\documentclass[11pt,a4paper,twoside,openany]{book}
\usepackage{amsfonts}
\usepackage{euscript}
\usepackage{calrsfs}
\usepackage{amsmath}
\usepackage{tabularx}
\usepackage{latexsym}
\usepackage{apacite}
\usepackage[dvips]{graphicx}
\usepackage{pstricks}
\usepackage{pst-node}
\usepackage[dvips]{rotating}
\usepackage{array}
\usepackage{accents}
\usepackage{mflogo}
\usepackage[labelfont=bf,font=small]{caption}

\usepackage{fancyhdr}
\renewcommand{\chaptermark}[1]{\markboth{#1}{}}

\fancypagestyle{plain}{\fancyhead{}}

\newcommand{\D}{\mathrm{d}}

\begin{document}

\frontmatter
\thispagestyle{empty}

\begin{center}
DEPARTMENT OF PHYSICS\\ UNIVERSITY OF JYV\"ASKYL\"A\\
RESEARCH REPORT No.~1/2008

\vspace{12mm}

{\bf\Large A METAGEOMETRIC ENQUIRY CONCERNING TIME, 
SPACE,\\ AND QUANTUM PHYSICS

\vspace{12mm}

BY \smallskip \\
DIEGO MESCHINI}\\

\vspace{12mm}

Academic Dissertation\\
for the Degree of\\
Doctor of Philosophy\\

\vspace{12mm}
\begin{center}
To be presented, by permission of the\\
Faculty of Mathematics and Natural Sciences of the\\
University of Jyv\"{a}skyl\"{a}, for public examination in\\
Auditorium FYS-1 of the University of Jyv\"{a}skyl\"{a} on\\
9th February, 2008, at 12 o'clock noon
\end{center}

\vfill
Jyv\"askyl\"a, Finland\\
February 2008\\
\end{center}
\hfill

\newpage 
\thispagestyle{empty}

\begin{figure*}[!h]
\begin{center}
\includegraphics[width=0.88\linewidth]{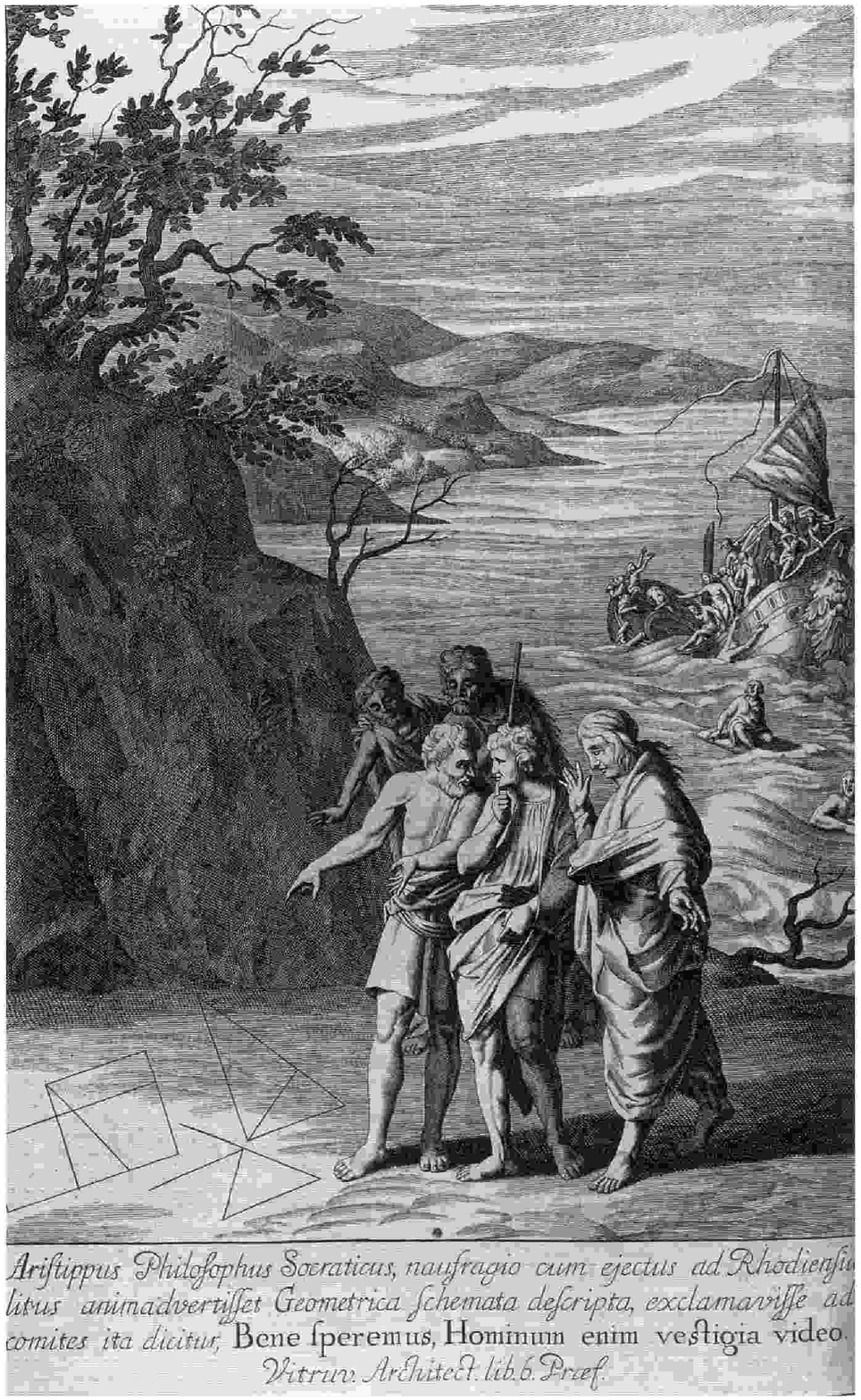}
\end{center}
\captionsetup{labelformat=empty}
\caption{``Aristippus, the Socratic philosopher, shipwrecked on the coast of Rhodes and having noticed sketches of geometric figures, it is said may have spoken thus to his companions: `We can hope for the best, for I see the signs of men.' (Vitruvius, \emph{De architectura}, Book 6, Preface)'' Frontispiece to \protect\cite{Euclid:1703}. Translation by H.~J.~Santecchia. Reproduced from \protect\cite{Heilbron:2000}.
}  
\end{figure*}

\newpage
\thispagestyle{empty}
\begin{flushright}
\begin{tabular}{p{7cm}} 
\emph{To Markku, who showed me the rare art of flying keeping one's feet on the ground}
\end{tabular}
\end{flushright}

\newpage
\thispagestyle{empty}
\vspace{5cm}
\begin{flushleft}
\begin{tabular}{p{8cm}}
\emph{Do space and time exist alone and}\\
\emph{As part and parcel of the world,}\\
\emph{Or are they merely illusions}\\
\emph{Our minds create and set unfurled?}\\
\emph{The very products of our brain}\\
\emph{Like, say, the feeling of disdain?}
\bigskip\\

\emph{Three hundred years of bitter history}\\
\emph{Have failed this issue to resolve:}\\
\emph{Cartesian vortices and whirlpools,}\\
\emph{Around which corpuscles revolve;}\\
\emph{Contraptions built out of material}\\
\emph{Not once upgraded from ethereal.}
\bigskip\\

\emph{Today the ether has not left us;}\\
\emph{It lives in new, geometric modes:}\\
\emph{As loops and foams, as fancy networks,}\\ 
\emph{As strings and membranes, links and nodes;}\\
\emph{For are these objects, say, these strings,}\\
\emph{The seat of any real things?}
\bigskip\\

\emph{Are even spacetime points ethereal?}\\
\emph{An argument about a hole:}\\
\emph{So wide to charm most varied thinkers,}\\  
\emph{But deep enough to lose your soul;}\\
\emph{Of which philosophers are fond,}\\
\emph{But which it's time to step beyond.}
\bigskip\\

\emph{Forgo the hole, the points, the ether.}\\
\emph{Instead use things you can detect;}\\
\emph{The simplest things than Man can think of,}\\
\emph{With sources Man can clearly affect.}\\
\emph{Find answers past the older helm}\\
\emph{Within a non-geometric realm.}
\bigskip\\

\emph{A basic secret of the cosmos}\\
\emph{It is our goal to better grasp.}\\
\emph{A highly ambitious undertaking;}\\
\emph{Might even make the Reader gasp,}\\
\emph{Or utter apprehensive shrieks:}\\
\emph{``A task to daunt the Ancient Greeks!''}
\bigskip\\

\emph{---Or utter cries of disbelief}\\
\emph{And hope this thesis to be brief.}\\
\emph{A wish that we cannot concede:}\\
\emph{Of much soul-searching there is need.}

\end{tabular}
\end{flushleft}

\chapter{Preface}
\begin{flushright}
\begin{tabular}{p{10cm}}
\emph{The prologue is the most emotive moment, where the tension accumulated during a year of work is released like a gunshot.}\smallskip \\
Miguel Indurain, \emph{Cycle Sport Magazine} 
\end{tabular}
\end{flushright}

\bigskip

The goal of this thesis is to contribute to the elucidation of an elemental, age-old problem, namely, the physical nature of space and time. The words of my supervisor, Dr.\ Markku Lehto, to whom I owe countless hours of illuminating revelations, perhaps describe the spirit of this thesis better than no other: 
\begin{quote}
We participate in one of mankind's most challenging intellectual efforts. The goal is to understand the biggest secrets of the cosmos: the structure of space and time as well as the connection of this structure with that of matter. One may even hold these secrets to be ``greater than life,'' when one thinks that life can be explained as an emergent phenomenon, i.e.\ a phenomenon which arises (probably very complicatedly) via evolution from the primitive structure of space and time and its interaction with matter. 

The first step is to explain whether space and time can have in general an own structure. Or are space and time in the end only human sensations like, for example, the feeling ``happy''?\footnote{Private communication, 2001.}
\end{quote}      

In the study of this problem, like in that of all problems enduring obstinately through the ages, it soon dawned on me that the question to focus on was not the immediately expected, ``How do we go about finding a solution?'' but rather the more careful and measured, ``What \emph{is} the problem?'' Moreover, it eventually became apparent that these two questions were one and the same; that an understanding of the essence of a problem was quite apt to reveal ill-founded attempts at solving it, as well as point implicitly to the path of a methodical resolution. 

This realization, together with mounting weariness of hackneyed, indulgent accounts of the nature and prospects of quantum gravity---the discipline which currently occupies itself with the issue of space and time---made an unsympathetic, bare-bones analysis of this field of research as indispensable as it was enlightening. It allowed me to disentangle few facts from a surfeit of clich\'{e}d beliefs (including even the occasional extravaganza) that populate the quantum-gravitational world---a referent-orphaned no-man's-land of theoretical research. In particular, I learnt from this analysis that the present work can be considered part of quantum gravity only in a very shallow, conventional sense, since it does not share with this field its standard motivations, methods of enquiry, or desired destination. If quantum gravity is to be reached by either one of the three geometric roads currently envisioned, this work treads none of them.  

After two and a half millennia of bitter and fruitless struggle---the last three and a half centuries of which within the context of modern science---the problem of the physical nature of space and time continues to be nothing short of an opinionated battlefield, a vicious display of parry and repost, a twisted labyrinth with no exit anywhere to be seen. We believe that unravelling this age-old problem requires a return to a physics based on observations, together with the conceptual overthrow of the geometric notions in terms of which the philosophically and mathematically minded debate continues exclusively to rage, i.e.\ via a thorough departure from all forms of geometric thought and language, an idea we condense in the name ``metageometry.''

Focused on metageometry, however, we notice with disquiet the ominous fact that human beings carry with them an innate craving for geometric modes of thought and expression. This is not only revealed by an overview of physical theories, or even of scientific theories, but, without fear of overstatement, one may even say of all human endeavour. Human beings' fascination for geometry easily grows to the point of enslavement; and sometimes even to the epitomical extreme of having Geometry become a modern religion---physicists' incarnation of God.
  
While a solution to the ontological spacetime problem cannot be found through philosophically minded, labyrinthine discussions void of fresh physical content, neither can it be found through the hit-and-miss invention of new spacetime structures off the top of one's head; and especially not when these procedures are based solely, or mostly, on pure mathematical insight. In physics, mathematics is a good servant but a bad master. Because it is a physical question that we have in our hands, its clarification requires strong physical intuition, the identification of relevant physical principles, and a firm anchorage in observations and experiments as an unconditional starting point. 

It is in view of all this that this work deals with the question, ``What do observations, quantum principles, and metageometric thought have to say about space and time?'' It consists primarily in the development of new, elemental physical ideas and concepts and, therefore, does not need to abound in complicated technicalities. Obstinate problems require conceptually fresh starts.

\begin{center}
$\Diamond \qquad \Diamond \qquad \Diamond$
\end{center}

\newpage

Literally speaking, this work would have been impossible without the wisdom and extraordinary guidance of my supervisor, Dr.~Markku Lehto. Not only did he gently and unselfishly show me the way through the mazelike passageways of quantum-gravity research, unobtrusively fostering at all times the independent development and expression of my own thoughts, but also showed me the difference between physics, mathematics, and philosophy. In so doing, he taught me by example (the one truly effective teaching method) the meaning of genuine physical thinking---a surprisingly difficult activity rarely witnessed in frontier theoretical physics. To him go my infinite gratitude and admiration.

From a financial point of view, the making of this thesis was assisted by the support of different organizations. I thank the Centre for International Mobility (CIMO) for sponsoring the beginning of my Ph.D.\ studies; and the Ellen and Artturi Nyyss\"{o}nen Foundation, the Finnish Cultural Foundation, and the Department of Physics, University of Jyv\"{a}skyl\"{a}, for allowing their continuation and completion. 

The research leading to this thesis was carried out during the period September 2001--July 2007 within the context of the last-mentioned Department, where I found ample freedom of action and better working conditions than I could have dreamt of. I particularly want to thank its administrative staff for its efficiency and helpfulness and my roommates, Ari Peltola, Johanna Piilonen, and Antti Tolvanen for pondering and assaying with me, sometimes lightheartedly, sometimes in all seriousness, the exactions of life in the brave new world of academia.  

Last but not least, the actual material creation of this thesis was made not only possible but, moreover, enjoyable by the typesetting system \LaTeX---an essential tool, almost on a par with the differential calculus, of the modern man of science; a tool which deserves more attention in our typeset-it-yourself world of amateur desktop publishing than is usually paid to it in the sciences and humanities curricula.

Man does not live by bread alone. I would also like to thank other people who have indirectly contributed to the fulfillment of this project; they are: my parents, Stella Maris and Omar, for bringing me up within the framework of an excellent family life and education---therein lie the seeds of all other achievements; Kaisuliina, for her companionship and warmth in the joys and commonplaces of everyday life alike; and Ulla and Teuvo, for unconditionally furnishing the context of a second family in a foreign, yet marvellous, land.

I bring this prelude to a close with some hope-ridden words\footnote{Paraphrase of Foss's \citeyear[p.~xiii]{Foss:2000}.} to the reader: if you should find this work simply readable, I shall be pleased; if it should give you a background on which to think new thoughts, I shall be delighted.

\bigskip

Diego Meschini

\smallskip

Jyv\"{a}skyl\"{a}, July 2007.

\chapter{Abstract}
Meschini, Diego\\
\noindent A metageometric enquiry concerning time, space, and quantum physics\\
\noindent Jyv\"{a}skyl\"{a}: University of Jyv\"{a}skyl\"{a}, 2008, xiv + 271 pp.\\
\noindent Department of Physics, University of Jyv\"{a}skyl\"{a},
Research Report 1/2008.\\
\noindent ISBN 978-951-39-3054-7\\ 
\noindent ISSN 0075-465X

\bigskip

This thesis consists of a enquiry into the physical nature of time and space and into the ontology of quantum mechanics from a metageometric perspective. This approach results from the belief that geometric thought and language are powerless to farther human understanding of these issues, their application continuing instead to restrict physical progress.

The nature and assumptions of quantum gravity are analysed critically throughout, including misgivings about the relevance of the Planck scale to it and its lack of observational referent in the natural world. 

The anthropic foundations of geometry as a tool of thought are investigated. The exclusive use of geometric thought from antiquity to present-day physics is highlighted and found to permeate all new attempts towards better theories, including quantum gravity and, within it, even pregeometry. Some early hints at the need to supersede geometry in physics are explored. 

The problem of the ether is studied and found to have perpetuated itself up to the present day by transmuting its form from mechanical, through metric, to geometric. Philosophically minded discussions of this issue (Einstein's hole argument) are found wanting from a physical perspective.

A clarification is made of the physical, mathematical, and psychological foundations of relativity and quantum theories. The former is founded geometrically on measurement-based clock-reading separations $\D s$. The latter is founded metageometrically on the experiment-based concepts of premeasurement and transition things, inspired in the physically unexplored aspect of time as a consciousness-related product. A concept of metageometric time is developed and coordinate time $t$ is recovered from it. Discovery of the connection between quantum-mechanical metageometric time elements and general-relativistic clock time elements $\D s$ (classical correlations) is deemed necessary for a combined understanding of time in the form of metageometric correlations of quantum-mechanical origin.

Time is conjectured to be the missing link between general relativity and quantum mechanics.

\newpage
\thispagestyle{empty}
\phantom{x} \vspace{2cm}

\noindent \textbf{OPPONENT}\smallskip\\ 
Dr.~Rosolino Buccheri\\ 
Institute for Educational Technology, National Research Council, Palermo \&\ 
Centro Interdipartimentale di Tecnologie della Conoscenza, University of Palermo

\vspace{1cm}

\noindent \textbf{SUPERVISOR}\smallskip\\
Dr.~Markku Lehto\\
Department of Physics, University of Jyv\"{a}skyl\"{a}

\vspace{1cm}

\noindent \textbf{EXAMINERS}\smallskip\\
Dr.~George Jaroszkiewicz\\
School of Mathematical Sciences, University of Nottingham \bigskip \\
Dr.~Metod Saniga\\
Astronomical Institute, Slovak Academy of Sciences

\chapter{Related publications}
\pagestyle{fancy}

\begin{flushright}
\begin{tabular}{p{10cm}} 
\emph{I am probably both dissatisfied with everything that I have written and proud of it.} \smallskip \\
Stanis{\l}aw Lem, \emph{Microworlds}
\end{tabular}
\end{flushright}

\bigskip

\begin{itemize}
\item[I.] Meschini, D., Lehto, M., \& Piilonen, J.\ (2005). Geometry, pregeometry and beyond. \emph{Studies in History and Philosophy of Modern Physics}, \emph{36}, 435--464. (arXiv:gr-qc/0411053v3)   

\item[II.] Meschini, D., \& Lehto, M.\ (2006).\ Is empty spacetime a physical thing? \emph{Foundations of Physics}, \emph{36}, 1193--1216. (arXiv:gr-qc/0506068v2)   

\item[III.] Meschini, D. (2006).\ Planck-scale physics: Facts and beliefs. To appear in \emph{Foundations of Science}. (arXiv:gr-qc/0601097v1)

\end{itemize}

Guided by his supervisor, Dr.~Markku Lehto, the author researched and wrote the vast majority of Article I and the totality of Articles II and III. 

The contents of this thesis consist of material partly drawn---with enhancements, abridgements, and amendments---from the above publications (Chapters \ref{ch:Planck-scale physics}, \ref{ch:Geometry in physics}, \ref{ch:Pregeometry}, \ref{ch:Beyond geometry}, and \ref{ch:ESE}), and partly previously unpublished (Chapters \ref{ch:Quantum gravity unobserved}, \ref{ch:Geometry}, \ref{ch:New geometries}, \ref{ch:Geometric anthropic principle}, and \ref{ch:Analysis of time}--\ref{ch:Time last taboo}). In particular, the contents of Chapters \ref{ch:Analysis of time} and \ref{ch:MQM} consist of enhanced versions by the author of public lectures given by Dr.~Lehto at the Department of Physics, University of Jyv\"{a}skyl\"{a}: Chapter \ref{ch:Analysis of time} is based on \emph{Relativity theory} (9th January--26th April, 2007); Chapter \ref{ch:MQM} is based on \emph{Physical, mathematical, and philosophical foundations of quantum theory} (12th April--29th June, 2005).

\tableofcontents
\newpage
\thispagestyle{empty}

\mainmatter
\pagestyle{fancy}

\chapter{Quantum gravity unobserved}
\label{ch:Quantum gravity unobserved}
\begin{flushright}
\begin{tabular}{p{10cm}}
\emph{If you don't know where you are going, any road will take you there.} 
\smallskip \\
Lewis Carroll, \emph{Alice's adventures in Wonderland}
\end{tabular}
\end{flushright}

\bigskip
The study of the physical nature of space and time falls currently under the conventional sphere of quantum gravity---a field of physics that has been abundantly described as an attempt to combine the clashing worldviews of general relativity and quantum mechanics, as though pieces of a faulty jigsaw puzzle.\footnote{See e.g.\ \cite{Butterfield/Isham:2000}.} It is not the intention of this thesis to go into yet one more of these stale descriptions. We are instead interested in analysing critically the mode of working and assumptions of quantum gravity, thereby to understand why it is not yet possible to know even superficially the answer to the question, ``What is quantum gravity?'' 

Paraphrasing Hardin \citeyear[p.~vii]{Hardin:1969} on population, the emerging history of quantum gravity is a story of confusion and denial---confusion experienced, but confusion psychologically denied. Quantum gravity is a field of physics whose subject matter has not been identified at the time of writing of this work. What is quantum gravity \emph{about}? This question has no real answer owing to the fact that no-one has as yet identified any relevant observations that could act to guide physics' next leap forward: either new, unaccounted-for observations involving interactions of some sort between gravitation and quantum systems or well-known ones that, when analysed from a new perspective, reveal previously hidden features of the natural world (cf.\ Einstein's analysis of time measurements as a source of special relativity). Because of this indefiniteness, there are, to varying degrees, almost as many meanings attached to quantum gravity as there are research groups of it. In other words, ignorance about what quantum gravity is \emph{about} leads to uncertainty about what quantum \mbox{gravity \emph{is}.}

In practice, however, this state of utter disorientation is not perceived to be as gloomy as all that. In quantum gravity's ivory tower, not only has research happily become highly speculative and ruled by matters of internal nature, but, much more worryingly, has to a large extent begun to display fictional features. In the self-assured expositions of quantum gravity, speculation is not only apt to supplant missing facts but is also capable of becoming fact itself: assumptions originally proposed as possibilities or beliefs have now, through concerted repetition, acquired the status of facts irrespective of the question of their unattested physical truth. As a result of this process, quantum gravity has become a clich\'{e}-ridden field. 

The most general of these clich\'{e}s holds quantum gravity to be basically the joint consideration of ``quantum'' and ``gravity.'' In other words, it holds quantum gravity to be essentially a mechanical theory of something physical (but for the moment unknown) of a certain typical size and energy evolving in time in typical periods (cf.\ the Planck scale). This unwarranted conclusion is either simply read off the label ``quantum gravity,'' or is based on a stronger, mostly unconscious human affinity to mechanical concepts to the detriment of other ideas farther removed from the mechanical sphere. In truth, however, one cannot tell at this stage what combination of known physics may be needed for a theory of quantum gravity; even less whether new, unforeseen physical ideas may be necessary instead.

Another clich\'{e} involves the belief that a theory of quantum gravity will (or worse, that it \emph{must}\footnote{See \cite[pp.~10--11]{Smolin:2003}.}) solve certain, most, or all of what are presently considered important open problems in frontier physics. For example, that quantum gravity will (or must) give a solution to the divergences of quantum field theory, unify the four forces of nature, explain the dimensionality of space, the value of the cosmological constant, the nature of quantum non-locality and of singularities, including black holes and the origin of the universe, recover general relativity as a low-energy limit, possibly rewrite the rules of quantum mechanics, etc. Such an almighty characterization of quantum gravity, however, seems to have more in common with expectations and desires than with physics realistically achievable at one fell swoop. 

Among these problems awaiting solution, the various forms of gravitational collapse predicted by general relativity and the confirmed existence of quantum-mechanical correlations figure most notably.\footnote{See e.g.\ \cite[pp.~3--9]{Monk:1997}.} To be sure, Misner, Thorne, and Wheeler \citeyear{Misner/Thorne/Wheeler:1973} called gravitational collapse ``the greatest crisis in physics of all time'' (p.~1196), while Stapp \citeyear{Stapp:1975} called Bell's theorem ``the most profound discovery of science'' (p.~271). Although one must not necessarily discard spacetime singularities or non-local correlations as motivations for the search for better theories, do we really need to \emph{increase our knowledge} of these issues as a conditional part of a theory of quantum gravity? This may or may not be the case. The case of Newtonian singularities, for example, was certainly not relevant to the invention of relativistic or quantum theories, while, on the other hand, the problem of electromagnetic singularities was relevant to the development of quantum mechanics, as evidenced by Bohr's efforts to solve the problem of the collapse of the atom. The same can be said of quantum non-locality. Here again one cannot tell whether our knowledge of these correlations that defy the ideas of spatial locality should be increased by a theory that will supersede the current ones, or whether they simply are a trouble-free characteristic of quantum theory whose expounding is irrelevant to any future developments. We simply do not know what current problems quantum gravity will solve, which it will declare non-issues, and which it will not even touch upon. What is certain, however, is that any future theory that may be called physical \emph{will} explain the origin and nature of the \emph{observables} it refers to.

A related clich\'{e} results from stretching the previous one beyond all measure. It involves the belief that a theory of quantum gravity will not only be a theory of gravitation and quantum phenomena, but a ``theory of everything'' that will put an end to research in elemental physics. This dubious idea, although nowadays not necessarily restricted to the particle-physics perspective, is presumably inspired in the worldview offered by this field of physics, according to which microscopic particles and their particle-mediated interactions are the most basic constituents of nature.\footnote{For approches from this direction, see e.g.\ \cite{Greene:1999,Weinberg:1993}; for approaches from a general direction, see e.g.\ \cite{Barrow:1992,Breuer:1997,Hartle:2003,Laughlin/Pines:2000,Tegmark:1998,Tegmark/Wheeler:2001}.} On this basis, it is believed that to account for all (known and presumed) basic microscopic constituents (particles, strings, or suchlike) and their mutual interactions is tantamount to accounting in principle for ``everything'' in the physical world. 

To say the least, this conception is at odds with another well-established field of physics, general relativity, from which we know that the matter-dependent geometric properties of spacetime are as physical and basic as the particles found embedded in its geometric network. If, as it seems to be, the suspect idea of a ``theory of everything'' was originally inspired by the optimistic prospect of a unification of the four known interactions on the basis of string-theory considerations,\footnote{The origin of claims regarding a ``theory of everything'' can be traced to a talk by Michael Green advertised at Oxford University in the mid-1980s. It is told that Brian Greene ``[w]alking home one winter day\ldots saw a poster for a lecture explaining a new-found `theory of everything.'~'' ``It made it sound dramatic,'' Greene adds. Dramatic, indeed, but to what purpose drama in science?  (Columbia College Today, http://www.college.columbia.edu/cct/sep99/12a.html. With variations also at Encyclopedia of World Biography, http://www.notablebiographies.com/news/Ge-La/Greene-Brian.html) In truth, however, we can trace the origin of the phrase ``theory of everything'' farther back in time. In a short story first published in 1966, Stanis{\l}aw Lem introduced the concept---as could not be otherwise---as part of harsh satire. His space-faring character Ijon Tichy relates that his grandfather
\begin{quote}
at the age of nine\ldots decided to create a General Theory of Everything and nothing could divert him from that goal. The tremendous difficulty he experienced, from the earliest years on, in formulating concepts, only intensified after a disastrous street accident (a steamroller flattened his head). \cite[p.~258]{Lem:1985} 
\end{quote}
This is an origin that does not bode well for the idea in question.} then a cursory glance at general relativity leaves us wondering, And what about space and time? In the last analysis, the particle-physics worldview originally inspiring a ``theory of everything'' is based on the presumption that to be more elemental means to be smaller; but whence should the elemental concepts of our most basic theories refer to small entities? This is also the presumption behind the Planck-scale considerations underlying quantum gravity.

Deeper reasons to doubt any claim about a ``theory of everything'' concern epistemological and ontological considerations. From an epistemological perspective, physical theories are the products of the human intellect, built with the conceptual tools of the mind. This, for one, entails that theories are built from limited experience out of the conventional (overwhelmingly geometric), volatile concepts of the physical tradition they belong to, and also that one can never be certain whether all aspects of nature---known and unforeseeable alike---can even in principle be reduced to those of any existing theory. As Stanis{\l}aw Lem \citeyear{Lem:1984}, that great discerner of the human spirit, put it, ``One has to look through the history of science to reach the most probable conclusion: that the shape of things to come is determined by what we do not know today and by what is unforeseeable'' (p.~124). 

But much more profoundly, the fact that our theories are nothing but conceptual tools of thought precludes the belief that a ``theory of everything'' is a reasonable or even meaningful human goal to pursue (or, worse still, just around the corner), because a tool of thought is a filter that selects according to a built-in bias. This must necessarily be so, because the jumble of impressions that make up the inner and outer world of conscious experience cannot be made sense of as an indivisible whole. Heed the pertinent words from a different context of another great discerner of the human spirit, biologist-ecologist Garrett Hardin:
\begin{quote}
We need to be acutely aware of the virtues and shortcomings of the tools of the mind. In the light of that awareness, many public controversies can be resolved.

[\ldots] Our intellectual tools are filters for reducing reality to a manageable simplicity. To assert that we have ever completely captured reality in an equation or a string of words would be extremely arrogant. Each filter captures only part of reality. Most expertise is single-filter expertise. We come closer to the truth when we compensate for the bias of one filter by using others (which have different biases).
\cite[pp.~20--21]{Hardin:1985}
\end{quote}
Any physical theory is such a filter, its mesh currently built out of geometry; indeed, a physical theory, as North \citeyear{North:1990} put it with a degree of modesty currently in short supply,\footnote{In a world where scientific theories have become marketable consumer goods, modesty is a luxury we cannot afford. This explains why drama is needed in science. For related ideas, see \cite{Lagendijk:2005}.} is ``merely an instrument of what passes for understanding'' (p.~407). ``Everything'' cannot be understood. Ultimately, the belief in a ``theory of everything'' is based on what Lem \citeyear{Lem:1984} has called ``the myth of our [and our scientific language's] cognitive universality'' (pp.~26--27)---in other words, on human folly. 

In connection with his quotation above, to better grasp the reality of a finite overpopulated world, Hardin proposed the application of a combination of \emph{several} filters: literacy, numeracy, and ecolacy. Hardin's various filters, because of being many and different, are filters \emph{against} folly---the folly of single-filter expertise. The single-filter explain-all schemes proposed by the advocates of a ``theory of everything,'' on the other hand, are then of necessity filters \emph{of} folly.

The only positive thing---yet far-fetched and, at any rate, of limited scope---that can be said for a ``theory of everything'' comes from the ontological perspective; in other words, from a differentiation between human knowledge and truth. Turning a blind eye to the built-in limitations of theories as filters for understanding, some have argued for a ``theory of everything'' as follows. As much as physical \emph{theories} may change, what is true of them as expressed in their \emph{laws} never changes beyond the degree of accuracy with which these laws have been corroborated. For example, as mechanical theories were successively improved from Galilei, through Kepler and Newton, to Einstein and the fathers of quantum mechanics, the laws of inertia, of equivalence between rest and gravitational mass, the inverse-square law of gravitation, etc., either remained untouched or were modified for new physical domains but never banished. What is once known is, as it were, known forever. Thus, a ``theory of everything'' could conceivably contain laws that are \emph{absolutely} and \emph{perfectly} true---whether we know it or not---of those aspects of nature, both known and to be disclosed in the future, that it as a filter manages to capture---not, however, of \emph{every} aspect of nature, which escapes any chosen mesh. This is the most a ``theory of everything'' \mbox{could be}.

Ontologically speaking, then, we see that a ``theory of everything'' in this severely restricted one-filter sense is not a logical impossibility, but only an utterly unreasonable expectation judging from the constant turmoil science has lived in during all of its history, including especially the (psychologically denied) utter disorientation of the present epoch. As Peacock \citeyear{Peacock:2006} enquires with regard to string theory, ``If the search for a unique and inevitable explanation of nature has proved illusory at every step, is it really plausible that suddenly string theory can make everything right at the last?'' (p.~170). 

Since when do scientists seriously concern themselves with the study of utter implausibilities? (Perpetual-motion machines and astrology come most readily to mind.) Is the enterprise of natural science about claiming non-impossibilities and asking others to demonstrate a ``no-go theorem'' to prove one to be logically wrong? This is, at any rate, the turf of mathematics or professional philosophy, but not a game that physics is ready to play. Or to borrow Hardin's \citeyear[pp.~40--43]{Hardin:1993} words, in science (as in ecology; but unlike in professional philosophy, as in economics) the ``default status'' is ``guilty until proven innocent,'' which means that the ``burden of proof'' lies on the proponent of extravagant claims (or ``progressive developments''), not on the receiver. Hardin asks, ``Aren't scientists supposed to be open-minded? Does not science progress by examining \emph{every} possibility?'' And forthrightly replies, ``The answer to the second question is \emph{no}. As for being open-minded, how much open-mindedness can we afford?'' (p.~40).  

A fourth and final clich\'{e} to be mentioned here consists in the generalized belief---regardless of research approach or investigator---that the Planck scale will be relevant to quantum gravity. This notion is without doubt the most cherished, overarching theme of quantum gravity. It goes unchallenged in virtually all researches, and has become such a central, yet worryingly little examined, assumption that we shall analyse it in detail in the following chapter. As we separate fact from belief in connection with the Planck scale, we give a new look at some of the other issues mentioned above.

Quantum-gravity clich\'{e}s, summarized in Table \ref{tabQG}, are fictional beliefs that have become facts through sheer repetition, in line with a reigning politics of publication that, as Lawrence \citeyear{Lawrence:2003} has remarked, uncritically favours fashionable ideas. This situation is fuelled by the fact that the standard accounts of quantum gravity are capable of offering alluring vistas to the human imagination, which are, in turn, invigorated by colourful popularizations in magazines, books, and visual media, for which the making of profit sets the main standard of action. Ultimately, these beliefs remain unchallenged due to the lack of a referent of quantum gravity in the natural world. Only the recognition of relevant observables, whether new ones yet unaccounted-for or old ones seen in a new light, will provide the clue as to what quantum gravity is about. And only knowledge of what quantum gravity is \emph{about} will spell out what quantum gravity \emph{is}. For this reason, the question one should worry about is not what quantum gravity is, but \emph{what are the relevant observables} on which to build a quantum theory of space and time.
 
\begin{table}
\centering
\setlength{\extrarowheight}{5pt}
\begin{tabular}{|p{5.8cm}|p{5.8cm}|}
\hline
\multicolumn{2}{|c|}{\textbf{What is quantum gravity?}}\\
\hline \hline
\underline{Clich\'{e} \#1}: Quantum gravity is quantum $\cap$ gravity. & 
\underline{Answer \#1}: Quantum gravity is known physics $\cup$ unforeseeable new physics.\\ \hline
\underline{Clich\'{e} \#2}: Quantum gravity to solve definite problems of current physics. &
\underline{Answer \#2}: Quantum gravity to explain origin and nature of its observables.\\ \hline
\underline{Clich\'{e} \#3}: Quantum gravity is the ``theory of everything.''&
\underline{Answer \#3}: A ``theory of everything'' is single-filter folly.\\ \hline
\underline{Clich\'{e} \#4}: Quantum gravity applies at the Planck scale.&
\underline{Answer \#4}: Quantum gravity applies at the scale of its observables.\\ 
\hline
\end{tabular}
\caption[Contrasting views on quantum gravity]{Contrast between the typical and our own views regarding the nature of quantum gravity. The symbol $\cap$ represents the logical operation AND, while the symbol $\cup$ represents the logical operation OR.}\label{tabQG} 
\end{table}

\chapter{The myth of Planck-scale physics}
\label{ch:Planck-scale physics}
\begin{flushright}
\begin{tabular}{p{10cm}}
\emph{There was a time when educated men supposed that myths were something of the past---quaint, irrational fairy tales believed in by our credulous and superstitious ancestors, but not by us, oh dear no!}
\smallskip \\
Garrett Hardin, \emph{Nature and Man's fate} 
\end{tabular}
\end{flushright}

\bigskip

An overview of the current confident status of so-called Planck-scale physics is enough to perplex. Research on quantum gravity has abundantly become synonymous with it, with the Planck length $l_P$, time $t_P$, and mass $m_P$ being unquestionably hailed as the scales of physical processes or things directly relevant to a theory of quantum gravity. And yet, in the last analysis, the only linking threads between Planck's natural units, $l_P$, $t_P$, and $m_P$, and quantum gravity ideas are (i) generally and predominantly, dimensional analysis---with all its vices and virtues---and, specific to some research programmes, (ii) joint considerations of gravitational (relativistic) and quantum theories, such as \emph{ad hoc} considerations about microscopic, Planck-sized black holes or the quantization of gravity---arenas in which the effects of both quantum theory and general relativity are to be relevant together.

Dimensional analysis is a surprisingly powerful method capable of providing great insight into physical situations without needing to work out or know the detailed principles underlying the problem in question. This (apparently) suits research on quantum gravity extremely well at present, since the physical mechanisms  to be involved in such a theory are unknown. However, dimensional analysis is not an all-powerful discipline: unless very judiciously used, the results it produces are not necessarily meaningful and, therefore, they should be interpreted with caution. So should, too, arguments concerning Planck-sized black holes or the quantization of gravity, considering, as we shall see, their suspect character.

In this chapter, we put forward several cautionary observations against the role uncritically bestowed currently on the Planck units as meaningful scales in a future physical theory of space and time. One remark consists in questioning the extent of the capabilities of dimensional analysis: although it possesses an almost miraculous ability to produce correct orders of magnitude for quantities from dimensionally appropriate combinations of physical constants and variables, it is not by itself a necessarily enlightening procedure. A conceptually deeper remark  has to do with the fact that, despite its name, a theory of quantum gravity need not be the outcome of ``quantum'' ($h$) and ``gravity'' ($G,c$): it may be required that one or more presently unknown natural constants be discovered and taken into account in order to picture\footnote{Do scientists \emph{picture} the natural world, or do they \emph{model} it? For a brief elucidation, see Appendix A.} the new physics correctly; or yet more unexpectedly, it may also turn out that quantum gravity has nothing to do with one or more of the above-mentioned constants. The meaning of these observations will be illustrated in the following with the help, where possible, of physical examples, through which we will cast doubt on the mostly undisputed significance of Planck-scale physics. We will argue that quantum gravity scholars, eager to embark on the details of their investigations, overlook the question of the likelihood of their assumptions regarding the Planck scale---thus creating seemingly indubitable facts out of merely plausible beliefs.

But before undertaking the critical analysis of these beliefs, let us first review the well-established \emph{facts} surrounding dimensional analysis and Planck's natural units.

\section{Facts} \label{Facts}
The Planck units were proposed for the first time by Max Planck \citeyear{Planck:1899} over a century ago. His original intention in the invention of these units was to provide a set of basic physical units that would be less arbitrary, i.e.\ less human-oriented, and more universally meaningful than metre, second, and kilogram units.\footnote{For an account of the early history of the Planck units, see \cite{Gorelik:1992}.} 

The Planck units can be calculated from dimensionally appropriate combinations of three universal natural constants, namely, Newton's gravitational constant $G$, Planck's constant $h$, and the speed of light $c$. The mathematical procedure through which Planck's units can be obtained is not guaranteed to succeed unless the reference set of constants $\{G, h, c\}$ is dimensionally independent with respect to the dimensions $M$ of mass, $L$ of length, and $T$ of time. This is to say that the $MLT$-dimensions of none of the above constants should be expressible in terms of combinations of the other two. 

The  reference set $\{G,h,c\}$ will be dimensionally independent with respect to $M$, $L$, and $T$ if the system of equations 
\begin{equation} \left\{
\begin{matrix}
[G] & = & M^{-1}L^3T^{-2}\\
[h] & = & M^1L^2T^{-1}\\
[c] & = & M^0L^1T^{-1}
\end{matrix} \right.
\end{equation}
is invertible. Following Bridgman \citeyear[pp.~31--34]{Bridgman:1963}, we take the logarithm of the equations to 
find
\begin{eqnarray} \left\{ \begin{matrix}
\ln([G])&=&-1\ln(M)+3\ln(L)-2\ln(T)\\
\ln([h])&=&1\ln(M)+2\ln(L)-1\ln(T)\\
\ln([c])&=&0\ln(M)+1\ln(L)-1\ln(T) \end{matrix} \right.,
\end{eqnarray}
which is now a linear system of equations in the logarithms of the variables of interest and can be 
solved applying Cramer's rule. For example, for $M$ we get
\begin{equation}
\ln(M)=\frac{1}{\Delta} \left|\begin{matrix}
\ln([G])&3&-2\\
\ln([h])&2&-1\\
\ln([c])&1&-1\\
\end{matrix}\right|,
\end{equation}
where 
\begin{equation} 
\Delta=\begin{vmatrix} 
-1 & 3 & -2 \\
1 & 2 &-1 \\
0 & 1 & -1 \\
\end{vmatrix}=2
\end{equation}
is the determinant of the coefficients of the system. We find
\begin{equation} 
M=[G]^{-\frac{1}{\Delta}}[h]^\frac{1}{\Delta}[c]^{\frac{1}{\Delta}},
\end{equation}
and similarly for $L$ and $T$. This means that the system of equations has a solution only if $\Delta$ is different from zero, which in fact holds in this case. Granted the possibility to construct Planck's units, we proceed to do so.\footnote{Although Cramer's rule could be used to actually solve for $l_P$, $t_P$, and $m_P$, in the following we shall use a less abstract, more intuitive method.}

The Planck length $l_P$ results as follows. We define $l_P$ by the equation 
\begin{equation}
l_P=K_l G^\alpha h^\beta c^\gamma, 
\end{equation} 
where $K_l$ is a dimensionless constant. In order to find $\alpha$, $\beta$, and $\gamma$, we subsequently write its dimensional counterpart
\begin{equation}
[l_P]=L=[G]^\alpha [h]^\beta [c]^\gamma.
\end{equation} 
In terms of $M$, $L$, and $T$, we get
\begin{eqnarray}
L&=&(M^{-1}L^3T^{-2})^\alpha (ML^2T^{-1})^\beta (LT^{-1})^\gamma.
\label{length}
\end{eqnarray} 
Next we solve for $\alpha$, $\beta$, and $\gamma$ by rearranging the right-hand side and comparing it with the left-hand side, to get
$\alpha=\beta=1/2$, and $\gamma=-3/2$. Thus, the Planck length is 
\begin{equation}
l_P=K_l \sqrt{\frac{Gh}{c^3}}=K_l 4.05 \times 10^{-35}\ \mathrm{m}. 
\end{equation}
The other two Planck units can be calculated by means of a totally analogous procedure. They are 
\begin{equation}
t_P=K_t \sqrt{\frac{Gh}{c^5}}=K_t 1.35 \times 10^{-43}\ \mathrm{s}
\end{equation}
and
\begin{equation}
m_P=K_m \sqrt{\frac{hc}{G}}=K_m 5.46 \times 10^{-8}\  \mathrm{kg}. 
\end{equation}

The crucial question now is: is there any physical significance in these natural units beyond Planck's original intentions of providing a less human-oriented set of reference units for length, time, and mass? Quantum-gravity research largely takes for granted a positive answer to this question, and confers on these units a completely different, and altogether loftier, role: Planck's units are to represent the physical scale of things relevant to a theory of quantum gravity, or at which processes relevant to such a theory occur.\footnote{See e.g.\ \cite{Amelino-Camelia:2003} for hypothetically possible roles of Planck's length.}

One may seek high and low for cogent justifications of this momentous claim, which is to be found in innumerable scientific publications; much as one might seek, however, one must always despairingly return to, by and large, the same explanation (when it is explicitly mentioned at all): dimensional analysis. For example, Butterfield and Isham \citeyear[p.~37]{Butterfield/Isham:2000} account for the appearance of the Planck units in passing---and quite accurately indeed---as a ``simple dimensional argument.'' Smolin \citeyear{Smolin:1997}, for his part, grants the suspect derivation of the Planck scale from dimensional analysis but goes on to justify it: ``This may seem like a bit of a trick, but in physics arguments like this [dimensional analytic] are usually reliable.'' He also assures the reader that ``several different approaches to quantum gravity do predict'' (p.~60) the relevance of this scale, but does not mention that all these different approaches use the \emph{same} trick. Is dimensional analysis such a trustworthy tool as to grant us definite information about unknown physics without needing to look into nature's inner workings, or has our \emph{faith} in it become exaggerated? 

Two uncommonly cautious statements regarding the meaning of the Planck units are those of Y.~J.~Ng's, and of J.~C.~Baez's. Ng \citeyear{Ng:2003} recognized that the Planck units are so extremely large or small relative to the scales we can explore today that ``it takes a certain amount of foolhardiness to even mention Planck-scale physics'' (p.~1, online ref.). Although this is a welcome observation, it appears to work only as a proviso, for Ng immediately moved on to a study of Planck-scale physics rather than to a criticism of it: ``But by extrapolating the well-known successes of quantum mechanics and general relativity in low energy, we believe one can still make predictions about certain phenomena involving Planck-scale physics, and check for consistency'' (p.~1).

Also Baez \citeyear{Baez:2000} made welcome critical observations against the hypothetical relevance of the Planck length in a theory of quantum gravity. Firstly, he mentioned that the dimensionless factor (here denoted $K_l$) might in fact turn out to be very large or very small, which means that the order of magnitude of the Planck length as is normally understood (i.e.\ with $K_l=1$) need not be meaningful at all. A moment's reflection shows that this problem does not really threaten the expected order of magnitude of the Planck units. An overview of physical formulas suggests that, in general, the dimensionless constant $K$ appearing in them tends to be very large or very small (e.g.\ $|K|\geq 10^2$ or $|K|\leq 10^{-2}$) only when the physical constants enter the equations to high powers. For example, consider a black body and ask what might be its radiation energy density $\rho_E$. Since the system in question involves a gas ($k_B\tau$) of photons ($c,h$), one proposes
\begin{equation}
\rho_E=K_\rho h^\alpha c^\beta (k_B\tau)^\gamma,    
\end{equation}
where $K_\rho$ is a dimensionless constant and $\tau$ is the absolute temperature of the black body. The application of dimensional analysis gives the result 
\begin{equation}
\rho_E=K_\rho \frac{(k_B\tau)^4}{h^3c^3}=K_\rho 4.64 \times 10^{-18}\tau^4 \frac{\mathrm{J}}{\mathrm{K}^4\mathrm{m}^3},
\end{equation}  
whereas the correct result is

\begin{equation}\label{energydensity}
\rho_E=7.57 \times 10^{-16}\tau^4\ \frac{\mathrm{J}}{\mathrm{K}^4\mathrm{m}^3}, 
\end{equation}
and thus $K_\rho=163$. The value of $K_\rho$ does not result from dimensional analysis but from detailed physical calculations; these values will from now on be given in parentheses for the purpose of assessing the accuracy of dimensional analysis. The result yielded by dimensional analysis, then, is two orders of magnitude lower than the correct value of $\rho_E$, which represents a meaningful difference for the physics of a black body. This discrepancy, however, could have been expected since it can be traced back to the constants $h$, $c$, and $k_B$ entering the equation to high powers (3, 3, and 4, respectively).\footnote{See \cite[pp.~88, 95]{Bridgman:1963} for related views.} In this respect, the situation appears quite safe for the Planck units, excepting perhaps the Planck time, in whose expression $c$ enters to the power $-5/2$.

More interestingly, Baez also recognized that ``a theory of quantum gravity might involve physical constants other than $c$, $G$, and $\hbar$'' (p.~180). This is indeed one important issue regarding the significance of Planck-scale physics---or lack thereof---which will be analysed in the next section. However, notwithstanding his cautionary observations to the effect that``we cannot prove that the Planck length is significant for quantum gravity,'' Baez also chose to try and ``glean some wisdom from pondering the constants $c$, $G$, and $\hbar$'' (p.~180).  

What these criticisms intend to point out is the curious fact that, within the widespread uncritical acceptance of the relevance of Planck-scale physics, even those who noticed the possible shortcomings of dimensional analysis in connection with quantum gravity decided to continue to pursue their studies in this direction---as if to seriously question the Planck scale was taboo. We enquire into the deeper reason behind this practice later on in Chapter \ref{ch:Geometric anthropic principle}. For now, we concentrate on the issue at hand and ask: where do the charms of dimensional analysis lie, then? 

``[D]imensional analysis is a by-way of physics that seldom fails to fascinate even the hardened practitioner,'' said Isaacson and Isaacson \citeyear[p.~vii]{Isaacson/Isaacson:1975}. Indeed, examples could be multiplied at will for the view that dimensional analysis is a trustworthy, almost magical, tool. Consider, for example, a hydrogen atom and ask what is the relevant physical scale of its radius $a$ and binding energy $E$. Dimensional analysis can readily provide an answer \emph{apparently} without much knowledge of the physics of atoms, only if one is capable of estimating correctly which natural constants are relevant to the problem. The general manner of this estimation will be considered in Section \ref{Appraisal}; for the moment, we assume that one at least knows what an atom \emph{is}, namely, a microscopic aggregate of lighter negative charges and heavier (neutral and) positive charges of equal strength, and where only the \emph{details} of their mutual interaction remain unknown. Without this kind of previous knowledge, in effect assuring us that we have a quantum-electrodynamic system in our hands, the application of dimensional analysis becomes \emph{blind}.\footnote{See \cite[pp.~50--51]{Bridgman:1963}.} Armed with previous physical experience, then, one notices that only quantum mechanics and the electrodynamics of the electron are involved, so that the constants to consider must be $h$, the electron charge $e$, the electron mass $m_e$, and the permittivity of vacuum $\epsilon_0$. Equipped with these constants, one proposes
\begin{equation}
a=K_a h^\alpha e^\beta m_e^\gamma \epsilon_0^\delta
\end{equation}
and
\begin{equation} 
E=K_E h^\lambda e^\mu m_e^\nu \epsilon_0^\xi,    
\end{equation}
where $K_a$ and $K_E$ are dimensionless constants. A procedure analogous to the one above for the Planck length gives the results 
\begin{equation}
a=K_a \frac{\epsilon_0 h^2}{m_e e^2}=K_a 1.66 \ \mathrm{\AA}=0.529 \ \mathrm{\AA} \qquad (K_a=1/\pi) 
\label{radius}
\end{equation}
and
\begin{equation}
E=K_E\frac{m_e e^4}{\epsilon_0^2 h^2}=K_E 108\ \mathrm{eV}=-13.60\ \mathrm{eV} \qquad (K_E=-1/8). 
\label{energy}
\end{equation}

As we can see, meaningful values of the order of magnitude of the characteristic radius and binding energy of the hydrogen atom can thus be found through the sheer power of dimensional analysis. We believe that it is on a largely unstated argument along these lines that quantum-gravity researchers' claims regarding the relevance of the Planck units rest. 

Regrettably, the physical world cannot be probed confidently by means of this tool alone, since it has its shortcomings. As we see in the next section, these are that dimensional analysis may altogether fail to inform us what physical quantity the obtained order of magnitude refers to; and more severely, much can go amiss in the initial process of estimating which natural constants must be consequential and which not. Therefore, any claims venturing beyond the limits imposed by these shortcomings, although plausible, must be based on \emph{faith}, \emph{hope}, or \emph{belief}, and should not in consequence be held as straightforward facts.
       
\section{Beliefs} \label{Beliefs}
Our first objection against a securely established significance of the Planck units in quantum gravity lies in the existence of cases in which dimensional analysis yields the order of magnitude of no clear physical thing or process at all. 

To illustrate this point, take the Compton effect. A photon collides with an electron, after which both particles are scattered. Assume---as is the case in quantum gravity---that the detailed physics behind the effect is unknown, and use dimensional analysis to predict the order of magnitude of the length of any thing or process involved in the effect. Since quantum mechanics and the dynamics of an electron and a photon are involved here, one assumes (quite correctly) that the natural constants to be considered are $h$, $c$, and $m_e$, where, like before, it is here essential to be assured that the physical system under study is quantum-electrodynamic. To find this Compton length $l_C$, we set 
\begin{equation}
l_C=K_Ch^\alpha c^\beta m_e^\gamma,
\end{equation}
where $K_C$ is a dimensionless constant, to find 
\begin{equation}
l_C=K_C\frac{h}{m_e c}=K_C 0.0243\ \mathrm{\AA}=0.0243\ \mathrm{\AA} \qquad (K_C=1).
\end{equation}
What is now $l_C$ the length of? In the words of Ohanian \citeyear{Ohanian:1989}, ``in spite of its name, this is not the wavelength of anything'' (p.~1039). 

Ohanian's statement is controversial and requires a qualification. In fact, one immediate meaning of the Compton length can be obtained from the physical equation in which it appears, namely, $\Delta \lambda=l_C [1-\cos(\theta)]$, where $\lambda$ is the wavelength of the photon and $\theta$ is the scattering angle. It is straightforward to see that $l_C$ must be the change of wavelength of a photon scattered \emph{at right angles} with the target electron.
In this manner, a meaning can be found to any constant appearing in a physical equation by looking at the special case when the functional dependence is set to unity. Perhaps this explains Ohanian's denial of any real meaning to $l_C$, since via this method meaning can be attached to any constant whatsoever.\footnote{Baez \citeyear{Baez:2000} explained another physical meaning of the Compton length as the distance at which quantum field theory becomes necessary for understanding the behaviour of a particle of mass $m$, since ``determining the position of a particle of mass $m$ to within one Compton wavelength requires enough energy to create another particle of that mass'' (p.~179). Here again, a new meaning can be found for $l_C$ based on already existing, detailed knowledge provided by quantum field theory. (But see observations on page \pageref{wave-packets}.)}

In this sense, the meaning of constants appearing in physical equations can be learnt from their very appearance in them. However, this takes all charm away from dimensional analysis itself, since such equations are not provided by it but become available only after well-understood physical theories containing them are known; i.e.\ after the content of the equations is related to actual observations. In consequence, the discovery of the meaning of the Planck units in a theory of quantum gravity depends on such equations being first discovered and interpreted against a background of actual observations.\footnote{Another possibility is that the meaning of physical constants could be obtained from simpler, theory-independent methods through which to measure them directly; however, that we have such a more direct method is certainly not true today of the Planck units. Neither is such an extreme operationalistic stance required to give meaning to them; it suffices to have a (Planck-scale) theory of quantum gravity making \emph{some} observable predictions. But at least this much is necessary in order to make quantum gravity a meaningful \emph{physical} theory, and thus disentangle the claims of this currently speculative and volatile field of research.}
And yet, what guarantees that the Planck units \emph{will} meaningfully appear in such a future theory? Only the very \emph{definition}---supported, as we shall see, by a hasty guess---of quantum gravity as a theory involving the constants $G$, $h$, and $c$. This brings us to our second objection. 

A more severe criticism of the widespread, unquestioned reign of the Planck units in quantum-gravity research results from conceptual considerations about quantum gravity itself, including an exemplifying look at the history of physics. 

As mentioned above, it is a largely unchallenged assumption that quantum gravity will involve precisely what its \emph{name} makes reference to---the quantum ($h$) and gravity ($G,c$)---and (i) nothing more or (ii) nothing less. The first alternative presupposes that in a future theory of spacetime, and any observations related to it, the combination of \emph{already known} physics---and nothing else---will prove to be significant. As correct as it might turn out to be, this is too restrictive a conjecture for it excludes the possibility of the need for \emph{truly new} physics (cf.\ Baez's observations). In particular, is our current physical knowledge so complete and final as to disregard the possibility of the existence of relevant natural constants yet undiscovered? 

To make matters worse, \emph{mainstream} quantum gravity only assumes $G$, $h$, and $c$ as relevant to its quest, while paying little heed to other readily available aspects of the Planck scale. For example, why not \emph{routinely} consider Boltzmann's constant $k_B$ and the permittivity of vacuum $\epsilon_0$ (or how about other suitable constants?) to get Planck's temperature 
\begin{equation}
\tau_P=K_\tau \sqrt{\frac{hc^5}{Gk_B^2}}=K_\tau 3.55 \times 10^{32}\ \mathrm{K},	
\end{equation}
and Planck's charge 
\begin{equation}
q_P=K_q\sqrt{hc\epsilon_0}=K_q 1.33 \times 10^{-18}\ \mathrm{C}	
\end{equation}
as relevant to quantum gravity as well? Just because only length, time, and mass appeal to our more primitive, mechanical intuition? Or is it perhaps because the notions of temperature and charge are not included in the \emph{label} ``quantum gravity''?\footnote{For approaches considering extended Planck scales, see e.g.\ \cite{Cooperstock/Faraoni:2003,Major/Setter:2001}.} 

The second alternative above takes for granted that \emph{at least} all three constants $G$, $h$, and $c$ must play a role in quantum gravity. Although this is a seemingly sensible expectation, it need not hold true either, for a theory of quantum gravity may also be understood in less conventional ways. For example, not as a quantum-mechanical theory of (general-relativistic) gravity but as a quantum-mechanical theory of empty spacetime, as we explain below.    
 
In order to present the import of the first view more vividly, we offer an illustration from the history of physics. Reconsider the hydrogen-atom problem before Bohr's solution was known and \emph{before} Planck's solution to the black-body problem was ever given, i.e.\ before any knowledge of the quantum, and therefore of $h$, was available.\footnote{This demand is, of course, anachronistic since $h$ \emph{was} known to Bohr. This notwithstanding, the hypothetical situation we propose serves as the basis for a completely plausible argument. It is only a historical accident that the problem of the collapse of the atom was first confronted with knowledge of the quantum, since the latter was discovered while studying black-body radiation, a problem independent of the atom. Since the black-body problem does not lend itself to the analysis we have in mind, we prefer to take the example of the hydrogen atom, even if we must consider it anachronistically.} With no solid previous experience about atomic physics, the atom would have been considered an electromagnetic-mechanical system, and dimensional analysis could have been (unjustifiably) used to find the order of magnitude of physical quantities significant to the problem. Confronted with this situation, a nineteenth-century quantum-gravity physicist may have decidedly assumed that no new constants need play a part in yet unknown physical phenomena, proceeding as follows. 

Since an electron and a (much more massive) proton of equal charge are concerned, the constants to be considered must be $e$, $m_e$, $\epsilon_0$, and the permeability of vacuum $\mu_0$---and nothing else. This is of course wrong, but only to the modern scientist who enjoys the benefit of hindsight. The inclusion of $\mu_0$ is not at all unreasonable since it could have been suspected that non-negligible magnetic effects played a part, too. 

In order to find the order of magnitude of the radius $a$ and binding energy $E$ of the hydrogen atom, one sets
\begin{equation}
a=K'_a e^\alpha m_e^\beta \epsilon_0^\gamma \mu_0^\delta
\end{equation}
and
\begin{equation}
E=K'_E e^\lambda m_e^\mu \epsilon_0^\nu \mu_0^\xi.
\end{equation}
Proceeding as before, one obtains
\begin{equation}
a=K'_a \frac{e^2\mu_0}{m_e}=K'_a 3.54 \times 10^{-4}\ \mathrm{\AA}	
\end{equation}
and
\begin{equation}
E=K'_E \frac{m_e}{\epsilon_0\mu_0}=K'_E 5.11 \times 10^5\ \mathrm{eV}.
\end{equation}
\emph{These expressions and estimates for $a$ and $E$ are completely wrong} (see Eqs.~(\ref{radius}) and (\ref{energy}) for the correct results and Table \ref{comparison} for a comparison). This comes as no surprise to today's physicist, who can tell at a glance that the constants assumed to be meaningful are wrong: $\mu_0$ should not be there at all, and $h$ has not been taken into account. How can then today's quantum-gravity physicists ignore the possibility that they, too, may be missing one or more yet unknown constants stemming from genuinely new physics? 

\begin{table}
\centering
\setlength{\extrarowheight}{5pt}
\begin{tabular}{|c|c|c|}
\hline
Hydrogen atom & Conscientious & Haphazard \\ \hline \hline
	Radius $a$ & $K_a\frac{\epsilon_0 h^2}{m_e e^2} \approx 1.7\ \mathrm{\AA}$ & $K'_a \frac{e^2\mu_0}{m_e} \approx 4 \times 10^{-4}\ \mathrm{\AA}$ \\ \hline
	Binding energy $E$ & $K_E \frac{m_e e^4}{\epsilon_0^2 h^2} \approx -108\ \mathrm{eV}$ & $K'_E \frac{m_e}{\epsilon_0\mu_0} \approx -5 \times 10^5\ \mathrm{eV}$\\ \hline	
\end{tabular}
\caption[Results of correct and incorrect application of dimensional analysis]{Contrasting results involving two characteristic features of the hydrogen atom obtained via dimensional analysis. A conscientious application takes into account the quantum-mechanical nature of the system, while a haphazard application does not and assumes the hydrogen atom to be an electromagnetic-mechanical system. Only the former gives reasonably approximate results.}
\label{comparison}
\end{table}

The second, less conventional view expressed above, namely, that quantum gravity need not be understood as a quantum-mechanical theory of (general-relativistic) gravity, is supported by the following reasoning. In view of the repeated difficulties and uncertainties encountered so far in attempts to uncover the quantum aspects of gravity, one may wonder whether the issue might not rather be whether spacetime \emph{beyond its geometric structure} may have quantum aspects. In particular and as we enquire further on, may the consideration of quantum theory reveal any physical reality (in the form of observables) that empty spacetime possesses but which classical general relativity denies to them (cf.\ hole argument)? From this perspective, the possible quantum features of gravity are not an issue and, if one stands beyond the geometric structure of spacetime---itself the bearer of its gravitational features ($G$) and causal structure ($c$)---the constants $G$ and $c$ have no reason to arise in the theory. 

\bigskip
 
As advanced in the introduction, there exist some approaches to quantum gravity that do not rely on dimensional analysis in order to reproduce Planck's length and mass. We start by analysing one type of these, namely, those involving the notion of \emph{Planck-scale black holes}. Baez \citeyear[pp.~179--180]{Baez:2000}, Thiemann \citeyear[pp.~8--9, online ref.]{Thiemann:2003}, and to some extent Saslow \citeyear{Saslow:1998} have all independently put forward the essence of the argument in question from different perspectives. In the language of wave mechanics, the gist of the idea is that in order for a wave packet of mass $m$ to be spread no more than one Compton length $l_C=h/mc$, its energy spectrum $\Delta E$ must be spread no less than $mc^2$, i.e.\ enough energy to create---according to quantum field theory---a particle similar to itself. On the other hand, in order for the same wave packet to create a non-negligible gravitational field (strong enough to interact with itself), it must be spread no more than one Schwarzschild radius $l_S=2Gm/c^2$, i.e.\ be concentrated enough to become---according to general relativity---a black hole. Both effects are deemed to become important together when $l_C=l_S$, i.e.\ when the particle becomes a black hole of mass $(hc/2G)^{1/2}=m_P/\sqrt{2}$ and radius $(2Gh/c^3)^{1/2}=\sqrt{2}l_P$---a so-called Planck black hole, the paradigmatic denizen of the quantum-gravitational world.\footnote{Planck's time is connected with this argument according to extrapolations of predictions from quantum field theory in curved spacetime, namely, that a Planck black hole is unstable with a lifetime of $t_P\approx 10^{-43}$ s. However, this theory, in which spacetime is taken to be classically curved but not ``quantized,'' is commonly referred to as a semiclassical theory (of quantum gravity) and paradoxically distrusted to hold at the Planck scale. See e.g.\ \cite[p.~164]{Repo:2001}.}  

Is this a sound argument or is it yet another \emph{ad hoc}, \emph{ad lib} armchair exercise devised only to quench our thirst for tangible ``quantum-gravitational things''? Regardless of the complicated and physically controversial question whether such black holes actually exist, let us concentrate here only on the conceptual aspect of the issue. Our criticism has three parts.

First and irrespective of quantum-gravity ideas, to give operational meaning to the above statements about wave packets and their spreads, one should talk about a set of identically prepared microscopic particles of mass $m$. Taking one microscopic system at a time and upon repeatedly measuring either its position or its momentum, the statistical results are found to be spread according to the uncertainty relation $4 \langle (\Delta \hat X)^2\rangle_{|\psi\rangle} \langle (\Delta \hat P)^2\rangle_{|\psi\rangle} \geq | \langle [\hat X,\hat P] \rangle_{|\psi\rangle}|^2=\hbar^2$.\footnote{In this context, the substitutions $\langle (\Delta \hat P)^2\rangle_{|\psi\rangle}\approx (\Delta E/c)^2$ (relativistic approximation) and $\langle (\Delta \hat X)^2\rangle_{|\psi\rangle}=(h/mc)^2$ (Compton spread) are used to get $\Delta E \geq mc^2/4\pi\approx mc^2$.} But spelt out this way, the argument loses its intuitive appeal and, with it, its intended meaning. We see that to prepare a system has nothing to do with our squeezing a particle into any sort of tight confinement, even less with our needing a certain minimum amount of energy to do so. Thus, neither the creation of new particles nor of black holes is naturally connected with the uncertainty relations at all.\label{wave-packets}

Second and regardless of the above, it is straightforward to see that this reasoning to the effect that Planck black holes must be paradigmatic of quantum gravity assumes the previously criticized view that such a theory must arise out of a combination of the quantum and gravity. It is really no surprise to find that, on this supposition, the Planck scale \emph{does} result even without the intervention of dimensional analysis; now instead of combining universal constants together, we combine \emph{theories} which contain them. 

In fact, a variation of this way of proceeding is found, for example, in loop quantum gravity and in the canonical quantization of Hamiltonian black-hole dynamics. In them, the appearance of the Planck length does not result from dimensional analysis\footnote{See \cite[p.~250]{Rovelli:2004} for an explicit mention of this point in regard to loop quantum gravity.} nor relying on the idea of Planck black holes. The Planck length arises instead from a different instance of the said physically uncalled-for procedure of now combining theories (instead of simply constants) whose fundamental constants are $G$, $c$, and $h$, i.e.\ from quantizing general relativity in one way or another. Why uncalled-for? Because no reasons are known as to why one should proceed thus---as if one were, say, to quantize the traffic on the streets or the solar system---nor are the consequences of quantization known to be attested in the physical world. All in all, what this analysis shows is something we already knew, namely, that the Planck scale must be relevant to a theory which glues the pieces of our current knowledge together. But is there reason to do so?

Third, the manner in which the black-hole argument is construed appears as notoriously haphazard as the quantum-gravity application of dimensional analysis also for reasons irrespective of the first criticism. In effect, general relativity is first assumed nonchalantly in order to implement the idea of a Planck black hole, but this concept is subsequently taken to belong in quantum gravity, which, according to the usual view, signals the breakdown of general relativity---thereby precluding the possibility of black holes!---at the Planck scale. (Or can there be black holes without general relativity? Does the mythical, unknown yet true, theory of quantum gravity allow for them?) Therefore, the reasoning here too seems to be conceptually flawed. 

The welcome view is sometimes expressed that the appropriateness of this notion is uncertain. For example, Thiemann shares (for his own reasons) our last point that the notion of Planck black holes must be conceptually flawed:
\begin{quote}
[A Planck black hole] is again [at] an energy regime at which quantum gravity must be important and these qualitative pictures must be fundamentally wrong\ldots \cite[p.~9, online ref.]{Thiemann:2003}
\end{quote}
Interestingly enough, Thiemann takes this \emph{negative} argument \emph{optimistically}, while we can only take it pessimistically---as a bad omen.

Given the conceptually uncertain nature and relevance of these ideas, we conclude that one should take them with a pinch of salt, and not anywhere more seriously than considerations arising from dimensional analysis.

\section{Appraisal} \label{Appraisal}
The state of affairs reviewed in the first part of the previous section leads to serious doubts regarding the applicability of dimensional analysis. How can we ever be sure to trust any results obtained by means of this method? Or as Bridgman \citeyear{Bridgman:1963} jocularly put it in his brilliant little book: ``We are afraid that\ldots we will get the incorrect answer, and not know it until a Quebec bridge falls down'' (p.~8). It is therefore worthwhile asking: where, in the last analysis, did we go wrong? 

In this respect, Bridgman enlightened us by explaining that a trustworthy application of dimensional analysis requires the benefit of \emph{extended experience} with the physical situation we are confronted with as well as \emph{careful thought}:

\begin{quote}
We shall thus ultimately be able to satisfy our critic of the correctness of our procedure, but to do so requires a considerable background of physical experience, and the exercise of a discreet judgment. The untutored savage in the bushes would probably not be able to apply the methods of dimensional analysis to this problem and obtain results which would satisfy us. \cite[p.~5]{Bridgman:1963}
\end{quote}
And further:
\begin{quote}
The problem [about what constants and variables are relevant] cannot be solved by the philosopher in his armchair, but the knowledge involved was gathered only by someone at some time soiling his hands with direct contact. \cite[\mbox{pp.~11--12}]{Bridgman:1963} 
\end{quote}

\begin{figure}
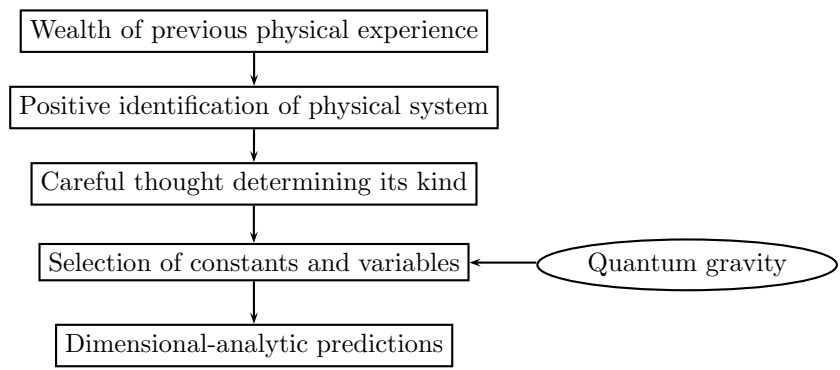

	\begin{psmatrix}[rowsep=0.4,colsep=0.5]
    \psframebox{{\small Wealth of previous physical experience}} \\
    \psframebox{{\small Positive identification of physical system}} \\
    \psframebox{{\small Careful thought determining its kind}} \\
    \psframebox{{\small Selection of constants and variables}} &
    \psovalbox{{\small Quantum gravity}} \\
    \psframebox{{\small Dimensional-analytic predictions}} 
    \ncline{->}{1,1}{2,1}
    \ncline{->}{2,1}{3,1}
    \ncline{->}{3,1}{4,1}
    \ncline{->}{4,1}{5,1}
    \ncline{->}{4,2}{4,1}
    \end{psmatrix}
\caption[Application of dimensional analysis]{Steps in the correct application of dimensional analysis. This is the only way to assure ourselves that the predictions obtained by means of this method will likely possess physical relevance. Dimensional analysis in quantum gravity becomes haphazard by omitting the first three steps of the procedure.}
\label{Application of dimensional analysis}
\end{figure}

In this light, a conscientious application of dimensional analysis to the affairs of quantum gravity is hopeless. There does not exist so far any realm of physical experience pertaining to interactions of gravitation and quantum mechanics, and therefore there is no previous physical knowledge available and no basis whatsoever on which to base our judgement. Thus, the quantum-gravity researcher proclaiming the relevance of the Planck scale resembles Bridgman's savage in the bushes or philosopher in the armchair, not because of being untutored or not wanting to soil his hands, but because there are no observational means available to obtain any substantial information about the situation of interest. In consequence, quantum gravity skips essential steps in its application of dimensional analysis, as shown in Figure \ref{Application of dimensional analysis}.   

Bridgman also dismissed any meaning to be found in Planck's units considering the manner in which they arise:
\begin{quote}
The attempt is sometimes made to go farther and see some absolute significance in the size of the [Planck] units thus determined, looking on them as in some way characteristic of a mechanism which is involved in the constants entering the definition. 

The mere fact that the dimensional formulas of the three constants used was such as to allow a determination of the new units in the way proposed seems to be the only fact of significance here, and this cannot be of much significance, because the chances are that any combination of three dimensional constants chosen at random would allow the same procedure. Until some essential \emph{connection} is discovered between the \emph{mechanisms} which are accountable for the gravitational constant, the velocity of light, and the quantum, it would seem that no significance whatever should be attached to the particular size of the units defined in this way, beyond the fact that the size of such units is determined by phenomena of universal occurrence. \cite[p.~101, italics added]{Bridgman:1963} 
\end{quote}
Gorelik \citeyear{Gorelik:1992}, and von Borzeszkowski and Treder \citeyear{vonBorzeszkowski/Treder:1988} have, in fact, criticized these remarks of Bridgman's on account that, indeed, the connection has now been found under the form of quantum gravity. A moment's reflection shows that this is a deceptive argument. One cannot legitimately contend that the new connection found is quantum gravity, and therefore that Planck's units have meaning because, with no phenomenological effects to support it, there is no physical substance to so-called quantum gravity besides (mostly Planck-scale based) theoretical speculations. Therefore, the argument is at best unfounded and at worst circular.

To sum up, the relevance of Planck-scale physics to quantum gravity rests on several uncritical assumptions, which we have here cast doubt upon. Most importantly, there is the question of whether quantum gravity can be straightforwardly decreed to be the simple combination of the quantum and gravity. This excluding attitude does not appear wise after realizing, on the one hand, that black-body physics did not turn out to be the combination of thermodynamics and relativity, nor atomic physics the combination of electromagnetism and mechanics; and on the other hand, that quantum gravity can also be interpreted in a less literal sense---for what's in a name after all?---and regarded to include, for instance, spacetime features beyond gravity. Dimensional, quantization, and \emph{ad hoc} analyses only yield meaningful results when the system studied and its physics are well known to start with. To presume knowledge of the right physics of an unknown system is for quantum gravity to presume too much.

Further, we argued that the physical meaning of the Planck units could only be known after the successful equations of the theory which assumes them---quantum gravity---were known. To achieve this, however, the recognition and observation of some phenomenological effects related to quantum mechanics and spacetime are essential. Without them, there can be no trustworthy guide to oversee the construction of, and give meaning to, the ``equations of quantum gravity,'' and one must then resort to wild guesses. Ultimately, herein lies the problem of dimensional, quantization, and \emph{ad hoc} analyses as applied to quantum gravity. The fact nevertheless remains that no such observables have as yet been identified. As we expressed earlier, it is only through the discovery and observation of such phenomena telling us what quantum gravity is \emph{about} that we will gradually unravel the question concerning what quantum gravity \emph{is}.  

In view of this uncertainty surrounding Planck's natural units, we believe it would be more appropriate to the honesty and prudence that typically characterize the scientific enterprise, not to abstain from their study if not so wished, but simply to express their relevance to quantum gravity as a humble belief, and not as an established fact, as is regrettably today's widespread practice.

\newpage
\thispagestyle{empty}

\chapter[The anthropic foundations of geometry]{The anthropic foundations\\ of geometry}
\label{ch:Geometry}
\begin{flushright}
\begin{tabular}{p{10cm}}
\emph{We can hope for the best, for I see the signs of men.}
\smallskip \\
Marcus Vitruvius Pollio, \emph{De architectura}, Book 6.
\end{tabular}
\end{flushright}
\bigskip

Leaving for now quantum gravity behind, we begin the study of geome\-try---a discipline which is not only ubiquitous in quantum gravity but, in fact, is basic to all of physics and even to a wealth of everyday cognitive tasks.

What are the nature and origins of geometry? How ubiquitous is it, really? Is it a tool of thought or a God-given feature of the world? Is its use beneficial to further human understanding? To what extent? How has geometry fared in physics so far in helping our understanding? And how does geometry fare as quantum gravity attempts to attain by its means a sounder comprehension of the physical meaning of space, time, and quantum theory? These questions will occupy us in the coming chapters.

\section{The rise of geometric thought}
The meaning of the word ``geometry'' is far wider than its etymological root suggests. That is to say, the meaning attributed to ``geometry'' is much more comprehensive and basic than the relatively advanced notions included in a ``measurement of the earth.''

To be sure, in order to imagine a triangle or a circle, no quantitative concept of length is needed, for such figures may well be thought of without any need to establish what the lengths of the sides of the triangle or what the circle radius might be. And yet, no-one would contend that a triangle or a circle is not a geometric object.

We could therefore lay down the conceptual foundations of geometry in the following way. Geometry is built out of two basic elements. The first and more primitive of them is that of \emph{geometric objects} such as, for example, point, line, arrow, plane, square, circle, sphere, cube, etc. Geometric objects stand for the idealized \emph{shape} or \emph{form} of the objects of the natural world. 

Geometric objects, in turn, display geometric properties. These are qualitative attributes abstracted from geometric objects and possibly shared by several of them. Examples of geometric properties are flatness, curvedness, straightness, roundness, concavity, convexity, closedness, openness, connectedness, disconnectedness, orientability, orthogonality, parallelism, etc. Geometric objects and some of their geometric properties (those invariant under continuous deformations) can be found at the root of the study of topology.  

\begin{figure}
\begin{center}
\includegraphics[width=\linewidth]{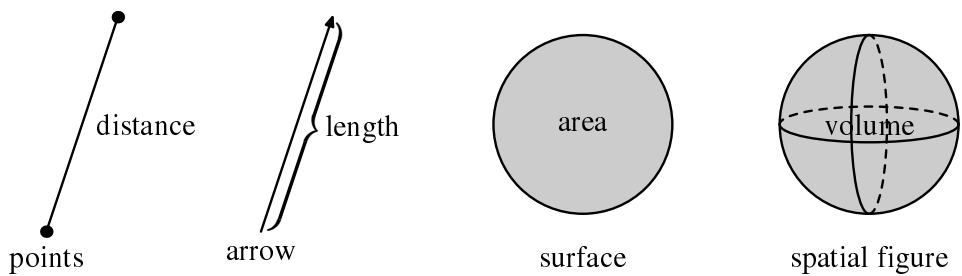}
\end{center}
\caption[Geometric objects and magnitudes]{Some geometric objects and associated geometric magnitudes.} \label{gogm}
\end{figure}

The second, much more advanced ingredient that geometry consists of is that of \emph{geometric magnitudes}. As its name suggests, these magnitudes involve the expression of a quantity, for which the concepts of \emph{number} and \emph{unit measure} are prerequisite. Geometric magnitudes arise as result of a process by means of which geometric objects can be attributed different notions of \emph{size} or \emph{measure}. The most basic and paradigmatic example of a geometric magnitude is \emph{length}; this applies to a geometric object (e.g.\ line, curve) or a part of it (e.g.\ edge of a polygon), and its original meaning is that of counting how many times a unit measure can be juxtaposed along the object in question. Other geometric magnitudes, shown with their corresponding geometric objects in Figure \ref{gogm}, are distance, area, and volume. 

\section{The origins of form}
According to Smith \citeyear[pp.~1--4]{Smith:1951}, the existence of geometric form can be traced back, as it were, to the mists of time, if one considers forms to be present in nature in themselves, and not as somebody's interpretation of nature. In this view, the galactic spirals, the ellipses of the planetary orbits, the sphere of the moon, the hexagons of snowflakes, etc., are all geometric displays of nature and antecede any presence of life on earth.

Is nature, then, intrinsically geometric? It is all too tempting to regard nature with human eyes, even when talking about a time when no sentient beings existed. Is it not rather a human prerogative to interpret the world geometrically? Or, paraphrasing Eddington \citeyear[p.~198]{Eddington:1920}, to filter out geometry from a jumble of qualities in the external world? Do not geometric objects actually arise through idealization of sensory perceptions by the mind to become geometric objects \emph{of human thought}? We believe that the history of geometric thought begins with the first conscious and purposeful \emph{appreciations} of shape or form. Geometry is to be equated with geometric thought rather than with geometric existence, and its origins must be located only after the advent of life. \emph{Geometry is a tool of thought}, not a property of the world.

Can we now consider, for example, the geometric patterns displayed on the surface of tree leaves or those of a sea-shell or spiderweb as primitive displays of geometric thought by living beings? Hardly. It is too far-fetched to say that a plant shows geometric patterns because \emph{it} thinks geometrically---plants have no brain. To a lesser degree, the same applies to spiders, which, although endowed with brains, appear as no great candidates for abstract thought either. As Smith \citeyear{Smith:1951} put it, ``Who can say when or where or under what influences the Epeira, ages ago, first learned to trace the logarithmic spiral in the weaving of her web?'' (p.~5). Who can say, indeed, that the Epeira ever learnt to do purposefully, insightfully, such a thing at all.

We reserve the capacity to think geometrically to humans, disregarding in so doing also mankind's closest cousins, the apes (or the more distant dolphins), as no other animal but Man engages in geometric thought in a deliberate, conscious, and purposeful way; no other animal is driven to geometric thought by the need to conceptualize the world around, nor psychologically rewarded in actually achieving the sought-for understanding after a geometric manner. Animals can certainly tell apart different \emph{man-made} (once again, for the purpose of understanding animals) geometric figures, but this does not qualify as an animal instance of geometric endeavour. Although most living creatures share with humans similar sense organs---especially, but not exclusively, those that allow vision---it is the workings of a particularly moulded \emph{brain} interpreting stimuli and striving to conceptualize the world that result in geometric understanding. The deeper reason why human brains can---or even must?---interpret the world geometrically is but an open question. As we proceed through the coming chapters, we shall analyse geometry as a human conceptual tool of thought further.

\section{The origins of number} \label{ON}
The first appreciations of form and symmetry must be as old as humanity itself. What about the first appreciations of number? In a rudimentary sense close to but not properly called counting, these may be taken to be even older than humanity. According to Smith \citeyear[p.~5]{Smith:1951}, some form of pseudo-counting exists among magpies, chimpanzees, and even some insects. More recently, also Hauser \citeyear{Hauser:2000} has described instances of apes, birds, and rats capable of some form of counting ``up to'' a small number. All these cases, however, only involve subitization, i.e.\ telling at a glance the number of objects present in a small group. The fact that animals cannot ``count'' beyond about four---the limit for accurate subitization---may in fact be seen as evidence that they are not actually counting. However, Hauser relates that, under \emph{human-induced} training, some animals can learn to count in exchange for food.

In this light, the view according to which the ability to count exists in animals in a way that precedes humanity is analogous to the previous one---that displays of geometric form are found in inert nature, plants, and animals---and similarly mistaken. The innate animal ability to recognize differences in the number of objects in a small group is far from what we mean by counting: an abstract process involving a deliberate act of enumeration. The ability to count, aided today by an advanced development of language and radix systems, comes second nature to us, and we are thus sometimes misled into seeing our own reflection in other animals.

The first human needs for counting might have arisen out of a desire to know the size of one's household, as well as that of hunted animals, or that of the members of the enemy. Originally, however, the members of large groups were not really counted, but simply considered as ``many'' and designated by some collective noun such as ``heap of stones,'' ``school of fish,'' and the like. Hauser has proposed that the development of a more advanced form of counting is to be found in the emergence of human trade.

\begin{figure} 
\begin{center}
\includegraphics[width=100mm]{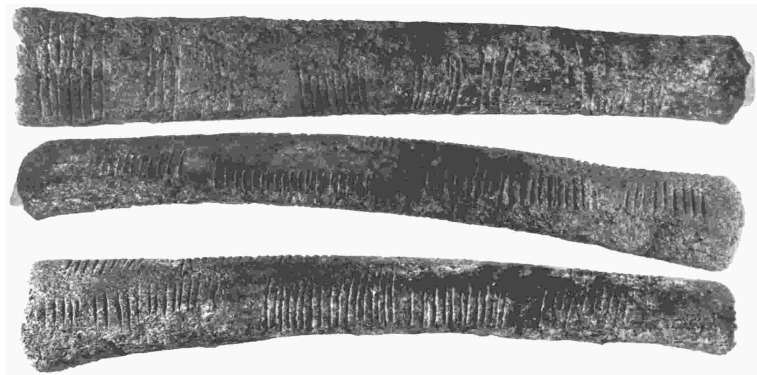}
\end{center}
\caption[Ishango bone]{Three views of the 10-centimetre-long Ishango bone dating back to about 18000 b.C. It displays notches in three rows, naively interpreted as evidence of Upper-Paleolithic knowledge of prime numbers, a decimal and sexagesimal number system, and duplication. According to A.~Marshack, the bone consists of a lunar-phase counter displaying a record of five and a half lunar months. Other interpretations, such as tallies of prey or mere decoration, are also possible. Reproduced from \cite{Marshack:1972}.}
\label{ishangobone}
\end{figure}

\begin{figure} \label{blanchardbone}
\begin{center}
\includegraphics[width=85mm]{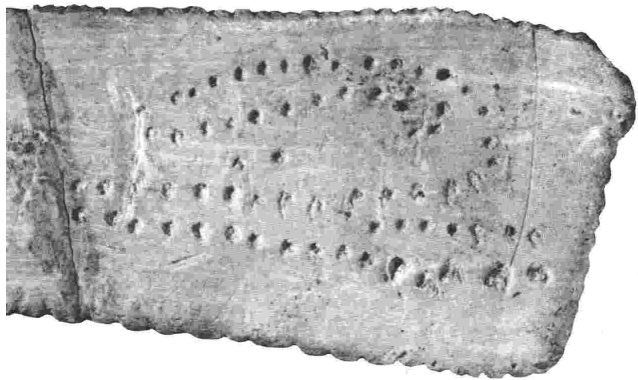}
\end{center}
\caption[Blanchard bone]{Partial view of the 11-centimetre-long Blanchard bone, found in Abri Blanchard, in the Dordogne region of France, and dating back to about 25000 b.C. in the Upper Paleolithic. It displays circular notches laid out in a meandering pattern, interpreted by A.~Marshack as a record of a two-and-a-quarter lunar month. Reproduced from \cite{Marshack:1972}.}
\end{figure}

The most famous---although not necessarily the oldest or most meaning\-ful---archaeological record of the human use of counting is the Ishango bone, found near Lake Edward, on the frontier of Uganda and Zaire, and dating back to about 18000 b.C.\ in the Upper Paleolithic. The 10-centimetre-long bone, shown in Figure \ref{ishangobone}, displays a series of marks carved in groups and placed in three rows. Naively, these markings have been interpreted as follows. Row (i) shows groups of 21 ($=20+1$), 19 ($=20-1$), 11 ($=10+1$), and 9 ($=10-1$) marks, which hints at a decimal number system; row (ii) contains groups of 11, 13, 17 and 19 notches, which are the prime numbers between 10 and 20; row (iii) shows a method of duplication. Furthermore, the markings on rows (i) and (ii) each add up to 60, and those on row (iii) to 48, possible evidence of knowledge of a sexagesimal number system. A speculative, yet painstakingly thorough and rather convincing study carried out by Marshack \citeyear{Marshack:1972}, however, revealed that it is more plausible for this bone---together with numerous other engraved specimens from the Upper Paleolithic, such as the Blanchard bone, shown in  Figure \ref{blanchardbone}---to be a lunar-phase counter displaying a record of five and a half lunar months.

The originally intended meaning of these carved objects remains nevertheless uncertain. In this respect, Flavin wrote:
\begin{quote}
Ancient straight lines, whether parietal or portable, defy forensic interpretation because we can't penetrate their constituent phenomenological simplicity. Marshack attempts to turn \emph{marques de chasse} (``hunting marks'') into lunar calendars, eschewing a tally of prey for a tally of days (or nights). As all of the artifacts presented by Marshack as evidence for Upper Paleolithic lunar calendars are individual, seeming to share no consistent notational pattern, every artifact remains uniquely problematic. Much like every generation receives a different interpretation of Stonehenge, perhaps years from now the engraved items from the Upper Paleolithic will be envisioned as something beyond our current imagining. \cite{Flavin:2004}
\end{quote}
In any case, as far as evidence of an ability to count is concerned, both markings representing a simple tally of prey and markings representing a more complex tally of days are equally signs of it; not so, however, if the markings in question were simply perfunctory decoration or the like.\footnote{When it comes to interpreting isolated pieces of archaeological evidence, a cautious attitude such as Flavin's has much to be recommended for. In March 2004, BBC News reported the finding, in the Kozarnika cave in northwest Bulgaria, of an animal bone over a million years old that displays several markings in the form of seemingly purposeful straight lines. Furthermore, ``another animal bone found at the site is incised with 27 marks along its edge'' (BBC News, http://news.bbc.co.uk/ 2/hi/science/nature/3512470.stm). Ought one to conclude that \emph{Homo erectus} already had the capacity to count and keep a tally of his findings, including records of the phases of the moon?}

Returning to the development of counting and following Smith \citeyear[pp.~7--14]{Smith:1951}, a primitive notion of counting was available much earlier than words existed for the numbers themselves. Primitively, counting consisted in explicitly associating the objects to be counted with, for example, the fingers of the hands or marks on the ground or on bone. Traces of such counting methods can still be found in the habits and languages of indigenous tribes living today. For example, the Andamans, an Oceanic tribe, only have words for ``one'' and ``two'' but manage to keep counting up to ten by tapping their nose with their fingers in succession and repeating ``and this one.'' The Zulu tribes express the number six by saying ``taking the thumb'' and the number seven by saying ``he pointed.'' Finally, in the Malay and Aztec tongues, in that of the Niu\`{e}s of the Southern Pacific, and in that of the Javans, the number names mean literally ``one-stone,'' ``two-stone,'' etc.; ``one-fruit,'' ``two-fruit,'' etc.; and ``one-grain,'' ``two-grain,'' etc., respectively.

Progressively, the ability to name the numbers arose. Since counting up to a considerably great number would have needed the invention of as many different words (and later symbols), some system was needed to handle and organize these numbers. Universally, this solution was found in the use of a radix system, i.e.\ using a certain amount of numbers and then repeating them over and again. 

\section{The origins of magnitude}
Like form and number, the idea of size must also be as old as Man himself, and as an animal-like rudimentary conception even older. A predatory animal, for example, must take into account the size of its prey before deciding whether it is sensible to attack it alone, or whether it should do so with the help of the pack. However, this is an instance of the qualitative (big, small) or even comparative (bigger, smaller) appreciation of size, but not of \emph{quantitative} size (three arms long), as is necessary for the birth of geometric magnitudes and the quantitative conception of geometry.

Thus, it was only after Man was armed with both his primitive, innate power to recognize form and symmetry in nature and a later, more involved capacity to count, that he had the necessary tools to develop the notion of geometric magnitude and measurement. The combination of these two skills proceeded through the realization that one need not only be satisfied with knowing whether a natural object was bigger or smaller than another (the enemy is taller than myself), but that their size could be \emph{quantified} and stated with a number. For the magnitude of length---the paradigm of a geometric magnitude---the process leading to it can be described as follows. If a natural object is geometrically idealized and thought of as a geometric object (the enemy as a straight line), one could then count how many times a comparatively smaller unit measure could be juxtaposed along the object in question. In addition to the natural object to be measured, also the successful implementation of a unit measure (a measuring rod) from a natural object (a tree branch) required the geometric idealization of the latter (a tree branch as a \emph{straight edge}). The quantitative size of the geometrically objectified natural object could then be given by means of this unit-measure-related number. Now---and only now---is geometry developed enough to live up to its etymological duty of ``measuring the earth.''

\section{The rise of abstract geometry} \label{OrAbGe}
The rise of more sophisticated geometric ideas beyond form and magnitude is recorded to have had its origin in the Neolithic Age in Babylonia and Egypt. It appears fair to say that the oldest geometric construction that rates higher in abstraction than other common uses of geometric thought is so-called Pythagoras' theorem. Evidence of knowledge of Pythagorean triples\footnote{A Pythagorean triple consists of an ordered triple $(x, y, z)$, where $x$, $y$, and $z$ satisfy the relation $x^2+y^2=z^2$.} and of the Pythagorean theorem can be found in the Babylonian cuneiform text Plimpton 322 and in the Yale tablet YBC 7289, which date from between 1900 and 1600 b.C.\ in the Late Neolithic or Early Bronze Age.

\begin{figure}
\begin{center}
\includegraphics[width=0.92\linewidth]{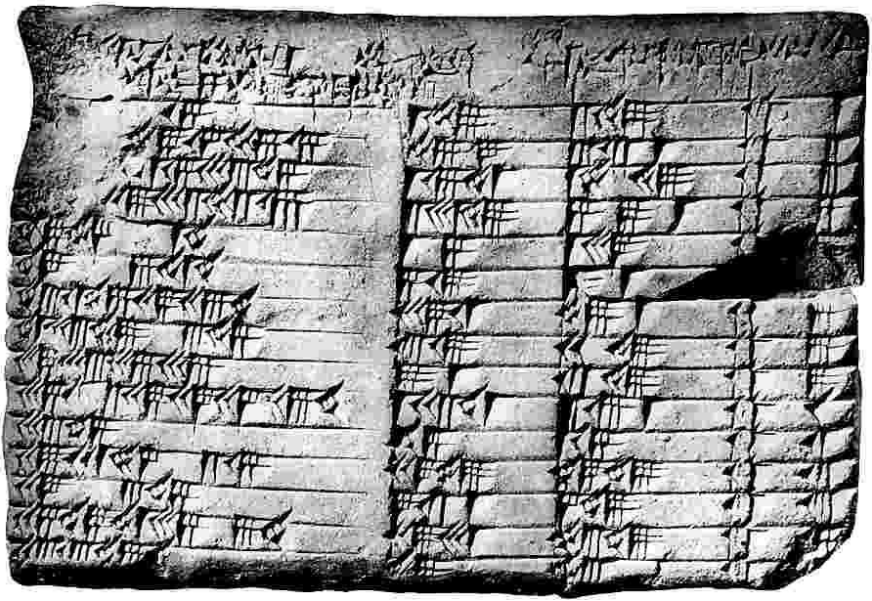}
\end{center}
\caption[Plimpton 322 tablet]{Plimpton 322 tablet at Columbia University. This Babylonian cuneiform tablet shows evidence of knowledge of Pythagorean triples as early as 1900--1600 b.C. Vertical strokes ``$\vee$'', alone or grouped, represent the numbers from 1 to 9; horizontal strokes ``$<$'', alone or grouped, represent 10, 20, 30, etc. Reproduced from http://www.math.ubc.ca/$\sim$cass/courses/m446-03/pl322/pl322.html} 
\label{plimpton322}
\end{figure}

In the preserved part of Plimpton 322, shown in  Figure \ref{plimpton322}, there are five columns of numbers in the sexagesimal system. Counting from the left, the fifth one shows the numbers 1 to 15, the fourth one repeats the word ``number'' (\emph{ki}), while the second and third show the width $y$ and the diagonal $z$ of a rectangle. If we calculate $z^2-y^2=x^2$, $x$ turns out to be (after some corrections) a whole number divisible by 2, 3 or 5, and a missing column for $x$ can be reconstructed with confidence. This helps explain the first (partly missing) column showing either $v=(y/x)^2$ or $w=(z/x)^2=1+v^2$. It has been inferred from this that the Babylonians must have calculated Pythagorean triples of the form $(1,v,w)$ satisfying the equation
\begin{equation} \label{Babylonia}
  w^2-v^2=1
\end{equation}
and then obtained triples $(x,y,z)$ multiplying by $x$. Finally, it has also been noticed that $w+v$ is in all cases a regular sexagesimal number, which means that the Babylonians may have written $(w+v)(w-v)=1$ and then found $w-v$ by inverting the regular number $w+v$.\footnote{See \cite[pp.~2--5]{vanderWaerden:1983}.}

\begin{figure}
\begin{center}
 \leavevmode
 \includegraphics[width=75mm]{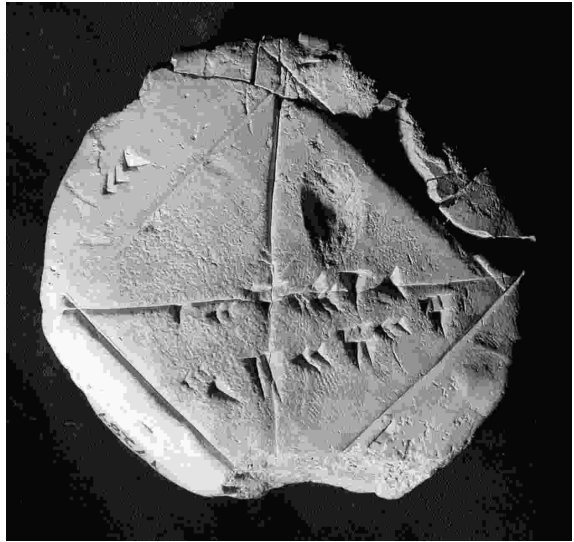}
\end{center}
\caption[Yale YBC 7298 tablet]{Yale YBC 7289 tablet at the Yale Babylonian Collection. This Babylonian cuneiform tablet shows incontestable evidence of knowledge of Pythagoras' theorem as early as 1900--1600 b.C., over a thousand years before Pythagoras. Reproduced from http://www.math.ubc.ca/people/faculty/cass/Euclid/ybc/ybc.html (Bill Casselman, photographer).} \label{yaletablet}
\end{figure}

The Yale tablet YBC 7289, shown in  Figure \ref{yaletablet}, shows a square whose side is marked off as 30 units long, and one of its diagonals attributed the numbers 1,24,51,10 and 42,25,35 (in base 60). If we fix the floating point and assume that the first number is 1;24,51,10, then its equivalent in the decimal system is 
\begin{equation}
1+24/60+51/60^2+10/60^3=1.414212963. 
\end{equation}
This is a remarkably good approximation of $\sqrt{2}=1.414213562\ldots$. The length of the diagonal 42;25,35 (42.42638889) can then be obtained as
\begin{equation}
42;25,35=30\times 1;24,51,10 \qquad (30\times \sqrt{2}). 
\end{equation}
This tablet constitutes archaeological proof beyond the shadow of a doubt that the Babylonians knew Pythagoras' theorem, not just Pythagorean triples. In particular, they knew that $\sqrt{2}$ is the length of the diagonal of a square of side 1, since they used this knowledge to solve for the diagonal of a square of side 30.

There exist records showing knowledge of Pythagorean triples by the Egyptians, too. The papyrus Berlin 6619 dating from around 1850 b.C.\ contains a mathematical problem in which the sides of two squares are to be found knowing that one side is
three-quarters of the other and that their total areas add up to 100 squared units; i.e.\ $y=3x/4$ and $x^2+y^2=100$. In the
solution, it is recognized that the total area is also that of a square of side 10 units, and therefore the triple (8,6,10) is recognized as well. However, no triangles are mentioned here. van der Waerden \citeyear[p.~24]{vanderWaerden:1983} suggested that the Egyptians may have learnt about Pythagorean triples from the Babylonians. Also in support of this view, Boyer and Merzbach \citeyear{Boyer/Merzbach:1991} wrote: ``It often is said that the ancient Egyptians were familiar with the Pythagorean theorem, but there is no hint of this in the papyri that have come down to us'' (p.~17).

All this being said, what purpose did the discovery and use of the Pythagorean theorem and its associated triangles and triples serve ancient Man? Boyer and Merzbach \citeyear{Boyer/Merzbach:1991} suggested that, although pre-Hellenistic mathematics was mostly of a practical sort and people ``were preoccupied with immediate problems of survival, even under this handicap there were in Egypt and Babylonia problems that have the earmarks of recreational mathematics'' (p.~42).
Asimov \citeyear[pp.~5--6]{Asimov:1972}, along similar lines, placed the rise of (the arts and of) pure knowledge in human curiosity or, as he eloquently put it,  ``the agony of boredom,'' stemming from the fact that any sufficiently advanced brain needs to be kept engaged in some activity. On the practical side, it is also
possible that the Pythagorean geometric construction could have originally represented something more practical to the ancients, either in connection with their religious beliefs, or as a calculational tool for their astronomical problems, or both.

To sum up, what the analysis of this chapter reveals is that nature is not geometric by essence, but that what seem to be nature's universal geometric features are, in fact, man-made, a direct consequence of Man's innate capacity to recognize form, number, and measure. This ability lies at the root of humans' past and present scientific achievements: these stem essentially from the relentless geometrization that humans have made and continue to make of the natural world, as we see in the next chapter.

\chapter{Geometry in physics}
\label{ch:Geometry in physics}
\begin{flushright}
\begin{tabular}{p{10cm}}
\emph{He deals the cards to find the answer,}\\
\emph{The sacred geometry of chance,}\\
\emph{The hidden law of a probable outcome;}\\
\emph{The numbers lead a dance.}
\smallskip \\
Gordon Sumner, \emph{Ten summoners' tales}
\end{tabular}
\end{flushright}

\bigskip

We presented a view of geometry that is founded on the human conceptions of geometric objects and geometric magnitudes attached to or between them. We exemplified this view of geometry by means of spatially extended geometric objects, and the attribution of quantitative size to them by means of rod measurements of length and its derivatives. In physics, however, we deal with various kinds of magnitude and their measurements, of which length and its measurement are but one case. What about these then? Is geometry entwined with all physical measurements? 

Giving a clear-cut answer to this question is difficult, because with it we enter a realm where the presence of geometry is tied not just to the way a measurement is made but also to how its results are interpreted theoretically. To begin with, rod measurements of length are an intrinsically geometry \emph{activity} because, as we said, a measuring rod must necessarily be a straight edge. This fact makes length and distance\footnote{Distance can be regarded as equivalent to length because it is in essence the same conception but applied to geometric objects differently. Length refers to the unidirectional size of a geometric object (e.g.\ a line, an edge), while distance refers to the size of the unidirectional space between the endpoints of the same object or between points as independent geometric objects (e.g.\ points of space).} paradigmatically geometric magnitudes. What about radar-based measurements of length? Are these geometric? We may now say that, \emph{as an activity}, they are not, because there is nothing essentially geometric about the instrument (e.g.\ a clock) needed to make the measurement. As a conception, however, length continues to be a geometric magnitude. How so?

In this case, geometry comes in as an immediate concomitant interpretation. The human mind visualizes length as spatial extension, and the mental picture of a geometric object (straight line) \emph{whose property} is the length just measured immediately arises. In a more abstract but equally vivid way, the same is done for the results of other measurements, now spatially unrelated, which as activities having nothing to do with geometry. For example, the observations of time via clock readings, probability via simple counting, speed via speedometer readings, spin via Stern-Gerlach device readings, etc., are again geometry-unrelated observations or procedures. The human mind, however, with its strong penchant for visualization in terms of extended objects in some form of space, takes the physical quantities resulting from these measuring activities and gives them a geometric theoretical background. In this way, observations of various natures are likened to the one which we experience as most familiar, namely, observations of length and its derivatives. Let us see some examples.

Although time intervals are measured with clocks and space intervals (customarily) with rods, physics likens time to space depicting it as a coordinate parameter along which material objects do not move but flow. Thus, measuring time intervals becomes theoretically equivalent to measuring stretches of space, and hence the phrase ``a length of time'' in the sense of a fourth spatial dimension. This geometric interpretation of time was only deepened by relativity theory, which, superficially at least (see Chapter \ref{ch:Analysis of time}), fused time and space into one geometric whole. Another example is furnished by probability. While this concept only has to do with counting positive and negative outcomes, it is understood in physical theories in one of two ways: as Isham \citeyear[p.~14--17]{Isham:1995} has explained, in classical physics, probability is likened to a ratio of volumes (the volume of successful outcomes divided by that of total outcomes), so that each outcome is represented by a unit volume;\footnote{In fact, one practical method to calculate $\pi$ through the measurement of probability consists in drawing a circle of unit-length radius on paper and subsequently enclosing it within four walls forming a square whose sides are two units of length. By randomly releasing a small object into the box, counting how many times it lands within the boundary of the circle, dividing by the total amount of trials, and multiplying by four (squared units of length), an approximate value of $\pi$ can be obtained.} in quantum physics, on the other hand, probability is likened to an area resulting from the squared overlap of two arrow state vectors. A third example is provided by velocity, which is depicted as an extended arrow tangent to the trajectory curve of a material particle, and whose magnitude is then interpreted as the length of this arrow. A fourth example is found in quantum-mechanical spin, which, like velocity, is mentally pictured as an arrow and its magnitude interpreted to be its length. 

A case farther detached from physics also provides an enlightening example. In the theory of coding, one is concerned with sending messages over a noisy information channel and with being able to recover the original input from a possibly distorted output. Evidently, no intrinsically geometric objects or magnitudes are involved in this activity. However, these are readily constructed to ease understanding of the situation. A word $w$ is thought of as an arrow vector $\vec w =(w_0,w_1,\ldots,w_{n-1})$ and a distance function, called Hamming metric, is defined for any pair of words $w$ and $x$ as $\max\{i\ |\ w_i\neq x_i, 0\leq i\leq n\}$, the maximum number of differing components between $\vec w$ and $\vec x$. After certain assumptions about the nature of the channel, a received word $w$ is corrected searching for the predefined codeword $c$ that is \emph{closest} to it in the above sense, i.e.\ differing from a codeword in the least number of elements.

For some physical magnitudes, however, this underlying process of geometric objectification is nearly nonexistent. Mass and charge, for example, appear as exceptions of physical magnitudes that our minds do not seem to find useful (at least not yet) to theoretically represent as the extension of any specific geometric object. This notwithstanding, one may (and in practice usually does) still view these magnitudes simply as the extension, or measure, of segments of an underlying metric line, in analogy with a timeline or graduated tape. From this perspective, any physical magnitude may be said to be a geometric magnitude. 

\begin{figure}
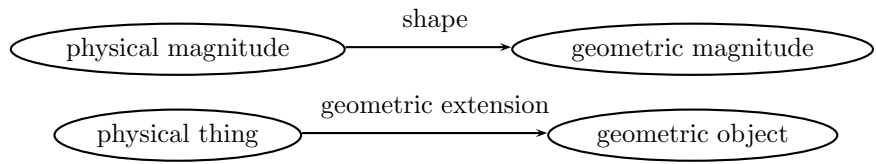

\centering
	\begin{psmatrix}[rowsep=0.4,colsep=2.2]
    \psovalbox{{\small physical magnitude}} & \psovalbox{{\small geometric magnitude}} \\
    \psovalbox{{\small physical thing}} & \psovalbox{{\small geometric object}}\\
    \ncline{->}{1,1}{1,2}^{\small {shape}}
    \ncline{->}{2,1}{2,2}^{\small{geometric extension}}
    \end{psmatrix}
\caption[Geometrization processes]{Two processes of geometrization in physics. A physical magnitude resulting from a measurement becomes a geometric magnitude when shape (geometric object) is attached to it; or a physical thing becomes a geometric object when geometric extension (length \& co.) is attached to it.}
\label{geometrization-process}
\end{figure}

In physical practice, geometrization can also take place in a different way. Instead of starting with a physical magnitude that becomes a geometric magnitude via the attachment of an underlying shape, it is sometimes the case that theoretical portrayal starts with an abstract object---possibly inspired directly in experience, in which case we have a physical thing---without any shape to speak of, but which is turned into a geometric object by attributing to it geometric extension in the form of length and its kin (distance, area, volume, overlap). We shall see examples of this shortly in connection with quantum theory, and later on in connection with pregeometry. Both types of process are schematized in Figure~\ref{geometrization-process}.

As a result of these processes of geometrization, we witness today a language of physics where geometric concepts are ubiquitous, as they have been since the birth of science, when Galilei \citeyear{Galilei:1957} expressed that without geometric concepts ``it is humanly impossible to understand a single word of [the book of nature]''; that without them ``one wanders about in a dark labyrinth'' (p.~238). Today as then, if one tried to find a physical theory that does not include geometric conceptions, one would be bound for certain failure.  

\begin{figure}
\begin{center}
\includegraphics[width=\linewidth]{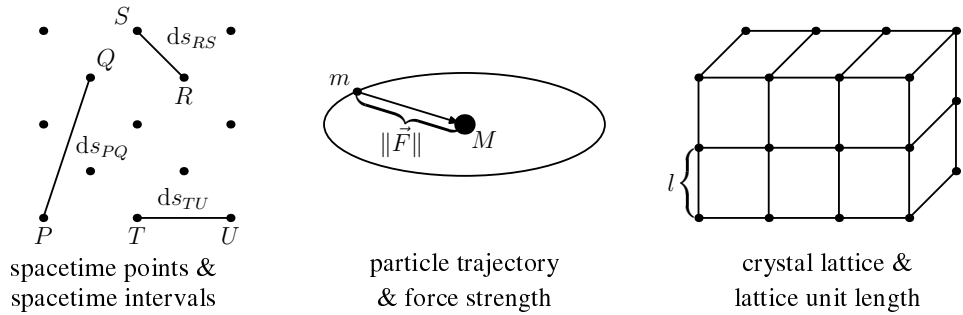}
 \end{center}
\caption[Geometric objects and magnitudes in physics]{Some geometric objects and associated geometric magnitudes used in physics.} \label{gogmip}
\end{figure}

For example, on the qualitative side, we find geometric objects such as spacetime \emph{points}, straight and circular \emph{trajectories}, \emph{arrow} state vectors, planetary \emph{surfaces}, crystal \emph{lattices}, and so forth. On the quantitative side, we find geometric magnitudes such as spacetime \emph{intervals}, \emph{lengths} of vectors (e.g.\ strength of a force), \emph{areas} of surfaces (e.g.\ probability of a quantum-mechanical outcome), volumes of spaces (e.g.\ probability of a classical outcome), and so on (Figure \ref{gogmip}). Although all the geometric concepts of physical science appear immediately familiar to us on account of our bend of mind, their level of detachment from the things we can perceive can range greatly. Some are developed from straightforward observations, whereas others bear a high level of abstraction. The farther detached the concept is from what we can directly observe and experiment with, the more troublesome the ascertainment of its physical meaning becomes (cf.\ spacetime points and state vectors). This, in turn, can lead to philosophical problems about their ontology---in other words, to considerable confusion.

\begin{figure}
\begin{center}
\includegraphics[width=90mm]{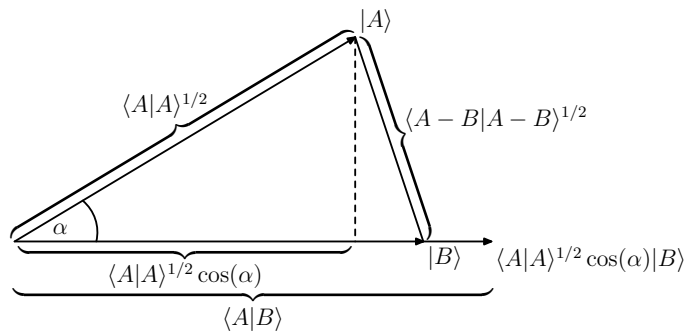}
 \end{center}
\caption[Length, distance, and inner product]{Four quantitative geometric notions: length of arrow vector $|A\rangle$, projection of $|A\rangle$ on arrow vector $|B\rangle$, distance between $|A\rangle$ and $|B\rangle$, and inner product between $|A\rangle$ and $|B\rangle$.} \label{ldip}
\end{figure}

Before moving on to the study of the geometric foundations of standard relativity and quantum theories, we note the existence of a geometric concept that is particularly noteworthy for its role in physics. We refer to the \emph{inner product}. In physical contexts, the inner product plays a much more essential role than the primitive idea of length or distance. This is partly because of a mathematical reason and partly because of a physical reason. Mathematically, the inner product provides a foundation for the emergence of three key quantitative geometric notions; in bracket notation, these are: the overlap of $|A\rangle$ and $|B\rangle$, $\langle A|B\rangle$ (with parallelism and orthogonality arising as special cases); the length of $|A\rangle$, $\langle A|A\rangle^{1/2}$; and the distance between $|A\rangle$ and $|B\rangle$, $\langle A-B|A-B\rangle^{1/2}$ (Figure \ref{ldip}). Physically, being itself an invariant scalar, the inner product provides a direct connection with measurements, the results of which are invariants as well. In physics, the representation of invariant scalars as inner products is always a sure sign of the introduction or existence of a geometric outlook upon the object of study. The advantages and pitfalls of this way of proceeding will come to light in later chapters.  

Next we analyse the geometric character of relativity and quantum theories as conventionally understood. We shall see that both theories are deeply geometric at their most elemental level, and even that the motivation behind their geometric character is similar. This view, however, seems to be at odds with common belief, as Ashtekar's statement reveals:
\begin{quote}
[W]e can happily maintain a \emph{schizophrenic} attitude and use the precise, \emph{geometric picture of reality} offered by
general relativity while dealing with cosmological and astrophysical phenomena, and the quantum-mechanical world of chance and intrinsic uncertainties while dealing with atomic and subatomic particles. \cite[p.~2, italics added]{Ashtekar:2005} 
\end{quote}
In the next two sections, we attempt to maintain an expository attitude towards the foundations of relativity and quantum theories as conventionally understood, as our present purpose is only to expose their geometric natures. In later chapters, we shall return to these topics and deal with them from a  more critical point of view. 

\section{Geometry in relativity theory} \label{GGRel}
Historical considerations aside, we can view relativity theory as essentially built upon several elemental geometric concepts. First and foremost, the theory deals with physical events, which are idealized as \emph{points} belonging to a four-dimensional spacetime continuum embracing the past, the present, and the future. A point-event in the four-dimensional spacetime continuum is characterized analytically by means of a set of four numbers $(x^1,x^2,x^3,x^4)$ called the point-event coordinates.     

The theory also deals with massive particles and photons, which it also idealizes as points; a particle is, moreover, a moving point that by its motion in the spacetime continuum determines a sequence of events called a \emph{worldline}, $x^\alpha=x^\alpha(u)$ $(\alpha=1,2,3,4)$, where $u$ is a one-dimensional parameter. The idea of the worldline of a massive particle also includes the idealization of the observer as a massive point that moves while making observations.

On these foundations, relativity theory builds its conceptual connection with physical measurements. This is done by means of the central concept of the differential spacetime interval $\D s$. The differential interval $\D s_{AB}$ gives a quantitative measure of the separation in spacetime between two neighbouring events $A$ and $B$. It is traditionally understood that $\D s$, being a spacetime interval, is measured with rods and clocks. Because time is thought as absorbed as a fourth coordinate $x^4$ on a par with the three spatial coordinates $x^1,x^2,x^3$ (with the speed of light $c$ as a transformation factor), the interval is conventionally regarded geometrically, as a \emph{generalized distance} between neighbouring events. The differential interval is also called the \emph{line} element; this is a denomination inspired in the geometric picture of two general events joined by a worldline composed of \emph{straight} pieces $\D x$, whose lengths $\D s$ add up to the finite separation $\Delta s$.

In relativity theory, the assumption is made that $\D s_{AB}$ depends analytically on the coordinates $x^\alpha$ of point-event $A$ and homogeneously (of first degree) on the displacement vector, or \emph{arrow}, $\D x^\alpha$ form $A$ to $B$; i.e.\ $\D s=f(x,\D x)$, where $f(x,\D x)$ is some function homogeneous in $\D x$, and $x$ denotes the set of four coordinates. Besides the demand of homogeneity, the form of $f(x,\D x)$ is not entirely free because the scalar $\D s$, being a measurable quantity, must be an invariant. As we mentioned previously, this situation suggests the use of an \emph{inner product}. The new assumption is now made that the relation between $\D s$, on the one hand, and $x$ and $\D x$, on the other, is an indefinite quadratic form, namely, $\D s=\sqrt{\epsilon g_{\alpha\beta}(x)\D x^\alpha \D x^\beta}$,\footnote{Here $\epsilon=\pm 1$ ensures that $\D s\geq 0$ always, since the quadratic form is not positive definite.} i.e.\ the line element is an invariant inner product between the displacement vectors and some function $g_{\alpha\beta}(x)$ of $x$. This quadratic form is in fact the foundation on which the rest of relativity theory is built. (But to what kind of theory would the simpler linear choice $\D s=g_\alpha(x)\D x^\alpha$ lead us? This is a question we shall ponder later on.)

A study of Riemannian space, in conjunction with its generalization to spacetime, allows one to determine that the function $g_{\alpha\beta}(x)$ is a second-order covariant tensor with a clear geometric meaning: it is the \emph{metric} of the spacetime continuum, and as such the formal (mathematical) source of the spacetime-continuum quantitative geometry. The metric is apt to inform us how much two vectors (in particular, two contravariant displacement vectors) at the same event $A(x)$ \emph{overlap}; on this basis, we can find the generalized length, or \emph{norm}, of a vector at $A(x)$, the generalized distance between two vectors at $A(x)$, and, via integration, the separation or generalized distance between any two events.

The Newtonian law of inertia is generalized in relativity theory via Einstein's geodesic hypothesis. According to it, a free (massive or massless) particle follows a \emph{geodesic} in spacetime, i.e.\ a \emph{straight line} in spacetime, such that the separation $\Delta s$ between two events joined by this line is maximized. On the other hand, the intrinsic shape of a spacetime geodesic followed by a free particle is given by an absolute property of spacetime, its \emph{curvature}. 

Curvature is set in a one-to-one correspondence with gravitation as a phenomenon as follows. Absolute spacetime curvature is characterized, not simply by some tensor (e.g.\ $g_{\alpha\beta}(x)$), but by some suitable tensor equation (e.g.\ $R^\delta_{\ \alpha\beta\gamma}(x)=0$) to be specified later on. When this equation vanishes identically ($R^\delta_{\ \alpha\beta\gamma}(x)\equiv 0$), i.e.\ when the tensor is null everywhere, spacetime is \emph{flat} and there are no gravitational effects; when this equation does not vanish identically ($R^\delta_{\ \alpha\beta\gamma}(x)\not\equiv 0$), i.e.\ when the tensor is not null everywhere, spacetime is \emph{curved} and there are gravitational effects.   

In view of the pervasiveness and essentialness of geometric notions in relativity theory, it seems understandable that this theory has come to be regarded as epitomically geometric, and ``more geometric'' than its Newtonian predecessor. But is this really so? Has general relativity truly geometrized gravitation? Is physics more geometric now after general relativity? Not at all. With general relativity, gravitation only acquires a \emph{new} geometric character. New because the Newtonian description of gravitation as a vectorial force was also a geometric concept, namely, a directed \emph{arrow} in Euclidean space with a notion of \emph{size} (strength) attached, acting between \emph{point-like} massive bodies a certain Euclidean \emph{distance} apart. 

In the current climate of misunderstanding regarding the essential lack of novelty introduced by general relativity as a geometric theory, it is not without a measure of surprise that we find that a correct assay of the unoriginal geometric status of relativity was made by its own creator over half a century ago. In a letter to Lincoln Barnett dated from 1948, Einstein considered as follows:
\begin{quote} \label{Einstein-field}
I do not agree with the idea that the general theory of relativity is geometrizing Physics or the gravitational field. The concepts of Physics have always been geometrical concepts and I cannot see why the $g_{ik}$ field should be called more geometrical than f.[or] i.[nstance] the electromagnetic field or the distance between bodies in Newtonian Mechanics. The notion comes probably from the fact that the mathematical origin of the $g_{ik}$ field is the Gauss-Riemann theory of the metrical continuum which we are wont to look at as a part of geometry. I am convinced, however, that the distinction between geometrical and other kinds of fields is not logically founded.\footnote{Quoted from \cite[p.~ix]{Earman/Glymour/Stachel:1977}.} 
\end{quote}

Having disclosed the geometric essence---but not geometric novelty---of relativity theory, but having left its physical foundations as yet unanalysed, we now turn our gaze to quantum theory and \emph{its} geometric essence.
 
\section{Geometry in quantum theory} \label{GQMec}
Quantum theory studies the measurable properties of microscopic physical systems, such as microscopic material particles. Unlike relativity theory, quantum theory does not find it essential to geometrically conceptualize particles as points, since the classical trajectory, or worldline, of a particle plays no role in it. Geometric concepts have in quantum theory a different physical source---yet the same formal expression motivated by needs similar to those of relativity theory.

In order to ascertain the properties of a microscopic system, one must set up measurements, which soon reveal that these systems display a non-classical feature of superposition. We study as an example the measurement of the polarization of photons. Following Dirac \citeyear[pp.~4--7]{Dirac:1958}, set up a plate which only transmits light polarized perpendicularly to its optical axis, and direct at it a low-intensity beam of light, polarized at an angle $\theta$ with the above axis, such that photons arrive one at a time. One finds that the intensity of transmitted light, and therefore of transmitted photons, is a fraction $\sin^2(\theta)$ of the total intensity, suggesting that polarization cannot be a classical property in the following sense: it cannot correspond to a tiny material stick attached to photons that is blocked by the plate unless aligned perpendicularly to its optical axis, for then no photons could pass through. Indeed, against classical expectations, some intensity is perceived, signalling the unimpeded passage of a fraction of the original photons: on arrival at the plate, each photon reorganizes itself polarization-wise as either compatible or incompatible with it.\footnote{A display of the non-classical nature of polarization can be produced in a purely phenomenological---and therefore much more striking---way. Place two plates with optical axes orthogonal to each other, so that no light will be perceived through both plates. Insert then a third plate obliquely between the previous two and observe now some light passing through the three plates. The addition of an \emph{extra filter} allows \emph{more light} to pass through---a phenomenon that goes against classical intuition.} Moreover, because the interaction at the polarization plate acts, to all intents and purposes, like a black box, one cannot predict what the fate of each individual photon will be with certainty, but only with a certain probability. This probability is an invariant scalar.

The task of quantum theory is to provide a link between this probabilistic experimental result and some theoretical feature attributable to each microscopic photon. On the basis of the said non-classical behaviour, quantum theory pictures each photon to be in a polarization state $|P\rangle$, such that it is a complex-number (see below) linear superposition  
\begin{equation}\label{eq:linear-superposition}
|P\rangle=c_t |P_t\rangle+c_a |P_a\rangle 
\end{equation}
of two polarization states, $|P_t\rangle$ and $|P_a\rangle$, associated with transmission and absorption with respect to a given plate. The superposition feature of $|P\rangle$ leads directly to the suggestion that quantum-mechanical states are vectors, since it holds physically that the linear combination of two states is also a state, but this is not to say (yet) that state vector $|P\rangle$ is a geometric object. This interpretation, however, provides a ready-made foundation for the introduction of geometry into quantum theory, which takes place in two steps.

For the construction of $|P\rangle$ as a linear superposition to make sense and be useful, the complex coefficients $c_t$ and $c_a$ should indicate (qualitatively at least) to what degree, on arrival at a given plate, the original state $|P\rangle$ is compatible with the one associated with transmission $|P_t\rangle$ and to what degree with the one associated with absorption $|P_a\rangle$, respectively. A geometric picture is immediately created in which state vectors are imagined as \emph{metric vectors} or \emph{extended arrows}, and their mutual compatibilities expressed in terms of their \emph{overlaps} or, in other words, the \emph{inner product}. The linear superposition of Eq.~(\ref{eq:linear-superposition}) is thus recast geometrically as 
\begin{equation}
|P\rangle=\langle P_t|P \rangle |P_t\rangle+\langle P_a|P \rangle |P_a\rangle,
\end{equation}
with $|P_t\rangle$ and $|P_a\rangle$ imagined as mutually \emph{orthonormal} and with unit arrow vector $|P\rangle$ imagined as being \emph{projected} or collapsed onto them. The inner product effectively turns in principle shapeless state vectors into extended arrows, because the idea of overlap induces a need for geometric objects in order for us to project onto one another (but why should we overlap arrows and not surfaces or spatial figures?). The inner product being a complex number, however, we cannot yet talk of geometric magnitudes in a physical sense.

\begin{figure}
\begin{center}
\includegraphics[height=66mm]{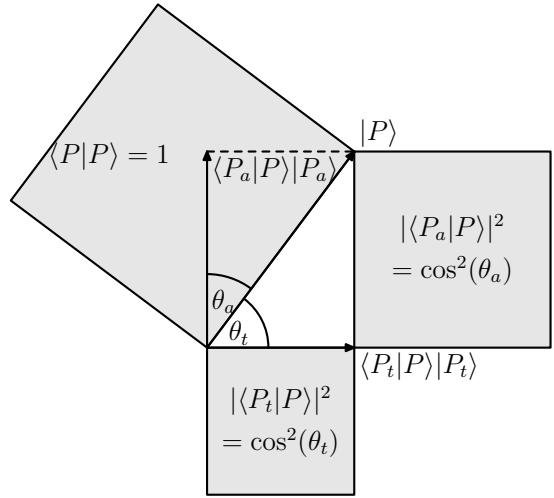}
 \end{center}
\caption[Probablity from Pythagoras' theorem]{Rise of quantum-mechanical probabilities from Pythagoras' theorem.} \label{ppt}
\end{figure}

The second step in the geometrization of quantum theory comes from the way it  is linked to the experimentally obtained probability. On the one hand, probability is a real scalar that should arise from some invariant function $f(|P\rangle,|P_t\rangle,|P_a\rangle)$ of state vectors; on the other hand, state vectors so far only combine to produce complex numbers\footnote{As justification for the significant but otherwise arbitrary requirement that the coefficients $\langle P_i|P \rangle$ be complex numbers, Dirac \citeyear[p.~18]{Dirac:1958} offered the following explanation. Within the context of the polarization experiment above, he argued that, given that the length of state vectors is irrelevant (normalization condition), in a linear superposition $c_1 |\psi\rangle +c_2|\phi\rangle$ of two states only the ratio $c_3=c_2/c_1$ carries a meaning, and so $c_3$ must be complex (two real parameters), i.e.\ $c_1$ and $c_2$ must be complex, since two independent real numbers are necessary to specify a general state of plane polarization. But why should two complex numbers (four real coefficients) be needed in a superposition of three linearly independent state vectors, etc.? Another proposal for the need of complex numbers was given by Penrose \citeyear[pp.~62--63]{Penrose:1998}. He stressed the feature of complex conjugation between a complex number $\alpha$ and its conjugate $\alpha^*$ in connection with quantum information and its relation to classical information. He argued that the existence of quantum information, which does not respect causality, is only possible via the existence of complex amplitudes and their conjugates, which can be interpreted as going forwards and backwards in time: $\langle P_i|P\rangle$ from the present state $|P\rangle$ to the future state $|P_i\rangle$ and $\langle P_i|P\rangle^\ast=\langle P|P_i \rangle$ from the future state $|P_i\rangle$ to the present state $|P\rangle$. One the other hand, classical information---in other words, probability---results from multiplying the complex amplitude $\alpha$ by its complex conjugate $\alpha^*$, obtaining thus a real number. In Chapter \ref{ch:MQM}, we give a deeper justification of why complex numbers are needed to describe quantum phenomena.} $\langle P_i|P \rangle$ ($i=t,a$). Further geometric analysis succeeds in finding common ground. On the basis of Pythagoras' theorem, one can see that the three state vectors $|P\rangle,|P_t\rangle,$ and $|P_a\rangle$ are connected by a real-number relation, 
\begin{equation}
\sum_i |\langle P_i|P \rangle|^2=\||P\rangle\|^2,
\end{equation} 
so long as we think them as arrows (but not surfaces or spatial figures). Now, because the \emph{length} of $|P\rangle$ is a geometric invariant and equal to unity, this relation becomes suitable for a probabilistic interpretation. The probability of each outcome $|P_i\rangle$ is given by the squared length, or \emph{area}, of the projection of $|P\rangle$ onto $|P_i\rangle$, and the probability of obtaining any outcome adds up to unity (i.e.\ a unit area), as shown in \mbox{Figure \ref{ppt}.}

It is in this manner that probability, an invariant\footnote{By quantum-mechanical probability being an invariant, we mean that it is an absolute property of a large ensemble of identically prepared microscopic systems (relative to the tests measuring apparatuses submit them to). The denomination ``invariant'' is used here in order to highlight the independence of measurable quantum-mechanical probability from the unobservable state vectors from which it traditionally arises. The name is also used to draw attention to the similar conceptual way in which both quantum and relativity theories introduce the invariant results of measurements via the inner product of entities deemed more elemental from a logico-axiomatic---but not physical---perspective. The denomination, however, is not intended to hint at a relativistic connection. For more on the relativistic interpretation of a single quantum-mechanical measurement, see \cite[pp.~4, 9]{Eakins/Jaroszkiewicz:2004}.} scalar that only has to do with counting positive and negative outcomes, is geometrized on the basis of the inner product and Pythagoras' theorem (in general, in $n$-dimensions) in order to provide the human mind with geometric understanding of, in principle, geometry-unrelated observations. Pythagoras' theorem comes thus from its humbler Neolithic uses to serve human understanding in yet another facet of Man's enquiries into nature---to instate, in this case, Sumner's ``sacred geometry of chance.''

\chapter{New geometries}
\label{ch:New geometries}
\begin{flushright}
\begin{tabular}{p{10cm}} 
\emph{It would be fascinating if, ultimately, we build all the fundamental things in the world out of geometry.} \smallskip \\
Interviewer to Abdus Salam, \emph{Superstrings: A theory of everything?} 
\end{tabular}
\end{flushright}

\bigskip

In the light of the previous chapter, it is not surprising to find that physics is not only built out of geometry, but also that it seeks progress through the invention of new geometries. Ptolemy's geocentric system of spheres was improved upon by Copernicus' heliocentric system of spheres, in turn replaced by Kepler's ellipses. Newton's Euclidean space and time structure were improved upon by Minkowski's flat spacetime structure, in turn replaced by Riemann and Einstein's curved one. In the maiden field of quantum mechanics, headway was made as the early collection of scattered results was set in a progressively more intuitive geometric framework: from Heisenberg-Born-Jordan matrix mechanics, passing through Schr\"{o}dinger's wave mechanics, and ending with Dirac's bracket algebra, whose geometric features we examined in the previous chapter.

In early attempts at unification of gravitation and electromagnetism, Weyl, Eddington, Kaluza, Klein, and Einstein and collaborators (Mayer, Bergmann, Bargmann, Strauss) attempted to improve on general relativity through the introduction of new geometries: generalized Riemannian geometry, affine geometry, five-dimensional spacetime, a non-symmetric metric tensor, teleparallelism, five-vectors on four-dimensional spacetime, bivector fields, a complex-valued metric tensor, etc.\footnote{See \cite[pp.~98--101, 115--148]{Goenner:2004,Kostro:2000}.} This fruitless quest for new geometries aiming to overthrow the reigning order in spacetime physics continues today unimpared in quantum-gravity research programmes, as follows.

According to unanimous agreement by decree in the physics community,\footnote{See e.g.\ \cite[pp.~10--11]{Chalmers:2003,Smolin:2001}.} there are three roads to quantum gravity. The first road is the road of string theory, which seeks progress through the geometric concepts of strings, membranes, deca- and endecadimensional metric spacetimes, etc. The second road is the road of loop quantum gravity, which seeks progress through the geometric concepts of loops of space, spin networks, spin foams, etc. And the third road to quantum gravity is the politely acknowledged road of everything else---a waste dump, as it were---that does not fit the first two characterizations and that, beyond politeness, one gets the impression,\footnote{See e.g.\ \cite{Chalmers:2003} and \cite[p.~37]{Rovelli:2003}.} can be safely ignored. Pursuing this less glittering third road, we can mention pregeometricians, who, inspired in Wheeler, also continue to further the same geometric trend (only now unwittingly) in the search for less common keys to a quantum theory of space and time.

It is also customary in any of these fields for yet \emph{newer} geometries to be readily invoked to explain any difficulties that their tentative \emph{new} geometries, attempting to explain well-established \emph{older} ones, might have given rise to. Woit \citeyear{Woit:2002} exemplifies this state of affairs when he writes, ``String theorists ask mathematicians to believe in the existence of some wonderful new sort of geometry that will eventually provide an explanation for M-theory'' (p.~110). That is to say, in the face of difficulties with the tentative new geometries of five distinct string theories, researchers naturally resort to a yet newer, more convoluted geometry on which to base so-called M-theory, and with the help of which problematic issues might still be resolved. The essence of the present situation is captured rather accurately by Brassard \citeyear{Brassard:1998}, when (for his own reasons) he remarks that ``The most difficult step in the creation of a science is\ldots the finding of an appropriate geometry. It is the difficulty that is facing the physicists who are trying to merge quantum physics with general relativity\ldots'' ($\S$ 3.4).

May this procedure of fitting nature into ever newer geometric moulds, essentially physically unenlightening since the establishment of general relativity,\footnote{The geometric formulation of quantum mechanics, which reached its full form with Dirac's bracket formulation, is not here considered precisely speaking enlightening, since the present controversies of quantum theory stem from a lack of physical meaning in its geometric concepts, as we shall see later on.} then reach a natural end? This question must be answered in the negative so long as the human mind continues to be exclusively gripped, as it is, by geometric modes of thought, remaining oblivious to other manners of theoretical reflection. But what would be the symptoms of a condition in which the physical problems that confront us---whether pertaining to quantum theory, spacetime theories, or both---were immune to our relentless geometric attacks? Trapped by our own conceptual tools of thought, we would witness (i) inexhaustible philosophical argument over old topics cast in the same recurring geometric language, and (ii) increasingly complicated, mathematically or philosophically inspired frameworks built out of geometrically ``elegant,'' ``self-consistent'' concepts yet without physical referents. Such symptoms are, in fact, already visible; the first category is populated by self-perpetuating philosophical discussions on Einstein's hole argument and spacetime ontology (substantivalism vs.\ relationalism, the meaning of spacetime points, etc.), as well as interpretational debate on the meaning of quantum-theoretical geometric concepts (the state vector and its collapse, probability, etc.); and the second category is inhabited by much of what presently goes by the name of quantum gravity. 

Is it reasonable to continue along a tangled geometric path when it gets more twisted the farther it is trodden? Are our developed geometric instincts leading us in the same ominous direction that the geometric instincts of the Aristotelian tradition led the Ptolemaic geocentric system of perfect spheres? For over a millennium since Ptolemy, an army of spheres interlocking with ever-mounting complexity was a price gladly paid to preserve an astronomy based on the philosophical opinion that spheres are perfect (and the earth all-important) yet ``fraught with cosmological absurdities'' \cite[p.~364]{Saliba:2002}. Confronted by so much absurd complexity, Ptolemy excused it as natural, considering that nothing short of understanding the mind of God was in question here:
\begin{quote}
Now let no-one, considering the complicated nature of our devices, judge such hypotheses to be over-elaborated. For it is not appropriate to compare human [constructions] with divine, nor to form one's beliefs about such great things on the basis of very dissimilar analogies.\footnote{Quoted from \cite[p.~366]{Saliba:2002}.}
\end{quote}
The layman's point of view, when he dares to voice it unintimidated by the proficiency of the expert, is usually closer to the truth. In the thirteenth century, King Alfonso X of Spain was much less impressed by Ptolemaic astronomy than Ptolemy himself. ``If the Lord Almighty had consulted me before embarking upon Creation,'' remarked the king (perhaps apocryphally), ``I should have recommended something simpler.''

The story of Aristotelian-Ptolemaic astronomy strikes a resonant chord with quantum gravity. Saving the distances, it warns against blindness towards our geometric biases, whether innate or inherited from a philosophy built upon opinion, as well as against the rationalization of complexity in the name of the fundamentality of the task at hand and the inexorability of geometry in dealing with it.

How far is quantum gravity ready to go, and how much to sacrifice, to save its geometric prejudices?  

\section{Three roads to more geometry}
String theory does not find fault with its thoroughly geometric character---and, in the context of a paradigmatically geometric science, why should it?---most elementally given by the very concept of a string and the multidimensional metric spacetimes to which they belong. What is more, its more vehement proponents predicate the correctness of the theory on a sense of ``elegance'' and ``beauty'' based on features of its geometric portrayal that, by their very geometric nature, tend to be psychologically appealing. In his revealing article, Gopakumar \citeyear{Gopakumar:2001} ponders the ``almost mystical'' (p.~1568) connection between geometry and the laws of nature and, embracing it, correctly describes progress in string theory as the evolution and growth of its geometric edifice, together with the geometric insights to be found in its newly built storeys. He also foretells---and we are likely to agree with him in view of current trends---more ``innovations in geometry'' (p.~1572) in the progressive reformulation of string theory. The desirability of furthering this geometric trend in string theory (and in physics as a whole) is perhaps best epitomized in the almost mystical words of an anonymous reporter during an interview to Abdus Salam: 
\begin{quote}
Clearly string theory is deeply rooted in geometry. I suppose one could say that science started with geometry (if one goes back to ancient Greece). It would be fascinating if, ultimately, we build all the fundamental things in the world out of geometry.\footnote{In \cite[p.~178]{Davies/Brown:1988}.} 
\end{quote}

These strong views notwithstanding, and as in other quantum-gravity endeavours (see below), even in expositions of an admittedly geometric theory such as this one, one puzzlingly witnesses isolated, casual allusions to the possible limitations of geometric theorizing. Gopakumar asks, ``Are there geometrical structures than can replace our conventional notions of space-time? Or perhaps geometry emerges only as an approximative notion at distances large compared to the Planck length?'' (p.~1568). Lacking appropriate context or further explanation, one can only attribute a vague and ambiguous meaning to ``geometry'' in this type of remarks. If by ``geometry'' is meant \emph{any} form of geometric concept, then quantum gravity answers the former question with a resounding yes, and the latter with a resounding no; but if by ``geometry'' is meant simply a continuous manifold, then quantum gravity answers the former question most probably with a no, and the latter most probably with a yes.

In loop quantum gravity, on the other hand, the standpoint regarding geometry is much more confusing. Its proponents not rarely but \emph{often} claim that ``the very notion of space-time geometry is most likely not defined in the deep quantum regime'' \cite[p.~17]{Perez:2006}, or that a ``quantum gravity theory would\ldots be free of a background space-time geometry'' \cite[p.~11]{Ashtekar:2005}, but we then repeatedly witness expositions of loop quantum gravity as a straightforward quantum theory of geometry.\footnote{See e.g.\ \cite[p.~13]{Ashtekar:2005}, \cite[pp.~75--191]{Smolin:2001} and \cite[p.~262]{Rovelli:2004}.} Despite their lack of clarity, what these phrases intend to mean is simply that, after the lessons of general relativity, one should not assume an absolutely fixed background spacetime in quantum gravity, but not that there is no \emph{background geometry} in the loop-quantum-gravity quantization of general relativity; else witness the thoroughly geometric formulation of loop quantum gravity in every exposition of it: from  elemental arrow-vector triads\footnote{Because of its inviability for quantization, talk of the metric field is at first banished and instead supplanted by talk of a (mathematically disembodied) ``gravitational field'' as an elemental physical concept. But because loop quantum gravity desperately needs a mathematical field in any case, a new mathematical concept---rechristened ``gravitational field''---appears at the starting point of the theory, namely, triads of arrow vectors $E_I^a$ ($a, I=1,2,3$) whose trace $\mathrm{Tr}(E_I^a E_I^b)=\det(g)g^{ab}$ (an \emph{inner product} on the internal indices $I$) mathematically recreates the previously banished metric field.} to the celebrated ``quantum geometry'' of spin networks in the form of their quantized area and volume. Like string theory, then, loop quantum gravity also progresses via the invention of new geometries. 

At this point, it is worthwhile noting that some loop-quantum-gravity scholars do not only deny the evident geometric foundations of their field but, more seriously, also those of general relativity (and by implication, should one understand, of loop quantum gravity?). Smolin \citeyear{Smolin:1997} explicitly claims that ``while the analogy between spacetime, as Einstein's theory describes it, and geometry is both beautiful and useful, it is only an analogy'' with ``little physical content'' (p.~333) and that
\begin{quote}
The metaphor in which space and time together have a geometry, called the spacetime geometry, is not actually very helpful in understanding the physical meaning of general relativity. The metaphor is based on a mathematical coincidence that is helpful only to those who know enough mathematics to make use of it\ldots \cite[p.~59]{Smolin:2001}
\end{quote}
Smolin's words are inspired in the general-relativistic feature of diffeomorphism invariance, according to which the spacetime metric $g_{\alpha\beta}(x)$ has no absolute meaning, as its values change from one coordinate system to another. Geometrically speaking, $g_{\alpha\beta}(x)$ has no absolute meaning because spacetime points $P$ (cf.\ $x$) are not physically real and fail to individuate the metric field. And this is correct, but to infer that general relativity is not essentially a geometric theory because the metric field has no absolute physical meaning is utterly mistaken---a confusion caused by the connotation of words. While tensors are relative to a coordinate system, tensor \emph{equations}, however, are not. The reason $g_{\alpha\beta}(x)$ plays no physical role in relativity theory is that it does not as such satisfy any tensor equation. Riemann's curvature tensor $R^\delta_{\ \alpha\beta\gamma}(x)$, on the other hand, does satisfy a tensor equation, Einstein's field equation, and thus has absolute (i.e.\ independent of coordinate system) meaning as a geometric property of spacetime, namely, its \emph{curvature}. 

We showed the geometric foundations of relativity in the previous chapter, and we  shall still say more about this issue later on, but for the moment let us stress that the role of geometry in general relativity is neither an analogy, nor a metaphor, nor a mathematical coincidence, nor devoid of physical content. Even at its more basic level, general relativity is essentially about a geometric network of invariant intervals $\mathrm{d}s_{AB}$ (\emph{norm} of the differential displacement \emph{arrows} $\D x_{AB}$) between \emph{point}-events $A(x)$ and $B(x+\D x)$, and, as for its physical content, we could not make any theoretical sense of the physical measurements $\D s$ without this underlying geometric structure. 

The helpfulness of the geometric foundations of the theory for understanding its physical meaning is therefore enormous; in this regard, we can perhaps do no better than heed the words of its creator, who, during an address to the Prussian Academy of Sciences in 1921, openly acknowledged:  
\begin{quote}
I attach special importance to the view of geometry which I have just set forth, because without it I should have been unable to formulate the theory of relativity. Without it the following reflection would have been impossible:---In a system of reference rotating relatively to an inert system, the laws of disposition of rigid bodies do not correspond to the rules of Euclidean geometry on account of the Lorentz contraction; thus if we admit non-inert systems we must abandon Euclidean geometry. \mbox{\cite[p.~33]{Einstein:1983b}}
\end{quote}
And during his Kyoto lecture a year later in 1922, Einstein \citeyear{Einstein:1982} expressed words to the same effect: ``I found that that the foundations of geometry had deep physical meaning in this problem'' (p.~47). Moreover, neither is it true that the geometric formulation of general relativity is ``helpful only to those who know enough mathematics'' since, as has been made plain time and again, Einstein was led to this geometric formulation \emph{despite} not knowing enough mathematics to express his physical insights in mathematical terms. 

Finally, not even Smolin's \citeyear{Smolin:1997} claim that he has himself ``described general relativity while making no mention of geometry'' (p.~333) is true, as his own description of this theory is in terms of ``the geometry of space and time'' (p.~234) and ``the metric of spacetime'' (pp.~235, 239). Smolin's remarks leave one feeling uneasy with a bitter aftertaste: if correct, why introduce such revolutionary ideas only as a short endnote or in passing in his books? Can one really do away with 90 years of physical understanding since Einstein, and with nearly 200 years of mathematical development relevant to relativity theory since Gauss and Riemann, \emph{in a footnote}? This can only be the benefit of speaking from an authoritative position.

Finally and most unexpectedly, also pregeometricians attempt to make progress via the invention of new geometries. By design, pregeometry is the attempt to explain the origin of geometry in physical theories from a conceptually different basis, like a material continuum finds explanation in terms of atoms. The use of geometric concepts is thus proscribed if pregeometry is to live up to its intended goal. Unwittingly, however, because physics knows of no other mode of operation than geometry, all ideas for pregeometries (excepting a few of Wheeler's) turned out to be fraught with geometric objects and magnitudes: graphs, networks, lattices, and causal sets with metrics determining the \emph{distances} of the \emph{edges} between their \emph{point}-like constituents. In the erroneous belief that mere engagement in pregeometry would turn its products pregeometric, the geometric nature of these building blocks went unnoticed, and the task of pregeometry was tacitly reinterpreted as the search for the origin of the geometric spacetime \emph{continuum} from a \emph{discrete} geometric foundation. The case of pregeometry is so exemplifying that it deserves a chapter of its own, to which we now turn. 

\begin{sidewaysfigure}[!h]\label{two-roads}
\begin{center} 
\includegraphics[width=0.94\linewidth]{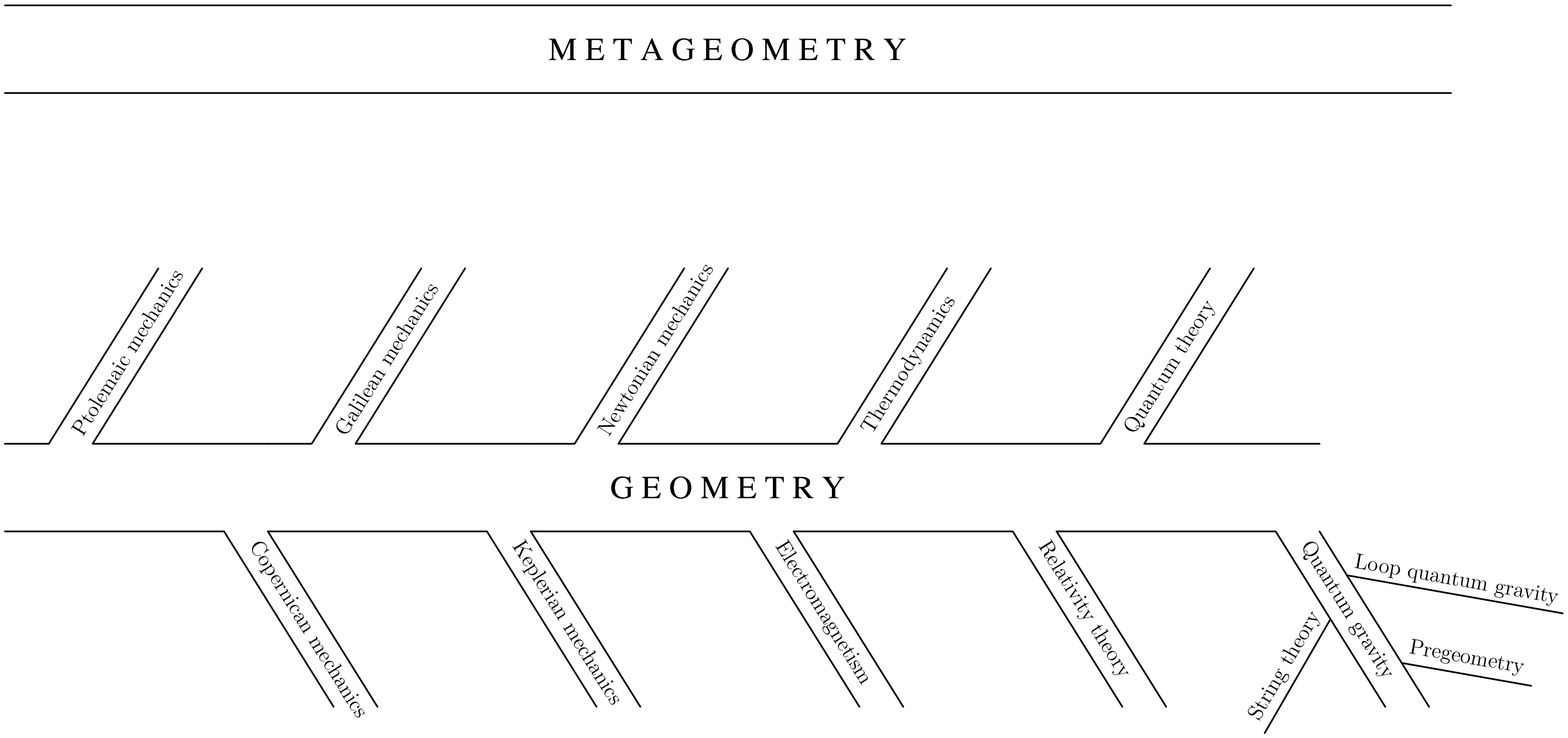}
\end{center}
\caption[Geometric and metageometric roads]{The language of physics is the language of geometry, and progress in physics has always been sought, whether successfully or not, via the invention of new geometries. This is still the case nowadays with the latest endeavours in the field of quantum gravity. We submit there exists a different, unexplored road, the treading of which can help us understand the physical world in ways deeper than any amount of further branching of the geometric road is capable of. This is the road of metageometry.}  
\end{sidewaysfigure}

We close this chapter with an illustration (Figure \ref{two-roads}) of the geometric road physics has followed so far, and suggest an unexplored alternative.
 
\chapter{Pregeometry}
\label{ch:Pregeometry}
\begin{flushright}
\begin{tabular}{p{10cm}}
\emph{Curious things, habits.\ People themselves never knew they had them.}
\smallskip \\
Agatha Christie, \emph{Witness for the prosecution}
\end{tabular}
\end{flushright}

\bigskip

After having dwelt in the meaning of geometry and witnessed its ubiquitous role in physics and attempts at new physics, we analyse those theoretical pictures that have explicitly proposed to forgo geometric ideas in their depiction of space and time, and which go by the name of \emph{pregeometry}---a very apt name indeed, for the word is evocative of positioning oneself before geometry. In pregeometry, the geometric spacetime continuum is considered a flawed physical picture, and it is attempted to replace it without assuming any other type of manifold for spacetime (in particular, without quantizing general relativity as a presupposed starting point); instead, spacetime is to be depicted via non-geometric constructs ontologically prior to the geometric ones of its present description. 

Indeed, if, like us, one is interested in going beyond geometry in physics, what better place than in pregeometry to search for the realization of this goal? Has not pregeometry, after all, already argued in earnest for the overthrow of geometric explanations, and even fulfilled the very goals that we have set ourselves in this work? Have we suddenly realized, on reaching Chapter \ref{ch:Pregeometry}, that the work we have embarked on is at best superfluous and at worst obsolete?

Hardly. A critical study soon reveals that the above preliminary characterization of what is \emph{called} pregeometry\footnote{See e.g.\ \cite[pp.~11--18]{Monk:1997}.} is for the most part incongruous with its actual archetypes. That is to say, despite its claims to the contrary, pregeometry surreptitiously and unavoidably fell prey to the very mode of description it endeavoured to evade, as evidenced by its all-pervading geometric pictures. Indeed, the distinctive feature of pregeometric approaches is that, as intended, they drop the assumption of spacetime being correctly described by a metric continuum, but \emph{only to replace it with some type or other of geometric-theoretical conception}.

Nothing supports this statement more dramatically and vividly than the words of Gibbs' \citeyear{Gibbs:1996}, who explains the mode of working of pregeometry thus: ``Once it has been decided which properties of space-time are to be discarded and which are to remain fundamental, the next step is to choose the appropriate \emph{geometric structures from which the pregeometry is to be built}'' (p.~18, italics added).  In this use of the term, pregeometry is synonymous with pre-continuum physics, although not synonymous with pregeometry in the proper semantic and historic sense of the word---the sense its inventor, J.~A.~Wheeler, endowed it with and which its current practitioners attempt to fulfill; in terms of Wheeler's \citeyear{Wheeler:1980} two most striking pronouncements: ``a concept of pregeometry that breaks loose at the start from all mention of geometry and distance'' and ``\ldots to admit distance at all is
to give up on the search for pregeometry'' (p.~4).\footnote{The criticism to pregeometry to be put forward here is not based merely on a verbal objection. It consists of an objection to the activities of pregeometricians proper, given that they explicitly set out to avoid geometry---and fail. Since this goal is depicted very well indeed by the name given to the field, the criticism applies to the word ``pregeometry'' just as well. It is only in this sense that it is \emph{also} a verbal objection.}

On the basis of the incongruity between Gibbs' and Wheeler's statements, it is striking to come across remarks such as Cahill and Klinger's \citeyear{Cahill/Klinger:1996}, who tell us that ``The need to construct a non-geometric theory to explain the time and space phenomena has been strongly argued by Wheeler, under the name of pregeometry. Gibbs has recently compiled a literature survey of such attempts'' (p.~314). At the reading of these words, one is at first quite pleased and subsequently nothing but disheartened. For, whereas it is true that Wheeler vowed for a non-geometric theory of space and time and named it pregeometry, it could not be farther from the truth to say that the pregeometric theories devised so far (for example, those in Gibbs' or Monk's survey) refrain from the use of geometric concepts; to the contrary, they use them extensively, as is our purpose to show.

At the risk of displeasing some authors, in the show-piece analysis that follows, attempts to transcend the spacetime continuum will be understood as pregeometric in tune with what is \emph{called} pregeometry, even when this designation is not used explicitly in every programme. In such cases, this way of proceeding is somewhat unfair, since the following works will be criticized chiefly as failed expressions of a \emph{genuine pregeometry} (but many also for lack of underlying physical principles). Where they exist, explicit mentions of pregeometric intentions will be highlighted; lack of any such mention will be acknowledged as well. 

\section{Discreteness}
One category could have embraced most of---if not all---the works
in this analysis. Without doubt, the best-favoured type of
programme to deal with new ways of looking at spacetime is, in
general terms, of a discrete nature. The exact meaning of
``discrete'' we will not attempt to specify; the simple intuitive
connotation of the word as something consisting of separate,
individually distinct entities\footnote{Merriam-Webster Online Dictionary, http://www.m-w.com} will do. A choice of this sort appears to be appealing to researchers; firstly, because having cast doubt upon the assumption of spacetime as a continuum, its reverse, discreteness of some form, is naturally envisaged; secondly, because quantum-theoretical considerations tend to suggest, at least on an intuitive level, that also spacetime must be built by discrete standards.

Last but not least, many find some of the special properties of discrete structures quite promising themselves. Among other things, such structures seem to be quite well-suited to reproduce a continuum in some limit, because they naturally support geometric relations as a built-in property. This was originally Riemann's realization. As to the possible need for such a type of structure, he wrote in an often-quoted passage:
\begin{quote}
Now it seems that the empirical notions on which the metrical determinations of space are founded, the notion of a solid body and of a ray of light, cease to be valid for the infinitely small. We are therefore quite at liberty to suppose that the metric relations of space in the infinitely small do not conform to the
hypothesis of geometry; and we ought in fact to suppose it, if we can thereby obtain a simpler explanation of phenomena.

The question of the validity of the hypotheses of geometry in the infinitely small is bound up with the question of the ground of the metric relations of space. In this last question, which we may still regard as belonging to the doctrine of space, is found the application of the remark made above; that in a discrete manifoldness, the ground of its metric relations is given in the notion of it, while in a continuous manifoldness, this ground must come from outside. Either therefore the reality which underlies space must form a discrete manifoldness, or we must seek the ground of its metric relations outside it, in the binding forces which act upon it. \cite[p.~11, online ref.]{Riemann:1873}
\end{quote}

Riemann can therefore be rightly considered the father of all discrete approaches to the study of spacetime structure. He should not be considered, however, the mentor of an authentic pregeometry since, as he himself stated, quantitatively geometric relations are given in the very notion of a discrete manifold, which, through these relations, becomes a geometric object itself.\footnote{In Article I, the sign of geometry in pregeometry was taken to be the appearance of the geometric magnitude length. Here, attention is also drawn to the presence of geometric objects. Although these appear sometimes in self-evident ways, at other times they appear at first as abstract objects, which \emph{become} geometric objects via the attachment of geometric magnitudes to them. For example, a ``set of objects with neighbouring relations'' becomes a graph, i.e.\ a set of \emph{points} with joining \emph{lines}, when a distance is prescribed on the abstract relations. In quantum mechanics, we saw that state vectors likewise become geometric objects (arrows) after the specification of their complex-number overlaps (inner product) and their squared real moduli (areas).} The absolute veracity of his statement can be witnessed in every attempt at pregeometry, not as originally designed by Wheeler, but as it of itself came to be. (Wheeler's pregeometry will be examined in Chapter~\ref{ch:Beyond geometry}.)

\section{Graph-theoretical pictures}
Graphs are a much-favoured choice in attempts at pregeometric schemes. Following Wilson \citeyear[pp.~8--10, 78]{Wilson:1985}, a graph of a quite general type consists of a pair $(V,E)$, where $V$ is a possibly infinite set of elements called vertices and $E$ is a possibly infinite family of ordered pairs, called edges, of not necessarily distinct elements of $V$. Infinite sets and families of elements allow for an infinite graph, which can be, however, locally-finite if each vertex has a finite number of edges incident on it. Families (instead of sets) of edges permit the appearance of the same pairs more than once, thus allowing multiple edges between the same pair of vertices. An edge consisting of a link of a single vertex to itself represents a loop. Finally, ordered pairs allow for the existence of directed edges in the graph.

A particular reason for the popularity of graphs in pregeometry is that, along the lines of Riemann's thoughts, they naturally support a metric. When a metric is introduced on a graph, all edges are taken to be of the same length (usually a unit) irrespective of their actual shapes and lengths as drawn on paper or as conjured up by the imagination. This distinguishes graphs from certain types of lattices, which have a more literal interpretation, as explained below.

\subsection{Dadi\'c and Pisk's discrete-space structure}
Dadi\'{c} and Pisk \citeyear{Dadic/Pisk:1979} went beyond the manifold assumption
by representing ``space as a set of abstract objects with certain relations of neighbourhood among them'' (p.~346), i.e.\ a graph; in it, the vertices appear to correspond to space points. This approach assumes a \emph{metric} as naturally inherited from the graph, and a definition of dimension---based on a modification of the Hausdorff dimension---that is scale-dependent. The graph
$|G\rangle$ representing spacetime is required to be unlabelled in its points and lines, and must be characterizable by its topological structure only. Operators $b^\dagger$ and $b$ are defined for the creation and annihilation of lines, such that
$|G\rangle$ can be constructed from the vacuum-state graph $|0\rangle$ by repeated application of $b^\dagger$. This is a Fock-space framework, which includes a \emph{second metric} (on state vectors) as part of its built-in structure.

What led Dadi\'c and Pisk to this choice of graph-theoretical approach? They argued that in analysing what is essential in the intuitive notion of space, they found this to be the existence of some objects and the relation of neighbourhood
between them. Based on this, they proceeded to construct their discrete-space structure. However, this motivational basis is not grounded on any further physical principles. Quantum-theoretical ideas---in particular, a Fock-space method---are used in this approach only as a formalism to deal with graphs, their states, and their evolution. 

Geometric concepts are here clearly assumed. They are typified by the natural metric defined on the graph on the one hand, that of Fock spaces on the other, and not least by a graph as a composite geometric object (vertices and edges) itself. This fact alone would make this approach a questionable case of an authentic pregeometry, although we acknowledge that its authors did not so argue.

\subsection{Antonsen's random graphs}
Antonsen \citeyear{Antonsen:1992} devised a graph-theoretical approach that is statistical in nature. Space is identified with a dynamical random graph, whose vertices appear to be associated with space points, and whose links have unit \emph{length}; time is given by the parameterization of these graphs (spaces) in a meta\-space. Primary point-creation and point-annihilation operators, $a^\dagger$ and $a$, are defined, as well as secondary corresponding notions for links, $b^\dagger$ and $b$; probabilities for the occurrence of any such case are either given or in principle calculable. According to Antonsen, this framework does not really assume traditional quantum mechanics although it might look like it. He claimed, echoing perhaps the way of proceeding of Dadi\'c and Pisk, that the operators and the Hilbert space introduced by him are only formal notions without direct physical interpretations besides that of being able to generate a geometric structure.

This approach is, according to its author, ``more directly in tune with Wheeler's original ideas'' (p.~8) and, in particular, similar to that of Wheeler's ``law without law'' (p.~9) in that the laws of nature are viewed as a statistical consequence of the truly random behaviour of a vast number of entities working at a deeper level. One must have reservations about this view, since Antonsen is already working in a geometric context (a graph with a notion of distance), which is not what Wheeler had in mind. Moreover, there does not appear to be any principle leading Antonsen to introduce this kind of graph structure in the first
place. He sometimes (pp.~9, 84) mentions the need to assume as little as possible, perhaps overlooking the fact that not only the quantity but also the nature of the assumptions made---as few as they might be---bears as much importance.

Antonsen is therefore the first claimant to pregeometry who calls for the use of geometric objects (a graph) and magnitudes (distance) in its construction; for this reason, his scheme is rather suspicious as an authentic expression of pregeometry.

\subsection{Requardt's cellular networks}
Requardt also went beyond the manifold assumption by means of a graph-theoretical approach. At a deeper level, space is associated with a graph whose nodes $n_i$ are to be found in a certain state $s_i$ (only differences of ``charge'' $s_i-s_k$ are meaningful), and its bonds $b_j$ in a state $J_{ik}$ that can be equal to 0, 1, or $-1$ (vanishing or directed bonds). In contrast to the preceding scheme, here spacetime points are not associated with the graph vertices but with ``densely entangled subclusters of \emph{nodes} and \emph{bonds} of the underlying network or graph'' \cite[p.~3040]{Requardt/Roy:2001}. A law for the evolution in the ``clock-time'' of such a graph is simply introduced (pp.~3042--3043) on the grounds that it provides, by trial and error, with some desired consequences. Requardt \citeyear[p.~1]{Requardt:2000} then assumed that this graph evolved from a chaotic initial phase or ``big bang'' in the distant past in such a way as to have reached a generally stable phase, associated with ordinary spacetime. 

A peculiarity of this scheme is that it seeks to understand the current structure of spacetime as emergent in time,\footnote{As is typical of arguments considering the dynamic emergence of time (also as in space\emph{time}), it is unclear in what dynamic time the emergent time is supposed to appear. This hypothetical, deeper dynamic time, when considered at all (discrete ticks of clock-time for Requardt), is nevertheless always the same kind of simple and intuitive external parameter as the notion of time it attempts to give rise to. The end result is then to trade one little-understood external-parameter time for another.} i.e.\ as a consequence of the primordial events that are believed to have given rise to it in the first place. In this respect, it could have also been classified as a cosmological scheme, although not as a quantum-cosmological one since quantum mechanics is not assumed: Requardt's \citeyear{Requardt:1996} goal is to ``identify both gravity and quantum theory as the two dominant but derived aspects of an underlying discrete and more primordial theory\ldots'' (pp.~1--2).

A lack of guiding principles must again be criticized. Requardt's \citeyear{Requardt:1995} choice of approach seems to rest on ``a certain suspicion in the scientific community that nature may be `discrete' on the Planck scale'' (p.~2). Therefore, he argued, cellular networks make a good choice because they are naturally discrete and can, moreover, generate complex behaviour and self-organize.
However, it must be said that basing a physical choice on a generally held suspicion whose precise meaning is not clear may not be very cautious.

Although Requardt \citeyear{Requardt:1995} intended to produce a pregeometric scheme (``What we are prepared to admit is some kind of \mbox{`\emph{pregeometry}'\ ''} (p.~6)) and strove for ``a substratum that does not support from the outset\ldots geometrical structures'' (p.~2), he soon showed---in total opposition to the earlier re\-marks---that his graphs ``have a natural metric
structure'' (p.~15). He introduced thus a very explicit notion of \emph{distance}. Requardt was concerned with avoiding the assumption of a continuum, and in this he succeeded. However, geometric notions are not merely continuum notions, as argued earlier. This attempt at pregeometry suffers therefore from the same geometric affliction as the previous one.

\section{Lattice pictures} \label{LP}
At an intuitive level, (i) a lattice consists of a regular geometric arrangement of points or objects over an area or in space.\footnote{Merriam-Webster Online Dictionary, http://www.m-w.com} From a mathematical point of view, (ii) a lattice consists of a partially ordered set $\langle L,\prec\rangle$ in which there exists an infimum and a supremum for every pair of its elements.

The first notion of lattice goes hand in hand with the same idea as used in Regge calculus. In it, an irregular lattice is used in the sense of an irregular mesh whose edges can have \emph{different} lengths. For this reason, such a lattice cannot be a graph, where all links are equivalent to one another. The second notion of lattice is, \emph{to a degree}, connected with the approach to spacetime structure called causal sets. Although, strictly speaking, a causal set is only a locally finite, partially ordered set without loops, when it also holds that there
exists a unique infimum and supremum for every pair of its elements (existence of a common past and a common future), the causal set becomes a lattice in the second sense of the term. Despite the fact that they are not rigorously always a lattice,
causal sets have been included here in order to also exemplify, in as close a fashion as possible, the use of a lattice of the second type in the construction of an alternative structure for spacetime.\footnote{Causal sets can also be classified as graph-theoretical, considering that they can also be thought of as
consisting of a locally finite, loopless, directed graph without circuits, in the sense that it is not possible to come back to a starting vertex while following the allowed directions in it.}

\subsection{Simplicial quantum gravity by Lehto et al.} 
\label{Simplicial quantum gravity}
Lehto \citeyear{Lehto:1988}, and Lehto, Nielsen, and Ninomiya \citeyear{Lehto/Nielsen/Ninomiya:1986b,Lehto/Nielsen/Ninomiya:1986a} attempted the construction of a pregeometric scheme in which it was conjectured that spacetime possesses a deeper, pregeometric structure described by three dynamical variables, namely, an abstract simplicial complex ASC, its number of vertices $n$, and a real-valued field $\ell$ associated with every one-simplex (pair of vertices) $Y$. It is significant that, up to this point, the approach has a chance of being truly pregeometric since the fields $\ell$ are not (yet) to be understood as lengths of any sort, and the vertices could also be thought as (or called) elements or abstract objects (cf.\ \emph{abstract} simplicial complex). However, the introduction of geometric concepts follows immediately.

Next, an abstract simplicial complex is set into a unique correspondence with a \emph{geometric} simplicial complex GSC, a scheme of \emph{vertices} and \emph{geometric simplices}. By piecing together these geometric simplices, a piecewise-linear space can be built so that, by construction, it admits a triangulation. The
crucial step is now taken to interpret field $\ell$ as the link \emph{distance} of the piecewise-linear manifold. The conditions are thus set for the introduction of a Regge-calculus lattice with \emph{metric} $g_{\mu\nu}^{\mathrm{RC}}$ given by the previously defined link lengths.

Traditionally, the primary idea of Regge calculus is to provide a piecewise-linear approximation to the spacetime manifolds of general relativity by means of the gluing together of four-dimensional simplices, with curvature concentrated only in
their two-dimensional faces. In this approach, however, the approximation goes in the opposite direction, since simplicial gravity is now seen as more fundamental, having the smooth manifolds of general relativity as a large-scale limit. Moreover,
whereas the basic concept of the Regge calculus consists in the link lengths, which once given are enough to determine the geometry of the lattice structure, Lehto et al.\ also introduced the number of vertices $n$ as a dynamical variable. 

A quantization of the Regge-calculus lattice is subsequently introduced by means of the Euclidean path-integral formalism. A pregeometric partition function $Z$ summing over ASC, fields $\ell$, and vertices $n$ is first proposed as a formal
abstraction to be geometrically realized by a corresponding Regge-calculus partition function $Z^\mathrm{RC}$ summing over piecewise-linear manifolds and link lengths $\ell$. Indeed, quantization can be viewed as one of the reasons why the notion of length must be introduced in order to move forwards.

Two reasons can be offered for the choice of the starting point of this framework. On the one hand, being concerned with the production of a genuine pregeometry, an abstract simplicial complex was chosen on the grounds that, as such, it is abstract
enough to be free from geometric notions intrinsic to it. On the other hand, an abstract simplicial complex with a variable set of elements was seen fit to provide an appropriate setting in which to study further and from a different point of view the character of diffeomorphism invariance---a useful concept to help us rethink the ontology of spacetime points (see Chapter \ref{ch:ESE}). The requirement of diffeomorphism symmetry of the functional integral measure with respect to a displacement of vertices resulted in a free-gas behaviour of the latter as elements of a geometric lattice. Any pair of vertices could then with high probability have any mutual distance, helping to avoid the rise of long-range correlations typical of fixed-vertex lattices.

In one sense, this work is no exception among all those analysed in this chapter. Geometric concepts such as link length have to be assumed to realize the abstract simplicial complex as a geometric object, a Regge-calculus lattice, bringing it thus closer to a more familiar geometric setting and rendering it fit for quantization. However, despite the assumption of geometry, a distinctive feature of this work is that it recognizes clearly pregeometric and geometric realms and, even though it cannot do without the latter, it keeps them clearly distinguished from one another rather than, as is more common, soon losing sight of all differences between them.

\subsection{Causal sets by Bombelli et al.}
Bombelli, Lee, Meyer, and Sorkin \citeyear{Bombelli/Lee/Meyer/Sorkin:1987} proposed that at the smallest scales spacetime is a causal set: a locally finite set of elements endowed with a partial order corresponding to the transitive and irreflexive macroscopic relation that relates past and future. 

In this framework, causal order is viewed as prior to metric relations and not the other way around: the differential structure and the conformal metric of a manifold are derived from causal order. Because only the conformal metric (it has no associated measure of length) can be obtained in this way, the transition is made to a non-continuous spacetime consisting of a finite but large number of ordered elements, a causal set. In such a space, size can be measured by counting. 

What considerations have led to the choice of a causal set? In this respect, Sorkin explains:
\begin{quote}
The insight underlying these proposals is that, in passing from
the continuous to the discrete, one actually \emph{gains} certain
information, because ``volume'' can now be assessed (as Riemann
said) \emph{by counting}; and with both order \emph{and} volume
information present, we have enough to recover geometry. \cite[p.~5, online ref.]{Sorkin:2005}
\end{quote}
This is a valid recognition resting on the fact that, given that the
general-relativistic metric is fixed by causal structure and conformal factor, a discrete causal set will be able to reproduce these features and also provide a way for the continuous manifold to emerge. But is Sorkin here following any physical principle that leads him to the conclusion that a causal-set structure underlies spacetime, or is the proposal only mathematically inspired?

As regards geometry, volume is already a geometric concept, so that ``to recover geometry'' above must mean ``to recover a continuous, metric spacetime,'' not ``geometry'' in the stricter sense of the word. The causal-set picture uses other geometric concepts as well. The correspondence principle between a manifold and a causal set from which it is said to emerge, for example, reads:
\begin{quote}
The manifold $(M,g)$ ``emerges from'' the causal set $C$ [if and only if] $C$
``could have come from sprinkling points into $M$ at unit density,
and endowing the sprinkled points with the order they inherit from
the light-cone structure of $g$.'' \cite[p.~156]{Sorkin:1991}
\end{quote}
The mentioned unit density means that there exists a \emph{fundamental unit of volume}, expected by Sorkin to be the Planck volume, \mbox{$10^{-105}$ m$^3$}. Furthermore, he also considers the \emph{distance} between two causal-set elements $x$ and $y$ as the number of elements in the longest chain joining them (p.~17), turning the ordering relations into lines and $x$ and $y$ into points. Thus, if a causal set is to be seen as a pregeometric framework, due to its geometric assumptions, it makes again a questionable case of it. However, we do not know  its proponents to have staked any such claim.

\section{Number-theoretical pictures}
Some investigators have identified the key to an advancement in the problem of spacetime structure to lie in the number fields used in current theories or to be used in future ones. Butterfield and Isham \citeyear[pp.~84--85]{Butterfield/Isham:2000}, for example, arrived at the conclusion that the use of real and complex numbers in quantum mechanics presupposes that space is a continuum. According to this view, standard quantum mechanics could not be used in the construction of any theory that attempts to go beyond such a characterization of space.\footnote{But do not real numbers simply arise as the probability of quantum-mechanical outcomes as the limiting relative frequency between successes and total trials? These are quite unproblematic observations made in the real world, so why should we fabricate problems worrying about what might go wrong in the ``quantum gravity regime'' where ``space and time might break down'' \cite[p.~6, online ref.]{Isham/Butterfield:2000}?}  

\subsection{Rational-number spacetime by Horzela et al.}
Horzela, Kapu\'{s}cik, Kempczy\'{n}ski, and Uzes \citeyear{Horzela/Kapuscik/Kempczynski/Uzes:1992} criticized those discrete representations of spacetime that assume an elementary length, and which may furthermore violate relativistic invariance, on the grounds that the former is experimentally not observed and
the latter is, in fact, experimentally verified.\footnote{See \cite{Hill:1955} for a precursor of this programme.} They proposed to start their analysis by studying the actual experimental method used to attribute coordinates to spacetime points, the radio-location method.

Their starting point is that the measured values of the coordinates \mbox{$t=(t_1+t_2)/2$} and $x=c(t_2-t_1)/2$ of an event in any system of reference are always rational (rather than real) numbers, since both the time of emission $t_1$ and reception $t_2$ of the radio signal are crude, straightforward measurements made
with a clock.\footnote{For this idea to work, it is also essential that the value $c$ of the speed of light be understood as the result of a ``crude'' measurement.} Subsequently, they showed that the property of being rational is preserved when calculating the coordinates of an event in another system, as well as in the calculation of the relative velocity of this system. Therefore, they maintained that Lorentz invariance holds in a spacetime pictured in terms of rational-number coordinates. In addition, since the rational numbers are dense in the real numbers, such a spacetime frees itself from any notion of elementary distance. If spacetime must be described in a discrete manner, the rational-number field is then seen as the key to a better understanding of it.

While Horzela et al.\ suggested that this line of investigation should be furthered by means of algebraic mathematical methods, they admitted not knowing at the time of their writing of any guiding physical principle to do this. As a
consequence, since they do not propose to go as far as building a more complete spacetime picture based on their preliminary findings, the usual criticisms being made in this chapter do not apply. Further physical guiding principles are lacking, but then this has been acknowledged and action taken in accord with it; geometric ideas are of course present, but then no real attempt to transcend the framework of relativity was made after all.

\subsection{Volovich's number theory}
Volovich \citeyear{Volovich:1987} argued that, since at the Planck scale the usual notion of spacetime is suspected to lose its meaning, the building blocks of the universe cannot be particles, fields, or strings, for these are defined on such a background; in their place, he proposed that the numbers be considered as the basic entities. At the same time, Volovich suggested that for lengths less than the Planck length, the Archimedean axiom\footnote{The Archimedean axiom states that any given segment of a straight line can be eventually surpassed by adding arbitrarily small segments of the same line.} may not hold. Putting the two ideas together, he asked why should one construct a non-Archimedean geometry over the field of real numbers and not over some other field? 

To make progress, Volovich proposed as a general principle that the number field on which a theory is developed undergoes quantum fluctuations (p.~14). He maintained that this means that the usual field of real numbers is apparently only realized with some probability but that, in principle, there also exists a probability for the occurrence of any other field. Volovich furthermore speculated that fundamental physical laws should be invariant under a change of the number field, in analogy with Einstein's principle of general invariance (p.~14).

It is difficult to see what principle led Volovich to propose this programme; it appears that only analogies have guided his choices. It is also difficult to come to terms with his opinion that ``number theory and the corresponding branches of
algebraic geometry are nothing else than the ultimate and unified physical theory'' (p.~15). Finally, the role of geometry in this framework is crystal-clear: instead of attempting to reduce ``all physics'' to Archimedean geometry over the real numbers, it is proposed that one should rather attempt to reduce ``all physics'' to non-Archimedean \emph{geometry} over a fluctuating number field.
For this reason, although Volovich did not so claim, his scheme could, again, hardly be considered a true instance of pregeometry.

\section{Relational or process-based pictures}
Having found inspiration in Leibniz's relational conception of space, some researchers have proposed that spacetime is not a thing itself but a resultant of a complex of relations among basic things. Others, on a more philosophical strand, have found inspiration in Heraclitus' principle that ``all is flux'' and have put forward pregeometric pictures in which, roughly speaking, processes are considered to be more fundamental than things. This view is sometimes linked with the currently much-disseminated idea of information, especially in its quantum-mechanical form.

\subsection{Axiomatic pregeometry by Perez Bergliaffa et al.}
Perez Bergliaffa, Romero, and Vucetich \citeyear{Perez/Romero/Vucetich:1998} presented an axiomatic framework that, via rules of correspondence, is to serve
as a ``pregeometry of space-time.'' This framework assumes the objective existence of basic physical entities called things and sees spacetime not as a thing itself but as a resultant of relations among those entities.

In order to derive from this more basic substratum the topological and metric properties of Minkowskian spacetime, a long list of concepts and axioms is presented. But for the said general requirements of objectivity and relationality, these ``ontological presuppositions'' (p.~2283) appear arbitrary and inspired in purely mathematical ideas rather than stemming from some physical principle concerning spacetime.

Among these concepts, we centre our attention on something called ontic space $E_{\mathrm{o}}$. Although its exact meaning cannot be conveyed without reviewing a great deal of the contents of the article in question, it could be intuitively taken to be a form of space connected with the basic entities, which is prior (in axiomatic development) to the more familiar geometric space $E_{\mathrm{G}}$ of physics---Minkowskian space, in this case. Perez Bergliaffa et al.\ proved (pp.~2290--2291) that the ontic space $E_{\mathrm{o}}$ is metrizable and gave an explicit metric $\mathrm{d}(x,y)$ for it, a \emph{distance} between things, turning, again, relations into lines and things into points. Finally, by means of an isomorphism between $E_{\mathrm{o}}$ and a subspace dense in a complete space, the geometric space $E_{\mathrm{G}}$ is obtained as this complete space itself, while at the same time it inherits the metric of $E_{\mathrm{o}}$.

One notices once again the stealthy introduction of a metric for pregeometry, with the ensuing geometric objectification of earlier abstract concepts, in order to derive from it further geometric notions. This procedure must be criticized in view that this approach explicitly claims to be an instance of pregeometry.

\subsection{Cahill and Klinger's bootstrap universe}
Cahill and Klinger \citeyear{Cahill/Klinger:1997} proposed to give an account of what they rather questionably called ``the `ultimate' modelling of reality''
(p.~2). This, they claimed, could only be done by means of pregeometric concepts that do not assume any notions of things but only those of processes since, according to them, the former notions (but not the latter) suffer from the problem of their always being capable of further explanation in terms of new things. One must wonder for what mysterious reason certain processes cannot be explained in terms of others.

As an embodiment of the concepts above, these authors used a ``bootstrapped self-referential system'' (p.~3) characterized by a certain iterative map that relies on the notion of relational information $B_{ij}$ between monads $i$ and $j$. In order to avoid the already rejected idea of things, these monads are said to have
no independent meaning but for the one they acquire via the relational information $B_{ij}$. Once again, it is evident that this mathematically inspired starting point does not rest on any physical foundation.

Due to reasons not to be reviewed here, the iterative map will allow the persistence of large $B_{ij}$. These, it is argued, will give rise to a tree-graph---with monads as nodes and $B_{ij}$ as links---as the most probable structure and as seen from the perspective of monad $i$. Subsequently, so as to obtain (by means of probabilistic mathematical tools) a persistent background
structure to be associated with some form of three-dimensional space, a \emph{distance} between any two monads is defined as the smallest number of links connecting them in the graph.

``This emergent 3-space,'' Cahill and Klinger argued, ``\ldots does not arise within any \emph{a priori} geometrical background structure'' (p.~6). Such a contention is clearly mistaken for a graph with its defined idea of distance is certainly a geometric background for the reasons we have pointed out earlier. Thus, yet another mathematically inspired pregeometric scheme falls prey to geometry.

\section{Quantum-cosmological pictures} \label{QCP}
Quantum-cosmological approaches tackle the problem of spacetime structure from the perspective of the universe as a whole and seen as a necessarily closed quantum system. For them, the problem of the constitution of spacetime is bound up with the problem of the constitution of the cosmos, customarily dealing as well with the
problem of how they interdependently originated.

\subsection{Eakins and Jaroszkiewicz's quantum universe} \label{EJ}
Jaroszkiewicz \citeyear{Jaroszkiewicz:2001} and Eakins and Jaroszkiewicz \citeyear{Eakins/Jaroszkiewicz:2002,Eakins/Jaroszkiewicz:2003} presented a theoretical picture for the quantum structure and running of the universe. Its first class of basic elements are event states; these may or may not be factored out as yet more
fundamental event states, depending on whether they are direct products (classicity) of yet more elementary event states $|\psi\rangle$, or whether they are entangled (non-separability) event states $|\Psi\rangle$. The second class of basic elements are the tests acting on the event states. These tests $\Sigma$, represented by Hermitian operators $\hat{\Sigma}$, provide the topological relationships between states, and endow the structure of states with an evolution (irreversible acquisition of information) via ``state collapse'' and with a quantum arrow of time. A third component is a certain information content of the universe---sometimes, although not necessarily, represented by the semiclassical observer---at each given step in its evolution (stage), which, together with the present state of the universe, determines the tests which are to follow. Thus, the universe is a self-testing machine, a quantum automaton, in which the traditional quantum-mechanical observer is accounted for but totally dispensed with.

Eakins and Jaroszkiewicz \citeyear[p.~5]{Eakins/Jaroszkiewicz:2002}, like Requardt but after their own fashion, also attempt to understand the present structure of spacetime as a consequence of primordial events, which in this case can be traced back to what they call the ``quantum big bang.'' In this case, however, the emergence of spacetime is more sophisticated than in Requardt's, since here it happens in the non-parameter intrinsic time given by state-vector collapse. 

The subject of pregeometry as pioneered by Wheeler is also touched upon \cite[p.~2]{Eakins/Jaroszkiewicz:2003}. The view that the characteristic feature of pregeometric approaches consists in ``avoiding any assumption of a pre-existing manifold'' (p.~2) is also stated there, although it is nowhere mentioned that the
notion of distance, according to Wheeler, should be avoided as well. In any case, these authors explain that their task is to reconcile pregeometric, bottom-up approaches to quantum gravity with other holistic, top-bottom ones as in quantum cosmology. Finally, since the traditional machinery of quantum mechanics is assumed in the approach, so is therefore geometry. Again, this fact undermines the status of this programme as a genuine representative of pregeometry.

Of all the above approaches, we find this one to be the most appealing. It gives clear explanations about the nature of its building blocks and puts forward judicious physical principles and mechanisms---as contrasted to purely mathematically inspired ones---by means of which the traditional notions of spacetime and quantum non-locality may emerge. 

In Table \ref{summary1}, we summarize the use of geometric concepts in all the pregeometric pictures analysed in this chapter, plus two more discussed later on in Chapter \ref{ch:Beyond geometry}.

\begin{table} 
\centering
\begin{tabular}{|l|c|c|} 
\hline
\textbf{Author} & \textbf{Geometric} & \textbf{Geometric}\\
					 &  \textbf{objects} & \textbf{magnitudes}\\
\hline \hline
Dadi\'{c} \& Pisk & vertices, edges & overlap, length, distance \\ \hline
Antonsen$^\ast$ & idem & length, distance \\ \hline
Requardt$^\ast$ & idem & idem \\ \hline
Lehto et al.$^\ast$ & idem & idem \\ \hline
Bombelli et al. & idem & length, distance, volume \\ \hline
Horzela et al. & as in sp.~rel. &  as in sp.~rel.\\ \hline
Volovich & as in mod.~phys. &   as in mod.~phys. \\ \hline
Perez Bergliaffa et al.$^\ast$ & points, lines &  length, distance \\ \hline
Cahill \& Klinger$^\ast$ & idem &  idem \\ \hline
Eakins \& Jaroszkiewicz$^\ast$ & as in q.~mech. &  as in q.~mech. \\ \hline
Nagels$^\ast$ & points, lines & length, distance \\ \hline
Stuckey \& Silberstein$^\ast$ & idem & idem \\ \hline
\end{tabular}
\caption[Geometric concepts in pregeometry]{Summary of the use of geometric concepts in the pregeometric approaches analysed in this chapter, plus two more discussed later in Section \ref{sec:bucket}. Approaches that mention pregeometry explicitly are marked with an asterisk ($\ast$). Abbreviations: Special relativity (sp.~rel.), quantum mechanics (q.~mech.), modern physics (mod.~phys.).}
\label{summary1} 
\end{table}

\section{The inexorability of geometric understanding?} \label{IGA}
By way of appraisal,  we take note of certain views expressed by Anandan, as they constitute the absolute epitome of the mode of working of pregeometry witnessed so far. Anandan suggests 
\begin{quote}
a philosophical principle which may be schematically \mbox{expressed as}
\begin{center} Ontology = Geometry = Physics. \end{center}
The last equality has not been achieved yet by physicists because
we do not have a quantum gravity. But it is here proposed as a
philosophical principle which should ultimately be satisfied by a
physical theory. \cite[p.~51]{Anandan:1997}
\end{quote}
Anandan appears to have taken the geometric essence of present physical theories as a ground to argue ahead that, correspondingly, any successful future physical theory must of necessity be geometric. (So much for Wheeler's pregeometry as a road to quantum gravity!) Such is his conviction that he raises this idea to the status of a principle that any physical theory should ``ultimately'' follow. But does it ensue from the fact that physics has so far described nature by means of geometry that it will continue to do so in the future? Physics \emph{must} describe nature geometrically come what may---\emph{whence such an inevitability?} Moreover, in what way has the first part of the quoted equality, as implicitly stated, already been achieved? Simply because it is a fact that reality itself is geometric? We are reminded of Smith's view, according to which form is an inherent part of nature. Why does this idea keep coming back to haunt us?

Thus, we witness here an overstated version of the spirit implicitly underlying the cases analysed through this section. If in the latter the use of geometry was, so to speak, accidental or non-intentional, in Anandan's statement we find the explicit claim materialized that geometry \emph{must} be the mode of description of physical science, in particular, concerning a solution to the problem of the quantum structure of spacetime. No stronger grip of geometry on physics---and on the human mind---could possibly be conceived. But is this conception justified? And if not, what is the source of this widespread misconception?

The connection underlying Chapters \ref{ch:Geometry}--\ref{ch:Pregeometry} now comes to the fore. From the mists of antiquity to the forefront of twenty-first century theoretical physics, geometry remains (for better or worse, successfully or vainly) the enduring, irreplaceable tool by means of which humans portray the world. Ironically enough, even in a field that sets itself the task of explaining the origin of geometry in physics and names itself pregeometry, one is to find the most explicit displays of geometric understanding. In the next chapter, we start to consider the reasons behind this peculiar state of affairs.

\chapter{The geometric anthropic principle and its consequences}
\label{ch:Geometric anthropic principle}
\begin{flushright}
\begin{tabular}{p{10cm}}
\emph{The more I study religions the more I am convinced that man never worshipped anything but himself.}\smallskip \\
Sir Richard Francis Burton, \emph{The book of the thousand nights and a night}, Vol.~10
\end{tabular}
\end{flushright}

\bigskip

J.~Ellis \citeyear{Ellis:2005} described the state of affairs documented in the two previous chapters rather insightfully when he said, ``Following Einstein, most theoretical physicists assign a central role to geometrical ideas'' (p.~57). \mbox{Ellis'} recognition is quite valid, but after pondering geometry to the extent that we have done, we are ready to make a few specifications. Although it is true that physicists currently draw their geometric inspiration to a certain extent from the success of Einstein's work---for it does \emph{look} more geometric than the physical theories preceding it---we have seen that the geometric drive in physics is rooted much deeper and much farther back in time than in any twentieth-century intellectual product. Moreover, one may even push Ellis' claim further and, without fear of error, say that, with the exceptions of the next chapter, it is essentially \emph{all} physicists that assign a central role to geometric ideas. How central is this role?

Pedersen \citeyear{Pedersen:1980}, for example, tells us that ``Geometry is the connection between the real world and the mathematics we use to solve problems about that real world,'' and reports on mathematician Ren\'e Thom's notion that ``Geometry is the most fundamental abstract representation of the real world'' \cite{Hilton/Pedersen:2004}. Pedersen's statement is an accurate description of the contingent manner in which geometry has so far happened to be used by humans, but it should not be understood as a description of the necessity of geometric theorizing. Thom's remark, on the other hand, errs in its conflation of contingency with necessity, namely, that the language of geometry is not simply a parochial picturing means of humans, but the universal tool by means of which any theorist (human or otherwise) at any time (now and forever) must represent any aspect (known and unforeseeable) of the physical world. What is the origin and reason for this exaltation of geometry?

\section{The geometric anthropic principle} 
The astounding effectiveness of geometry in the portrayal of the natural world has since antiquity been interpreted by many to signify that the natural world is geometric in itself. The roots of this persuasion are found in ancient Greece and are represented by Plato's thought; in particular, they are best epitomized by a notable phrase attributed to him, namely, ``God eternally geometrizes.'' 

Instead of Plato, Foss \citeyear[p.~36]{Foss:2000} attributes the origin of this belief to Pythagoras, and calls it the ``Pythagorean Intuition.'' However, Pythagoras' view was that ``reality is number'' or that ``at its deepest level, reality is mathematical in nature,''\footnote{Encyclopaedia Britannica, http://www.britannica.com} which is not equivalent to its being essentially \emph{geometric}. Jammer \citeyear{Jammer:1969}, on the other hand, does capture the distinction, together with the Platonic origins of the geometric physical tradition, when he writes, ``With Plato physics becomes geometry, just as with the Pythagoreans it became arithmetic'' (p.~14).\footnote{The conflation of mathematics with geometry is an error commonly incurred in. For example, in connection with the ``Pythagorean Intuition,'' Foss \citeyear{Foss:2000} writes: ``\emph{why} has geometry become \emph{the} modeling tool of science? Is it because Pythagoras was right in thinking that reality itself is somehow deeply mathematical by nature?'' (p.~87). For a similar error, see \cite[p.~5]{Stewart:2002}.} 

Two millennia later, hardly anything had changed. The same opinion as Plato's but without its religious overtones was articulated by Galilei, for whom the book of nature \begin{quote} is written in the language of mathematics, and its characters are triangles, circles, and other geometric figures without which it is humanly
impossible to understand a single word of it; without these, one wanders about in a dark labyrinth. \cite[p.~238]{Galilei:1957}
\end{quote}
Whence so much emphasis on the essentialness of geometry?

These statements propounding the fundamentality of geometry can be understood as originating from a geometric form of the notorious \emph{anthropic principle}. Ordinarily, this principle asserts that ``the universe must be as it is because, otherwise, we would not be here to observe it.'' In a similar vein, Plato's words epitomize a strong geometric version of the anthropic principle, which could be phrased thus: ``God must have made the universe geometrically because, otherwise, we would not be able to describe it as we do.'' Galilei's words embody a weaker version of the same principle: ``The universe must be geometric because, otherwise, we would not be able to describe it as we do.'' As common sense is apt to reveal, the reasoning expressed by anthropic principles is nevertheless
inadequate, considering that the fact that Man can do something---in the former case, simply exist; in the latter, devise successful geometric theories---is a contingent consequence, not a cause, of the features of the world.

Despite its ineffectualness and basically flawed nature, the anthropic principle continues to be widely upheld today in both its geometric varieties. Its weaker
version finds expression in statements such as Brassard's \citeyear{Brassard:1998} ``A science has to be based on a geometry'' ($\S$ 3.4), whereby physics is posited as a branch of geometry---a view that puts the cart before the horse. It also finds expression in Anandan's \citeyear{Anandan:1997} previously analysed principle ``Reality = Geometry = Physics'' (p.~57) that any physical theory should ultimately follow. Its strong version is in turn exemplified by straightout religious assertions such as ``God is a geometer.''\footnote{See e.g.\ \cite{Stewart/Golubitsky:1993} and \cite[p.~5]{Stewart:2002}; see also \cite[p.~881]{Pollard:1984}.} The strong version of the geometric anthropic principle has even been spelt out in its full explicit form---and shockingly, too---by Stewart \citeyear{Stewart:2006}: ``Only a Geometer God can create a mind that has the capacity to delude itself that a Geometer God exists'' (p.~203).

Has Geometry perchance become a modern religion---physicists' latest incarnation of God? Unexpectedly, this rhetorical question finds a willing reply from Smolin \citeyear{Smolin:1997}, who answers forthrightly in the positive: ``the belief that at its deepest level reality may be captured by an equation or a geometrical construction is the private religion of the theoretical physicist'' (p.~178); he
even goes on to say that ``an education in physics or mathematics is a little like an induction into a mystical order'' (p.~179). But if this is a statement of fact (and it may well be), then it is of a sad fact, and not of one we should be sympathetic of or mention lightly.

Not even twenty-first century physics---much though it may secretly presume---has ridden itself of the religious mindset originally part of natural philosophy. Peacock \citeyear{Peacock:2006}, for example, rightly criticized physicists in general for subscribing to the religious view that ``God wrote the equations'' of nature revealed (not made) by our theories, and Susskind in particular for transferring ``the quasi-religious awe [from intelligent design] to string theory, whose mathematical results he repeatedly describes as `miraculous'~'' (p.~170). And Horgan \citeyear{Horgan:1996} hit the nail nearly on the head when he attributed physicists' calling God a geometer to their being ``intoxicated by the power of their mathematical [geometric] theories'' (p.~77; see also p.~136).

Far from God or nature, we place the origin of geometry in human nature; more specifically, in the character of the human brain. It is not because the world is geometric itself (a cryptic notion, at any rate), or because God made it so, that our theories of it speak the language of geometry; it is because evolutionary vicissitudes have moulded the human brain such that, for better or worse, successfully or vainly, \emph{we write} geometric theories of the world. Indeed, it appears rather more evident that Galilei's geometric book of nature is not pre-existent but written by Man---who eternally geometrizes but, ingrained as his ways of thinking are, is unable to recognize the handwriting as his own.

Without hesitation, we place the source of human geometric abilities---or even, compul\-sions---in the character of the brain and not elsewhere, such as in that of the sense organs, especially eyesight. It is quite possible that if we were still endowed with the familiar five senses but our brains were essentially different, we would not develop any geometric concepts and insights into nature. An acquaintance with the self-related life experiences of Helen Keller, the famous deaf-blind woman who overcame her sensory handicaps to become a full human being, gives observational support to this view. 

Keller \citeyear{Keller:1933} tells us about her mind's natural ability to grasp all the established, conventional thoughts and concepts of seeing and hearing people despite her own, practically life-long, visual and hearing disabilities. She wrote: ``Can it be that the brain is so constituted that it will continue the activity which animates the sight and the hearing, after the eye and the ear have been destroyed?'' (p.~63). And further reflected:
\begin{quote}
The blind child---the deaf-blind child---has inherited the mind of seeing-and-hearing ancestors---a mind measured to five senses. Therefore he must be influenced, even if it is unknown to himself, by the light, colour, song which have been transmitted through the language he is taught, for the chambers of the mind are ready to receive that language. \cite[p.~91]{Keller:1933}
\end{quote}
Keller discovers in her personal experience that her mind naturally bridges the gap to that of her seeing-and-hearing peers by producing appropriate conceptual correspondences, such as between a ``lightning flash,'' which she cannot experience, and a ``flash of thought and its swiftness'' (p.~92), which she can; and she notes that no matter how far she will exercise this method, it will not break down. Keller tells us that she conceives of the world in essentially the same manner as do her normal peers, and that she knows this from the fact that she has no difficulty in comprehending other people; she distinctly knows that her consciousness---primarily a product of her brain---is in essence the same as that of sensorially unimpaired people. ``Blindness,'' she metaphorically decrees from her vantage position, ``has no limiting effect upon mental vision'' (p.~95). Keller's awe-inspiring words, then, bear out the point in question in first-hand manner, namely, that it is primarily the brain, not the senses, that leads us to our particular description and understanding of the world. Change a man's senses, or to a certain extent deprive him of them, and his worldview will not change; but change a man's mind and witness his vision, hearing, and touch acquire unsuspected new meanings.

Geometric thought is no exception to this rule. Shape and extension are not properties we abstract any more from the sight of objects than from the sounds they make or the way they feel. As one can infer from the above account, the nature of the mind's means of contact with the outer world is to a large extent irrelevant to the formation of geometric conceptions---the task of creating these rests essentially upon the brain.

It is sometimes told that a fisherman whose fishing net is loosely knit will only catch big fish. But while we intellectually understand the meaning of this statement, we fail to feel its meaning in our bones. After all, we \emph{can} imagine a more tightly knit net that would catch the smaller fish in the ocean; we can easily see on the other side of this artificially imposed limitation; we can easily see how mistaken it is to infer an ocean populated exclusively by big fish from the character of the net---and so the moral eludes us. The extent to which it eludes us is directly proportional to the difficulty we experience in trying to come to terms with a statement of the type, ``Snowflakes are not inherently hexagonal,'' because snowflakes look invariably hexagonal to us, and we find it hard---and pointless---to conceive of them otherwise. This is not to suggest that the hexagonality of snowflakes is an illusion, but that it is one possible true description of them, just like bigness is one true description of fish in the ocean---the net does not lie, only possibly misleads.

When the geometric worldview, instead of stemming from an outside filter such as a fishing net, stems from the innermost recesses of our own selves, it is little wonder that it will go unrecognized as the conceptual tool it is, and be instead conflated with God, nature, and reality themselves. What is more, it will continue to pass unrecognized as a non-essential tool even after we have caught a glimpse of its possible shortcomings. Foss \citeyear{Foss:2000} makes a striking case of this situation in his book on consciousness. After insightfully recognizing the exclusively geometric mode of working of science, he goes on to note that qualia (i.e.\ the quality of conscious experience) have escaped scientific understanding because science deals with the world only geometrically. But such is the grip of the geometric worldview, and such is the strength of the taboo on any thought that would cast doubt upon it, that Foss's valuable realization soon regresses and becomes a victim of itself. Instead of concluding that scientific descriptions need not then always be geometric, he writes: ``The crucial insight is that even though science must model things geometric\emph{ally}, this does not entail that it can only model geometric things, geometric properties, or geometric aspects of reality'' (p.~70). (We return to Foss's ideas later on in Chapter \ref{ch:MQM}.)

But whence \emph{must} science picture things geometrically, beyond the contingency that it presently \emph{happens} to do so? Is it not more natural to conclude that some things (e.g.\ qualia)---\emph{which are neither geometric nor non-geometric in themselves}---defy geometric representation simply because their physical nature is such that it does not allow for its casting into geometric moulds? It is gripped by an indelible geometric imperative as they probe the physical world that, no matter how speculative and non-conservative today's
scientists might be, they continue to pound away with the only artillery they can conceive---geometry, more geometries, new geometries.

Returning once again to quantum gravity, we ask: would it be fair to say that, after three-quarters of a century of disorientation and frustration in quantum gravity, the disclosure of a deeper layer of the nature of things through new geometries has reached a dead-end? That, after nearly a century of philosophical puzzles concerning the ontology of state vectors and spacetime points, more and new geometric deliberations on quantum and spacetime ontology can take us no further? That with more and ever-renewed geometric modes of thought we can only continue to imprison ourselves in a labyrinth of our own making? 

In the rest of this chapter, we expound some of the consequences for theoretical physics of belief in the geometric anthropic principle.

\section[Ontological difficulties]{Ontological difficulties: State vectors and\\ spacetime constituents}
Attitudes towards the state vector and its collapse are as varied as human imagination. The ontological dependence of quantum theory on states, combined with the human yearning to gain insight exclusively through geometric visualization, has led to a frantic search for the physical meaning of the arrow state vector. Isham \citeyear[p.~83]{Isham:1995} submits some of the many possibilities: does the state vector refer to an individual system, a collection of identical systems, our knowledge of the system(s), the result of a measurement on a system, the results of repeated measurements on a collection of them, the definiteness with which a system possesses a value, etc.? Further, to make sense of the collapse of the geometric state vector, potentially physical mechanisms have been proposed, such as gravity or consciousness, that might cause it; others have in denial engineered a metaphysical escape to a myriad of parallel universes; etc.

Not only does one universe suffice to give answers to the ontological questions of quantum theory but, we submit, not even physical influences beyond quantum mechanics need be invoked for the task, so long as we are willing to reexamine the geometric foundations of the portrayal of phenomena and, in line with a long physical tradition, are ready to start the development of theoretical concepts from physical things we are familiar with (observations, experiments) instead of, as is today more common, mathematical contrivances without physical referents.

Attitudes towards the ontology of empty spacetime are in no short supply either. These arise primarily as a sequel of Einstein's hole argument, which, looking at what is left beyond the metric structure of spacetime (i.e.\ its quantitatively geometric scaffold) concluded that what remains, spacetime points (i.e.\ its qualitatively geometric constituents), have no physical meaning. As a fallout of this argument, professional philosophers and philosophically minded physicists continue to devise, as far as logic will allow, ingenious stratagems to furnish the geometric objectification of spacetime with new meanings (substantivalism); or then to reassure us that Einstein was correct in the first place, that spacetime points have in effect no physical meaning, and that relations between particles and fields in spacetime is all there is to it (relationalism).\footnote{For a review, see e.g.\ \cite[esp.~pp.~175--208]{Earman:1989}.} But is a solution to be sought by means of an increase in our knowledge within the very geometric tradition that gave rise to the problem?  If these philosophical battles have no end in sight, it is because, in all probability, the concepts used to seek solutions lie at the root of the difficulties.

Addressing the problem of spacetime ontology from a different per\-spec\-tive---yet with equally doubtful physical relevance---we find quantum-gravity endeavours. Here, due to conceptual discontent with the general-relativistic spacetime picture, complicated geometric edifices representing new spacetime structures are industriously built in an almost exclusively mathematical spirit; only subsequently is physical meaning sought for these psychologically compelling geometric creations, but with starting points based on observationally disembodied concepts, physical meaning is later hard to find regardless of how ``beautiful'' or ``self-consistent'' these geometric creations may be. As we said in Section \ref{Beliefs}, not even the quantization of general relativity, when mathematically viable, is a reliable starting point, so long as the consequences of quantization fail to be attested in the physical world.

\section[Psychological difficulties]{Psychological difficulties: Time and space}
Besides finding inspiration in the physical world rather than in purely mathematical realms, quantum gravity may also benefit from reconsidering the possibilities contained in the elemental question that drives it, ``What are space and time?'' This question is always tacitly assumed to be synonymous with ``What is the geometry of space and time?'' but this reading only circumscribes research to a narrow context. Is it thoroughly unthinkable that there may be more aspects to the physical world than those that can be captured in geometric language? We see in this light that ``What are space and time?'' is neither, as we know, a simple question to answer but not even, as Smolin \citeyear{Smolin:2001} suggests, ``the simplest question to ask'' (p.~1). Research still uniquely focuses on  the \emph{geometry} of space and time, while remaining oblivious to any alternatives.

As for the geometric answers offered, it is noteworthy that the treatments given to space and time are in general disunited, and that, moreover, neither treatment is as radically new, unconventional, or different as the characterizations made by quantum-gravity researchers would have one believe. Although quantum gravity endeavours for an elucidation of the quantum-mechanical structure of space \emph{and time}, the classical concept of time as an external one-parameter is so deep-rooted in physics, and so much more so than the classical concept of space as a three-dimensional continuum of points, that it is common for quantum-gravity ventures to invent notions of discrete or quantized space that are non-classical to varying degrees but, when it comes to ``dynamics,'' some form of external time marking parametric evolution is surreptitiously added to the picture. The psychological difficulty to think of space other than geometrically (as a metric network of points), and of time in ways other than geometrically (as a parametric line) \emph{and} classically (as an external parameter) reveals itself, in different ways, in all geometric roads to quantum gravity.

In string theory and most pregeometries, space is depicted less classically than time, yet classically enough. Whether described as a wrapped-up multidimensional metric manifold, or as a graph, a network, a lattice, or a causal set, the difference with the classical notion of space is only one of adding dimensions, changing the topology and metric, or making a continuum discrete in literal-minded ways. As for time, dynamic evolution either still takes place along a continuous parameter or, when any new treatment is given to time at all, this is at the outside postulated to be discrete (in steps of Planck's time, naturally), but still an external parameter nonetheless. The discretization of both space and time normally comes from forcing the classical notions to comply with common preconceptions in simple-minded manners, rather than as the result of a process (e.g.\ quantization) or, better still, some observationally inspired idea that would naturally lead to it.

In loop quantum gravity, on the other hand, space is represented by a quantum-mechanical state $|s\rangle$, called a spin network, which is not put in by hand but arises from a quantization of general relativity.\footnote{The question remains, however, to what degree the state vector itself is ``put in by hand'' (or better: by mind) in a quantum theory. We give an answer in Chapter \ref{ch:MQM}.} In this sense and unlike in the instances mentioned above, space as per loop quantum gravity may be called a genuinely non-classical space.\footnote{Examples of less literal-minded pregeometries in this sense are provided by Dadi\'{c} and Pisk \citeyear{Dadic/Pisk:1979} and Antonsen \citeyear{Antonsen:1992}, who view graphs themselves as quantum states $|G\rangle$; and by Jaroszkiewicz \citeyear{Jaroszkiewicz:2001}, who considers a topological network whose elements are quantum states $|\psi\rangle$ and test operators $\hat\Sigma$.}  It is here, then, that the contrast with the treatment of time stands out in sharpest relief.  Although Rovelli \citeyear{Rovelli:2004} insists on presenting a ``timeless'' loop quantum gravity, this is far from true. In his book \emph{Quantum gravity}, a \emph{magnum opus} born ahead of time, loop quantum gravity\footnote{Presumably \emph{the} quantum theory of gravity? So much for the three roads to quantum gravity---or even two! (Cf.\ Rovelli's quotation on page \pageref{Rovelli-quote}.)} is presented in terms of a mathematical, unobservable parameter $\tau$ which acts to order spin networks and along which these states of space evolve dynamically.\footnote{Most notably, the probability amplitude $W(s|s')$ of a transition  from $|s'\rangle$ to $|s\rangle$ is given in terms of $\tau$ as $\langle s| \int \exp(i\tau H)\mathrm{d}\tau|s'\rangle$, with suitable integration limits in $\tau$, and where $H$ is the Hamiltonian.} One may call parameter $\tau$ time, or one may call it what one will; Rovelli, determined to do without time come what may, calls it ``an artifact of this technique'' (p.~112) referring to the Hamiltonian formalism. Time as an unobservable parameter also appears---now without any proviso---as a coordinate variable $t$ when dynamic evolution is represented by means of the spin-foam (i.e.\ sum-over-histories) formalism; spin networks are said to be embedded in a four-dimensional spacetime and to move ``upward along a `time' coordinate'' (p.~325). Finally, time also appears as a partial observable $q$ (also $t$); of all of the above, this is the only type of time parameter that can be measured directly by macroscopic systems called clocks but, being external, it is useless as an intrinsic measure of time for any ``quantum-gravitational system'' purportedly at the Planck scale. (In the macroscopic theory of relativity, clock measurements \emph{do} play an essential physical role, as we see in Chapter \ref{ch:Analysis of time}.)

Despite the psychological difficulties involved in dealing with the concept of time other than classically, it is common for researchers to ignore the matter (whether purposefully or unwittingly), presenting their theories as if in them space and time were on equally non-classical footings. Witness the following comments going back and forth without ryhme or reason between ``space'' and ``space and time,'' or ``space'' and ``spacetime'': Smolin \citeyear{Smolin:2001} writes, ``Is there an atomic structure to the geometry of \emph{space and time}\ldots? [\ldots] Both loop quantum gravity and string theory assert that there is an atomic structure to \emph{space}'' (pp.~102--103, italics added; see also p.~170); in the same vein, Rovelli \citeyear{Rovelli:1998} says, ``A knot state is an elementary quantum of \emph{space}. In this manner, loop quantum gravity ties the new notion of \emph{space and time} introduced by general relativity with quantum mechanics'' (p.~19, italics added); last but not least, Weinberg \citeyear{Weinberg:1999} writes, ``The nature of \emph{space and time} must be dealt with in a unified theory. At the shortest distance scales, \emph{space} may be replaced by a continually reconnecting structure of strings and membranes---or by something stranger still'' (p.~36, italics added). \emph{And time?} Is time, in the end, just a classical parameter even in concept-sweeping quantum gravity? Or is there any non-classical, non-geometric concept of time intrinsic to quantum-mechanical systems?

A step away from this conceptual confinement was taken by Jaroszkiewicz \citeyear{Jaroszkiewicz:2001}. He pictured the passage of time in a non-parameter and intrinsically quantum-mechanical way, as quantum ticks determined by the irreversible acquisition of information at every quantum-mechanical measurement. Although attractive, this idea is potentially treacherous as long as it is connected with the geometric formulation of quantum theory. For if a quantum tick corresponds to the collapse of the state vector, one could---enslaved to geometric thinking and parameter time---immedi\-ately ask: how, where, and \emph{when} does $|\psi(t)\rangle$ collapse when a measurement occurs? Ultimately, the idea of a quantum tick and the geometric state-vector formalism are mutually uncongenial. In Chapter \ref{ch:MQM}, we leave geometry behind in order to find a non-parameter concept of time that is intrinsic to quantum-mechanical systems and from which the origin of time $t$ can be understood.

To pose the question, ``What are space and time?'' and, withheld by the shackles of the geometric anthropic principle, attempt a thousand and one geometric answers to it reminds one of the irony in Shakespeare's Hamlet words: we could be bounded in a theoretical nutshell and count ourselves kings of an infinite space of theoretical possibilities. 

\section[Experimental difficulties]{Experimental difficulties: The Planck scale}
Another instance of the grip of geometric thought on quantum-gravity theorizing is revealed by the notorious issue of the Planck scale. As many and varied as the geometric approaches to quantum gravity may be, a detached observer is greatly surprised to find that all theoretical pictures nevertheless agree on a seemingly elemental point: the relevance of the Planck scale. The observer's surprise mounts as he hears of the pettiness of this scale, and wonders why all these endeavours into the unknown should impose such stringent limits on their area of relevance, awkwardly cornering their experimental hopes to the probing of the secluded, far-off depths of space and time (and heights of energy). Does this agreement stem from a shred of dependable \emph{shared knowledge} obtained via observation of the physical systems of interest? Or does it rather stem---in the absence of any guiding observations or even positive identification of the physical system under study---from \emph{shared ignorance} combined with the desire to yet make a statement? As we learnt in Chapter \ref{ch:Planck-scale physics}, the relevance of the Planck scale to quantum gravity is not justified by the use of dimensional, quantization, and \emph{ad hoc} analyses: their application is meaningful only \emph{after} the system studied is well known, but rather hopeless in the absence of previous knowledge of these systems.

In view of its shaky foundations, why is yet so much thrust put into the essentialness of this scale? A deeper answer may be here considered by taking into account the psychological make-up of the human researcher. The human mind, as it seeks to reach out beyond the limits of knowledge, craves nonetheless the familiarity of mechanical concepts and intuition, the ease of geometric tools, and the comfort and safety of geometric containment. In this particular instance, this led to its trapping itself in the marginality of a $10^{-105}$-cubic-metre cage---and intellectual sanctuary, as it were---from which it is unwilling to peek out, much less escape. In this light, the Planck scale stands today as the \emph{leitmotiv} of human beings' psychological needs as they attempt new physics. It is the scale that came most handy to do a job already in dire need of being filled---if the gravitational constant $G$, Planck's constant $h$, and the speed of light $c$ had not combined to produce it, another scale would have been ordained to take the vacant role.

But, alas! So long as we continue to be ushered to ``get in lane now'' by following either one of the ``two obvious routes to quantum gravity'' \cite{Chalmers:2003} hand in hand with the Planck scale and its presumed phenomenology, and so long as we rejoice in not being ``lost in a multitude of alternative theories'' because we have ``two well developed, tentative theories of quantum gravity'' \cite[p.~37]{Rovelli:2003}; \label{Rovelli-quote} in other words, so long as reiteration and endorsement of conventional ideas by large communities continues to be taken as a certificate of innovativeness and worthiness, and so long as self-assured exposition of physically uncertain ideas continues to be mistaken for relevance and correctness, we shall remain far from the genuine revolution in physical thought whose absolute necessity is yet the inclination of main-road Planck-scale-bound researchers to insistently proclaim.

But while the value of this attitude to physics is doubtful, its extra-physical values are clearer, for, as Hardin \citeyear{Hardin:1993} has pointedly put it, ``In the learned world, as in others, repetition and self-advertizement contribute mightily to the amassing of a reputation'' (p.~223). After all, as another critic of science said, ``in science,\ldots despite appearances, a discovery needs credentials louder than its own merit'' \cite[p.~50]{Lem:1984}.

\chapter{Beyond geometry}
\label{ch:Beyond geometry}
\begin{flushright}
\begin{tabular}{p{10cm}} 
\emph{Describing the physical laws without reference to geometry is similar to describing our thought without words.}\smallskip \\
Albert Einstein, ``How I created the theory of relativity''
\end{tabular}
\end{flushright}

\bigskip

Throughout the preceding chapters, we witnessed how essential and ubiquitous geometry is in human theorizing. This notwithstanding, it is even this cherished realm that, as early as over a century ago, some proposed may need to be transcended in the search for new physics---whether it be a better theory of spacetime structure or, more generally, the uncovering of a deeper layer of the nature of things. To this rare attempt to go beyond geometry, stated in each case with varying degrees of explicitness, belong the views espoused by Clifford in 1875, by Eddington in 1920, by Wheeler in the decades between 1960 and 1980, and by Einstein in the latter part of his life. 

Of these four scholars, only Wheeler makes an explicit case for the superseding of geometry in physics with his well-known proposals for a pregeometry. Clifford and Eddington, on the other hand, do not make a very explicit case for transcending geometric thought, although one can see the seeds of this idea in their works, too. Einstein, for his part, does not take issue with geometry in particular, but worries in general about the nefarious effects of being unconsciously ruled by our well-established conceptual tools of thought, whatever these may be.

Einstein's reflections bring up the question of the influence of language on thought, a notorious issue that has been most famously dealt with by linguist Benjamin L.\ Whorf. His views, when conservatively interpreted in the light of the unchallenged reign of geometry in physics, provide a new angle to understand the orthodox behaviour of physicists currently engaged in the creation of new physics and, in particular, to uncover another reason behind the failure of pregeometry to follow the directions marked out by its own road map.    

\section{Clifford's elements of feeling} \label{CEF}
Views that seem to advocate the provisional character of geometric explanations in physical theories date at least as far back as 1875. For then, not only did Clifford express his better-known belief that matter and its motion are in fact nothing but the curvature of space---and thus reducible to geometry---but also lesser-known ideas about matter and its motion---and so perhaps, indirectly, geometry---being, in turn, only a partial aspect of ``the complex thing we call feeling.''

Within the context of a general investigation in terms of vortex-atoms, the mechanical ether, and other speculative ideas of the day---ideas which the modern physicist feels arcane but which, at the same time, give rise in him to the ominous thought whether this is the way most of today's speculation will look a century from now---we rescue an idea that we still find meaningful today. Clifford \citeyear[pp.~172--173]{Clifford:1886a} asked about the existence of something that is not part of the ``material or phenomenal world'' such as matter and its motion, but that is its ``non-phenomenal counterpart,'' something to which one can ascribe existence independently of our perceptions. Clifford wrote:
\begin{quote}
The answer to this question is to be found in the theory of sensation; which tells us not merely that there is a non-phenomenal counterpart of the material or phenomenal world, but also in some measure what it is made of. Namely, the reality corresponding to our perception of the motion of matter is an element of the complex thing we call feeling. What we might perceive as a plexus of nerve-disturbances is really in itself a feeling; and the succession of feelings which constitutes
a man's consciousness is the reality which produces in our minds the perception of the motions of his brain. These elements of feeling have relations of nextness or contiguity in space, which are exemplified by the sight-perceptions of contiguous points; and relations of succession in time which are exemplified by all perceptions. Out of these two relations the future theorist has to build up the world as best he may. \cite[p.~173]{Clifford:1886a}
\end{quote}
In an essay published three years later, Clifford \citeyear{Clifford:1886b} spells out the meaning he attributes to the phrase ``elementary feeling'' and the logical inference leading him to see them as the primordial constituents of the world. He argues that as the complex feelings that constitute our consciousness always occur in parallel with material nerve-disturbances in the brain, and because the evolution of life forms a continuum from inorganic matter to the highest life forms, a non-arbitrary line cannot be drawn only past which one is justified in inferring the existence of conscious feelings in other organisms---i.e.\ ejects, facts ejected from one's consciousness and attributed to others. Clifford concludes that ``every motion of matter is simultaneous with some ejective fact or event which might be part of a consciousness'' (p.~284), but that in its most elementary form is not and does not need a consciousness to support it. He arrives thus at the idea of an elementary feeling as something non-phenomenal, something ``whose existence is not relative to anything else'' (p.~284).       

Clifford pursued after this fashion the conviction that matter and motion do not lie at the bottom of things, but that, on the contrary, they are themselves an aspect of different entities---elementary feelings---and their mutual relations, contiguity and succession, of which the contiguity and succession of matter are an image. If one so chooses, both the said basic constituents and their two relations may be understood to be geometry-unrelated, as neither form nor size is attributed to the former or to the latter. In fact, this we shall do when, in Chapter \ref{ch:MQM}, we take Clifford's dictum in the last sentence quoted above in earnest, and try ``to build the quantum-mechanical world as best we may'' out of metageometric physical things, bearing a resemblance to conscious feelings, and the two relations of nextness and succession reinterpreted in a strictly metageometric sense; in particular, without any reference to space or time. It is in these elements of feeling, then, in these things beyond matter and possibly geometry, that Clifford believed one can catch a glimpse of a deeper layer of the constitution of the world. 

When in addition we consider Clifford's \citeyear{Clifford:1886a} further words indicating a possible way to apply the idea of ``contiguity of feelings in space,'' namely, ``There are lines of mathematical thought which indicate that distance or quantity may come to be expressed in terms of \emph{position} in the wide sense of \emph{analysis situ}'' (p.~173), the allusion to dispensing with typically geometric concepts such as distance becomes clearer still. One may then say that in Clifford we find the (perhaps unwitting) germ of the idea to move away from geometric concepts in physics---indeed, a sort of forefather of the later idea of pregeometry.

\section{Eddington's nature of things}
Forty-five years later, Eddington \citeyear[pp.~180--201]{Eddington:1920} expressed views in a similar spirit as above regarding what he called the nature of things, while, at the same time, more explicitly casting doubt on the extent of the powers of geometric theorizing. In analysing general relativity, he identified the point-events as its basic elements, and the interval as an elementary relation between them. As to the nature of the latter, Eddington wrote:
\begin{quote}
Its [the interval's] geometrical properties\ldots can only represent one aspect of the relation. It may have other aspects associated with features of the world outside the scope of physics. But in physics we are concerned not with the nature of the relation but with the number assigned to express its intensity; and this suggests a graphical representation, leading to a geometrical theory of the world of physics. \cite[p.~187]{Eddington:1920}
\end{quote}
Along these lines (in agreement with the idea that size induces shape), he suggested that the \emph{individual} intervals between point-events probably escape today's scales and clocks, these being too rudimentary to capture them. As a consequence, in general relativity one only deals with macroscopic values composed of many individual intervals. ``Perhaps,'' Eddington ventured further in an allusion to transcending geometric magnitudes,
\begin{quote}
even the primitive interval is not quantitative, but simply 1 for certain pairs of point-events and 0 for others. The formula given [$\mathrm{d}s^2=g_{\mu\nu}\mathrm{d}x^\mu \mathrm{d}x^\nu$] is just an average summary which suffices for our coarse methods of investigation, and holds true only statistically. \cite[p.~188]{Eddington:1920}
\end{quote}
One should not be confused by the association of the numbers 1 and 0 to the primitive intervals and think that such numbers represent their lengths. In the passage, the \emph{non-quantitativeness} of the primitive intervals is clearly stated. What the 1 and 0 represent is merely whether the intervals exist or not, any other designation being equally satisfactory for the purpose. Eddington's insight into the need to transcend geometric notions is reinforced by his further remarks: ``[W]e can scarcely hope to build up a theory of the nature of things if we take a scale and a clock as the simplest unanalysable concepts'' (p.~191). It is nothing short of remarkable to find expressed such views, on the one hand, inspired in the general theory of relativity while, on the other, in a sense contrary to its spirit, only five years after the publication of the latter. Eddington's avowal, however, only seems to be for the overthrow of geometric magnitudes but not of geometric objects, since he does not seem to find fault with the concepts of point-events and intervals as lines.

Eddington also stressed the role of the human mind in the construction of physical theories. He argued that, out of the above primitive intervals (taken as elements of reality and not of a theory of it) between point-events, a vast number of more complicated qualities can arise; as a matter of fact, however, only certain qualities of all the possible ones do arise. Which qualities are to become apparent (in one's theories) and which not depends, according to Eddington, on which aspects of these elementary constituents of the world (and not of a theory of it) the mind singles out for recognition. ``Mind filters out matter from the meaningless jumble of qualities,'' he said,
\begin{quote}
as the prism filters out the colours of the rainbow from the chaotic pulsations of white light. Mind exalts the permanent and ignores the transitory\ldots\ Is it too much to say that mind's search for permanence has created the world of physics? So that the world we perceive around us could scarcely have been other than it is? \cite[p.~198]{Eddington:1920}\footnote{Eddington talks about the mind creating the world of physics, but one finds such a thing a daring exploit, for how could the mind create the physical world? Or to put it in the words of Devitt and Sterelny \citeyear{Devitt/Sterelny:1987}, ``how could we, literally, have made the stars?'' (p.~200). Rather, what the mind creates is \emph{theories of} the world and into these inventions it puts, naturally, all its partialities and prejudices.}
\end{quote}

In these words, we find the echo of a provoking statement we made earlier on, namely, that ``snowflakes are not inherently hexagonal,'' but that it is our minds which, to use now Eddington's phrasing, filter out their geometry from what are to us a remaining jumble of indiscernible qualities. Mind exalts geometry and ignores the non-geometric---so much so that no word is apt to describe that which is ignored other than by clumsy negation. Mind exalts and filters out geometry because it is what it knows how to do, just as a prism naturally decomposes white light according to the frequency of its component parts. White light, however, may by the use of other suitable instruments be understood not in terms of component frequencies, but in terms of its intensity or its polarization. Since we only have one kind of intellect at our disposal, in order to find out what may be the possible counterparts of intensity and polarization in this comparison, we must learn to apply the mind to the physical world in new ways. 

As for the way things stand today, paraphrasing Eddington's last-quoted question, we ask: is it too much to say that \emph{mind's geometric instinct} has created all the current theories of the physical world? So that our \emph{theories of the world} could scarcely have been other than they are?

\section{Wheeler's pregeometry} \label{WP}
Wheeler \citeyear{Wheeler:1964,Wheeler:1980}, Misner et al.\ \citeyear{Misner/Thorne/Wheeler:1973}, and Patton and
Wheeler \citeyear{Patton/Wheeler:1975} expressed pioneering ideas on what they called pregeometry already four decades ago. Since we understand Wheeler to be by far the main contributor to this idea, the kind of pregeometry analysed in this section is called Wheeler's pregeometry. In order to avoid confusion, we remark yet again---in addition to the early comments of Chapter \ref{ch:Pregeometry}---that Wheeler's pregeometry is not really what later on came to be known by the same name, i.e.\ pre-continuum physics, but the more radical stance of going beyond geometry in a proper sense.

His basic demand amounts to the rejection of geometric concepts in order to explain geometric structure. Indeed, Wheeler \citeyear{Wheeler:1980} advocated ``a concept of pregeometry that breaks loose at the start from all mention of geometry and distance,'' he was wary of schemes in which ``too much geometric structure is presupposed to lead to a believable theory of geometric structure,'' and he clearly recognized that ``to admit distance at all is to give up on the search for pregeometry'' (pp.~3--4). This means that one surprisingly finds the very inventor of pregeometry generally invalidating all the schemes analysed in Chapter \ref{ch:Pregeometry}, as far as they are to be an expression of his original pregeometry.

What are the grounds for these strong pronouncements of Wheeler's? Firstly, he envisaged,\footnote{See \cite[p.~547]{Patton/Wheeler:1975} and \cite[p.~1201]{Misner/Thorne/Wheeler:1973}.} among other things, spacetime collapse in the form of the presumed big bang and big crunch, black holes, and a supposed foam-like structure of spacetime on small scales, as indicators that spacetime cannot be a continuous manifold, since it has the ability to become singular. Feeling for these reasons that the spacetime continuum needed rethinking, he posed the question:
\begin{quote}
If the elastic medium is built out of electrons and nuclei and
nothing more, if cloth is built out of thread and nothing more, we
are led to ask out of what ``pregeometry'' the geometry of space
and spacetime are built. \cite[p.~1]{Wheeler:1980}
\end{quote}
A reading of Wheeler suggests that the reason why he seeks a ``pregeometric building material'' \cite[p.~1203]{Misner/Thorne/Wheeler:1973} in order to account for the spacetime continuum instead of simply a ``discrete building material'' is that he believes that genuine explanations about the nature of something do not come about by explicating a concept in terms of similar ones, but by reducing it to a different, more basic kind of object. This is evidenced in the passage: ``And must not this something, this `pregeometry,' be as far removed from geometry as the quantum mechanics of electrons is far removed from elasticity?'' (p.~1207).

What were Wheeler's proposals to implement this programme to go beyond geometry? His attempts include pregeometry as binary-choice logic, pregeometry as a self-referential universe, and pregeometry as a Borel set or bucket of dust.\footnote{See \cite[pp.~157--161]{Demaret/Heller/Lambert:1997} for an independent review of Wheeler's pregeometry.}

An early attempt at pregeometry based on binary choice was Wheeler's ``sewing machine,'' into which rings or loops of space were fed and connected or left unconnected by the machine according to encoded binary information (yes or no) for each possible pair of rings (pp.~1209--1210). Wheeler considered the coded instructions to be given from without either deterministically or probabilistically, and then asked whether fabrics of different dimensionality would arise, and whether there would be a higher weight for any one dimensionality to appear. As a design for pregeometry, this idea succeeds in staying clear of geometric magnitudes but fails, at least superficially, to avoid geometric objects: Wheeler's loops of space have no size and yet are \emph{rings}. We say this failure is superficial because these rings or loops, so long as they have no size, could as well be called abstract elements. If nothing else, the mention of rings is witness to the fact that geometric modes of thought are always lurking in the background.

A subsequent idea still based on binary-choice logic was pregeometry as the calculus of propositions (pp.~1209, 1211--1212). This conception was much less picturesque and rather more abstract. Wheeler exploited the isomorphism between the truth-values of a proposition and the state of a switching circuit to toy with the idea that `` `physics' automatically emerges from the statistics of very long propositions, and very many propositions'' \cite[p.~598]{Patton/Wheeler:1975} in thermodynamic analogy. In that reference and in \cite{Wheeler:1980}, however, the expectation that pure mathematical logic \emph{alone} could have anything whatsoever to do with providing a foundation for physics was sensibly acknowledged as misguided. In this case, perhaps due to its abstractness and total divorcement with physics, this conception of pregeometry is truly free of all geometry in a strict sense. For all this scheme was worth, it is at least rewarding to find it has this feature, instead of quickly proceeding to negate its intended pregeometric nature.

Later on, inspired in self-referential propositions, Wheeler conceived of the idea of pregeometry as a self-referential universe. As with the previous approaches, Wheeler \citeyear{Wheeler:1980} admitted to having no more than a vision of how this understanding of geometry in terms of (his) pregeometry might come about. Two basic ingredients in this vision appear to be (i) events consisting of a primitive form of quantum principle: investigate nature and create reality in so doing (see below) and, more intelligibly, (ii) stochastic processes among these events producing form and dependability out of randomness. His vision reads thus:
\begin{quote}
(1) Law without law with no before before the big bang and no after after collapse. The universe and the laws that guide it
could not have existed from everlasting to everlasting. Law must have come into being\ldots\ Moreover, there could have been no message engraved in advance on a tablet of stone to tell them how to come into being. They had to come into being in a
higgledy-piggledy way, as the order of genera and species came into being by the blind accidents of billions upon billions of
mutations, and as the second law of thermodynamics with all its dependability and precision comes into being out of the blind
accidents of motion of molecules who would have laughed at the second law if they had ever heard of it. (2) Individual events. Events beyond law. Events so numerous and so uncoordinated that flaunting their freedom from formula, they yet fabricate firm form. (3) These events, not of some new kind, but the elementary act of question to nature and a probability guided answer given by nature, the familiar everyday elementary quantum act of observer-participancy. (4) Billions upon billions of such acts giving rise, via an overpowering statistics, to the regularities of physical law and to the appearance of continuous spacetime. \cite[pp.~5--6]{Wheeler:1980}
\end{quote}
Earlier, Patton and Wheeler \citeyear[pp.~556--562]{Patton/Wheeler:1975} had elaborated further on this. They considered two alternatives to the problem
of spacetime collapse, namely: (a) taking into account quantum mechanics, the event of universal collapse can be viewed as a ``probabilistic scattering in superspace.'' In this way, the universe is reprocessed at each cycle, but the spacetime manifold retains its fundamental mode of description in physics. This alternative was not favoured by these authors, but instead (b) considering as an overarching guiding principle that the universe must have a way to come into being (cosmogony), they suggested that such a requirement can be fulfilled by a ``quantum principle'' of ``observer-participator.''\footnote{This principle is tantamount to the first ingredient (i) above, and perhaps helps to clarify it.} According to it, the universe goes through one cycle only, but only if in it a ``communicating community'' arises that can give meaning to it. This universe is thus self-referential, which was the standpoint favoured by these authors.\footnote{This latter proposal seems to confuse nature with our theories of it. It is indeed correct to say that only a communicating community of beings can give meaning to the displays of nature, but it is rather something else to contend that a communicating community creates the world in so doing; once again: ``how could we, literally, have made the stars?''} As far as pregeometric ideas are concerned, this vision of a self-referential cosmology succeeds in eschewing all kinds of geometric concepts again due to its being conceived in such broad and general terms. Regardless of its worth and in the general context of paradoxical geometric pregeometry, let this be a welcome feature of Wheeler's effort.

Regardless of Wheeler's actual implementations of his pregeometry and any criticisms thereof, the imperative demand that the foundations of such a theory should be completely free of geometric concepts remains unaltered. The fact that so many (including, to the smallest extent, Wheeler himself) have misconstrued his idea of a programme for pregeometry is a tell-tale sign of something beyond simple carelessness in the reading of his views. Indeed, the large-scale misinterpretation witnessed in Chapter \ref{ch:Pregeometry} may just as well stem from the lack of all background for thought that one encounters as soon as one attempts to dispense with geometry. Physics knows of no other mode of working than geometry, and thus physicists reinterpreted Wheeler's new idea in the only way that made sense to them---\emph{setting out to find geometry in pregeometry.}

\subsection{More buckets of dust} \label{sec:bucket}
As two final claimants to pregeometry, we turn to the works of Nagels and of Stuckey and Silberstein in the light of their close connection with another, partly new approach to pregeometry by Wheeler. In the works of these authors, two independent attempts are made to build again ``space as a bucket of dust'' after Wheeler's \citeyear[pp.~495--499]{Wheeler:1964} earlier effort. This earlier effort, clearly connected with the above-mentioned ``sewing machine'' in character and purpose, consisted in starting with a Borel set of points without any mutual relations whatsoever and assembling them into structures of different dimensionality on the basis of different quantum-mechanical probability amplitudes attributed to the
relations of nearest neighbour between the points. Wheeler dismissed his own trial because, among other reasons, the same quantum-mechanical principles used to define adjacency rendered the idea untenable, since points that had once been neighbours would remain correlated after departing from each other (p.~498). We observe once more that this idea, in its use of the concept of points (cf.\ earlier rings), is not free from geometric objects, but that this flaw is minimized by Wheeler's avoidance of the introduction of any quantitative geometry. As for the mentioned quantum amplitudes, these need not have geometric meaning as long as they do not arise from an inner product of state vectors. In Table \ref{summary2}, we summarize the use of basic objects and relations in the non-geometric and partly non-geometric approaches we have discussed.

\begin{table} 
\centering
\begin{tabular}{|l|c|c|} 
\hline
\textbf{Author} & \textbf{Basic objects} & \textbf{Basic relations}\\
\hline \hline
Clifford & feelings & nextness, succession \\ \hline
Eddington & \emph{points}, \emph{intervals} & statistical interaction \\ \hline
\multicolumn{3}{|l|}{Wheeler}\\ \hline
\quad Sewing machine & \emph{rings} & nextness (bin.~choice) \\ \hline
\quad Borel set & \emph{points} & nextness (prob.~amp.) \\ \hline
\quad Calculus of propositions & propositions & statistical interaction \\ \hline
\quad Self-referential universe & act of observation & idem \\ \hline
\end{tabular}
\caption[Basic objects and relations in beyond-geometry proposals]{Summary of the use of basic objects and relations in the non-geometric or partly non-geometric approaches analysed. Geometric concepts appear in italics. Abbreviations: Binary (bin.), probability amplitude (prob.~amp.).}
\label{summary2}
\end{table}

Nagels' \citeyear{Nagels:1985} attempt started off with a scheme similar to Wheeler's, where ``the only structure imposed on individual points is a uniform probability of adjacency between two arbitrarily chosen points'' (p.~545). He only assumed that these probabilities are very small and that the total number of points is very large (and possibly infinite). He thus believed his proposal to satisfy the always desired requirement of ``a bare minimum of assumptions'' and very natural ones at that. However, much too soon he fell prey to quantitative geometry. Given this is a pregeometric framework directly inspired in Wheeler, it is startling to come across the following early remarks:
\begin{quote}
Without a background geometry, the simplest way to \emph{introduce distance} is to specify whether or not two given points are ``adjacent.'' [\ldots] We may then say that two ``adjacent'' points are a distance of 1 unit of length apart. \cite[p.~546, italics added]{Nagels:1985}
\end{quote}
For his part, Stuckey introduced in his attempt 
\begin{quote}
a pregeometry that provides a metric and dimensionality over a Borel set (Wheeler's ``bucket of dust'') without assuming probability amplitudes for adjacency. Rather, a non-trivial \emph{metric} is produced over a Borel set $X$ per a uniformity base generated via the discrete topological group structures over $X$. \cite[p.~1, italics added]{Stuckey:2001}
\end{quote}

Thus, both authors managed to create pregeometric metrics, i.e.\ metrics that will give a \emph{notion of distance for pregeometry} and by means of which they hoped to obtain the usual spacetime metric. This assertion is supported by Stuckey and Silberstein's \citeyear{Stuckey/Silberstein:2000} words: ``our pregeometric notion of distance'' (p.~9) and ``the process by which this metric yields a spacetime metric with Lorentz signature must be obtained'' (p.~13). Stuckey \citeyear{Stuckey:2001} moreover asserted that the pregeometries of Nagels, Antonsen, and Requardt have, arguably, overcome the difficulties represented by the presupposition of ``too much geometric structure'' as previously brought up by Wheeler, since their assumptions are minimal and ``the notion of length per graph theory is virtually innate'' (p.~2). Arguably, indeed, because the assumption of length is totally contrary to the spirit of pregeometry according to its very inventor, and its introduction acts to geometrically objectify what could have otherwise been pregeometric schemes.

Alas, what has become of Wheeler's reasonable dictum: ``to admit distance at all is to give up on the search for pregeometry,'' or his demands for ``break[ing] loose at the start from all mention of geometry and distance,'' and for pregeometry to be ``as far removed from geometry as the quantum mechanics of electrons is far removed from elasticity?'' Or yet to quote Wheeler in two other early illuminating passages:
\begin{quote}
One might also wish to accept to begin with the idea of a distance, or edge length, associated with a pair of these points,
even though this idea is already a very great leap, and one that one can conceive of later supplying with a natural foundation of its own.

[\ldots] [T]he use of the concept of distance between pairs of points seems unreasonable\ldots [L]ength is anyway not a natural idea with which to start. The subject of analysis here is ``pregeometry,'' so the concept of length should be derived, not assumed \emph{ab initio}. \cite[pp.~497, 499]{Wheeler:1964}
\end{quote}
Has all this simply gone into oblivion?

The question, indeed, remains---why the need of a metric for pregeometry? Should not pregeometry produce the traditional spacetime metric by non-geometric means alone? Should it not go even ``beyond Wheeler'' and do without any sort of geometric objects too? This state of affairs comes to show to what extent the craving of the human mind for geometric explanations goes; at the same time, one catches a glimpse of the psychological difficulties that might be encountered in any attempt that genuinely attempts to go beyond such explanations, so deeply rooted in the mind.

\section{Einstein's conceptual tools of thought}
Einstein, alongside his contributions to physics, also entertained thoughts about the validity of the ``conceptual tools of thought'' we use to portray the world. He begins the foreword to Max Jammer's \emph{Concepts of space} with a crisp and clear reflection, which we here quote at length:
\begin{quote}
In the attempt to achieve a conceptual formulation of the confusingly immense body of observational data, the scientist makes use of a whole arsenal of concepts which he imbibed practically with his mother's milk; and seldom if ever is he aware of the eternally problematic character of his concepts. He uses this conceptual material, or, speaking more exactly, these conceptual tools of thought, as something obviously, immutably given; something having an objective value of truth which is hardly ever, and in any case not seriously, to be doubted. How could he do otherwise? [\ldots] And yet in the interests of science it is necessary over and over again to engage in the critique of these fundamental concepts, in order that we may not unconsciously be ruled by them. This becomes evident especially in those situations involving development of ideas in which the consistent use of the traditional fundamental concepts leads us to paradoxes difficult to resolve.\footnote{In \cite[pp.~xi--xii]{Jammer:1969}.}
\end{quote}
We sympathize deeply with these words, which eloquently bring forth one of the underlying themes of this work when we identify the conceptual tools of thought referred to with our ingrained and ubiquitous geometric concepts.
  
Einstein indirectly questioned certain geometric concepts when, as Pais \citeyear[p.~347]{Pais:1982} recounts, in the early 1940s he considered whether partial differential equations may prove unsuitable for a more elemental description of nature. This idea appears to have been inspired in the face of quantum mechanics. Einstein \citeyear[pp.~148, 156]{Einstein:1935} notes that quantum theory does away with the possibility of describing the world fundamentally in terms of spatiotemporal fields subject to causal laws given by differential equations, since quantum-mechanical systems are described by state functions that are not fields in spacetime and whose behaviour is not strictly causal. This view is relevant in this connection when we consider that transcending the field concept together with differential equations means leaving behind the qualitative geometry inherent in spatiotemporal points $P$ (locality and causality; cf.\ field as $f(P)$) and the quantitative geometry inherent in the values (physical quantities) of the field.\footnote{See also Einstein's quotation on page \pageref{Einstein-field}.} Although J.~Ellis \citeyear{Ellis:2005}, referring to Einstein's idea, correctly asserts that ``Modern theorists can hardly be accused of excessive conservatism, but even they have not revived this startling speculation!''(p.~56), it is even this startling notion of leaving fields and differential equations behind that should be part of a plan to do without geometry entirely. 

Finally, in this context of reservation towards the limitations imposed by our tools of thought, it is difficult to know how to interpret a deeper pronouncement of Einstein's \citeyear{Einstein:1982} (which opens this chapter) now in direct connection with geometric concepts in general. Explaining how he had created the theory of relativity, he said: ``Describing the physical laws without reference to geometry is similar to describing our thought without words'' (p.~47). Is Einstein here suggesting that, despite being possibly ruled by it, doing physics without the assistance of a geometric language is as hopelessly inviable as describing our thoughts without words, or does he mean to say that it would be as hard, yet not necessarily fruitless, to do so? Perhaps it is wisest not to seek too much meaning in these otherwise remarkable words and simply take them at face value, as a correct indication of the fact that geometry is indeed the only physical language so far available---dispensing with it, therefore, leaves us wordless.   

However, this circumstance need not be as inescapable as it seems. Einstein's comparison only holds true until the non-geometric words of a non-geometric language are invented and become available, allowing our physical thoughts to be expressed anew in terms of them. Further on in this thesis, we shall see that the task of going beyond geometry, although mentally strenuous, is not only feasible but moreover essential to better comprehend some aspects of the physical world.

\section{Whorf's linguistic relativity}
Linguist Benjamin L.\ Whorf is best known for his now notorious principle of linguistic relativity. A judicious interpretation of this principle provides us with quite a straightforward framework to understand the source of Einstein's apprehensions regarding our conceptual tools of thought and our own remarks that thought beyond geometry can be buttressed by a supporting language: language, whether it be scientific or otherwise, is apt to greatly influence thought both positively and negatively. Whorf wrote:
\begin{quote}
It was found that the background linguistic system\ldots of each language is not merely a reproducing instrument for voicing ideas but rather is itself the shaper of ideas\ldots We dissect nature along lines laid down by our native languages. The categories and types that we isolate from the world of phenomena we do not find there because they stare every observer in the face; on the contrary, the world is presented in a kaleidoscopic flux of impressions which has to be organized by our minds---and this means largely by the linguistic systems in our minds.

This fact is very significant for modern science, for it means that no individual is free to describe nature with absolute
impartiality but is constrained to certain modes of interpretation even while he thinks himself most free\ldots We are thus
introduced to a new principle of relativity, which holds that all observers are not led by the same physical evidence to the same picture of the universe, unless their linguistic backgrounds are similar, or can in some way be calibrated. \cite[pp.~212--214]{Whorf:1956}
\end{quote}
One ought to be careful as to what to make of Whorf's remarks, as their significance could go from (i) an implication of an
\emph{immutable constraint} of language upon thought with the consequent fabrication of an unavoidable world picture to which
one is led by the language used, to (ii) milder insinuations about language being ``only'' a \emph{shaper of ideas} rather than a tyrannical master. How far is the influence of language on thought to be taken to go?

In this, we share the views of Devitt and Sterelny's. In their book, they  explained the circular---although not viciously so---process in which thought and language interact; they wrote:
\begin{quote}
We feel a pressing need to understand our environment in order to manipulate and control it. This drive led our early ancestors, in time, to express a primitive thought or two. They grunted or gestured, \emph{meaning something by} such actions. There was speaker meaning without conventional meaning. Over time the grunts and gestures caught on: linguistic conventions were born. As a result of this trail blazing it is much easier for others to have those primitive thoughts, for they can learn to have them from the conventional ways of expressing them. Further, they have available an easy way of representing the world, a way based on those conventional gestures and grunts. They borrow their capacity to think about things from those who created the conventions. With primitive thought made easy, the drive to understand leads to more
complicated thoughts, hence more complicated speaker meanings, hence more complicated conventions. \cite[p.~127]{Devitt/Sterelny:1987}
\end{quote}
In the first place, this means that, as one could have already guessed, thought must precede any form of language, or else how could the latter have come into being? Secondly, it shows that thought, as a source of linguistic conventions, is not in any way restricted by language, although the characteristics of the latter do, in fact, \emph{facilitate} certain forms of thought by making them readily available, in the sense that the thought processes of many can now benefit from existing concepts already made by a few others. This is especially the case in science, which abounds in instances of this kind. For example, thought about complex numbers is facilitated by the already invented imaginary-number language convention ``$i=\sqrt{-1}$,'' just like the thought of a particle being at different places at the same time is facilitated by the already invented state-vector-related conventions of quantum mechanics.

At the same time, an existing language can also \emph{discourage}
certain thoughts by making them abstruse and recondite to express \cite[p.~174]{Devitt/Sterelny:1987}. This does not mean that certain things cannot be thought---they eventually can be---but only that their expression does not come easily as it is not straightforwardly supported by the existing language. As an example of this instance, we can mention the reverse cases of the above examples. That is to say, the evident difficulty of conjuring up any thoughts about complex numbers and delocalized particles \emph{before} the formal concepts above supporting these thoughts were introduced into the language of science by some specific individuals. The fact, however, that such linguistic conventions were created eventually shows that thought beyond linguistic conventions is possible---essential, moreover, for the evolution of science.

One is thus led to the view that the picture that physics makes of nature is not only guided by the physicist's imagination, but also by the physicist's language---with its particular vices and virtues. This language, due to its constitution attained after a natural development, favours geometric modes of thought, while it appears to dissuade any sort of non-geometric contemplation. Pregeometric schemes have shown this fact clearly: imagination was able to produce varied creative frameworks, but they all spoke the same geometric language despite the attempt to avoid it. Only against this \emph{language-favoured background} did physicists' minds roam effortlessly.

The reason why humans have naturally developed primitive geometric thoughts, leading to geometric modes of expression, in turn reinforcing more geometric-like thinking, and so forth---in conclusion, developing a geometric understanding---can only be guessed at. It is reasonable to suppose that geometric insight should be connected with the natural provision of an early evolutionary advantage, rather than with a specially designed tool to devise physical theories. But as little as geometric insight may have been designed for high forms of abstract conceptualization, it is the only insight we can \emph{effortlessly} avail ourselves of, and so the prospect of dispensing with geometric thought in physical science is not encouraging, for when geometry is lost, much is lost with it. We close this section, then, with the encouraging words of others who also believed that, mentally strenuous as it might be, this endeavour is worth pursuing:
\begin{quote}
And is not the source of any dismay the apparent loss of guidance that one experiences in giving up geometrodynamics---and not only geometrodynamics but geometry itself---as a crutch to lean on as one hobbles forward? Yet there is so much chance that this view of nature is right that one must take it seriously and explore its consequences. Never more than today does one have the incentive to explore [Wheeler's] pregeometry. \cite[p.~1208]{Misner/Thorne/Wheeler:1973}
\end{quote}

\section{Appraisal} 
A long way has so far been travelled. Starting off in the realm of geometry, we laid down its foundations in the form of geometric objects and magnitudes, and concluded that physical theories are woven out of these conceptions because of a basic human predilection to conceptualize the world after a geometric fashion. Indeed, we witnessed the essentially geometric nature of all physical theorizing, from Babylonian clay diagrams through Ptolemaic astronomy to modern science, and continuing with the attempted improvements of twenty-first-century frontier physics despite the lack of physical counterparts for its latest geometric entities, and despite the mounting complexity of the geometric edifices built upon them. 

Furnishing a particularly conspicuous instance of quantum-gravity research, we visited the abode of pregeometry. Although judging by its name and goals, pregeometry was the last place in which one would have expected to find geometry, the former nevertheless fell prey to the latter. Pregeometry, in its need for its own geometric means to explain the origin of geometry in physics, cannot but raise the disquieting question: is the human mind so dependent upon geometric means of description that they cannot be avoided? Or has pregeometry not tried hard enough? In any case, we concluded that pregeometry has failed to live up to the semantic connotations of its name and to the original intentions of its creator, as well as to the intentions of its present-day practitioners, only to become a considerable incongruity.

Subsequently, we surveyed the vistas of a land farther beyond. Different older suggestions for the need to overcome geometry in physics were analysed, including those views pertaining to the father of pregeometry proper. In general, this proposal stemmed from the desire to reach into yet another layer of the nature of things, in conjunction with the realization that everyday and scientific normal language---i.e.\ geometric language---were likely to stand in our way. 

After this critical general survey of the geometric physical tradition---from the mists of prehistory, ancient Babylonia and Greece, through the rise and golden age of classical physics, to contemporary modern physics up to its pushing frontiers---it is now time to take a deeper look at space and time according to our best available knowledge of them. In other words, it is now time to take a step, as it were, backwards and leave the speculative doctrines of quantum gravity behind to study what classical and modern physics have to say about the physical nature of space and time. This inexorably leads us to the problem of the ether.

\chapter{The tragedy of the ether} 
\label{ch:ESE}
\begin{flushright}
\begin{tabular}{p{10cm}} 
\emph{The essence of dramatic tragedy is not unhappiness. It resides in the solemnity of the remorseless working of things.}\smallskip \\
Alfred N.~Whitehead, \emph{Science and the modern world}
\end{tabular}
\end{flushright}

\bigskip

Upon hearing the word ``ether,'' the modern physicist finds his lips almost involuntarily forming a knowing smile. The vision of nineteenth-century physicists with their convoluted theoretical pictures of this all-pervading luminiferous medium and their experimentally vain attempts to observe the motion of the earth through it comes to his mind, and he cannot but feel sorry for his precursors embarking on such a wild-goose chase. For they were, after all, only chasing a ghost: what is the material medium in which light, gravity, heat, and the electric and magnetic forces propagate? Today we know better. The luminiferous ether simply does not exist. Electromagnetic and gravitational interactions need no medium to propagate, and so we can put this notorious idea, greatly troubled since its very inception, to rest. Or can we? 

Not really. Contrary to common belief, the ether still haunts physics at the turn of a new century. Not in the form of a mechanical medium---that idea \emph{is} dead and buried---but in a different, transmuted guise. That this is so is not readily acknowledged, because the belief that only our ancestors were capable of falling prey to their own conventional conceptions overpowers all capability to look at our own position with anything resembling objectivity. Our knowing smiles, however, slowly give way to chagrin as we start to recognize ourselves in the reflection of the past. What kind of ether still haunts physics? A quick preliminary survey of the history of this concept is apt to naturally lead us to an outline of the present predicament.
 
The need for an ether as a material medium with mechanical properties first became apparent to Descartes in the first half of the seventeenth century in an attempt to avoid any actions that would propagate, through nothing, from a distance. During its history of roughly three centuries in its original conception, the idea of the ether managed to materialize in endless forms via the works of countless investigators. Its main purpose of providing a medium through which interactions could propagate remained untouched, but the actual properties with which it was endowed in order to account for and unify an ever-increasing range of phenomena were mutually incongruous and dissimilar. Never yielding to observation and constantly confronted by gruelling difficulties, the mechanical ether had to reinvent itself continually, but its very notion staggered not a bit.

After almost 300 years of bitter struggle, the mechanical ether eventually gave in. The first step of this change took place in the hands of Lorentz, for whom the ether was a sort of substantial medium that affected bodies not mechanically but only dynamically, i.e.\ due to the fact that bodies moved through it. Drude and Larmor further declared that the ether need not actually be substantial at all but simply space with physical properties. The second and decisive step in this demechanization of the ether was brought about by Poincar\'{e} and Einstein through ideas that eventually took the shape of the special theory of relativity. In Einstein's \citeyear{Einstein:1983a} own words, this change ``consisted in taking away from the ether its last mechanical quality, namely, its immobility'' (p.~11).

Far from being dead, however, the ether had only transmuted its charac\-ter---from a mechanical substance to an inertial spacetime characterized by an absolute metric structure. The need for regarding inertial spacetime as an ether came after noticing that empty spacetime, despite being unobservable and unalterable, displayed physical properties: it provided an absolute reference not for velocity but for acceleration, and it determined the behaviour of measuring rods and clocks, similar to the function Newton's ethereal absolute space had had before.

The nature of this new, special-relativistic ether underwent yet another change with the general theory of relativity, as the ether became the dynamic and intrinsic metric structure of spacetime \cite[p.~176]{Einstein:1961}. This was so significant a change that it modified the very ideas of ether and empty spacetime. By making the metric structure alterable, it actually put an end to its status as a genuine ether. And by making the metric field a content of spacetime, it did away with empty spacetime, since now to vacate spacetime meant to be left with no metric structure and, therefore, with nothing physical at all: only an amorphous, unobservable substratum of spacetime points, whose physical reality was denied by Einstein's hole argument. From this standpoint, Einstein concluded that empty spacetime cannot possess any physical properties, i.e.\ that empty spacetime does not exist.

What degree of certainty can we attach to this conclusion? Are spacetime points truly superfluous entities, such that they neither act nor are acted upon? Far from it. Spacetime points $P$ act by performing the localization of fields $f(P)$, i.e.\ by giving the location $P$ of physical properties $f$; in other words, spacetime points are part and parcel of the \emph{field conception} in physics, on which the elemental physical categories of \emph{locality} and \emph{causality} rest. And yet spacetime points cannot be acted upon such that their displacement would produce a measurable effect. So did the banishment of the second ether---the metric ether---give way to empty spacetime and the birth of a third ether---the geometric ether. 

Einstein's conclusion as to the physical unreality of empty spacetime must be viewed with caution also as a result of more general considerations. This conclusion rests solely on the classical, geometric framework of the general theory of relativity, and this is all very well, but if we seek better physics we should not restrict ourselves unduly. May quantum theory have anything new to add to the concept of spacetime points and empty spacetime? May, moreover, the adoption of a metageometric outlook reveal a deeper layer of this ontological problem, currently hiding beneath its geometric face value? Be this as it may, when heeding the history of the ether, one thing is at any rate certain: for empty spacetime to lay any claim on physical reality, the recognition of observables is an unconditional must.

\section{The mechanical ether}\label{ME}
We start by tracing the rich history of the ether, starting here from its older conception as a material substance, passing through its virtual disappearance after the progressive removal of its mechanical attributes, and ending with its new form of an immaterial metric substratum, analysed in the next section. For the historical review of this section, the very comprehensive work of Whittaker \citeyear{Whittaker:1951} will be followed as a guideline.\footnote{Page numbers in parentheses in this section are to Vol.~1 of this reference unless otherwise stated.}

Ren\'{e} Descartes (1596--1650) was the first to introduce the conception of an ether as a mechanical medium. Given the belief that action could only be transmitted by means of pressure and impact, he considered that the effects at a distance between bodies could only be explained by assuming the existence of a medium filling up space---an ether. He gave thus a new meaning to this name, which in its original Greek ($\alpha\accentset{,}{\iota}\theta\acute{\eta}\rho$) had meant the blue sky or the upper air; however, the elemental qualities the Greeks had endowed this concept with, namely, universal pervasiveness (above the lunar sphere) and indestructibility (cf.\ unaffectability),\footnote{Kostro \citeyear{Kostro:2000} writes: ``The ether filled and made up the supra-lunar universe. Earth, water, air, and fire were composed of primary matter and form, thus being capable of transformations from one into the others, whereas the ether was so perfect that it was not capable of being transformed, and was thus indestructible'' (p.~3).} remained untouched throughout its modern history. Descartes' ether was unobservable and yet it was needed to account for his mechanistic view of the universe, given his assumptions.\footnote{With the invention of the coordinate system, Descartes was, at the same time, the unwitting precursor of the later conception of the ether as a form of space.} At the same time, the notion of an ether was right from its inception entwined with considerations about the theory of light. Descartes himself explained the propagation of light as a transmission of pressure from a first type of matter to be found in vortices around stars to a second type of matter, that of which he believed the ether to be constituted (pp.~5--9).

The history of the ether continued tied to the theory of light with Robert Hooke's (1635--1703) work. In an improvement with
respect to Descartes' view, he conceived of light as a wave motion, an exceedingly small vibration of luminous bodies that
propagated through a transparent and homogeneous ether in a spherical manner. Hooke also introduced thus the fruitful idea of a wave front (pp.~14--15).

Isaac Newton (1642--1727) rejected Hooke's wave theory of light on the grounds that it could not explain the rectilinear propagation of light or its polarization (see below). In its place, Newton proposed that light consisted of rays that interacted with the ether to produce the phenomena of reflection, refraction, and diffraction, but that did not depend on it for their propagation. He gave several options as to what the true nature of light might be, one of which was that it consisted of particles---a view that later on would be associated with his name; nevertheless, as to the nature of light, he ``let every man here take his fancy.'' Newton also considered it possible for the ether to consist of different ``ethereal spirits,'' each separately suited for the propagation of a different interaction (pp.~18--20).

Regarding gravitation in the context of the universal law of attraction, Newton did not want to pronounce himself as to its nature. He nonetheless conjectured that it would be absurd to suppose that gravitational effects could propagate without the mediation of an ether. However, Newton's eighteenth-century followers gave a twist to his views; antagonizing with Cartesians due to their rejection of Newton's gravitational law, they went as far as denying the existence of the ether---originally Descartes' concept---and attempted to account for all contact interactions as actions at a distance between particles (pp.~30--31).

Christiaan Huygens (1629--1695) was also a supporter of the wave theory of light after observing that light rays that cross each other do not interact with one another as would be expected of them if they were particles. Like Hooke, he also believed that light consisted of waves propagating in an ether that penetrated vacuum and matter alike. He managed to explain reflection and refraction by means of the principle that carries his name, which introduced the concept of a wave front as the emitter of secondary waves. As to gravitation, Huygens' idea of a Cartesian ether led him to account for it as a vortex around the earth (pp.~23--28).

An actual observation that would later have a bearing on the notions of the nature of light and of the ether was that of
Huygens' regarding the polarization of light. He observed that light refracted once through a so-called Icelandic crystal, when refracted through a second such crystal, could or could not be seen depending on the orientation of the latter. Newton correctly understood this result as the first light ray being polarized, i.e.\ having properties dependent on the directions perpendicular to its direction of propagation. He then concluded that this was incompatible with light being a (longitudinal) wave, which could not carry such properties (pp.~\mbox{27--28}).

Another thoroughly Cartesian account of the ether was presented by John Bernoulli Jr.\ (1710--1790) in an attempt to provide a mechanical basis for his father's ideas on the refraction of light. Bernoulli's ether consisted of tiny whirlpools and was interspersed with solid corpuscles that were pushed about by the whirlpools but could not astray far from their average locations. A source of light would temporarily condense the whirlpools nearest to it, diminishing thus their centrifugal effects and displacing the said corpuscles; in this manner, a longitudinal wave would be started (pp.~95--96).

In the midst of a general acceptance of the corpuscular theory of light in the eighteenth century, also Leonhard Euler (1707--1783) supported the view of an ether in connection with a wave theory of light after noticing that light could not consist of the emission of particles from a source since no diminution of mass was observed. Most remarkably, Euler suggested that, in fact, the same ether served as a medium for both electricity and light, hinting for the first time at a unification of these two phenomena. Finally, he also attempted to explain gravitation in terms of the ether, which he assumed to have more pressure the farther from the earth, so that the resulting net balance of ether pressure on a body would push it towards the centre of the earth (pp.~97--99).

At the turn of the century, the wave theory of light received new support in the hands of Thomas Young (1773--1829). Within this theory, Young explained reflection and refraction in a more natural manner than the corpuscular theory and, more importantly, also accounted successfully for the phenomena of Newton's rings (and hinted at the cause of diffraction) by introducing an interference principle for light waves. It was Augustin Fresnel (1788--1827) who, in 1816 and amidst an atmosphere of hostility towards the wave theory, managed to explain diffraction in terms of Huygens' and Young's earlier findings (pp.~100--108).

Young and Fresnel also provided an alternative explanation of stellar aberration, which had been first observed by James Bradley (1692--1762) in 1728 while searching to measure stellar parallax and which had so far been explained in terms of the corpuscular theory of light. Young first proposed that such effect could be explained assuming that the earth did not drag the ether with it, so that the earth's motion with respect to it was the cause of aberration. Subsequently, Fresnel provided a fuller explanation that could also account for aberration being the same when observed through refractive media. Following Young, Fresnel suggested that material media partially dragged the ether with them in such a way that the latter would pick a fraction $1-1/n^2$ (where $n$ is the medium's refractive index) of the medium's velocity. So far the ether was viewed as a somewhat non-viscous fluid that could be dragged along in the inside of refractive media in proportion to their refractive index, and whose longitudinal excitations described light (pp.~108--113).

Considerations about the polarization of light would bring along fundamental changes to the conception of the ether. As Newton had previously observed, the properties of polarized light did not favour a longitudinal-wave theory of light. Inspired by the results of an experiment performed by Fran\c{c}ois Arago (1786--1853) and Fresnel, Young hit on the solution to the problem of polarization by proposing that light was a \emph{transverse} wave propagating in a medium. Fresnel further hypothesized that the ether must then be akin to a solid and display rigidity so as to sustain such waves \mbox{(pp.~114--117).}

The fact that only a rigid ether could support transverse waves robbed the idea of an immobile, undragged ether of much of its plausibility, since it was hard to imagine a solid medium of some sort that would not be, at least, partially dragged by bodies moving through it. George Stokes (1819--1903) rose up to this challenge by providing a picture of the ether as a medium that behaved like a solid for high-frequency waves and as a fluid for slow-moving bodies. As a fluid, Stokes' ether was dragged by material bodies such that, in particular, it was at rest relative to the earth surface \mbox{(pp.~128, 386--387).}

Michael Faraday (1791--1867) gave a new dimension to the ether conception by introducing the notion of \emph{field}, which in hindsight was the most important concept to be invented in this connection.\footnote{And this not without a sense of irony. At first, it was on the field, a physical entity existing on its own and needing no medium to propagate, that the overthrow of the mechanical ether would rest; however, later on Einstein would reinstate the ether as a (metric) field itself.} In studies of the induction of currents, of the relation of electricity and chemistry, and of polarization in insulators, he put forward the concepts of magnetic and electric lines of force permeating space. He introduced thus the concept of a field as a stress in the ether and present where its effects took place. Faraday went on to suggest---prophesying the relativistic development---that an ether may not be needed if one were to think of these lines of force, which he understood as extensions of material bodies, as the actual carriers of transverse vibrations, including light and radiant heat as well. Or then that, if there was an luminiferous ether, it might also carry magnetic forces and ``should have other uses than simply the conveyance of radiations.'' By including also the magnetic field as being carried by the ether, Faraday added now to Euler's earlier prophecy, hinting for the first time at the conception of light as an electromagnetic wave (pp.~170--197).

Another unifying association of this type was made by William Thomson (1824--1907), who in 1846 compared heat and electricity in that the isotherms of the former corresponded to the equipotentials of the latter. He suggested furthermore that electric and magnetic forces might propagate as elastic displacements in a solid. James Clerk Maxwell (1831--1879), inspired by Faraday's and W.\ Thomson's ideas, strove to make a mechanical picture of the electromagnetic field by identifying static fields with displacements of the ether (for him equivalent to displacements of material media) and currents with their variations. At the same time, Maxwell, like Gustav Kirchhoff (1824--1887) before him, was impressed by the equality of the measured velocity of light and that of the disturbances of the electromagnetic theory and suggested that light and electromagnetic waves must be waves of the same medium (pp.~242--254).

So far, the theories of Maxwell and Heinrich Hertz (1857--1894) had not made any distinction between ether and matter, with the former considered as totally carried along by the latter. These theories were still in disagreement with Fresnel's successful explanation of aberration in moving refractive media, which postulated a partial ether drag by such bodies. However, experiments to detect any motion of the earth with respect to the
ether, such as those by Albert Michelson (1852--1931) and Edward Morley (1838--1923), had been negative and lent support to Stokes' theory of an ether totally dragged at the surface of the earth (pp.~386--392).

Not content with Stokes' picture, in 1892 Hendrik Lorentz (1853--1928) proposed an alternative explanation with a theory of electrons, which reconciled electromagnetic theory with Fresnel's law. However, Lorentz's picture of the ether was that of an electron-populated medium whose parts were mutually at rest; Fresnel's partial drag was therefore not allowed by it. Lorentz's theory denied the ether mechanical properties and considered it only space with dynamic properties (i.e.\ affecting bodies because they moved through it), although still endowed it with a degree of substantiality.\footnote{See \cite[p.~18]{Kostro:2000} and \cite[pp.~201, 207]{Kox:1989}.} The negative results of the Michelson-Morley experiment were then explained by Lorentz by means of the existence of FitzGerald's contraction, which consisted in a shortening of material bodies by a fraction $v^2/2c^2$ of their lengths in the direction of motion relative to the ether. Thus, the ether would cease being a mechanical medium to become a sort of substantial dynamic space (pp.~392--405).

Near the end of the nineteenth century, the conception of the ether would complete the turn initiated by Lorentz with the views of Paul Drude (1863--1906) and Joseph Larmor (1857--1942), which entirely took away from the ether its substantiality. Drude \citeyear[p.~9]{Drude:1894} declared:
\begin{quote}
Just as one can attribute to a specific medium, which fills space
everywhere, the role of intermediary of the action of forces, one
could do without it and attribute to space itself those physical
characteristics which are now attributed to the ether.\footnote{Quoted from \cite[p.~20]{Kostro:2000}.}
\end{quote}
Also Larmor claimed that the ether should be conceived as an immaterial medium, not a mechanical one; that one should not attempt to explain the dynamic relations so far found in terms of
\begin{quote}
mechanical consequences of concealed structure in that medium; we should rather rest satisfied with having attained to their exact dynamical correlation, just as geometry explores or correlates, without explaining, the descriptive and metric properties of space.\footnote{Quoted from \cite[Vol.~1, p.~303]{Whittaker:1951}.}
\end{quote}
Larmor's statement is so remarkable that it will receive more attention later on in Section \ref{BGE}.

Despite the seeming superfluousness of the ether even taken as a fixed dynamic space, Lorentz held fast to the ether until his death, hoping perhaps that motion relative to it could still somehow be detected \cite{Kox:1989}. Others, like Poincar\'{e} and Einstein, understood the repeated failed attempts to measure velocities with respect to the ether as a clear sign that the ether, in fact, did not exist. Henri Poincar\'{e} (1854--1912) was the first to reach such conclusion; in 1899 he asserted that absolute motion with respect to the ether was undetectable by any means, and that optical experiments depended only on the relative motions of bodies; in 1900 he openly distrusted the existence of the ether with the words, ``Our ether, does it really exist?''; and in 1904 he proposed a principle of relativity (Vol.~2, pp.~30--31). 

It was Albert Einstein (1879--1955) who in 1905 provided a theory where he reinstated these earlier claims but with a new, lucid interpretational basis. In particular, Einstein considered
\begin{quote}
The introduction of a ``luminiferous ether''\ldots to be
superfluous inasmuch as the view here to be developed will not
require an ``absolutely stationary space'' provided with special
properties\ldots \cite[p.~38]{Einstein:1952a} 
\end{quote}
Thus, in the hands of Poincar\'{e} and Einstein, \emph{the ether had died}.

\section{The metric ethers} \label{EE}
Belief in the nonexistence of the ether would not last very long, for it would soon rise from its ashes---transmuted. A rebirth of the ether as an inertial medium was advocated by Einstein about a decade after 1905 on the grounds that, without it, empty space could not have any physical properties; yet it displayed them through the effects of absolute acceleration, as well as effects on measuring rods and clocks.

It is a well-known fact that all motion simply cannot be reduced to the symmetric relationship between any two reference systems. Newton \citeyear[pp.~10--12]{Newton:1962}, through the rotating-bucket and revolving-globes (thought) experiments, realized that the effect of acceleration is not relative, and that rotation in empty space is meaningful. As a result, he held on to the need for a space absolutely at rest with respect to which this absolute acceleration could be properly defined.\footnote{In fact, only a family of inertial spaces linked by Galilean transformations would have sufficed.} Einstein \citeyear[pp.~112--113]{Einstein:1952b} was impressed by the same fact, which he brought up by means of the rotating-spheres thought experiment. The conclusion forces itself upon us that some reference frames are privileged. In Newtonian space and special-relativistic spacetime, these are the inertial frames; in general relativity, these are the freely-falling frames. 

However, regarding the classical and special-relativistic cases, Einstein \citeyear[p.~113]{Einstein:1952b} complains that no ``epistemologically satisfactory'' reason that is ``an observable fact of experience'' can be given as to why or how space singles out these frames and makes them special. Speaking for a layman whose natural intelligence has not been corrupted by a study of geometry and mechanics, he says in blunt reply to the mechanist: 
\begin{quote}
You may indeed be incomparably well educated. But just as I could never be made to believe in ghosts, so I cannot believe in the gigantic thing of which you speak to me and which you call space. I can neither see such a thing nor imagine it. \cite[p.~312]{Einstein:1996a}
\end{quote}

Inertial Newtonian space and inertial special-relativistic spacetime are metric substrata, each characterized by a metric structure determined by an \emph{invariant quadratic form}\footnote{For more on the physical and mathematical meaning of this concept, see Section~\ref{sec:Invariance-line-element}.} (not just a metric). The Newtonian metric substratum is determined by 
\begin{equation}
\D l^2=E_{ab}\D x^a\D x^b \qquad (a,b=1,2,3),
\end{equation}
where $E_{ab}=\mathrm{diag}(1,1,1)$, and the special-relativistic metric substratum by 
\begin{equation}
\D s^2=\epsilon \eta_{\alpha\beta}\D x^\alpha\D x^\beta \qquad (\alpha,\beta=1,2,3,4),
\end{equation}
where $\eta_{\alpha\beta}=\mathrm{diag}(1,1,1,-1)$ and $\epsilon=\pm 1$ ensures that $\D s^2\geq 0$. In both cases, the metrics $E_{ab}$ and $\eta_{\alpha\beta}$ are constant, in the sense that coordinates can always be found such that their values do not depend on the space or spacetime coordinates $x$. Do Newtonian space and special-relativistic spacetime have anything in common with the luminiferous ether? 

Just like the luminiferous ether never displayed any property intrinsic to its material nature---most notably, the ether wind always had null relative velocity---so do these two metric substrata fail to display any non-null or non-trivial properties intrinsic to their own metric nature---most notably, they are flat and, therefore, structurally (metrically) sterile. For example, gently release two massive particles and observe their mutual separation remain constant as a result of inertial media having no \emph{active} effect on them; inertial media only act dynamically, by having bodies move through them, but not statically, as a result of only being placed in them. Further, just like the luminiferous ether acted as a medium that made the propagation of light and other interactions possible, yet had an enigmatic origin and could not be altered in any way, so do these two metric substrata act as the source of physical effects by determining inertial frames; however, because they have no sources, they cannot be acted upon in any way, and so their reason of being cannot be determined in a physical sense. The Newtonian and special-relativistic metric substrata are thus in precise correlation with the original characterization of the older mechanical ether, and we are compelled to hold them as a new, metric embodiment of the ether.

Not content with the absolute nature of special-relativistic spacetime  insofar as it could not be affected, Einstein looked for sources via which the physical properties displayed by spacetime could be influenced and thus no longer fixed and beyond reach. While dealing with the rotating-spheres thought experiment, he wrote:
\begin{quote}
What is the reason for this difference [spherical and ellipsoidal] in the two bodies? No answer can be admitted as epistemologically satisfactory, unless the reason given is an \emph{observable fact of experience}\ldots [T]he privileged space $R_1$ [inertial] of Galileo\ldots is merely a \emph{factitious} cause, and not a thing that can be observed\ldots\ The cause must therefore lie \emph{outside} this system\ldots [T]he distant masses and their motions relative to $S_1$ and $S_2$ [the spheres] must then be regarded as the seat of the causes (which must be susceptible to observation) of the different behaviour of our two bodies $S_1$ and $S_2$. They take over the role of the factitious cause $R_1$. \cite[pp.~112--113]{Einstein:1952b} 
\end{quote}
In the metric context of relativity theory, this meant searching for a spacetime whose metric structure could be changed by the distribution of matter and which had gravitation as its \emph{active} metric effect. On accomplishing this, Einstein was the first physicist since the conception of the ether to bring it to full physical accountability, i.e.\ not only to  observe its active effects but also to control its structure. In other words, after two conceptual revolutions, Einstein was the first physicist to \emph{observe and control the ether}. But this is surely a contradiction in terms, and so the ether died again. Einstein had uprooted it for the second time, but this time with the right tool---not denial, observation. 

Einstein nevertheless continued to call the general-relativistic metric substratum an ether, because he seemingly meant the word primarily in its primitive sense of a ubiquitous substratum with physical properties regardless of whether its effects had now turned from passive to active and its nature changed from absolute to alterable.\footnote{See \cite{Kostro:2000} for a useful source of historical material on Einstein and the ether.} 
 
When we say that the metric ether became fully accounted for as a consequence of the second relativistic revolution, we generally mean that the metric ether (i) became observable as a result of its non-trivial metric structure, displaying now an active geometric effect in the form of gravitation; and (ii) lost its absolute character when, through the field equation, its intimate linkage to matter as its source was established. But did the dynamic, alterable nature of the general-relativistic metric substratum, now determined by the invariant quadratic form 
\begin{equation}
	\D s^2=\epsilon g_{\alpha\beta}(x)\D x^\alpha\D x^\beta,
\end{equation}
put the metric field $g_{\alpha\beta}(x)$ on an equal footing with other physical fields? Is $g_{\alpha\beta}(x)$, in fact, nothing else than the new gravitational field? The answers to these two questions are not as straightforward as we might wish, because mathematical concepts, otherwise of aid, also blur our physical vision.
 
On the surface, the metric field $g_{\alpha\beta}(x)$ is akin to other physical fields, like the electric field $E^a(x)$ or magnetic field $B^a(x)$---or, better, to the electromagnetic tensor field $F_{\alpha\beta}(x)$. In effect, whereas matter acts as the source of the former, charge and current act as the sources of the latter. But this comparison is misleading. How could the metric tensor be the gravitational field if it is not even null in the special-relativistic case, i.e.\ when $g_{\alpha\beta}(x)=\eta_{\alpha\beta}$? Should we perhaps concede that one should not take the correspondence so literally but continue to claim that it holds, for example, by claiming something like ``the `gravitational field' is null when the `gravitational field' $g_{\alpha\beta}(x)$ is constant''? Even if we accepted this contradiction in terms, it would not be workable. The values of $g_{\alpha\beta}(x)$ change from one coordinate system to another, and even when there exists a coordinate system in flat spacetime with respect to which $g_{\alpha\beta}(x)=\mathrm{constant}$ holds, there are many other coordinate systems in which it does not hold. In other words, there is no \emph{tensor equation}, meaningful in \emph{all} coordinate systems, linking the metric tensor as such to the stress-energy tensor $T_{\alpha\beta}$, which characterizes the matter distribution in spacetime; there is \emph{no straightforward invariant connection between $g_{\alpha\beta}(x)$ and gravitation}. The same analysis applies to the view that, if not the metric field, then the Christoffel symbol $\left\{\begin{smallmatrix}\alpha\\\beta\gamma\end{smallmatrix}\right\}$ must be the gravitational field; it can be actually null (not just constant) in gravitation-free spacetime, but this value too changes from one coordinate system to another. In consequence, equations $g_{\alpha\beta}(x)\neq\mathrm{constant}$ and $\left\{\begin{smallmatrix}\alpha\\\beta\gamma\end{smallmatrix}\right\}\neq 0$ tell us nothing about the existence or nonexistence of a gravitational field. 

By looking at the actual field equation, 
\begin{equation}
	R_{\alpha\beta}(x)-\frac{1}{2} g_{\alpha\beta}(x)R(x)=-\frac{8\pi G}{c^4}
	T_{\alpha\beta}(x),
\end{equation} 
where $R_{\alpha\beta}$ is the Ricci tensor and $R$ the Ricci scalar, we notice that, if a correspondence with a ``gravitational field'' must be made, it is in general the \emph{curvature tensor} $R^\delta_{\ \alpha\beta\gamma}$ (of which the Ricci tensor and curvature scalar are contractions) that should be identified with it. It is only the curvature tensor, a complicated combination of second derivatives and products of first derivatives of the metric tensor,\footnote{The physical dimensions $[x^\alpha]^{-2}$ of curvature arise from the dimensionless metric tensor in this way.} that (i) can be set in direct correspondence with the existence of gravitational effects in all coordinate systems, and (ii) has matter as its source by means of which it can be altered. Compare with the case of electromagnetism, where the electromagnetic tensor field $F_{\alpha\beta}(x)$ (i) can be set in a one-to-one correspondence with the existence of electromagnetic effects in all inertial coordinate systems, and (ii) has the four-current $J^\alpha(x)$ (charge and current) as its source according to the inhomogeneous Maxwell equations 
\begin{equation}
	\frac{\partial F^{\alpha\beta}(x)}{\partial x^\alpha}=\mu_0 J^\beta(x), 
\end{equation}
where $\mu_0$ is the permeability of vacuum. 

The confusion of the metric field $g_{\alpha\beta}(x)$ with the ``new gravitational field'' is rooted in the fact that, in general relativity, $g_{\alpha\beta}(x)$ is the most fundamental field in \emph{mathematical development}. Forgetting that mathematics is in physics only a useful tool to aid and guide our thought, we fall prey to the belief that what is mathematically fundamental must also be physically fundamental. This is not so. The gravitational effects that we experience and observe are not in direct correspondence with the metric field but with the curvature of spacetime. In the next chapter, we shall see that the metric field is, in fact, closely connected with (but not equivalent to) an observable fact of experience, but not in connection with curvature---in connection with clocks. Finally, a supplementary cause of misdirection, whose psychological influence should not perhaps be underestimated, is that the metric field is denoted with the letter ``$g$'' for ``geometry,'' suggesting at the same time ``gravitation.'' 

Can now the analogy of the metric and the luminiferous ether be furthered in such a way that, by considering cases that are not actually the case in the physical world---cases intermediate between the fixed, flat special-relativistic metric substratum and the alterable, curved general-relativistic one---we may gain better understanding of the ether concept? What properties are characteristic of an ether, and what meaning do we attach to them? 

If instead of having a fixed, flat spacetime, we had a fixed spacetime of everywhere-constant curvature such as, for example, a spherical spacetime, would we consider this a metric ether, and what would be its analogous mechanical counterpart? A curved spacetime possesses a property that a flat spacetime lacks, namely, it displays an \emph{active} physical property on account of its non-trivial metric structure, i.e.\ of its curvature, which we experience as gravitational effects: bodies are affected not dynamically, because they move through this substratum, but statically, because they are placed in this substratum. This would be analogous to having had a luminiferous ether with an active mechanical property, namely, an ether wind constant everywhere. Since this is precisely what was sought as evidence of the existence of the old ether, we may as well conclude that such a metric substratum could not be an ether at all (in the negative sense of the word). And yet, are we not nonetheless troubled by the fact that gravitation as the result of a fixed spacetime curvature has a physically unexplained origin? Would not nineteenth-century physicists have been likewise disturbed at having attained to the measurement of an immutable property (ether wind) of physically unaccountable origin? In both cases, after momentary exultation, the need to know the origin of these otherwise God-given media would take over. (But see page \pageref{lambda-solution} for a possible way to view this issue as no puzzle.) 

What would happen if, generalizing the above, we were confronted by a  spacetime of variable curvature, but that is fixed once given, such as, for example, a spherical spacetime glued smoothly to a hyperbolic one? Here gravitation would vary in different regions of space; bodies released at rest would gradually converge towards each other at some places and gradually diverge from each other at other places. Finding ourselves unable to modify this variable active effect of the metric spacetime substratum, we would in this case be even more puzzled than before, and the need to account for its physical source (whether it had any or not) would be even more compelling. Analogously, if the luminiferous ether had displayed a wind velocity that varied in speed and direction at different locations of space, the need to account for the origin of these ``upper winds of the Gods'' would have become even more pressing. 

Now, do these these two kinds of observable yet absolute media qualify as ethers or not? We see that they are midway between the idea of a genuine ether, which is unobservable, homogeneous, and absolute, and a non-ether, which is observable, inhomogeneous, and malleable. The first of these fictitious constructions is, in fact, observable, homogeneous, and absolute, while the second is observable, inhomogeneous, and absolute. Because they share part of the full characterization of ether, we call them \emph{partial ethers} or \emph{demiethers}.\footnote{Demi, from Latin \emph{dimidius}, \emph{dis-} ``apart'' + \emph{medius} ``middle'': one that partly belongs to a specified type or class. (Merriam-Webster Online Dictionary, http://www.m-w.com)}

We move forwards in the analysis of the ether conception by asking next: did the metric ether die a \emph{clean} death in the hands of observation as gravitation and affectability via its two-way interaction with matter, or did it leave behind a difficult legacy in the emptiness lying beneath it? 

General relativity changed not only the notion of ether, but also that of empty spacetime. Another feature of $g_{\alpha\beta}(x)$ is that it depends on the spacetime coordinates $x$ (geometrically speaking, on its points), so that it does not give rise to an immutable metric \emph{background} that can be associated with empty spacetime, as $\eta_{\alpha\beta}$ did in special relativity. On the contrary, the metric field $g_{\alpha\beta}(x)$ appears now as an \emph{intrinsic content}\footnote{Einstein \citeyear{Einstein:1961} distinguished space from its contents with the words: ``[S]pace as opposed to `what fills space,' \emph{which is dependent on the coordinates}\ldots'' (p.~176, italics added). This distinction is based on the idea that anything that depends on the spacetime coordinates (or points) must be \emph{in} spacetime.} of spacetime. The removal of this special field means that the metric structure is gone, so that not an empty spacetime with physical properties but, rather, nothing physical remains without it \cite[p.~176]{Einstein:1961}.

Upon removal of the metric content of spacetime, what remains, what we can still call empty spacetime, are the spacetime points $P$ (cf.\ $x$). Now, are we certain that the amorphous substratum formed by these geometric objects displays no physical properties at all? If this were so, the ether problem in physics would be definitely solved, perhaps not forever but at least so far as we can see, i.e.\ within the reigning physical worldview. 

However, luck does not seem to be on our side---faithful to its history, the ether rises from its second ashes transmuted yet once more, as if enacting its tragic destiny of immortality. Why tragic?

In the classic essay, ``The tragedy of the commons,'' Hardin \citeyear{Hardin:1968} brings our attention to the meaning of ``tragedy'' as something beyond simple unhappiness. He bases the use of the word ``tragedy'' in Whitehead's observation that
\begin{quote}
The essence of dramatic tragedy is not unhappiness. It resides in the solemnity of the remorseless working of things\ldots This inevitableness of destiny can only be illustrated in terms of human life by incidents which in fact involve unhappiness. For it is only by them that the futility of escape can be made evident in the drama. \cite[p.~17]{Whitehead:1948} 
\end{quote}
Like Hardin's grazing commons but in reverse, so does the ether remorselessly follow an inevitable destiny---endless life. 

What passive physical properties do spacetime points display that we must still call them an ether? And, if endowed with physical properties, why can spacetime points not be observed? What is this third rise of the ether all about? 

\section{The geometric ether} \label{GE}
We call the substratum of spacetime points the geometric ether. But before we can move on, the reader asks: are not the metric ethers also geometric? They are indeed, but in different ways. Revealing the ways in which they are similar and the ways in which they differ can help us see what the third rise of the ether shares with its ancestors and in what it is different.

We have characterized geometry as comprising two essential parts, geometric objects and geometric magnitudes attached to them. What we called metric ethers each possessed an inherent quantitative geometric structure given by its invariant quadratic form. This metric structure determined spatial or spatiotemporal \emph{geometric extension}, $\D l^2$ or $\D s^2$, in space or spacetime. A metric ether is for this reason endowed with a quantitatively geometric structure. 

Spacetime points, on the other hand, are geometric objects but, after the removal of $g_{\alpha\beta}(x)$, no quantitatively geometric structure is any longer placed on them. Spacetime points consist only of a qualitatively geometric medium whose elements bear no mutual quantitative relations. We acknowledge, then, that the metric ethers are also geometric ethers, and so that the classification of spacetime points as ``geometric ether'' is, strictly speaking, too inclusive. However, the more accurate denominations ``qualitatively geometric ether'' or ``geometric-object ether'' do not have a good ring about them, and considering that ``metric ether'' brings out the essential characteristic of the ethers of the previous section, we shall value naturalness of expression over taxonomic precision.

Spacetime points are actors not acted upon. What is the physical role they play? To find it, we should not look for passive or active metric effects, even less for mechanical ones, because these geometric objects form neither a metric nor a mechanical substratum. The physical properties displayed by spacetime points are as far removed from the ones attributed to the inertial ethers as these were from those attributed to the mechanical ether. To become aware of them, we must learn to see this new geometric ether in its proper light. 

Spacetime points are inherent in the physical concept of field $f(P)$ by performing their localization---or, more accurately, by performing the localization of physical quantities. Regarding the intrinsic connection that holds between spacetime points and the field, Auyang writes:
\begin{quote}
The spatiotemporal structure is an integral aspect of the field\ldots\ We can theoretically abstract it and think about it while ignoring the dynamical aspect of the field, but our thinking does not create things of independent existence; we are not God. \cite[p.~214]{Auyang:2001}
\end{quote}
She further criticizes Earman's \citeyear{Earman:1989} remarks that ``When relativity theory banished the ether, the space-time manifold $M$ began to function as a kind of dematerialized ether needed to support fields'' (p.~155). 

We agree with Auyang's view that ``spacetime'' (spacetime points) is not merely a substratum on which to \emph{mathematically} define fields. Spacetime points are inherent in fields inasmuch as they perform the \emph{physical} task of localizing physical quantities. This is what, after Weyl \citeyear{Weyl:1949}, Auyang \citeyear[p.~209]{Auyang:2001} called the ``this'' or ``here-now'' aspect of a field, additional to its ``thus'' or qualitative aspect. We also concur with Auyang in that isolated points are the illusive creations of our geometric thinking, but should we offhand renounce the possibility that empty spacetime---\emph{beyond its geometric description}---be real on its own, and not simply an illusion created by the mind? Might we not still discover that it has active observable effects, if we learnt to probe it with different conceptual tools?

In performing the localization of fields, spacetime points give foundation to the concept of locality (here-now) of physical quantities. This is, furthermore, the foundation on which the causality (from here-now smoothly to there-then) of all physical laws is grounded, and which also goes hand in hand with the notion of time as a one-parameter along which things flow causally. Without the field as a local concept, no theoretical explanation can be given for the primary \emph{physical properties} of \emph{locality} and \emph{causality}, whose effects we \emph{passively observe} in the workings of the world. In current theories, we make sense of these perceived properties by attributing to empty spacetime a point-like constitution, and then holding locality and causality to be consequences of it.

In question here is not just an ``ether of mechanics'' or an ``ether of spacetime theories'' but an ether of field-based physics, of all local and causal physics. There is, then, no part of current physics that the geometric ether does not touch, with the exception---not surprisingly---of that part of quantum mechanics that cannot be understood within the field conception. In quantum physics, the field concept survives partly (although not as a spacetime field) in the state vector $|\psi(x,t)\rangle$ and its causal Schr\"{o}dinger equation of motion, but it cannot be applied at the global, non-causal circumstance of its preparation or measurement, when it said to ``inexplicably collapse.'' And conversely, if we felt that locality and causality are not good moulds to describe the natural world, we could not move past them without giving up on the local field.

On the other hand, however, active spacetime points cannot be acted upon such that their displacement would cause any observable effects. This conclusion  forced itself upon Einstein \citeyear{Einstein:1996b} as a consequence of an analysis he called the \emph{hole argument}. This analysis was brought back into the spotlight by professional philosophers of science after about six decades of inconspicuousness, and is today the object of luxuriant philosophical debate---so luxuriant, in fact, that the argument has ramified beyond recognition, and one wonders whether more harm than good has been done by resurrecting the matter. Originally, Einstein offered this argument in 1914 as proof that the equations of general relativity could \emph{not} be generally covariant,\footnote{Ultimately, the equations of all spacetime theories can be expressed in generally covariant form, but general relativity alone admits \emph{only} a generally covariant formulation. See \cite[pp.~46--61]{Friedman:1983}.} because accepting the alternative consequence of the hole argument was at first unthinkable.  However, this path led to difficulties, and about two years later Einstein \citeyear{Einstein:1952b} re-embraced general covariance while at the same time making his peace with the hole argument by accepting the unthinkable, namely, that general covariance ``takes away from space and time the last remnant of physical objectivity'' (p.~117). In other words, that empty spacetime has no physical meaning.\footnote{For a historically complete review of the hole argument, see e.g.\ \cite[pp.~71--81]{Stachel:1989}, where relevant quotations and a full list of early references can be found.}

\begin{figure} 
\centering
\includegraphics[width=85mm]{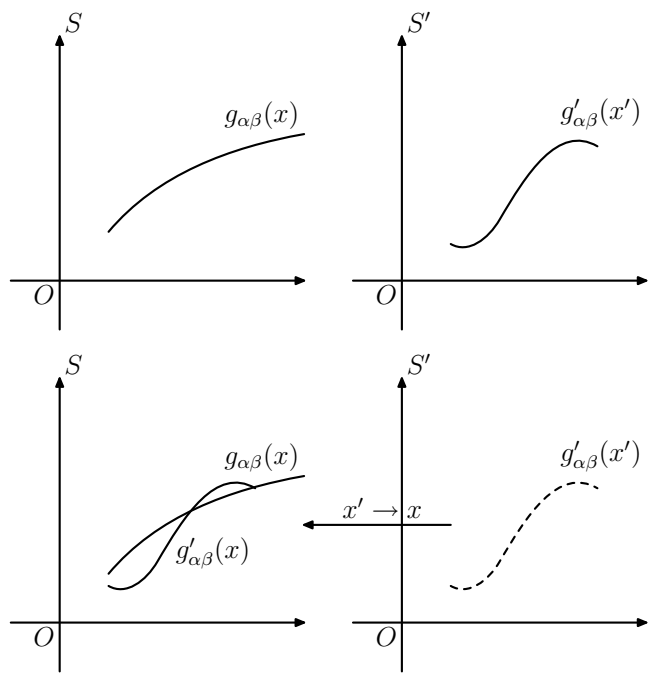} 
\caption[Passive and active coordinate transformations]{Same metric field seen from $S$ and $S'$ (\emph{top}). $S'$ is dropped; points $x'$ are moved to $x$ to express the metric field in $S$ as previously expressed in $S'$ (\emph{bottom}).} 
\label{atoc}
\end{figure}

We review the train of thought leading Einstein to this conclusion. Because the field equation is expressed in generally covariant form, if $g_{\alpha\beta}(x)$ is a solution to it corresponding to the matter source $T_{\alpha\beta}(x)$, then so is $g'_{\alpha\beta}(x')$ with corresponding matter source $T'_{\alpha\beta}(x')$ for any continuous coordinate transformation $x'=\hat{x}'(x)$. Here the same metric and matter fields are being described in two different coordinate systems, both of which are viable on account of the field equation being expressed in generally covariant form. Dropping the primed coordinate system $S'$ completely, how to express $g'_{\alpha\beta}(x')$ and $T'_{\alpha\beta}(x')$ in the unprimed coordinate system $S$? This is done by replacing $x'$ by $x$, so that $g'_{\alpha\beta}(x)$ and $T'_{\alpha\beta}(x)$ describe now with respect to $S$ the fields as were earlier described with respect to $S'$. This transformation is the active equivalent of the original coordinate transformation. It works by replacing $x'$ by $x$, as if, within $S$, point $x'$ was displaced to point $x$ via the transformation $\hat x(x')=x$ (here $\hat x=\hat x'^{-1}$), i.e.\ as if the backdrop of spacetime points was moved as a whole (Figure \ref{atoc}). 

For an \emph{arbitrary} transformation $\hat x(x')=x$ as here considered, a new valid solution $g'_{\alpha\beta}(x)$ can be obtained in this active way only because, in general relativity, $g_{\alpha\beta}(x)$ is not absolute but varies with $T_{\alpha\beta}(x)$ (see below). The result is that both $g_{\alpha\beta}(x)$ with source $T_{\alpha\beta}(x)$ and $g'_{\alpha\beta}(x)$ with source $T'_{\alpha\beta}(x)$ are solutions to the field equation in the same coordinate system. Here are two different metric fields at the same point $x$, which is acceptable so long as each is attached to a different matter source and, therefore, are not simultaneously valid at $x$.

\begin{figure} 
\centering
\includegraphics[width=70mm]{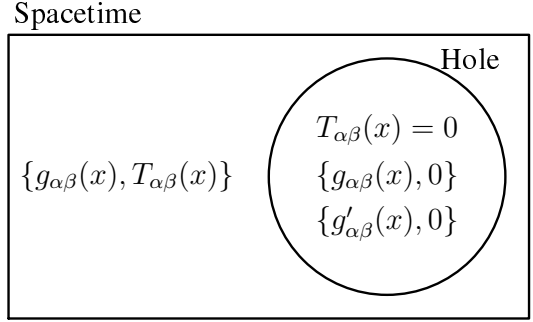} 
\caption[Hole argument in coordinate language]{Schematic representation of the hole argument in coordinate language.} 
\label{hacl}
\end{figure}

Now $H$ is a so-called hole in spacetime in the sense that within this region no matter is present, i.e.\ $T_{\alpha\beta}(x)=0$; outside $H$, on the other hand, $T_{\alpha\beta}(x)$ is non-null. Furthermore, $g_{\alpha\beta}(x)$ is the spacetime metric, necessarily non-null both inside and outside the hole, and $\phi$ is an active coordinate transformation (akin to $\hat x$ before) with the property that it is equal to the identity outside $H$, different from the identity inside $H$, and continuous at the boundary. The unchanged $g_{\alpha\beta}(x)$, with unchanged source $T_{\alpha\beta}(x)$, is now a solution outside $H$, but inside $H$, where no matter is present, both $g_{\alpha\beta}(x)$ and $g'_{\alpha\beta}(x)$ are \emph{two different} and \emph{coexisting} solutions at the same point $x$ (Figure \ref{hacl}). Einstein reasoned that this implies the field equation is not causal, since both coexisting metric fields in $H$ are produced by the \emph{same} source outside $H$. To avoid this, Einstein postulated that, their mathematical differences notwithstanding, $g_{\alpha\beta}(x)$ and $g'_{\alpha\beta}(x)$ must be physically the same. In other words, Einstein postulated that the point transformation $\hat x(x')=x$ leading from $g'_{\alpha\beta}(x')$ to $g'_{\alpha\beta}(x)$ is physically empty, i.e.\ there is no physical meaning to be attached to (the displacement of) spacetime points.

We add parenthetically that only the hole argument applied to general relativity leads to this problem. In special relativity, for example, where the metric field $\eta_{\alpha\beta}$ is absolute and does not result as a solution to a field equation, a new field $\eta'_{\alpha\beta}(x)$ can be obtained from $\eta'_{\alpha\beta}(x')$  by the \emph{active} method $\hat x(x')=x$ of moving the spacetime points \emph{only} if $x'$ and $x$ are otherwise connected by special transformations $\hat x_L$ that leave $\eta_{\alpha\beta}$ invariant: these are the Lorentz transformations. As a consequence, the quadratic form remains invariant:
\begin{equation}
	\D s^2=\epsilon\eta_{\alpha\beta}(x)\D x^\alpha \D x^\beta = \epsilon\eta'_{\alpha\beta}(x)\D x^\alpha \D x^\beta = \D s'^2.
\end{equation} 
In the general-relativistic case, where no special transformations exist that leave $g_{\alpha\beta}(x)$ invariant, 
\begin{equation}
	\D s^2=\epsilon g_{\alpha\beta}(x)\D x^\alpha \D x^\beta \neq \epsilon g'_{\alpha\beta}(x)\D x^\alpha \D x^\beta = \D s'^2, 
\end{equation}
but then again this is no violation of the invariance of $\D s^2$, because the source of geometric structure has changed too, from $T_{\alpha\beta}(x)$ to $T'_{\alpha\beta}(x)$.\footnote{For more details, see \cite[pp.~46--61]{Friedman:1983} and \cite[p.~78]{Stachel:1989}.}

Einstein was therefore right to declare that empty spacetime does not have physical properties---but only if by ``physical properties'' we understand observable (active) effects arising from their displacement. Locality and causal evolution in parameter time, on the other hand, are nothing but the \emph{passive physical effects} that we attribute to the point-like constitution of empty spacetime and to the behaviour of the local fields the points are part of. Does not all this ring a bell? The ether repeats its history---tragically---once again.

Despite the odds against them, Friedman \citeyear[pp.~216--263]{Friedman:1983} nonetheless argued for the physical reality of bare and isolated spacetime points. According to him, the core of an explanation of natural phenomena is to be able to reduce a wide variety of them to a single framework \cite{Friedman:1974}, so that what one is required to believe is not a vast range of isolated representational structures, but a single, all-encompassing construct. However, in order to provide scientific understanding, theoretical entities with \emph{sufficient unifying power} must be taken to be of a literal kind. Thus, Friedman argued that denying the existence of spacetime points---themselves essential for geodesics to exist---can only lead to a loss of unifying and explanatory power in spacetime theories.

As logically impeccable as this typically philosophical argument may be, it is at odds with history itself and with the interests of physical science. If anything has been learnt from the narration of the history of the mechanical ether, it should be this: no other conception gripped the minds of so many illustrious men of science for longer and more strongly than that of the luminiferous ether. It was always held to be physically real and unified a wide range of phenomena (light, heat, gravitation, electricity, and magnetism) despite its relentlessly unobservable existence. And yet, at the opening of the twentieth century, a scientific revolution rendered it superfluous and useless, and so, with no further ado and hardly enough mourning to do justice to its nearly three hundred years of age, it was declared nonexistent and duly buried. Science quickly forgets its waste products and moves on.

Evidently, no spell of time during which a conception proves to be successful is long enough to declare it real because of its utility or unifying power. What is to guarantee that today's vastly successful spacetime points---the geometric ether---will not run, in their own due time, the same fate as their mechanical and inertial ancestors? If any concept that is of aid in physics is to be held real, nothing more and nothing less should be demanded of it that it be observable, i.e.\ that it have active physical effects and be open to human manipulation and control.

While professional philosophy does not try to overcome the geometric ether, neither does quantum gravity. A study of this discipline reveals that it does not try to determine from where observable locality and causal evolution in parameter time may come by attempts to find observables related to empty spacetime. Instead, quantum gravity focuses on the creation of ever more involved geometric structures for space and time, leaving the qualitative problem of the geometric ether untouched. When an attempt at observation is at all made in quantum gravity, it is only to link its proposed, so far unobservable, geometric structures with some hypothetical effect, but never giving second thoughts to the observability of the underlying geometric objects---spacetime points among them.

In this respect, Brans \citeyear[pp.~597--602]{Brans:1999} argues that much by way of unobservable structure is taken for granted in current spacetime theories, such as the existence of a point set, a topology, smoothness, a metric, etc. He compares the vortices and atoms of the mechanical ether with these unobservable building blocks of spacetime and, moreover, with the yet considerably more complex spacetime structures and ``superstructures'' devised more recently. We interpret Brans to be asking: will strings and membranes, spin networks and foams, nodes and links---to name but a few---appear a hundred years from now like the mechanical ether does today? Are the above the new geometric counterparts of the old mechanical contraptions?\footnote{For a recent explicit revival even of the long-dead mechanical ether, see \cite{Jacobson/Parentani:2005}, where the authors ask: ``Could spacetime literally be a kind of fluid, like the ether of pre-Einsteinian physics?'' (p.~69).} In other words, will the phrase ``quantum theory of gravity'' come to be regarded, a hundred years from now, as we regard today the phrase ``vortex-atom theory of the ether''? It is likely that a new name become necessary for a theory that, by being both physical and faithful to nature, succeeds in differentiating itself from the failed attempts that preceded it, and that the abused term ``quantum gravity'' only remain as a useful handle to denote the historical debacle of another dark age of physics. 

Be all this as it may, even if we turned a blind eye to their physical uncertainty and assumed that some of these more recently proposed structures were not to follow such a fate and could eventually be observed, the observations relating to them would at any rate give evidence of a new metric spacetime structure. But if the goal is to advance the understanding of the geometric ether (locality, causality, parameter time, and fields) then we must move away from geometry---forwards on a new path.

For an appraisal of the different concepts of ether analysed, turn to the summary table at the end of this chapter on page \pageref{ether-summary}.

\section{Beyond the geometric ether} \label{BGE}
The luminiferous ether was superseded by stripping it of all its intrinsic mechanical properties and rendering them superfluous; there was no luminiferous ether or any such thing at all. The metric ether, on the other hand, did not have its metric properties rendered superfluous but instead accounted for once their connection with matter was found. The story of the mechanical and inertial ethers is repeating itself today in a geometric, instead of a metric or a mechanical, guise---but which of these two fates (denial or observation), if any, may the future hold in store for the geometric ether? Both, we hope. We seek for the geometric ether both a conceptual change and the recognition of observables linked to its new nature, a change comparable to the shift from the \emph{unobservable} \emph{mechanical} ether to the \emph{observable} \emph{metric} structure of general-relativistic spacetime.

As we seek to learn what lies beneath locality, causality, parametric time, and fields---as we seek to find, as it were, the \emph{source} of these passive physical properties inherited from empty spacetime---the geometric ether must, like its mechanical ancestor, have another layer of its nature revealed. This is to be done, once again, by stripping the ether of its intrinsic properties, but this time they are \emph{geometric} ones. We believe that, taking a conceptual step beyond the geometric ether, i.e.\ beyond geometry \emph{entirely}, may be the key to identifying observables related to empty spacetime. These hypothetical observables, however, would not be directly related to spacetime points, which are geometric objects, but rather to ``metageometric things,'' just like spacetime curvature has little to do with the mechanical properties of the earlier-hypothesized material medium.  

To put it differently, we are here proposing an updated version of Larmor's centenary words:
\begin{quote}
We should not be tempted towards explaining the simple group of
relations which have been found to define the activity of the
aether by treating them as mechanical consequences of concealed
structure in that medium; we should rather rest satisfied with
having attained to their exact dynamical correlation, just as
geometry explores or correlates, without explaining, the
descriptive and metric properties of space.\footnote{Quoted from \cite[Vol.~1, p.~303]{Whittaker:1951}.}
\end{quote}
Larmor's statement is remarkable for its correctness, and all the more remarkable for its incorrectness. Its first half (up to the semicolon) is a correct testimony of what soon would prove to be the way out of the mechanical-ether problem: its denial by special relativity. Its second half \emph{was}---at the time of its utterance---a correct comparison between the way Larmor thought the mechanical ether should be considered and the way geometry was then regarded, namely, not as background for the explanation of phenomena. However, 16 years later in 1916 the ether would be
overtly reinstated as a dynamic metric structure, and the physical properties of spacetime would be \emph{explained} by it in the sense of a metric geometric substratum.\footnote{In fact, Newton's space had already assumed this role over 200 hundred years before, and Einstein's special-relativistic inertial spacetime 11 years before, but neither of them had been openly considered as substrata with physical geometric properties until after 1916. Then Einstein identified $\eta_{\alpha\beta}$ and $g_{\alpha\beta}(x)$ with metric ethers. Breaking with history, we chose to identify the \emph{invariant quadratic forms} derived therefrom with the metric ethers due to their closer connection with actual physical observables.} This trend of attempting to explain phenomena in terms of geometry, and in particular metric geometry, did not stop with general relativity but, as we saw, continues with the efforts of quantum gravity. Therefore, nowadays geometry does explain, as a substratum, the properties of spacetime,\footnote{Friedman \citeyear{Friedman:1983}, for example, writes that ``the development of relativity appears to lead\ldots to the view that the geometrical structure of space or space-time is an object of empirical investigations in just the same way as are atoms, molecules, and the electromagnetic field'' (p.~xi).} and the second part of Larmor's view is no longer correct. 

The updated version of Larmor's words proposed here reads thus:
\begin{quote}
We should not be tempted towards explaining the simple group of relations (\emph{locality, causality, parametric time}) which have been found to define the activity of the \emph{geometric ether} (\emph{spacetime points}) by treating them as consequences of concealed structure in that \emph{geometric medium}; we should rather \emph{seek to explain the activity of the geometric ether beyond its geometric nature, searching for empty-spacetime observables metageometrically}.	
\end{quote}
Unlike Larmor, we do not renounce an explanation of the geometric ether,
but we do not attempt to find it at the same conceptual level as this ether finds itself.

The search for observables related to empty spacetime requires taking a
genuine conceptual step beyond the state of affairs as left by Einstein's hole argument. We believe that the current philosophical literature on this problem, as reviewed e.g.\ in \cite[pp.~175--208]{Earman:1989}, has not been able to take this step by adding something \emph{physically new} to the discussion. On the contrary, the philosophical debate appears to function in the spirit of Earman's words, which measure the fruitfulness of a work by asking ``How many discussions does it engender?'' (p.~xi). What is needed is a new physical insight by means of which the present philosophical debate (like every other) may be rendered inconsequential, much like the older disputes as to the shape and position of the earth or the nature of the heavenly bodies were only settled by new physical investigations. Indeed, what is needed is a discovery in the sense of Oppenheimer's words:
\begin{quote}
All history teaches us that these questions that we think the pressing ones will be transmuted before they are answered, that they will be replaced by others, and that the very process of discovery will shatter the concepts that we today use to describe our puzzlement. \cite[p.~642]{Oppenheimer:1999}
\end{quote}

\begin{figure} 
\centering
\includegraphics[width=70mm]{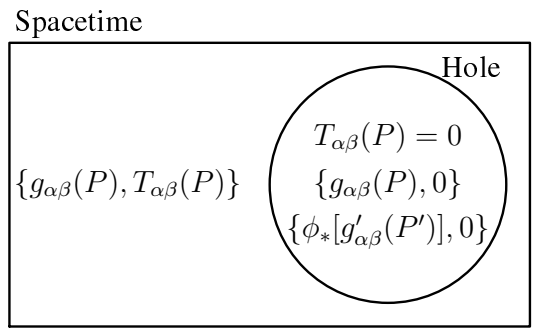} 
\caption[Hole argument in geometric language]{Schematic representation of the hole argument in geometric language in tune with the coordinate notation of the previous section. Here $\phi_\ast[g'_{\alpha\beta}(P')]=g'_{\alpha\beta}(P)$. To match the geometric notation just introduced, replace: $g_{\alpha\beta}(P)$ by $g'_{\alpha\beta}(Q)$, $T_{\alpha\beta}(P)$ by $T'_{\alpha\beta}(Q)$, and $\phi_\ast[g'_{\alpha\beta}(P')]$ by $\phi^\ast[g_{\alpha\beta}(P)]$.} 
\label{hagl}
\end{figure}

To understand how a possible way to go beyond the hole argument is related to the hole argument itself, we now re-rehearse it in geometric, rather than coordinate, language and for a general field $f(P)$. We start with two points, $P$ and $Q$, inside $H$ linked by a diffeomorphism $\phi(P)=Q$, and we have these two points be the local aspect or location of fields $f(P)$ and $f'(Q)$, each solution to the field equation with the same source. The demand that
\begin{equation}\label{coordinate}
f(P)=f'(Q) 
\end{equation}
(cf.\ $g_{\alpha\beta}(x)=g'_{\alpha\beta}(x')$) reproduces geometrically
the requirement that, after an arbitrary coordinate transformation $x'=\hat x'(x)$, a physical situation remains unchanged. In terms of an active
transformation, point $P$ (cf.\ $x$) is dropped in solution $f(P)$ (cf.\ $g_{\alpha\beta}(x)$), and a new solution is obtained by displacing $P$ to $Q$ (cf.\ $x'$) as follows:
\begin{equation}\label{active}
f(P) \rightarrow f(Q)=f(\phi P)=: \phi^\ast[f(P)],
\end{equation}
where $\phi^\ast[f(P)]$ is induced by $\phi$ in the way shown and is called the push-forward of $f(P)$. The new field $\phi^\ast[f(P)]=f(Q)$ is also a solution, but
\begin{equation}\label{different}
f(Q)\neq f'(Q) 
\end{equation}
(cf.\ $g_{\alpha\beta}(x')\neq g'_{\alpha\beta}(x')$) (Figure \ref{hagl}).\footnote{It would have been more natural to denote $Q$ by $P'$ and later to perform an active transformation dropping $P'$ (cf.\ $x'$) instead of $P$
(cf.\ $x$), in which case one would have obtained the expressions
$f(P)=f'(P')$ instead of (\ref{coordinate}), $f'(P')\rightarrow f'(P)=f'(\phi^{-1}P') =: \phi_\ast[f'(P')]$ instead of (\ref{active}), and $f(P)\neq f'(P)$ instead of (\ref{different}), in closer parallel with Einstein's original coordinate notation and with the one of the previous section. However, remaining faithful to this starting point would have complicated the notation and obscured the intuitive appeal of the investigation that follows at the end of this section.} Again, to preserve the causality of the field equation one postulates that displacements of points lead to no observable effects and that, therefore, (displacements of) points have no physical meaning.

In order to move forward and catch a glimpse of what form empty-spacetime observables may take, we reason \emph{heuristically} as follows. We note that the postulate above appears to go farther in its claim than it really needs to. The required property of diffeomorphism invariance with respect to an arbitrary displacement of points is satisfied by simply having spacetime points be all alike, i.e.\ have no physical identity and form a perfectly homogeneous substratum. Thereby, the problem posed by the hole argument is avoided, since having physically indistinguishable spacetime points makes the two solutions $g_{\alpha\beta}(P)$ and $\phi_\ast[g'_{\alpha\beta}(P')]$ physically equivalent.\footnote{Rewrite $\phi_\ast[g'_{\alpha\beta}(P')]$ as $g'_{\alpha\beta}(P)$ and replace $P$ by $P'$ to get $g'_{\alpha\beta}(P')=g_{\alpha\beta}(P)$.}

What do we mean by physically indistinguishable points? We can see this by recognizing the difference between mathematical and physical points. Mathematical points may well be labelled points but, as Stachel remarked,
\begin{quote}
No mathematical coordinate system is \emph{physically} distinguished per se; and without such a distinction there is no justification for physically identifying the points of a [mathematical] manifold\ldots as physical events in space-time. Thus, the mathematician will always correctly regard the original and the dragged-along fields as distinct from each other. But the physicist must examine this question in a different light\ldots \cite[p.~75]{Stachel:1989} 
\end{quote}
The physicist must, in this case, rather ask whether in a spacetime empty of matter and metric structure there is anything by means of which its points can be told apart from one another. Finding there are no such means, the physicist is forced to hold spacetime points physically identical to one another. 

However, having a multitude of indistinguishable spacetime points does not necessarily mean that they must be physically meaningless \emph{in every other way}. How so? Think, for example, of hydrogen atoms, which are exact physical copies of each other and yet display physical properties. Which type of property displayed by hydrogen atoms may serve as a useful analogy for the case of spacetime points? 

The hydrogen atoms conforming a gas of this element are, like spacetime points, physically identical, but denying the reality of hydrogen atoms on these grounds does not seem at all reasonable.\footnote{See \cite[p.~409]{Horwich:1978} and \cite[p.~241]{Friedman:1983}.} Hydrogen atoms, despite being identical, possess a property that spacetime points do not seem to have: they interact with other matter and correlate with one another creating bonds, hydrogen molecules. In empty spacetime, there is nothing else for points to interact with, and so the interaction of the hydrogen atoms with, for instance, their container walls seems to be an unworkable analogy. However, hydrogen atoms also interact among themselves---and this is the only analogy with spacetime points that remains open for us to pursue. Could spacetime points show observable properties---active physical effects---due to \emph{mutual correlations}, just like hydrogen molecules display observable properties that give evidence of their constituent identical atoms? 

Forty years ago, Wheeler touched upon an idea somewhat similar to this one within the context of the so-called ``bucket of dust.'' Pondering the consequences of changes in the topology of space, he wrote:
\begin{quote}
Two points between which any journey was previously very long have suddenly found themselves very close together. However sudden the change is in classical theory, in quantum theory there is a probability amplitude function which falls off in the classically forbidden domain. In other words, there is some residual connection between points which are ostensibly very far apart. \cite[p.~498]{Wheeler:1964} 
\end{quote}
However, Wheeler did not develop this idea, but only used it in order to reject the quantum-theoretical concept of nearest neighbour, according to the manner in which he had previously defined it.

A more thorough attempt in this direction was carried out by Lehto \citeyear{Lehto:1988} and collaborators \cite{Lehto/Nielsen/Ninomiya:1986b,Lehto/Nielsen/Ninomiya:1986a}. In the context of the geometrized pregeometric lattice analysed in Section \ref{Simplicial quantum gravity}, they found that when the principle of diffeomorphism invariance was considered in connection with quantum theory, fields on the said lattice displayed correlations of quantum-mechanical origin.

Like the analogy between identical spacetime points and hydrogen at\-oms---themselves identical on account of their quantum nature---and like Wheeler's conjectured consequence of a change in the topology of space in a quantum-mechanical world, these results insinuate the same conclusion. When the physical identity of spacetime points is considered in conjunction with quantum theory---the only branch of natural science, as Isham \citeyear[p.~65]{Isham:1995} remarked, so far forced to confront the problem of existence---we are led to results that differ from the conclusions drawn by classical general relativity alone, namely, to the concept of \emph{quantum-mechanical spacetime correlations}. Could they hold a key to the identification of empty-spacetime observables?

In order to display the idea of correlations within the existing context of the hole argument---but moving, at the same time, beyond it by introducing a physically new idea into the classical setting---we consider the following. We saw that, in an otherwise empty spacetime, a field $f(P)$ at point $P$ is also an acceptable field when dragged onto another point $Q$,
\begin{equation}
f(P) \rightarrow \phi^\ast [f(P)]=f(Q).
\end{equation}
But what happens if we now \emph{pull it back again} to get $\phi_\ast [f(Q)]$? The field could carry properties pertaining to $Q$ back with it, so that a comparison of the original field $f(P)$ and the restored field $\phi_\ast [f(Q)]$ could yield that they are physically different. 

This is not to say that we expect to find new physics by means of a purely mathematical operation plus its inverse. It means the reverse: \emph{given} the privately gathered (not logically deduced) insight that indistinguishable points may nonetheless display physical effects by correlating with each other quantum-mechanically, then the above construction constitutes a \emph{geometric means to represent} the ensuing broken physical symmetry of diffeomorphism invariance. Moreover, when we speak of ``moving points,'' no system is actually being displaced in the physical world; the physical meaning of the mathematical expressions above (and below) ultimately falls back on spacetime correlations of quantum origin, the existence of which is here \emph{conjectured}.

In other words, this heuristic analysis does not try to prove, by mathematical manipulation or deduction, that such quantum correlations exist, like some (as criticized in Chapter \ref{ch:Planck-scale physics}) infer the existence of Planck black holes---and, therefore, the relevance of the Planck scale---from a heuristic analysis of the relations between mass, energy, and spatial localization. We only use heuristics to inspire our physical thought and to guide its mathematical expression, but not to establish the physical substance of the ideas thus arrived at. The ideas presented in this section, then, are part of an exploratory approach to the problem of the geometric ether. Its true solution, however, must be rooted at its origin in familiar observations of the physical world; it must come from simple experiments that humans can understand and affect and not from disembodied mathematical push-forwards and pull-backs. These ideas should therefore be interpreted from this exploratory perspective. 

\begin{figure} 
\centering
\includegraphics[width=80mm]{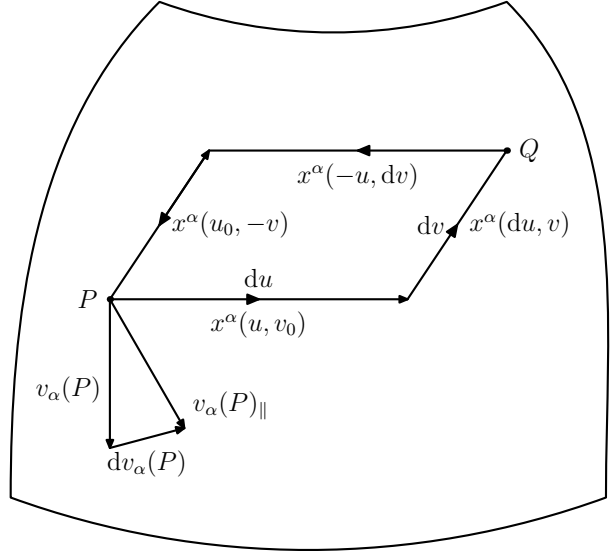} 
\caption[Parallel transport]{Parallel transport of a vector $v_\alpha(P)$ around a closed loop in curved spacetime.} 
\label{ptgr}
\end{figure}

Given $f(P)$, the appearance of the pull-back $\phi_\ast [f(Q)]$ rests on the need to compare locally the original field and the original field restored. This need arises from the fact that physical experiments are not performed globally but locally. An analogy close at hand is that of parallel transport in general relativity. Given two (covariant) vector fields $v_\alpha(P)$ and $v_\alpha(Q)$ in curved spacetime, they can only be compared by computing $\Delta v_\alpha$ at one point $P$ by parallel-transporting the latter field: 
\begin{equation}
\Delta v_\alpha(P)=v_\alpha(Q)_\parallel - v_\alpha(P).
\end{equation} 
The analogy works best for the case of a small closed loop traced starting at $P(u_0,v_0)$ by following a worldline $x^\alpha=x^\alpha(u,v_0)$ until we reach a separation $\D u$, and then a worldline $x^\alpha=x^\alpha(\D u,v)$ until we reach a separation $\D v$, thus arriving at $Q$; we then come back\footnote{We cannot actually come back to $P$ in the literal sense of the word because we can only traverse spacetime forwards, in the direction of the tangent vectors to the worldlines. That is to say, the separation $\D u$ alluded to is given by the values of a parameter $u$ that increases monotonically on the worldline, e.g.\ a clock reading. What we can in fact do is travel from $P$ to $Q$ following two different routes and then mathematically form a loop by subtracting one change in the transported vector from the other. See Section~\ref{sec:curvature-tensor} for details.} from $Q$ to $P$ following worldline \mbox{$x^\alpha=x^\alpha(-u,\D v)$} until we reach a separation $\D u$ and then worldline \mbox{$x^\alpha=x^\alpha(u_0,-v)$} until we reach a separation $\D v$. Field $v_\alpha(P)$ is compared to itself after having travelled the loop and restored to its original position. We find that 
\begin{equation}
\D v_\alpha(P)=R^\delta_{\ \alpha\beta\gamma}(P) v_\delta 
\left(\frac{\partial x^\beta}{\partial u}\D u\right) \left(\frac{\partial x^\gamma}{\partial v}\D v \right), 
\end{equation}
where $R^\delta_{\ \alpha\beta\gamma}$ is the (mixed) Riemann tensor, quantifies how much the components, but not the size, of $v_\alpha(P)$ have been changed by the experience (Figure \ref{ptgr}). 

Similarly, if instead of thinking of a field as a mere value, we visualize it as a \emph{vector} $f$, and $f(P)$ as a \emph{component} of $f$, then a new (geometric) dimension to the hole argument is revealed: the local comparison of $\phi_\ast [f(Q)]$ and $f(P)$ quantifies how much the field component $f(P)$, but not the field value, changes due to the possible quantum correlation of $P$ with $Q$; in short, $f$ as a vector would behave like $v_\alpha$. This change would stem from any physical reality of quantum-mechanical origin that the points may have. Pictorially speaking, $P$ would behave as if it remembered having interacted with $Q$, remaining ``entangled'' with it. 

Let us now push the parallel-transport analogy further. The parallel-transport procedure constitutes the mathematical representation of a physical property of spacetime. This property is curvature and is unheard-of in flat spacetime, which is sterile with respect to active metric effects. A curved spacetime, on the other hand, is not. The parallel transport of a field around a closed loop uncovers curvature as a \emph{previously hidden} active metric effect. Now, the description above intends to point to the appearance of quantum-related spacetime correlations in a corresponding manner. They are a hypothetical property completely foreign to the homogeneous substratum of spacetime points, which is sterile with respect to active geometric effects (diffeomorphism invariance)---but could such correlations arise as the active effect of a different kind of medium? And if so, what kind of medium? 

We believe that we should not seek the source of quantum spacetime correlations in \label{inner-structure-1} extra hidden geometric properties of points but somewhere else, within a deeper layer of the nature of things: as an active effect of the quantum-mechanical connection between \emph{metageometric things}. Indeed, the search for observables related to empty spacetime should be accompanied by the conceptual overthrow of the geometric ether as well. If nothing else, the fact that the present geometric description of spacetime leads again to the futile problem of an ether is hint enough that there must be something amiss with this form of description. 

Where shall we look to find observables that capture the above-described behaviour of field $f(P)$ and that are, at the same time, in principle free of geometric denotation? These cannot certainly be the fields themselves, since they are purely geometric concepts, nor are they observable as such. 

Sometimes the answers to the hardest questions are staring one in the face. What is the most fundamental physical concept in the present understanding of spacetime? What is general relativity essentially \emph{about}? Not the metric field but an \emph{observable} network of invariant intervals $\D s$ between events; it is on the measurable intervals that the metric description of spacetime ultimately rests. The metric field, as we mentioned earlier and will show in more detail in the next chapter, comes \emph{after} the directly measurable intervals, and endows them with geometric meaning as part of a differential-geometric theoretical description. By picturing in our minds the measurable separation $\D s$ between two events as the extension of a spacetime vector $\D \vec r(x)$ parametrized with coordinates $x$, we find
\begin{equation}
\mathrm{d}s^2= \epsilon \mathrm{d}\vec r \cdot \mathrm{d}\vec r = \epsilon \left( \frac{\partial \vec r}{\partial x^\alpha}\Big\vert_P\mathrm{d}x^\alpha\right)\cdot \left( \frac{\partial \vec r}{\partial x^\beta}\Big\vert_P\mathrm{d}x^\beta\right)
=\epsilon g_{\alpha\beta}(P) \mathrm{d}x^\alpha \mathrm{d}x^\beta.
\end{equation}
The realization that $\D s$, as a direct observable, takes physical precedence over its geometric theoretical interpretation is useful to this search, because it helps us disentangle all posterior geometric baggage (full differential geometry) from the physically more essential measurement result itself. Recalling the ideas of Chapter~\ref{ch:Geometry in physics}, we are then left with $\D s$ as a minimal, bare geometric notion, namely, as the extension of some underlying metric line. This is not yet a metageometric realm, but it places us one step closer to it. It suggests that we may look at the physical essence behind $\D s$---a classical link correlating two events as phenomena---to inspire and guide the search for \emph{metageometric correlations of quantum origin}.

\begin{figure} 
\centering
\includegraphics[width=75mm]{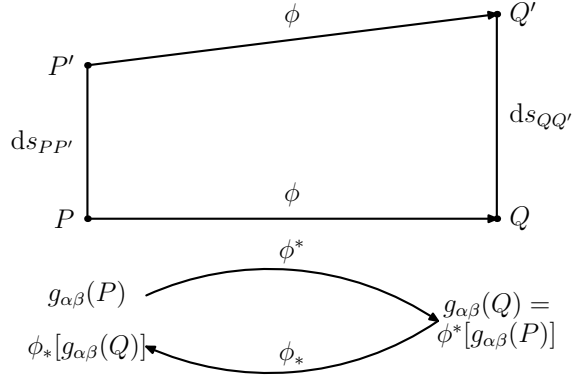} 
\caption[Push-forward and pull-back of metric field]{Push-forward and successive pull-back of the metric field $g_{\alpha\beta}(P)$ in an otherwise empty spacetime. If spacetime points had any physical reality of quantum-mechanical origin, long-range correlations would been seen in the line element $\D s$.} 
\label{pfpb}
\end{figure}

Once again in geometric language, we could now explore\footnote{Here $P'$ and $Q'$ are points in the neighbourhoods of $P$ and $Q$ respectively, and are linked by a diffeomorphism $\phi(P')=Q'$, in the same way that $P$ and $Q$ are.} whether the line element
\begin{equation}
\mathrm{d}s_{PP'}=\sqrt{\epsilon g_{\alpha\beta}(P)\mathrm{d}x^\alpha \mathrm{d}x^\beta} 
\label{intial}
\end{equation}
between two points remains unchanged after $g_{\alpha\beta}(P)$ is pushed forward, $\phi^\ast [g_{\alpha\beta}(P)]$, to get
\begin{equation}\label{dragged}
\mathrm{d}s_{QQ'}=\sqrt{\epsilon \phi^\ast [g_{\alpha\beta}(P)]\mathrm{d}x^\alpha \mathrm{d}x^\beta} 
=\sqrt{\epsilon g_{\alpha\beta}(Q)\mathrm{d}x^\alpha \mathrm{d}x^\beta}, 
\end{equation}
and subsequently pulled back, $\phi_\ast [g_{\alpha\beta}(Q)]$, to get for the original line element
\begin{equation}
\sqrt{\epsilon \phi_\ast [g_{\alpha\beta}(Q)] \mathrm{d}x^\alpha \mathrm{d}x^\beta}. \label{restored}
\end{equation}
If spacetime points had any physical reality of quantum-mechanical origin, the final expression (\ref{restored}) could not be equal to the initial one (\ref{intial}), and there would be some long-range correlations seen in the line element $\D s$ (Figure \ref{pfpb}).

This would not upset general relativity, because the envisioned correlations represent a new kind of effect rather than corrections to physical magnitudes predicted by the theory. Moreover, at a conceptual level, the diffeomorphism-invariance principle and the hole-argument conclusion are not challenged either, since these belong in the general-relativistic \emph{classical}, \emph{geometric} description of spacetime, which does not consider quantum theory at all and, therefore, must view quantum-mechanical correlations as an element foreign to its framework. In particular and furthering the previous analogy, we may say that the local comparison between $\phi_\ast[f(Q)]$ and $f(P)$ deals with changes in the components of a field, whereas the hole argument refers to (lack of) changes in the value of a field. Thus, the \emph{classical}, \emph{geometric} theory of general relativity remains untouched from this perspective. 

Strictly speaking and consistent with the general-relativistic picture, then, it is not possible for the correlations envisioned to be displayed by $\mathrm{d}s_{PP'}$ \emph{as a classical geometric notion}, i.e.\ as the \emph{distance} between \emph{points} of a \emph{non-quantum} theory. Building beyond the raw material provided by $\D s$ with the aid of metageometric thought and quantum-mechanical ideas, we seek to learn what lies behind it as a measurement value. Our expectation is for the classical geometric interval $\mathrm{d}s_{PP'}$ to result as a geometric remnant or trace left behind by metageometric things. It is through these \emph{things beyond geometry} that the conceptual overthrow of the geometric ether can be pursued. In this light, one should regard points the way one would Wittgenstein's ladder: as concepts to be discarded after one ``has climbed out through them, on them, over them'' \cite[$\S$~6.54]{Wittgenstein:1922}.

But the road to this goal is conceptually demanding; much physical insight needs first to be gained and brought to light before the metageometric path involving $\D s$ can be pursued. Why is $\D s$ so important and, especially, more important than the metric field? How is all relativity to be built upon it? And what is its physical meaning, what kind of measurement is it? Does it really involve the proverbial \emph{rods and clocks}? These queries lead us to an analysis of the concept of time.  

\begin{sidewaystable}
\centering
\setlength{\extrarowheight}{8pt}
\begin{tabularx}{\linewidth}{|X|X|X|X|}\hline
	\textbf{Ether} & \textbf{Action} & \textbf{Observability} & \textbf{Affectability} \\ \hline \hline
	\textbf{Mechanical}. \hspace{2cm} Material medium & Medium for the propagation of light, gravitation, electricity, magnetism, heat, etc. & 
	Mechanically sterile homogeneous substratum. No observable (active) mechanical properties (e.g.\ velocity, density) & 
	Its effect on light, gravitation, electricity, magnetism, heat, etc.\ cannot be changed \\ \hline
	\textbf{Inertial}. \hspace{3cm} Absolute\phantom{-}space\phantom{-}and\phantom{-}absolute spacetime 	&  Media for the determination of acceleration and spatial, or spatiotemporal, extension 
      & Geometrically sterile homogeneous substrata. No observable (active) geometric properties & 
      Their effect on particles, or particles and clocks, cannot be changed\\ \hline
	\textbf{Gravitational}. \hspace{2cm} Dynamic spacetime
	 & Medium for the determination of acceleration and spatiotemporal extension & 
	Geometrically rich inhomogeneous substratum. Observable geometric property: curvature & 
	Has matter as source. Its effect on particles and clocks can be changed \\ \hline
	\textbf{Geometric}. \hspace{2cm} Spacetime points  &
      Medium for the localization of fields $f(P)$. Embodiment of locality and causal evolution in parametric time  & 
      Qualitatively\phantom{-}homogeneous, quantitatively\phantom{-}structureless geometric substratum. No observable (active) geometric properties & 
      Its localizing effect on fields cannot be changed by moving the points ($f(P)=\phi_\ast[f(Q)]$). Not even self-interacting due to geometric amorphousness\\ \hline
\end{tabularx}
\caption{A summary of the different concepts of the ether in physics, from the mechanical luminiferous ether to the geometric spacetime points. Of the four, only the gravitational ``ether'' was brought to full physical accountability: it is observable and open to manipulation and control.}
\label{ether-summary}
\end{sidewaystable}

\chapter{A geometric analysis of time}
\label{ch:Analysis of time}
\begin{flushright}
\begin{tabular}{p{10cm}}
\emph{An analysis of the concept of time was my solution.} \smallskip \\
Albert Einstein, ``How I created the theory of relativity''
\end{tabular}
\end{flushright}

\bigskip

Philosophers of science thrive in the analysis of scientific concepts. If we want to draw a lesson on how to make a good analysis of the physical concept of time, we should take a look at the works of (broadly speaking) natural philosophers---from V b.C.\ to XXI a.D.---\emph{or should we?} What do philosophers of science do when they make an analysis?

A conceptual analysis can be interpreted variously. It may refer to the meticulous \emph{logical dissection} of a concept into constituent parts or to the \emph{physical examination} of a concept regarding its physical origin and nature. If, as we suggest, these two activities are not one and the same, how do they differ? And what is their connection with the activities of scientists and philosophers of science?

In his enlightening book, Reichenbach \citeyear{Reichenbach:1951} finds the source of ``philosophical error'' in speculative philosophy in ``certain \emph{extralogical motives}'' (p.~27). These are compounded in the speculative philosopher's psychological make-up, ``a problem which deserves more attention than is usually paid to it'' (p.~36). Reichenbach describes the fall of speculative (i.e.\ traditional) philosophy and the rise of scientific philosophy, a philosophy to be guided by modern science and its tools. Unlike speculative philosophy, the goal of scientific philosophy is not a one-person explanation of the natural world but the elucidation of scientific thought; its object of study is science itself. Or in the words of Reichenbach, ``Scientific philosophy\ldots leaves the explanation of the universe entirely to the scientist; it constructs the theory of knowledge by the analysis of the results of science'' (p.~303). Identifying scientific
philosophy with modern philosophy of science, we can preliminarily conclude that, whereas the \emph{raw material} of science is the natural world, the \emph{raw material} of philosophy of science is (as its name correctly suggests) the scientific enterprise.

As distinct as science and philosophy of science seem to be, we still find ourselves at a loss to tell which of the two disciplines we should pursue if we want to better understand the physical concept of time. A reading of Reichen\-bach suggests that we should embrace philosophy of science, because it is through it, and not normal science, that we gain \emph{clarification} of scientific concepts (p.~123). Indeed, Reichenbach traces the rise of the full-blown ``professional philosopher of science'' to the incipient attempts of scientists who would unsystematically ponder the nature of their fields between the lines of their technical works. But has this evolutionary development from enquiring scientist to professional philosopher of science truly yielded a species better adapted to the clarification of conceptual scientific puzzles? Or has the speciation product, becoming overspecialized, shifted instead to a separate and altogether different habitat, a new place where it can not only survive but even thrive?

If by ``philosophy of science'' we understand the natural extension of the ordinary work of the scientist that comes from asking not only ``how do I calculate this?'' but also---or instead---``what is this, and why should I calculate it?'' then philosophy of science is a useful and integral part of science. Progress in the scientific worldview would not be possible without it. But philosophy of science as a discipline in its own right, carried out ``professionally'' by philosophers because, as Reichenbach put it, ``scientific research does not leave a man time enough to do the work of logical analysis, and\ldots conversely logical analysis demands a concentration which does not leave time for scientific work'' (p.~123)---this is another matter altogether.

Does the meticulous logical dissection of scientific concepts produce useful results without limit? Or is there instead a fuzzy boundary carried beyond which logical analysis stops clarifying anything at all? Reichenbach goes on to admit that philosophy of science, ``aiming at clarification rather than discovery may even impede scientific productivity.'' May it not impede, when carried too far, not just scientific productivity but even clarity itself, thus turning against its original goal?

With the professionalization of the conceptual examination naturally falling on the enquiring scientist, the link with the natural world as the motivator of scientific concepts and \emph{informer of their analysis} was severed in professional philosophy of science, for, as Reichenbach put it, the professional philosopher of science \emph{has no time to deal with nature}. What is then his working method? His starting point is established science and his methods scientific, but his logical analyses roam largely unconstrained by feedback from the physical world. That is to say, philosophy of science freezes the science of the day and, armed with the tools of the same science, looks away from the physical world and proceeds to dissect the concepts of science in as many different ways, into as many possible constituents, as \emph{logic and frozen science} will allow. Thus, philosophers of science become---in the words of one of them---``professional distinction drawers'' \cite[p.~2]{Earman:1989}, because, without \emph{fresh feedback} from the natural world, progress (appropriately understood) must necessarily be made from the differentiation of frozen scientific material into new, abstract forms.

Whether these new forms are useful to the scientific enquiry, whether they actually clarify it or only succeed in making of it a worse conceptual shambles, is not the measure by which philosophical fruitfulness is assayed. Inevitably, having severed the link with nature in the name of professionalism, philosophers judge their works, \emph{de facto} as well as in the words of the same renowned philosopher of science above, ``by one of the most reliable yardsticks of fruitful philosophizing, namely, How many discussions does it engender? How many dissertation topics does it spin off?'' \cite[p.~xi]{Earman:1989}. To a non-philosopher, this is an astounding confession, for one would have thought that philosophers were not just after \emph{more discussion} but \emph{solutions}.

In this way, the discussions of scientific philosophy become similar to those of speculative philosophy, which it thought to have superseded. Stuck in a field of hardened scientific lava and without any objective, natural standard to assay their merits and discard the unfit, the debates of scientific philosophy too are apt to perpetuate themselves through millennia, to last from everlasting to everlasting. To be sure, the debates of scientific philosophy, unlike those of traditional philosophy, are capable of growing in complexity on a par with the development of scientific tools---but whether this is for better or for worse, one does not dare say.

Can we expect the resolution of a conceptual scientific problem out of the arguments of professional scientific philosophy? Hardly, because its yardstick of success inevitably demands of it that it become argument for argument's sake (cf.\ Earman's remarks). Scientific puzzles have always been solved by more and better scientific investigations, by individuals who dared to give a fresh new look at the physical world around them. For this to be possible, the link with the natural world must not be severed. The establishment of new scientific knowledge is then apt to banish some philosophical debates that look rather rusted and obsolete in the light of the new knowledge. (Are spheres the only acceptable shape for the locus of the heavenly bodies? Is the centre of the universe the only acceptable position of the earth? How can Achilles ever overtake the tortoise?) But scientific philosophy drinks in the waters of science---insatiably: for every banished debate, an untold number of others will arise on the basis of the new science, and so on forevermore.

A telltale indication of scientific philosophy as argument for argument's sake can be found in its conspicuously bellicose language, i.e.\ the language typically used by philosophers as they, intellectual enemies, battle each other---till they draw blood (see below) but not to death: it is essential for the well-being of philosophy that all arguments be kept alive. The scientific philosopher's everyday vocabulary is fraught with all sorts of words allusive of violent conflict, so much so that at times one gets the impression to be reading an account of war. One hears\footnote{All
italicized words from genuine texts by professional philosophers of science.} of \emph{battlegrounds}, where \emph{enemies}, unfurling myriad \emph{pro-} and \emph{anti-ism} flags, \emph{deploy} \emph{attacks} on the \emph{opponent's} \emph{position}, and \emph{defend} their own with \emph{counterattacks}; of \emph{struggles} in which, amidst the \emph{smoke of battle}, \emph{rivals} seek \emph{victory} over \emph{defeat} by dealing each other \emph{blows} and \emph{undermining} each other's \emph{ground}. Not without a shock, we catch an epitomic glimpse of the scientific philosopher's bellicose mindset in Earman's \citeyear{Earman:1989} book on spacetime ontology, where he seeks to ``\emph{revitalize} what has become an insular and \emph{bloodless} philosophical discussion'' (p.~3, italics added). One may be inclined to excuse such expressions as mere figures of speech, but when these repeat themselves constantly, one is led to presume an underlying reason of being.

The reason behind this custom is now apparent: locked within the inert confines of a frozen science, yet seeking development and progress, the focus of scientific philosophy shifts in practice to the dynamic and constantly evolving views of other philosophers. As a result, argument is built upon argument in a viciously upward spiral, which only a fresh confrontation with the \emph{bare bones of nature's raw material} can collapse back to the ground. A sustained reading of scientific philosophy is thus apt to lead a scientist (besides despair) to the following reflection: philosophy---or \emph{philoversy}? The love of knowledge, or the love of war/confusion?\footnote{The etymological association is particularly apt. \emph{Wers-}, meaning ``to mix up, to lead to confusion,'' is the original Indo-European root of the modern English ``war.'' Cf.\ German \emph{verwirren}, meaning ``to confuse, to perplex,'' originally from the Proto-Germanic root \emph{werra-}. From the first reference below we learn that ``Cognates suggest the original sense was `to bring into confusion.' There was no common Germanic word for `war' at the dawn of historical times.'' (See ``war'' in Online Etymology Dictionary, http://www.etymonline.com; and American Heritage Dictionary, http://www.bartleby.com/61)}

This situation reminds us of Wittgenstein's words: 
\begin{quote} A person caught in a philosophical confusion is like a man in a room
who wants to get out but doesn't know how. He tries the window but it is too high. He tries the chimney but it is too narrow. And if he would only \emph{turn around}, he would see that the door has been open all the time!\footnote{Quoted from \cite[p.~51]{Malcolm:1972}.} 
\end{quote} 
The symbolic door is in actuality not so easy to find, but the direction in which we should look to find it is clear: the way out of the philosophical house is the way that originally brought us in, namely, the study of nature. Wittgenstein's metaphorical ``turn around'' turns out to be singularly apt.

To sum up with an exemplifying question: if Einstein had been a philosopher, as he is sometimes called, would he have discovered special relativity through his physical analysis of time, or would he have instead remained scientifically frozen in an endless logical analysis of Newtonian absolute time and space?

All this is not to say that professional philosophy of science \emph{never} helps science or clarifies anything. At times, in the hands of exceptional, scientifically inspired philosophers, professional though they may be, it does; but on a whole, in the hands of the average professional philosopher, it does not.

If the analyses of scientific philosophy go astray due to being logical dissections uninformed by fresh experience, scientists should know better. Scientists, after all, deal with the physical world. This is at any rate true of the ideal scientist, for it is not today uncommon to come across the variety that finds more inspiration in ``self-consistent'' mathematical abstractions than in nature, and has thus more in common with scientific philosophers than scientists. But what about the scientist who investigates the physical world and works in the spirit in which physics has been traditionally carried out for most of its history?\footnote{In connection with space and time, most notably by Riemann and Einstein. In his famous essay, Riemann \citeyear{Riemann:1873}, a mathematician, talks as a physicist when he points out that ``the properties which distinguish [physical] space from other conceivable triply extended magnitudes are only to be deduced from experience,'' and is even wary of assuming the validity of experience ``beyond the limits of observation, on the side both of the infinitely great and of the infinitely small'' (pp.~1--2, online ref.). As for Einstein, his continual use of simple thought experiments during the first half of his life as a starting point for the theoretical concepts of physics is well known.} What does an analysis mean to him? To him, an analysis does not mean definition, postulation, or logical dissection (into what?) but examination. Examination of a concept's \emph{physical} meaning, of its \emph{physical} root and nature. The content of his analysis is not to be found primarily in his head, but in the world around him. This is not to say that an analysis of this type does not rely on the scientist's conceptual tools of thought. It necessarily does, but only in conjunction with a keen observation of nature: his elemental theoretical concepts are
developed directly from observations and experiments---science's raw
material---not off the top of his secluded head. One particular advantage of proceeding thus is that the scientist in question knows from the start what he is talking \emph{about}: the ontology of his foundational concepts is, by and large, uncontroversial.

This being said, we proceed to a geometric analysis of the concept of time. What can be expected from such an analysis? First and foremost, an understanding of relativity theory on the basis of the raw material of simple measurements and on physical (as opposed to mathematical or philosophical) ideas. Secondly, an understanding of relativity built progressively as a natural concatenation of steps, where by ``natural'' we mean that, at every roadblock, we let physics (and, when necessary, psychology) suggest the way forward. In particular, we let only our physical needs suggest what kind of mathematical concepts we need, and we develop these naturally, not as formal, hard-to-swallow definitions without
any apparent reason of being.

An expected upside of this way of proceeding is a minimization of mathematical baggage, but---as every upside surely must have a downside---the sceptical reader may ask, are we perhaps trading depth of understanding for mathematical lightness? Breaking the above-insinuated conservation law of upsides and downsides, we are here unexpectedly confronted by an upside without a downside, but instead conducive to a further upside: it is \emph{because} mathematics is used only in its due measure, as physics demands and no more, that we end up gaining understanding of relativity theory
instead of, as is more common, ending up clouded in confusion. The fact that mathematics is in physics only a tool can be ignored only at the physicist's peril.\footnote{Mathematics as practiced by the mathematician, i.e.\ as a formal system empty of semantic meaning, is another matter altogether and not the subject of this criticism.}

The dire consequences of the abuse of mathematical baggage in physics have also been noticed by other people pondering these issues. Physicist Robert Geroch writes: 
\begin{quote} 
One sometimes hears expressed the view that some sort of uncertainty principle operates in the interaction between mathematics and physics: the greater the mathematical care used to formulate a concept, the less the physical insight to be gained from that formulation. It is not difficult to imagine how such a  viewpoint could come to be popular. It is often the case that the essential physical ideas of a discussion are smothered by
mathematics through excessive definitions, concern over irrelevant generality, etc. Nonetheless, one can make a case that mathematics as mathematics, if used thoughtfully, is almost always useful---and occasionally essential---to progress in theoretical physics.
\cite[p.~vii]{Geroch:1985} 
\end{quote}

In conclusion, for an insightful physical analysis, mathematics in
due measure and professional philosophy not at all.

\section{Clock readings as spatiotemporal basis}

\subsection{Events and particles as geometric objects}\label{Events-and-particles}

As we survey the macroscopic world in connection with classical and relativistic theories, two key concepts come to our attention: that of an event and that of a particle. An event is, quite simply, a phenomenon, anything that we  register as happening: a flash of lightning, the sound of a horn, the touch of a
hand on the shoulder, etc.

Is an event as such a geometric concept? It certainly is a human-made concept, for what we call an event never appears alone and differentiated from the rest of a jumble of impressions. But it certainly is not a geometric concept. We can ask of it a pair of test questions: does an event have a shape? Does the concept of size apply to it? The answer to both questions is no. Not before we idealize it.

The idealization of an event made in physics is one of the most basic yet extreme cases of geometric idealization. Going much farther than having, for example, the more tangible earth crust idealized as an even spherical surface, here we have an indefinite physical phenomenon squeezed into the tightest of all geometric objects, a point.

Does now a point-event have a shape, and can we speak of its size? The first question is the trickier of the two because, of all geometric objects, a point is by far the most abstract---an extreme case. Unlike all other geometric objects, the shape of a point is \emph{irrelevant} because its size (understood in any way) is zero. This answers the second question above: the concept of size does apply to a point, namely, it is zero. This is not at all the same as saying that it is meaningless to speak of its size, as it is to speak of the size of a raw event. Due to their utmost geometric abstractness, points are strictly speaking impossible to visualize, although they are normally thought of and depicted as dots of some irrelevant shape (circular, square, spherical, etc.).

How does this translate to the physically inspired point-events? By geometrically idealizing physical phenomena as points, the former become endowed with two seemingly physical properties: a point-event is something that does not take space and does not have duration. But with such extreme features, how can point-events be measured (or localized) as such? After all, if they do not take space and do not have duration, they must not exist.

At this difficult junction, counting comes to our aid. The quality of being point-like can be characterized analytically by means of numbers: a point is simply a set of real numbers. In classical physics, a point-event is identified with the set of four numbers $(x,y,z;t)$, and the whole of all possible point-events forms a four-dimensional continuum called space-time---with a hyphen, since $t$ is special and given separately. The same idea can be transferred to relativity theory by loosening the restriction above, according to which time $t$ is given independently of the other three numbers. In relativity theory, a point-event is identified with the set of four numbers $(x^1,x^2,x^3,x^4)$, and the whole of all possible point-events forms a four-dimensional continuum called spacetime---without a hyphen, since no element of the set $(x^1,x^2,x^3,x^4)$ can be separately singled out as special and, in particular, representing time.

The sets of four numbers $(x,y,z;t)$ and $(x^1,x^2,x^3,x^4)$ are the
coordinates of a point-event, but what physical meaning shall we attribute to them? If coordinates are to have any meaning, it must be given them operationally: \emph{how} are these numbers attributed to a point-event?

Since the space-time coordinate $t$, symbolizing absolute time in $(x,y,z;t)$, has in fact no operational meaning, we focus only on the general spacetime coordinates $(x^1,x^2,x^3,x^4)$. To localize a spacetime event, we resort to an arbitrary operational method, whereby we specify the way in which the four numbers are measured. Consider as an example Ohanian's \citeyear[pp.~222--223]{Ohanian:1976} astrogator coordinates used  by a hypothetical astronavigator. To determine the spacetime coordinates of his spaceship, he identifies four different stars (red, yellow, blue, and white), measures the angles $\theta^\alpha$ between any four different pairs of stars, and associates each number with a spacetime coordinate, $\theta^\alpha =: x^\alpha$ ($\alpha=1,2,3,4$). 

In spacetime, no coordinate is differentiated from the others; each shares the same status with the other three. But so that two distinct events (e.g.\ two distinct phenomena inside the spaceship) will be tagged with different numbers, at least one of the stars (but possibly all four) must be in relative motion with respect to any of the other three. Ohanian calls this coordinate the ``time coordinate.'' However, this name may be misleading for, if all the stars are moving relative to one another, which one is to be the ``time coordinate''? If the four coordinates are operationally equivalent and any of them can be taken as the ``time coordinate,'' then why give one a special name without reason? The ghost of Newtonian time keeps haunting relativity. 

The establishment of a connection between time and a coordinate, say, $x^4$, is only possible in special cases. For example, if spacetime is separable into a spatial part and a temporal part, i.e.\ if its quadratic form $Q$ can be written as
\begin{equation}
	Q=g_{ab}(x^1,x^2,x^3,x^4)\D x^a \D x^b + g_{44}(x^1,x^2,x^3) (\D x^4)^2, 
\end{equation}
and if we choose comoving coordinates such that $x^1=x^2=x^3=\mathrm{constant}$ (only one moving star with respect to the ship), then $\D s^2=\mathrm{constant}\ (\D x^4)^2$. When these requirements are fulfilled, the differential $\D x^4$ of the fourth coordinate is directly proportional to the clock separation $\D s$ (see Sections \ref{Clocks-and-line-element} and \ref{Problem-cosmology}). 

Whenever spacetime coordinates are chosen without establishing at the same time their operational meaning (e.g.\ spherical coordinates), these remain physically meaningless until a connection is established between them and a measurable parameter such as $s$. For example, once the metric field is known and the geodesic equation solved, one can find a relation $x^\alpha=x^\alpha(s)$.

Given two sets of coordinates $(x^1,x^2,x^3,x^4)$ and $(y^1,y^2,y^3,y^4)$, we can find analytic transformations $x^\alpha=x^\alpha(y)$ and $y^\alpha=y^\alpha(x)$ that take one set of coordinate values to another ($x$ represents the whole set of coordinate values, and likewise for $y$).

The next concept essential to classical and relativistic theories is that of a (massive or massless) particle. Like events but less dramatically so, particles are also idealized geometrically as points. Unlike events, however, particles are taken to be moving points both in classical and relativistic theories. An event and a particle are connected as follows: the history of a particle is a sequence of events.

In classical physics, the history of a particle is given by the set of equations $\{x=x(t),y=y(t),z=z(t)\}$, where $t$ has the role of a (coordinate) parameter. These three equations determine a particle's classical worldline. In relativistic physics, on the other hand, the history of a particle is given by the set of equations $x^\alpha=x^\alpha(u)$, where $u$ is a one-dimensional (non-coordinate) parameter. These four equations determine a particle's relativistic worldline. Now, if possible and if so wanted, one could solve for $u$ (e.g.\ $u=u(x^4)$) and use this to rewrite the above set of four equations as a set of three, $x^a=x^a(x^4)$ ($a=1,2,3$), but in so doing the separation between space and time has not taken place---there is no reason to take in general $x^4$ as a time coordinate in relativity theory.

In classical physics, a particle of mass $m$ satisfies the equations of motion
\begin{equation}
m\frac{\D^2 x^a}{\D t^2}=F^a,
\end{equation}
where the classical force $F^a$ is of the form $F^a(x,\mathrm{d}x/\mathrm{d}t)$. These three equations determine the history of a particle, or its worldline, if one is given (i) an event $(x^1,x^2,x^3;t)$ that belongs to it and (ii) a direction in space-time at the said event, i.e.\ the ratios between the differentials $\mathrm{d}x^a$ and $\mathrm{d}t$. Can this idea of an equation of motion also be applied in relativity theory? We now demand that the history of a massive particle, or its worldline, be determined once one is given (i) an event $(x^1,x^2,x^3,x^4)$ that belongs to it and (ii) a direction in spacetime at the said event, i.e.\ the ratios between the differentials $\mathrm{d}x^\alpha$ and $\D u$. This demand can be met if the differential equations for the worldline of a particle are of the form 
\begin{equation}\label{equation-of-motion}
\frac{\mathrm{d}^2x^\alpha}{\mathrm{d}u^2}=X^\alpha,	
\end{equation}
where $X^\alpha=X^\alpha(x,\mathrm{d}x/\mathrm{d}u)$. Here $\mathrm{d}u$ appears only as a differential. What is $X^\alpha$ equal to and what is its physical meaning? We return to this issue later on, when the time is ripe for it.

\subsection{The observer as psychological agent}
We saw that physical phenomena are geometrically idealized as point-events, and that matter is geometrically idealized as point-particles. In particular, now a human observer is geometrically idealized as a moving point-particle that makes observations along its worldline. Conventionally, the so-called observer does not need to be human, but is embodied by a set of measuring devices moving along a worldline. However, there exists one essential side to the observer that only a human observer can provide, namely, its subjective experience. According to this experience, events can be laid down in a certain order along the human observer's worldline, who thus creates in his mind the \emph{past}, the \emph{present}, and the \emph{future}.

We saw earlier that none of the four coordinates of a relativistic event $(x^1,x^2,x^3,x^4)$ could in general be isolated as time. The isolation of one of these four coordinates as universal time would unavoidably lead to an absolute ordering of events as in classical physics (classical absolute time), but relativity theory does not claim the existence of a natural order for an arbitrary pair of point-events. When at all possible, a pair of events can only be ordered with respect to the worldline of a given observer---he is actually the one who does the ordering. For those pairs of events that cannot possibly be contained on the observer's worldline due to restrictions of his motion, no ordering relation is possible at all. All these events belong together in the relativistic \emph{four-dimensional present}, whereas those than can be ordered along the observer's worldline belong in the relativistic \emph{four-dimensional past} or \emph{future}. 

\begin{figure}
\centering
\includegraphics[width=\linewidth]{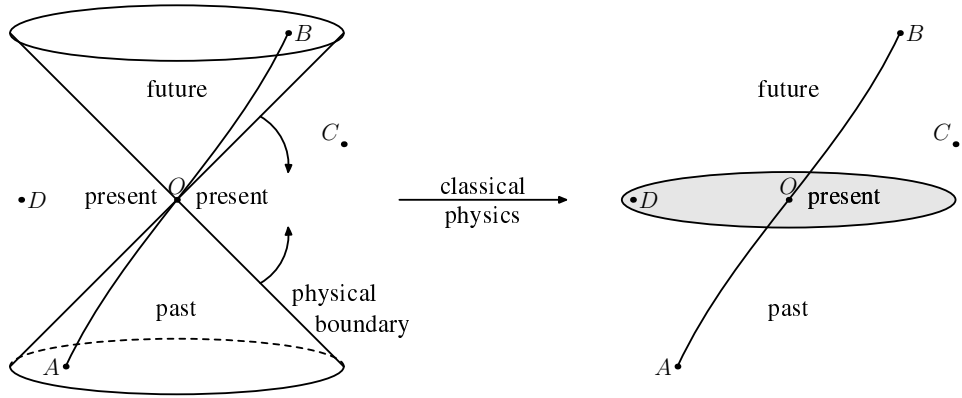}
\caption[Past, present, and future]{In a relativistic world, an
observer determines the past, the present, and the future, as he orders events on his worldline into these three categories according to his subjective experience. The present is formed by the set of all events that lie outside the physical boundary imposed by restrictions on his motion. The classical conception of an absolute present arises when the boundary is collapsed onto a three-dimensional plane \mbox{$t=\mathrm{constant}$.} Event $D$ belongs to both the relativistic and classical present, while event $C$ belongs to the relativistic present but to the classical future.} 
\label{past-present-future}
\end{figure}

Also the classical worldview can be understood from this temporal perspective. As the constraint imposed by the three-dimensional hypersurface limiting the motion of the observer is relaxed, the forbidden four-dimensional present becomes an allowed three-dimensional present, together with a wider four-dimensional past and future (Figure \ref{past-present-future}). Classical physics, then, appears to be a special case of relativistic physics as far as the dimensions of the present are concerned, but, on the other hand, relativistic physics is also more limiting than classical physics, because in it the present is a forbidden area. The restrictive nature of the relativistic present is one of the reasons why classical physics continues to be preferred for most of our daily needs, since in it the allowed present is understood more widely.

\subsection{Clocks and the line element} \label{Clocks-and-line-element}
The foundations of classical and relativistic physics can be laid down on the basis of \emph{clock measurements}. Ideally, a clock should progress at a constant pace; this requirement is realized in practice by searching for natural processes with respect to which a great many other natural processes are in synchronicity, thereby simplifying their study. As an ideal clock can only be determined by comparison with other clocks, its basic property may be given as follows: two ideal clocks carried along by an observer along his worldline remain synchronous---their ticking ratio remains constant---for any worldline $\mathcal{C}$ joining any pair of events $A$ and $B$.

Consider now an observer who carries an ideal clock along his worldline $\mathcal{C}$ from point-event $A$ to point-event $B$. At $A$ he records the \emph{clock reading} $s_A$ and on reaching $B$ records the new \emph{clock reading} $s_B$. These clock readings (not time instants) are the results of physical measurements. We now say that the \emph{separation} between events $A$ and $B$ along worldline $\mathcal{C}$ is \mbox{$\Delta s=s_B-s_A$.} When event $B$ is close to event $A$, by which we mean operationally that the observer ``stops his watch quickly after $A$,''\footnote{We may also say that $A$ and $B$ are close if their coordinate values differ by a small amount.} a good approximation to $\Delta s$ is furnished by the separation $\D s$ along a straight worldline element; in symbols, $\Delta s \approx \D s$.

The worldline-element separation $\mathrm{d}s$ as here presented (in terms of clock measurements) gives enough foundation to lead, via different choices, to the concepts of either Newton's classical physics, Einstein's special- and general-relativistic physics, and presumably to any other geometric theory of space and time. Here we shall deal with classical and relativistic theories on the basis of $\mathrm{d}s$. 

\begin{figure}
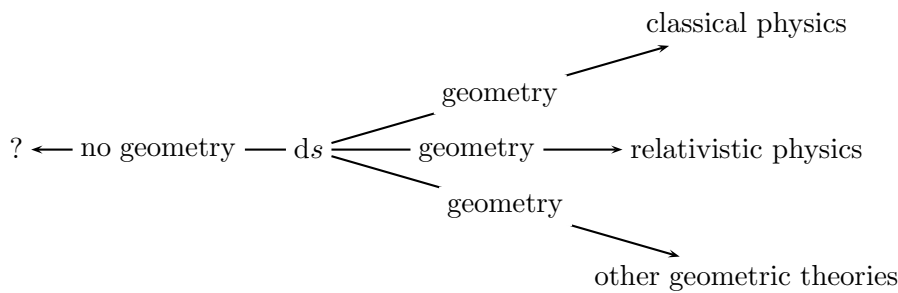

\centering
\[
\begin{psmatrix}[colsep=3.6cm,rowsep=1.2cm]
       &               & {\small \mathrm{classical\ physics}} \\
     {\small?} & {\small\mathrm{d}s}   & {\small \mathrm{relativistic\ physics}} \\
       &               & {\small \mathrm{other\ geometric\ theories}}
\psset{arrows=->,labelsep=3pt,nodesep=3pt}
\ncline{2,2}{2,1}\ncput*{{\small \mathrm{no\ geometry}}}
\ncline{2,2}{2,3}\ncput*{{\small \mathrm{geometry}}}
\ncline{2,2}{1,3}\ncput*{{\small \mathrm{geometry}}}
\ncline{2,2}{3,3}\ncput*{{\small \mathrm{geometry}}} \end{psmatrix} \]
\caption[Line element as starting point]{The line element
$\mathrm{d}s$ can act as a starting point for the development of
classical and relativistic theories, and presumably for any other
geometric theory of space and time, when we choose to geometrize it
according to tradition. But what kind of metageometric theory, describing what physics, would one arrive at if one followed the opposite, no-geometry direction?}
\label{line-element}
\end{figure}

It is noteworthy, however, that $\mathrm{d}s$, as a temporal correlation between two events, or two ``nows,'' is not a geometric concept. There is nothing geometric in the (psychological) identification of the present (when the initial and final observations are made), and there is nothing geometric in an event, since, as we saw, these are not originally points but raw phenomena. The activity comprised by a clock measurement and the values it produces, however, \emph{can} lead to a geometric representation. Such a geometrization and its main results are in fact the essence of this chapter. These observations lead then to a provoking thought: what kind of theory may be arrived at by following an opposite, no-geometry direction from $\mathrm{d}s$? What kind of metageometric theory, describing what physics of time, would this be? (Figure~\ref{line-element}).

A geometric picture emerges when events become \emph{point}-events, an observer a \emph{point}-particle, an observer's history a world-\emph{line}, $\Delta s$ the \emph{length} of the worldline between the joined point-events, and $\mathrm{d}s$ the length of the differential \emph{arrow} $\D x$ between point-events $A$ at $x$ and $B$ at $x+\mathrm{d}x$, i.e.\ when we start to view $\mathrm{d}s$ in terms of coordinate points and coordinate arrows. In fact, the geometrization of the differential $\D s$ leads us---naturally enough---to differential geometry. What else do we know about $\D s$? How does it depend on the coordinates of point-events $A$ and $B$? 

To move forward, some choices must be made. The main stopovers on the road to Einstein's geometric theory of gravitation have, in fact, already been called into attention in two public lectures: by Riemann in 1854, ``On the foundations which lie at the bases of geometry,'' and by Einstein in 1922, ``How I created the theory of relativity''---it is through their steps that we shall progress. 

The first clue is older than relativity itself and comes from Riemann's famous 1854 lecture. As he sets himself the task of determining the length of a line in a manifold, he writes:
\begin{quote} The problem consists then in establishing a
mathematical expression for the length of a line\ldots I shall
confine myself in the first place to lines in which the ratios of
the quantities $\mathrm{d}x$ of the respective variables vary
continuously. We may then conceive these lines broken up into
elements, within which the ratios of the quantities $\mathrm{d}x$
may be regarded as constant; and the problem is then reduced to
establishing for each point a general expression for the linear
element $\mathrm{d}s$ starting from that point, an expression which
will thus contain the quantities $x$ and the quantities
$\mathrm{d}x$. I shall suppose, secondly, that the length of the
linear element, to the first order, is unaltered when all the points
of this element undergo the same infinitesimal displacement, which
implies at the same time that if all the quantities $\mathrm{d}x$
are increased in the same ratio, the linear element will also vary
in the same ratio. On these assumptions, \emph{the linear element may be
any homogeneous function of the first degree of the quantities $\mathrm{d}x$}\ldots \cite[p.~5, online ref., italics added]{Riemann:1873}
\end{quote} 
Following Riemann, we set the line element $\mathrm{d}s$ to be a function of the coordinates of $A$ and $B$,
\begin{equation}
    \mathrm{d}s=f(x,\mathrm{d}x),
\end{equation} 
where $f(x,\mathrm{d}x)$ is a homogeneous function of first degree in the differentials $\mathrm{d}x$ , i.e.\ $f(x,k\mathrm{d}x)=kf(x,\mathrm{d}x)$ ($k>0$ can be a function of $x$).\footnote{We have found that also Synge \citeyear{Synge:1964} has understood relativity chronometrically based on the Riemannian hypothesis.}  When the equation of worldline $\mathcal{C}$ is given by $x^\alpha=x^\alpha(u)$, where $u$ is a parameter whose value grows from the past to the future, we can then write
\begin{equation}
    \mathrm{d}s=f\left(x,\frac{\mathrm{d}x}{\D u}\D u\right)=
    f\left(x,\frac{\mathrm{d}x}{\D u}\right)\D u,
\end{equation}
and therefore the separation $\Delta s$ between $A$ and $B$ is given by
\begin{equation}
    \Delta s=\int^B_{\begin{smallmatrix}A\\\mathcal{C}
    \end{smallmatrix}}\mathrm{d}s=
    \int_{u_A}^{u_B}f\left(x,\frac{\mathrm{d}x}{\mathrm{d}u}\right)\mathrm{d}u.
\end{equation}

\subsection{Invariance of the line element}\label{sec:Invariance-line-element}
Talking about his personal struggle to understand gravitation geometrically, Einstein \citeyear{Einstein:1982} relates during his Kyoto lecture discussing with Marcel Grossmann whether ``the problem could be solved using Riemann theory, in other words, by using the concept of the invariance of the line elements'' (p.~47). This is a key issue, whose study now allows us to move on.

Event separation $\Delta s$ and its differential approximation $\D s$ are the results of physical measurements with a clock made by the observer along his worldline; $\D s$ is an absolute spacetime property of this worldline, namely, its extension between events $A$ and $B$. Because $\D s$ is here meant only in this classical sense, its invariance is a simple matter.\footnote{In the present context, the physical scale (seconds, minutes, hours) or dimension used to express $\D s$ is inconsequential to its invariance. A measurement result expressed as $60\ \mathrm{s}$, $1\ \mathrm{min}$, etc., or in a different dimension, say $2\pi\ \mathrm{rad}$, makes reference to the same objective physical extension---this is something we can metaphorically point at, in the same way that ``book'' and ``Buch'' refer to the same physical object. (Cf.\ comments by Eakins and Jaroszkiewicz \citeyear[p.~10]{Eakins/Jaroszkiewicz:2004}). Beyond the tentative, exploratory remarks at the end of Chapter~\ref{ch:ESE}, a quantum-mechanical discussion of invariance (in particular, that of $\D s$) falls outside the scope of this thesis. For a discussion of quantum experiments involving two different inertial frames, see \cite[pp.~50--54]{Jaroszkiewicz:2007}.} Having proposed that the line element is a function of the coordinates, $\D s=f(x,\D x)$, the \emph{physical} raw material encoded as $\D s$ now demands that its \emph{mathematical} expression $f(x,\D x)$ be an invariant in the mathematical sense of the word. That is, the value of $f(x,\D x)$ must be independent of the set of coordinates chosen by the observer on his worldline, i.e.\ $f(x,\D x)=f'(y,\D y)$, where $y^\alpha=y^\alpha(x)$.

The need to deal with physical invariance leads us to search for mathematical objects that are meaningful regardless of the coordinate system in which they may be expressed. \emph{Tensor equations} (not just tensors) are such objects; if a tensor equation holds in one coordinate system, it holds in any coordinate system. Tensor equations have this property because tensor transformations between coordinate systems are linear and homogeneous.

Events $A$ at $x$ and $B$ at $x+\mathrm{d}x$ determine a differential displacement vector whose components are $\mathrm{d}x^\alpha$ and whose tail is fixed at $A$. In another coordinate system, the components of this differential displacement vector are $\mathrm{d}y^\alpha$ and can be obtained by differentiation of the coordinate transformation $y^\alpha=y^\alpha(x)$. We find
\begin{equation}
\mathrm{d}y^\alpha=\frac{\partial y^\alpha}{\partial x^\beta}\mathrm{d}x^\beta,
\end{equation}
which is a linear and homogeneous coordinate transformation of the differential with coefficients $\partial y^\alpha/\partial x^\beta$. The set of differentials $\mathrm{d}x^\alpha$, together with the way they transform from one coordinate system to another, gives us a prototype for a first type of vector (and, in general, tensor) called \emph{contravariant}.\footnote{Although contravariant vectors and tensors are inspired in differential displacement vectors (their components are small), the components of contravariant vectors and tensors do not in general need to be small. For example, the components $\mathrm{d}x^\alpha/\mathrm{d}u$ of the derivative along a worldline represent a unbounded contravariant vector.}

Following the contravariant vector prototype, we can build other sets of magnitudes attached to $A$ that transform in the same way. The magnitudes $T^{\alpha\beta}$, transforming linearly and homogeneously in $T^{\alpha\beta}$ according to
\begin{equation}
    T'^{\alpha\beta}=\frac{\partial x'^\alpha}{\partial x^\gamma}\frac{\partial x'^\beta}{\partial x^\delta} T^{\gamma\delta},
\end{equation}
constitute a second-order \emph{contravariant tensor}. And similarly for higher-order contravariant tensors.

A zeroth-order tensor is a scalar that transforms according to the identity relation, $T'=T$, i.e.\ it is an invariant. If we now take a zeroth-order tensor to be a function of the coordinates $f(x)$, we must have that \mbox{$f(x)=f'(x')$.} How does the gradient of $f(x)$ transform when expressed in another coordinate system? Using the chain rule, we have
\begin{equation}
    \frac{\partial f'(x')}{\partial x'^\alpha}=
    \frac{\partial f(x)}{\partial x'^\alpha}=
    \frac{\partial x^\beta}{\partial x'^\alpha} \frac{\partial f(x)}{\partial x^\beta}.
\end{equation}
It turns out that a gradient transforms in a manner opposite to that of contravariant vectors (and tensors). The gradient then serves as a prototype for a second type of vector (and, in general, tensor) called \emph{covariant}. Historically, this name actually comes from a shortening and reshuffling of ``variation co-gradient,'' i.e.\ variation in the same manner as the gradient, and the name contravariant comes from ``variation contra-gradient,'' i.e.\ variation against the manner of the gradient.

Like before, following the covariant-gradient prototype, we can have other sets of magnitudes attached to $A$ that transform in the same way. The magnitudes $T_ {\alpha\beta}$, transforming linearly and homogeneously in $T_{\alpha\beta}$ according to
\begin{equation}
    T'_{\alpha\beta}=\frac{\partial x^\gamma}{\partial x'^\alpha}\frac{\partial x^\delta}{\partial x'^\beta} T_{\gamma\delta},
\end{equation}
constitute a second-order \emph{covariant tensor}. And similarly for higher-order covariant tensors, and also for mixtures of both types. Finally, when a tensor is attached not only to a single event, but to every event along a worldline or to every event in the four-dimensional spacetime continuum, we have a spacetime field called \emph{tensor field}.

\subsection{The Newtonian line element}
So far, we know two chief properties of $\D s$. One is that it represents a classical measurement of an absolute spacetime property and is therefore a physical invariant; this means that $f(x,\D x)$ must be a mathematical invariant. The other is that $f(x,\D x)$ is homogeneous with respect to all components $\D x^\alpha$. Within these requirements there is still freedom to choose the form of $f(x,\D x)$, but what choices lead to physically relevant results? We proceed in what one may call a natural way. Let us try to separate the dependence of $f(x,\D x)$ on $x$ and $\D x$ using the mathematical invariance and homogeneity of this function. Both requirements can be fulfilled by considering an \emph{inner product}. 

We propose as a first case to be analysed an inner product of two
vectors, a \emph{linear form} 
\begin{equation}\label{linear-form}
    \D s=f(x,\D x)=g_\alpha(x)\D x^\alpha.
\end{equation}
This function is clearly homogeneous of first degree with respect to $\D x^\alpha$ and produces an invariant number. Moreover, because $g_\alpha(x)\D x^\alpha$ is invariant and $\D x^\alpha$ is a general contravariant vector, we must have that $g_\alpha(x)$ is a covariant vector.\footnote{The invariance of the inner product means that $g'_\alpha\D x'^\alpha=g_\beta \D x^\beta$, and the fact that $\D x^\alpha$ is a contravariant vector means that it transforms as $\D x'^\alpha=(\partial x'^\alpha/\partial x^\beta)\D x^\beta$. Together these two relations lead to $[(\partial x'^\alpha/\partial x^\beta)g'_\alpha - g_\beta]\D x^\beta=0$. On account of $\D x^\beta$ being an arbitrary contravariant vector, the bracketed expression must be null, which means that $g_\alpha$ transforms as a covariant vector.} But what is the physical meaning of $g_\alpha(x)$?

The simplest option is to choose the covariant vector after the covariant prototype,  
\begin{equation}\label{covariant-prototype}
g_\alpha(x)=\frac{\partial t(x)}{\partial x^\alpha},
\end{equation}
where $t(x)$ is an invariant function of the spacetime coordinates $x$. The separation between events $A$ and $B$ becomes
\begin{equation}
    s_B-s_A =
    \int_A^B \D s = \int^B_{\begin{smallmatrix}A\\\mathcal{C}
    \end{smallmatrix}} \frac{\partial t(x)}{\partial x^\alpha}\D x^\alpha = \int^B_{\begin{smallmatrix}A\\\mathcal{C}
    \end{smallmatrix}} \D t(x) = t_B-t_A.
\end{equation}
In other words, because $\D t(x)$ is an exact differential, the separation between events $A$ and $B$ does not depend on the worldline $\mathcal{C}$ travelled. Because we know experimentally that this result is false, we can conclude that there must be something wrong with our choice of Eq.~(\ref{covariant-prototype}), and presumably even with the linear form (\ref{linear-form}). On the positive side, we learn that this choice leads to the interpretation of $s$ as \emph{absolute classical time} $t$. This is one of the essential differences between classical and relativistic physics; in the latter, $\D s$ is not an exact differential and the separation $\Delta s$ between events depends on the worldline travelled.

\subsection{The Riemann-Einstein line element}
The second simplest choice for $f(x,\D x)$ satisfying the same two requirements set above consists of an inner product of a second-order covariant tensor and two differential displacement vectors, that is, a \emph{quadratic form} 
\begin{equation}\label{line-element-eq}
    \D s= f(x,\D x)=\sqrt{-g_{\alpha\beta}(x)\D x^\alpha \D x^\beta}.
\end{equation}
The square root is necessary to meet the requirement of homogeneity of first degree with respect to the differential displacements, whereas the reason for the minus sign will become apparent later on.

On the basis of a similar procedure as above (see previous footnote), we can now deduce something about the tensorial nature of $g_{\alpha\beta}(x)$, but not as much as before. Although $\D x^\alpha$ is a general contravariant tensor, since it appears in the symmetric product $\D x^\alpha \D x^\beta$, we can now only deduce that $g_{\alpha\beta}+g_{\beta\alpha}$ must be a second-order covariant tensor, but we cannot conclude anything about the tensorial nature of $g_{\alpha\beta}$ itself on the basis of the invariance of the inner product. However, if we assume that $g_{\alpha\beta}(x)$ is symmetric---and this we shall do---then we can conclude that $g_{\alpha\beta}(x)$ is a second-order covariant tensor.\footnote{Einstein considered the possibility of an asymmetric metric tensor in the latter part of his life in the context of attempts at a unified theory of electromagnetism and gravity.}

Given then a symmetric second-order covariant tensor $g_{\alpha\beta}(x)$ that is, moreover, non-singular ($\mathrm{det}(g_{\alpha\beta})\neq 0$), the invariant quadratic form 
\begin{equation}
Q=g_{\alpha\beta}(x)\D x^\alpha \D x^\beta	
\end{equation}
determines a four-dimensional continuum called \emph{Riemann-Einstein spacetime}. The line element (\ref{line-element-eq}) of Riemann-Einstein spacetime, which is based on this quadratic form, is the foundation of relativity theory.

A challenging question, however, remains: what is the physical meaning of $g_{\alpha\beta}(x)$? To sketch a solution, we now need to take a look at the foundations of Riemannian metric geometry. In connection with the search for a new description of gravitation, the necessity of this stopover was related by Einstein during his Kyoto lecture:
\begin{quote}
What should we look for to describe our problem? This problem was unsolved  until 1912, when I hit upon the idea that the surface theory of Karl Friedrich Gauss might be the key to this mystery. I found that Gauss' surface coordinates were very meaningful for understanding this problem. Until then I did not know that Bernhard Riemann [who was a student of Gauss] had discussed the foundation of geometry deeply\ldots I found that the foundations of geometry had deep physical meaning. \cite[p.~47]{Einstein:1982}
\end{quote}

\subsection{From Gaussian surfaces to Riemannian space}
Metric geometry---geometry that deals essentially with magnitudes---has length as its basic concept. In Euclidean metric geometry, this concept always refers to the finite distance between two space points. In Riemannian metric geometry, on the other hand, it refers to the distance between two very (not necessarily infinitesimally) close points. In other words, the passage from finite Euclidean geometry to differential Riemannian geometry is achieved by approximating the finite distance $\Delta l$ between two points with the differential distance $\D l$.

We start by examining Cartesian coordinates (straight lines meeting at right angles) in a three-dimensional Euclidean space. On the basis of Pythagoras' theorem, the squared distance between the points $(y^1,y^2,y^3)$ and $(y^1+\D y^1,y^2+\D y^2,y^3+\D y^3)$ is given by 
\begin{equation}
      \D l^2=(\D y^1)^2+(\D y^2)^2+(\D y^3)^2.	
\end{equation}
In terms of general coordinates $x^1,x^2,x^3$ (curved lines meeting at oblique angles), because $\D y^i=(\partial y^i/\partial x^a)\D x^a$ ($i,a=1,2,3)$, we obtain 
\begin{equation}
      \D l^2=h_{ab}(x)\D x^a \D x^b,
\end{equation}
where
\begin{equation}
    h_{ab}(x)=\sum_{i=1}^3 \frac{\partial y^i}{\partial x^a}
    \frac{\partial y^i}{\partial x^b}
\end{equation}
is symmetric with respect to $a$ and $b$ by inspection. Again, because $h_{ab}(x)$ is assuredly symmetric, the squared length $\D l^2$ is an invariant, and $\D x^a$ is a general vector, we can conclude (based on the results of the previous section) that $h_{ab}(x)$ is a (symmetric) second-order covariant tensor.\footnote{If $h_{ab}(x)=0$ when $a\neq b$, the coordinate lines meet at right angles. If, in addition, $h_{ab}(x)=1$ when $a=b$, the coordinates are Cartesian.} This tensor determines the metric structure, or metric, of three-dimensional Euclidean space, and is therefore called the \emph{metric tensor}.

The next step towards Riemannian space is to consider with Gauss a two-dimensional \emph{surface} $y^i=y^i(x^1,x^2)$ ($i=1,2,3$), where $x^1$ and $x^2$ are parameters, \emph{embedded} in a three-dimensional Euclidean space. Now the squared distance between neighbouring points on the surface is given by 
\begin{equation}
\D l^2=(\D y^1)^2+(\D y^2)^2+(\D y^3)^2=\gamma_{ab}(x)\D x^a \D x^b,	
\end{equation}
where
\begin{equation}
    \gamma_{ab}(x)=\sum_{i=1}^3 \frac{\partial y^i}{\partial x^a}
    \frac{\partial y^i}{\partial x^b} \qquad (a,b=1,2).
\end{equation}
This symmetric second-order covariant tensor is called the \emph{surface-induced metric}, i.e.\ the metric that the surrounding three-dimensional Euclidean space induces on the surface. What happens if we now \emph{remove} the space surrounding the surface altogether?

Although it was Gauss who first developed the idea of a surface embedded in a higher-dimensional Euclidean space, it was Riemann who took the seemingly nonsensical step of removing the embedding space and taking what remained seriously: a two-dimensional \emph{non-Euclidean space}, a two-dimensional continuum determined by the invariant quadratic form 
\begin{equation}\label{two-dim-Q}
Q=\gamma_{ab}(x)\D x^a \D x^b \qquad (a,b=1,2),	
\end{equation}
where $x^1$ and $x^2$ are Gauss's surface coordinates and $\gamma_{ab}(x)$ is a second-order covariant tensor. We assume this new tensor inherits its metric role in this new space from Gauss's case, and that it, too, is symmetric. What remains, then, is a two-dimensional Riemannian \emph{space}, a space one can inhabit but not look at from the outside because it has no outside. For example, what is normally thought as a spherical surface is, in fact, only a Gaussian picture (embedded in Euclidean space) of what is truly a non-Euclidean elliptical plane.

If $\gamma_{ab}(x)$ is the (symmetric) metric tensor of two-dimensional Riemannian space, does this mean that its quadratic form gives a meaningful idea of the squared distance between two neighbouring points of this space? As it stands, it does not, because there is no reason to assume that the quadratic form (\ref{two-dim-Q}) is positive definite,\footnote{The quadratic form is positive definite if it is always positive except when all the components of the differential displacement $\D x^a$ are null, i.e.\ when the two points in question coincide with each other.} as it was for Euclidean space and Gaussian surfaces embedded in it. In Riemannian space, the quadratic form can be negative for some differential displacements $\D x^a$ (i.e.\ in certain directions from a point), positive for others, and null for others; in consequence, the concept of distance must be introduced separately. This can be done with the help of an indicator $\epsilon$, which can have the values $+1$ or $-1$, such that 
\begin{equation}
\D l^2=\epsilon \gamma_{ab}(x)\D x^a \D x^b	\geq 0
\end{equation}
for all $\D x^a$. The equality holds for a null displacement, but can also hold for a non-null one, i.e.\ the distance between two points can be null even when they are not coincident.

Indicators can also be used in a different way. The coordinates can be so chosen that the quadratic form $\gamma_{ab}(x)\D x^a \D x^b$ can be brought into the simpler one
\begin{equation}
Q=\epsilon_1(\D x'^1)^2+\epsilon_2(\D x'^2)^2, 	
\end{equation}
\emph{at one point at a time} but not globally.\footnote{To do this, write $\gamma_{ab}(x)\D x^a \D x^b$ ($a=1,2$) in explicit form, complete squares, and make the change of variables $ x'^1=\sqrt{\epsilon_1\gamma_{11}}x^1+(\gamma_{12}/\sqrt{\epsilon_1\gamma_{11}}x^2)$, $x'^2=\sqrt{\epsilon_2[\gamma_{22}-(\gamma_{12})^2/\gamma_{11}]}x^2$ to obtain $\epsilon_1(\D x'^1)^2+\epsilon_2(\D x'^2)^2$. The quadratic form thus obtained is valid only at a point at a time, because for this method to work $\gamma_{ab}(x)$ must be kept constant. The indicators should be chosen so as to avoid the appearance of imaginary roots, thus keeping the transformations real.} For a Riemannian space with a positive-definite quadratic form, we have the case $\epsilon_1=\epsilon_2=+1$, leading to a locally Cartesian quadratic form; it  describes a space whose geometry is locally Euclidean. For a Riemannian space with an indefinite quadratic form, we have the rest of the cases, leading to a pseudo-Cartesian quadratic form; they describe a space whose geometry is locally pseudo-Euclidean. The pair $(\epsilon_1,\epsilon_2)$ is called the space signature and it is an invariant, i.e.\ it cannot be changed by means of a (real) coordinate transformation.

With the introduction of a stand-alone two-dimensional Riemannian space, we have banished three-dimensional Euclidean space. But as we next add a dimension to take into account a three-dimensional space, should Euclidean space be resurrected again? If we want to conserve the chance of describing new physics through the idea of Riemannian space, we should not. We should now consider a three-dimensional Riemannian space (i) which is not embedded in any higher-dimensional Euclidean space, (ii) whose geometry is by nature non-Euclidean, and (iii) which is determined by the invariant indefinite quadratic form $h_{ab}(x)\D x^a \D x^b$ ($a,b=1,2,3$). Here $h_{ab}(x)$ is a symmetric second-order covariant metric tensor, and the quadratic form in question is locally Cartesian or locally pseudo-Cartesian, i.e.\ locally of the form 
\begin{equation}
Q=\epsilon_1(\D x^1)^2+\epsilon_2(\D x^2)^2+\epsilon_3(\D x^3)^2,	
\end{equation}
with $\epsilon_1,\epsilon_2,\epsilon_3$ each being $+1$ or $-1$. Finally, the concept of distance is attached to this three-dimensional space via the use of indicator $\epsilon$ as before:\footnote{In physical spacetime, $\D l^2$ is not an invariant, but the spatial part of the invariant $\D s^2$ (when it is possible to make the space-time split).} 
\begin{equation}
\D l^2=\epsilon h_{ab}(x)\D x^a \D x^b.
\end{equation}

\section{Riemann-Einstein physical spacetime}

\subsection{Towards local physical geometry}
Having studied Gaussian surfaces and Riemannian space to suit our needs, we now turn to physical spacetime, resuming where we left earlier.

We proposed the form of the line element $\D s$ of Riemann-Einstein spacetime in Eq.~(\ref{line-element-eq}) to be
\begin{equation}
\D s=\sqrt{-g_{\alpha\beta}(x)\D x^\alpha \D x^\beta}. \nonumber	
\end{equation}
We can now make physical sense of this expression, including an elucidation of the minus sign, by approaching the line element from the perspective of Riemannian geometry. Because $\D s$ represents clock-reading separations and these are positive real numbers, $\D s$ is always real and positive. Its square $\D s^2$ must then also be real and positive.\footnote{The case of a photon is dealt with later.} On the other hand, we saw that the quadratic form $g_{\alpha\beta}(x)\D x^\alpha \D x^\beta$ of a Riemannian (now four-dimensional) space is not positive definite. However, we learnt how to solve this problem (in trying to get a non-negative length $\D l^2$) using an indicator $\epsilon$. It is then natural to propose
\begin{equation}\label{squared-line-element}
\D s^2=\epsilon g_{\alpha\beta}(x)\D x^\alpha \D x^\beta
\end{equation}
in order to ensure that $\D s^2$ is never negative.

As an aside, we also find the geometric meaning of $g_{\alpha\beta}(x)$: $g_{\alpha\beta}(x)$ is the \emph{metric tensor}, or metric, of Riemann-Einstein four-dimensional spacetime. As a Riemannian metric, $g_{\alpha\beta}$ links contravariant and covariant vectors, $T^\alpha$ and $T_\alpha$, which have otherwise no connection; the invariant inner product between two such vectors $T^\alpha T_\alpha$ is the same as the invariant quadratic form $g_{\alpha\beta}T^\alpha T^\beta$, whereby we obtain the identification 
\begin{equation}
g_{\alpha\beta}T^\beta=:T_\alpha.	
\end{equation}
We conclude that, from the point of view of Riemannian geometry, contravariant and covariant vectors (and, in general, tensors) are essentially the same. 

This interpretation of $g_{\alpha\beta}(x)$ comes as an anticlimactic answer to the challenging question we posed earlier, for at the back of our minds we already knew that $g_{\alpha\beta}(x)$ was the metric of spacetime. This mathematical, geometric meaning, then, is not really what we were pursuing when we asked for the meaning of $g_{\alpha\beta}(x)$; it is a more physical meaning, if any, that we are still after.

As we now compare Eq.~(\ref{line-element-eq}) with Eq.~(\ref{squared-line-element}), we must conclude that for any observer moving along his worldline and actually measuring the \emph{positive} real-number separation between two events with his clock, $\epsilon$ must be equal to $-1$. In other words, an observer with a clock can only move in directions determined by differential displacements $\D x^\alpha$ such that $g_{\alpha\beta}(x)\D x^\alpha \D x^\beta$ is negative; other directions are physically forbidden.

According to the direction of a differential displacement $\D x^\alpha$, and in keeping with the usual names, we can now say that\footnote{The same classification can be generalized to include any contravariant vector $T^\alpha$. An example of another timelike (i.e.\ allowed) contravariant vector is the tangent vector $\D x^\alpha/\D u$ to a worldline.}
\begin{itemize}
	\item[(i)] $\D x^\alpha$ is timelike (i.e.\ allowed) if $g_{\alpha\beta}(x)\D x^\alpha \D x^\beta < 0$ ($\epsilon = -1)$,
	\item[(ii)] $\D x^\alpha$ is spacelike (i.e.\ forbidden) if $g_{\alpha\beta}(x)\D x^\alpha \D x^\beta > 0$ ($\epsilon = +1)$, and 
	\item[(iii)]$\D x^\alpha$ is lightlike (i.e.\ allowed only for photons) if $g_{\alpha\beta}(x)\D x^\alpha \D x^\beta = 0$.
\end{itemize}
Regarding item (iii), in relativity theory the hypothesis is made that the differential displacements characterizing the history of a photon satisfy the relation $\D s=0$. Strictly speaking, this hypothesis is outside the boundaries of the operational meaning of $\D s$, since the clock to be carried along by a photon to make the necessary measurements is a material object which can never reach the speed of light. One may, however, understand this photon-history hypothesis as the limiting case of increasing velocities of material bodies (see Section \ref{absolute-derivative-sec}).

Because Riemann-Einstein spacetime is locally pseudo-Cartesian, coordinates can be found such that its indefinite quadratic form takes the form 
\begin{equation}
Q=\epsilon_1(\D x^1)^2+\epsilon_2(\D x^2)^2+\epsilon_3(\D x^3)^2+\epsilon_4(\D x^4)^2	
\end{equation}
at one point at a time. What are all the possible different combinations of $\epsilon_i$? Because no separation between three space coordinates and one time coordinate has as yet been made---and any such separation would at this point be artificial---all four coordinates stand on an equal footing. It is then immaterial the signs of which $\epsilon_i$ are being changed in any one combination. The five different possible combinations are listed in Table \ref{epsilon-table}.

\begin{table}
\centering
\begin{tabular}{|c|c|c|c|c|}
\hline
Signature & $\epsilon_1$ & $\epsilon_2$ & $\epsilon_3$ & $\epsilon_4$ \\
\hline \hline
$(+,+,+,+)$  & $+1$ & $+1$ & $+1$ & $+1$ \\ \hline
$(+,+,+,-)$  & $+1$ & $+1$ & $+1$ & $-1$ \\ \hline
$(+,+,-,-)$  & $+1$ & $+1$ & $-1$ & $-1$ \\ \hline
$(+,-,-,-)$  & $+1$ & $-1$ & $-1$ & $-1$ \\ \hline
$(-,-,-,-)$  & $-1$ & $-1$ & $-1$ & $-1$ \\ \hline
\end{tabular}
\caption[Choices of local spacetime geometry]{Five different possibilities for the signature of Riemann-Einstein spacetime. Each of these choices leads to a different local physical geometry for spacetime.}
\label{epsilon-table}
\end{table}

Which of the five possibilities should we choose? And on what grounds? No answer makes itself naturally available to us---we have reached a roadblock. \emph{This is a critical stage of this analysis}. It is time to recapitulate.

\subsection{Local physical geometry from psychology}
The present analysis of time has so far been based, first and foremost, on a \emph{physical component} consisting of clock measurements $\D s$; and secondly, on the other side of the equation, on a \emph{mathematical component} consisting of the geometric description of spacetime via its characteristic invariant quadratic form $g_{\alpha\beta}(x)\D x^\alpha \D x^\beta$. Is there anything missing from this analysis of time? Should we consider something else besides physics and mathematics? We must; if we are to move forwards in a natural, non-axiomatic way, we are forced to also take into account a \emph{psychological component} regarding the conception of time: the human notions of past, present (or being), and future.

The past and the future include all the allowed directions that an observer can follow, and so, recalling that we can only measure positive values of $\D s$ and that Eq.~(\ref{line-element-eq}) carries a minus sign, we have that
\begin{itemize}
	\item[(i)] $g_{\alpha\beta}(x)\D x^\alpha \D x^\beta < 0$ ($\epsilon=-1$) for the past and the future, 
	\item[(ii)] $g_{\alpha\beta}(x)\D x^\alpha \D x^\beta > 0$ ($\epsilon=+1$) for the present, and 
	\item[(iii)] $g_{\alpha\beta}(x)\D x^\alpha \D x^\beta = 0$ for the boundary surface between the present and the past and future, i.e.\ the lightcone (see below).
\end{itemize}

What choice of indicators $\epsilon_i$ do the conceptions of past, present, and future lead to? Choices $(+,+,+,+)$ and $(-,-,-,-)$ can be discarded as soon as we notice that they lead to quadratic forms of the type
\begin{equation}
Q=\pm[(\D x^1)^2+(\D x^2)^2+(\D x^3)^2+(\D x^4)^2],	
\end{equation}
which are always positive and do not allow the conceptions of the past and the future, or always negative and do not allow the conception of the present. Of the remaining options, which should we choose? And again, on what grounds?

While the existence of the present on the one hand, and the past and future on the other, acted as a first guideline, it is now the distinction between the past and the future that acts as the next one. We perceive the past becoming the future as a serial, unidirectional, or one-dimensional affair through the present, and so it appears that this process of becoming should be pictured locally by means of the progression of a single coordinate parameter and not of several of them. Given one coordinate parameter $x$ that we may locally associate with the psychological conception of the unidirectional flow of time, we may then connect the past with decreasing values of $x$ and the future with increasing values of $x$. 

Because we associated the past and the future with negative values of the invariant quadratic form, we may now discard options $(+,+,-,-)$ and $(+,-,-,-)$, which introduce, respectively, two and three (coordinate) parameters locally associated with the past and the future (prefixed with negative signs)---a so-called embarrassment of riches. Why? Because there do not exist two or three different ways in which we feel it is possible to go from the past to the future---e.g.\ is there a way through the present but perhaps also another that shortcuts it, as if past and future shared one or more direct connections? This idea does not conform with the seriality of the psychological conception of time, and we therefore discard these two options. We are left with option $(+,+,+,-)$, which we adopt.\footnote{If we had not prefixed Eq.~(\ref{line-element-eq}) with a minus sign, we would have been forced to choose option $(+,-,-,-)$, with the positive coordinate now being associated with the past and the future. This would have been a workable alternative too, although we would now have to carry along three minus signs instead of one.}

The choice of the invariant quadratic form with signature $(+,+,+,-)$ leads to the following classification:
\begin{itemize}
	\item[(i)] $(\D x^1)^2+(\D x^2)^2+(\D x^3)^2-(\D x^4)^2 < 0$ for the past and future,
	\item[(ii)] $(\D x^1)^2+(\D x^2)^2+(\D x^3)^2-(\D x^4)^2 > 0$ for the present, and 
	\item[(iii)] $(\D x^1)^2+(\D x^2)^2+(\D x^3)^2-(\D x^4)^2 = 0$ for the lightcone.
\end{itemize}
Moreover, through the increase and decrease of coordinate $x^4$, we can locally differentiate the past from the future, thus: 
\begin{itemize}
	\item[(i-a)] $(\D x^1)^2+(\D x^2)^2+(\D x^3)^2-(\D x^4)^2 < 0 \quad \& \quad \D x^4<0$ for the past, and
	\item[(i-b)] $(\D x^1)^2+(\D x^2)^2+(\D x^3)^2-(\D x^4)^2 < 0 \quad \& \quad \D x^4>0$ for the future.
\end{itemize}
In addition, we can also identify 
\begin{itemize}
	\item[(iii-a)] $(\D x^1)^2+(\D x^2)^2+(\D x^3)^2-(\D x^4)^2 = 0 \quad \&\quad \D x^4 < 0$ with the branch of the lightcone that differentiates the past from the present, and 
      \item[(iii-b)] $(\D x^1)^2+(\D x^2)^2+(\D x^3)^2-(\D x^4)^2 = 0 \quad \& \quad \D x^4 > 0$ with the branch of the lightcone that differentiates the future from the present.
\end{itemize}

Now, why should physics take into account these aspects of the psychological notion of time? Why should physics be partly based on them? Should physics also take into account the properties of other psychological conceptions, such as hunger, happiness, pain, and disdain? Not really, not in any specific way. What is then special about time? It is this: the psychological notion of time as a stream that drags us from past to future through the ubiquitous present is the deepest rooted feature of human consciousness and, as a result, physics has since its beginnings described the changing world guided by this form of human experience, yet without succeeding (or even trying) to explain it. From Galilean dynamics to quantum gravity, a picture of time as ``$t$''is all physics has so far managed: an abstract mathematical parameter that by itself cannot tell us why the past is gone, the present all-pervading, and the future unknown and out of reach.  

In classical physics, the mathematical depiction of time as ``$t$'' is simple enough for the psychological property of things-in-flow to be superimposed on it in a straightforward way: past, present, and future are taken as universal notions, each associated with different positions along the absolute parametric timeline. This is not to say that the human experience of time was \emph{explained} by $t$, but that it was easy enough to project ourselves onto that timeline and imagine ourselves to flow along it. Classically, then, as well as in conventional expositions of relativity theory with its block-universe picture, the human categories of past, present, and future remain physically ignored.

In this presentation of relativity, however, we notice that in this theory an allowance must be made for human temporal notions in a somewhat more explicit way. On the one hand, from the subjective point of view of the ordering of events along an observer's worldline, we again find that past, present, and future are ignored by physics, if we now replace the absolute parametric timeline given by $t$ by the relative parametric timeline given by clock readings $s$. On the other hand, however, relativistic physics is not just about the history of \emph{one} observer's worldline; it is a description of any and every worldline. In other words, relativistic physics is about \emph{absolute spacetime properties}, and we find that in the construction of the absolute property of spacetime signature $(+,+,+,-)$, it needs to explicitly acknowledge, though minimally, the psychological conceptions of past, present, and future for the problem to be unlocked in a meaningful way. 

The sceptical reader, having heard that physics ought to be independent of humans and external to them, may ask: but if the conceptions of past, present, and future are psychological, why make them a part of physics? Or then, if they are part of physics, why stress they are of psychological origin? Because we want psychology, \emph{where relevant}, to inform physics, so that physics, \emph{where possible}, will be reconciled with psychology. The pursuit of this twofold connection, which one may call \emph{psychophysics}, requires that we seriously acknowledge the factualness of conscious experience and that we identify those features of it that are relevant to physics. In order to do this, we must start by taking ourselves seriously and not dismiss, in the pursuit of the ideal of ``objectivity,'' human experience as an irrelevant delusion.

When it comes to time, that we should take ourselves seriously means that we should not consider ourselves the victims of an illusion in which \emph{we feel} that time flows irreversibly, with the past known and irretrievable, the present a permanent and all-pervading ``now,'' and the future unknown and out of reach, while \emph{in fact} time is reversible because the deterministic equations of physics make no dynamical distinction between $t$ and $-t$, and past, present, and future are all on an equal footing because spacetime theories lay them out as a homogeneous block. If physics says the earthly carousel and the heavenly bodies in their orbits can equally well run backwards in time, but our raw experience of time says that this is impossible, which should we trust? Is the physics of time correct as it is and we delusional, or could we learn a better physics of time by taking ourselves more seriously? 

That we should do so is not a forgone conclusion but an attitude we propose. Against such an attitude, one may observe with Dennett \citeyear{Dennett:1991} that ``nature doesn't build epistemic engines'' (p.~382). That is, why should we trust our minds to capture something essential about the nature of time that physical theories have not yet grasped, if our minds did not arise to know the world but only to improve our chances of staying alive? Regarding colour perception, for example, the context in which Dennett makes his observation, this is quite true; the perception of red does not represent any basic property of the world. But when it comes to time, we are inclined to give the mind the benefit of the doubt. Unlike any other feeling, time is an inalienable experiential product of human consciousness and, while physics desperately needs \emph{some} concept of time, so far it has only managed to offer a caricature of it, grasping barely its serial aspect in a mathematical parameter and no more. Because the conception of time is a consciousness-related idea, to understand it, we need to pay heed to ourselves.

This presentation of relativity takes a small step in a positive direction. Through it we learn that, despite appearances, relativity does not straightout hold the spatiotemporal universe to be a completely undifferentiated block, because in its choice of local physical geometry (or global signature) there is a mild recognition that (i) past and future are different from the present (first discard of signatures), and (ii) echoing ``$t$'' in classical physics, that past and future are serially connected (second discard of signatures).

However, this understanding of relativity does not achieve nearly enough. How are past, present, and future qualitatively different from each other? Why is past-present-future a serial chain? And, more significantly, what about the present, that permanent now in which we constantly find ourselves? How does physics explain the \emph{present moment}? Can it single it as special in any way? Our consciousness does, and so distinctly so that we feel we are nothing but the present. If we decide to heed ourselves when it comes to time, we should consider this an enigma. An enigma neither classical nor relativistic physics can solve. But what about quantum physics? Can it succeed where its predecessor theories fail? Can quantum physics do better in accounting for what is otherwise the human delusion that the present is paramount? And what is the role of geometric tools of thought in this search? We return to this issue in Chapter~\ref{ch:MQM}.

\subsection{Spacelike separation and orthogonality from clocks}
Up to now, we have used an ideal clock to measure separations between events that can be reached by an observer travelling along his worldline, i.e.\ the separations measured have always been in directions given by timelike displacement vectors $\D x^\alpha$. Can we also measure separations between events along directions given by spacelike displacement vectors? And if so, what kind of experimental setup do we need to do this?

\begin{figure}
\centering
\includegraphics[width=70mm]{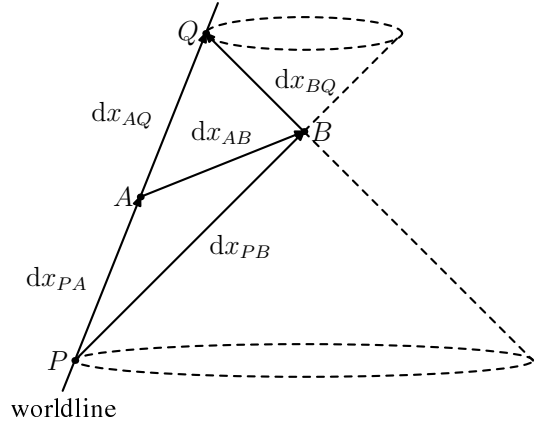}
\caption{Measurement of a spacelike separation by means of a clock
and a reflected photon. Displacement vectors can be detached from
their original positions due to the fact that they are small.}
\label{spacelike-separation}
\end{figure}

We examine two neighbouring events $A$ and $B$ joined by a spacelike displacement vector $\D x_{AB}$, with $A$ situated between events $P$ and $Q$ and all three events on the observer's timelike worldline. Events $P$ and $Q$ are determined experimentally by the application of the radar method: $P$ is an event such that a light signal (a photon) emitted from it is reflected at $B$ and received by the observer at $Q$; both timelike separations $\D s_{PA}$ and $\D s_{AQ}$ are experimentally determined with the observer's clock. Looking at the diagram of Figure \ref{spacelike-separation}, we can write down the system of equations
\begin{equation}\left\{ \begin{array}{rclc}
	   \D x_{AQ}&=&k \D x_{PA}& \qquad (k>0) \\
   \D x_{PB}-\D x_{PA}&=&\D x_{AB}&\\
   \D x_{AQ}-\D x_{AB}&=&\D x_{BQ}& 
   \end{array}\right..
\end{equation}
Using the last two equations to express the lightlike
separations $\D s^2_{PB}=0$ and $\D s^2_{BQ}=0$ in terms of the two
timelike and one spacelike displacement vectors, and subsequently
using the first equation, we find 
\begin{equation}
\D s^2_{AB}=\D s_{PA}\D s_{AQ}.\footnote{In more detail, we have (1) $\D s^2_{PB}=-\D s^2_{PA}+\D s^2_{AB}+2g_{\alpha\beta}\D x^\alpha_{PA} d x^\beta_{AB}=0$ and (2) $\D s^2_{BQ}=-k^2\D s^2_{PA}+\D s^2_{AB}-2g_{\alpha\beta}(k\D x^\alpha_{PA}) d x^\beta_{AB}=0$. Now $k$(1)+(2) gives $\D s^2_{AB}=k\D s^2_{PA}$, from which follows
the desired result.}	
\end{equation}
Spacelike separations can then be measured with
the simple addition of emission, reflection, and reception of
photons to the original clock-on-worldline type of measurements, i.e.\
spacelike separations can be expressed in terms of timelike ones.

We find, moreover, that if events $A$, $B$, $P$, and $Q$ can be experimentally so arranged as to have $\D s_{PA}=\D s_{AQ}$, then
the $k$-factor is unity and $\D x_{AQ}=\D x_{PA}$. In that case, $\D s^2_{BQ}=0$ becomes the bilinear form 
\begin{equation}\label{orthogonality-eq}
g_{\alpha\beta}\D x^\alpha_{AB}\D x^\beta_{AQ}=0,	
\end{equation}
where $\D x^\alpha_{AB}$ is spacelike and $\D x^\alpha_{AQ}$ is timelike. What is the geometric meaning of this result? 

\begin{figure} 
\centering
\includegraphics[height=30mm]{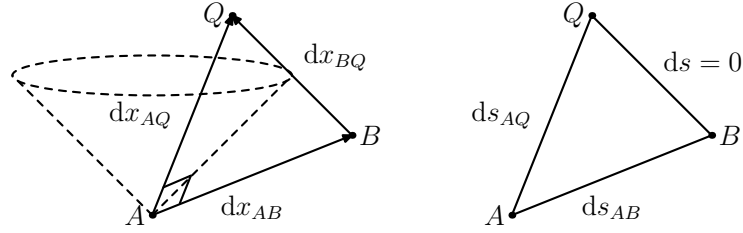}
\caption{Experimental test to determine the orthogonality of a
spacelike vector $\D x_{AB}$ and a timelike vector $\D
x_{AQ}$. The two vectors are orthogonal if $\D
s_{AB}=\D s_{AQ}$, where events $B$ and $Q$ are joined
by the worldline of a photon.} 
\label{orthogonality-time-space}
\end{figure}

We know that, in Euclidean space, whose quadratic form is positive definite, the cosine of the angle between vectors $\overrightarrow{AB}$ and $\overrightarrow{AQ}$ is given by the dot product of the corresponding unit vectors:
\begin{equation}
\cos(\theta)=\hat u_{AB}\cdot \hat u_{AQ}=u_{AB}^1u_{AQ}^1+u_{AB}^2
u_{AQ}^2+u_{AB}^3 u_{AQ}^3.	
\end{equation}
Generalizing for a Riemannian space with a \emph{positive-definite} quadratic form, we have 
\begin{equation}
\cos(\theta)=h_{ab}\left(\frac{\D x^a_{AB}}{\D l_{AB}}\right)\left(\frac{\D x^b_{AQ}}{\D l_{AQ}}\right).	
\end{equation}
In consequence, also in Riemannian space we say that two vectors are orthogonal when their inner product is null. Although in the case of an indefinite quadratic form it is not sensible to define an acute or obtuse angle with a cosine-type relation, we do nevertheless accept the concept of a straight angle, i.e.\ of orthogonality, from the positive definite case, and characterize two orthogonal vectors in Riemann-Einstein spacetime by the relation of Eq.~(\ref{orthogonality-eq}). As a result, the experimental determination of spacelike separations supplies a method for the experimental determination of orthogonality between displacement vectors: Eq.~(\ref{orthogonality-eq}) means that spacelike displacement vector $\D x^\alpha_{AB}$ and timelike displacement vector $\D x^\beta_{AQ}$ are \emph{orthogonal} in an \emph{experimental test} where $\D s_{AB}=\D s_{AQ}$ and where events $B$ and $Q$ are joined by the worldline of a photon (Figure \ref{orthogonality-time-space}). This concept of orthogonality can be generalized for any pair of contravariant vectors $T^\alpha$ \mbox{and $U^\alpha$.}

\begin{figure} 
\centering
\includegraphics[height=27mm]{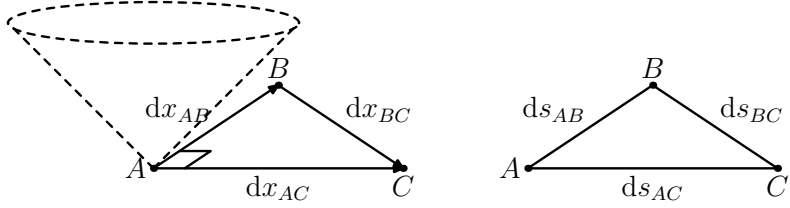}
\caption{Experimental test to determine the orthogonality of two
spacelike vectors $\D x_{AB}$ and $\D x_{AC}$. The two
vectors are orthogonal if they are joined by a spacelike vector $\D
x_{BC}$ and $\D s^2_{BC}=\D s^2_{AC}+\D s^2_{AB}$, that is,
when $\D x_{BC}$, $\D x_{AC}$, and $\D x_{AB}$
satisfy Pythagoras' theorem in spacetime.} 
\label{orthogonality-space-space}
\end{figure}

Can now two spacelike displacement vectors $\D x_{AB}$ and $\D
x_{AC}$ be orthogonal? And if so, under what conditions? By means of
what experimental test can we verify their orthogonality? Because now we have two spacelike displacement vectors, we use the radar method twice to measure $\D s^2_{AB}$ and $\D s^2_{AC}$ in terms of timelike separations. Noting that $\D x_{BC}=\D x_{AC}-\D x_{AB}$ (Figure \ref{orthogonality-space-space}), we find 
\begin{equation}
      g_{\alpha\beta}\D x^\alpha_{BC}\D x^\beta_{BC}-\D s^2_{AC}-\D s^2_{AB} =
      -2 g_{\alpha\beta}\D x^\alpha_{AC}\D x^\beta_{AB}.	
\end{equation}
The left-hand side can only be null and orthogonality possible when $\D x_{BC}$ is spacelike and thus $g_{\alpha\beta}\D x^\alpha_{BC}\D x^\beta_{BC}=\D s^2_{BC}>0$. In
that case, the condition for the orthogonality of $\D x_{AB}$ and $\D x_{AC}$ is 
\begin{equation}
\D s^2_{BC}=\D s^2_{AC}+\D s^2_{AB}.	
\end{equation}
This is Pythagoras' theorem in spacetime!

Two timelike vectors, however, can never be orthogonal. This fact leads to a counterintuitive (from a Euclidean point of view) triangle inequality in spacetime, with consequences reaching farther than one would expect. To show first that two timelike vectors $T^\alpha$ and $U^\alpha$ cannot be orthogonal, we investigate their behaviour at a point (chosen to be the origin) after recasting them in pseudospherical coordinates $r,\theta,\phi,\chi$ ($r>0$):
\begin{equation} \left\{
\begin{array}{ccl}
    T^1&=&r\sin(\theta)\cos(\phi)\sinh(\chi)\\
    T^2&=&r\sin(\theta)\sin(\phi)\sinh(\chi)\\
    T^3&=&r\cos(\theta)\sinh(\chi)\\
    T^4&=&r\cosh(\chi)
\end{array} \right.
\end{equation}
and
\begin{equation}\left\{
\begin{array}{ccl}
    U^1&=&\tilde r\sin(\tilde\theta)\cos(\tilde\phi)\sinh(\tilde\chi)\\
    U^2&=&\tilde r\sin(\tilde\theta)\sin(\tilde\phi)\sinh(\tilde\chi)\\
    U^3&=&\tilde r\cos(\tilde\theta)\sinh(\tilde\chi)\\
    U^4&=&\tilde r\cosh(\tilde\chi)
\end{array}\right.
\end{equation}
The quadratic form of $T^\alpha$ at the origin gives 
\begin{equation}
(T^1)^2+(T^2)^2+(T^3)^2-(T^4)^2=-r^2<0,	
\end{equation}
which assures us that $T^\alpha$ is timelike at that event; and similarly for $U^\alpha$. After using several trigonometric- and hyperbolic-function identities,\footnote{The trigonometric identities needed are $2\sin(x)\sin(y)=\cos(x-y)-\cos(x+y)$, $2\cos(x)\cos(y)=\cos(x-y)+\cos(x+y)$, and $\cos(2x)=\cos^2(x)-\sin^2(x)$. The hyperbolic identities needed are $2\sinh(x)\sinh(y)=\cosh(x+y)-\cosh(x-y)$ and $2\cosh(x)\cosh(y)=\cosh(x+y)+\cosh(x-y)$.} we find that the inner product of these two vectors at the origin is
\begin{multline}
T^1U^1+T^2U^2+T^3U^3-T^4U^4=-r\tilde r[\cosh(\chi-\tilde\chi)\cos^2(\omega/2)-\\ \cosh(\chi+\tilde\chi)\sin^2(\omega/2)]\leq -r\tilde r <0,	
\end{multline}
where 
\begin{equation}
\cos(\omega)=\cos(\theta)\cos(\tilde\theta)+\sin(\theta)\sin(\tilde\theta)\cos(\phi-\tilde\phi).
\end{equation}
This means that 
\begin{equation}
g_{\alpha\beta}T^\alpha U^\beta \leq -r\tilde r \neq 0,	
\end{equation}
i.e.\ timelike vectors $T^\alpha$ and $U^\alpha$ cannot be orthogonal.

The triangle inequality in spacetime results now as follows. In Euclidean space, the triangle inequality for the sides of a triangle with vertices $A$, $B$, and $C$ reads $|\overrightarrow{AC}|\leq |\overrightarrow{AB}|+|\overrightarrow{BC}|$, which says that the sum of the lengths of any two sides of a triangle is greater than the length of the remaining side. In spacetime, things are otherwise.

\begin{figure}
\centering
\includegraphics[width=42mm]{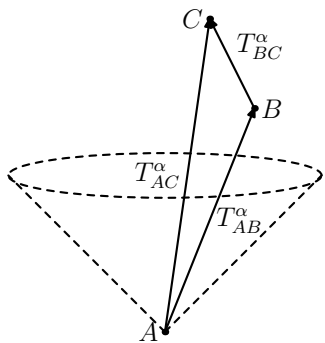} \caption{The
counterintuitive triangle inequality $T_{AC}\geq T_{AB}+T_{BC}$ for
three timelike vectors in spacetime.} 
\label{timelike-triangle}
\end{figure}

Let $T^\alpha_{AB}$, $T^\alpha_{BC}$, and $T^\alpha_{AC}$ be three timelike vectors forming a triangle with vertices at events $A$, $B$, and $C$. We assume that the three vectors point towards the future and that the vectors characterizing the displacements are small. On the basis of Figure \ref{timelike-triangle}, we have
\begin{equation}
T^\alpha_{AC}=T^\alpha_{AB}+T^\alpha_{BC}.	
\end{equation}
We express these vectors using their norms
\begin{equation}
T=\sqrt{\epsilon g_{\alpha\beta}T^\alpha T^\beta} \qquad (\epsilon=-1)	
\end{equation}
and unit vectors $\hat u$ to get 
\begin{equation}
T_{AC}u^\alpha_{AC}=T_{AB}u^\alpha_{AB}+T_{BC}u^\alpha_{BC}.	
\end{equation}
Because the unit vectors are timelike, we have
\begin{equation}
      g_{\alpha\beta}u^\alpha_{AB} u^\beta_{AB}=
      g_{\alpha\beta}u^\alpha_{BC} u^\beta_{BC}=-1	
      \quad \mathrm{and} \quad
      g_{\alpha\beta}u^\alpha_{AB} u^\beta_{BC}\leq -1. 
\end{equation}
As a result, we find
\begin{equation}
T_{AC}^2=T_{AB}^2+T_{BC}^2-2T_{AB}T_{BC}g_{\alpha\beta}u^\alpha_{AB}u^\beta_{BC},	
\end{equation}
from where 
\begin{equation}
T_{AC}^2\geq (T_{AB}+T_{BC})^2. 	
\end{equation}
It follows that
\begin{equation}
T_{AC}\geq T_{AB}+T_{BC}	
\end{equation}
is the \emph{triangle inequality} in spacetime, with equality holding when
\mbox{$T^\alpha_{AC}=T^\alpha_{AB}$} or \mbox{$T^\alpha_{AC}=T^\alpha_{BC}$} (no actual triangle).

The triangle inequality for timelike vectors has consequences that extend beyond the character of the geometric structure of spacetime. Notably, it has a bearing on the features of material processes that take place in spacetime, such as the spontaneous decay or induced fission of a massive particle into decay or fission products with an ensuing release of energy.

In classical physics, the collision of material particles is understood in terms of the conservation of two quantities, the total quantity of motion $\sum_i m_iv_i$ and the total kinetic energy $\sum_i m_iv_i^2/2$ of the system of particles. From a strictly mechanical point of view, this is only true of elastic collisions, in which no energy is lost by the colliding bodies in the form of heat. Classical mechanics needs to be complemented by thermodynamics in order to make full sense of energy-dissipating (i.e.\ inelastic) collisions. On the other hand, relativity theory is in this respect self-contained; every aspect of the interaction of massive particles can be explained based on the geometric structure of spacetime where the processes occur.

Taking now the natural step of using clock readings $s$ as a worldline parameter, we recast the worldline equation as $x^\alpha=x^\alpha(s)$. Because $\D s$ is always positive, the covariant vector $\D x^\alpha/\D s$ points in the same direction as $\D x^\alpha$, and so
\begin{equation}
\frac{\D x^\alpha}{\D s} =: V^\alpha 	
\end{equation}
is a good representation of the worldline tangent vector, which is called the \emph{four-velocity}. The four-velocity $V^\alpha$ is, in fact, a unit vector. Writing $\D x^\alpha=V^\alpha \D s$ in the quadratic form for $\D s^2$, we find 
\begin{equation}
V^2=\epsilon g_{\alpha\beta} V^\alpha V^\beta=1.      
\end{equation}
Because the worldline is that of a material particle, then $\epsilon=-1$, and so
\begin{equation}
g_{\alpha\beta} V^\alpha V^\beta=-1.	
\end{equation}
Finally, we attach the mass of the particle $m$ (an invariant scalar) to the four-velocity $V^\alpha$ to get the \emph{four-momentum}
\begin{equation}
P^\alpha=m V^\alpha.	
\end{equation}

\begin{figure}
\centering
\includegraphics[width=42mm]{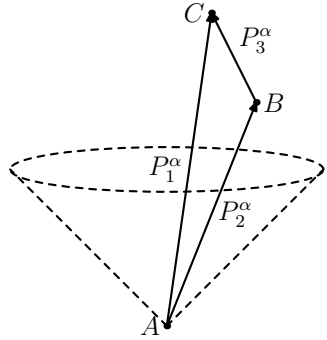} \caption{Particle decay
($A$) into two subproducts ($B$ and $C$). Energy is released in these processes as a
result of the spacetime triangle inequality.} 
\label{particle-decay}
\end{figure}

In the event of a particle decaying into two subproducts, in relativity theory we only require the conservation of the four-momentum of the total system, namely, 
\begin{equation}
P^\alpha_1=P^\alpha_2+P^\alpha_3,	
\end{equation}
which we can graphically represent with three timelike vectors in the form 
\begin{equation}
P^\alpha_{AC}=P^\alpha_{AB}+P^\alpha_{BC},	
\end{equation}
as shown in Figure \ref{particle-decay}. Because $V^\alpha$ is a unit vector, $m$ acts as the norm of $P^\alpha$, and we can apply the results of the triangle inequality we found for $T^\alpha$, to get 
\begin{equation}
m_1>m_2+m_3.	
\end{equation}
We find that the conservation of the four-momentum in the decay of a particle involves the release of energy, $m_1c^2>m_2c^2+m_3c^2$, as an \emph{inescapable consequence} of the non-Euclidean geometric structure of Riemann-Einstein spacetime, namely, the non-orthogonality of timelike vectors. This positive energy deficit $m_1c^2-(m_2c^2+m_3c^2)$ is normally denoted as $Q$ in textbooks on the topic, but its origin is left unexplained.

\section{Gravitation as spacetime curvature}
In his Kyoto lecture, Einstein describes the train of thought leading him to the general theory of relativity: 
\begin{quote}
If a man falls freely, he would not feel his weight\ldots A falling man is accelerated\ldots In the accelerated frame of reference, we need a new gravitational field\ldots [I]n the frame of reference with acceleration Euclidean geometry cannot be applied. \cite[p.~47]{Einstein:1982}
\end{quote}

If a falling man does not feel his weight yet is accelerated, it seems he must not be subjected to any gravitational \emph{force} or \emph{field} but to some form of interaction (whose effects we see in his acceleration) nonetheless---he must be subjected to a ``new gravitational field.'' What is this new interaction that induces gravitational effects yet is not a force? An answer can be sketched as follows. It has been long since realized that the gravitational force is a special interaction. Galilei and Newton observed that inertial masses $m_i$ and gravitational masses $m_g$ were the same (Newton's principle) and that, therefore, all massive bodies in free fall in a vacuum suffer the same acceleration (Galilei's principle) regardless of their constitution. This pulling on all bodies and light in like fashion insinuates that the ``new gravitational field'' we are after could be some property of space or spacetime in which massive bodies and light move.

Could gravitation, then, be conceived as a property of \emph{space}? Intuitively, as Rindler \citeyear[p.~115]{Rindler:1977} explains, this cannot be just so because the future motion of a body under gravitational interaction depends on its initial velocity, but this information is not included in the description of a \emph{space} geodesic, which is determined only by initial position and direction.\footnote{For example, a free particle confined to a surface follows a geodesic that depends only on its initial position and direction, but is independent of its initial velocity.} However, an account of all necessary initial conditions is included in the specification of a \emph{spacetime} geodesic. The validity of this heuristic reasoning is strengthened when we observe with Einstein that, in the man's accelerated frame of reference, Euclidean geometry cannot be applied beyond a point-event,\footnote{This means that, because the velocity of the (rigid) frame changes constantly, we must attach a different inertial system to each \emph{spacetime} point or event. However, Einstein \citeyear{Einstein:1983b} sometimes reached the same conclusion in a more intuitive way by looking at a rotating frame; here an inhomogeneous distortion of ``rigid bodies'' (measuring rods) ensues in a more easily visualizable way, as even each \emph{space} point becomes an own, separate inertial frame ``on account of the Lorentz contraction'' (p.~33).} and note at the same time the essential role that the geometry of the \emph{four-dimensional}, \emph{non-Euclidean} Riemann-Einstein spacetime has had so far in the development of relativity theory.

Given the geometric setting of relativity theory, we suspect then that the ``new gravitational field'' must be somehow included in the geometry of Riemann-Einstein spacetime. Gravitation, moreover, must be an \emph{absolute property of spacetime}, in the sense that either gravitational \emph{effects} are present or they are not. As has been often remarked, when it comes to \emph{spacetime} properties, there is nothing relative in relativity theory. The prime example of this truth is the spacetime interval $\D s$. In absolute classical physics, on the other hand, gravitation is, in fact, a relative concept: an observer on earth deems it to be a force acting on the falling man, while the falling man deems it as no force at all. Einstein's classical thought experiment would teach us nothing were we not willing to also take into account the relative point of view of the man on earth.

What spacetime property shall we now consider to represent gravitation as an absolute interaction? And what is the role of measurements in this search? On account of a long acquaintance with it (and perhaps also because of the suggestive notation), one's first intention is to take the metric tensor $g_{\alpha\beta}(x)$ to play the vacant role of ``new gravitational field,'' while at the same time finally giving an answer to the still-standing question of its physical meaning. However, the metric tensor is a \emph{rather bad candidate} to act as ``gravitation,'' the absolute spacetime property we are seeking. For, if truth be told, there is nothing absolute in $g_{\alpha\beta}(x)$: it is a tensor field, and that is all very well, but what absolute spacetime property does it represent? In one coordinate system, $g_{\alpha\beta}(x)$ takes certain values; in another, it takes different ones. 

When we introduced tensors at the beginning of this chapter, we observed that their reason of being was to provide mathematical objects meaningful in all coordinate systems. We then identified such mathematical objects, not with the tensors themselves, but with \emph{tensor equations}. Here is then the key to this search. Is there any tensor equation that $g_{\alpha\beta}(x)$ satisfies as such? So far the only absolute equation connected with the metric field has been $\D s^2-\epsilon g_{\alpha\beta}(x)\D x^\alpha \D x^\beta=0$, but this is an equation for the whole quadratic form and for the clock separation $\D s$. Evidently, this elemental relation is tightly connected with clock measurements but has no straightforward connection with gravitation. It is time to give up the illusion that, in relativity theory, it is simply the metric field that has taken up the role of the ``new gravitational field.'' If a name with a strong physical connotation must still be given to the metric field, one should then perhaps call $g_{\alpha\beta}(x)$ the \emph{chronometric field}---and stop at that.

\subsection{Einstein's geodesic hypothesis}
Newton's first law of motion, or the law of inertia, states that a free massive particle follows a uniform and rectilinear trajectory. Here ``free'' means free of all external interactions, including the gravitational force.

In relativity, we proceed to include a description of gravitation by means of an extension of Newton's first law of motion called Einstein's \emph{geodesic hypothesis}. According to it, a free massive particle has a timelike worldline that is a geodetic curve, or geodesic, in Riemann-Einstein spacetime, where by ``free'' we now mean free of all external \emph{forces} but not of gravitation, which we want to describe not as an external force but as an absolute geometric property of spacetime. As such, it is impossible to be free of its effects. Einstein's geodesic hypothesis also applies to a free photon, in which case it says that a free photon follows a null geodesic in Riemann-Einstein spacetime.

A geodesic is a line whose extension (length) between two given points is stationary, i.e.\ has an extremal value. This can be a maximum or a minimum. When we consider two points in three-dimensional Euclidean space, a geodesic refers to a spatial path of minimal length, i.e.\ the shortest distance between the two points. In Riemann-Einstein spacetime, on the other hand, our Euclidean intuition fails again: where we expect less we get more (cf.\ triangle inequality and energy surplus in particle decay). In spacetime, a geodesic represents a worldline of maximal separation as measured by clocks. 

Incidentally, this is not, strictly speaking, a maximal \emph{length} in any literal sense, although our geometric intuition leads us to think of it that way. Insightfully, in this regard, Synge writes:
\begin{quote}
It is indeed a Riemannian \emph{chronometry} rather than a \emph{geometry}, and the word \emph{geometry}, with its dangerous suggestion that we should go
about measuring \emph{lengths} with \emph{yardsticks}, might well be
abandoned altogether in the present connection were it not for the
fact that the crude literal meaning of the word geometry has been
transmuted into the abstract mathematical concept of a ``space'' with
a ``metric'' in it. \cite[p.~109]{Synge:1964}
\end{quote}
Here is yet another instance of a rule we have repeatedly observed: the mind turns everything it beholds into its golden standard---geometry.\footnote{What mythical character does this behaviour remind us of? (See Epilogue.)} 

In order to find the equation $x^\alpha=x^\alpha(u)$ satisfied by the geodetic worldline with extremes at events $A$ and $B$, we calculate the variation of the event separation $\Delta s$ and equate it to zero, 
\begin{equation}
\delta \int_A^B \D s=\delta \int_{u_A}^{u_B} f\left(x,\frac{\D x}{\D u}\right)\D u=0.	
\end{equation}
Taking the variations of $x^\alpha$ and $x'^\alpha=:\D x^\alpha/\D u$, we find
\begin{equation}
    \int_{u_A}^{u_B} \left(\frac{\partial f}{\partial x^\alpha}\delta x^\alpha+
    \frac{\partial f}{\partial x'^\alpha}\delta x'^\alpha \right) \D u=0.
\end{equation}
Rewriting $\delta x'^\alpha=\D (\delta x^\alpha)/\D u$ and integrating the second term by parts (\mbox{$\tilde u_\alpha=\partial f/\partial x'^\alpha$,} $\D \tilde v=\D(\delta x^\alpha)$), we get
\begin{equation}
    \Big\vert_{u_A}^{u_B}\frac{\partial f}{\partial x'^\alpha}\delta x^\alpha(u)+
    \int_{u_A}^{u_B} \left[\frac{\partial f}{\partial x^\alpha}-
    \frac{\D}{\D u}\left(\frac{\partial f}{\partial x'^\alpha}\right)\right] \delta x^\alpha \D u =0.
\end{equation}
Because the first term is null on account of $\delta x(u_B)=\delta x(u_A)=0$ (the extremes are fixed), and because $\delta x^\alpha$ is an arbitrary small variation, we find that, at every point of the geodesic, $f(x,x')$ must satisfy the Euler-Lagrange equations
\begin{equation}\label{Euler-Lagrange-I}
\frac{\D}{\D u}\left(\frac{\partial f}{\partial x'^\alpha} \right)-
\frac{\partial f}{\partial x^\alpha}=0,
\end{equation}
where $f(x,x')=\sqrt{-g_{\alpha\beta} x'^\alpha x'^\beta}$. To avoid having to differentiate a square-root expression, we rewrite Eq.~(\ref{Euler-Lagrange-I}) in the equivalent form
\begin{equation}\label{Euler-Lagrange-II}
\frac{\D}{\D u}\left(\frac{\partial f^2}{\partial x'^\alpha} \right)-
\frac{\partial f^2}{\partial x^\alpha}-
\frac{1}{2f^2}\frac{\D f^2}{\D u}
\frac{\partial f^2}{\partial x'^\alpha}=0.
\end{equation}
Replacing $u$ by the natural worldline parameter $s$, we find 
\begin{equation}
f^2(x,x')=-g_{\alpha\beta} V^\alpha V^\beta=+1 \qquad \mathrm{and} \qquad \frac{\D f^2}{\D s}=0, 
\end{equation}
to get
\begin{equation}\label{Euler-Lagrange-III}
\frac{\D}{\D s}\left(\frac{\partial f^2}{\partial x'^\alpha} \right)-
\frac{\partial f^2}{\partial x^\alpha}=0.
\end{equation}
Replacing $f^2(x,x')$ by $-g_{\alpha\beta} x'^\alpha x'^\beta$, we obtain the geodesic equation in the more explicit form
\begin{equation}\label{Euler-Lagrange-IV}
g_{\delta\beta}\frac{\D x'^\beta}{\D s}+[\alpha\beta,\delta] x'^\alpha x'^\beta=0,
\end{equation}
where
\begin{equation}\label{Christoffel-first-kind}
[\alpha\beta,\delta]=\frac{1}{2}\left(
\frac{\partial g_{\alpha\delta}}{\partial x^\beta} +
\frac{\partial g_{\beta\delta}}{\partial x^\alpha} -
\frac{\partial g_{\alpha\beta}}{\partial x^\delta} \right)
\end{equation}
is called \emph{Christoffel's symbol of the first kind}, which is symmetric with respect to its first two indices. Finally, multiplying Eq.~(\ref{Euler-Lagrange-IV}) by $g^{\gamma\delta}$, we obtain a yet more explicit form of the geodesic equation:
\begin{equation}\label{Euler-Lagrange-V}
\frac{\D^2 x^\gamma}{\D s^2}+\big\{\begin{smallmatrix}\gamma\\\alpha\beta
\end{smallmatrix}\big\} \frac{\D x^\alpha}{\D s} \frac{\D x^\beta}{\D s}=0,
\end{equation}
where
$\big\{\begin{smallmatrix}\gamma\\\alpha\beta\end{smallmatrix}\big\}=g^{\gamma\delta}[\alpha\beta,\delta]$ is called \emph{Christoffel's symbol of the second kind}. In addition, the timelike condition 
\begin{equation}
g_{\alpha\beta}\frac{\D x^\alpha}{\D s}\frac{\D x^\beta}{\D s}=-1	
\end{equation}
holds for all events on the geodesic. The solution $x^\alpha=x^\alpha(s)$ to this second-order differential equation can be determined unequivocally when we know $x^\alpha(s)$ and $\D x^\alpha$ at some event on the timelike geodesic, i.e.\ for some clock-reading value $s$.

The geodesic equation (\ref{Euler-Lagrange-V}) gives us now the answer to a question we asked earlier on. On page~\pageref{equation-of-motion}, we had expected that the equation of motion~(\ref{equation-of-motion}) of a particle would be of the form $\D^2 x^\alpha/\D u^2=X^\alpha(x,\D x/\D u)$. Now we have found  that this is indeed the case, with
\begin{equation}\label{form-of-X}
X^\alpha \left( x,\frac{\D x}{\D s}\right)=
-\big\{\begin{smallmatrix}\alpha\\ \beta\gamma\end{smallmatrix}\big\} \frac{\D x^\beta}{\D s} \frac{\D x^\gamma}{\D s},
\end{equation}
where the $x$-dependence is contained in Christoffel's symbol.

As an example, we can apply the geodesic equation to the case of a spacetime whose quadratic form is 
\begin{equation}
Q=(\D x^1)^2+(\D x^2)^2+(\D x^3)^2-(\D x^4)^2	
\end{equation}
everywhere and its metric, therefore, $g_{\alpha\beta}=\mathrm{diag}(1,1,1,-1)=\eta_{\alpha\beta}$. As a result, all Christoffel symbols are null and the geodesic equation reduces to
\begin{equation}
\frac{\D^2 x^\alpha}{\D s^2}=0,
\end{equation}
with timelike condition
\begin{equation}\label{timelike-condition}
\left(\frac{\D x^1}{\D s}\right)^2+\left(\frac{\D x^2}{\D s}\right)^2+
\left(\frac{\D x^3}{\D s}\right)^2-\left(\frac{\D x^4}{\D s}\right)^2=-1.
\end{equation}
The solutions are \emph{straight lines in spacetime} 
\begin{equation}
x^\alpha(s)=A^\alpha s+B^\alpha,	
\end{equation}
where $A^\alpha$ and $B^\alpha$ are constants, and $A^\alpha$ satisfies timelike condition (\ref{timelike-condition}). Excepting this last restriction, \label{flat-space} the equation of motion and its solution have the same form as those for a free particle in classical physics, namely, $\D^2 x^a/\D t^2=0$ with solution $x^a(t)=A^a t+B^a$. The correspondence suggests that a spacetime whose quadratic form is pseudo-Cartesian everywhere resembles flat Euclidean space.

So far we have dealt with geodesics of material particles, for which \mbox{$\D s>0$}. How shall we deal with geodesics of photons, for which $\D s=0$? In this case, the use of $s$ as a geodesic parameter is impossible, and we must use a different parameter $u$ together with its corresponding lightlike condition 
\begin{equation}
	f^2(x,x')=g_{\alpha\beta}U^\alpha U^\beta=0,
\end{equation} 
where $U^\alpha=\D x^\alpha/\D u$. On account of it, the simplification $\D f^2/\D s=0$ introduced earlier now becomes $\D f^2/\D u=0$. From here on, the procedure is analogous to the one for a timelike geodesic. The equation of a lightlike geodesic is then
\begin{equation}\label{null-geodesic-equation}
\frac{\D^2 x^\gamma}{\D u^2}+\big\{\begin{smallmatrix}\gamma\\\alpha\beta\end{smallmatrix}\big\} \frac{\D x^\alpha}{\D u} \frac{\D x^\beta}{\D u}=0,
\end{equation}
with lightlike condition
\begin{equation}
g_{\alpha\beta}\frac{\D x^\alpha}{\D u} \frac{\D x^\beta}{\D u}=0.
\end{equation}
This way of proceeding, however, begs the question as to what kind of parameter $u$ is. How general can it be and what is its relation to $s$? We defer the answer to this question until we have studied the absolute derivative in connection with the generalized law of inertia of relativity theory.

The geodesic equation for material particles and photons can also be derived by means of a different method. We return to the original Euler-Lagrange equation (\ref{Euler-Lagrange-I}) and proceed to find its first integral. In order to avoid confusion with the preceding analysis, we now rename $f$ as $\phi$ in this equation.

Multiplying Eq.~(\ref{Euler-Lagrange-I}) by $x'^\alpha=\D x^\alpha/\D u$ in the sense of an inner product, we find
\begin{equation}
    x'^\alpha \frac{\D}{\D u}\left(\frac{\partial \phi}{\partial x'^\alpha} \right)-
    x'^\alpha \frac{\partial \phi}{\partial x^\alpha}=0,
\end{equation}
which can be rewritten as
\begin{equation}
    \frac{\D}{\D u}\left( x'^\alpha \frac{\partial \phi}{\partial x'^\alpha} \right) -
    \left( \frac{\partial \phi}{\partial x^\alpha} \frac{\D x^\alpha}{\D u} +
           \frac{\partial \phi}{\partial x'^\alpha} \frac{\D x'^\alpha}{\D u}\right)=0.
\end{equation}
Because the second term is $\D \phi/\D u$, the whole expression can be recast as a total derivative,
\begin{equation}
    \frac{\D}{\D u}\left( x'^\alpha \frac{\partial \phi}{\partial x'^\alpha}-\phi \right)=0.
\end{equation}
Integrating once with respect to $u$, we get
\begin{equation}\label{first-integral}
    x'^\alpha \frac{\partial \phi}{\partial x'^\alpha}-\phi = C\ (\mathrm{constant}).
\end{equation}
Proposing the solution $\phi=g_{\alpha\beta}x'^\alpha x'^\beta$, we find $\phi=C$. Choosing $u=s$ leads to the earlier results for a \emph{timelike geodesic}; in that case, $C=-1$ (timelike condition). When $C=+1$, we find the spacelike condition $g_{\alpha\beta}x'^\alpha x'^\beta=+1$, and we are led to a ``spacelike geodesic,'' a mathematical concept that has no correspondent in the physical world and we can forthwith forget. When $C=0$, we find the lightlike condition $g_{\alpha\beta}x'^\alpha x'^\beta=0$, and we are led to a \emph{null geodesic}, the worldline followed by a free photon. Finally, because we now have $\phi=-f^2(x,x')$,\footnote{The extra minus sign does not change the final result.} the earlier derivation of the geodesic equation leading to Eq.~(\ref{null-geodesic-equation}) remains the same.

\subsection{The absolute derivative}\label{absolute-derivative-sec}
Given the geodesic equation (\ref{null-geodesic-equation}) (in principle applicable to both material particles and photons), let us perform a general parameter transformation $u\mapsto v(u) $ and rewrite the geodesic equation in terms of the new parameter $v$. Using the chain rule to go from $u$ to $v$, we find
\begin{equation}
\frac{\D^2 x^\gamma}{\D v}+\big\{\begin{smallmatrix}\gamma\\\alpha\beta\end{smallmatrix}\big\} \frac{\D x^\alpha}{\D v} \frac{\D x^\beta}{\D v}=
-\frac{\D^2 v}{\D u^2}\left(\frac{\D v}{\D u}\right)^{-2} \frac{\D x^\gamma}{\D v}.
\end{equation}
This result tells us two things. First, that the geodesic equation retains its simple form only for linear (also called affine) parametric transformations. Returning to the possible relation between the null parameter $u$ and timelike parameter $s$, we learn that, given $s$ is a naturally valid parameter, in order for the geodesic equation of a photon to conserve its simplest form, we must have that $u=as+b$, where $a$ and $b$ are yet physically undetermined constants. Second, it tells us that because the right-hand side is a contravariant vector $U^\gamma(v)$ of the form $f(v)\D x^\gamma/\D v$, then so must be the left-hand side, despite the fact that neither its first term nor its second term in isolation are vectors themselves (their sum must be). As we see next, the second observation opens up a natural (i.e.\ physically inspired) road towards the concept of absolute derivative. We pursue the implications of the first observation later on.

What we wish to do is find a form for the geodesic equation that is explicitly tensorial; to achieve it, we need to remove from sight both the total second derivative of the coordinates (which is not a tensor) and Christoffel's symbol (which is not a tensor either) and recast the equation in terms of a single, explicitly contravariant vector.

Returning to a notation in terms of $u$, in a change of coordinates $x\mapsto x'$ we should then have $U'^\gamma(u)=(\partial x'^\gamma/\partial x^\delta)U^\delta$; explicitly, this is
\begin{equation}\label{coordinate-change}
\frac{\D^2 x'^\gamma}{\D u}+\big\{\begin{smallmatrix}\gamma\\\alpha\beta\end{smallmatrix}\big\}' \frac{\D x'^\alpha}{\D u} \frac{\D x'^\beta}{\D u}=
\frac{\partial x'^\gamma}{\partial x^\delta}\frac{\D^2 x^\delta}{\D u}+\big\{\begin{smallmatrix}\delta\\\alpha\beta\end{smallmatrix}\big\} \frac{\partial x'^\gamma}{\partial x^\delta} \frac{\D x^\alpha}{\D u} \frac{\D x^\beta}{\D u}.
\end{equation}
Applying the chain rule repeatedly to express the right-hand side in terms of partial derivatives with respect to $x'$ and total derivatives of $x'$ with respect to $u$, and subsequently comparing with the left-hand side, we get
\begin{equation}
\left( \big\{\begin{smallmatrix}\gamma\\\alpha\beta\end{smallmatrix}\big\}'-
\big\{\begin{smallmatrix}\delta\\\epsilon\zeta\end{smallmatrix}\big\}
\frac{\partial x'^\gamma}{\partial x^\delta}\frac{\partial x^\epsilon}{\partial x'^\alpha}\frac{\partial x^\zeta}{\partial x'^\beta}-
\frac{\partial x'^\gamma}{\partial x^\delta}
\frac{\partial^2 x^\delta}{\partial x'^\alpha \partial x'^\beta}\right)
\frac{\D x'^\alpha}{\D u}\frac{\D x'^\beta}{\D u}=0.
\end{equation}
Therefore, Christoffel's symbol of the second kind transforms as follows:
\begin{equation}\label{Christoffel-symbol-transformation}
\big\{\begin{smallmatrix}\gamma\\\alpha\beta\end{smallmatrix}\big\}'=
\big\{\begin{smallmatrix}\delta\\\epsilon\zeta\end{smallmatrix}\big\}
\frac{\partial x'^\gamma}{\partial x^\delta}\frac{\partial x^\epsilon}{\partial x'^\alpha}\frac{\partial x^\zeta}{\partial x'^\beta}+
\frac{\partial x'^\gamma}{\partial x^\delta}
\frac{\partial^2 x^\delta}{\partial x'^\alpha \partial x'^\beta};
\end{equation}
the appearance of the second term on the right-hand side shows that Christoffel's symbol of the second kind is not a tensor.\footnote{With a similar procedure, one can also show that Christoffel's symbol of the first kind is not a tensor either.}

Next, we take a general contravariant vector $T^\alpha$, which is attached to every point of a worldline $x^\alpha(u)$; this is, in other words, a vector field $T^\alpha[x(u)]$, although its being attached only to a worldline is, at the moment, a more realistic (i.e.\ physical) situation than its being attached to the whole spacetime continuum. Inspired by the   present form of the geodesic equation and our goal of expressing it tensorially, we propose a \emph{derivative extension} of it of the form
\begin{equation}\label{derivative-extension-eq}
\frac{\delta T^\alpha}{\delta u}=:\frac{\D T^\alpha}{\D u} + \big\{\begin{smallmatrix}\alpha\\\beta\gamma\end{smallmatrix}\big\} T^\beta \frac{\D x^\gamma}{\D u}.
\end{equation}
Is this derivative extension of $T^\alpha$ a contravariant vector?

To check the possible tensorial nature of $\delta T^\alpha/\delta u$, we must verify whether
\begin{equation}
\frac{\delta T'^\alpha}{\delta u} - \frac{\partial x'^\alpha}{\partial x^\beta} \frac{\delta T^\beta}{\delta u} \nonumber
\end{equation}
is null. Written explicitly, the expression becomes
\begin{equation}
\frac{\D T'^\alpha}{\D u} - \frac{\partial x'^\alpha}{\partial x^\beta}
\frac{\D T^\beta}{\D u} +
\big\{\begin{smallmatrix}\alpha\\\beta\gamma\end{smallmatrix}\big\}' T'^\beta \frac{\D x'^\gamma}{\D u} -
\frac{\partial x'^\alpha}{\partial x^\beta}
\big\{\begin{smallmatrix}\beta\\\gamma\delta\end{smallmatrix}\big\}
T^\gamma \frac{\D x^\delta}{\D u}.
\end{equation}
The first two terms taken together reduce to
\begin{equation}
\frac{\partial^2 x'^\alpha}{\partial x^\beta \partial x^\gamma}T^\beta
\frac{\D x^\gamma}{\D u}\neq 0,
\end{equation}
showing that $\D T^\alpha/\D u$ alone is not a vector. Using Eq.~(\ref{Christoffel-symbol-transformation}), the third and fourth terms taken together reduce to
\begin{equation}
\frac{\partial^2 x^\delta}{\partial x'^\epsilon \partial x'^\zeta}
\frac{\partial x'^\alpha}{\partial x^\delta}
\frac{\partial x'^\epsilon}{\partial x^\beta}
\frac{\partial x'^\zeta}{\partial x^\gamma} T^\beta
\frac{\D x^\gamma}{\D u}\neq 0,
\end{equation}
showing that the second term of the derivative extension (\ref{derivative-extension-eq}) is not a vector either. However, when both results are taken together, we find
\begin{equation}
\left( \frac{\partial^2 x'^\alpha}{\partial x^\beta \partial x^\gamma} +
\frac{\partial^2 x^\delta}{\partial x'^\epsilon \partial x'^\zeta}
\frac{\partial x'^\alpha}{\partial x^\delta}
\frac{\partial x'^\epsilon}{\partial x^\beta}
\frac{\partial x'^\zeta}{\partial x^\gamma}\right) T^\beta
\frac{\D x^\gamma}{\D u}=0
\end{equation}
on account of the parenthesized expression being null.\footnote{This can be proved by differentiating both sides of $(\partial x'^\alpha/\partial x^\gamma)
(\partial x^\beta/\partial x'^\alpha)=\delta^\beta_{\ \gamma}$ with respect to $x^\delta$.} We conclude that $\delta T^\alpha/\delta u$ is an explicitly contravariant vector and call it the \emph{absolute derivative} of $T^\alpha$.

\begin{figure}
\centering
\includegraphics[width=60mm]{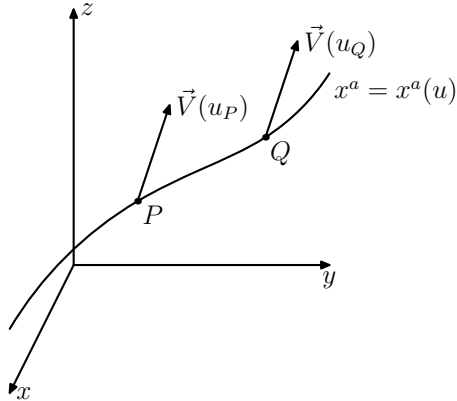} 
\caption{Euclidean
vizualization via space vector $\vec V$ of the parallel
transport of a spacetime vector $T^\alpha$ along a worldline. The
parallel-transport condition on $\vec V$ is given by the null ordinary
derivative, $\D \vec V/\D u=0$, at each point of a space trajectory,
while the same condition on $T^\alpha$ is given by a null absolute
derivative, $\delta T^\alpha/\delta u=0$, at each point of the
wordline.} 
\label{parallel-transport}
\end{figure}

We can now put the absolute derivative of a vector to good use. We observe that the condition $\delta T^\alpha/\delta u=0$ pictures the \emph{absolute constancy} of the vector $T^\alpha$; $T^\alpha$ is said to be \emph{parallel-transported} along a worldline. These names are inspired in the Euclidean picture, shown in Figure \ref{parallel-transport}, that results from visualizing this situation: a curve in space and an arrow vector $\vec V$ attached to two points $P$ and $Q$ on the curve, such that $\vec V(P)$ points in the same direction as $\vec V(Q)$. This Euclidean image is given by the condition $\D \vec V/\D u=0$, which is only a special case of $\delta T^\alpha/\delta u=0$. The image also holds under conditions of pseudo-Euclidicity as in special relativity, where the Christoffel symbol is null and the absolute derivative reduces to the usual total derivative, $\delta T^\alpha/\delta u=\D T^\alpha/\D u$. The null absolute derivative pictures the parallel transport of a vector along a worldline in a spacetime that is not in general flat.

Because the form of the absolute derivative was inspired in that of the geodesic equation, we should not be surprised to find that, when parallel-transported, $T^\alpha$ satisfies an equation remarkably similar to the geodesic equation, namely,
\begin{equation}
\frac{\delta T^\alpha}{\delta u}=:\frac{\D T^\alpha}{\D u} + \big\{\begin{smallmatrix}\alpha\\\beta\gamma\end{smallmatrix}\big\} T^\beta \frac{\D x^\gamma}{\D u}=0.
\end{equation}
It is now only a small step to recast the geodesic equation in the form we were seeking, one that is explicitly covariant---and not only that but, moreover, one that has both an elegant form and interpretation:
\begin{equation}\label{geodesic-equation-final}
\frac{\delta}{\delta u}\frac{\D x^\alpha}{\D u}=0.
\end{equation}
While the elegant form is self-evident, the elegant interpretation is the following: a geodesic tangent vector progresses along its worldline in the direction which it itself points to, and so, necessarily, without changing direction. This is a generalization of the case of a geodesic in the Euclidean plane, which, being a straight line, is traced by the unidirectional, unperturbed progression of its tangent vector. In consequence, a geodesic in Riemann-Einstein spacetime is a straight line which progresses, as it were, by following the natural \emph{intrinsic} shape of spacetime. The usual Euclidean visualization of this situation is, however, somewhat misleading, because by picturing Riemann-Einstein spacetime as an \emph{extrinsically} curved surface (e.g.\ a sphere embedded in three-dimensional Euclidean space), a geodesic tangent vector (the tangent to a maximal circle) sticks out of the surface into the embedding space, thus giving the impression that it does not progress in the direction that it points to but that, rather, some external pressure forces it to turn with the sphere.

We find, furthermore, that the above metaphorical allusions to ``unperturbed progression'' of and ``external pressure'' on a geodesic tangent vector are particularly apt to describe its behaviour, since in question here is a (new) physical picture of gravitation. A free particle (in the general-relativistic sense, i.e.\ not gravitation-free) follows a geodesic in spacetime. Our classical intuition does not mislead us when we expect the free particle to continue along its worldline unperturbed, since there are no external forces to change the direction or rate of its progression; and so it does, as it is carried along in the intrinsic spacetime direction of least resistance on account of its inertia.

This is a generalization of Newton's first law of motion. Because the mass $m$ of a particle is a constant, we can rewrite Eq.~(\ref{geodesic-equation-final}) in terms of its four-momentum $P^\alpha=mV^\alpha$ and timelike parameter $s$ to get
\begin{equation}\label{generalized-first-law}
\frac{\delta P^\alpha}{\delta s}=0.
\end{equation}
This means that the four-momentum of a free massive particle progresses unperturbed (i.e.\ is parallel-transported) in spacetime, of which a special case is
\begin{equation}
\frac{\D \vec p}{\D t}=\vec 0,
\end{equation}
namely, Newton's first law of motion, or the law of inertia (with the proviso that we interpret $s$ as absolute time $t$).

Finally, we examine the special case of a photon. This is a massless particle whose four-momentum $P^\alpha$ is nevertheless not null, and so for a photon we may simply write\footnote{Choosing $P^\alpha=\theta \D x^\alpha/\D u$, we find that $\theta$ must be absolutely constant, $\delta \theta/\delta u=0$, on account of $\delta P^\alpha/\delta u=0$ and $\delta U^\alpha/\delta u=0$, where $U^\alpha=\D x^\alpha/\D u$. The simplest choice is $\theta=1$. In general, Eq.~(\ref{particle-photon-relation}) becomes $u=(\theta/m)s$.} 
\begin{equation}
P^\alpha=\frac{\D x^\alpha}{\D u}.	
\end{equation}
Comparing with the four-momentum $P^\alpha=m \D x^\alpha/\D s$ of a massive particle leads to the suggestion that a photon can be understood as a limiting case of a massive particle whose mass $m$ tends to zero as the interval separation $\D s$ measured by a clock on its worldline tends to zero with it, such that the quotient of $\D s$ to $m$ converges to a non-null limit; this limit is $\D u$. The linear connection between $u$ and $s$ envisioned earlier is thus 
\begin{equation}\label{particle-photon-relation}
u=\frac{1}{m}s.	
\end{equation}

So far, the absolute derivative was developed only for a contravariant vector, but because in the context of Riemannian geometry contravariant and covariant vectors are fundamentally the same, it is reasonable to expect that the absolute derivative can also be a derivative extension of covariant vectors. In order to find its form, we resort once more to the concept of invariance. Let $T_\alpha$ be a covariant vector attached to the worldline \mbox{$x^\alpha=x^\alpha(u)$}, and let $U^\alpha$ be an arbitrary contravariant vector which is parallel-transported along the same worldline, and therefore 
\begin{equation}\label{parallel-transport-u}
\frac{\D U^\alpha}{\D u}=-\big\{\begin{smallmatrix}\alpha\\\beta\gamma\end{smallmatrix}\big\} U^\beta \frac{\D x^\gamma}{\D u}.	
\end{equation}
Comparing the ordinary rate of change of the invariant overlap between $T_\alpha$ and $U^\alpha$, after applying the chain rule and Eq.~(\ref{parallel-transport-u}), we find
\begin{equation}
\frac{\D (T_\alpha U^\alpha)}{\D u}= \left( \frac{\D T_\alpha}{\D u}-
\big\{\begin{smallmatrix}\beta\\\alpha\gamma\end{smallmatrix}\big\} T_\beta
\frac{\D x^\gamma}{\D u} \right) U^\alpha.
\end{equation}
Because the left-hand side is an invariant and $U^\alpha$ on the right-hand side is an arbitrary contravariant vector, the parenthesized expression, which is an extension of vector $T_\alpha$, must then be a covariant vector. We call 
\begin{equation}
      \frac{\delta T_\alpha}{\delta u}=:
      \frac{\D T_\alpha}{\D u}-\big\{\begin{smallmatrix}\beta\\\alpha\gamma\end{smallmatrix}\big\} T_\beta \frac{\D x^\gamma}{\D u}
\end{equation}
the absolute derivative of $T_\alpha$.

The same heuristic process that led us to the absolute derivative of a covariant vector can also lead us to the absolute derivative of a tensor of any type and order. For example, by setting up the ordinary derivative of the invariant $T^{\alpha\beta}U_\alpha W_\beta$, where $U_\alpha$ and $W_\beta$ are parallel-transported along the worldline, we obtain, proceeding as before, the absolute derivative of a second-order contravariant tensor
\begin{equation}
\frac{\delta T^{\alpha\beta}}{\delta u}= 
\frac{\D T^{\alpha\beta}}{\D u}+\big\{\begin{smallmatrix}\alpha\\\gamma\delta\end{smallmatrix}\big\} T^{\gamma\beta}\frac{\D x^\delta}{\D u}+
\big\{\begin{smallmatrix}\beta\\\gamma\delta\end{smallmatrix}\big\} T^{\alpha\gamma}\frac{\D x^\delta}{\D u}.
\end{equation}
The manner in which this result is obtained allows us to conclude that the absolute derivative of a general tensor is built out of its ordinary derivative plus one positive term for each of its contravariant indices and one negative term for each of its covariant indices, each index being summed over one at a time.

From its general form follows that the absolute derivative satisfies the linear-combination rule and Leibniz's chain rule. Another useful result also follows, namely, that the absolute derivative of the metric field is null in both its covariant and contravariant versions,\footnote{This can be shown by rewriting $\D g_{\alpha\beta}/\D u$ as $(\partial g_{\alpha\beta}/\partial x^\delta) (\D x^\delta/\D u)$ and considering the identity $[\delta\alpha,\beta]+[\delta\beta,\alpha]=0$ in the expression for the absolute derivative of the metric tensor.} 
\begin{equation}
\frac{\delta g_{\alpha\beta}}{\delta u}=0 \qquad \mathrm{and} \qquad \frac{\delta g^{\alpha\beta}}{\delta u}=0.	
\end{equation}
This means that, although the metric field is not a constant in the usual sense, it is a constant in the absolute sense of Riemann-Einstein spacetime. Finally, because an invariant scalar has no indices, we add the convention that its absolute derivative corresponds to its ordinary derivative, 
\begin{equation}
\frac{\delta T}{\delta u}=:\frac{\D T}{\D u}.	
\end{equation}

At this point, the reader might object that our way of proceeding in order to find the absolute derivative as an extension of vectors and general tensors has no real justification behind it. Why should it be what it is and not otherwise? To this we reply simply that there is no certified, royal road to mathematical physics. By pretending that there is and defining mathematical concepts in an axiomatic spirit, we trade learning the informal, freewheeling, and privately conceived physical motivation behind mathematical concepts for the illusion of security offered by authoritative disembodied definitions. Would defining the absolute derivative in a dry axiom, while leaving out all mention of the heuristic method by which we obtained it, have made this concept clearer or more obscure in the present physical context? Let us leave rigorously stated definitions and axioms to pure mathematicians; these are their rightful turf---but physics is not mathematics nor should it strive to be.

\subsection{The covariant derivative}
Absolute differentiation applies to tensors that are attached to a worldline in Riemann-Einstein spacetime. And this is all very well because the object of our examination has so far been material particles and photons moving along geodesics in spacetime. However, because the geodesic equation does not give us knowledge of the spacetime geometry but only informs us how particles will move given the metric tensor, it needs the partnership of a field equation. A field equation can fill this gap by providing knowledge of the metric tensor given the distribution of matter in spacetime, which thus acts as the source of spacetime geometry, akin to the way charge and current act as the sources of the electric and magnetic forces.

A field equation linking matter distribution to curvature distribution in spacetime differs from the geodesic equation in that the tensors involved in the former need to be attached to whole regions of spacetime and not just to the worldlines of particles. The absolute derivative, then, will not do as a physically meaningful derivative extension of fields, since the term $\D x^\alpha/\D u$ appearing in it does not have any meaning in the absence of a worldline. This is enough motivation for the development of a different kind of extension of a tensor---for a \emph{spacetime derivative}---as closely related to the absolute derivative as possible, yet without reference to a worldline.

What this spacetime derivative should be is insinuated after a simple rewriting of the absolute derivative as follows:
\begin{equation}
\frac{\delta T^\alpha}{\delta u}=\left( \frac{\partial T^\alpha}{\partial x^\gamma}+\big\{\begin{smallmatrix}\alpha\\\beta\gamma\end{smallmatrix}\big\} T^\beta\right) \frac{\D x^\gamma}{\D u}.
\end{equation}
Again, because the absolute derivative of $T^\alpha$ on the left-hand side is a contravariant vector and the worldline derivative on the right-hand side is as well, the parenthesized expression---which does not depend on the worldline---must now be a second-order once-covariant once-contravariant tensor. As this expression satisfies all we asked from a derivative extension applicable to fields in spacetime, we take
\begin{equation}
T^\alpha_{\ \ ;\gamma}=: \frac{\partial T^\alpha}{\partial x^\gamma}
+\big\{\begin{smallmatrix}\alpha\\\beta\gamma\end{smallmatrix}\big\} T^\beta 
\end{equation}
as the spacetime derivative we seek. We see that, just like the absolute derivative was a generalization of the total derivative in flat space or flat spacetime, so is the spacetime derivative a generalization of the partial derivative in flat space or flat spacetime. From now on, we call $T^\alpha_{\ \ ;\gamma}$ the \emph{covariant derivative} of $T^\alpha$ in line with a long physical tradition. As justification for this name---less descriptive of its physical function than the name spacetime derivative---it is noted that this derivative always has the effect of adding one extra covariant index to the covariantly differentiated tensor. As before, because the covariant derivative comes from the absolute derivative in a straightforward way, the covariant derivative of a tensor of general type follows the same form as that of its absolute derivative. Finally, also the covariant derivative satisfies the linear-combination rule and Leibniz's rule, and the metric tensor is also covariantly constant, $g_{\alpha\beta ;\gamma}=g^{\alpha\beta}_{\ \ \ ;\gamma}=0$.

Once the covariant derivative has been developed, we can recreate with it generalizations of well-known concepts from vector analysis to be useful in the following pages:
\begin{itemize}
\item[(i)] The \emph{covariant gradient} $T_{;\alpha}$ of an invariant scalar $T$ is by convention (because it has no indices) taken to correspond with its ordinary gradient (ordinary partial derivative), 
\begin{equation}
T_{;\alpha}=: \frac{\partial T}{\partial x^\alpha} =: T_{,\alpha},	
\end{equation}
where from now on we also denote partial derivatives with a comma. 

\item[(ii) ]The \emph{covariant divergence} $T^\alpha_{\ \ ;\alpha}$ of a contravariant vector $T^\alpha$ is\footnote{This is a consequence of the simplified form of the appearing Christoffel symbol $\big\{\begin{smallmatrix}\alpha\\\beta\alpha\end{smallmatrix}\big\}$, when this form is made explicit.}
\begin{equation}
T^\alpha_{\ \ ;\alpha}=T^\alpha_{\ \ ,\alpha}+\frac{1}{2}g^{\alpha\gamma} \frac{\partial g_{\alpha\gamma}}{\partial x^\beta} T^\beta.
\end{equation}

\item[(iii)] The \emph{covariant curl} $T_{\alpha ;\beta}-T_{\beta ;\alpha}$ of a covariant vector $T_\gamma$ is its ordinary curl, 
\begin{equation}
T_{\alpha ;\beta}-T_{\beta ;\alpha}=T_{\alpha ,\beta}-T_{\beta ,\alpha}	
\end{equation}
on account of the symmetry property of the Christoffel symbol. As a consequence, when $T_\alpha$ is some invariant gradient $U_{,\alpha}$, we recover the spacetime (i.e.\ covariant) version of the well-known result that a gradient is a \emph{conservative vector field}, 
\begin{equation}
U_{,\alpha ,\beta}-U_{,\beta ,\alpha}=0,	
\end{equation}
since ordinary partial derivatives commute. From now on, we denote consecutive ordinary partial derivatives of tensors with a single comma; e.g.\ $U_{,\alpha ,\beta}=: U_{,\alpha\beta}$.
\end{itemize}

\subsection{The curvature tensor}\label{sec:curvature-tensor}
We saw on page \pageref{flat-space} that the geodesics of a free particle in a spacetime whose quadratic form is everywhere pseudo-Cartesian are straight lines resembling those of flat Euclidean space of classical mechanics. This motivates us to characterize flat spacetime by the condition that it be possible to choose coordinates such that the quadratic form be everywhere pseudo-Cartesian, i.e.\ \mbox{$Q=(\D x^1)^2+(\D x^2)^2+(\D x^3)^2-(\D x^4)^2$}; in other words, that $g_{\alpha\beta}$ can be transformed into $\eta_{\alpha\beta}=\mathrm{diag}(1,1,1,-1)$ at every point. If this were not possible, we take spacetime to be curved.

This characterization of flat and curved spacetime is clear and intuitive, but is it practical? Given a quadratic form, how shall we proceed to show that it can or cannot be recast as an everywhere pseudo-Cartesian one? Can we devise an easier way of testing whether spacetime is flat or curved? To do this, we take advantage of the intuitive image given by the spacetime-related concepts inherited from ordinary vector calculus, and think in terms of an analogy. It may be objected that the analogy is far-fetched, but we should not worry about this so long as it naturally guides us to the solution of our problem.

We take a covariant vector field $T_\alpha$ and use its second covariant derivative
\begin{equation}
T_{\alpha;\beta\gamma}=: (T_{\alpha;\beta})_{;\gamma} = T_{\alpha;\beta,\gamma}-
\big\{\begin{smallmatrix}\delta\\\alpha\gamma\end{smallmatrix}\big\} T_{\delta;\beta}-
\big\{\begin{smallmatrix}\delta\\\beta\gamma\end{smallmatrix}\big\} T_{\alpha;\delta}
\end{equation}
to construct \emph{a kind of covariant curl} (third-order tensor) $T_{\alpha;\beta\gamma}-T_{\alpha;\gamma\beta}$ of its first covariant derivative $T_{\alpha;\beta}$ (second-order tensor). This curl can be written out explicitly to find
\begin{multline}
T_{\alpha;\beta\gamma}-T_{\alpha;\gamma\beta}=
T_{\alpha;\beta,\gamma}-\big\{\begin{smallmatrix}\delta\\\alpha\gamma\end{smallmatrix}\big\} T_{\delta;\beta}-
\big\{\begin{smallmatrix}\delta\\\beta\gamma\end{smallmatrix}\big\} T_{\alpha;\delta}-
T_{\alpha;\gamma,\beta}+\\
\big\{\begin{smallmatrix}\delta\\\alpha\beta\end{smallmatrix}\big\} T_{\delta;\gamma}+
\big\{\begin{smallmatrix}\delta\\\gamma\beta\end{smallmatrix}\big\} T_{\alpha;\delta}.
\end{multline}
The third and sixth terms on the right-hand side cancel out. Writing out the remaining covariant derivatives, we find
\begin{multline}
T_{\alpha;\beta\gamma}-T_{\alpha;\gamma\beta}=
(T_{\alpha,\beta}-\big\{\begin{smallmatrix}\delta\\\alpha\beta\end{smallmatrix}\big\}T_\delta)_{,\gamma}-
\big\{\begin{smallmatrix}\delta\\\alpha\gamma\end{smallmatrix}\big\}
(T_{\delta,\beta}-\big\{\begin{smallmatrix}\epsilon\\\delta\beta\end{smallmatrix}\big\}T_\epsilon)-\\
(T_{\alpha,\gamma}-\big\{\begin{smallmatrix}\delta\\\alpha\gamma\end{smallmatrix}\big\}T_\delta)_{,\beta}+
\big\{\begin{smallmatrix}\delta\\\alpha\beta\end{smallmatrix}\big\}
(T_{\delta,\gamma}-\big\{\begin{smallmatrix}\epsilon\\\delta\gamma\end{smallmatrix}\big\}T_\epsilon).
\end{multline}
The first and fifth terms, part of the second term and the seventh term, and the third term and part of the sixth term cancel out in pairs. Finally, we find
\begin{equation}\label{covariant-rotor-eq}
T_{\alpha;\beta\gamma}-T_{\alpha;\gamma\beta}= R^\delta_{\ \alpha\beta\gamma}T_\delta,
\end{equation}
where
\begin{equation}\label{curvature-tensor}
R^\delta_{\ \alpha\beta\gamma}=
\big\{\begin{smallmatrix}\delta\\\alpha\gamma\end{smallmatrix}\big\}_{,\beta}-
\big\{\begin{smallmatrix}\delta\\\alpha\beta\end{smallmatrix}\big\}_{,\gamma}+
\big\{\begin{smallmatrix}\epsilon\\\alpha\gamma\end{smallmatrix}\big\}
\big\{\begin{smallmatrix}\delta\\\epsilon\beta\end{smallmatrix}\big\}-
\big\{\begin{smallmatrix}\epsilon\\\alpha\beta\end{smallmatrix}\big\}
\big\{\begin{smallmatrix}\delta\\\epsilon\gamma\end{smallmatrix}\big\}
\end{equation}
must be a fourth-order mixed tensor, since the left-hand side of Eq.~(\ref{covariant-rotor-eq}) is a third-order covariant tensor and $T_\delta$ on the right-hand side is a covariant vector. We call $R^\delta_{\ \alpha\beta\gamma}$ the \emph{mixed curvature tensor} or the \emph{mixed Riemann tensor}. In general, $R^\delta_{\ \alpha\beta\gamma}T_\delta\neq 0$, suggesting that the discrepancy arising from the order of spacetime directions in which a vector is differentiated is the result of a \emph{non-conservative property of spacetime} captured by the tensor field $R^\delta_{\ \alpha\beta\gamma}$. But what is the connection of all this with curvature?

If spacetime is flat, there exist coordinates such that the metric field is $\eta_{\alpha\beta}=\mathrm{diag}(1,1,1,-1)$ at every point. In that case, all Christoffel symbols and their ordinary partial derivatives in Eq.~(\ref{curvature-tensor}) are null, and so 
\begin{equation}
R^\delta_{\ \alpha\beta\gamma}T_\delta(x) \equiv 0 	
\end{equation}
at every spacetime point $x$. Since this is a \emph{tensor equation}, its validity is independent of the coordinate system used; if it holds in one coordinate system, it holds in all of them. Now, while this is a necessary condition for the flatness of spacetime, is it also a sufficient condition, such that we may say that there is a one-to-one correspondence between the curvature tensor and curvature? A reading of Einstein's original paper on the general theory of relativity is not auspicious regarding the level of complexity required to answer this question. Einstein \citeyear{Einstein:1952b} prefers not to tackle this problem and, in a footnote, writes: ``The mathematicians have proved that this is also a \emph{sufficient} condition'' (p.~141). Let us nonetheless attempt this demonstration---the physicist's way.

We start with an arbitrary covariant vector $T_\alpha(\tilde x)$ at an arbitrary spacetime point, and parallel-transport it first along a worldline $x^\alpha=x^\alpha(u,v_0)$ a separation $\D u$, and subsequently along another worldline $x^\alpha=x^\alpha(\D u,v)$ a separation $\D v$, where $u_0$ and $v_0$ are constants, thus reaching spacetime point $x$ (see Figure \ref{ptgr} on page \pageref{ptgr}). Alternatively, we can reach $x$ from $\tilde x$ by parallel-transporting $T_\alpha(\tilde x)$ first along worldline $x^\alpha=x^\alpha(u_0,v)$ a separation $\D v$, and subsequently along worldline $x^\alpha=x^\alpha(u,\D v)$ a separation $\D u$. The two resulting vectors at $x$ are the same only if the difference
\begin{equation}
\frac{\delta^2T_\alpha}{\delta v \delta u}-
\frac{\delta^2T_\alpha}{\delta u \delta v}=
\left( T_{\alpha;\beta} \frac{\partial x^\beta}{\partial u} \right)_{;\gamma}
\frac{\partial x^\gamma}{\partial v}-
\left( T_{\alpha;\beta} \frac{\partial x^\beta}{\partial v} \right)_{;\gamma}
\frac{\partial x^\gamma}{\partial u}
\end{equation}
is null. Because $\partial x^\beta/\partial u$ and $\partial x^\beta/\partial v$ on the right-hand side are the vectors tangent to the worldlines, they are absolutely constant and can be taken out of the covariant derivative. We find
\begin{align}
\frac{\delta^2T_\alpha}{\delta v \delta u}-
\frac{\delta^2T_\alpha}{\delta u \delta v}&=
T_{\alpha;\beta\gamma}\frac{\partial x^\beta}{\partial u}\frac{\partial x^\gamma}{\partial v}-
T_{\alpha;\beta\gamma}\frac{\partial x^\beta}{\partial v}\frac{\partial x^\gamma}{\partial u}\nonumber\\
&=(T_{\alpha;\beta\gamma}-T_{\alpha;\gamma\beta})\frac{\partial x^\beta}{\partial u}\frac{\partial x^\gamma}{\partial v},
\end{align}
and we now recognize the Riemann tensor in the parenthesized expression:
\begin{equation}
\frac{\delta^2T_\alpha}{\delta v \delta u}-
\frac{\delta^2T_\alpha}{\delta u \delta v}=
R^\delta_{\ \alpha\beta\gamma}T_\delta \frac{\partial x^\beta}{\partial u}
\frac{\partial x^\gamma}{\partial v},
\end{equation}
where the worldlines are arbitrary and so is $T_\alpha$. 

Therefore, \emph{if} the Riemann tensor is identically null, then the result of parallel-transporting $T_\alpha(\tilde x)$ to $x$ is a \emph{unequivocal} vector $T_\alpha(x)$ independently of the worldline followed. But then, its covariant derivative must be null, $T_{\alpha;\beta}(x)=0$. Since spacetime point $x$ is arbitrary, given $R^\delta_{\ \alpha\beta\gamma}\equiv 0$ (sufficient condition),\footnote{This is also a necessary condition. The system of equations $T_{\alpha,\beta}=\big\{\begin{smallmatrix}\delta\\\alpha\beta\end{smallmatrix}\big\}T_\delta$ is integrable only if $T_{\alpha,\beta\gamma}=T_{\alpha,\gamma\beta}$. This gives $
\left(\big\{\begin{smallmatrix}\delta\\\alpha\beta\end{smallmatrix}\big\} T_\delta \right)_{,\gamma}=\left(\big\{\begin{smallmatrix}\delta\\\alpha\gamma\end{smallmatrix}\big\} T_\delta\right)_{,\beta}$, implying $R^\delta_{\ \alpha\beta\gamma}\equiv 0$.} we can create a spacetime vector \emph{field} at any spacetime point $x$ via parallel-transport of an original vector. Based on our intuition, we now expect that the unequivocal existence of $T_\alpha(x)$ will imply that spacetime is flat. For example, in the intuitively curved case of a spherical surface, the parallel-transport of a vector from the equator to the north pole produces different vectors at the north pole depending on the path chosen, but the same vector on a flat surface.

We now take the covariant vector field $T_\alpha(x)$ to be the gradient $S_{,\alpha}(x)$ of an arbitrary invariant scalar field $S(x)$. We take, moreover, four independent scalars fields, which we choose to be the coordinates $x'^1(x),x'^2(x),x'^3(x)$, and $x'^4(x)$. Now $T_{\alpha;\beta}=0$ translates into $x'^\gamma_{,\alpha;\beta}(x)=0$, which, in turn, gives 
\begin{equation}\label{constant-x}
\frac{\partial^2 x'^\gamma}{\partial x^\alpha\partial x^\beta}=\big\{\begin{smallmatrix}\delta\\ \alpha\beta\end{smallmatrix}\big\}\frac{\partial x'^\gamma}{\partial x^\delta}.
\end{equation}
We now show that in the primed coordinates obtained through parallel transport, the metric tensor $g'_{\alpha\beta}(x')$ is constant.

Because 
\begin{equation}
g_{\alpha\beta}=\frac{\partial x'^\gamma}{\partial x^\alpha} \frac{\partial x'^\delta}{\partial x^\beta} g'_{\gamma\delta},	
\end{equation}
we find
\begin{equation}
\frac{\partial g_{\alpha\beta}}{\partial x^\epsilon}=
\frac{\partial^2 x'^\gamma}{\partial x^\alpha\partial x^\epsilon} \frac{\partial x'^\delta}{\partial x^\beta} g'_{\gamma\delta}+
\frac{\partial x'^\gamma}{\partial x^\alpha} \frac{\partial^2 x'^\delta}{\partial x^\beta \partial x^\epsilon} g'_{\gamma\delta}+
\frac{\partial x'^\gamma}{\partial x^\alpha}\frac{\partial x'^\delta}{\partial x^\beta} \frac{\partial g'_{\gamma\delta}}{\partial x^\epsilon},
\end{equation}
which, on account of Eq.~({\ref{constant-x}}), becomes
\begin{equation}
\frac{\partial g_{\alpha\beta}}{\partial x^\epsilon}=
\big\{\begin{smallmatrix}\zeta\\\epsilon\alpha\end{smallmatrix}\big\}
\frac{\partial x'^\gamma}{\partial x^\zeta} \frac{\partial x'^\delta}{\partial x^\beta} g'_{\gamma\delta}+
\big\{\begin{smallmatrix}\zeta\\\epsilon\beta\end{smallmatrix}\big\}
\frac{\partial x'^\gamma}{\partial x^\alpha} \frac{\partial x'^\delta}{\partial x^\zeta} g'_{\gamma\delta}+
\frac{\partial x'^\gamma}{\partial x^\alpha} \frac{\partial x'^\delta}{\partial x^\beta} \frac{\partial g'_{\gamma\delta}}{\partial x^\epsilon}.
\end{equation}
The first and second terms can be simplified to get
\begin{align}
\frac{\partial g_{\alpha\beta}}{\partial x^\epsilon}&=
\big\{\begin{smallmatrix}\zeta\\\epsilon\alpha\end{smallmatrix}\big\}
g_{\zeta\beta}+
\big\{\begin{smallmatrix}\zeta\\\epsilon\beta\end{smallmatrix}\big\}
g_{\alpha\zeta}+
\frac{\partial x'^\gamma}{\partial x^\alpha} \frac{\partial x'^\delta}{\partial x^\beta} \frac{\partial g'_{\gamma\delta}}{\partial x^\epsilon}\nonumber\\
&=[\epsilon\alpha,\beta]+[\epsilon\beta,\alpha]+\frac{\partial x'\gamma}{\partial x^\alpha} \frac{\partial x'^\delta}{\partial x^\beta} \frac{\partial g'_{\gamma\delta}}{\partial x^\epsilon},
\end{align}
finally giving
\begin{equation}
\frac{\partial g_{\alpha\beta}}{\partial x^\epsilon}=
\frac{\partial g_{\alpha\beta}}{\partial x^\epsilon}+ \frac{\partial x'^\gamma}{\partial x^\alpha} \frac{\partial x'^\delta}{\partial x^\beta}\frac{\partial g'_{\gamma\delta}}{\partial x^\epsilon}.
\end{equation}
It follows that $\partial g'_{\gamma\delta}/\partial x^\epsilon=0$, and therefore $\partial g'_{\gamma\delta}/\partial x'^\zeta=0$. In conclusion, if the Riemann curvature tensor is identically null, then coordinates $x'$ exist such that $g'_{\alpha\beta}(x')$ is constant everywhere in spacetime.\footnote{We have not proved, however, that $g'_{\alpha\beta}(x')=\eta_{\alpha\beta}$.}

The mixed Riemann tensor $R^\delta_{\ \alpha\beta\gamma}$, its covariant version $R_{\alpha\beta\gamma\delta}$, and its contractions, $R^{\gamma}_{\ \alpha\beta\gamma}$ and $R^\alpha_{\ \alpha}$, satisfy several useful identities. We can see straight from Eq.~(\ref{curvature-tensor}) that the mixed Riemann tensor is antisymmetric on its last two covariant indices, 
\begin{equation}
R^\delta_{\ \alpha\beta\gamma}=-R^\delta_{\ \alpha\gamma\beta},	
\end{equation}
and that the sum of its cyclic permutations is null, 
\begin{equation}
R^\delta_{\ \alpha\beta\gamma}+R^\delta_{\ \gamma\alpha\beta}+R^\delta_{\ \beta\gamma\alpha}=0. 	
\end{equation}
The covariant Riemann tensor can be obtained from the mixed one lowering the contravariant index, $R_{\alpha\beta\gamma\delta}=g_{\alpha\epsilon}R^\epsilon_{\ \beta\gamma\delta}$, which gives\footnote{Express $g_{\alpha\epsilon}\big\{\begin{smallmatrix}\epsilon\\\beta\delta\end{smallmatrix}\big\}_{,\gamma}$ as the derivative of the product minus an extra term, and rewrite both in terms of Christoffel symbols of the first kind and their derivatives.}
\begin{equation}
    R_{\alpha\beta\gamma\delta}=
    [\beta\delta,\alpha]_{,\gamma}-[\beta\gamma,\alpha]_{,\delta}+
    g^{\epsilon\zeta}                         \left(
    [\alpha\delta,\epsilon][\beta\gamma,\zeta]-
    [\alpha\gamma,\epsilon][\beta\delta,\zeta]\right).
\end{equation}
A more useful expression of the covariant Riemann tensor can be obtained by writing the Christoffel symbols of the first kind in terms of those of the second kind, the metric tensor, and its derivatives. We find
\begin{multline}
    R_{\alpha\beta\gamma\delta}=
    \frac{1}{2}(g_{\alpha\delta,\beta\gamma}+g_{\beta\gamma,\alpha\delta}-
      g_{\alpha\gamma,\beta\delta}-g_{\beta\delta,\alpha\gamma})+\\
      g_{\epsilon\zeta}\left(
      \big\{\begin{smallmatrix}\epsilon\\\alpha\delta\end{smallmatrix}\big\}
      \big\{\begin{smallmatrix}\zeta\\\beta\gamma\end{smallmatrix}\big\}-
      \big\{\begin{smallmatrix}\epsilon\\\alpha\gamma\end{smallmatrix}\big\}
      \big\{\begin{smallmatrix}\zeta\\\beta\delta\end{smallmatrix}\big\}
                        \right).
\end{multline}
By direct inspection, we see that the covariant Riemann tensor is antisymmetric on its first two and last two indices, 
\begin{equation}
R_{\alpha\beta\gamma\delta}=-R_{\beta\alpha\gamma\delta} 
\qquad \mathrm{and} \qquad 
R_{\alpha\beta\gamma\delta}=-R_{\alpha\beta\delta\gamma},
\end{equation}
and it is symmetric with respect to a swap of the first and second pairs of indices taken as a unit,
\begin{equation}
R_{\alpha\beta\gamma\delta}=R_{\gamma\delta\alpha\beta}. 		
\end{equation}
Applying all these results together, we also find that the covariant Riemann tensor is symmetric with respect to an opposite reordering of its indices,
\begin{equation}
R_{\alpha\beta\gamma\delta}=R_{\delta\gamma\beta\alpha}.
\end{equation}

In addition, the covariant derivatives of the mixed and covariant Riemann tensors each satisfy an identity called the \emph{Bianchi identity}, which says that the sum of the cyclic permutations on the last three covariant indices of the derivative tensor is null. For the mixed Riemann tensor, it reads\footnote{To prove the Bianchi identity, calculate the ``covariant curl'' of the third-order tensor $(A_\alpha B_\beta)_{;\gamma}$ to find $(A_\alpha B_\beta)_{;\gamma\delta}-(A_\alpha B_\beta)_{;\delta\gamma}=A_\alpha R^\epsilon_{\ \beta\gamma\delta}B_\epsilon+B_\beta R^\epsilon_{\ \alpha\gamma\delta}A_\epsilon$. Next identify $A_\alpha B_\beta$ with $T_{\alpha;\beta}$ and form the sum of its three cyclic permutations of $\beta$, $\gamma$, and $\delta$. Express this sum, on the one hand, in terms of the mixed Riemann tensor and first covariant derivatives of $T_\alpha$ and, on the other hand, in terms of covariant derivatives of the product of the mixed Riemann tensor and $T_\alpha$. Use $R^\delta_{\ \alpha\beta\gamma}+R^\delta_{\ \gamma\alpha\beta}+R^\delta_{\ \beta\gamma\alpha}=0$ and simplify.} 
\begin{equation}
R^\epsilon_{\ \alpha\beta\gamma;\delta}+R^\epsilon_{\ \alpha\delta\beta;\gamma}+R^\epsilon_{\ \alpha\gamma\delta;\beta}=0.	
\end{equation}
The corresponding Bianchi identity for the covariant Riemann tensor can be obtained from the former by lowering the contravariant index; it reads 
\begin{equation}\label{Bianchi-II}
R_{\alpha\beta\gamma\delta;\epsilon}+R_{\alpha\beta\epsilon\gamma;\delta}+R_{\alpha\beta\delta\epsilon;\gamma}=0.	
\end{equation}

The \emph{Ricci tensor} $R_{\alpha\beta}$ is a useful tensor derived from the covariant Riemann tensor $R_{\delta\alpha\beta\gamma}$ by contracting the first and last indices, 
\begin{equation}
g^{\gamma\delta} R_{\delta\alpha\beta\gamma}=R^\gamma_{\ \alpha\beta\gamma}=: R_{\alpha\beta}.	
\end{equation}
The Ricci tensor is symmetric\footnote{To show this result, write the Ricci tensor in terms of the covariant Riemann tensor and use the symmetry properties of the latter.} and has only 10 independent components, which is the same amount of independent components as that of the metric tensor. This property becomes essential when we form Einstein's field equation linking gravitation with its sources. Another useful quantity, the invariant \emph{curvature scalar} $R$, can be derived from the Ricci tensor by contracting its two indices, 
\begin{equation}
g^{\alpha\gamma}R_{\gamma\alpha}=R^\alpha_{\ \alpha}=: R.	
\end{equation}

Do the Ricci tensor and curvature scalar have any noteworthy properties? Taking the second Bianchi identity (\ref{Bianchi-II}) and contracting its indices in a suitable way, we find
\begin{multline}
g^{\beta\gamma}g^{\alpha\delta}R_{\alpha\beta\gamma\delta;\epsilon}+
g^{\alpha\delta}g^{\beta\gamma}R_{\alpha\beta\epsilon\gamma;\delta}+
g^{\beta\gamma}g^{\alpha\delta}R_{\alpha\beta\delta\epsilon;\gamma}=\\
g^{\beta\gamma}R_{\beta\gamma;\epsilon}-g^{\alpha\delta}R_{\alpha\epsilon;\delta}-g^{\beta\gamma}R_{\beta\epsilon;\gamma}=
R_{;\epsilon}-2R^\delta_{\ \epsilon;\delta}=0.
\end{multline}
Multiplying by $g^{\epsilon\zeta}$, we get
\begin{equation}
g^{\epsilon\zeta}R_{;\epsilon}-2R^{\delta\zeta}_{\ \ ;\delta}=
\left( R^{\epsilon\zeta}-\frac{1}{2}g^{\epsilon\zeta}R\right)_{;\epsilon}=0.
\end{equation}
We have found that the symmetric tensor 
\begin{equation}
G^{\alpha\beta} =: R^{\alpha\beta}-\frac{1}{2}g^{\alpha\beta}R, 	
\end{equation}
called the \emph{Einstein tensor}, has the special property that its covariant divergence is null, 
\begin{equation}\label{Einstein-divergence-free}
G^{\alpha\beta}_{\ \ \ ;\beta}=0.	
\end{equation}

Finally, noticing that the Einstein tensor $G^{\alpha\beta}$ consists of a second-order tensor $R^{\alpha\beta}$ and a scalar $R$, we consider whether any further, still simpler, term could be added to it such that its covariant divergence continues to be null. Thinking in purely mathematical terms, a term of the form $\Lambda g^{\alpha\beta}$, where $\Lambda$ is some constant unrelated to the Ricci tensor, satisfies $\Lambda g^{\alpha\beta}_{\ \ \ ;\beta}=0$. Adding it to the Einstein tensor for now only with this purely mathematical motivation, we get the \emph{generalized Einstein tensor} 
\begin{equation}
\mathcal{G}^{\alpha\beta}=:R^{\alpha\beta}-\frac{1}{2}g^{\alpha\beta}R+\Lambda g^{\alpha\beta},	
\end{equation}
where 
\begin{equation}
\mathcal{G}^{\alpha\beta}_{\ \ \ ;\beta}=0.	
\end{equation}
Is the inclusion of $\Lambda$ in the Einstein tensor also justified on physical grounds by the display of relevant observable effects? And if so, what is its physical interpretation? Not dark and mysterious, one should hope. We return to this issue later on.

\subsection{Geodesic deviation and the rise of rod length}
How do we observe the active effects of the curvature tensor? In a curved spacetime, we expect that two particles that follow at first parallel geodesics will not remain on parallel spacetime tracks but will start to converge or diverge, i.e.\ two test particles released at rest will either move towards or away from each other (discounting their mutual gravitational interaction). This two-geodesic method of probing curvature sounds promising, because the test in question involves only local information.

\begin{figure}
\centering
\includegraphics[width=60mm]{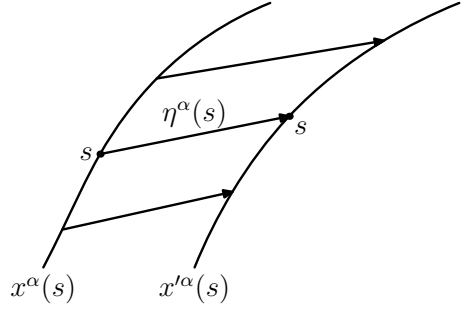} 
\caption{Deviation of two timelike geodesics as a local active effect of spacetime curvature. When spacetime is flat, two geodesics deviate from each other linearly. When the size $\eta$ of $\eta^\alpha$ does not depend on $s$, \mbox{$\D \eta/\D s=0$}, $\eta^\alpha$ acts like an idealized measuring rod.}
\label{geodesic-deviation}  
\end{figure}

To inspect the behaviour of the separation between two particle geodesics in
relation to curvature, we examine two close-lying geodesics \mbox{$x^\alpha=x^\alpha(u)$} and \mbox{$x'^\alpha=x'^\alpha(u)$} (not its derivative) that are nearly parallel to each other, as shown in Figure \ref{geodesic-deviation} for the case of two massive particles. The small difference \mbox{$\eta^\alpha(u)=x'^\alpha(u)-x^\alpha(u)$} is a differential vector that connects those points in the geodesics that have the same value of parameter $u$. Since the two geodesics are nearly parallel, the rate of change $\D \eta^\alpha/\D u$ of their mutual separation is also very small.

In order to obtain an equation for $\eta^\alpha$, we subtract the
geodesic equation of $x^\alpha(u)$ from that of $x'^\alpha(u)$.
Using the first-order approximation to the primed Christoffel symbol,
\begin{equation}
\big\{\begin{smallmatrix}\alpha\\\beta\gamma\end{smallmatrix}\big\}'=
\big\{\begin{smallmatrix}\alpha\\\beta\gamma\end{smallmatrix}\big\}+\big\{\begin{smallmatrix}\alpha\\\beta\gamma\end{smallmatrix}\big\}_{,\delta}\eta^\delta,
\end{equation}
on account of $x'^\alpha(u)=x^\alpha(u)+\eta^\alpha(u)$, we find
\begin{equation} 
\frac{\D^2 \eta^\alpha}{\D u^2}+
\big\{\begin{smallmatrix}\alpha\\\beta\gamma\end{smallmatrix}\big\}_{,\delta}
\frac{\D x^\beta}{\D u} \frac{\D x^\gamma}{\D u}\eta^\delta+
\big\{\begin{smallmatrix}\alpha\\\beta\gamma\end{smallmatrix}\big\}
\frac{\D x^\beta}{\D u} \frac{\D \eta^\gamma}{\D u}+
\big\{\begin{smallmatrix}\alpha\\\beta\gamma\end{smallmatrix}\big\}
\frac{\D \eta^\beta}{\D u} \frac{\D x^\gamma}{\D u}=0. 
\end{equation}
Introducing the derivative of a sum and subtracting two extra
terms, we get
\begin{multline} 
\frac{\D}{\D u}\left( \frac{\D \eta^\alpha}{\D u} +
\big\{\begin{smallmatrix}\alpha\\\beta\gamma\end{smallmatrix}\big\}\eta^\beta
\frac{\D x^\gamma}{\D u} \right) -
\big\{\begin{smallmatrix}\alpha\\\beta\gamma\end{smallmatrix}\big\}_{,\delta}
\eta^\beta \frac{\D x^\gamma}{\D u} \frac{\D x^\delta}{\D u} -
\big\{\begin{smallmatrix}\alpha\\\beta\gamma\end{smallmatrix}\big\}\eta^\beta
\frac{\D^2 x^\gamma}{\D u^2}+\\
\big\{\begin{smallmatrix}\alpha\\\beta\gamma\end{smallmatrix}\big\}_{,\delta}
\frac{\D x^\beta}{\D u} \frac{\D x^\gamma}{\D u}\eta^\delta+
\big\{\begin{smallmatrix}\alpha\\\beta\gamma\end{smallmatrix}\big\}
\frac{\D x^\beta}{\D u} \frac{\D \eta^\gamma}{\D u}=0. 
\end{multline}
Finally, using the geodesic equation to rewrite $\D^2x^\gamma/\D
u^2$, adding and subtracting a term (fourth and last below) to form an extra $\delta \eta^\delta/\delta u$, and rearranging the indices, we find 
\begin{multline} 
\frac{\D}{\D u}\left( \frac{\D \eta^\alpha}{\D u} +
\big\{\begin{smallmatrix}\alpha\\\beta\gamma\end{smallmatrix}\big\}\eta^\beta
\frac{\D x^\gamma}{\D u} \right) +
\big\{\begin{smallmatrix}\alpha\\\delta\epsilon\end{smallmatrix}\big\}
\left( \frac{\D \eta^\delta}{\D u} +
\big\{\begin{smallmatrix}\delta\\\beta\gamma\end{smallmatrix}\big\}\eta^\beta
\frac{\D x^\gamma}{\D u} \right)\frac{\D x^\epsilon}{\D u}+\\ \left(
\big\{\begin{smallmatrix}\alpha\\\gamma\delta\end{smallmatrix}\big\}_{,\beta}-
\big\{\begin{smallmatrix}\alpha\\\beta\gamma\end{smallmatrix}\big\}_{,\delta}+
\big\{\begin{smallmatrix}\epsilon\\\gamma\delta\end{smallmatrix}\big\}
\big\{\begin{smallmatrix}\alpha\\\beta\epsilon\end{smallmatrix}\big\}-
\big\{\begin{smallmatrix}\epsilon\\\beta\gamma\end{smallmatrix}\big\}
\big\{\begin{smallmatrix}\alpha\\\delta\epsilon\end{smallmatrix}\big\}
\right)\eta^\beta \frac{\D x^\gamma}{\D u}\frac{\D x^\delta}{\D
u}=0. 
\end{multline} 
In the first two parenthesized expressions, we recognize the absolute derivative of the geodesic separation and, in the last one, the Riemann tensor. The equation satisfied by the geodesic separation, which connects spacetime curvature with the way the two geodesics deviate from one another, can then be expressed in simpler terms as
\begin{equation} 
\frac{\delta^2\eta^\alpha}{\delta u^2}+R^\alpha_{\
\gamma\beta\delta}\eta^\beta \frac{\D x^\gamma}{\D u} \frac{\D
x^\delta}{\D u}=0. 
\end{equation}
The solution is completely determined when we know $\eta^\alpha$ and $\D
\eta^\alpha/\D u$ for some value of the parameter $u$. In flat
spacetime, where the curvature tensor is identically null, the
equation reduces to 
\begin{equation}
\frac{\D^2 \eta^\alpha}{\D u^2}=0.	
\end{equation}
Its solutions are $\eta^\alpha(u)= A^\alpha u +B^\alpha$, which means that in flat spacetime the separation vector between two geodesics changes linearly with $u$---the two geodesics are straight lines in spacetime and, therefore, converge or diverge linearly.

The separation between timelike spacetime geodesics gives us now the
possibility to recover the concept of \emph{rod distance} $\eta$ from
chronometric foundations. In relativity theory, the concept of
length between two particles is not  straightforward because, in 
Riemann-Einstein spacetime, the present---that is, when the length measurement should be made---is tied to each observer's worldline; in classical physics, on the other hand, the present is an absolute notion. This difficulty
notwithstanding, the rod length between two particles can be determined
for special cases in relativity theory too, as follows.

The geodesic separation $\eta^\alpha(s)$ is a spacelike vector,
whose size 
\begin{equation}
\eta(s) =: \D s_{\eta(s)}=\sqrt{g_{\alpha\beta}[x(s)]\eta^\alpha(s)\eta^\beta(s)}	
\end{equation}
depends on the value of the geodesic parameter $s$. This size, indicating the separation between two neighbouring free massive particles, is measured with a clock and a photon; showing the dependence on $s$ explicitly, it is 
\begin{equation}
\eta(s)=\D s_{s_P s_A} \D s_{s_A s_Q},	
\end{equation}
where $P$ and $Q$ are the events of photon emission and reception and $A$ marks the tail of $\eta^\alpha(s)$ (cf.\ Figure \ref{spacelike-separation} on page \pageref{spacelike-separation}). If $\eta(s)$ does not depend on $s$, the spacelike separation must be constant, 
\begin{equation}
\frac{\D \eta}{\D s}=0.	
\end{equation}
In that case, we can identify the two particles joined by $\eta^\alpha$ with the end-points of an idealized rigid measuring rod and $\eta$ with its length. In practice, this condition is obtained when all the photons sent out from the first particle and bouncing back from the second particle arrive at constant clock intervals. In this roundabout manner, the concept of rod length arises as a particular case of measurements of spacetime intervals with clocks and photons.

\section{The sources of gravitation}\label{Sources-of-gravitation}

\subsection{The vacuum}
We start the search for the different instances of Einstein's field equation, which connects spacetime curvature with sources in spacetime, in each case looking for guidance in Newton's classical theory of gravitation. At the outset with the case of the vacuum, Newtonian theory helps us realize that there is no \emph{straightforward} general-relativistic counterpart of the classical gravitational potential $\Phi(\vec r)$, but it also provides a useful analogical picture on which to base the search for the different instances of the general-relativistic field equation.

In the vacuum outside a spherically symmetric mass distribution, the Newtonian gravitational potential $\Phi(\vec r)$, connected with the universal force of gravitation according to $\vec \nabla \Phi(\vec r)=-\vec g(\vec r)$, satisfies Laplace's equation 
\begin{equation}
\nabla^2\Phi(\vec r)=0.	
\end{equation}
This follows from the fact that $\vec g(\vec r)$ is divergence-free, $\vec\nabla\cdot \vec g(\vec r)=0$, where $\vec g(\vec r)=-GM\hat r/r^2$. In Cartesian coordinates, Laplace's equation becomes 
\begin{equation}
\sum_{i=1}^3 \frac{\partial^2 \Phi(\vec r)}{(\partial x^i)^2}=0.	
\end{equation}

In special relativity, an equation corresponding to the classical Laplace equation is 
\begin{equation}
\square \phi(x)=:\eta^{\alpha\beta}\phi_{,\alpha\beta}(x)=0.	
\end{equation}
This equation has the form of the Klein-Gordon equation for a scalar field associated with a massless particle; expressed in the form of a wave equation, it is 
\begin{equation}
\sum_{i=1}^3 \frac{\partial^2 \phi(x)}{(\partial x^i)^2}-\frac{\partial^2 \phi(x)}{(\partial x^4)^2}=0.
\end{equation}

Could we now generalize Laplace's equation for the general-relativistic case to find the equation of gravitation in the vacuum? Should we perhaps, in analogy, take our equation to be $g^{\alpha\beta}\phi_{;\alpha\beta}(x)=0$? But what is now the meaning of field $\phi(x)$? Having come this far in the formulation of general relativity, we know it for a fact that its foundation is given by the quadratic form $\D s^2=\epsilon g_{\alpha\beta}(x)\D x^\alpha \D x^\beta$, so why should all of a sudden an extraneous field like $\phi(x)$ be needed? How about $g^{\alpha\beta}g_{\gamma\delta;\alpha\beta}(x)=0$ then? Not really, because the covariant derivative of the metric tensor is identically null, $g_{\gamma\delta;\alpha}(x)\equiv 0$, and so no real information can be extracted from such an equation. Once again, it seems that simply $g_{\alpha\beta}(x)$ is not enough to serve as the ``new gravitational field.''

As we suspected earlier, gravitation should instead be related to spacetime curvature, and we know that curvature is characterized by the Riemann tensor. But how should one take advantage of this knowledge to build the field equation in vacuum? If we say that ``spacetime curvature is null in vacuum,'' would this mean that spacetime is flat, and thus gravitation-free, in vacuum even when there is matter elsewhere and not necessarily very far away? Such an interpretation goes against our intuition, because gravitational effects  exist in the vacuum near massive objects; spacetime cannot therefore be flat there.

When we say, as we will, that spacetime curvature is null in vacuum, we take the meaning of this assertion in direct analogy with Newtonian theory, by identifying a null curvature tensor with the null Laplacian of the Newtonian potential. And so, just like in classical physics the null Laplacian can give the solution to the non-null gravitational potential \emph{outside} a material distribution, so can a null curvature tensor give the non-null curvature solution \emph{outside} a material distribution. Only when the vacuum extends thoroughly everywhere are gravitational effects nonexistent and, in relativistic physics, we can say that spacetime is flat.

So far we have developed several versions of the curvature tensor. Which of the following five alternative field equations: (i) $R_{\alpha\beta\gamma\delta}(x)=0$, (ii) $R^\delta_{\ \alpha\beta\gamma}(x)=0$, (iii) $R_{\alpha\beta}(x)=0$, 
(iv) $R(x)=0$, and (v) $G_{\alpha\beta}(x)=0$ should we now choose as the analogue of $\nabla^2\Phi(\vec r)=0$? The field equation we search should be capable of unequivocally
determining the 10 independent components $g_{\alpha\beta}$ of the
metric tensor, and so we can narrow down the possibilities to
expressions (iii) and (v)---a null Ricci or Einstein tensor,
respectively---the only ones suitable from this perspective, because only they have no more nor less than 10 independent components. But which of the two should we choose?

For the purposes of this and the next two sections, that is, for the
study of the vacuum, which of the two tensors we choose is irrelevant, because in the vacuum, as we see below, the Einstein tensor reduces to the Ricci tensor. However, when we later on consider solutions inside a distribution of matter (i.e.\ matter as an explicit source), only the choice of the Einstein tensor will be suitable, because only it is capable of leading us to a law for the local conservation of the momentum and energy of the matter source, i.e.\ to the generalization of the classical law of
continuity of matter. This is made possible by the extra mathematical property of Eq.~(\ref{Einstein-divergence-free}) seen earlier, namely, that the Einstein tensor is divergence-free (in the covariant sense). When, after establishing a connection between curvature and matter, we demand that the matter source be divergence-free for conservation purposes, this mathematical property of $G^{\alpha\beta}$ becomes physically relevant.

In general, we should then choose the field equation in vacuum to be
\begin{equation} 
G_{\alpha\beta}(x)=0. 
\end{equation} 
In practice, however, the curvature scalar $R$ is null in vacuum,\footnote{Express $G_{\alpha\beta}(x)=0$ in explicit form and contract its indices with the metric tensor to find $-R=0$.} and in that case we can use $R_{\alpha\beta}(x)=0$ for simplicity.

The simplest solution to the field equation in vacuum is obtained when the vacuum extends everywhere. Then the curvature tensor is identically null, $R_{\alpha\beta}(x)\equiv 0$, and coordinates can be found such that the solution to the field equation is $g_{\alpha\beta}(x)=\eta_{\alpha\beta}$, that is, flat spacetime.

Another exact solution to the spacetime metric can be found outside a spherically symmetric matter distribution of mass $M$. This solution was found by Karl Schwarzschild in 1916. Considerations of spherical symmetry lead to constraints in the possible form that the metric field can take. Using spherical coordinates, the metric can only be of the form 
\begin{equation}
g_{\alpha\beta}(x)=\mathrm{diag}\left( e^{\lambda(x^1)},(x^1)^2, (x^1)^2\sin^2(x^2),-e^{\nu(x^1)}\right),	
\end{equation}
with asymptotic flatness at infinity, i.e.\ with boundary conditions $\lambda(x^1)\rightarrow 0$ and $\nu(x^1)\rightarrow 0$ when $x^1\rightarrow \infty$.

The calculation of the 13 non-null Christoffel symbols gives
\begin{align}
\big\{\begin{smallmatrix}1\\11\end{smallmatrix}\big\}&=\frac{1}{2}\lambda', &\qquad \big\{\begin{smallmatrix}1\\22\end{smallmatrix}\big\}&=-x^1e^{-\lambda},\nonumber\\
\big\{\begin{smallmatrix}1\\33\end{smallmatrix}\big\}&=-x^1\sin^2(x^2)e^{-\lambda}, &\qquad \big\{\begin{smallmatrix}1\\44\end{smallmatrix}\big\}&=\frac{1}{2}e^{\nu-\lambda}\nu', \nonumber\\
\big\{\begin{smallmatrix}2\\12\end{smallmatrix}\big\}&=\big\{\begin{smallmatrix}2\\21\end{smallmatrix}\big\}=\frac{1}{x^1}, & \qquad 
\big\{\begin{smallmatrix}2\\33\end{smallmatrix}\big\}&=-\sin(x^2)\cos(x^2),\\
\big\{\begin{smallmatrix}3\\13\end{smallmatrix}\big\}&=\big\{\begin{smallmatrix}3\\31\end{smallmatrix}\big\}=\frac{1}{x^1}, \qquad &
\big\{\begin{smallmatrix}3\\23\end{smallmatrix}\big\}&=\big\{\begin{smallmatrix}3\\32\end{smallmatrix}\big\}=\cot(x^2),\nonumber\\
\big\{\begin{smallmatrix}4\\14\end{smallmatrix}\big\}& =\big\{\begin{smallmatrix}4\\41\end{smallmatrix}\big\}=\frac{1}{2}\nu'. && \nonumber
\end{align}
As a result, the Ricci tensor is diagonal, with components given by
\begin{align}
R_{11}&=\frac{1}{2}\left( \nu''-\frac{1}{2}\lambda'\nu'+\frac{1}{2}\nu'^2-\frac{2}{x^1}\lambda' \right),\\
R_{22}&=-1+e^{-\lambda}\left[1+\frac{1}{2}x^1(\nu'-\lambda') \right],\\
R_{33}&=\sin^2(x^2)R_{22},\\
R_{44}&=-\frac{1}{2}e^{\nu-\lambda}\left( \nu''-\frac{1}{2}\lambda'\nu'+\frac{1}{2}\nu'^2+\frac{2}{x^1}\nu' \right).
\end{align}
To solve the system of four equations $R_{\alpha\alpha}=0$, we consider first
$R_{11}=0$ and $R_{44}=0$ together with the condition of asymptotic flatness. We find straightforwardly that $\nu(x^1)=-\lambda(x^1)$. Next we use this result to simplify and solve $R_{22}=0$. Grouping the variables $x^1$ and $\lambda$ with their respective differentials and integrating once, we find
\begin{equation}
e^\lambda=\left(1-\frac{C}{x^1} \right)^{-1}
\end{equation}
and, therefore,
\begin{equation}
e^\nu=1-\frac{C}{x^1}.
\end{equation}

The value of the integration constant $C$ is found by comparing with the Newtonian limit (expressed in natural units). Because\footnote{See next section, Eq.~(\ref{g44}).} 
\begin{equation}
-\left(1-\frac{C}{x^1}\right)=g_{44}\overset{\mathrm{c}}{=}-(1+2\Phi)=-\left(1-\frac{2GM}{c^2r}\right), 
\end{equation}
then $C=2GM/c^2$, where $x^1$ can be identified with the Euclidean distance $r$ only when $x^1 \rightarrow \infty$. The quadratic form in the vacuum outside a spherical matter distribution then is 
\begin{multline}
\D s^2=\epsilon\Bigg\{ \left(1-\frac{2GM}{c^2x^1}\right)^{-1}(\D x^1)^2+
(x^1)^2[(\D x^2)^2+\sin^2(x^2)(\D x^3)^2]-\\\left(1-\frac{2GM}{c^2x^1}\right)(\D x^4)^2 \Bigg\},
\end{multline}
where $0\leq x^2\leq \pi$, $0\leq x^3\leq 2\pi$, and $-\infty< x^4< \infty$. This solution is valid either for (i) $x^1_B<x^1<\infty$, where $x^1_B$ is the radius of the material edge of the matter distribution or (ii) $x^1_S<x^1<\infty$, where $x^1_S=2GM/c^2$ is the Schwarzschild radius of the matter distribution, depending on which of the two values is reached first as we approach from infinity. If $x^1_S=2GM/c^2$ is reached first, the spacetime region in question is called a (non-rotating) \emph{black hole}.

\subsection{Tremors in the vacuum}
If not in the general way described (and rejected) at the beginning of the previous section, do the classical gravitational potential $\Phi(\vec r)$ and the Laplace equation $\nabla^2\Phi(\vec r)=0$ have some kind of relativistic counterpart in some special case? 

To find out, we consider physically weak tremors (from the spacetime perspective) induced by a small deviation from metric flatness, 
\begin{equation}\label{tremors}
g_{\alpha\beta}(x)=\eta_{\alpha\beta}+\phi_{\alpha\beta},	
\end{equation}
where all $\phi_{\alpha\beta}$ and their derivatives are very small. In this case, it is accurate enough to solve the equation $R_{\alpha\beta}(x)=0$ to first order in $\phi_{\alpha\beta}$. What kind of weak ``gravitational field'' $\phi_{\alpha\beta}(x)$ results from this approximation?

To first order ($\sim$) in $\phi_{\alpha\beta}$, we have that the inverse metric tensor is 
\begin{equation}\label{tremors-inverse}
g^{\alpha\beta}(x)=\eta^{\alpha\beta}-\phi^{\alpha\beta}	
\end{equation}
because 
\begin{equation}
g^{\alpha\beta}g_{\beta\gamma} \sim (\eta^{\alpha\beta}-\phi^{\alpha\beta})(\eta_{\beta\gamma}-\phi_{\beta\gamma})=\delta^\alpha_{\ \gamma}+\phi^\alpha_{\ \gamma}-\phi^\alpha_{\ \gamma}=\delta^\alpha_{\ \gamma}.	
\end{equation}
In the following, the metric tensor $g_{\alpha\beta}$ is the only tensor whose indices will be raised and lowered with itself; for any other field, we use the flat-spacetime metric $\eta_{\alpha\beta}$ or $\eta^{\alpha\beta}$.

The next step is to find the first-order approximation to the Ricci tensor. From expression (\ref{curvature-tensor}) of the mixed Riemann tensor, we know that the exact Ricci tensor is
\begin{equation}\label{Ricci-tensor}
R_{\alpha\beta}=
\big\{\begin{smallmatrix}\gamma\\\alpha\gamma\end{smallmatrix}\big\}_{,\beta}-
\big\{\begin{smallmatrix}\gamma\\\alpha\beta\end{smallmatrix}\big\}_{,\gamma}+
\big\{\begin{smallmatrix}\epsilon\\\alpha\gamma\end{smallmatrix}\big\}
\big\{\begin{smallmatrix}\gamma\\\epsilon\beta\end{smallmatrix}\big\}-
\big\{\begin{smallmatrix}\epsilon\\\alpha\beta\end{smallmatrix}\big\}
\big\{\begin{smallmatrix}\gamma\\\epsilon\gamma\end{smallmatrix}\big\}.
\end{equation}
Because to first order the Christoffel symbol is
\begin{equation}
\big\{\begin{smallmatrix}\gamma\\\alpha\beta\end{smallmatrix}\big\}\sim
\frac{1}{2}\eta^{\gamma\delta}(\phi_{\alpha\delta,\beta}+\phi_{\beta\delta,\alpha}-\phi_{\alpha\beta,\delta}),
\end{equation}
we then have 
\begin{equation}
\big\{\begin{smallmatrix}\gamma\\\alpha\beta\end{smallmatrix}\big\}_{,\gamma}\sim
\frac{1}{2}(\phi_{\alpha\ ,\beta\gamma}^{\ \gamma}+\phi_{\beta\ ,\alpha\gamma}^{\ \gamma}-\square \phi_{\alpha\beta}) 
\end{equation}
and
\begin{equation}
\big\{\begin{smallmatrix}\gamma\\\alpha\gamma\end{smallmatrix}\big\}_{,\beta} \sim \frac{1}{2} \phi_{,\alpha\beta}.
\end{equation}
To first order, disregarding products of $\phi_{\alpha\beta}$ or its derivatives, the Ricci tensor is
\begin{equation}
R_{\alpha\beta}\sim
\big\{\begin{smallmatrix}\gamma\\\alpha\gamma\end{smallmatrix}\big\}_{,\beta}-
\big\{\begin{smallmatrix}\gamma\\\alpha\beta\end{smallmatrix}\big\}_{,\gamma}\sim
\frac{1}{2}(\phi_{,\alpha\beta}-\phi_{\alpha\ ,\beta\gamma}^{\ \gamma}-\phi_{\beta\ ,\alpha\gamma}^{\ \gamma}+\square \phi_{\alpha\beta}).
\end{equation}
Because we find ourselves outside any material distribution, we set \mbox{$R_{\alpha\beta}=0$} to get
\begin{equation}
\square \phi_{\alpha\beta}+\phi_{,\alpha\beta}-\phi_{\alpha\ ,\beta\gamma}^{\ \gamma}-\phi_{\beta\ ,\alpha\gamma}^{\ \gamma}=0.\footnote{Taking advantage of the gauge freedom of this equation, it can be recast simply as $\square \phi_{\alpha\beta}=0$ using the gauge $\chi_{\alpha\ ,\gamma}^{\ \gamma}=0$, where $\chi_{\alpha\beta}=\phi_{\alpha\beta}-\eta_{\alpha\beta}\phi/2$. In some more detail, the linearized equation is invariant with respect to the gauge transformation $\phi_{\alpha\beta} \rightarrow \tilde\phi_{\alpha\beta}=\phi_{\alpha\beta}+(\lambda_{\alpha,\beta}+\lambda_{\beta,\alpha})/2$, where $\lambda_\alpha(x)$ is an arbitrary vector field. When $\square \lambda_{\alpha}=2\chi_{\alpha\ \beta}^{\ \beta}$, the gauge condition is satisfied (i.e.\ $\tilde\chi_{\alpha\ ,\gamma}^{\ \gamma}=0$) and the linearized equation simplified with respect to the new field $\tilde\phi_{\alpha\beta}$. This is the starting point of the study of gravitational waves.}
\end{equation}

This is the linearized general-relativistic equation for a perturbation $\phi_{\alpha\beta}(x)$ of the metric field. Its form coincides with the Bargmann-Wigner equation for a field in flat spacetime associated with a massless spin-2 particle. On account of its formal correspondence with the linearized general-relativistic equation for the metric perturbation $\phi_{\alpha\beta}(x)$, this particle is sometimes called a ``graviton.'' However, the correspondence appears to be only formal, because the Bargmann-Wigner field equation was derived for a field in Minkowski spacetime, which is flat and has therefore nothing to do with gravitation. As hard as one might try to establish the resemblance between a ``graviton'' and other quantum particles, a ``graviton'' does not stand on the same conceptual footing as quantum particles, because the field associated with it is not an \emph{ordinary field on} flat spacetime but is itself (part of) the \emph{metric field of} spacetime.

To see what $\phi_{\alpha\beta}(x)$ may have to do with the Newtonian potential $\Phi(\vec r)$, we start by considering a ``slow particle'' such that the components $V^a$ ($a=1,2,3$) of its four-velocity $V^\alpha=\D x^\alpha/\D s$ are very small and, therefore, $g_{\alpha\beta}V^\alpha V^\beta=-1$ becomes 
\begin{equation}
g_{44}(V_4)^2 \sim -1,	
\end{equation}
from where 
\begin{equation}\label{V4}
V^4\sim \frac{1}{\sqrt{-g_{44}}}.	
\end{equation}
When the slow-particle condition is applied to the geodesic equation for $\alpha=a$, we find 
\begin{equation}
\frac{\D V^a}{\D s}\sim -\big\{\begin{smallmatrix}a\\44\end{smallmatrix}\big\}(V^4)^2. 	
\end{equation}
But because 
\begin{equation}
\frac{\D V^a}{\D s}=\frac{\D V^a}{\D x^\alpha} V^\alpha \sim \frac{\D V^a}{\D x^4}V^4, 	
\end{equation}
we get 
\begin{equation}
\frac{\D V^a}{\D x^4}\sim -\big\{\begin{smallmatrix}a\\44\end{smallmatrix}\big\}V^4. 
\end{equation}
The left-hand side of this equation is comparable with the Newtonian acceleration $\D v_a/\D t$.

Assuming next that the full spacetime metric is ``quasistatic,'' i.e.\ that $g_{\alpha\beta,4}$ is much smaller than $g_{\alpha\beta,a}$, we get
\begin{equation}
\big\{\begin{smallmatrix}a\\44\end{smallmatrix}\big\} \sim -\frac{1}{2}g^{a\alpha}\frac{\partial g_{44}}{\partial x^\alpha} \sim -\frac{1}{2}g^{ab} \frac{\partial g_{44}}{\partial x^b}.	
\end{equation}
Using this result, we find 
\begin{equation}
\frac{\D V^a}{\D x^4}\sim  \frac{1}{2}g^{ab} \frac{\partial g_{44}}{\partial  x^b} V^4.	
\end{equation}
Using Eq.~(\ref{V4}), we come to the expression 
\begin{equation}
\frac{\D V^a}{\D x^4} \sim -g^{ab} \frac{\partial \sqrt{-g_{44}}}{\partial x^b}.	
\end{equation}

Finally, because we are dealing only with tremors in the vacuum, 
\begin{equation}
g^{44}=1-\phi^{44}	
\end{equation}
on account of Eq.~(\ref{tremors-inverse}). The counterpart of the Newtonian acceleration thus becomes 
\begin{equation}
\frac{\D V^a}{\D x^4} \sim -(\eta^{ab}-\phi^{ab}) \frac{\partial (1-\phi_{44}/2)}{\partial x^b} \sim \frac{1}{2} \eta^{ab} \frac{\partial\phi_{44}}{\partial x^b};	
\end{equation}
that is
\begin{equation}
\frac{\D V_a}{\D x^4}\sim \frac{1}{2}\frac{\partial \phi_{44}}{\partial x^a}.
\end{equation}
Comparing this expression with the Newtonian acceleration \begin{equation}
	\frac{\D v_a}{\D t}=g_a(\vec r)=-\frac{\partial \Phi(\vec r)}{\partial x^a},
\end{equation} 
we discover the correspondences (valid only in the context of the approximations made)
\begin{equation}
\Phi \overset{\mathrm{c}}{=} -\frac{1}{2}\phi_{44} 
\end{equation}
and
\begin{equation}\label{g44}
	g_{44} \overset{\mathrm{c}}{=} -(1+2\Phi).
\end{equation}

A rough estimate of how $\Phi$ compares with unity in Eq.~(\ref{g44}) at the surface of the earth, where $\Phi(r)=-GM/c^2r$,\footnote{In natural units, as used so far, $\Phi$ is a dimensionless number. On the other hand, $-GM/r$ has SI units $L^2T^{-2}$. To convert a quantity from SI units to natural units ($c=1$), we must divide it by $c^2$ as we just did; to convert a quantity from natural units to SI units, we must multiply it by $c^2$.} gives 
\begin{equation}
	\Phi \sim -10^{-9} 
	\qquad \mathrm{and} \qquad
	\frac{\partial \Phi}{\partial x^a} \sim 10^{-16}\ \mathrm{m}^{-1}.
\end{equation} 
From this, the acceleration at the surface of the earth turns out to be the familiar result
\begin{equation}
c^2 \frac{\partial \Phi}{\partial x^a} = 10\ \mathrm{ms}^{-2}.
\end{equation}
The relative size of the gravitational tremors with respect to the flat-spacetime background is comparable with ripples rising 1 centimetre above the surface of an (unearthly) lake 10,000 kilometres in depth.

We find, moreover, that in this correspondence the Laplacian $\nabla^2 \Phi$ of the Newtonian potential is the 44-component of the Ricci tensor,
\begin{multline}
R_{44}\sim \frac{1}{2}(\phi_{,44}-2\phi_{4\ ,4\gamma}^{\ \gamma}+\square \phi_{44})\sim
\frac{1}{2}\eta^{\alpha\beta}\phi_{44,\alpha\beta}\sim \frac{1}{2}\eta^{ab}\phi_{44,ab}=\\
\frac{1}{2}\sum_{a=1}^3 \phi_{44,aa}=\frac{1}{2}\nabla^2\phi_{44}\overset{\mathrm{c}}{=}-\nabla^2\Phi.
\end{multline}
Thus, the Laplace equation $\nabla^2\Phi=0$ for the Newtonian potential in vacuum corresponds to $R_{44}=0$.

The special role of coordinate $x^4$ in the way the correspondence with classical physics was established deserves special mention. Early on, $x^4$ played a crucial part in the local characterization of the past, the present, and the future in relativity theory. Now it seems that, in this approximation, $x^4$ has become the absolute time $t$ of classical physics. The likelihood of this conclusion is strengthened by a look at Einstein's \citeyear{Einstein:1952b} original paper, where, on making this approximation, he writes that ``we have set $\D s=\D x_4=\D t$'' (p.~158). But is this true \emph{to first order in} $\Phi$? It is not, because we have \mbox{$\Delta s=\Delta t$} in the Newtonian case, but
\begin{equation}
	\frac{\D x^4}{\D s}=V^4 \sim \frac{1}{\sqrt{1-\phi_{44}}} \sim 1+\frac{1}{2}\phi_{44} \overset{\mathrm{c}}{=} 1-\Phi \neq 1.
\end{equation} 
We conclude that the identification of the fourth coordinate $x^4$ with absolute time $t$ is not possible even in the relativistic approximation to Newtonian physics. If, in spite of this, the association $x^4=ct$ still wants to be made, one must remember that in this case $t$ is just a notation---the misleading name of a spacetime coordinate, not absolute time.

\subsection{The cosmic vacuum}
So far we have studied the field equations $G_{\alpha\beta}=0$ or $R_{\alpha\beta}=0$, where the constant $\Lambda$ is null. In neglecting this constant, we have assumed that $|\Lambda|$ is small enough, but now we would like to make sure that this is correct by setting an upper bound for $|\Lambda|$. 

We start by noting that, because
\begin{equation}
	g^{\alpha\beta}\mathcal{G}_{\alpha\beta}=R-2R+4\Lambda=0, 
\end{equation}
then $R=4\Lambda$, and the field equation reduces to 
\begin{equation}\label{Einstein-lambda}
	R_{\alpha\beta}-\Lambda g_{\alpha\beta}=0. 
\end{equation}
Using again the correspondence $R_{44}\overset{\mathrm{c}}{=}-\nabla^2\Phi$ in the context of the Newtonian approximation, to first order in $\Lambda$ we have 
\begin{equation}
	\Lambda g_{44}\overset{\mathrm{c}}{=}-\Lambda(1+2\Phi)\sim -\Lambda.
\end{equation} 
If we now take $\Lambda g_{44}$ \emph{as if} it was a source of curvature (instead of being a curvature term itself), we can write the Poisson equation 
\begin{equation}\label{Poisson-lambda}
	\nabla^2\Phi=\frac{4\pi G\rho}{c^2}+\Lambda 
\end{equation}
for the Newtonian potential \emph{inside} a medium of density $\rho+\Lambda c^2/4\pi G$ (expressed in SI units). Here $\rho$ stands for the ordinary density of matter, while $\Lambda c^2/4\pi G$ stands for the mass density of a new source of gravitation pervading all space in the manner of an ubiquitous ether, present also inside matter. 

We see from Eq.~(\ref{Poisson-lambda}) that, if the classical dynamics inside a system of density $\rho$ is well described by Newtonian theory, then $|\Lambda|\ll 4\pi G \rho/c^2$, obtaining an upper bound for this new constant. The more dilute a system inside of which Newtonian theory holds, the more stringent the upper bound set on $|\Lambda|$ is and, consequently, the smaller it must be. 

For a conservative estimate, we take our galaxy as a system inside which Newtonian mechanics can be applied successfully. Given that the average density of the Milky Way is approximately $\rho= 1.7 \times 10^{-21}\ \mathrm{kg/m}^3$,\footnote{Expressed differently, $\rho= 10^6 \times 1.7 \times 10^{-27}\ \mathrm{kg/m}^3$, that is, $10^6$ hydrogen-atom masses per cubic meter. This is one million times the density of the intergalactic ``vacuum.''} we find, $|\Lambda|\ll 1.4 \times 10^{-30}\ \mathrm{s}^{-2}$. To express this bound in a more familiar way in terms of natural units, we divide by $c^2$ to get $|\Lambda|\ll 1.6 \times 10^{-47} \mathrm{m}^{-2}$.  

It has been conjectured by other methods that a more accurate but less conservative estimate is $|\Lambda|\leq 10^{-35}\ \mathrm{s}^{-2}$ or, more commonly in natural units, $|\Lambda|\leq 10^{-52}\ \mathrm{m}^{-2}$. On account of Eq.~(\ref{Poisson-lambda}), the gravitational effect of a constant $\Lambda$ of this magnitude would begin to be felt inside a material medium of density $\rho\leq 1.7 \times 10^{-26}\ \mathrm{kg/m}^3$, i.e.\ 10 hydrogen-atom masses per cubic metre. This value corresponds to the density of the \emph{intergalactic medium}. Thus, in order for $\Lambda$ to have non-negligible effects, we must consider cosmological distances---hence its name, the \emph{cosmological constant}.

To see what kind of gravitational effects are caused by the cosmological constant---whose so far enigmatic source, we recall, pervades all space---in the absence of any other material medium (i.e.\ $\rho=0$), we consider a point-like\footnote{Or spherical, in which case we are only interested to find the gravitational potential in the vacuum outside it.} mass $M$ surrounded by a perfect vacuum in the ordinary sense of the word. In the Newtonian approximation under consideration, we should then solve Eq.~(\ref{Poisson-lambda}).

Since $\Lambda$ is constant and the central mass point-like (or spherical), we solve for $\Phi(r)$ in a spherically symmetric context:
\begin{equation}
\frac{1}{r^2}\frac{\D}{\D r}\left( r^2\frac{\D \Phi}{\D r} \right)=\Lambda.
\end{equation}
Integrating twice and absorbing one constant into the potential, we find
\begin{equation}
\Phi(r)=-\frac{C'}{r}+\frac{1}{6}\Lambda r^2,
\end{equation}
which leads to a gravitational force on a particle of mass $m$ of the type
\begin{equation}
\vec F_g=-m\frac{\D \Phi}{\D r}\hat r=-m\left( \frac{C'}{r^2}+\frac{1}{3}\Lambda r\right) \hat r,
\end{equation}
where $C'=GM/c^2$. If $\Lambda$ is positive, it generates an extra attractive gravitational force between the central mass and a test mass. More interestingly, if it is negative, it generates an extra repulsive gravitational force between the two masses. In both cases, the extra force increases linearly as we recede from the central mass. For negative $\Lambda$, there exists then a distance at which the attraction from the centre is balanced by the repulsion caused by the so far enigmatic source of the cosmological constant. Given the upper bound $|\Lambda|\leq 10^{-52}\ \mathrm{m}^{-2}$, we can get a feel for the distance involved when the central mass is the sun ($M\approx 2\times 10^{30}\ \mathrm{kg}$). We find that, in an otherwise perfect universal vacuum, the cosmological constant is so small that its repulsive force cancels out the attraction of the sun at a distance 
\begin{equation}
r=\sqrt[3]{\frac{3GM}{\Lambda c^2}}\approx 3.5 \times 10^{18}\ \mathrm{m} \approx 375\ \mathrm{ly}. 	
\end{equation}
The value of this result is only pedagogical. Nothing near the perfect universal vacuum here assumed exists either in the solar system or in the galaxy. We should go as far out as intergalactic space to find a medium of low enough density that the cosmological constant can exert its influence at all. This is a distance of millions or tens of millions of light-years---four to five orders of magnitude greater than the mere 375 light-years just obtained.

Leaving classical theory behind, we can now consider $\Lambda$ and find an exact solution to the generalized Einstein field equation $\mathcal{G}_{\alpha\beta}=0$, which, outside an ordinary material distribution, we saw reduces to Eq.~(\ref{Einstein-lambda}). Following essentially the same method as for the Schwarzschild solution with the added extra source terms,\footnote{To solve the resulting differential equation by integration, rewrite $(1+x^1 \nu')e^\nu$ as the derivative of a product.} we get
\begin{equation}
e^\nu=e^{-\lambda}=\left[1-\frac{C}{x^1}+\frac{1}{3}\Lambda (x^1)^2\right].
\end{equation}
To consider a cosmos everywhere void of matter, we set $C=0$, since the origin of this constant is a localized matter distribution. As a result, we get
\begin{multline}\label{de-Sitter}
\D s^2=\epsilon \Bigg\{ \left[1+\frac{1}{3}\Lambda (x^1)^2\right]^{-1}(\D x^1)^2+
(x^1)^2[(\D x^2)^2+\sin^2(x^2)(\D x^3)^2]-\\\left[ 1+\frac{1}{3}\Lambda (x^1)^2\right](\D x^4)^2 \Bigg\}.
\end{multline}
This solution was found by Willem de Sitter in 1917. For $\Lambda < 0$, the value $x^1_H=\sqrt{-3/\Lambda}$ characterizes the radius of the \emph{event horizon} of the spacetime in question, beyond which no information can reach an observer.

When all is said and done, what are we to make of the cosmological constant? What physical meaning are we to attach to it? When the generalized field equation is written in the form $R_{\alpha\beta}=\Lambda g_{\alpha\beta}$, one gets the impression that the cosmological constant is a material source of spacetime curvature. But since it certainly is not related to matter, we appear to be confronted by a mysterious kind of stuff that is all-pervading and whose origin we know nothing of, but which has active physical effects---a demiether. Evoking this sense of wonder, the presumed physical stuff behind the cosmological constant has been named ``dark energy.'' But is in question here a physical mystery or only a mysterious physical interpretation? The power of psychological inducement perhaps deserves more attention that is usually paid to it. If the generalized field equation is instead simply written in the form $R_{\alpha\beta}-\Lambda g_{\alpha\beta}=0$, the impression vanishes, and one is free to see the cosmological constant simply as an \emph{extra term of spacetime curvature} possibly relevant in a cosmic near-vacuum. \label{lambda-solution}

In the context of present theories, consideration of a negative cosmological constant is deemed justified by the observed cosmological redshift of distant galaxies. If this interpretation of the cosmological redshift stood the test of time in the light of hypothetical better cosmological theories to come, and the need for a cosmological constant became thus better secured, we could then, in the light of the previous paragraph, regard it as nothing more than a \emph{new fundamental constant} related to the curvature of spacetime. If we do not worry  and try to explain the origin of the gravitational constant $G$, the speed of light $c$, and Planck's constant $h$, why should we any more worry about the origin of the \emph{curvature constant} $\Lambda$?

\subsection{The dust-filled cosmos}

The next and final step is to generalize the field equation for the
case of a non-empty cosmos. Now, instead of the homogeneous field
equation $G_{\alpha\beta}(x)=0$ in vacuum, we expect an
inhomogeneous equation with a source term characterizing the matter
distribution in analogy with the classical Poisson equation
\begin{equation}
      \nabla^2\Phi=4\pi G\rho.
\end{equation} 
The source term is usually expressed as
$\kappa  T_{\alpha\beta}$, where $\kappa $ is a constant. Further,
because $G_{\alpha\beta}$ is symmetric and divergence-free (a choice based on a mathematical property so far without physical relevance), $T_{\alpha\beta}$ should be so too.

A simple case is that of a cosmos filled with matter resembling an
ideal fluid whose pressure is null (or approximately null) and its constituent
particles are free, interacting with each other only gravitationally. This kind of fluid is usually called dust. When we identify the fluid particles with galaxies, we obtain a useful application of this idea for the study of the cosmos as a whole. Since we are dealing with a continuous material distribution, instead of considering its four-momentum \mbox{$P^\alpha(x)=mV^\alpha(x)$}, it is more appropriate to characterize this fluid by means of its invariant \emph{matter density} $\rho(x)$ and four-velocity $V^\alpha(x)$. Together they form a field $\rho(x)V^\alpha(x)$, called the \emph{four-momentum-density field}, the natural counterpart of the four-momentum field $mV^\alpha(x)$.

However, $\rho V^\alpha$ is not yet a viable choice for the source
term $T^{\alpha\beta}$ we seek, because this should be a second-order tensor. Since we are looking for a symmetric second-order tensor and the only fields
characterizing the ideal pressureless fluid are its density and its
four-velocity, we do not seem to have any other options but to
propose that 
\begin{equation}
	T^{\alpha\beta}=\rho V^\alpha V^\beta
\end{equation}
is the source term we are looking for. If it is a bad choice, it should
next lead us to physically unpleasant results.

Because $T^{\alpha\beta}$ should be divergence-free, we have
\begin{equation}\label{divergence-free-matter} 
T^{\alpha\beta}_{\ \ \ ;\beta}= (\rho V^\beta)_{;\beta}V^\alpha+ \rho V^\beta
V^\alpha_{\ \ ;\beta}=0. 
\end{equation} 
Multiplying by $V_\alpha$ on both sides, we get 
\begin{equation} 
(\rho V^\beta)_{;\beta}V_\alpha V^\alpha+\rho V^\beta V_\alpha V^\alpha_{\ \ ;\beta}=0. \end{equation}
The second term is null\footnote{Writing
$V_\alpha V^\alpha_{\ \ ;\beta}=(V_\alpha V^\alpha)_{;\beta}-V^\alpha
V_{\alpha;\beta}$, we find that the first term is null because
$V^\alpha V_\alpha=-1$ and that the second is
$-g^{\alpha\gamma}V_\gamma V_{\alpha;\beta}=-V_\gamma V^\gamma_{\ \ ;\beta}$. Therefore, $V_\alpha V^\alpha_{\ \ ;\beta}=0$.} and, because
$V^\alpha V_\alpha=-1$, the equation reduces to 
\begin{equation}
(\rho V^\beta)_{;\beta}=0. 
\end{equation} 
This is a conservation equation for the four-momentum density, and it is the
general-relativistic generalization of the classical continuity equation 
\begin{equation}
	\vec\nabla\cdot (\rho\vec v)+\frac{\partial \rho}{\partial t}=0, 
\end{equation}
expressing the local conservation of matter. This is a physically pleasant result. Furthermore, it justifies the choice of $G_{\alpha\beta}$ over $R_{\alpha\beta}$ from a physical point of view.

Returning to Eq.~(\ref{divergence-free-matter}), since the first term is null, it follows that so must be the second term. Because neither $\rho$ nor $V^\beta$ are in general null, then 
\begin{equation}
	V^\alpha_{\ \ ;\beta}=0. 
\end{equation}
In words, the \emph{free particles} composing the ideal pressureless fluid follow spacetime geodesics. And this is as it should be---another physically pleasant result. Our presumption about the form of $T^{\alpha\beta}$ is now confirmed.

In this ideal-fluid picture, the field equation is then
\begin{equation}
	R^{\alpha\beta}-\frac{1}{2}g^{\alpha\beta}R=\kappa \rho V^\alpha V^\beta.
\end{equation}
Contracting the indices with $g_{\alpha\beta}$, we find that
$R=\kappa \rho$. The field equation is thus simplified to
\begin{equation}
	R^{\alpha\beta}=\kappa \rho\left[V^\alpha V^\beta+\frac{1}{2}g^{\alpha\beta}\right].
\end{equation}
To determine $\kappa $ we resort again to the correspondence
with Newtonian theory. The 44-component of the Ricci
tensor gives 
\begin{equation}
	-\nabla^2\Phi \overset{\mathrm{c}}{=}R^{44}=\kappa \rho \left[(V^4)^2+\frac{1}{2}g^{44}\right]. 
\end{equation}
To the lowest non-trivial order in $\kappa \rho$, $(V^4)^2\sim
1$ and $g^{44}\sim -1$, and so 
\begin{equation}
	\nabla^2\Phi=-\frac{1}{2}\kappa \rho.
\end{equation}
Comparison with Poisson's equation $\nabla^2\Phi=4\pi G\rho$ leads
finally to the result\footnote{In SI units, the constant of proportionality is $\kappa =-8\pi G/c^4$.} 
\begin{equation}
	\kappa =-8\pi G.
\end{equation} 
In the pressureless-fluid picture here considered, the field equation finally becomes
\begin{equation} 
G^{\alpha\beta}=-8\pi G\rho V^\alpha V^\beta.
\end{equation}

In his original paper, Einstein \citeyear{Einstein:1952b} holds that the general expression of this equation, namely,
\begin{equation}
G^{\alpha\beta}=-8\pi G T^{\alpha\beta}
\end{equation}
is indeed the field equation describing ``the influence of the gravitational field on all processes, without our having to introduce any new hypothesis whatever'' (pp.~151--152). This is to say that the equation describes the gravitational interaction of \emph{all} fields related to matter and radiation (velocity field, mass-density field, pressure field, electromagnetic field, etc.), so long as they are described by a matter-radiation tensor $T_{\alpha\beta}$ that is divergence-free. Finally, when we consider the cosmological constant as well, the most general field equation for gravitational interactions is
\begin{equation}
\mathcal{G}^{\alpha\beta}=-8\pi G T^{\alpha\beta}.
\end{equation}
The curvature of spacetime becomes in this way connected to the distribution of matter and radiation in it.

\subsection{The problem of cosmology}\label{Problem-cosmology}
``The problem of cosmology,'' said Einstein \citeyear{Einstein:1982} in his Kyoto lecture, ``is related to the geometry of the universe and to time'' (p.~47). We finish this chapter with a brief analysis of the insight contained in this deceptively simple statement.

In order for the study of the cosmos as a whole to be practically possible, standard cosmology is forced to make several assumptions. Firstly, the assumption is made that galaxies, conceived as material particles, are in free fall, interacting with each other only gravitationally (cf.\ previous section). Next follows a much less innocuous supposition, according to which cosmic space is locally isotropic, i.e.\ looks the same in all directions from anywhere at any one time. Against this supposition, it might be argued that an approximation in which the cosmos is materially smooth is too unrealistic, and that, moreover, the cosmos may only look approximately smooth from our own vantage point, which is both severely restricted spatially to the Milky Way and temporally to a few thousand years of astronomical observations. Martin \citeyear{Martin:1996}, for example, readily admits that ``we have no way of telling that the Universe is at all homogeneous and isotropic when seen from other places or at other times'' (p.~143). 

Indeed, all this and more could be argued against the cosmological premise of local isotropy. But it is the \emph{unspoken} presumption lying beneath this cosmological premise that must be regarded with a much higher degree of suspicion. For what is ``cosmic space at one time'' supposed to mean in general relativity? How shall we separate the whole universal four-dimensional spacetime into universal spaces at a given time, when this cannot even be done for smaller spacetime regions? What kind of time is it that the existence of cosmic space implies?  

In cosmology, the separation of spacetime into a spatial part and a temporal part is translated into geometric language by expressing the line element in the form
\begin{equation}\label{space-time}
	\D s^2=\epsilon\left[g_{ab}(x)\D x^a\D x^b-(\D x^4)^2\right]. 
\end{equation}
Once this split is made, the local isotropy of cosmic space is expressed geometrically by recasting the (non-invariant) spatial part $Q_s=:g_{ab}(x)\D x^a\D x^b$ of this equation in spherical coordinates as 
\begin{equation}
	Q_s=e^{A(x^4,x^1)} \left[ (\D x^1)^2+(x^1)^2 (\D x^2)^2+
	(x^1)^2\sin^2(x^2)(\D x^3)^2 \right].
\end{equation}

How are the coordinates interpreted in this space-time setting? Coordinate $x^4$ is taken to represent ``cosmic time'' and denoted $x^4=c\tau$. ``Cosmic time'' $\tau$ works as a ``universal time,'' equal for all ``cosmic observers'' travelling along galactic worldlines. This means that ``cosmic observers'' throughout the cosmos measure the same \emph{absolute} ``cosmic separation'' 
\begin{equation}\label{cosmic-time}
	\D s^2=c^2 \D\tau^2. 
\end{equation}
Cosmic space is then nothing but the three-dimensional hypersurfaces orthogonal to the galactic worldlines at $\tau=\tau_{\mathrm{now}}$; i.e.\ cosmic space is an instantaneous slice of spacetime at absolute ``cosmic time'' $\tau_{\mathrm{now}}$. But is the assumption of ``cosmic time'' realistic in a general-relativistic world? 

As for $x^1$, it is interpreted as the radial distance to an event from a cosmic observer and denoted $x^1=\rho$. The supposition is then made that the function $A(\tau,\rho)$ can be separated into a part $B(\tau)$ connected only with ``cosmic time'' and to be related to the expansion of cosmic space, and a part $C(\rho)$ connected only with matter and to be related to the curvature of cosmic space, i.e.\ 
\begin{equation}
	A(\tau,\rho)=B(\tau)+C(\rho).
\end{equation}
Since the cosmos is spatially homogeneous at a given ``cosmic instant,'' the spatial part of the tensor describing its matter and radiation content must be of the form \mbox{$T^a_{\ b}=\mathrm{constant}\ \delta^a_{\ b}$}.

On all these assumptions and with suitable integration constants, the field equation leads to the solution
\begin{equation}
\D s^2=\epsilon \left[ e^{B(\tau)}\left( 1+\frac{k\rho^2}{4\rho_0^2}\right)^{-2}(\D\rho^2+\rho^2\D\theta^2+\rho^2\sin^2(\theta)\D\phi^2)-\\
c^2\D\tau^2\right],
\end{equation}
where $k=0,1$ or $-1$. An equivalent solution was found by Howard Robertson in 1935 and by Arthur Walker in 1936, and it is the geometric foundation of relativistic cosmology. It is interpreted to picture a flat, positively, or negatively curved locally isotropic space---pictured by $\exp[C(\rho)]$---``expanding'' in ``cosmic time''---pictured by $\exp[B(\tau)]$.

A similar interpretation is usually made of the de Sitter solution (\ref{de-Sitter}) for the cosmic vacuum we saw earlier. By means of a change of coordinates, 
\begin{equation}\left\{
\begin{array}{lll}
	x^1&=&\rho e^{c\tau\sqrt{-\Lambda/3}} \\
      x^4&=&c\tau+\frac{1}{2}\sqrt{-3/\Lambda}\ln\left[1+(\Lambda/3)\rho^2 e^{2c\tau\sqrt{-\Lambda/3}}\right]
\end{array} \right.,
\end{equation}
the de Sitter line element can be expressed as a spatially isotropic part that, for $\Lambda<0$, ``expands'' in ``cosmic time'' $\tau$:
\begin{equation}
\D s^2=\epsilon \left\{ e^{2c\tau\sqrt{-\Lambda/3}}[\D\rho^2+\rho^2\D\theta^2+\rho^2\sin^2(\theta)\D\phi^2]-c^2\D\tau^2\right\}.
\end{equation}
This pictures, as it were, an expanding empty cosmos or ``motion without matter.'' It stands in contrast to Einstein's proposed solution of a matter-filled static universe or ``matter without motion.''

We can now see that when Einstein deemed the problem of cosmology to be tied to the geometry of the universe (i.e.\ of space) and to time, he was dead right. Moreover, realizing now that the former problem depends on the latter, we recognize that the problem of cosmology really boils down to its presumptions about the \emph{nature of time}.

The problem of space refers to the question of the spatial curvature of the locally isotropic cosmos. But, as we just saw, in order to be able to even talk about spatial properties, spacetime should be separable into a spatial part $g_{ab}(x)\D x^a\D x^b$ and a temporal part $g_{44}(x^1,x^2,x^3)(\D x^4)^2$. This is feasible, for example, for flat spacetime or for the spacetime around an isolated spherical mass distribution, but it is not a general property of spacetime. In fact, it seems \emph{least} circumspect to assume the validity of the space-time split in the cosmological setting, since we are here dealing with the \emph{totality} of spatiotemporal events, which is therefore the \emph{most general} case there is. 

This unwarranted space-time split of cosmology signals a regress to a Newtonian worldview: an absolute cosmic space whose existence requires that of absolute time. Through ``cosmic time,'' the universe becomes endowed with an absolute present (absolute space) and a history. The validity of the space-time split in cosmology, however, runs counter to the basic lessons of general relativity, and it is a presumption whose truth can only be hoped for and guessed at. The psychologically suggestive notation $x^4=c\tau$ does not by itself turn $\tau$ into a physically meaningful absolute time.

What motivates this cosmological regress to the classical view of time? Ultimately, only our human quest for understanding. The problem of cosmology lies essentially in the suppositions it must make about the nature of time in order to create a theoretical picture of something as vast, complex, and remote as the cosmos simple enough to be humanly manageable and understandable. It lies in the reversed logic of a situation in which we fit the great wild cosmos to our existing, familiar tools of thought instead of trying to develop new tools of thought that suit and do justice to the cosmos. 

It seems ironic---although on second thoughts only natural---that the more we have learnt, the deeper and farther we have probed the natural world in pursuit of the disclosure of one more of its secrets, the closer we have come to experience the deficiencies of our own cognitive tools. As if the deeper we sink in an oceanic expedition, the more the dimming light and building pressure interfere with our intended objective study of the ocean's fauna---until, in the utmost darkness of the inhospitable depths, we become oblivious to all fauna, only remaining aware of the feeble light our torch dispenses and the low oxygen supply in our tank. Dimming light leads to virtual blindness and oxygen deficiency to brain damage. The first symptoms of these conditions are easy for a diver to self-diagnose. The same, regrettably, cannot be said of their counterparts in the human expedition to the intellectually inhospitable depths of the cosmos. 

In an often-quoted passage, Eddington captured, with rare insight, this human dilemma. Concluding his book on spacetime and gravitation, he wrote:
\begin{quote}
[W]e have found that where science has progressed the farthest, the mind has but regained from nature that which the mind has put into nature.

We have found a strange foot-print on the shores of the unknown. We have devised profound theories, one after another, to account for its origin. At last, we have succeeded in reconstructing the creature  that made the foot-print. And Lo! it is our own. \mbox{\cite[pp.~200--201]{Eddington:1920}}
\end{quote}

The geometric analysis of clock time, leading us to the classical and relativistic worldviews, is now complete.

What next?

\chapter[Interlude: Shall we quantize time?]{Interlude: Shall we\\ quantize time?}
\label{ch:Quantize time}
\begin{flushright}
\begin{tabular}{p{10cm}}
\emph{A picture held us captive. And we could not get outside it, for it
lay in our language and language seemed to repeat it to us inexorably.} 
\smallskip \\
Ludwig Wittgenstein, \emph{Philosophical investigations}
\end{tabular}
\end{flushright}
\bigskip

The main lesson to be drawn from the previous chapter is the identification of clock time as the key physical concept in relativity theory. Our goal being a ``quantum theory of gravity,'' this seems an excellent opportunity to find it. Having uncovered the confusion that holds the metric field $g_{\alpha\beta}(x)$ as the physical essence of relativity, and having identified the worldline separation $\D s$ in its stead as the rightful holder of this place, what else is there left for us to do than quantize not $g_{\alpha\beta}(x)$ but $\D s$? What else is there left for us to do than \emph{quantize time}?

Right at the outset, we can consider two alternatives. If we believe that the smooth background of Riemann-Einstein spacetime is a suitable starting point, we may quantize $\D s$ and introduce a quantum field of the form $\widehat{\D s}$. However, if for some reason we prefer to give up the comfort of differential geometry and return to finite Euclidean geometry, we may quantize $\Delta s$ and introduce a quantum field of the form $\widehat{\Delta s}$. In the following, we use the differential version $\widehat{\D s}$; which version we choose, however, is irrelevant to our ulterior aim.

Is $\widehat{\D s}$ an observable? We know that clock readings $\D s$ are positive real numbers; they are the results of the experiments made with clocks as we measure the interval between two events, $A$ and $B$. It is then natural to assume that $\D s$ are the eigenvalues of $\widehat{\D s}$. On this condition, $\widehat{\D s}$ is indeed an observable, i.e.\ Hermitian. In effect, introducing a Hilbert space of quantum-mechanical eigenvectors $|\psi_i\rangle$, we have that
\begin{equation} \left\{
\begin{matrix}
\widehat{\D s}| \psi_i\rangle  & = & \D s_{|\psi_i\rangle}|\psi_i\rangle  \\
\langle \psi_i| \widehat{\D s}^\dagger  & = & \D s_{|\psi_i\rangle} \langle \psi_i|
\end{matrix} \right..
\end{equation}
The second equation is obtained as the dual correspondent of the first on account of $\D s^\ast=\D s$. Multiplying the first equation by $\langle \psi_i|$ on the left and the second by $|\psi_i\rangle$ on the right, we find 
\begin{equation}
      \langle \psi_i|\widehat{\D s}|\psi_i\rangle = \langle \psi_i|\widehat{\D s}^\dagger|\psi_i\rangle.
\end{equation}
Since $|\psi_i\rangle$ is an arbitrary eigenvector of $\widehat{\D s}$, it follows that $\widehat{\D s}$ is Hermitian, 
\begin{equation}
\widehat{\D s}=\widehat{\D s}^\dagger.
\end{equation} 

For a general state vector $|\psi\rangle$, we also find that the expression $\langle \psi|\widehat{\D s}|\psi\rangle$ can be attributed a special meaning. Rewriting it in terms of the orthonormal eigenstates $|\psi_i\rangle$ of $\widehat{\D s}$, we find
\begin{equation}\label{expected-value}
	\langle \psi|\widehat{\D s}|\psi\rangle = 
	\sum_{i,j}\langle\psi |\psi_i \rangle \langle \psi_i|\widehat{\D s}|\psi_j \rangle \langle \psi_j |\psi \rangle = 
	\sum_i |\langle \psi_i|\psi \rangle|^2 \D s_{|\psi_i\rangle},
\end{equation}
which suggests that $|\langle \psi_i|\psi \rangle|^2$ can be interpreted as the probability that, upon measuring with a clock the spacetime interval between events $A$ and $B$, the value $\D s_{|\psi_i\rangle}$ will be obtained. The measurement of this value is, at the same time, linked with an underlying quantum-mechanical transition from spacetime state $|\psi\rangle$ to spacetime state $|\psi_i\rangle$. Thus, $\langle \psi|\widehat{\D s}|\psi\rangle $ is the average obtained after a series of measurements of the separation operator $\widehat{\D s}$ in state $|\psi\rangle$; in other words, the expected value. 

These ideas could be developed further, but even at this early stage it is  appropriate to \emph{stop, think, and ask}: is this the physical foundation of a quantum theory of gravity? Or perhaps the foundation of a quantum theory of time? Hardly. At best, the above only teaches us a lesson about how one should \emph{not} go about doing new physics, and about how easy it is to be held captive of the conventional pictures of a preestablished physical tradition. Let us see why.

The only positive thing to be said about the quantization of the spacetime interval is that $\D s$ is indeed a real number resulting from a simple physical measurement that we know how to perform: stand at event $A$, take a clock, read its dial, record the displayed number, travel to event $B$, read the dial, and record the displayed number. But as soon as we enquire into the operational meaning of the quantum-theoretical framework introduced around $\D s$, we are in deep trouble. 

Essential to the quantum-theoretical picture developed is the state vector $|\psi\rangle$. When we introduced it, we did so non-committally; when we later on referred to it, we called it a quantum-mechanical spacetime state. This is indeed what $|\psi\rangle$ must denote in theory, because the intervals $\D s$ between spacetime events are determined by the geometric structure of spacetime, but is there \emph{in practice} a denotation behind the words ``spacetime state'' and its mathematical symbol ``$|\psi\rangle$''? 

Talk about the expected value $\langle \psi|\widehat{\D s}|\psi\rangle$ of observable $\widehat{\D s}$ in state $|\psi\rangle$ implies that we can make a (large enough) set of measurements of $\widehat{\D s}$ starting each time with a system in the same ``spacetime state'' $|\psi\rangle$, and then proceed to calculate the average of the results. But what is here the system to be prepared into state $|\psi\rangle$? The quantum-mechanical system must surely be spacetime, but how are we supposed to prepare it into state $|\psi\rangle$? What are the \emph{humanly feasible} operations involved? What is a spacetime state, after all? How do we gain access to it, affect it, modify it?

The same problem affects the concept of probability. What operational meaning do we attribute to the probability of obtaining value $\D s_{|\psi_i\rangle}$ upon a measurement of $\D s$ when spacetime is in state $|\psi\rangle$, when we do not know what a spacetime state means physically? And, further, how plausible is it for a clock measurement to exert an influence and modify the state of spacetime, inducing the transition of $|\psi\rangle$ into $|\psi_i\rangle$? 

In this light, the quantization of the spacetime interval appears as a mimicry---a caricature---of the quantum mechanics of material systems. The operationally well-determined notion of quantum-mechanical state\footnote{But not state \emph{vector}; see next chapter.} of a material system (such as a particle shower) is uncritically transferred to spacetime, arriving at the idea of a quantum-mechanical state of spacetime. But just because we know how to prepare material particles into states, it does not follow by the magic of analogy that we know how, or even what it means, to prepare spacetime into states. In the context of the quantum mechanics of material systems, the state of a particle is induced to change by the direct action of an external device upon the particle, e.g.\ by the magnetic field which in a Stern-Gerlach device changes the spin state of a particle shower. If this were to translate to the quantum mechanics of spacetime, the making of a clock measurement should have the capacity to directly modify the state of spacetime. Unlikely? Far-fetched?

The same criticism applies to the uncritical importation of other operationally well-determined key concepts of the quantum mechanics of material particles, such as expected values and probabilities, that rely on the idea of an \emph{accessible}, \emph{modifiable}, and \emph{reproducible} quantum-mechanical state of a system to have physical meaning.  

Some of these and other similar problems beset an attempt along these lines carried out by Isham, Kubyshin, and Renteln \citeyear{Isham/Kubyshin/Renteln:1991}. A noteworthy remark by these authors deserves comment. Regarding how their proposal should be interpreted, they wrote:
\begin{quote}
Since the metric [distance function $\mathrm{d}(x,y)$; cf.\ $\Delta s$]\ldots is non-negative it is not clear how the pseudo-Riemannian structure of spacetime of general relativity and the notion of time could emerge in quantum metric topology. Thus the approach can be considered either as a generalization of the Euclidean version of gravity, or as a description of the spatial part of physical spacetime. In the latter case time remains an additional parameter labelling the metric in the quantum metric topology approach. \cite[p.~163]{Isham/Kubyshin/Renteln:1991}
\end{quote} 
First of all, there is, in fact, no problem here---only a long-standing confusion. As we saw in the previous chapter, the non-negativity of $\Delta s$ is perfectly compatible with the positive, negative, or null quadratic form $g_{\alpha\beta}(x)\D x^\alpha \D x^\beta$, because their connection takes place via the $\epsilon$ indicator,
\begin{equation}
	\Delta s =\int_{s_A}^{s_B} \sqrt{\epsilon g_{\alpha\beta}(x) \frac{\D x^\alpha}{\D s} \frac{\D x^\beta}{\D s}}\D s,
\end{equation}
where $\epsilon=\pm 1$. For timelike displacements, $\epsilon=-1$ and $\Delta s$ remains positive, as it should. But what is here remarkable is not the confusion---which is common---but the attitude taken in the face of seeming difficulties. If the framework chosen will not describe spacetime and gravitation as intended, then it might as well describe something else. In the face of adversity, even the essential sacrifice is made to leave the description of time out of the originally intended quantum-theoretical picture of space\emph{time}, so as to thus rescue and continue to cling to an idea that started off with the wrong foot. 

This is only an allegorical example of the widespread working philosophy of frontier theoretical physics. If the mathematics we come up with will not conform to the physical world, then the physical world had better conform to the mathematics we come up with. Is this the path of physical discovery? \emph{Is this the road to quantum gravity?} 

What spacetime may be quantum-mechanically---if anything at all beneath the classical geometric structure so far uncovered---we do not know. But our ignorance in this regard should not justify the desperate invention of quantum-theoretical frameworks without a shred of physical meaning. One may play with the transmuted mathematical symbols borrowed from a successful physical tradition, but to presume that this, too, is physics because of mere resemblance is to presume too much. 

Physics must be done from the bottom up,  building upon nature's raw material, explicitly, in terms of things that human beings have access to and can affect. In the context of the quantum mechanics of particles, we turn to such things next.

\chapter{The metageometric nature of quantum-mechanical things and the rise of time}
\chaptermark{The metageometric nature of quantum-mechanical things}
\label{ch:MQM}
\begin{flushright} 
\begin{tabular}{p{10cm}}
\emph{Groping, uncertain, I at last found my identity, and after seeing my thoughts and feelings repeated in others, I gradually constructed my world of men and God. As I read and study, I find that this is what the rest of the race has done. Man looks within himself and in time finds the measure and the meaning of the universe.} \smallskip \\
Helen Keller, \emph{The world I live in}
\end{tabular} 
\end{flushright}

\bigskip

Passages from famous physicists about the difficulties involved in understanding quantum mechanics abound. Speaking of it, Erwin Schr\"{o}dinger is reported to have said, ``I do not like it, and I am sorry I ever had anything to do with it.'' Niels Bohr, for his part, stated that ``those who are not shocked when they first come across quantum mechanics cannot possibly have understood it.'' Murray Gell-Mann described quantum mechanics as ``that mysterious, confusing discipline which none of us really understands but which we know how to use.'' And Richard Feynman decreed, ``I think it is safe to say that no one understands quantum mechanics.''

Although these views were sometimes expressed only lightheartedly, one can yet grasp the gist of these physicists' complaints---and the seriousness thereof. It is not only that quantum-theoretical descriptions cannot be mechanically visualized; more worrying is the fact that the foundational concepts of quantum theory have \emph{physically unexplained origins}. In other words, it is not just the \emph{mechanical-how} but, more importantly, the \emph{physical-why} of quantum mechanics that remains elusive. On the other hand, the theory affords us a geometric visualization of the processes it
describes; its \emph{geometric-how} is readily available---but may this be more of a hindrance than a help?

\begin{figure}
\centering
\includegraphics[width=32mm]{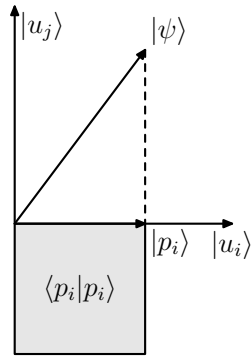}
\caption[Probability as projection]{Quantum-mechanical probability
as the squared length of the projection $|p_i\rangle$ of
$|\psi\rangle$ on $|u_i\rangle$.} 
\label{pfp} 
\end{figure}

Among quantum theory's long-standing enigmas are the following: what is the physical meaning of the arrow state \emph{vector} $|\psi\rangle$? Why are vectors needed at all to formulate quantum mechanics? How and why do they collapse when a measurement or state preparation occurs? Why is it possible to picture such a collapse through a \emph{projection} by means of an \emph{inner product}? Why is the probability of an outcome the squared \emph{length} $|\langle u_i|\psi \rangle|^2 = \langle p_i|p_i\rangle$ of the complex-valued projection vector (cf.\ collapse) $|p_i\rangle=\langle u_i|\psi\rangle |u_i\rangle$ of an initial state vector $|\psi\rangle$ onto another $|u_i\rangle$ (Figure \ref{pfp})? Why do probabilities arise \emph{geometrically} from Pythagoras' theorem? What is the physical meaning of complex numbers and complex conjugation? Why is it always possible to find Hermitian operators (with real eigenvalues) for any observable? And why is it always possible to attach \emph{orthogonal} eigenvectors to an observable?

These enigmas, some more pressing and disquieting than others, are directly related to the postulates of quantum theory; and all of them are, in turn, founded on the \emph{geometric rendering} of the theory (cf.\ italized words above), which we analysed in Section \ref{GQMec}. Further thought, however, reveals that it is precisely in these unquestioned (and unquestionable) geometry-ridden postulates that our problems comprehending the physical roots of quantum mechanics lie.

Indeed, geometry resembles a double-edge sword. Whereas it greatly aids at first our comprehension of the world, it eventually brings up problems of its own, i.e.\ problems stemming from the particular way in which it forces us to view the states of affairs it describes. On the one hand, it allows for a geometrically intuitive formulation of quantum theory that is consistent with observations but, on the other, its very geometric formulation turns out to be the bearer of its least understood concepts. In other words, although we only seem to be able to formulate and understand quantum mechanics geometrically, it is precisely its geometric notions whose physical underpinnings are elusive.

The need to understand the physical origin of the basic, yet controversial, geometric concepts of quantum mechanics, and thereby clarify the above enigmas, suggests itself as the immediate motivation to go beyond geometry in the formulation of this theory. But the motivation of the metageometric analysis that follows does not end at this, as it also has a less immediate goal. This is to lay the foundation for a metageometric quantum-mechanical picture of time, which we intend to connect with the clock-based formulation of general relativity of Chapter \ref{ch:Analysis of time}. To achieve this goal, we will seek inspiration for the present analysis of quantum mechanics in so far unaccounted-for aspects of time as a conscious experience.

\section{Quasigeometry and metageometry}\label{GQM}
As explained in Chapter \ref{ch:Geometry}, we equate geometry, not with
geometric existence, but with geometric thought and language. Neither geometric objects (shape) nor geometric magnitudes (size) are present in nature, but are a conceptual tool by means of which humans picture nature. In this chapter, we introduce the notion of quasigeometry and, by example, elucidate the meaning of the anticipated notion of metageometry.

By \emph{quasigeometry} we refer to a realm between metageometry and geometry that helps bridge the gap between these two in connection with the foundations of quantum theory. The word ``quasigeometry''\footnote{Quasi, from Latin \emph{quam} ``as much as'' + \emph{si} ``if'': as if.} means ``having some resemblance to geometry but not being really geometric,'' and is attributed to this realm due to its similarity and connection with geometry. We shall find the main quasigeometric entities to be the familiar real and complex numbers. How they
appear and help bridge the gap between metageometric quantum-mechanical things and the usual geometric representations will be one of the issues of this chapter.

Real numbers can be thought as extended arrows laid along a graduated one-dimensional line. In spite of this, they are neither genuine geometric objects nor magnitudes for the elementary reason that they are themselves the very \emph{carriers of geometric size}. As we saw in Chapter \ref{ch:Geometry}, real numbers act, in a geometric setting, as the go-betweens between geometric quality and geometric quantity. In other words, a real number is not size but \emph{gives size}. If thought of as arrows, they can, in particular, \emph{give size to themselves} (that which they inherit from their moduli $|a|$), in which case they become \emph{degenerate}\footnote{Degenerate, from Latin \emph{degenerare}, \emph{de-} ``down from, off'' + \emph{genus} ``birth, race, kind'': to depart from one's kind, fall from ancestral quality (Online Etymology Dictionary, http://www.etymonline.com); having declined or become less specialized from an ancestral or former state; having sunk to a condition below that which is normal to a type (Merriam-Webster Online Dictionary, http://www.m-w.com).} geometric concepts. Due to their degenerate geometric nature, we give the name quasigeometry to the realm they characterize. 

\begin{figure}
\begin{center}
\includegraphics[width=75mm]{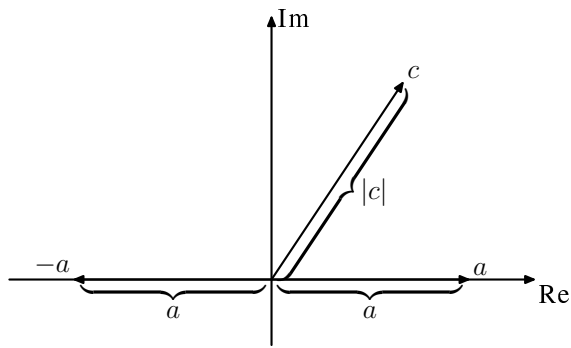}
\end{center}
\caption[Geometric visualization of real and complex numbers]{Geometric visualization of real and complex numbers. When real numbers $a$, originally the carriers of size, are thought of as arrows of length $|a|$, they become degenerate geometric concepts. Complex numbers $c$, on the other hand, have a genuine geometric interpretation as arrows of length $|c|$, but to conserve our chances of a deeper understanding of quantum ontology, it is essential to stay clear of this geometrization of the complex number.} 
\label{naqgc}
\end{figure}

Complex numbers, too, can be thought of as arrows with a definite length only now laid on the (Argand) plane. In this case, however, the situation is different. Now it is the real modulus $|c|$ of a complex number that represents its length, while the complex number $c$ itself can be viewed as a non-degenerate geometric object (arrow). For this reason, if we want to use complex numbers as quasigeometric concepts---concepts truly devoid of geometric meaning---we must be careful never to regard them as extended arrows. We must eschew all geometric visualization of complex numbers and, instead, seek their physical meaning other than as \emph{merely conducive} to probability as the size ($|c|^2$) of a geometric object (arrow $c$): we must strive to find the physical meaning of $c$ itself, which does not as such refer to the shape or length of anything. To conserve our chances of a deeper understanding of quantum ontology, it is essential to stay clear of this geometrization of the complex number. As we progress in this chapter, so far the only known role of complex numbers as descriptive of something physical will be uncovered, namely, the characterization of the physical process of a \emph{directional} quantum-mechanical transition. This will give us a different reason to call complex numbers quasigeometric (Figure \ref{naqgc}).

By metageometry\footnote{Meta, from Greek preposition \emph{meta}: in the midst of, among, with, after; from Proto-Indo-European *\emph{me-}: in the middle. The original sense of the Greek word is ``after'' rather than ``beyond,'' but it was misinterpreted when Aristotle's treatises were discussed by Latin writers. From the reference below, we learn that, originally, \emph{metaphysika} meant ``~`the (works) after the Physics' [in the] title of the 13 treatises which traditionally were arranged after those on physics and natural sciences in Aristotle's writings. The name was given circa 70 B.C.E. by Andronicus of Rhodes, and was a reference to the customary ordering of the books, but it was misinterpreted by Latin writers as meaning `the science of what is beyond the physical.'~'' (See ``meta'' and ``metaphysics'' in Online Etymology Dictionary, http://www.etymonline.com)} we mean ``beyond geometry,'' and therefore refer to a theoretical picture---a human scientific language---that does not use geometric concepts of any sort. This is the sense in which one should understand the reference to the ``metageometric nature of quantum-mechanical things'' in the title of this chapter. That is, not as hinting at the metageometric essence of the physical world, as if to say that ``God is a metageometrician'' in neo-Platonic fashion, or that ``the Book of Nature is written in the language of metageometry'' in neo-Galilean fashion, but rather meaning that some aspects of nature---quantum-mechanical ones, in this case---may not be open to human understanding through geometric tools of thought. We see that geometric modes of thinking and talking have not only been capable of leading to myriad physical triumphs ranging from Newton to Einstein, but also to possible philosophical and physical blind alleys; for example, spacetime ontology, quantum ontology, and the geometric roads to quantum gravity. 

We suggest that, if progress concerning the conceptual difficulties of quantum theory wants to be made, the theory should be written after a metageometric fashion, i.e.\ in terms of metageometric things. Now this may seem like a long shot indeed. Given that humans by nature think geometrically, and that thus all physical theories so far speak the language of geometry, what chance is there that any metageometric theory could be understood---even less developed---by us? Would not a metageometric theory be hopelessly abstruse? Would not metageometric things, devoid of all shape and size, be hopelessly unimaginable, meaningless, and abstract? 

Not so. We shall see that metageometric things can be found such that, far from abstruse, are actually based on physical experiments and are rather close to human experience. Indeed, we shall find that, besides the well-established, unquestionably geometric aspects of our practical thinking, a deeper realm of human conscious experience---closely connected with the feeling of time---is naturally captured metageometrically. By identifying its constituents and their mutual relations, and by having these be \emph{inspirational sources} of the experiment-based physical things of a metageometric quantum theory, we will lay the foundation for a theory that is transparent to human cognition, is open to human access and control, and provides a setting for a new understanding of time.

\section{Towards metageometric quantum-mechanical things}
\label{CQT}
Our search for a metageometric, but at the same time familiar and comprehensible, foundation of quantum mechanics is guided by the following working principle:

\begin{quote}
A profound and trustworthy knowledge of the ontology of quantum mechanics can only be obtained from the results of the activities of Man, from such things whose origins Man can understand, have access to, and affect. Our description of quantum mechanics must be expressible with the help of simple physical things because, otherwise, the workings of quantum mechanics are not knowable or understandable. These things must be such that we know the physical methods through which they arise.
\end{quote}

This means that we do not start the search for metageometric quantum-mechanical things in an implicit way, e.g.\ with wave functions as in wave mechanics, with canonical matrix pairs as in matrix mechanics, or with state vectors as in Dirac's algebra. On the contrary, if quantum theory is to be laid plain before our eyes and its enigmatic features comprehended, we should start the metageometric search explicitly, with concepts close to both human psychological and physical experience.

The pursuit of the \emph{psychological connection} means that we would like quantum-mechanical things to be closely connected with the realm of experienced thoughts and feelings. This is not to say that the intention here is to develop a new theory of consciousness; this requirement rather aims to start with a setting closely connected with the \emph{raw feeling of time}, in which the present moment is paramount, because in this way the quantum-mechanical theory will afford the possibility to picture time in a physically new way. In turn, the pursuit of the \emph{physical connection} means that we require that quantum-mechanical things be based upon an experimentally explicit basis, because this is the only way to make sure that the elemental ontological constituents of the resulting theory will be \emph{observationally concrete}, i.e.\ wide open to human manipulation and control.

The discovery, within a metageometric theoretical setting, of quantum-mechanical things that are related to consciousness and are experimentally explicit is the task of what may be called \emph{psychophysics}. In the rest of this section, we pursue the path of psychophysics along two complementary directions. The first involves the investigation of possible phenomenological similarities between conscious and quantum-mechanical phenomena in search for an \emph{experimental link} between psychology and physics, whereas the second involves the investigation of the metageometric quality of conscious phenomena and their relations in search for an extra \emph{metageometric link} between psychology and physics. In Section \ref{Metageometry}, these two paths converge on the sought-for metageometric psychophysical thing,\footnote{In his book, Isham \citeyear[pp.~63--67]{Isham:1995} ponders the question, ``What is a thing?'' from a perspective that takes into account ontology and epistemology---the existence of things and our knowledge of things---and the degree to which ontology may be the projection of our own epistemological shadows. For us, a thing will be essentially a simple psychological abstraction of a physical experiment; insofar as this is so, the ontology of things is as uncontroversial (or controversial, according to philosophical taste) as the physical existence of our minds and the experiment itself, while the epistemology of things becomes conflated with their ontology, for our knowledge of things is tantamount to our first-hand experience of things themselves. As for the metageometric language in which we shall picture physical things, into it go, naturally and unavoidably, our methodological biases and prejudices.} on which to build the foundations of an experimentally overt metageometric formulation of quantum theory that can throw new light on quantum ontology and the notion of time.

\subsection{Experimental facet of psychophysics} 
Do quantum-mechanical phenomena share any similarities with phenomena related to consciousness that may serve as inspiration in this physical search? Dennett's \citeyear{Dennett:1991}\footnote{Consult this reference (especially, pp.~101--137, 209--226, 275--282) for more details on the consciousness-related topics discussed below. Page numbers in parentheses in this section and the next refer to this reference unless otherwise stated.} theory of consciousness, with its unorthodox approach to conscious phenomena, can help us give a positive answer to this question.

A paradigmatic experiment that is apt to reveal several peculiar features of human consciousness is the \emph{colour-phi phenomenon} studied by Kolers and von Gr\"{u}nau \citeyear{Kolers/vonGrunau:1976}, but originally studied by
Wertheimer \citeyear{Wertheimer:1912} nearly a hundred years ago in its single-colour version. A red spot of light is flashed on a screen and 200 milliseconds later a green one is flashed within 4 degrees of angle from the first spot. Subjects report not only that the original red spot follows a continuous trajectory\footnote{This is the result of Wertheimer's experiment with a single-colour spot and is the working principle behind motion pictures.} but also that
it changes colour (from red to green) \emph{midway} between the endpoints of the continuous trajectory. How should we make sense of this phenomenological observation, and what does it reveal about the mechanisms of consciousness?

The explanation that comes most readily to mind is that, since in order for the red spot to change colour seemingly before the green spot has even been flashed, there must be an intrinsic delay in consciousness itself of at least 200 milliseconds. However, there is enough evidence to the effect that conscious responses (e.g.\ button pressing) to stimuli are not as sluggish as this hypothesis requires (pp.~121--122, 150). If conscious experience is not being put temporarily on hold (certainly not an evolutionary advantage), is the colour-phi phenomenon, then, an indication that conscious events violate causality in time and space? This alternative seems even less palatable. We seem to be trapped, here as in frontier physics, in a deep conceptual confusion.

To help disentangle it, Dennett urges us to rid ourselves of the natural idea that to become conscious of something is tantamount to certain (e.g.\ electrochemical) processes occurring at some specific point in the brain; to reject the notion that what we experience as the stream of consciousness has an exact analogue at some singular location in the brain. Dennett asks, ``Where does it all come together?'' and replies, ``Nowhere\ldots [T]here is no place in the brain through which all these causal trains must pass in order to deposit their content `in consciousness' '' (pp.~134--135). On the assumption of a brain singularity---and without the benefit of long delays or violations of causality---the colour-phi phenomenon defies explanation.

It seems remarkable that, in a field of science as removed from cosmology and spacetime physics as can be, one should also be forced to come to terms with a singularity---a point at which our understanding becomes feeble and powerless---\emph{if} unable to refrain from the application of geometric thinking beyond its
ostensible capabilities. On second thoughts, however, there is nothing remarkable about hitting upon a new singularity, as both psychological and cosmological theories are made by humans with the same geometric tools of thought. Dennett, also noticing the reminiscence with singularities in physics, criticized the \emph{seemingly natural geometric method} by which one arrives at the belief in a singularity in the brain: 

\begin{quote} Doesn't it follow \emph{as a matter of geometric necessity} that our conscious minds are located at the \emph{termination} of all the \emph{in}bound processes, just before the \emph{initiation} of all the \emph{out}bound processes that implement our actions? \cite[p.~108]{Dennett:1991} 
\end{quote} 
He subsequently argued that the geometric force of this contention is misleading: inbound processes need not end at, and outbound processes need not leave from, a physiological point: the brain works largely as a unit.

The neuronal activity giving rise to consciousness is extended over the brain as a whole. Dennett suggests that, as a result, we cannot say where or when a conscious experience occurs (p.~107); that although 

\begin{quote} 
Every event in your brain has a definite spatio-temporal location\ldots asking ``Exactly when do \emph{you} become conscious of the stimulus?'' assumes that some one of these events is, or amounts to, your becoming conscious of the stimulus. [\ldots] Since cognition and control---and hence consciousness---is distributed around in the brain, no moment [or place] can count as the precise moment [or place] at which each conscious event happens. \cite[pp.~168--169]{Dennett:1991} 
\end{quote}

Thus, the colour-phi phenomenon (and other tests to the same effect) leads to the need to differentiate between a representation---consciousness---and the medium of the representation---the spatiotemporal brain events: the brain does not represent space (e.g.\ the image of a house) with space in the
brain, and, to a limited extent, neither does it represent time (e.g.\ the sequential motion in the colour-phi test) with time in the brain. Viewed in this way, consciousness is a representational abstraction of the internal and external world, whose conveying medium is the spatiotemporal events in the brain. It is, among other things, a \emph{representation of} space and time, but it is not itself located anywhere in the space and operating time of the
brain in any literal sense. Consciousness is, in this sense, a \emph{spatially} and \emph{temporally delocalized}, \emph{global}, and \emph{locally irreducible} property of the brain. How come, then, that conscious experience resembles so closely a linear sequence of spatiotemporal events?

Dennett elaborates as follows. Neurons work together multifunctionally and globally within the space and time of the brain to produce numerous, parallel, undecanted streams of consciousness, which he calls ``multiple drafts.'' It is parts of these drafts that can project themselves as a single \emph{serial} sequence of narrative. This is achieved by means of probes such as self-examination and external questioning, which act to decant certain draft contents of consciousness to a different level. These probes are as essential to the determination of what one is conscious of as are the external stimuli themselves (pp.~113, 135--136, 169). This loosely suggests also an \emph{indeterministic} picture of consciousness, since there is no single, actual stream of consciousness unequivocally determined by external stimuli independently of the random probes that originate it.

According to Dennett, it is the decanted stream of consciousness that we experience as a serial chain of events and readily associate with conscious experience. It is this stream that provides a subjective worldline of events as reported by the conscious subject. This worldline can be subsequently compared with other worldlines of objective events as elements of the external world (p.~136). The comparison of two such worldlines gives meaning to the idea of the projection of consciousness, as a sequential stream outside spacetime, onto the spatiotemporal world.

A better understanding of the colour-phi phenomenon is now closer at hand. If the phenomenon is thought of as occurring in consciousness in a mirror-like, linear sequence as the light stimuli occur in spacetime, it leads to the mistaken belief that the brain must house a consciousness singularity capable of projecting certain events (green flash) backwards to times earlier than their actual occurrence in the brain (p.~127). However, if, as Dennett proposes, the colour-phi phenomenon is thought of as occurring in consciousness as a multifaceted, higher-order representation of stimuli-in-spacetime that does not nevertheless use space in the brain to picture space nor time in the brain to picture time (p.~131), a new conception arises. Because consciousness only projects itself onto spacetime in the way just explained, there is no question of the later appearance of the green spot being projected backwards in time or anything of the sort, but rather of a multitude of brain events working in parallel to yield a biased representation of the outer world. The projection of this representation onto spacetime is fuzzy within the time scale (of the order of half a second) of the said brain workings.

The phenomenological properties of consciousness just revealed display some similarity with quantum-mechanical ideas---in particular, with the preparation of physical systems into quantum-mechanical ``states.'' But before proceeding, a warning must be sounded. So far as we can see, the resemblances to be uncovered between the two phenomenological realms are superficial; or to use the precisely suited words of Klaus Gottstein \citeyear{Gottstein:2003} in a letter to Carl Weizs\"{a}cker dealing with a different issue, ``It appears to me that this is not a real connection, but rather a kind of parallelism'' (p.~2, online ref.) What is then the value of the parallelism we intend to draw? If nothing else, it is its heuristic power, as it is apt to guide thought towards physically useful concepts. Ideas, after all, must come from somewhere. As such, the following parallelism stakes no claim other than to be part of our private train of thought; the reader might as well ignore it if personally unconcerned with the private aspect of public science.\footnote{In particular, the parallelism to
be drawn does not mean to suggest any microscopic role for quantum mechanics in the brain, nor any connections between the mechanisms of consciousness and state-vector collapse. In fact, the concepts of state vector and collapse will occupy no place at all in the metageometric formulation. This should have been already clear---but, in an atmosphere as prone to misunderstanding as is that of science, better safe than sorry.}

We start by noting a reminiscence of the first, erroneous interpretation of the colour-phi phenomenon with equally misleading interpretations of quantum phenomena. For example, like the phenomenon of consciousness, so has the phenomenon of quantum correlations (or quantum entanglement) global, non-causal, non-reductionistic, and indeterministic characteristics, leading sometimes to talk about particles or information travelling backwards in time.

The first to propose this idea was Costa de Beauregard \citeyear{Costa:1977}, originally in 1953; talking of a correlated, widely-separated system of two photons, he wrote: ``Einstein's prohibition to `telegraph into the past' does \emph{not} hold at the level of the \emph{quantal stochastic event} (the wave collapse)'' (p.~43). Davidon \citeyear{Davidon:1976} also expressed similar views: ``In this formulation of quantum physics, effects of interactions with a previously closed system `propagate backwards in time' '' (p.~39); and so did Stapp \citeyear{Stapp:1975}: ``This information flows from an event both forward to its potential successors and backward to its antecedents\ldots'' (p.~275). These interpretations were devised on the erroneous assumption that the quantum-mechanical preparation of systems is a spatiotemporal process and that, therefore, it gainsays spacetime causal structure. These interpretations, then, stem from a desire not to relinquish this local, causal spacetime structure, and so trade it for a new, anomalous way in which causal connections can still be achieved locally. As Stapp put it: 

\begin{quote} 
Bohr has insistently maintained that quantum phenomena are incompatible with causal space-time description. The model of reality proposed here conforms to Bohr's dicta, in so far as classical causal space-time description is concerned, but circumvents it by introducing a nonclassical space-time structure of causal connection. \cite[p.~276]{Stapp:1975} 
\end{quote}

How do the colour-phi phenomenon and that of quantum correlations measure up to each other within the troubled framework of this local spacetime picture? On the one hand, the consciousness singularity in the brain needs to know that a green flash has
occurred before it has had time to acquire it (before the green flash has had time to travel past the singularity). This is solved by thinking that the green flash travels backwards in time and inserts itself at an earlier time of the stream of consciousness. On the other hand, a quantum-mechanical subsystem $S_2$ needs to know the result of an experiment $E_1$ performed on subsystem $S_1$ before it has had time to acquire it. This is solved by thinking that information about the result of $E_1$ travels backwards in time to a time when $S_1$ and $S_2$ were together and gets transferred to $S_2$. Both problems are created by thinking locally about consciousness and quantum-mechanical information, namely, (i) consciousness is a local process, and (ii) the transfer of quantum-mechanical information is a local process. However, by regarding consciousness and the transfer of quantum-mechanical information as intrinsically global phenomena that are \emph{ontologically prior to spacetime}, neither needs to challenge the notion of causal spacetime structure and thus lead to far-fetched explanations. 

Finally, when we picture the transfer of quantum-mechanical information in the usual way, namely, as the preparation of a quantum-mechanical system into a certain ``state'' (not a state vector), the parallelism can be brought even further. We find that preparation, a global and spatiotemporally delocalized phenomenon, has, like consciousness, the ability to impress itself onto particular physical systems (i.e.\ matter). It can thus create events that are localized in spacetime, providing a worldline in analogy with the stream of consciousness.

The experimental approach to psychophysics, understood as a heuristic discipline, has led to the physical notion of quantum-mechanical preparation. Is this an adequate or useful starting point for a metageometric quantum theory?

\subsection{Metageometric facet of psychophysics} \label{MetagConsc}
Despite our first-hand acquaintance with the subjective quality of our conscious experience, a systematic account of this realm---a theory of conscious\-ness---is lacking. Not only does conscious experience presently elude quantitative but even qualitative scientific description; in other words, not only do we fail to grasp the size of thoughts and feelings, but also fail to picture their forms, shapes, outlines, or structures in any sensible way. On the other hand, however, thoughts and feelings can be \emph{about} objects with shape and size. In particular, as we just learnt, conscious events have no proper spatial location, but they can be \emph{about} spatial objects. In this light, Pylyshyn \citeyear{Pylyshyn:1979} has made the further point that it should strike us as curious to ``speak so freely of the \emph{duration} of
a mental event,'' when we ``would not properly say of a thought (or image) that it was large or red'' (p.~278). Thus, in the same token, conscious events have no temporal location either, although they can be \emph{about} temporal happenings.

While this failure to picture conscious events geometrically need not be a problem in itself, it is nevertheless perceived as a problem in the context of our typically geometric ways of thinking---scientifically or otherwise. To us, the fact that the readily available ideas of geometric objects and geometric magnitudes are inapplicable to conscious events, instead of suggesting that consciousness is a dead-end problem, points to a world of possibilities: may the phenomena of consciousness be properly captured in a metageometric theoretical picture?

Remarkably, in a scientific context in which geometry is taken for granted as the God-given essence of the physical world, Foss \citeyear{Foss:2000} held the geometric non-representability of conscious experience responsible for the fact that consciousness has remained a scientific mystery up until now. ``Since science deals only with geometric properties,'' he wrote, ``and the sensual qualities are not geometric, they cannot even be expressed, let alone explained, in scientific terms\ldots Flavors have no edges, tingles have no curvature'' (p.~59). Foss furthermore made an enlightening historical analysis of the process that, through what he called the Pythagorean Intuition\footnote{As noted in Chapter \ref{ch:Geometric anthropic principle}, ``Platonic Intuition'' may be more accurate.} that ``reality is geometrical at its heart'' and the Galilean Intuition that ``the sensuous qualities are essentially non-geometric''\footnote{For example, Galilei \citeyear{Galilei:1957} wrote: ``Now I say that whenever I conceive any material or corporeal substance, I immediately feel the need to think of it as bounded, and as having this or that shape; as being large or small in relation to other things, and in some specific place at any given time; as being in motion or at rest; as touching or not touching some other body; and as being one in number, or few, or many. From these conditions I cannot separate such a substance by any stretch of my imagination. But that it must be white or red, bitter or sweet, noisy or silent, and of sweet or foul odor, my mind does not feel compelled to bring in as necessary accompaniments. [\ldots] Hence I think that tastes, odors, colors, and so on are no more than mere names so far as the object in which we place them is concerned, and that they reside only in the consciousness'' (p.~274).} (pp.~40, 66), banned consciousness from a paradigmatically geometric-minded science. Foss argued that this way of thinking stems from a confusion between the geometric vs.\
the non-geometric essence of things---which, understood in a metaphysical sense, science does not concern itself with---and the geometric pictures that science makes of the phenomena it studies.

Foss's insight about the geometric nature of scientific description simply as a successful \emph{tool} is a true gem, valuable as it is rare, but did he manage to stare at the taboo on thought about geometry being a \emph{limited} tool straight in the eye? Not quite, because his assessment of the ``riddle of consciousness'' comes down to this:

\begin{quote} 
Science does not require that the things it models \emph{be} geometric, but only that they can be modeled geometric\emph{ally}. And there is no reason to think, no immovable intuition indicating, that consciousness, including its many-colored
qualia, cannot be just as successfully modeled by science as any other natural phenomenon. \cite[pp.~94--95]{Foss:2000} 
\end{quote}
Pictured by science---how? Geometrically?

We do not see any reason why the study of consciousness in the bosom of a geometric theory, as implied by Foss, should not be possible. But having looked throughout this work at the reigning thought-taboo on geometric limitations in the eye, we are interested in a different line of thought. May these facts be saying that consciousness also has ontological aspects that can only be captured by going beyond geometry as a tool of thought in science? Based on the heuristic parallelism with quantum-mechanical phenomena drawn earlier, we should be inclined to at least consider the possibility. Why? Because it is on the other hand the case with quantum phenomena that geometric methods succeed in capturing their superficial features, but that a description of their heart and nature continues to elude such methods; because the heart and nature of quantum phenomena, as we see in this chapter, \emph{can} in fact be disclosed beyond the bounds of geometric thought.

Understood from this metageometric point of view, Foss's description of the current climate of thought that ``it is intuitively obvious that the sensuous qualities cannot be conceived in geometric terms, hence cannot be convincingly modeled by science, therefore cannot find a place in its ontology'' (p.~66) demands what seems to be an unforeseen qualification: the sensuous qualities cannot find a place in the ontology of science \emph{unless} this ontology was expanded beyond the geometric realm. Understood thus, the following words from Dennett could be given unsuspected meaning: 

\begin{quote} 
[We should not be dissuaded] from asking\ldots whether mind stuff
might indeed be something above and beyond the atoms and molecules
that compose the brain, but still a scientifically investigatable
kind of matter\ldots Perhaps some basic enlargement of the
ontology of the physical sciences is called for in order to account for
the phenomena of consciousness. \mbox{\cite[p.~36]{Dennett:1991}} 
\end{quote}

If we take the geometric non-representability of conscious phenomena in earnest, how can it help to develop a metageometric formulation of quantum theory and time based on the previously arrived-at experimental notion of quantum-mechanical preparation? Perhaps we should leave the \emph{details} of that myriad of sensations, perceptions, emotions, thoughts, and general mental activities that we call conscious feelings alone. Perhaps, instead of nearsightedly enquiring into their \emph{inner structure} (even worse, into their inner \emph{geometric} structure) after the prevailing winds of a science that progresses through dissection into smaller (but not necessarily simpler) constituents, one should take conscious feelings as primitive notions, and enquire then into their mutual relationships.

At this point, part of Clifford's words, quoted and analysed in Chapter~\ref{ch:Beyond geometry}, on the role of the \emph{relations} between feelings in science strike a resonant chord:

\begin{quote} 
These elements of feeling have relations of nextness or contiguity in space, which are exemplified by the sight-perceptions of contiguous points; and relations of succession in time which are exemplified by all perceptions. Out of these two relations the future theorist has to build up the world as best he may. \cite[p.~173]{Clifford:1886a}
\end{quote} 
Given our purposes, we do not wish to take Clifford's account in a literal sense; that is, we do not want to concern ourselves with relations between feelings (i.e.\ between metageometric things) \emph{in space and time}. But we do nevertheless relate to Clifford's general exhortation to the future theorist, namely, to build up the world as best he may out of the two relations between feelings, \emph{nextness} and \emph{succession}, when the two relations are understood outside a geometric context.

We interpret nextness between conscious feelings to picture the capacity to recognize, distinguish, or discriminate among the contents of consciousness, to the degree that it is possible to tell one conscious feeling from another. According to Dennett (pp.~173--175), the ulterior reason behind this is evolutionary: our unicellular ancestors, in order to survive, had to distinguish their own bodies from the rest of the world, developing a primitive form of selfishness. Nextness, then, does not picture the geometric idea of how close (in space or elsewhere) two conscious feelings are, but tells whether or not they are distinct from each other.

In turn, we take succession between conscious feelings to picture the stream of consciousness, which we experience as a linear, narrative-like sequence of conscious feelings. The inspirational role of the stream of consciousness in the search for knowledge and understanding is not unprecedented. Famously, Turing \citeyear{Turing:1950} inspired himself in his own stream of consciousness to understand how a machine could solve a mathematical problem. In Dennett's words, Turing 

\begin{quote} 
took the important step of trying to break down the \emph{sequence} of his
\emph{mental acts} into their primitive components. ``What do I do,'' he must have asked himself, ``when I perform a computation? Well, first I ask myself which rule applies, and then I apply the rule, and then write down the result, and then I ask myself what to do next, and\ldots'' \cite[p.~212]{Dennett:1991}
\end{quote}
Succession, then, does not picture the geometric idea of how close (in time or elsewhere) two conscious feelings are, but tells whether or not one follows the other.

These two relations between conscious feelings are by nature metageometric. Neither nextness nor succession involve any geometric objects or magnitudes in their depiction; the attribution of shapes and sizes is utterly foreign to these ideas, and we should not hesitate to eschew their introduction if an ill-advised future theorist were to attempt it---such geometric tampering would undoubtedly result in his psychological satisfaction, but most likely not in any gain in physical understanding.

Nextness and succession of feelings form the essence of the psychological conception of time. Time is linked to the sensitivity of consciousness to changes; it arises as the serial chaining (succession) of similar feelings that are yet distinct enough to be discriminated (nextness) from each other. This serial chaining of discriminated feelings---psychological time---constitutes the permanent content of our consciousness, i.e.\ the \emph{permanent present}.

Has psychophysics as heuristics served its purpose? It has. By first bringing to our attention some basic features of our psychological experience, it led to the physical idea of quantum-mechanical preparation; and by alerting us now to the metageometric nature and mutual relations between feelings, it suggests the physical notion of preparation as a metageometric physical concept. It even gives us a clue about what useful metageometric relations may hold among such physical things, and gives us hope that a new quantum-mechanical picture of time may be achievable on their basis. The development of quantum theory along these lines begins in the next main section.

Incidentally, we now see that if, as we suggested in Chapter~\ref{ch:Pregeometry}, Riemann is the father of discrete spacetime and Wheeler is the father of pregeometry, then Clifford is the father of psychophysics.

\subsection{Appraisal}
How necessary is, in general, a psychological approach to physics? At the end of Chapter \ref{ch:Analysis of time}, we criticized cosmology for falling victim to the earthly necessities of human cognitive tools, evidenced especially in its treatment of the concept of time. But we are who we are, the makers of our own tools of thought, so how could the essential limitation imposed by human cognitive tools be overcome? Strictly speaking, it cannot; the ``objective study of the ocean's fauna'' referred to in that chapter is practically unattainable. A theory, any one theory, is (and must necessarily be) a biased filter.\footnote{Cf.\ comments about a ``theory of everything'' in Chapter~\ref{ch:Quantum gravity unobserved}.} But the exploration of our cognitive tools, not in search for absolute ones but better ones, \emph{is} open to enquiry, and it is in it that the chance of progress in science lies.

The clich\'{e} that ``to know the world we must first know ourselves'' becomes especially relevant at the edge of our theoretical picturing abilities. A sign of arrival at this limit can be recognized by the endless, repetitive application of the same traditional concepts to the fields of human enquiry regardless of the results obtained, i.e.\ by the onset of what Lem has called an age of epigonism (see Epilogue). If, as is our contention, science has reached today such a state with its geometric worldview, to overcome it we should look within ourselves in search for better tools of the mind. As we do so, and furthering the previous parable, we may discover, in the utmost darkness and pressure of the inhospitable oceanic depths, that we have a budding ability for photoluminescence, the vestiges of a swim bladder, or even the traces of gills.

No-one has been forced to first look within herself before she could make sense of her surroundings more than the extraordinary deaf-blind woman, Helen Keller. If we heed her remarkable words, perhaps then we shall conserve a greater-than-nil chance to grow humanly wiser:

\begin{quote} 
Groping, uncertain, I at last found my identity, and after seeing my thoughts and feelings repeated in others, I gradually constructed my world of men and God. As I read and study, I find that this is what the rest of the race has done. Man looks within himself and in time finds the measure and the meaning of the universe. \cite[pp.~88--89]{Keller:1933} 
\end{quote}

\section{Metageometric realm} \label{Metageometry}

\subsection{Premeasurement things} \label{Premthings} 
We take preparation or \emph{premeasurement} as the physical basis of a metageometric thing.

We picture a quantum-mechanical, measurable magnitude by means of an observable $A$ (not an operator). The possible results of measuring $A$ are $a_1, a_2, a_3, \ldots$, such that $a_i$ is a real number; these constitute the spectrum $\{a_i\}$ of $A$. Likewise for other observables $B$, $C$, etc.\footnote{In this investigation, the spectrum of observables need not be discriminated into discrete and continuous; all that matters is that the measurement result $a_i$
exist. Both possibilities will be taken into account simultaneously by the metageometric formalism, although the traditional distinction can, if so wished, be made once we have proceeded to the geometric realm later on.} Now a \emph{premeasurement} refers to a selective process which consists of preparation and filtering of parts of a physical system, such that, after it, only certain measurement results are possible. 

\begin{figure}
\begin{center}
\includegraphics[width=90mm]{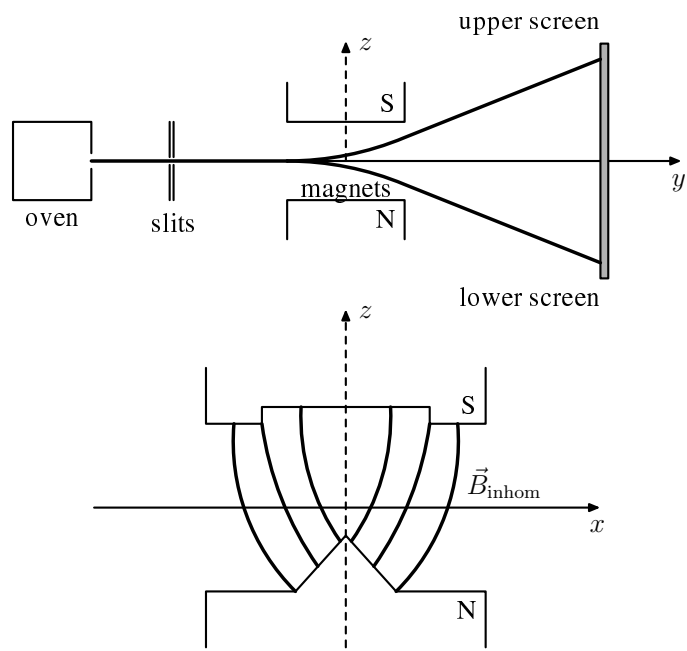}
 \end{center}
\caption[Stern-Gerlach experiment]{Schematic setup of the Stern-Gerlach experiment. Side view (\emph{top}). Front view (\emph{bottom}). For a non-null premeasurement, at least one of the screens (upper or lower) must be removed.} 
\label{sgd}
\end{figure}

The basic idea of a premeasurement is exemplified by (but not limited to) a
Stern-Gerlach experimental setup, shown in Figure \ref{sgd}. If a (lower) screen is placed blocking only those silver atoms with spin value $s_2=-\hbar/2$, while those silver atoms with spin value $s_1=\hbar/2$ are allowed to pass through, then this constitutes a premeasurement of spin magnitude $s_1$ (normally referred to as ``spin up''). Another example of a quantum-mechanical premeasurement is the following. If a beam of light is made to pass through a polarizer, the photons of the outgoing beam will all be polarized perpendicular
to the polarizer's optical axis, while all others will be blocked.

\begin{figure}
\begin{center}
\includegraphics[width=85mm]{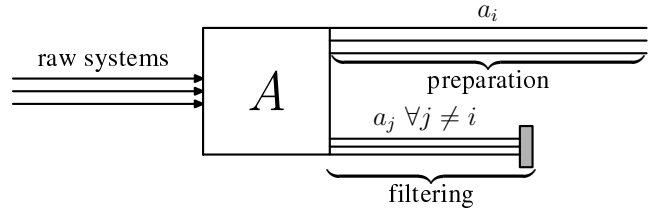}
 \end{center}
\caption[A premeasurement]{Schematic representation of a quantum-mechanical premeasurement thing $\mathcal{P}(a_i)$.} 
\label{pb}
\end{figure}

Premeasurement is here taken as a primitive and irreducible---but under\-standable---concept in quantum ontology on which to build the metageometric edifice of quantum theory. In analogy with a conscious feeling, we picture the basic physical idea in question by means of a metageometric \emph{premeasurement thing} $\mathcal{P}(a)$, characterizing a premeasurement associated with an observable $A$. This consists in the filtering of all parts of a physical system (e.g.\ a beam of particles) except those that succeed in being prepared into $a$-systems, i.e.\ systems on which the result of a possible, later measurement of $A$ gives the real value $a$ (Figure \ref{pb}).

A quantum-mechanical thing is inherently global; in resemblance with conscious events produced by the brain, it is inadequate to ask where and when within the premeasurement device the premeasurement takes place. The global ontology in question here does not respect the spatiotemporal laws of causal processes. Premeasurement things, however, can create prepared physical systems (matter) that are localized in spacetime. 

Furthermore, like a conscious feeling, a premeasurement thing calls for a metageometric depiction. The reason why we should regard a premeasurement thing $\mathcal{P}(a)$ as a metageometric concept is, quite simply, that it is not a geometric object of any sort---it really has no shape for our mind's eye to grab onto---and has no geometric magnitudes attached, such as length or area or volume; neither do premeasurement things have mutual geometric relationships such as overlaps or distances defined between them; expressions such as ``is $\mathcal{P}(a)$ round or square?'', as well as $\| \mathcal{P}(a) \|$, $\langle
\mathcal{P}(a)|\mathcal{P}(b)\rangle$, and $\mathrm{dist}[\mathcal{P}(a),\mathcal{P}(b)]$ are \emph{meaningless} and \emph{utterly foreign} to them. There is nothing in the idea of premeasurement that requires any form of
geometric portrayal; and we should, in fact, reject its geometrization if anyone were to attempt it to satiate his geometric instinct.

Now, what relationships hold among different premeasurement things, and what meaning should we attach to them? Clifford's ideas of nextness and succession from the previous section can now help us uncover these relationships, leading to the mathematical algebra obeyed by premeasurement things, while at the same time pointing at the explicit physical meaning of the operations of this algebra.\footnote{The experimental foundation behind what we have called a premeasurement (and later also a transition) thing has been discussed by Schwinger \citeyear{Schwinger:1959} under the name of measurement symbols, including the algebraic relationships holding between them. The metageometric interpretation of premeasurement things, their relationships to conscious experience and psychological time, the elucidation of the need for complex numbers, the differentiation between meta-, quasi-, and geometric realms, the origin of the state vector, and the rise of metageometric and preparation time, however, are our original contributions. This chapter, then, may be seen as a fresh, physically and psychologically inspired conceptual elaboration on Schwinger's previous ideas.}

\begin{figure}
\begin{center}
\includegraphics[width=85mm]{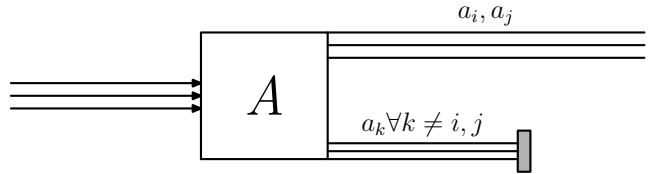}
 \end{center}
\caption[Sum of premeasurement things]{Schematic representation of the sum $\mathcal{P}(a_i)+\mathcal{P}(a_j)$ of two premeasurement things.} 
\label{spt}
\end{figure}

First, we centre our attention on the relation of nextness between feelings. If we take the different premeasurement things, $\mathcal{P}(a_i)$, $\mathcal{P}(a_j)$, etc., associated with the same observable $A$ to be related by the metageometric relation of nextness, we are led to the idea of \emph{adjustable discrimination} (within a single and indivisible premeasurement process) among the different measurement values $a_i$, $a_j$, etc. We embody such a discrimination in the mathematical operation of a sum of premeasurement things as follows. The sum $\mathcal{P}(a_i)+\mathcal{P}(a_j)$ of two premeasurement things describes a selective premeasurement process that discriminates $a_i$- and $a_j$-systems (taken together) from the totality of all possible systems associated with $A$. It works by allowing only $a_i$- or\footnote{Understood as the logical operation OR.} $a_j$-systems to pass through as prepared systems, while blocking all other $a_k$-systems, $k\neq i,j$ (Figure \ref{spt}). Therefore, the more terms in the sum, the less discriminatory the indivisible premeasurement pictured by the whole sum is.

\begin{figure}
\begin{center}
\includegraphics[width=\linewidth]{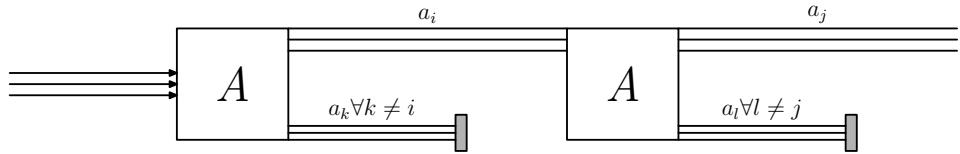}
 \end{center}
\caption[Product of premeasurement things]{Schematic representation of the product $\mathcal{P}(a_j)\mathcal{P}(a_i)$ of two premeasurement things.} 
\label{pfpt}
\end{figure}

Second, we focus on the relation of succession between feelings. It serves as straightforward inspiration for the idea of \emph{linear succession} or \emph{consecutiveness} of premeasurement things, which we embody in the mathematical operation of a product of premeasurement things as follows. The product $\mathcal{P}(a_j)\mathcal{P}(a_i)$ of two premeasurement things describes the succession of two premeasurements, such that some fraction of physical systems is first prepared into $a_i$-systems, and these are consecutively fed into premeasurement $\mathcal{P}(a_j)$ (Figure \ref{pfpt}).

Having established this physical foundation, we proceed to develop the metageometric theory. We start by uncovering the properties of the sum and the product of premeasurement things.

It is straightforward to see that the sum of two premeasurement things is commutative,
\begin{equation} \mathcal{P}(a_i)+\mathcal{P}(a_j)=\mathcal{P}(a_j)+\mathcal{P}(a_i),
\end{equation}
because in the premeasurement $\mathcal{P}(a_i)+\mathcal{P}(a_j)$, $a_i$- or $a_j$-systems pass through \emph{together} irrespective of the order in which they appear. Further, the sum of three premeasurement things is associative,
\begin{equation}
[\mathcal{P}(a_i)+\mathcal{P}(a_j)]+\mathcal{P}(a_k)=
\mathcal{P}(a_i)+[\mathcal{P}(a_j)+\mathcal{P}(a_k)],
\end{equation}
for similar reasons.

\begin{figure}
\begin{center}
\includegraphics[width=80mm]{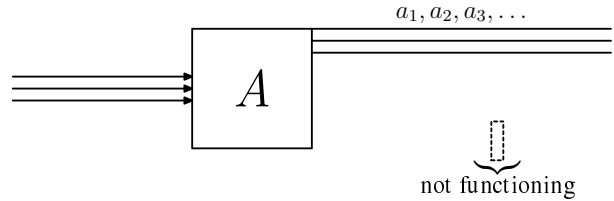}
 \end{center}
\caption[Identity premeasurement]{Schematic representation of the identity premeasurement \mbox{thing $\mathcal{I}$.}} 
\label{identity}
\end{figure}

The identity thing $\mathcal{I}$ appears now naturally as the premeasurement which allows all (prepared) systems to pass through (Figure \ref{identity}). Therefore, the identity thing must be physically equal to the sum of all possible premeasurement things for a given observable: 
\begin{equation} \label{CR}
\sum_i\mathcal{P}(a_i)=\mathcal{I}. 
\end{equation}
We call this the (metageometric) \emph{completeness relation}. By ``all possible
premeasurement things,'' we mean as many as there are different measurement results $a_i$. Thus, the above sum $\sum_i$ includes as many terms $\mathcal{P}(a_i)$, with as closely spaced values $a_i$, as made necessary by observations. It is of no consequence here whether the possible values resulting from a measurement are believed to form a continuous spectrum or not, i.e.\ whether they must be real or may only be rational. It is enough for us to be assured that, no matter what value $a_i$ results from a measurement, its premeasurement appears expressed via $\mathcal{P}(a_i)$.

\begin{figure}
\begin{center}
\includegraphics[width=80mm]{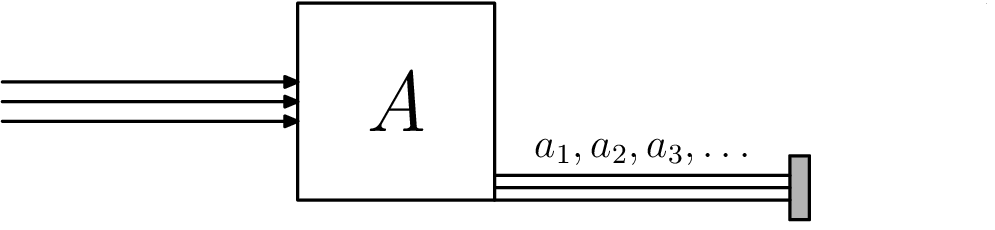}
 \end{center}
\caption[Null premeasurement]{Schematic representation of the null premeasurement thing $\mathcal{O}$.} 
\label{null}
\end{figure}

Similarly, the null thing $\mathcal{O}$ describes the situation in which all systems are filtered and none pass through (Figure \ref{null}).

In the product $\mathcal{P}(a_j)\mathcal{P}(a_i)$ of two premeasurement things associated with the \emph{single} observable $A$, the latter premeasurement $\mathcal{P}(a_j)$ takes in $a_i$-systems and filters all except $a_j$-systems, and so all systems will be filtered unless $i=j$. This product can then be understood as allowing the passage of $a_i$- and\footnote{Understood as  the logical operation AND.} $a_j$-systems. Thus, we obtain the results
\begin{equation}
\mathcal{P}(a_j)\mathcal{P}(a_i)=\mathcal{O} \qquad \mathrm{for}\ i\neq j
\label{product0}
\end{equation}
and
\begin{equation}
\mathcal{P}(a_j)\mathcal{P}(a_i)=\mathcal{P}(a_i) \qquad
\mathrm{for}\ i=j, \label{product1}
\end{equation}
representing the fact that the repeated succession of the same premeasurement is physically (and therefore mathematically) equivalent to itself, i.e.\ $\mathcal{P}(a_i)$ is an idempotent thing. The two results (\ref{product0}) and (\ref{product1}) can be connected with the help of the delta thing $\mathcal{D}_{ij}$:
\begin{equation}
\mathcal{P}(a_j)\mathcal{P}(a_i)=\mathcal{D}_{ij}\mathcal{P}(a_i),
\end{equation}
where
\begin{equation}
\mathcal{D}_{ij}=\left\{ \begin{matrix}
  \mathcal{O}& \mathrm{for}\  i\neq j\\
  \mathcal{I}& \mathrm{for}\  i=j
\end{matrix} \right..
\end{equation}
From this result follows (now even without recourse to physical thinking) that the product of premeasurement things associated with the same observable is both commutative and associative:\footnote{The product is associative in general, i.e.\ $[\mathcal{P}(c)\mathcal{P}(b)]\mathcal{P}(a)=\mathcal{P}(c)[\mathcal{P}(b)\mathcal{P}(a)]$. To show this, the further results (\ref{PremTrans}) and (\ref{TP}) are needed.}
\begin{equation}
\mathcal{P}(a_j)\mathcal{P}(a_i)=\mathcal{P}(a_i)
\mathcal{P}(a_j),
\end{equation}
\begin{equation}
[\mathcal{P}(a_k)\mathcal{P}(a_j)]\mathcal{P}(a_i)=\mathcal{P}(a_k)
[\mathcal{P}(a_j)\mathcal{P}(a_i)].
\end{equation}

We note finally that $\mathcal{I}$ and $\mathcal{O}$ \emph{behave as if they were numbers} with respect to the sum and the product:
\begin{align}
\mathcal{I}+\mathcal{O}&=\mathcal{I},\nonumber\\
\mathcal{P}(a)+\mathcal{O}&=\mathcal{P}(a),\nonumber\\
\mathcal{I}\mathcal{I}&=\mathcal{I},\nonumber\\ 
\mathcal{O}\mathcal{O}&=\mathcal{O},\\
\mathcal{I}\mathcal{P}(a)&=\mathcal{P}(a)=\mathcal{P}(a)\mathcal{I},\nonumber\\
\mathcal{O}\mathcal{P}(a)&=\mathcal{O}=\mathcal{P}(a)\mathcal{O},\nonumber\\
\mathcal{I}\mathcal{O}&=\mathcal{O}=\mathcal{O}\mathcal{I}.\nonumber 
\end{align}
Could, under appropriate circumstances, the things $\mathcal{O}$ and $\mathcal{I}$ be replaced by the numbers 0 and 1?

\subsection{Transition things}\label{TransThings}
We examine next the products $\mathcal{P}(b)\mathcal{P}(a)$ and $\mathcal{P}(a)\mathcal{P}(b)$ associated with two \emph{different} observables $A$ and $B$. In general,
\begin{equation}
\mathcal{P}(b)\mathcal{P}(a)\neq \mathcal{P}(a)\mathcal{P}(b)
\end{equation}
because the first product prepares $a$-systems into $b$-systems, while the second product prepares $b$-systems into $a$-systems. Both situations are equivalent only if $A$ and $B$ are compatible observables (see Section \ref{Mentalities} for details on compatibility).

\begin{figure}
\begin{center}
\includegraphics[width=\linewidth]{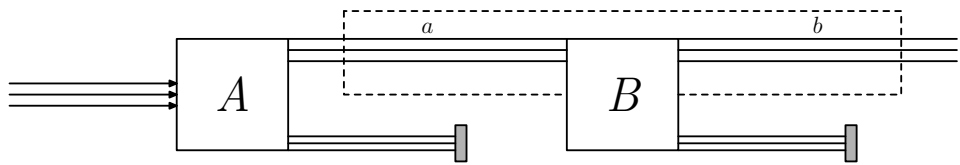}
 \end{center}
\caption[Transition for different observables]{Schematic representation of a quantum-mechanical transition thing $\mathcal{P}(b|a)$ for the case of different observables $A$ and $B$. The transition thing is represented by the region enclosed by the dashed line (the whole of transition box $B$ is included but not its filtering action).} 
\label{tbdo}
\end{figure}

The product of premeasurement things gives motivation for a new type of metageometric thing, namely, transition things. A \emph{transition thing} $\mathcal{P}(b|a)$ represents a \emph{pure} quantum-mechanical transition from $a$-systems to $b$-systems (Figure \ref{tbdo}). We notice that $\mathcal{P}(b|a)$ is not equal to $\mathcal{P}(b)\mathcal{P}(a)$ but a part of it; a transition thing $\mathcal{P}(b|a)$ makes reference to a pure transition in which every $a$-system becomes a $b$-system without any filtering loss, while a product $\mathcal{P}(b)\mathcal{P}(a)$ of premeasurements involves (i) the preparation of some raw systems into $a$-systems with the loss of some, (ii) the transformation of some $a$-systems into $b$-systems, and (iii) the filtering of $a$-systems that failed to become $b$-systems. What we identify as a pure transition (and mark off in dashed lines in the pictures), then, is item (ii) within the three-part process constituting a product of premeasurements.\footnote{The exact  correspondence between the pictorial representation of a transition as part of the product of premeasurements and its physico-mathematical representation will become clearer as we move along; see Eq.~(\ref{connection}).}

\begin{figure}
\begin{center}
\includegraphics[width=95mm]{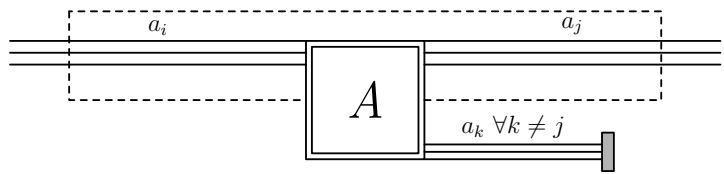}
 \end{center}
\caption[Transition for single observable]{Schematic representation of a quantum-mechanical transition thing $\mathcal{P}(a_j|a_i)$ for the case of a single observable $A$. The double box represents the fact that such a transition thing cannot be implemented with a single premeasurement device (see p.~\pageref{double-box-exp}). The transition thing is represented by the dashed-off region (the whole double box $A$ is included); the filtering action of the box is not a proper part of it.} 
\label{tbso}
\end{figure}

The first feature to notice about a transition thing is that it is in general asymmetric,
\begin{equation}
\mathcal{P}(a|b)\neq \mathcal{P}(b|a),
\end{equation}
since a transition from $b$-systems to $a$-systems is physically different from a transition from $a$-systems to $b$-systems. (When $A$ and $B$ are compatible observables, we shall see that $\mathcal{P}(a|b)$ and $\mathcal{P}(b|a)$ commute.) Asymmetry holds in particular for transition things associated with a single observable,
\begin{equation}
\mathcal{P}(a_i|a_j)\neq \mathcal{P}(a_j|a_i),
\end{equation}
since $a_i$- and $a_j$-systems are necessarily mutually exclusive (for $i\neq j$) (Figure~\ref{tbso}).

\begin{figure}
\begin{center}
\includegraphics[width=\linewidth]{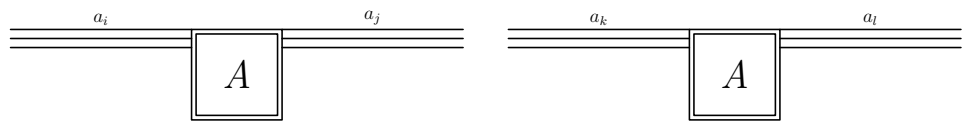}
\end{center}
\caption[Product of transitions for a single obserable]{Schematic representation of the product $\mathcal{P}(a_l|a_k)\mathcal{P}(a_j|a_i)$ of two quantum-mechanical transition things for the case of a single observable $A$. The filtering action not belonging to the transition proper has not been drawn.} 
\label{ptso}
\end{figure}

More detailed knowledge about the nature of transition things can be obtained by analysing the product $\mathcal{P}(a_l|a_k)\mathcal{P}(a_j|a_i)$ for the case of a single observable $A$ (Figure \ref{ptso}). After Schwinger \citeyear[p.~1544]{Schwinger:1959}, since a product describes the outcome of the first process being fed into the second, and the systems associated with a single observable are mutually exclusive, the only way for the above product to be non-null is to have the outcome of the first transition match the input of the second:
\begin{equation} \label{delta}
\mathcal{P}(a_l|a_k)\mathcal{P}(a_j|a_i)=\mathcal{D}_{jk}\mathcal{P}(a_l|a_i).
\end{equation}
From this result follows that the product of transition things associated with a single observable is not commutative except in the mathematically trivial cases $j\neq k$ and $i\neq l$ (products reduce to the null thing), and $i=j=k=l$ (products reduce to the same transition thing).

Physical reasoning can also convince us at this early stage that the product $\mathcal{P}(d|c)\mathcal{P}(b|a)$ of transition things associated with different observables is not commutative either. Despite a lack of proper tools at the moment, one might wish to speculate on the nature of this product. In principle, it should work akin to $\mathcal{P}(d|a)$, although in general $\mathcal{P}(d|c)\mathcal{P}(b|a)\neq \mathcal{P}(d|a)$. To find the manner in which they are connected, we must know \emph{to what degree} $b$- and $c$-systems are compatible (cf.\ the case of a single observable $A$, where different systems are totally incompatible).

\begin{figure} 
\begin{center}
\includegraphics[width=\linewidth]{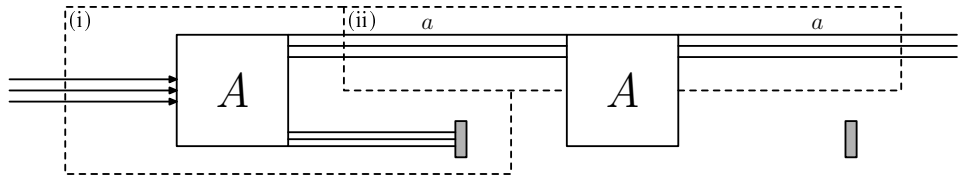} 
\end{center}
\caption[Idempotence]{Schematic representation of the product
$\mathcal{P}(a)\mathcal{P}(a)$ of the same premeasurement thing in
succession for observable $A$. The overall process is presented as
the succession of two parts: (i) the premeasurement of $a$-systems
and (ii) the superfluous transition of $a$-systems into
$a$-systems.} 
\label{prodpapa} 
\end{figure}

Finally, reasoning on the basis of a schematic picture is apt to lead us to a noteworthy connection between a premeasurement thing $\mathcal{P}(a)$ and a transition thing $\mathcal{P}(a|a)$, namely, 
\begin{equation} \label{PremTrans} 
\mathcal{P}(a)=\mathcal{P}(a|a). 
\end{equation} 
In effect, consider the representation of the product $\mathcal{P}(a)\mathcal{P}(a)$ (Figure \ref{prodpapa}), and analyse it in two parts: (i) the premeasurement of raw systems into $a$-systems implemented via preparation and filtering, and (ii) the subsequent preparation of $a$-systems into $a$-systems. We can see that the first part acts as a regular premeasurement $\mathcal{P}(a)$, while the second part---implemented with an \emph{identical device} as the first---acts as a superfluous transition $\mathcal{P}(a|a)$ that outputs its full input; its implementation requires no filtering action, although the filter is present. From this we learn intuitively that
\begin{equation}\label{premtrans2}
\mathcal{P}(a)\mathcal{P}(a)=\mathcal{P}(a|a)\mathcal{P}(a)
\end{equation} 
holds, i.e.\ that once $a$-systems are given, $\mathcal{P}(a)$ and $\mathcal{P}(a|a)$ are equivalent.

On the other hand, a similar reasoning starting from the representation of $\mathcal{P}(b)\mathcal{P}(a)$ (see Figure \ref{tbdo}) does not lead to a similar conclusion. From the picture, we have $\mathcal{P}(b)\mathcal{P}(a)\neq \mathcal{P}(b|a)\mathcal{P}(a)$, because in the implementation of $\mathcal{P}(b|a)$ (second box) an extra filtering mechanism is needed to get rid of some $a$-systems.

What about then those situations in which we do not enjoy the benefit of having $a$-systems as given? Can $\mathcal{P}(a)$ and $\mathcal{P}(a|a)$ be equivalent also then? The equivalence continues to seem plausible. A premeasurement thing $\mathcal{P}(a)$ prepares some raw systems into $a$-systems, while the rest of raw systems are filtered out. On the other hand, the special transition thing $\mathcal{P}(a|a)$ implements a superfluous
transition from $a$-systems to $a$-systems. The physical processes represented by these two things are not now identical, but they appear to be equivalent: to prepare $a$-systems from raw ones with filtering loss is equivalent to having $a$-systems and superfluously preparing them into $a$-systems without any filtering loss. Both processes amount to the same thing: they have the \emph{same output} and are \emph{implemented with the same device}---with respect to $\mathcal{P}(a|a)$, the filtering action of $\mathcal{P}(a)$ balances or neutralizes what is in excess in its raw input.

\begin{figure} 
\begin{center} 
\includegraphics[width=\linewidth]{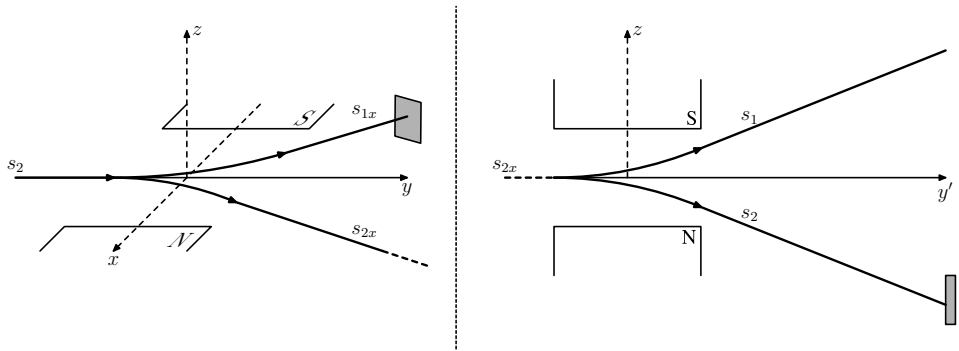}
\end{center}
\caption[Transition with Stern-Gerlach setup]{Realistic
representation of the Stern-Gerlach setup for the performance of a
transition $\mathcal{P}(s_1|s_2)$ corresponding to a single spin
observable $S$. Symbolized as a double box in schematic diagrams
such as Figure \ref{tbso}.} 
\label{rssgd} 
\end{figure}

The meaning of ``the same output with the same device'' can be illustrated as follows. Imagine we want to perform $\mathcal{P}(s_1)$ and $\mathcal{P}(s_1|s_1)$ with a Stern-Gerlach setup. Both can be performed using one single Stern-Gerlach device set to filter out all $s_2$-systems by means of an appropriately placed screen; further, although the input systems of the premeasurement and transition things are different (a raw beam vs.\ a prepared beam), the output produced by the \emph{same} device is the same. On the other hand, it is impossible to perform $\mathcal{P}(s_1|s_2)$ with a \emph{single same} device because $s_1$- and $s_2$-systems are mutually incompatible: given an input of $s_2$-systems, the above device would output no systems at all, i.e.\ it would act as the null thing $\mathcal{O}$. What is required to perform $\mathcal{P}(s_1|s_2)$ is a compound device (drawn as a double box)\label{double-box-exp}; for example, a succession of two Stern-Gerlach devices, with the first one oriented along an axis perpendicular to that of the second such that it will output, say, $s_{1x}$- and $s_{2x}$-systems in equal quantities. Subsequently, either of these, say, $s_{1x}$-systems, can be filtered and the output of $s_{2x}$-systems transformed into $s_1$- and $s_2$-systems by a second device (oriented perpendicularly to the first), which could filter out the unwanted $s_2$-systems and output only $s_1$-systems\footnote{Because transition things represent pure transitions, all systems filtered out in their implementation should not be considered part of the transition thing itself.} (Figure~\ref{rssgd}). Therefore, $\mathcal{P}(s_1|s_1)$ is the only single-observable\footnote{Transitions involving different observables \emph{can} be performed with the same devices used for premeasurements since, in this case, the degree of compatibility is not either all or nothing.} transition that can be performed with the same device used for $\mathcal{P}(s_1)$.

We decide to trust this physical intuition and accept Eq.~(\ref{PremTrans}) as an equivalence valid in general for any situation. We remain alert, however, to the possibility that it may lead to inconsistencies later on in the development of the theory. What we shall find is that, after the introduction of the transition postulate in the next section, the origin of $a$-systems becomes in each and every case irrelevant, and that any situation can be recast in the spirit of Eq.~(\ref{premtrans2}).

With this, the metageometric foundation of quantum theory is ready. Ahead lies the task of showing how a description of quantum phenomena arises from it, thereby revealing the metageometric origin of their characteristic features. However, quantum theory cannot be derived on the basis of metageometric things and their sums and products alone: where would numbers come from given that quantum-mechanical things have nothing to do with them? The introduction of a main postulate will therefore be needed, but its being a postulate does not mean that its physical meaning must remain unclear.

\section{Quasigeometric realm} \label{Quasigeometry}

\subsection{The transition postulate and the transition function}
We take as a starting point relation (\ref{delta}), since it appears that the only way in which numbers could be introduced in a physically reasonable way is via the delta thing. Let us replace the delta thing $\mathcal{D}_{jk}$ by Kronecker's delta $\delta_{jk}$ to get
\begin{equation}\label{delta-replacement}
\mathcal{P}(a_l|a_k)\mathcal{P}(a_j|a_i)=\delta_{jk}\mathcal{P}(a_l|a_i),
\end{equation} 
thus starting with the two most natural numbers there can be, 0 and 1. The former number represents the complete absence of an overall transition (total incompatibility between different systems associated with $A$), while the latter represents the existence of an overall transition in which no systems are lost.\footnote{As we saw in Chapter~\ref{ch:Beyond geometry}, Eddington's \citeyear[p.~188]{Eddington:1920} own quest for the nature of things led him to an analysis of spacetime point-events and the intervals between them. He argued that the intervals might have \emph{non-quantitative} aspects; that the \emph{individual} intervals probably cannot be captured by rulers and clocks, but that they could simply exist (1) or not (0). Thus, Eddington's manner of introducing numbers into an outspokenly non-quantitative framework is, at least superficially, akin to the present one. Eddington's either existent or non-existent primitive intervals are analogous to the transition functions $\langle a_j|a_i\rangle=\delta_{ij}$ (to be introduced) for the case of a single observable $A$. These transition functions tell whether something occurs or not---in a premeasurement of $A$, its own systems are accepted completely (1) or not at all (0).}

This replacement is quite significant since here a thing is being replaced by a number, which is an object of a totally different nature. However, the replacement is acceptable and does not lead to inconsistencies because the delta thing always appears multiplying other things; therefore, the nature of all equations is preserved when we consider that, in all cases, things continue to be equated to things: 
\begin{eqnarray} 0\mathcal{P}(a_l|a_i)&=&\mathcal{O},\\ 1\mathcal{P}(a_l|a_i)&=&\mathcal{P}(a_l|a_i). 
\end{eqnarray} 
A question, however, remains. Where could all other numbers come from? This is a critical stage of this investigation.

Generalizing the idea behind Eq.~(\ref{delta-replacement}), we postulate after Schwinger \citeyear[p.~1545]{Schwinger:1959} that 
\begin{equation} \label{TP}
\mathcal{P}(d|c)\mathcal{P}(b|a)=\langle c|b\rangle \mathcal{P}(d|a), 
\end{equation} 
where the \emph{notation} $\langle c|b\rangle$ represents a number. We call this the \emph{transition postulate} and name $\langle c|b \rangle$ the \emph{transition function} (or transformation function).\footnote{Function, from Latin \emph{fungi}: to perform, to execute an action. The transition function is a function in this etymological sense; it is not to be interpreted as an explicit function $f(b,c)$ of the real numbers $b$ and $c$.} If we are to throw new light on quantum ontology, we \emph{must not} interpret the notation $\langle c|b\rangle$ as an inner product---not just yet and without further qualification---but simply as a manner of marking a physical action involving $b$- and $c$-systems. The transition function $\langle c|b\rangle$ is a number that is taken to commute with premeasurement and transition things, since these are objects of completely different natures.

But what \emph{kind} of number is the transition function? Natural, rational, real, complex, hypercomplex? This is not something that we can postulate without a reason; to say what kind of number the transition function may be, we must let the physics involved naturally lead to an answer. We make and justify this choice in Section \ref{Transition-function postulate}.

Physically, the transition postulate describes what part of the group of $b$-systems associated with observable $B$ is accepted in the premeasurement associated with observable $C$, succeeding to become $c$-systems. This fraction of systems, described by a number, is the share of systems that premeasurements $\mathcal{P}(b)$ and $\mathcal{P}(c)$ have in common. One can see from this more general perspective that the earlier result that $a_i$- and $a_j$-systems are either fully compatible or fully incompatible follows now as the special case of a single observable in the form $\langle
a_j|a_i\rangle=\delta_{ij}$.

We remark that the characterization in question here is \emph{qualitative}---``part'' is a number. We would only be forced to call this characterization quantitative if the transition function turned out to be a non-negative real (or natural or rational) number but not otherwise, because other kinds of numbers, e.g.\ complex numbers, do not picture size---only their moduli do.

The above wording ``have in common'' suggests the possibility of establishing a connection---if so we wished to do---between the transition function and the inner product of two arrow-like vectors. The way in which this geometric interpretation of the transition function arises will teach us how state vectors are born---and reveal them as physically insubstantial concepts. At this point, however, we have not yet introduced any geometric concepts, and the transition function must be understood simply as a concept---a
number---in the quasigeometric realm that serves to bridge the chasm between a genuinely metageometric realm and a genuinely geometric one.

Is now the transition postulate in harmony with the equality we introduced in the previous section, namely, $\mathcal{P}(a)=\mathcal{P}(a|a)$? It is indeed, because, on the basis of the transition postulate, the method by which $a$-systems are first obtained becomes now irrelevant for any situation. In effect, using both Eq.~(\ref{PremTrans}) and Eq.~(\ref{TP}), we have $\mathcal{P}(b|a)=\mathcal{P}(b|a)\mathcal{P}(a)$ because $\mathcal{P}(a)=\mathcal{P}(a|a)$ and $\langle a|a\rangle=1$. As a result, we can now write 
\begin{equation}\label{connection}
\mathcal{P}(b)\mathcal{P}(a)= \langle b|a\rangle
\mathcal{P}(b|a)\mathcal{P}(a), 
\end{equation} 
whereby we obtain a complete correspondence, element by element, between the pictorial and the physico-mathematical representation of the product of two different premeasurements. Comparing with Figure \ref{tbdo}, we see that each element on the right-hand side of Eq.~(\ref{connection}) is pictured in it: (i) a premeasurement $\mathcal{P}(a)$ of $a$-systems including filtering, (ii) a pure transition $\mathcal{P}(b|a)$ from $a$-systems to $b$-systems, and (iii) a qualification $\langle b|a\rangle$ of the part of $a$-systems that manage to become $b$-systems symbolized by the second filtering action. The pictorial-algebraic parallelism is now complete.

Before concluding this section, we mention two noteworthy properties
of the transition function. The first one results as follows. On the
one hand, 
\begin{equation}\label{tfprop1}
\mathcal{P}(b)\mathcal{P}(a)=\mathcal{P}(b|b)\mathcal{P}(a|a)=\langle b|a\rangle \mathcal{P}(b|a); 
\end{equation} 
on the other hand, introducing the identity thing $\sum_i \mathcal{P}(c_i)$, we
get 
\begin{align} \mathcal{P}(b)\left[\sum_i \mathcal{P}(c_i)\right]\mathcal{P}(a)&=\mathcal{P}(b|b)\left[\sum_i \mathcal{P}(c_i|c_i)\right]\mathcal{P}(a|a) \nonumber\\
&=\sum_i \langle b|c_i\rangle \langle c_i|a\rangle \mathcal{P}(b|a).
\end{align}
Therefore, 
\begin{equation} \label{DR}
\langle b|a\rangle=\sum_i \langle b|c_i\rangle \langle c_i|a\rangle. 
\end{equation} 
We call this the (quasigeometric) \emph{decomposition relation}. It tells us that the part of $a$-systems that succeeds in being transformed into $b$-systems can be expressed via transitions to and from \emph{all possible} $c_i$-systems.\footnote{This proof assumes that the sum of things is distributive with respect to the product. Since the expression $\mathcal{P}(c_i)+\mathcal{P}(c_j)$ is irreducible, this result cannot be proved \emph{mathematically}; however, we do no find this problematic at all. That the sum is distributive with the product can be seen to hold by considering the meaning of the
expression \emph{physically}. We should not demand anything else from a physical theory. Alternatively, one may choose to impose this as a mathematical axiom, but we consider this to be a pointless procedure in the present physical context.}

A similar procedure yields a second property, namely, \begin{equation} \mathcal{P}(b|a)=\sum_{i,j}\langle d_j|b\rangle \langle a|c_i\rangle \mathcal{P}(d_j|c_i), \end{equation} which tells that a transition from $b$-systems to $a$-systems can be built out of transitions from $c_i$-systems to $d_j$-systems, so long as we know what part of $c_i$-systems is compatible with $a$-systems and what part of $b$-systems is compatible with $d_j$-systems. We see, then, how the transition functions have the role of building a linear combination of transition things; transition functions tell what qualitative weight each transition thing possesses in its combination with other such things, depending on what part every pair of physical systems have in common.

\subsection{A lead on probability}
We found previously that, in the special case of a single observable, $\langle a_j|a_i\rangle$ is either 0 or 1. One may be tempted by this fact and  simple-mindedly interpret the transition function to be the probability for the occurrence of a quantum-mechanical transition, with each $\langle b_i|a\rangle$ assumed to be a real number between 0 and 1, and $\sum_i \langle b_i|a\rangle=1$. This interpretation, however, is not based on any physical reasoning and is, in fact, completely misguided.

In order to obtain a sensible interpretation of the transition function, and one that is also in harmony with property (\ref{DR}), we must not blindly assume a physical meaning for it, but rather find it as it naturally arises from a further analysis of premeasurement things---especially, of a succession of them after Schwinger \citeyear[p.~1547]{Schwinger:1959}.

In the physical situation described by the product $\mathcal{P}(b)\mathcal{P}(a)$, premeasurement $\mathcal{P}(b)$ in general destroys a part of the information obtained from premeasurement $\mathcal{P}(a)$ because all those $a$-systems that fail to perform a transition into $b$-systems are lost. The part of systems that is not lost, or the degree of compatibility between $a$- and $b$-systems, is described by the transition function $\langle
b|a\rangle$. The expression \begin{equation} \label{transitionf} \mathcal{P}(b)\mathcal{P}(a)=\langle b|a\rangle \mathcal{P}(b|a) \end{equation} tells us precisely this, namely, that the succession of the two premeasurements $\mathcal{P}(a)$ and $\mathcal{P}(b)$ is equivalent to a (pure) transition $\mathcal{P}(b|a)$ from $a$-systems to $b$-systems qualified by $\langle b|a\rangle$, the degree of compatibility between the systems in question.

\begin{figure} 
\begin{center}
\includegraphics[width=\linewidth]{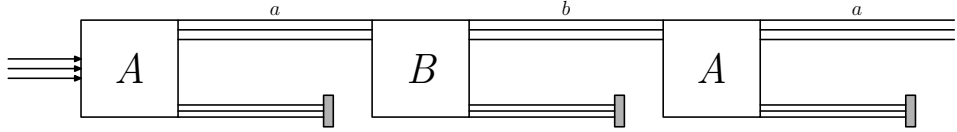} 
\end{center}
\caption[Probability from premeasurement things]{Schematic
representation of the succession
$\mathcal{P}(a)\mathcal{P}(b)\mathcal{P}(a)$ of premeasurement
things suggesting the concept of quantum-mechanical probability.}
\label{probability} 
\end{figure}

If now we perform premeasurement $\mathcal{P}(a)$ again in order to regain $a$-systems (Figure \ref{probability}), we can see what kind of disturbance premeasurement $\mathcal{P}(b)$ has caused; we find 
\begin{equation} 
\mathcal{P}(a)\mathcal{P}(b)\mathcal{P}(a)=\langle a|b\rangle
\langle b|a\rangle \mathcal{P}(a). 
\end{equation} 
\emph{If} $\langle a|b\rangle \langle b|a\rangle$ was a non-negative real number, this result could be physically interpreted to \emph{quantify} the disturbance caused by $\mathcal{P}(b)$, i.e.\ to be telling us \emph{what the chances are} that a given, individual $a$-system will survive its becoming a $b$-system and be recovered again as an $a$-system. In other words, $\langle a|b\rangle \langle b|a\rangle$ could quantify the equivalence between a premeasurement of $a$-systems having passed through $b$-systems, and a single
premeasurement of $a$-systems. These two different situations are physically the same only in those cases when the disturbance introduced by $\mathcal{P}(b)$ is cleared successfully.

In the special case of a single observable, the results obtained are
in harmony with this possible interpretation. In effect,
\begin{equation}\label{SC} 
\mathcal{P}(a_i)\mathcal{P}(a_j)\mathcal{P}(a_i)= \langle a_i|a_j\rangle \langle a_j|a_i\rangle \mathcal{P}(a_i) =\delta_{ij}\mathcal{P}(a_i)
\end{equation} 
would tell us that the chances of an $a_i$-system surviving a disturbance introduced by $\mathcal{P}(a_j)$ is null unless $i=j$, in which case the chances of survival are guaranteed.

The role of the product of metageometric things as a lead on transition probabilities cannot be overemphasized. It is our intuitive grasp of the situation represented by $\mathcal{P}(a)\mathcal{P}(b)\mathcal{P}(a)$ as a
recovery operation, and our intuition that the recovery of an $a$-system will sometimes fail and sometimes succeed, that gives us reason to think that, perhaps, $\langle a|b\rangle \langle b|a\rangle$ can be interpreted as a probability. 

The analysis of the role of the sum of premeasurement things in the possible birth of probability adds more justification to this interpretation. What results would we obtain if we allowed for the possibility of $a$-systems being disturbed through \emph{all} possible (mutually exclusive) $b_k$-systems? We find that the supposed probability of an $a$-system surviving this least selective disturbance introduced by $\sum_k \mathcal{P}(b_k)$ adds up to unity. In effect, because we have $\sum_k \mathcal{P}(b_k)=\mathcal{I}$ and $\mathcal{P}(a)$ is idempotent, we find 
\begin{equation} 
\mathcal{P}(a)=\mathcal{P}(a)\left[\sum_k \mathcal{P}(b_k)\right]\mathcal{P}(a) 
=\sum_k \langle a|b_k\rangle \langle b_k|a\rangle \mathcal{P}(a), 
\end{equation} 
so that 
\begin{equation}
\label{normalization} 
\sum_k \langle a|b_k\rangle \langle b_k|a\rangle =1. 
\end{equation} 
This result is consistent with property (\ref{DR}). In this manner, the product of metageometric things is compatible with the idea of probability for single transitions, and the sum of products with the idea of probability for a set of mutually exclusive transitions. Unexpectedly, the interpretation of the product and sum of \emph{things} agrees with the interpretation of the product and sum of \emph{numbers} in probability theory.

If this interpretation is workable and its lead on probability correct, then there is no need to postulate controversial state vectors---with their mysterious collapses---here. Probability would arise not mysteriously but naturally, from the very physical meaning of the situation under examination.

\subsection{The transition-function postulate}
\label{Transition-function postulate}
The hurdles that must still be overcome on the way to probability are (i) to reveal and justify the nature of the number representing the transition function $\langle b|a\rangle$ guided not only by the picture of the recovery of an $a$-system and its lead on probability, but also by the physical meaning of the transition function and a deeper insight into the significance and utility of numbers in physics; and (ii) to make sure that the product $\langle a|b\rangle \langle b|a\rangle$ we have hypothetically identified with a probability is indeed a real number between \mbox{0 and 1}.

To move forwards, we ask whether any kind of relation could exist between the two numbers $\langle b|a\rangle$ and $\langle a|b\rangle$. In other words, considering that they both picture related processes, namely, a transition from $a$-systems to $b$-systems and viceversa, could one be obtained from the other? And if so, on what grounds?

Consider the following reasoning within the realm of real numbers. Let $r$ be a real number, and let $r^\star=-r$ denote the \emph{dual counterpart} of $r$. Next we ask (i) what is the physical meaning of operation ``$\star$''? and (ii) what special properties does it inherit from its physical meaning? The first question is straightforward to answer. Operation ``$\star$'' describes the \emph{reversal} of a \emph{simple physical process}. For example, if $r$ represents gaining $r$ apples, $r^\star$ represents losing or owing $r$ apples; if $r$ symbolizes a force of intensity $r$ pulling in a certain direction, $r^\star$ symbolizes a force of the same intensity acting in the opposite direction; etc. As we can see, the physical meaning of operation ``$\star$'' is very close to everyday experience. 

Since ``$\star$'' represents the reversal of a physical process, and since to reverse a physical process twice brings us back to the starting point, this operation must be its own inverse, $r^{\star\star}=r$ (recovery of original transaction). Moreover, the conflation of a process in one direction with another in the opposite direction must produce no physical process at all; this finds expression in the property $r^\star r=0$ (null net transaction),\footnote{Expression $r^\star r$ is to be read $-r+r$.} where $0$ is a special real number with the property $0^\star=0$ because, roughly speaking, $0$ points in no direction.

The previously unearthed physical meaning of $\langle b|a\rangle$ as giving the \emph{direction} of a quantum-mechanical transition from $a$- to $b$-systems gives us now an essential clue, because the transition function, through its explicit expression, likewise tells us that $\langle a|b\rangle$ represents a quantum-mechanical transition in a \emph{direction opposite} to that of $\langle b|a\rangle$.

This fits the meaning of the $\star$ operation, and so we ask: could the transition function, then, be a real number and ``$\star$'' the relation we are looking for? We see immediately that it cannot, because we would then have that $\langle a|b\rangle \langle b|a\rangle=-r+r \equiv 0$, i.e.\ always null.\footnote{We read $ \langle b|a\rangle^\star \langle b|a\rangle$ as $-\langle b|a\rangle + \langle b|a\rangle$ according to the meaning of the $\star$-operation.} Returning to the product $\mathcal{P}(a)\mathcal{P}(b)\mathcal{P}(a)$, this would mean that the original $a$-system is always irrecoverable for arbitrary observables $A$ and $B$. But this runs counter to our intuition of the original physical situation. And so would a similar use of natural and rational numbers. We must try harder.

A clue to this problem comes from realizing that the kind of physical process represented by the transition function is farther removed from everyday experience than those characterized by real numbers. The quantum-mechanical process in question simply cannot be grasped by real numbers, which have been designed and introduced into physics to picture classical phenomena, phenomena
whose physical nature is more intuitive than that of the quantum world. Real numbers have been designed to picture gaining and losing, coming and going, pulling and pushing, and other such classical transactions. Quantum transitions have nothing to do with these actions; they are transactions of a different kind, which must, accordingly, be grasped in terms of a different kind of numbers. If simpler numbers also fall short of the task at hand, then how about the next kind of number in line, complex numbers?

Save for an increment in the complexity of the picturing number, let us conjecture that the underlying logic of quantum-mechanical transactions is nonetheless the same as that of classical ones and see where it leads. Let then the transition function be a complex number, and let each pair of complex numbers representing opposite transitions be \emph{dual counterparts}, linked by a new operation of dual correspondence ``$\ast$,'' $\langle a|b\rangle=\langle b|a\rangle^\ast$. Operation ``$\ast$,'' as before, must symbolize the reversal of the physical process in question, and must therefore share its properties with ``$\star$.'' A double reversal of a quantum-mechanical transition must bring us back to the original transition, $\langle b|a\rangle^{\ast\ast}=\langle b|a\rangle$, i.e. ``$\ast$'' is its
own inverse; and the conflation of a quantum-mechanical transition from $a$- to $b$-systems with another from $b$- to $a$-systems must produce no transition at all, $\langle b|a\rangle^\ast \langle b|a\rangle= r \in \mathbb{R}$. 

\begin{figure} 
\begin{center}
\includegraphics[width=80mm]{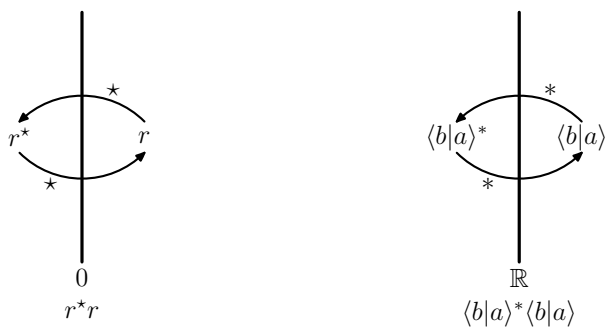} 
\end{center}
\caption[Dual counterparts]{The operation ``$\star$'' generates the dual counterpart $r^\star$ of a real number $r$ (\emph{left}). The operation ``$\ast$'' generates the dual counterpart $\langle b|a\rangle^\ast$ of a complex number $\langle b|a\rangle$ (\emph{right}). The meaning of both is to reverse a physical action; the former a classical, the latter a quantum action. Both are, as they should, their own inverses, and the conflation of a physical action with its reverse results in no action at all.}
\label{complexconj} 
\end{figure}

Why a real number? Because, like 0 satisfies $0^\star=0$ and fails to picture a classical transaction, real numbers have the property $r^\ast=r$, as they fail to represent any quantum-mechanical transition. We know this fact from two separate sources. We just learnt heuristically that real numbers are inadequate to represent quantum-mechanical transactions; and we also know from the earlier result $\langle a_j|a_i\rangle= \delta_{ij} \in \mathbb{R}$ that transition functions that reduce to real numbers represent no proper transition: in those cases, $a_i$-systems either remained $a_i$-systems (1) or they failed to become $a_j$-systems (0) (Figure \ref{complexconj}). Moreover, because $\langle a_j|a_i\rangle^\ast \langle a_j|a_i\rangle = 0$ (cf.\ Eq.~(\ref{SC})) represents the extreme case in which the systems involved in the recovery operation have nothing in common ($i\neq j$), we conjecture that this is the lower bound of the more general real number $\langle b|a\rangle^\ast \langle b|a\rangle$, characterizing a recovery operation where the systems involved are partially compatible; in symbols, $\langle b|a\rangle^\ast \langle b|a\rangle \geq 0$. 

With their underlying physical meaning clarified, complex numbers and complex conjugation lose their aura of mystery, while at the same time a new dimension is added to the concept of isolated complex numbers, namely, that every meaningful complex number has an equally physically meaningful dual counterpart characterizing an opposite quantum-mechanical transition.\footnote{We saw in Section \ref{GQMec} that Penrose understood complex conjugation as reversing the direction \emph{in time} of the flow of quantum-mechanical information. The present findings go deeper into the nature of complex conjugation, as they do not rely on any idea of time, nor on the inessential geometric decomposition---and ensuing geometrization as an element of Argand's plane---of a complex number $\langle b|a\rangle$ into $x+iy$.} Furthermore, with this, the quasigeometric realm of the transition function re-earns its name, since it introduces the notion of direction---a look-alike of geometric thought---without yet, however, introducing any notions of
shape or size.

Now how about numbers more complicated than complex ones? Could these be suitable or even more suitable? Taking $\langle b|a\rangle$ to be a more complicated number such as a quaternion,\footnote{A quaternion is a number $q$ of the form $x_0+ix_1+jx_2+kx_3$, where $i^2=j^2=k^2=-1$, $ij=-ji=k$, $jk=-kj=i$, and $ki=-ik=j$.} would not lead to difficulties as far as the rise of probability is concerned, because quaternions behave like complex numbers with respect to complex conjugation: $qq^\ast=q^\ast q \geq 0$. However, their adoption leads to difficulties in other areas. At the meta- and quasigeometric level, the product of three premeasurement things fails to be associative,\footnote{Taking the commutation of transition functions (in this case, quaternions) and things to be a reasonable assumption, we find:
\begin{multline*}
[\mathcal{P}(c)\mathcal{P}(b)]\mathcal{P}(a)=\langle c|b\rangle
\mathcal{P}(c|b)\mathcal{P}(a)=\langle c|b\rangle \langle
b|a\rangle\mathcal{P}(c|a)\neq\\
\mathcal{P}(c)[\mathcal{P}(b)\mathcal{P}(a)]= \mathcal{P}(c)\langle
b|a\rangle\mathcal{P}(b|a)=\langle b|a\rangle \langle c|b\rangle
\mathcal{P}(c|a),
\end{multline*}
since quaternions do not commute. At this point, one might want to invent an extra ``chaining rule'' stating that the only correct order of multiplication for transition functions is one in which the left-hand system on the right matches the right-hand system on the left (cf.\ $b$ in first product). This fix appears artificial and, although it does not conflict with most of the forthcoming calculations at the meta- and quasigeometric level, it does make the findings of Section \ref{Geometric traces} impossible to obtain.} contrary to the results established in Section \ref{Premthings}. And, although inessential to our main concerns, quaternions do not work at the traditional geometric level either because, their product being non-commutative, they do not form a field and cannot therefore act as the scalars of a (Hilbert) vector space. The choice of other hypercomplex numbers, such as biquaternions, octonions, hypercomplex numbers proper, etc., is equally troublesome.

However, this is not say that the need for complex numbers, to the exclusion of any other types of more general number, remains thus proved. These findings rather suggest that only complex numbers are adequate for the \emph{simple} framework of this metageometric formulation of quantum theory, as well as for that of the
traditional formulation. One suspects that if either of these formulations could be modified into more complex alternatives, quaternions (or other hypercomplex numbers) might be suitable for a quantum theory just as well. The difficulty lies in fathoming out such a theory; and the question remains as to why attempt this feat when a simpler theory that uses complex numbers already achieves the desired explanatory goals.

Postulates and hypotheses should not be feared in physics---so long as we understand their physical origin and meaning (cf.\ Einstein's geodesic hypothesis). Resting assured that we now understand the underlying reason of being of complex numbers, we postulate that the transition function is a complex number, and that the relation between $\langle a|b\rangle$ and $\langle b|a\rangle$ is 
\begin{equation}\label{TFP} 
\langle b|a\rangle^\ast=\langle a|b\rangle. 
\end{equation} We call this the \emph{transition-function postulate}. It attributes to complex conjugation (later on: dual correspondence in general) the role of \emph{reversing the direction of a quantum-mechanical transition}.\footnote{Dr.~Jaroszkiewicz has brought up the issue of the role of the observer; in particular, in relation to reversals of quantum processes: if the observer is an inalienable part of a quantum system (involving also systems under observation and apparatuses), how could he (or his state), too, be said to be reversed? From the metageometric standpoint, however, the role of human beings is of an epistemological rather than ontological nature. Human beings are essential to quantum physics because quantum theory must be tailored to do justice to both quantum phenomena and human cognition and methodological capabilities. Quantum theory is then to be an explicit explanation of quantum phenomena by humans for humans but, in the metageometric setting, their role does not extend to an essential modification of microscopic systems via ``acts of observation.'' Here humans do not act as observers in the usual quantum-mechanical sense of the word; that is, a human looking or not looking has no effect upon outcomes---only their preparing microscopic systems with macroscopic devices does.}

The interpretation of $\langle a|b\rangle \langle b|a\rangle$ as a
probability is now fully justified. We have that
\begin{equation}\label{nonnegprob} 
\mathrm{Pr}(b|a)=\langle
a|b\rangle\langle b|a\rangle=\langle b|a\rangle^\ast \langle
b|a\rangle =:|\langle b|a\rangle|^2 \geq 0,\footnote{We have now identified the real product of a transition function and its dual counterpart with the squared modulus of a complex number, turning it into an arrow. This is done at this late stage partly for notational convenience, and partly to give a familiar look to the traditional probability results. So far as this section is concerned, nothing else should be read into this mere notation: we initially set out to avoid the geometrization of complex numbers---and so we did, with positive new results. This geometric-ridden notation, however, becomes essential in the next section, because it is it that inspires the rise of the geometric state vector.}
\end{equation}
where the probability of each individual transition is symmetric: \begin{equation} \mathrm{Pr}(b|a)=\mathrm{Pr}(a|b), \end{equation} i.e.\ the probability of an $a$-system becoming a $b$-system is equal to that of a $b$-system becoming an $a$-system. Moreover, for an exhaustive set of mutually exclusive outcomes---a single $a$-system can transform into any of all possible
$b_i$-systems---we recover from Eq.~(\ref{normalization}) the usual result \begin{equation} \sum_i \mathrm{Pr}(b_i|a)=\sum_i |\langle b_i|a\rangle|^2 =1. \end{equation} This, together with Eq.~(\ref{nonnegprob}), finally implies 
\begin{equation} 
0\leq \mathrm{Pr}(b|a)\leq 1. 
\end{equation}

Before closing this section and moving onto the geometric realm, it
is worthwhile to analyse some properties of probabilities, seen as
direct consequences of the product of premeasurement things. We
study three different cases:
\begin{enumerate} 
\item The series of premeasurements $\mathcal{P}(c)\mathcal{P}(b)\mathcal{P}(a)$ (here the filter associated with $\mathcal{P}(b)$ is on) involves a
transformation of $a$-systems into $c$-systems going through a
determined $b$-system (Figure \ref{probability1}). Because the
transition from $a$-systems to $b$-systems is independent of the
second transition from $b$-systems to $c$-systems, the probability
$\mathrm{Pr}(c|b|a)$ of the whole event can be expressed as
$\mathrm{Pr}(c|b)\mathrm{Pr}(b|a)$. This gives 
\begin{equation}
\mathrm{Pr}(c|b|a)=\langle b|c\rangle \langle c|b\rangle \langle
a|b\rangle \langle b|a\rangle =|\langle c|b\rangle\langle b|a\rangle|^2. 
\end{equation}

\begin{figure}
\begin{center}
\includegraphics[width=\linewidth]{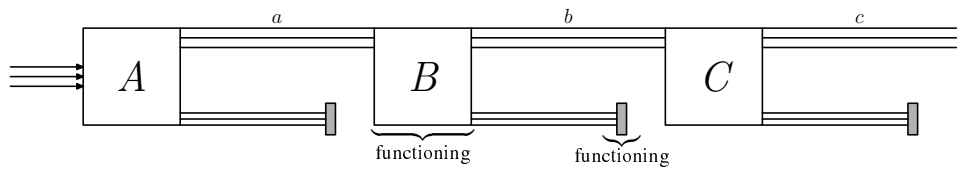}
\end{center}
\caption[Probability (case 1)]{Schematic representation of the succession $\mathcal{P}(c)\mathcal{P}(b)\mathcal{P}(a)$ of premeasurement things transforming $a$-systems into $c$-systems passing through a certain $b$-system.} 
\label{probability1}
\end{figure}

\item The series of premeasurements $\mathcal{P}(c)\mathcal{I}\mathcal{P}(a)$ (here the filter associated with $\mathcal{I}$ is off) involves a transformation of $a$-systems into $c$-systems going through either one of all possible $b_i$-systems (Figure \ref{probability2}). Because the different possibilities of passing through each $b_i$-system are mutually exclusive, we find
\begin{align}
\sum_i \mathrm{Pr}(c|b_i|a)&=\sum_i \langle b_i|c\rangle \langle c|b_i\rangle \langle a|b_i\rangle \langle b_i|a\rangle \nonumber\\
&=\sum_i |\langle c|b_i\rangle\langle b_i|a\rangle|^2.
\end{align}

\begin{figure}
\begin{center}
\includegraphics[width=\linewidth]{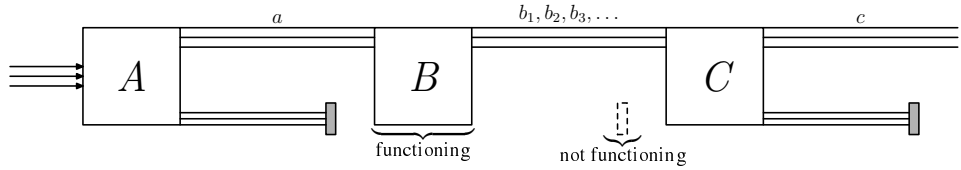}
\end{center}
\caption[Probability (case 2)]{Schematic representation of the succession $\mathcal{P}(c)\mathcal{I}\mathcal{P}(a)$ of premeasurement things transforming $a$-systems into $c$-systems passing through all possible $b_i$-systems.} 
\label{probability2}
\end{figure}

\item The series of premeasurements $\mathcal{P}(a)\mathcal{P}(c)$ (here the premeasurement associated with $B$ is totally absent) involves a straight transformation of $a$-systems into $c$-systems, apparently without going through any $b_i$-system (Figure \ref{probability3}). However, using the quasigeometric decomposition relation (\ref{DR}), we find that the probability
\begin{equation}
\mathrm{Pr}(c|a)=\langle a|c\rangle \langle c|a \rangle
\end{equation}
is also equal to
\begin{eqnarray}
\mathrm{Pr}(c|a)&=&\sum_i\sum_j  \langle a|b_i\rangle \langle b_i|c\rangle \langle c|b_j \rangle \langle b_j|a \rangle \nonumber\\
&=&\sum_i  \langle a|b_i\rangle \langle b_i|c\rangle \langle c|b_i \rangle \langle b_i|a \rangle + \sum_i\sum_{j\neq i}  \langle a|b_i\rangle \langle b_i|c\rangle \langle c|b_j \rangle \langle b_j|a \rangle \nonumber\\
&=&\sum_i \mathrm{Pr}(c|b_i|a)+ \mathrm{interference\ terms}.
\end{eqnarray}

\begin{figure}
\begin{center}
\includegraphics[width=\linewidth]{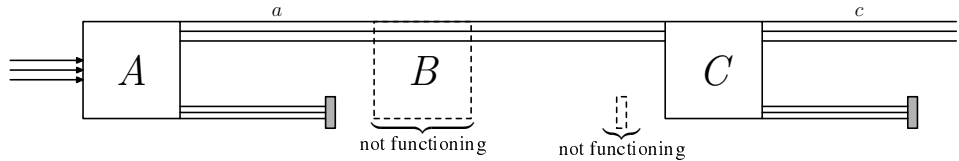}
\end{center}
\caption[Probability (case 3)]{Schematic representation of the succession $\mathcal{P}(a)\mathcal{P}(c)$ of premeasurement things transforming $a$-systems into $c$-systems without passing through any $b_i$-system.} 
\label{probability3}
\end{figure}
\end{enumerate}
The extra cross terms that appear with respect to situation (2.) represent the interference of all possible ways of passing through $b_i$-systems taken in pairs. This is as if the systems, besides being able to pass through any $b_i$-system separately, were also able to pass through all of them at once when no premeasurement associated with observable $B$ is present. This is a phenomenon that is classically forbidden. This comparison tells us that even the identity premeasurement thing $\mathcal{I}$ (which has the weakest action as far as premeasurement is concerned) has physical consequences, since the mediation of $\mathcal{I}$ associated with $B$ in case (2.) leads to different probabilistic predictions for the measurement of $c$-systems than in case (3.), where no premeasurement associated with $B$ is present at all. This is a positive result as far as the experimental concreteness of quantum-mechanical things is concerned.

This concludes the characterization of the quasigeometric realm made possible by
the introduction of numbers, midway between the metageometric and geometric realms. How do the traditional geometric concepts of quantum theory arise from the meta- and quasigeometric foundations we have just laid?

\section{Geometric realm} \label{Geometry}

\subsection{State vectors from geometric decomposition} \label{SVMD}
Whereas traditionally state vectors constitute the starting point of quantum theory, they arise here as derived---and physically insubstantial---objects. 
The quasigeometric probability relation $\sum_i |\langle b_i|a\rangle|^2=1$ is the key to understand how the introduction of state vectors and the geometrization of quantum theory takes place.\footnote{See Schwinger's \citeyear{Schwinger:1960} follow-up article for his way of introducing state vectors. He then suggests that ``the geometry of states'' thus introduced is more fundamental than ``the measurement algebra,'' which ``can be derived'' from the former (p.~260). This is true from a mathematical but not physical point of view, where observations always come first; we keep metageometry as ontologically primary.}  

\begin{figure}
\begin{center}
\includegraphics[width=65mm]{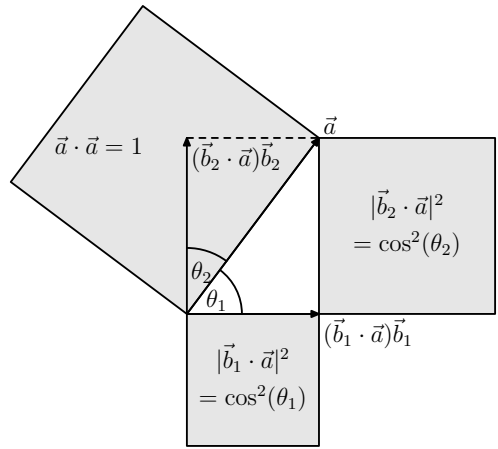}
\end{center}
\caption[Pythagoras' theorem in $n$ dimensions]{Two-dimensional representation of an $n$-dimensional real Euclidean vector space, in which a unit arrow vector $\vec a$ can be decomposed into projections along orthonormal axes $\vec b_i$ with components $\vec b_i\cdot\vec a$ satisfying $\sum_i(\vec b_i\cdot \vec a)^2=|\vec a|^2=1$ by virtue of Pythagoras' theorem in $n$ dimensions.} 
\label{mentaldec}
\end{figure}

\begin{figure}
\begin{center}
\includegraphics[width=65mm]{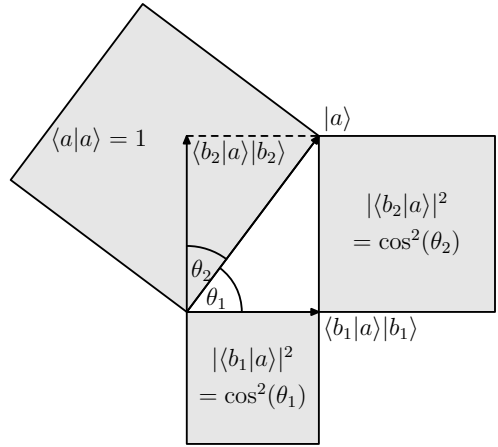}
\end{center}
\caption[State vectors from mental decomposition]{Rise of quantum-mechanical arrow state vectors from the geometric decomposition of quasigeometric property $\sum_i |\langle b_i|a\rangle|^2=1$.} 
\label{mentaldec2}
\end{figure}

Let us use an analogy and examine a real $n$-dimensional Euclidean space to which belong the general unit vector $\vec a$ and the vectors $\vec
b_i$ \mbox{($i=1,2,\ldots,n$)}, which form an orthonormal basis of this space: \mbox{$\vec b_i\cdot \vec b_j=\delta_{ij}$} (Figure \ref{mentaldec}). Thus, if $\theta_i$ is the \emph{angle} between $\vec a$ and $\vec b_i$, by virtue of Pythagoras' theorem in $n$ dimensions, we have 
\begin{equation}
\sum_i\cos^2(\theta_i)=\sum_i(\vec b_i\cdot \vec a)^2=1,
\end{equation}
an analogical geometric version of quasigeometric property (\ref{normalization}). This result tells us that the  transition function $\langle b|a\rangle$ \emph{can be geometrically decomposed} and \emph{thought as} the \emph{inner product}, or overlap, $\langle b| \cdot |a\rangle$ of two different arrow ``state vectors'' to yield now the quantum-mechanical geometric version of property (\ref{normalization}) (Figure \ref{mentaldec2}). Because, as we have seen, the physically more primitive concept is that of the transition function, and because its geometric decomposition is only a human prerogative---only a connotation\footnote{Connotation: something suggested (two overlapping arrows, $\langle b| \cdot |a\rangle$) by a word or thing (``$\langle b|a\rangle$'') apart from the thing it explicitly names or describes (common part between the players in a quantum-mechanical transition). (Adapted from Merriam-Webster Online Dictionary, http://www.m-w.com)}---state vectors need not be associated with any concrete, genuinely physical entities. In this light, state vectors appear as the artificial constructions of our minds---which cling to the intuitive picture given by arrow-like vectors out of their built-in habit to seek visualization in terms of familiar geometric concepts. In the lecture course on which this chapter is based, Dr.~Lehto described the predicament in which we have put ourselves as a result of this habit in this way:
\begin{quote}
Philosophical problems in quantum physics arise from our simplifying mental pictures, which we create out of our habit to use familiar concepts and words. These problems cannot be solved by increasing our knowledge---we have imprisoned ourselves in a labyrinth of our own making, and we cannot find the way out.\footnote{\emph{Physical, mathematical, and philosophical foundations of
quantum theory} (2005), Lecture notes.}
\end{quote}

This finding about the origin of the state vector is reminiscent of Auyang's \citeyear{Auyang:2001} view regarding what we might call the connotative decomposition of physical fields $f(P)$ into a local aspect (``spacetime'' $P$) and a qualitative aspect (``quality'' $f$). Auyang argued, in a manner similar to the present one regarding the transition function, that fields are physically indivisible entities, and that it is only our prerogative to decompose them into two constituent parts. However, she maintained, spacetime (points) need not have any independent physical meaning just because we can abstract them from the field---``our thinking does not create things of independent existence;'' she wrote, ``we are not God'' (p.~214). We notice then that, also in the quantum-mechanical case, our mental activities work towards the creation of originally absent \emph{geometric} concepts with unclear physical meaning. 

It is also noteworthy that the result about the transition function suggests, contrary to common belief, that from a \emph{physical} (not logical) point of view the metric (probability) comes first and vectors arise only \emph{after} the metric, when this is split up in order to satisfy the human need for geometric visualization.\footnote{In Article I (see Related publications), we presented an idea concerning the possible metageometric foundations of quantum theory which is contradicted by the present findings. We speculated then that the geometric formulation of quantum theory could arise from the ``overlap of geometry-free things $|\psi\rangle$ and their dual counterparts $\langle \psi|$'' (p.~437). By this we meant that, if \emph{metric-free} vectors could be found to be physically meaningful, the geometric concepts of projection and probability would come second as a consequence. Here the sense of vectors as non-geometric concepts is the following. Vectors are not geometric concepts in general, but \emph{become} geometric objects with associated geometric magnitudes after being endowed with a metric structure. In the light of the present findings, state vectors are geometric concepts \emph{from the start} because they arise from geometric decomposition of the transition function; at no time are there vectors \emph{without} a metric. For this reason, we now leave the older idea behind.} This idea can be related to similar concepts in spacetime physics. May it be the case that, from a physical standpoint, the worldline separation $\mathrm{d}s$ comes first and that spacetime points follow only as mentally helpful appendages? \label{inner-structure-2}

From their role as geometric parts of the transition function, we gather that kets $|a\rangle$ could be associated with incoming systems, and bras $\langle a|$ with outgoing systems. By analogy with the transition-function postulate, we can say that state vectors are linked by some form of dual correspondence, $|a\rangle \sim \langle a|$. This is, however, only a transmogrification of the physical dual correspondence between complex numbers (transition functions), and need not carry with it an autonomous physical meaning.

From the metageometric perspective, the following two geometric properties follow automatically. Interpreting \mbox{$\langle b|a\rangle^\ast=\langle a|b\rangle $} geometrically, and associating ket $|a\rangle$ with $\vec a$ and bra $\langle a|$ with $\vec a^\ast$, we recover a well-known property of complex vectors, namely, that their inner product satisfies \mbox{$(\vec b^\ast \cdot \vec a)^\ast=\vec a^\ast \cdot \vec b$}. Also the orthonormality condition $\langle a_j|a_i\rangle=\delta_{ij}$ for vectors follows now automatically from earlier considerations of compatibility of $a_i$- and $a_j$-systems.

\subsection{Derivative geometric connotations} \label{Mentalities}
If the geometric counterpart of the transition function $\langle b|a\rangle$  is the inner product of arrow vectors $\langle b| \cdot |a\rangle$, what may be the geometric counterparts of the metageometric premeasurement and transition things $\mathcal{P}(a)$ and $\mathcal{P}(b|a)$? To answer this question, we proceed as we did with the transition function, namely, by finding a correspondence with the geometric realm through geometric decomposition.

A transition thing $\mathcal{P}(b|a)$ has the capacity to cleanly turn incoming $a$-systems into $b$-systems. We notice that, geometrically, this action is displayed by the operator $|b\rangle \cdot \langle a|$, which takes in ``$a$-systems'' $|a\rangle$ and turns them into ``$b$-systems'' $|b\rangle$, thus: $(|b\rangle \cdot \langle a|)|a\rangle = |b\rangle$. The \emph{outer product}, then, arises from the geometric decomposition of a transition thing into the product of two separate state vectors, a bra and a ket. We denote this correspondence as follows: 
\begin{equation}
\mathcal{P}(b|a)\overset{\mathrm{g}}{=}|b\rangle\langle a|.
\end{equation}
In particular and in agreement with earlier remarks, 
\begin{equation}
\mathcal{P}(a)\overset{\mathrm{g}}{=} |a\rangle\langle a|
\end{equation} 
represents no transition at all, since $(|a\rangle\langle a|)|a\rangle=|a\rangle$.

When the transition function is interpreted as an inner product of arrow vectors, and premeasurement and transition things as an outer product of arrow vectors, the whole premeasurement process can be reinterpreted geometrically in two parts: an \emph{incoming} one related to the \emph{inner} product and an \emph{outgoing} one related to the \emph{outer} product. It is not essential to do so, but it is comforting to be able to recover the traditional geometric results from metageometric foundations.

\begin{figure}
\begin{center}
\includegraphics[width=90mm]{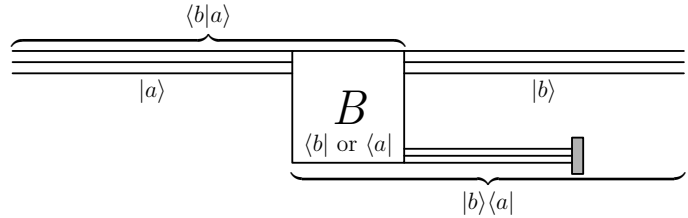}
 \end{center}
\caption[Inner and outer products]{Geometric representation of the premeasurement process in terms of the inner and outer products, $\langle b|a\rangle$ and $|b\rangle\langle a|$, of arrow vectors $|a\rangle$ and $|b\rangle$.} 
\label{inoutprod}
\end{figure}

In the incoming part, we start with systems in state $|a\rangle$ before their preparation into $b$-systems, and with no systems in the preparation box itself. Some systems are then ``absorbed'' in the preparation box into state $\langle b|$. In the outgoing part, we have systems in the preparation box ``emitted'' from state $\langle a|$ into state $|b\rangle$, which leave the preparation box---now empty of systems---behind. In this geometric setting, it is then natural to connect the inner product $\langle b|a\rangle$ with the internal rearrangement (``absorbtion'') of systems at preparation, and the outer product $|b\rangle\langle a|$ with the external rearrangement (``emission'') of systems at preparation (Figure \ref{inoutprod}).

At this point of the geometric development, and if so wished, the correspondence between premeasurement and transition things  and the hypothetical continuum of measurement results $a$ can be introduced. This can be done by assuming that an infinitesimally small variation $\delta a$ of a measurement result $a$ is physically meaningful, and thus that it is significant to write
\begin{equation}
\mathcal{P}(a)\overset{\mathrm{g}}{=}\int_{a-\delta a}^{a+\delta a}|a'\rangle\langle a'|\mathrm{d}a',
\end{equation}
as well as
\begin{equation}
\mathcal{P}(b|a)\overset{\mathrm{g}}{=}\int_{a-\delta a}^{a+\delta a}\int_{b-\delta b}^{b+\delta b}|b'\rangle\langle a'|\mathrm{d}b'\mathrm{d}a'.
\end{equation}
However, geometric expressions such as $\int_{a_1}^{a_2}|a\rangle\langle a| \mathrm{d}a$ have no metageometric counterparts at all.

As expected, the transition postulate (\ref{TP}) also holds in geometric form:
\begin{equation}
\mathcal{P}(d|c)\mathcal{P}(b|a)\overset{\mathrm{g}}{=}(|d\rangle\langle c|)(|b\rangle\langle a|)=\langle c|b\rangle |d\rangle\langle a|
\overset{\mathrm{g}}{=} \langle c|b\rangle \mathcal{P}(d|a).
\end{equation}

The outer products $|a\rangle\langle a|$ and $|b\rangle\langle a|$, geometric connotations of the premeasurement and transition things $\mathcal{P}(a)$ and $\mathcal{P}(b|a)$, transform a vector into another vector. Therefore, they are (linear) operators. All other operators come from these basic operators, which are, in turn, connected with measurement values.

We notice next that Hermitian conjugation generalizes the concept of complex conjugation in a genuinely physical way because, unlike the case of dual correspondence between state vectors, it retains the previous physical meaning. In effect, if we compare the relation $\langle b|a\rangle^\ast = \langle a|b\rangle$ for the inner product with the relation $(|b\rangle\langle a|)^\dag =|a\rangle\langle b|$ for the outer product (operator), we see that in both cases the action of ``$\ast$'' and ``$\dag$,'' respectively, consists in inverting the direction of the quantum-mechanical transition: \mbox{$(|b\rangle\langle a|)|a\rangle=|b\rangle$} and $(|b\rangle\langle a|)^\dagger|b\rangle=|a\rangle$. Further, if we accept the notation
\begin{equation}
\mathcal{P}(b|a)^\dagger \overset{\mathrm{g}}{=} (|b\rangle\langle a|)^\dag,
\end{equation}
we then have the metageometric relation
\begin{equation}\label{hermconj}
\mathcal{P}(b|a)^\dag =\mathcal{P}(a|b),
\end{equation}
and, in particular, 
\begin{equation}
\mathcal{P}(a)^\dag=\mathcal{P}(a).
\end{equation}
Also this latter result is physically reasonable, considering that there is no transition involved in a premeasurement. If we take Hermitian conjugation (including complex conjugation) as a form of dual correspondence, then \emph{the physical meaning of dual correspondence is to invert the direction of a quantum-mechanical transition}. This interpretation applies to metageometric things, quasigeometric complex numbers (transition function), and geometric operators, but not to isolated state vectors. 

We now have enough knowledge to analyse the compatibility of observables. So far little has been said about the possibility of $\mathcal{P}(a)$ and $\mathcal{P}(b)$ commuting; what physical results do we obtain in such a case? Using the transition postulate, we find
\begin{eqnarray}\label{commutation}
\mathcal{P}(b)\mathcal{P}(a)&=&\mathcal{P}(a)\mathcal{P}(b) \nonumber\\
\mathcal{P}(b|b)\mathcal{P}(a|a)&=&\mathcal{P}(a|a)\mathcal{P}(b|b) 
\nonumber\\
\langle b|a\rangle \mathcal{P}(b|a)&=&\langle a|b\rangle \mathcal{P}(a|b)
\end{eqnarray}
Using Eq.~(\ref{hermconj}) to obtain the Hermitian conjugate of Eq.~(\ref{commutation}), and multiplying term by term, the transition functions vanish to get
\begin{equation}
\mathcal{P}(b|a)\mathcal{P}(a|b)=\mathcal{P}(a|b)\mathcal{P}(b|a),
\end{equation}
or equivalently
\begin{equation}\label{equalthings}
\mathcal{P}(b)=\mathcal{P}(a).
\end{equation}
Furthermore, we can now use Eq.~(\ref{equalthings}) to find
\begin{equation}
\mathcal{P}(a)\mathcal{P}(b)\mathcal{P}(a)=\mathcal{P}(a)=|\langle b|a\rangle|^2 \mathcal{P}(a),
\end{equation}
so that
\begin{equation}
\mathrm{Pr}(b|a)=\mathrm{Pr}(a|b)=|\langle b|a\rangle|^2=1.
\end{equation}
To sum up, if $\mathcal{P}(a)$ and $\mathcal{P}(b)$ commute, transition things $\mathcal{P}(b|a)$ and $\mathcal{P}(a|b)$ also commute, $\mathcal{P}(a)$ and $\mathcal{P}(b)$ are themselves equal to each other, and in transitions involving observables $A$ and $B$, any $b$-system becomes with certainty an $a$-system and viceversa. In other words, $A$ and $B$ are compatible observables.

Finally, since the geometric counterpart of the identity thing $\mathcal{I}$ must be the identity operator $\hat 1$, $\mathcal{I}\overset{\mathrm{g}}{=} \hat 1$, and since $\sum_i\mathcal{P}(a_i)\overset{\mathrm{g}}{=} \sum_i |a_i\rangle\langle a_i|$, we recover the geometric completeness relation
\begin{equation}
\sum_i |a_i\rangle\langle a_i|=\hat 1.
\end{equation}
With its help, a general vector $|\psi\rangle$ and a general operator $\hat X$ can be built in terms of a vector basis $\{ |a_i\rangle \}$ or operator basis
$\{ |a_i\rangle \langle a_i| \}$, respectively.

\subsection{Geometric traces}\label{Geometric traces}
The next noteworthy concept is that of the trace. For an operator, the trace consists of the sum of its diagonal elements,
\begin{equation}
\mathrm{Tr}(\hat X)=\sum_i \langle a_i|\hat X|a_i\rangle,
\end{equation}
which is in general a complex number. The trace, together with the decomposition relation (\ref{DR}) and the commutativity of complex numbers,\footnote{Cf. non-commutativity of quaternions.} now allows us to find enlightening connections between the metageometric and geometric realms:
\begin{enumerate}
\item The geometric trace left behind by the metageometric transition thing $\mathcal{P}(b|a)$ is the inner product (geometrized transition function) $\langle a|b\rangle$:
\begin{equation}
\mathrm{Tr}[\mathcal{P}(b|a)]\overset{\mathrm{g}}{=}\mathrm{Tr}(|b\rangle\langle a|)=\sum_i \langle c_i|b\rangle\langle a|c_i\rangle =\langle a|b\rangle.
\end{equation}

\item In particular,
\begin{equation}
\mathrm{Tr}[\mathcal{P}(a_j|a_i)]\overset{\mathrm{g}}{=}\mathrm{Tr}(|a_j\rangle\langle a_i|)=\langle a_i|a_j\rangle=\delta_{ij}.
\end{equation}

\item The geometric trace left behind by the metageometric premeasurement thing $\mathcal{P}(a)$ is
\begin{equation}
\mathrm{Tr}[\mathcal{P}(a)]\overset{\mathrm{g}}{=}\mathrm{Tr}(|a\rangle\langle a|)=\langle a|a\rangle=1.
\end{equation}

\item The geometric trace left behind by the product of metageometric transition things $\mathcal{P}(d|c)$ and $\mathcal{P}(b|a)$ is
\begin{align}
\mathrm{Tr}[\mathcal{P}(b|a)\mathcal{P}(d|c)]&=\mathrm{Tr}[\mathcal{P}(d|c)\mathcal{P}(b|a)]\nonumber\\
&\overset{\mathrm{g}}{=}\mathrm{Tr}[(|d\rangle\langle c|)(|b\rangle\langle a|)]=\langle c|b\rangle \langle a|d\rangle.
\end{align}
From this follows a noteworthy result:

\item The geometric trace left behind by the product of premeasurement things $\mathcal{P}(b)$ and $\mathcal{P}(a)$ is the probability of a transition from $a$-systems to $b$-systems or viceversa:
 \begin{equation}
\mathrm{Tr}[\mathcal{P}(a)\mathcal{P}(b)]=
\mathrm{Tr}[\mathcal{P}(b)\mathcal{P}(a)]\overset{\mathrm{g}}{=}
\langle b|a\rangle \langle a|b\rangle =\mathrm{Pr}(b|a).
\end{equation}
In other words, \emph{the geometric fingerprint left by metageometric things is quantum-mechanical probability}.
\end{enumerate}

\subsection{Observables}
We finish this section---and complete the geometrization of the theory---with the geometric and metageometric interpretation of observables. We start with the weighted average of measurement results $\{b_i\}$,
\begin{equation}
\sum_i \mathrm{Pr}(b_i|a)b_i=\sum_i \langle a|b_i\rangle \langle b_i|a\rangle b_i = \langle a|\left(\sum_i  b_i |b_i\rangle \langle b_i|\right) |a\rangle,
\end{equation}
where we geometrically identify observable $B$ with an operator $\hat B$ such that
\begin{equation}
B \overset{\mathrm{g}}{=}
\hat B=\sum_i b_i |b_i\rangle \langle b_i|. 
\label{GRO}
\end{equation}
Then
\begin{equation}
\langle a|\hat B |a\rangle=:\langle \hat B\rangle_{|a\rangle}
\end{equation}
is the \emph{expected value} of operator $\hat B$ in state $|a\rangle$ (i.e.\ when measurements are performed on a sample of $a$-systems), and $\langle(\Delta\hat B)^2\rangle_{|a\rangle}$ is its variance, where \mbox{$\Delta\hat B=\hat B-\langle \hat B \rangle\hat 1$}.

From the geometric representation (\ref{GRO}) of an observable, the eigenvalue equation follows \emph{automatically}:
\begin{equation}
\hat B|b_i\rangle=\left(\sum_j b_j|b_j\rangle \langle b_j|\right) |b_i\rangle = \sum_j b_j|b_j\rangle \delta_{ij}=b_i|b_i\rangle.
\end{equation}
In contrast, for a general operator $\hat X=\sum_{ij} x_{ij}|b_j\rangle\langle b_i|$ which is not an observable, the eigenvalue equation does not follow. The Hermiticity of observables also follows \emph{automatically} from Eq.~(\ref{GRO}):
\begin{equation}
\hat B^\dag=\sum_i b_i^\ast(|b_i\rangle\langle b_i|)^\dag=\hat B.
\end{equation}
Thus, an observable $B$ is always represented geometrically by a Hermitian operator, and we do not need to postulate this separately. Particularly, the eigenvalues of operators associated with observables, coming from measurement results $b_i$, are \emph{automatically} real; and different eigenvalues \emph{automatically} have orthogonal eigenvectors associated with them ($\langle b_j|b_i\rangle=0$, $i\neq j$) since $b_i$-systems and $b_j$-systems are mutually incompatible unless $i=j$. These are the benefits of basing physics on observations instead of disembodied mathematics.

It is now only a small step to find that a \emph{pure observable}, i.e.\ the metageometric counterpart of $\hat B$, is 
\begin{equation} 
B=\sum_i b_i \mathcal{P}(b_i). 
\end{equation}
How does it differ from $\hat B$? When we began building the theory, we introduced observables $A,B,C$, etc., which we assumed to represent physical magnitudes. But we saw in Chapter~\ref{ch:Geometry in physics} that all physical magnitudes are underlain by geometric associations to become geometric magnitudes. These are now pictured by geometric operator $\hat B$.\footnote{We may complement the geometric picture underlying $\hat B$, and discover some more connections, by finding out what geometric object represents this operator and how size is attached to it. Operator $\hat B$ is associated with a matrix related to a linear combination of outer products of arrow vectors. In general, if the matrix operator \mbox{$\hat X=|Z\rangle\langle Y|$} comes from the outer product of \mbox{$|Y\rangle=\sum_i y_i|y_i\rangle$} and \mbox{$|Z\rangle=\sum_j z_j |z_j\rangle$}, $\hat X$ is taken as an \emph{arrow} vector with components $x_{ij}=y_i z^\ast_j$ and \emph{length} 
\begin{equation}
\|\hat X \| = \sqrt{\sum_{ij}|x_{ij}|^2}= \sqrt{\mathrm{Tr}(\hat X^\dagger \hat X)}.  	
\end{equation}
A geometric observable $\hat B$ is then pictured as an arrow vector with one component $b_i|b_i\rangle$ for each measurable value $b_i$, and its length is given by 
\begin{equation}
\| \hat B \|= \sqrt{\sum_i b_i^2} = \sqrt{\mathrm{Tr}(\hat B^2)}. 	
\end{equation}
We now notice how the trace plays the role of an \emph{inner product} in order to produce the size of $\hat B$. We also learn that the geometric traces of the previous section are generalizations (because they do not necessarily produce a non-negative or even real number) of the norm of matrix operators. Cases (3.) and (5.), however, can be reinterpreted as squared operator norms if rewritten as $\mathrm{Tr}[\mathcal{P}(a)^\dagger\mathcal{P}(a)]$ and $\mathrm{Tr}\{[\mathcal{P}(b)\mathcal{P}(a)]^\dagger \mathcal{P}(b)\mathcal{P}(a)\}$.} Observable $B$, on the other hand, is a linear combination of metageometric things. It involves not measurements---activities producing physical \emph{values}---but premeasure\-ments---activities producing prepared \emph{systems}.
In fact, this is a helpful distinction to visualize how the starting points of a geometric and a metageometric theory differ.  

We call the geometry-free measurable observable represented by $B$ a quantum-mechanical, pure \emph{physical property}. We can then say that \emph{a quan\-tum-mechanical observable is by nature metageometric}. In other words, pure metageometric physical properties lie at the ontological root of the geometric physical magnitudes we are used to dealing with in everyday physics. 

Finally, what geometric traces do metageometric observables, i.e.\ pure physical properties, leave behind? The trace of the product of $B$ and $\mathcal{P}(a)$ is
\begin{equation}
\mathrm{Tr}[B\mathcal{P}(a)]\overset{\mathrm{g}}{=}\langle \hat B\rangle_{|a\rangle},
\end{equation}
which tells us that \emph{the geometric fingerprint left by a metageometric observable is the quantum-mechanical expected value}, i.e.\ statistics.

\section{Time} \label{MTS} 
Do metageometric things and the constructions derived therefrom also have something to tell us about time as a concept inherent in quantum-mechanical systems? Or is time, after all, just an external parameter also according to quantum mechanics? And how can the original psychophysical parallelism between quantum-mechanical things and conscious feelings guide the answers to these questions? 

In this section, we study the rise of \emph{metageometric time}---itself not a coordinate or parameter of any kind---purely from metageometric quantum-mechanical things. Metageometric time, in turn, can lead to the rise of coordinate time intervals $\delta t$ and coordinate time $t$ appearing in the equations of motion of classical physics and non-relativistic quantum mechanics, thus revealing its metageometric origin.  

Besides metageometric time, we have also found another intrinsically quantum-mechanical concept of time but now at the geometric level. This time is liked to the preparation of quantum-mechanical systems and called \emph{preparation time}. We introduce it separately in Appendix B because it has no direct application in this thesis but may nonetheless prove useful for future research.

\subsection{Metageometric time}
We are interested in a description of time that is connected to physics through quantum mechanics but that, at the same time, captures so far disregarded aspects of conscious experience---most notably, the permanent present. Quantum-mechanical things were introduced with this goal in mind. They are global and belong outside conventional time and space; their product pictures succession (yet not conventional temporal evolution) and their sum pictures discrimination, both in analogy with the succession and discrimination of conscious feelings forming the basis of psychological time as an experience in a permanent now. 

Quantum-mechanical things have the potential for the new description of time we seek, but not simply as such. Instead of a premeasurement or a transition thing alone, we would like a simple construction out of them that evokes some properties of the time of conscious experience somewhat more closely. For example, we would like, first and foremost, a metageometric time that does not evolve in the conventional sense of physics, that determines no past or future nor, therefore, a present in the customary sense of a time instant. We would also like a metageometric time in which the event of no (quantum-mechanical) change translates into an empty, or better, ineffectual ``element of time,'' in analogy with the lack of feeling of time when consciousness is insensitive to change (e.g.\ during a dreamless sleep). In addition, metageometric time should also allow for a succession of its elements to be reversed and thus recover its beginning, in analogy with the exact repetition of a conscious feeling. And finally, a metageometric time whose building blocks succeed each other smoothly, in analogy with the continuity of conscious experience. 

We arrive at a metageometric concept of time having all these properties by searching for an arrangement of quantum-mechanical things such that (i)~only a pure transition be involved and thus leave no place for any filtering action (cf.\ preparation time), and such that (ii) each and every $a_i$-system associated with an observable $A$ be transformed into a $b_i$-system associated with another observable $B$ in a one-to-one correspondence.

\begin{figure}
\begin{center}
\includegraphics[width=90mm]{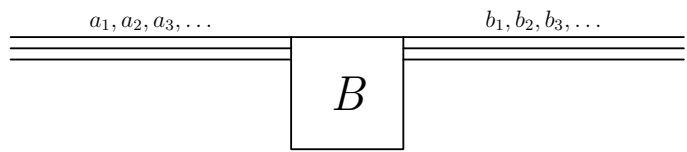}
 \end{center}
\caption[Metageometric time element]{Schematic representation of a metageometric time element $\sum_i \mathcal{P}(b_i|a_i)$.}
\label{primtime} 
\end{figure}

The transition thing $\mathcal{P}(b_i|a_i)$ is almost what we want. It satisfies the first requirement because, as we saw earlier on, it involves only a pure transition and does not contain any filtering action. However, the following geometric test shows that the said one-to-one correspondence does not hold. In effect, 
\begin{equation}
(|b_i\rangle\langle a_i|)|a_j\rangle = \delta_{ij}|b_i\rangle\neq
|b_j\rangle, 
\end{equation} 
which means that any $a_j$-system where $j\neq i$ is left without a counterpart into which to undergo a transition. Now a small modification gives us what we seek, as the sum $\sum_i \mathcal{P}(b_i|a_i)$ of transition things satisfies both the said requirements. It does not contain any filtering
action, and it attaches a counterpart to each and every $a_i$-system, as guaranteed by the same geometric test as before: 
\begin{equation} \left(\sum_i|b_i\rangle\langle
a_i|\right)|a_j\rangle = \sum_i\delta_{ij}|b_i\rangle= |b_j\rangle.
\end{equation} 
We call $\sum_i \mathcal{P}(b_i|a_i)$ a \emph{metageometric time element}\footnote{Here ``metageometric time element'' is favoured over ``metageometric time instant'' because the latter refers to an infinitesimal interval of time, whereas the former is free from any geometric connotations. A metageometric time element is a metageometric concept, to which it is meaningless to apply the geometric concepts of point or length (be it null,
infinitesimal, or finite).} (Figure \ref{primtime}).

We can justify this provocative name from two different perspectives. From a psychological point of view, we can do so through the consciousness-related properties it displays, shown immediately below. From a physical point of view, we can retrospectively justify $\sum_i \mathcal{P}(b_i|a_i)$ as metageometric time by noticing that it \emph{can} explain the origin of time as an \emph{absolute} parametric (coordinate) interval $\delta t$ as used in physics. However, we would still like to justify it as an expression of quantum-mechanical time by linking it to the \emph{relativistic} separation $\D s$, directly connected with clock time as a genuinely measurable magnitude (absolute time intervals $\delta t$ are not). Finding the link between clock time $\D s$ and metageometric time $\sum_i \mathcal{P}(b_i|a_i)$ is nevertheless a task demanding enough to fall outside the scope of this thesis; we must, for that reason, leave it for the future, and rest content with giving an outlook on how to approach this problem at the end of the next and last chapter.

\begin{figure} 
\centering 
\includegraphics[width=\linewidth]{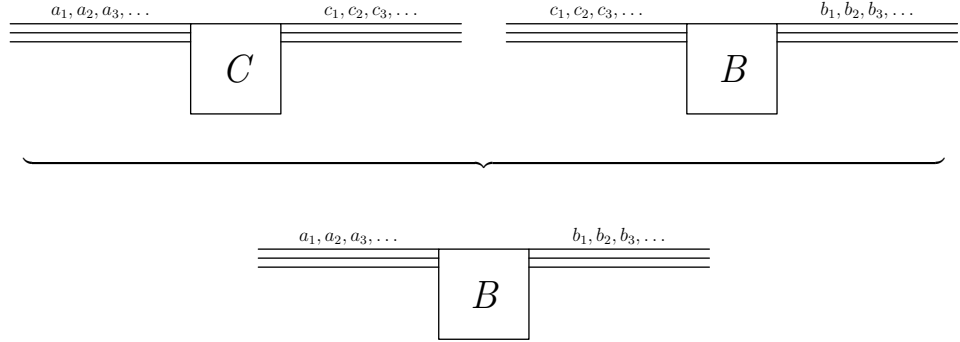}
\caption[Physically possible product]{Schematic representation of a physically possible succession $\sum_i \mathcal{P}(b_j|c_j)\sum_i \mathcal{P}(c_i|a_i)$ of two
metageometric time elements resulting in one time element $\sum_i \mathcal{P}(b_i|a_i)$.} 
\label{prodpti} 
\end{figure}

A metageometric time element satisfies four essential properties:
\begin{enumerate} 
\item Two successive metageometric time elements lead to one metageometric time element (Figure \ref{prodpti}):
\begin{equation} 
\sum_j \mathcal{P}(b_j|c_j)\sum_i \mathcal{P}(c_i|a_i)=\sum_{ij}\langle c_j|c_i\rangle \mathcal{P}(b_j|a_i)=\sum_i \mathcal{P}(b_i|a_i). 
\end{equation}
This says that, in the evolution of metageometric time, the succession of two time elements is one time element. In other words, from the perspective of coordinate time, \emph{metageometric time does not evolve}---it happens, so to speak, always and never. (Cf.\ permanent present of conscious experience.)

\item The succession of three metageometric time elements is
associative: \begin{multline} \sum_k
\mathcal{P}(b_k|d_k)\left[\sum_j \mathcal{P}(d_j|c_j) \sum_i
\mathcal{P}(c_i|a_i)\right]=\\\left[\sum_k
\mathcal{P}(b_k|d_k)\sum_j \mathcal{P}(d_j|c_j)\right] \sum_i
\mathcal{P}(c_i|a_i)=\sum_i \mathcal{P}(b_i|a_i). \end{multline}

\item There exists a metageometric time element
$\mathcal{I}=\sum_i\mathcal{P}(x_i|x_i)$, with $x_i$ associated with
any \emph{suitable} observable, which does not change any preceding
or succeeding metageometric time element: \begin{equation}
\sum_j\mathcal{P}(b_j|b_j)\sum_i\mathcal{P}(b_i|a_i)=\sum_i\mathcal{P}(b_i|a_i)
=\sum_i\mathcal{P}(b_i|a_i)\sum_j\mathcal{P}(a_j|a_j).
\end{equation}
(Cf.\ insensitivity of consciousness to change.)

\item Any metageometric time element can be turned backwards,
cancelled, or undone with another preceding or succeeding
metageometric time element: \begin{equation}
\sum_j\mathcal{P}(b_j|a_j)\sum_i\mathcal{P}(a_i|b_i)=\mathcal{I}=
\sum_i\mathcal{P}(a_i|b_i)\sum_j\mathcal{P}(b_j|a_j).
\end{equation} 
(Cf.\ repetition of conscious feeling or experience.)
\end{enumerate}

We can immediately see that the above four properties constitute the
four axioms that define a group $G$ with the product as the group
operation. The four group axioms are: 
\begin{enumerate} 
\item Closure: if $A, B$ belong to $G$, $AB$ belongs to $G$; 
\item Associativity: if $A, B, C$ belong to $G$, $A(BC)=(AB)C$; 
\item Existence of an identity: if $A$ belongs to $G$, there exists $I$ such that $AI=A=IA$;  
\item Existence of an inverse: if $A$ belongs to $G$, there exists $A^{-1}$ such that $AA^{-1}=I=A^{-1}A$.
\end{enumerate} 
However, we notice that there exists one significant difference between this traditional group structure and that of the metageometric time elements when it comes to the group operation. On the one hand, the elements of the traditional group are, so to speak, structureless, while, on the other, the group of metageometric time elements possesses internal constitution: a metageometric time element consists of a sum of transitions $\mathcal{P}(b_i|a_i)$ from certain $a_i$-systems to certain $b_i$-systems (cf.\ element $A$). As a result, the group operation consisting of the product of \emph{successive} time elements is not free but instead \emph{physically restricted}: from a physical standpoint, it is only possible to have products in which the outgoing system in the factor on the right is associated with the
same observable as the incoming system in the factor on the left. We
call the group of metageometric time elements a \emph{physically restricted group}.\footnote{In a conventional group, the identity is unique. Assuming $I_1$ and $I_2$ are two different identities, we find $I_1I_2=I_1$ and $I_1I_2=I_2$, and therefore $I_1=I_2$, which contradicts the hypothesis. In this respect, the existence of myriads of \emph{different} identities in the group of metageometric time elements does not lead to any inconsistencies
because the physical restriction on what is a possible product only allows the same identity to be multiplied by itself.}

\begin{figure}
\begin{center} 
\includegraphics[width=\linewidth]{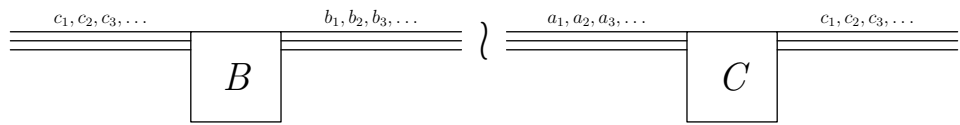}
\end{center} 
\caption[Physically impossible product]{Schematic representation of a physically impossible succession $\sum_i\mathcal{P}(c_i|a_i)\sum_j\mathcal{P}(b_j|c_j)$ of two
metageometric time elements.} 
\label{impprod} 
\end{figure}

In particular, as a consequence of its inner structure, the question of whether the physically restricted group of metageometric time elements is Abelian or non-Abelian is now not a reasonable question to ask. In general, whenever a succession of two elements is physically possible, e.g.\
\begin{equation}
\sum_j\mathcal{P}(b_j|c_j)\sum_i\mathcal{P}(c_i|a_i),	
\end{equation}
the succession determined by the inverse order,
\begin{equation}
\sum_i\mathcal{P}(c_i|a_i)\sum_j\mathcal{P}(b_j|c_j),	
\end{equation}
does not make physical sense as a metageometric time element, as shown in Figure \ref{impprod}. (Cf.\ continuity of conscious experience.)

Far from finding these incongruities with the traditional concept of group distressing, we believe that the distinctiveness displayed by the group of metageometric time elements makes them even more appealing from a physical point of view. Here we witness yet another example of physics giving rise, for its own independent reasons, to a new bit of mathematics---mathematics does not rule physics but is only its compliant tool.

Next we pursue the geometric interpretation of metageometric time, leading to an understanding of time intervals $\delta t$ and, in general, time $t$. The key to this pursuit will lie in the physical restriction imposed on the group structure of the metageometric time elements and in the circumstances of its disappearance.

\subsection{The rise of coordinate time}\label{Virtual}
Coordinate time intervals $\delta t$ and coordinate time $t$ can be understood as a particular outgrowth of metageometric time. To achieve this, we start by finding the geometric counterpart of a metageometric time element, namely, 
\begin{equation} 
\sum_i \mathcal{P}(b_i|a_i)\overset{\mathrm{g}}{=}\sum_i |b_i\rangle\langle a_i|=:\hat T_{ba}. 
\end{equation} 
We call operator $\hat T_{ba}$ the \emph{time-element operator}. The well-known claim that in quantum mechanics there exists no universal time operator\footnote{The widespread belief that the time coordinate constitutes a special
kind of problem in quantum mechanics, after Pauli's famous argument,
has been uprooted, for example, by Hilgevoord \citeyear{Hilgevoord:2002}. He
accounts for this false impression as a confusion between spacetime coordinates $x,t$ and the dynamic variables $\hat q,\hat\theta$ of systems: whereas in quantum mechanics coordinate time $t$ is a parameter exactly as in Newtonian mechanics---and should therefore not be attempted to turn into an operator---particular \emph{systems} (clocks) may admit particular dynamic time variables $\hat\theta$, in analogy with position $\hat q$. The properties of $\hat\theta$, or whether it exists at all, are not universal but dependent on the system of which $\hat\theta$ is a variable.} is therefore true for infinitesimal \emph{intervals}, or instants, $\delta t$ of coordinate time, but not for time \emph{elements}. Operator $\hat T_{ba}$, however, is not Hermitian in any case. In fact, we find
\begin{equation} 
\label{conjugate} 
\hat T_{ba}^\dag=\sum_i (|b_i\rangle\langle a_i|)^\dag=\sum_i |a_i\rangle\langle b_i|=\hat T_{ab}. 
\end{equation}

By interpreting geometrically all earlier results applying to metageometric time, we find that what holds for metageometric time elements also holds for geometric time-element operators. The geometric evolution of time elements is given by the product $\hat T_{bc}\hat T_{ca}$. Based on this, we can now see that the properties (1.)--(4.) above for metageometric time elements hold for time-element operators: 
\begin{enumerate} 
\item Closure: $\hat T_{bc}\hat T_{ca}=\hat T_{ba}$; 
\item Associativity: $\hat T_{bd}(\hat T_{dc}\hat T_{ca})=(\hat T_{bd}\hat T_{dc})\hat T_{ca}$; 
\item Existence of an identity $\hat T_{xx}$: $\hat T_{bb}\hat T_{ba}=\hat T_{ba} =\hat T_{ba}\hat T_{aa}$; 
\item Existence of an inverse: $\hat T_{ab}\hat T_{ba}=\hat 1=\hat T_{ba}\hat T_{ab}$. 
\end{enumerate} 
Therefore, time-element operators also form a physically restricted group. Moreover, from property (4.) and Eq.~(\ref{conjugate}) follows the new result that time-element operators $\hat T_{ba}$ are \emph{unitary}: 
\begin{equation} 
\hat T_{ba}^\dag\hat T_{ba}=\hat 1=\hat T_{ba}\hat T_{ba}^\dag.
\end{equation} 
Knowing, as we do, that the physical meaning of Hermitian conjugation is to reverse the direction of a transition, this result is telling us something that is perfectly reasonable, namely, that \emph{to undo a time element} ($\hat
T_{ba}^{-1}$) \emph{is to reverse the transitions} it involves ($\hat T_{ba}^\dagger$). Here, then, are the seeds of unitary time evolution in coordinate time $t$ in quantum mechanics.

How should we proceed to recover coordinate time from metageometric time? The first thing to note is that this recovery, leading to all the usual equations of quantum mechanics, is not a \emph{necessary} but a \emph{possible} consequence. This is enough and all that we require.

Coordinate time can be recovered \emph{if} the physically restricted group of unitary time-element operators $\hat T_{ba}$ is replaced by a new group of likewise unitary operators $\hat U$, in which the discrete inner structure ($ba$) of the time elements is lost and replaced by some set of continuous coordinate parameters $x^1,x^2,x^3,\ldots$ on which $\hat U$ depends, $\hat U(x^1,x^2,x^3,\ldots)$. How many coordinate parameters should we choose, and on what physical grounds? Nothing prevents us from choosing, for example, $x^1,x^2$, and $x^3$ and associating them with the three spatial coordinates. But on what grounds would we do this? Why choose three coordinate parameters and not four, five, or just one?

The only physical idea behind the operators $\hat U$ left to guide us is that they originate from the operators $\hat T_{ba}$, which are connected with quantum-mechanical transitions. We proposed that these do not belong in spacetime, but that at preparation they project themselves onto spacetime as a \emph{linear} sequence of events in analogy with the projection of the stream of consciousness. The heuristic value of this psychophysical parallelism is now as follows. As a geometric spatiotemporal concept (as we now examine), a transition takes place instantaneously in an infinitesimal instant of (unidimensional) time. This suggests that we should choose $\hat U$ to depend only on one coordinate parameter, call it an infinitesimal (cf.\ instantaneity) coordinate interval $\delta t$, and associate it with the usual physical concept of \emph{time instant}.

\begin{figure} 
\begin{center} 
\includegraphics[width=70mm]{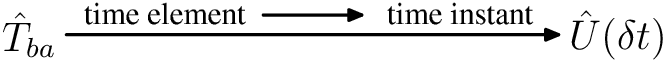}
\end{center} 
\caption[Coordinate time from metageometric
time]{Replacement leading to the obtention of evolution in
coordinate time from time-element operators, in turn originating from metageometric time elements.} 
\label{eleminst}
\end{figure}

We call the unitary continuous one-parameter operators $\hat U(\delta t)$ \emph{time-instant operators}, also known as time-evolution operators. The replacement leading from time-element operators to time-instant operators, schematized in Figure \ref{eleminst}, is \emph{seemingly} small, but in fact it is dramatic. Seemingly small because a time instant (as small as we wish yet non-null) has substituted an extensionless time element; in fact dramatic because in question is not the difference between an infinitesimal interval and an extensionless point, but between the attribute of temporal extension (be it null, infinitesimal, or finite) and the meaninglessness of the application of this attribute to time elements. In this dramatic passage, the inner structure of the time elements is lost. As a result, the operators $\hat U(\delta t_i)$ become free of physical restrictions concerning multiplication and form now an unrestricted group.

To recover from this basis the equations of motion of non-relativistic quantum mechanics, we start by noting that the closure property of the unrestricted group of unitary operators $\hat U(\delta t_i)$ demands that 
\begin{equation} 
\label{virtualtimedevelop} 
\hat U(\delta t_1) \hat U(\delta t_2)=\hat U(\delta t_1+\delta t_2). 
\end{equation} 
Furthermore, assuming evolution in coordinate time to be continuous,
we have \begin{equation} \label{timedevcontinuity} \hat U(\delta
t)=\hat 1+ \delta\hat U(\delta t). \end{equation} What is now the
form of $\delta\hat U(\delta t)$? Because $\hat U$ is unitary, then
\begin{equation} \hat 1=\hat U^\dag \hat U=(\hat 1+\delta\hat
U^\dag)(\hat 1+\delta\hat U)=\hat 1+\delta \hat U^\dag+\delta \hat
U+O[(\delta \hat U)^2], \end{equation} which implies
\begin{equation}
 \delta\hat U^\dag=-\delta\hat U.
\end{equation} 
Therefore, $\delta \hat U$ is anti-Hermitian and we can write $\delta\hat U=-i\delta\hat V$, where $\delta\hat V$ is Hermitian. We obtain 
\begin{equation} \hat U(\delta t)=\hat 1-i\delta\hat V(\delta t). 
\end{equation}

The fact that the group property (\ref{virtualtimedevelop}) holds for time-evolution operators $\hat U(\delta t)$, together with (\ref{timedevcontinuity}), now requires that 
\begin{eqnarray} 
\delta \hat V(\delta t_1)+  \delta \hat V(\delta t_2)=\delta \hat V(\delta
t_1+\delta t_2). 
\end{eqnarray} 
This means that it is possible for us to take $\delta \hat V(\delta t)$ to be of the form $\delta t \hat\omega$, where $\hat\omega$ is Hermitian, does not depend on $\delta t$ anymore, and is not itself an infinitesimal, although it can depend on $t$. We obtain 
\begin{equation} 
\hat U(\delta t)=\hat 1-i\delta t\hat\omega. 
\end{equation} 
In view that our context is purely quantum-mechanical, we can justifiably\footnote{What cannot be justified without previous knowledge of quantum theory is the choice of $\hbar$ over $h$.} use dimensional analysis to find that $[\hat\omega]=[\mathrm{time}]^{-1}=[\mathrm{energy}][\hbar]^{-1}$, so that we can take $\omega=\hat H/\hbar$, where the Hamiltonian operator $\hat H$ works as the generator of time translations.
Finally, we obtain \begin{equation} \label{timedevelp} \hat U(\delta
t)=\hat 1-\frac{i}{\hbar}\delta t\hat H, \end{equation} where $\hat
H$ is Hermitian and can depend on $t$.

All the equations of motion of quantum mechanics follow from
Eq.~(\ref{timedevelp}). Because 
\begin{equation} 
\hat U(t+\delta t)-\hat U(t)=[\hat U(\delta t)-\hat 1]\hat U(t)=-\frac{i}{\hbar}\delta t\hat H\hat U(t),
\end{equation} 
we get 
\begin{equation} 
\frac{\hat U(t+\delta t)-\hat U(t)}{\delta t}=:\frac{\mathrm{d}\hat
U(t)}{\mathrm{d}t}=-\frac{i}{\hbar}\hat H\hat U(t), 
\end{equation}
which is \emph{Schr\"{o}dinger's equation of motion} for the time
evolution operator: 
\begin{equation} \label{Udevelop}
i\hbar\frac{\mathrm{d}\hat U(t)}{\mathrm{d}t}=\hat H\hat U(t).
\end{equation} 
Notice that the linearity of this differential equation need not be assumed, but follows directly from the continuous group structure and unitarity of the evolution operators $\hat U(\delta t_i)$.

Schr\"{o}dinger's equation for a geometric state vector results as follows. Since $|a;t+\delta t\rangle=\hat U(\delta t)|a;t\rangle$, we have
\begin{equation}
|a;t+\delta t\rangle-|a;t\rangle=[\hat U(\delta t)-\hat 1]|a;t\rangle=-\frac{i}{\hbar}\delta t\hat H |a;t\rangle.
\end{equation}
Dividing on both sides by the infinitesimal $\delta t$, we find \emph{Schr\"{o}dinger's equation of motion} for a state vector, namely,
\begin{equation} \label{vectorSdevelop}
i\hbar\frac{\mathrm{d} |a;t\rangle}{\mathrm{d}t}=\hat H |a;t\rangle.
\end{equation}

Finally, we obtain Heisengberg's equation of motion for a time-indepen\-dent
operator $\hat A$ (in the Schr\"{o}dinger picture). Because we can use $\hat U(t)$ to make a ket evolve in coordinate time, and therefore $\hat U^\dag(t)$ to make a bra evolve in coordinate time, we have
\begin{equation} 
\langle b;t|\hat A_S|b;t\rangle=\langle b;0|\hat U^\dag(t)\hat A_S\hat U(t)|b;0\rangle, 
\end{equation} 
where we identify 
\begin{equation}\label{eq:Heisenberg-Schroedinger}
\hat U^\dag(t)\hat A_S\hat U(t)=:\hat A_H(t)
\end{equation}
with the Heisenberg operator $\hat A_H(t)$, where $\hat A_S=\hat A_H(0)$. Using Eq.~(\ref{Udevelop}) to differentiate Eq.~(\ref{eq:Heisenberg-Schroedinger}), we find \emph{Heisenberg's equation of motion} for $\hat A_H(t)$: 
\begin{equation} 
i\hbar\frac{\mathrm{d} \hat A_H(t)}{\mathrm{d}t}=-[\hat H, \hat A_H(t)].
\end{equation} 
Thus, all major traditional equations of motion in coordinate time can be
explained on the basis of metageometric time elements which, in turn, rest solely on metageometric things.

Several layers of the veil obscuring the physical origin and meaning of quantum physics have now been lifted from in front our eyes. We can now see more clearly into the heart of quantum physics---and as we stare the old quantum mysteries in the eye with the fresh vision metageometry affords us, we see complexity and philosophical confusion turn to simplicity and physical understanding. 

For a conceptual appraisal of the metageometric picture of time presented here, we turn to the next and final chapter.

\chapter{Time, the last taboo}
\label{ch:Time last taboo}
\begin{flushright}
\begin{tabularx}{10cm}{X}
\emph{El tiempo es la sustancia de que estoy hecho. El tiempo es un r\'io que me arrebata, pero yo soy el r\'io; es un tigre que me destroza, pero yo soy el tigre; es un fuego que me consume, pero yo soy el fuego.}\footnote{``Time is the substance I am made of. Time is a river that grabs me, but I am the river; a tiger that tears me apart, but I am the tiger; a fire that consumes me, but I am the fire.'' (Jorge Luis Borges, A new refutation of time)}\smallskip \\
Jorge Luis Borges, \emph{Nueva refutaci\'{o}n del tiempo}
\end{tabularx}
\end{flushright}

\bigskip

\section{The enigma of time}
The metageometric idea of time as the succession of complexes of quantum-mechanical things outside conventional spacetime that we arrived at in the previous chapter is inspired in an essential aspect of raw conscious experience. At the same time, however, it runs counter to both our belief of what time is, could be, or should be, as well as what physics tells us that time is. 

We came to a metageometric quantum-mechanical time that does not flow, that ``happens always and never''---the physical correspondent of the permanent present of human experience. And yet, we cannot let go of the feeling that also the past, which we remember, and the future, which we anticipate, are as real and basic as the present. This belief is partly innate, stemming from the fact that we have memory and the ability to plan and anticipate, and partly imbibed, stemming from the complementary fact that we are educated (perhaps against experience) to regard time as an \emph{equipartition} past-present-future. Moreover, as we look to the higher court of physics for an answer, we only get from it an indifferent reply: time is only a parameter, $t$, and all three, past, present, and future, are mere illusions. 

Like Dennett said, the mind is not an epistemic engine. This is clearly true. What is not so clear is what practical value to attach to this truth. Shall we let it guide us negatively, by distrusting all our intuitions, or shall we let it guide us positively, by distrusting only some of our intuitions but not all of them? 

The first alternative is unfeasible. To doubt the reality of everything in our conscious experience and dismiss it all,\footnote{Excepting, that is, that ``there is a thought.'' Not much help in that, though.} like a Descartes reborn, can lead us nowhere useful. On what basis would we then build and test scientific theories? The second is a more reasonable and measured option. It tells us that the mind is not built to grasp the ontological essence of the world, but that neither must it be constantly misleading us for that reason. 

Now, what conscious intuitions should we take seriously as we attempt to picture the world, and which should we cast aside? Our perceptions of colour and of hot and cold, for example, belong to the latter group; these are primitive and simple human notions evolved to keep us alive, but they are complicated constructions viewed in terms of basic physical constituents: combinations of wavelength, surface reflectivity, lighting conditions, etc.\ in the former case; combinations of temperature, conductivity, specific heat, heat flow, etc.\ in the latter.\footnote{See \cite[pp.~375--383]{Dennett:1991} for colour and \cite[pp.~80--82]{Foss:2000} for hot and cold.} 

On the other hand, our perceptions of shape, of depth (close and far), and of light and heavy match more closely, under normal circumstances, the scientific concepts of geometric object, distance, and weight inspired in them. That is, science has not yet seen any need to relegate these human feelings to secondary roles---as it has done with red, hot, and the sound of violins. What we perceive as straight, the straight-edges of physics verify (save for specially framed optical illusions); what we perceive as close, the measuring rods of physics tell us is a short distance away; what we perceive as heavy, the scales of physics tell us possesses considerable weight. In these cases, the mind does a good job as an epistemic engine.     

What about time? Is it a conscious experience that we should dismiss in the formulation of a physical theory because it is ostensibly without a simple counterpart in the world, or is it an intuition of something basic enough to be physically relevant and useful? With this we come back full circle to the early question of the Preface: does time have an own structure, or it is only a human sensation like the feelings ``happy,'' ``red,'' and ``hot''?  

We do not know. But at least this much can be said for psychological time. As mentioned in Chapter~\ref{ch:Analysis of time}, this form of experience is so ingrained in human thought that, since its beginnings, physics has described the changing world guided by it. The feeling of serial change is pictured---albeit poorly---by a background mathematical parameter $t$ along which we imagine things to flow; a parameter in terms of which we write equations of things in change, but for which physics has no equation describing \emph{its} own properties and behaviour. We see, then, that in a sense physics takes psychological time in earnest but, unable to picture it in any depth, remains content with a minimalistic account of it as ``$t$.'' As a result of this mixture of close acquaintanceship and impotence for further elucidation, rare is the theory in which the innocent-looking symbol ``$t$'' does not appear, but just as rare is encounter with a questioner who demands, ``What does `$t$' stand for?'' We are left to gather its meaning from our psychological experience and, physics being unable to say more, we are kindly asked to refrain from asking awkward questions.

Besides this scientific denial of any substance to any aspect of time as psychologically experienced, there exists another sense in which even our common-sense belief of what time is differs from our feeling of it. Perhaps as a result of the physical worldview, we are educated to regard time as an equal division into past, present, and future, but the conscious feeling of time is not precisely along these lines. We certainly remember the past, we certainly anticipate the future, but does not the present---this conspicuous ``now'' in which we \emph{exist}---at any rate \emph{feel} infinitely more especial and distinct than the past we remember (in the present) and the future we anticipate (in the present)? Why is the ``now'' a special time? Is this a sign of something physically essential or just a psychological accident?

Natural science is silent on these mysteries---on the \emph{enigma of time}. Is that because physics does not need to explain time? Because, perhaps, time is only a human illusion without any bearing on the foundations of physics? Baffled by the elusive nature of this psychological concept, some have opted to outright deny the relevance of time to elemental physics, telling us that ``In quantum gravity, [they] see no reason to expect a fundamental notion of time to play any role'' despite the fact that ``the \emph{nostalgia for time} is hard to resist;'' they tell us that ``At the fundamental level, we should, simply, forget time'' \cite[p.~114]{Rovelli:2000}.\footnote{See also \cite[p.~84]{Rovelli:2004} and \cite[p.~4]{Rovelli:2006}.} If all there is to time is nostalgia but no physics, then physics should at least be capable of telling us that the ``now,'' the past, and the future are such and such complicated constructions of such and such elemental entities, like it tells us that ``hot'' is a combination of temperature, conductivity, specific heat, and heat flow. One way or another, \emph{the enigma of time remains}. 

Other physicists, however, far from dismissing time so lightly, have taken human psychological experience in earnest as a path towards better physics. Buccheri, Jaroszkiewicz, Saniga, and others have vowed for an endophysical perspective---i.e.\ taking into account human experience---on what has so far been an exophysics---i.e.\ physics built solely upon an ``objective reality'' outside and independent of human experience \cite{Buccheri/Buccheri:2005}. They, too, have deemed the dichotomy between psychological and physical time not a failure of psychology but of physics, and have expected that, if time as an experiential product is not an illusion, then physics should be able to account for it \cite[p.~3, online ref.]{Buccheri/Saniga:2003}.
  
In particular, and indirectly making a connection with the idea of metageometry, Buccheri and Saniga \citeyear{Buccheri/Saniga:2003} criticized normal physics for striving exclusively towards a \emph{quantitative} account of nature, and deemed it ``necessary to go beyond quantifying'' (p.~4) to grasp the qualitative aspects of psychological time.\footnote{Echoing J.~Sanfey, Buccheri \citeyear[pp.~2--3, online ref.]{Buccheri:2002} expressed that the physical problem of psychological time may even be related to a so far elusive understanding of consciousness. Extending this idea, closing the 2002 NATO Workshop, Jaroszkiewicz pondered the future of physics in historical perspective with these thought-provoking words: ``If the eighteenth century may be called the century of the classical particle, the nineteenth century the century of the classical field, the twentieth century the century of the quantum, then the twenty-first century may turn out to be the century of the mind'' (http://www.pa.itd.cnr.it/buccheri).} This point had been previously brought up by Shallis \citeyear[pp.~153--154]{Shallis:1982}, who wrote that 
\begin{quote}
the fact that the experience of time is not quantifiable puts it into an arena of human perceptions that are at once richer and more meaningful than those things that are merely quantifiable\ldots The lack of quantification of temporal experiences is not something that should stand them in low stead, to be dismissed as nothing more than fleeting perceptions or as merely anecdotal; rather that lack should be seen as their strength. It is because the experience of time is not quantifiable and not subject to numerical comparison that makes it something of quality, \emph{something containing the essence of being}\ldots (italics added)\footnote{Quoted from \cite[pp.~5--6]{Buccheri/Saniga:2003}.} 
\end{quote}

To get an edge on this problem and break free from the restricted conception of time only as a background parameter, we took the permanent-now aspect of psychological experience in earnest. Metageometric quantum-mechanical time is a step in this direction. Coming to grips with a new view of time, however, is a tall order, because the current view of time is deeply ingrained in the scientific and common-sense views of the world. A look at an abnormal mind, however, may help us see time in this new light.\footnote{For a review of anomalous psychological experiences concerning time (and space), though not necessarily arising from brain damage, see \cite{Saniga:2005}.}

Tulving \citeyear{Tulving:1989} reports on the case of a man called K.C., who, upon suffering brain damage in a motorcycle accident, became amnesiac in a rather unique way: K.C.\ knows the past but does not remember it. What does this mean? K.C.\ has impersonal knowledge about himself and the world, but he does not remember any episode from his past. K.C.\ can drive a car and play chess, but has no memory of ever driving a car or playing chess with anyone. What is more, K.C.\ not only lacks conception of the past but also of the future; as far as he is concerned, there is only the present. Tulving writes:
\begin{quote}
He cannot conjure up images about his future in his mind's eye any more than he can do so about his past. Without the ability to remember what he has done or to contemplate what the future might bring, K.C.\ is destined to spend the remainder of his life in a permanent present. \cite[p.~364]{Tulving:1989}
\end{quote}
This sounds tragic, but is it not there---in the permanent present---where in a less dramatic sense we \emph{all} spend our lives after all? 

In the context of the physical enigma of time, K.C.'s case raises an intriguing question: what would physics look like if normal human beings knew the past but did not remember it and neither had conception of the future? In principle, since we would \emph{know} the past as a set of recorded facts, physics could continue to work in the same local and causal way as we know it. But would we, \emph{in practice}, without the suggestive encouragement of our episodic memory and our anticipation of the future, have come to develop the same theories? Would then any roles be attributed to locality and causality? Would ``$t$'' be a part of physics? Would there be differential equations for things changing \emph{in} time? If we were all like K.C., how would we have managed? One can only wonder.

All this is not to say that the past and future of common sense are nothing but insubstantial delusions. The purpose of this probing is rather to find a clue, an edge we can grab onto, for a deeper understanding of time. The question we intend, the seed of doubt we mean to plant, is this: is the present \emph{more basic} than the past and future from a physical point of view? And is this hypothetical physical fact actually being mirrored by aspects of the way our brains capture the world? Pondering K.C.'s case, Tulving hazards a promising interpretation in this regard. He writes: ``Remembering one's past is a different and perhaps \emph{more advanced} achievement of the brain than simply knowing about it'' (p.~361, italics added). If this were so, the psychological present would be more elemental than the psychological past and future. And if the mind has any role in informing physics in this connection, then perhaps a kind of permanent present is what physical time is made of.

\section{The taboo of time}
As we presented the enigma of time, we remarked that a conception of time other than as a featureless parameter runs counter to physical custom, and that a conception of time other than as an equipartition past-present-future runs counter to our partly innate, partly imbibed belief of what time is. In the former case, we violate scientific custom; in the latter, we provoke the emotional aversion of common sense. 

We remarked that the question, ``What is time?'' is not posed in physics, as if the tools of science were helpless to answer it; we behave as if only speculative philosophers had anything (useless) to say about it. To ask the question of a professional physicist, it is most likely to be met with annoyance. ``Well, time is \emph{time}, of course!'' replies the scholar, ``a parameter along which things flow. What else could it be?''

An annoyed reaction to the probing question, ``What is time?'' is a sign that we have trodden upon a taboo. Of the varying definitions of taboo, in this context a taboo is best described by the first reference in the footnote below as ``a ban or an inhibition resulting from social custom or emotional aversion.''\footnote{Taboo, from Tongan adjective \emph{tabu}: under prohibition. The English word first appeared in Captain James Cook's logbook ``A voyage to the Pacific ocean'' from 1777. During his stay in the Polynesian island of Tonga, he wrote that the word ``has a very comprehensive meaning; but, in general, signifies that a thing is forbidden\ldots When any thing is forbidden to be eat, or made use of, they say, that it is taboo.'' (The Free Dictionary, http://www.tfd.com) In this sense, the Western word originates from Tongan \emph{tabu}, but similar words also exist in other Pacific languages: ``[L]inguists in the Pacific have reconstructed an irreducible Proto-Polynesian *\emph{tapu}, from Proto-Oceanic *\emph{tabu} `sacred, forbidden' (cf.\ Hawaiian \emph{kapu} `taboo, prohibition, sacred, holy, consecrated;' Tahitian \emph{tapu} `restriction, sacred;' Maori \emph{tapu} `be under ritual restriction, prohibited'). The noun and verb are English innovations first recorded in Cook's book.'' (Online Etymology Dictionary, http://www.etymonline.com)}

Hardin \citeyear[pp.~xi--xiii]{Hardin:1973}, a self-professed stalker of taboos, noted that, while we like to believe that only primitive cultures have taboos, we too in sophisticated societies have our own taboos. He explains that taboos on objects and actions, although present, do not offer any intellectual challenges because, although we observe them, we can see right through them---they are rules of good taste, not sacred cultural elements. Word- and thought-taboos, however, are quite a different matter.

A word-taboo makes reference to a subject that is outside the range of topics we are ready or willing to discuss. Two excellent, complementary examples of word-taboos in current society are ``overpopulation'' and ``the desirability of economical growth.'' These are tabooed topics whose mere mention is off-bounds. They lie at the root of mankind's mounting ecological woes (e.g.\ poverty; land, air, water, and sound pollution; traffic congestion; deforestation; nuclear waste; species extinction; global warming; etc.), but nowhere besides in ignored publications are these words dared to be uttered. Instead, we dodge taboo by believing that technology will solve our ecological problems too, but in a finite world with unrestrained population growth, this hope can only be illusive.\footnote{See \cite{Hardin:1968,Hardin:1972}.}

From a taboo on expression to a taboo on thought, i.e.\ from word- to thought-taboo, it is only a small step. The less we hear a word or question uttered, the less capable we become to think about the problem, falling victim of a thought-taboo:  ``A word-taboo held inviolate for a long time becomes a taboo on thinking itself (for how can we think of things we hear no words for?)'' (p.~xi). 
 
In physics, the question about the nature of time, in the sense that it could be anything else than an unexplainable background parameter, is taboo. But, we believe, the nature of this thought-taboo on time is special and unlike any other we may be victims of. As the provocative title of this chapter declares, time is the last taboo we must deal with---last because of being rooted deepest, to different degrees, in both scientific and common-sense thought and language; last because the received view of time is knit into the fabric of science and society, and to change it is to change both the scientific and common-sense views of the world.\footnote{Time as last taboo in the sense of one coming at the end of a series of recognizable taboos we may be in need of confronting, but not, however, in any way implying that with its overthrow ``the series [of taboos] is completed or stopped'' (Cf.\ ``last'' in Merriam-Webster Online Dictionary, http://www.m-w.com). What new taboos the scientific and common-sense worldviews of the more distant future may have in store for us, there is no way of telling now.}  

How shall one proceed to being to break the spell of taboo? Not without a hint of humour, Hardin suggests as follows:
\begin{quote}
Violating a word-taboo is bad taste, but he who is an \emph{eccentric}---i.e., one who is out of the center of things---may ultimately hear a call to break the taboo lest his silence permit the final metamorphosis of word-taboo into thought-taboo---of repressed state into non-existence. When the vocation of the eccentric reaches this painfully insistent point, he shouts the word from the rooftops. This is a dangerous moment; the time must be chosen carefully. \cite[pp.~xi--xii]{Hardin:1973} 
\end{quote}
We do not know how well we may have chosen the moment and place to put forth this unorthodox view of what time could be---the serial chaining of metageometric quantum-mechanical things outside spacetime mirrored in the permanent now of human consciousness. But while we do not know how well we may have observed this rule, in the course of this work we have with certainty infringed upon another tactics of Hardin's never to ``be forgotten by the ambitious stalker of taboos: Never tackle more than one taboo at a time'' (p.~30). For, together with the assumed nature of time, we have also questioned the minor taboo on the relevance of the Planck scale to quantum gravity and, more significantly, the taboo on the nature and relevance in physical science of the sacred discipline of geometry. We believe that at least the thought-taboo on time simply had to be tackled together with the one on geometry---but then again, perhaps we have been too ambitious stalkers of the taboos of physics.

The deeply felt human dichotomy between the life-as-irretrievable-flow and permanent-present aspects of conscious experience has been captured masterfully and with rare insight by the Argentinian writer Jorge Luis Borges. He concludes his essay ``A new refutation of time'' in an attempt to come to terms with this seeming contradiction, thus: 
\begin{quote}
Time is the substance I am made of. Time is a river that grabs me, but I am the river; a tiger that tears me apart, but I am the tiger; a fire that consumes me, but I am the fire. \cite[p.~187]{Borges:1976}
\end{quote}
Which of the two feelings shall we hang on to?

\section{Overview and outlook}
Physics misrepresents time and fails to picture it because it views it as a property superimposed on the existence of things---for physics, things exist \emph{in} time. But time is not a property of the existence of things; it is instead an experiential product related to consciousness. This is to say that there is no actual ``observation of time'' or ``perception of time,'' but only the observation or perception of \emph{phenomena} or \emph{events}. How does time enter the picture, then? These events---psychologically, complexes of elementary feelings---are chained serially into the stream of consciousness. It is this flow of events in the permanent present that we call time.

To give a physical explanation of time, we need to know what plays the part of these three key concepts of things, events, and the flow of events just mentioned in general relativity and in quantum mechanics, as well as find the way the  theoretical pictures from both sides are interconnected.  

In general relativity, the role of things is so far vacant, and we cannot say whether the players of this part should be first recognized in order to progress, or whether this is inessential. Neither can we say whether these missing general-relativistic things are to be metageometric, nor what a metageometric formulation of general relativity may look like. 

At any rate, we intuitively seek for relativity theory a foundation similar to the metageometric one we gave to quantum theory. In the latter, we explained the appearance of state vectors, $|a\rangle$ and $|b\rangle$, and the full \emph{geometric background} traditionally built upon them out of the decomposition of the physically more essential probability \emph{measurement values} $|\langle b|a\rangle|^2$; and the rise of these values were, in turn, explained in terms of physically yet more elemental metageometric things, $\mathcal{P}(a)$ and $\mathcal{P}(b)$, associated with \emph{physical activities} of premeasurements. Likewise, in relativity theory, we explained the appearance of the metric field $g_{\alpha\beta}(x)$, differential arrow vectors $\D x^\alpha$, and the full \emph{geometric background} built upon them out of the decomposition of the physically more essential \emph{clock measurement} $\D s=\sqrt{-g_{\alpha\beta}(x) \D x^\alpha \D x^\beta}$. But from what metageometric, physically yet more elemental foundation do these clock-measurement values arise?

The role of events, on the other hand, is crystal-clear: it is taken up by spacetime events, understood not as spacetime points but as phenomena. Better still, to bring forth the connection with time, events can be pictured by clock readings $s_A$ at the event $A$ in question. Finally, the flow of events also finds a natural characterization in terms of the clock separation $\Delta s=s_B-s_A$, \emph{correlating} the pair of events $A$ and $B$. Central to this idea of classical event correlations $\Delta s$ is that the focus is placed on the shared connection between events instead of on the events themselves. 

In quantum mechanics, we identified things as metageometric things, for example, $\mathcal{P}(b|a)$. The role of events is now naturally filled by the metageometric time elements $\sum_i \mathcal{P}(b_i|a_i)$, which we can view as complexes of metageometric things. How can we then give a still missing characterization of the flow of these events? What shall be the metageometric tool of thought that allows us to move forward on this flank? A tool well-suited  to the task of describing metageometric quantum-mechanical correlations associated with time elements is category theory (Table~\ref{outlook-comparison}). 

\begin{table}
\centering
\setlength{\extrarowheight}{6pt}
\begin{tabular}{|p{.18\linewidth}|p{.15\linewidth}|p{.27\linewidth}|p{.26\linewidth}|}
\hline
         & Psychology &Quantum  mechanics & General relativity \\ \hline \hline
things & feelings & metageometric things $\mathcal{P}(b|a)$ & 	\qquad \qquad ? \\ \hline
events & complexes of feelings & metageometric time\phantom{-}elements $\sum_i\mathcal{P}(b_i|a_i)$  & spacetime events $A$; clock readings $s_A$ \\ \hline
flow of events & stream\phantom{-}of consciousness & category-theoretical relations\phantom{-}$r$\phantom{-}between groups\phantom{-}of\phantom{-}time elements & correlations $\Delta s$ between events\\ 
\hline	
\end{tabular}
\caption{Comparison between the three key concepts of things, events, and the flow of events for a physical description of time from the point of view of psychology, quantum mechanics, and general relativity. The task remains to provide the missing links between the three fields; in particular, to fathom a metageometric formulation of general relativity and characterize metageometric quantum-mechanical correlations with the help of category theory.}	
\label{outlook-comparison}
\end{table}

The basic components of a category are objects and relations between them. Denoting objects as $\Omega,$ $\Omega'$, and a relation $r$ between them as $\Omega \overset{r}{\longrightarrow} \Omega'$, a category is determined as follows:
\begin{enumerate}
	\item Closure: if $\Omega \overset{r}{\longrightarrow} \Omega'$ and $\Omega' \overset{s}{\longrightarrow} \Omega''$ are relations, their composition $\Omega \overset{s\circ r}{\longrightarrow} \Omega''$ is also a relation;
	\item Associativity: if $\Omega \overset{r}{\longrightarrow} \Omega'$, $\Omega' \overset{s}{\longrightarrow} \Omega''$, and $\Omega'' \overset{t}{\longrightarrow} \Omega'''$ are relations, $\Omega \overset{s\circ r}{\longrightarrow} \Omega''\overset{t}{\longrightarrow}\Omega'''$ =  $\Omega \overset{r}{\longrightarrow} \Omega'\overset{t\circ s}{\longrightarrow}\Omega'''$;
	\item Existence of an identity: if $r$ is a relation, there exists relation $i$ such that $\Omega \overset{r\circ i}{\longrightarrow} \Omega'= \Omega \overset{r}{\longrightarrow} \Omega' = \Omega \overset{i\circ r}{\longrightarrow} \Omega' $;
	\item Existence of an inverse: if $r$ is a relation, there exists relation $r^\ast$ such that $\Omega \overset{ r^\ast \circ r}{\longrightarrow} \Omega =	\Omega \overset{i}{\longrightarrow} \Omega = 	\Omega \overset{r\circ r^\ast}{\longrightarrow} \Omega $.\footnote{The existence of an inverse relation is not an axiom of category theory.}
\end{enumerate}
These four properties look conspicuously familiar. They are precisely the same ones satisfied by a group, but with a difference. Now the emphasis has shifted from the objects to the \emph{relations} between the objects. Indeed, in a category, the inner structure of the objects is not a concern; on the contrary, in a category it is the existence of relations that carries essential significance. What is more, a category---unlike its natural geometric counterpart, a graph---is abstract enough to be naturally free of geometric denotation: its objects are not points, nor are its relations extended arrows. 

We mentioned in passing on pages \pageref{inner-structure-1} and \pageref{inner-structure-2} that a key to the problem of the geometric ether could be to look not into the inner structure or physical reality of the spacetime points themselves but instead at the classical correlations given by the intervals $\Delta s$ between them; in non-geometric language, to look at the classical temporal correlations $\Delta s$ between events as phenomena for a physical description of the \emph{flow} of events from the point of view of general relativity.

Could now category theory, with its shift of focus from objects to relations, be meaningfully used as a suitable setting for a similar study of co\emph{relations} between metageometric time elements $\sum_i\mathcal{P}(b_i|a_i)$? Not really; the connection need not be so straightforward. For a physical description of metageometric flow from the point of view of quantum mechanics, instead of simply taking isolated objects (cf.\ time elements) or sets of them, the most natural choice of category is the \emph{category of groups}: its objects are the groups formed by metageometric time elements and its relations are group homomorphisms.

So far, regarding a connection between ``spacetime and the quantum,'' this is what we have achieved. After reducing the clock time of relativistic physics to the absolute time of classical physics, we found a link between the latter and metageometric quantum-mechanical time:
\begin{equation}
\Delta s = \Delta t \longrightarrow \hat U(\delta t) \longrightarrow \hat T_{ba} \longrightarrow \sum_i \mathcal{P}(b_i|a_i). \nonumber
\end{equation}
The challenge that remains is to find the link between the full-fledged general-relativistic clock time element and the quantum-mechanical metageometric time element for a combined understanding of time.

We believe that it is through the concept of time that quantum mechanics and general relativity are connected---that it is the concept of time that has the potential to give physical meaning to the presently disoriented field of quantum gravity. But rather than ask \emph{what time is}, we should ask \mbox{\emph{why time is}}.

\backmatter

\chapter{Epilogue}
\begin{flushright}
\begin{tabular}{p{10cm}}
\emph{If you pay a man a salary for doing research, he and you will want to have something to point to at the end of the year to show that the money has not been wasted. In promising work of the highest class, however, results do no come in this regular fashion, in fact years may pass without any tangible result being obtained, and the position of the paid worker would be very embarrassing and he would naturally take to work on a lower, or at any rate a different plane where he could be sure of getting year by year tangible results which would justify his salary. The position is this: You want this kind of research, but, if you pay a man to do it, it will drive him to research of a different kind. The only thing to do is to pay him for doing something else and give him enough leisure to do research for the love of it.}
\smallskip \\
Sir Joseph J.~Thomson, \emph{The life of Sir J.~J.~Thomson} 
\end{tabular}
\end{flushright}

\bigskip

In this thesis, we brought to attention the tight relationship existing between geometric thought and the human mind. Resembling the remarkable gift given King Midas by Dionysus, the mind has greatly profited from its Midas touch of turning into geometry everything it beholds, as a world of understanding opened up through it; but like King Midas' gift, geometry too turned from gift to curse by enslaving us. Our exclusive use of geometry is to blame for the vicious perpetuation of several mysteries in our physical worldviews, as well as for restricting thought towards new and better physics. As King Midas discovered upon his first meal, turning everything we touch into gold---gold though it is---is not as desirable as we might have thought. But where shall we find our river Pactolus to divest the mind of its King-Midas touch, of this superlative gift turned like a curse upon ourselves?

Although far from finding river Pactolus, at least partly through metageometric thinking and theorizing, and based on existing observations, we managed to clarify the experimental, mathematical, and psychological foundations of relativity and quantum theories. We founded the former geometrically on clock-reading separations $\D s$, and the latter metageometrically on the experiment-based concepts of premeasurement and transition things. Metageometric things were inspired in the physically unexplored aspect of time as a consciousness-related product, which allowed us to develop the concept of metageometric time. We deemed it necessary to discover the connection between quantum-mechanical metageometric time elements and general-relativistic clock time elements $\D s$ (classical correlations) in order to attain a combined understanding of time in the form of metageometric correlations of quantum-mechanical origin.

In both cases, clarification was achieved through the identification of suitable experiments and measurements that Man has access to and can affect. Physics must, after all, be based not on mathematics or professional philosophy but on physics---a triviality that seems to have been largely forgotten. Experiment is, in this light, both the beginning and the end of physical theorizing. In the field of quantum gravity, however, the physical referent---that which it is about---has not been identified. The present quest of frontier physics, an orphan without a model in the  world, is then a search for its own identity. For this reason, the identification of observables remains a central goal.

The kind of metageometric endeavours contemplated here, however, are no easy task. Human geometric compulsions are significant, and our imagination easily succumbs to its geometric overlords. Consider the following analogy. Imagine, as part of the normal universe we know, an accidentally static world populated by motionless inhabitants, who are nonetheless endowed with a good understanding of length. What kind of toiling against mental barriers might they experience, not naturally able to conceive of motion (for no evolutionary advantage attached to this ability), in order to discover and formulate the special-relativistic effect of length contraction? Length contraction itself would be perfectly comprehensible and measurable for these beings, but its discovery could be impossible and its cause would remain an enigma without an understanding of motion. In our case, the discovery of quantum-mechanical effects related to empty spacetime may be impossible, and an understanding of their cause an enigma, without previous comprehension of the metageometric things from which they arise.

In spite of the evident difficulties to forgo geometric thinking in human theorizing, the situation is not hopeless. Geometric thinking is a deep-rooted human habit, but a habit nonetheless; an extrinsic limitation forced on the plastic structure of the human brain by evolutionary processes, but not an intrinsic one impossible to overcome. A clue supporting this view is found in human language---itself a more recent evolutionary product---which, as we showed with this thesis, is capable of supporting geometric thought as well as metageometric ruminations. Language, as it were, succeeds in aiding our metageometric thought processes right where our picturing abilities leave us high and dry. Whether, in the space of eons, the brain of \emph{Homo sapiens} could evolve to naturally support metageometric thought in full depends on whether or not, as the needs imposed by the environments in which we live gradually change, any evolutionary advantage would attach itself to this unexplored manner of psychological portrayal. At such a time, ``describing the physical laws without reference to geometry'' (like Einstein said) would be much more effortless and natural than it has been for me to describe the metageometric thoughts poured in these pages with these words. It is preeminently the possibility of human change---coupled with the recognition that the human mind is not, neither now nor ever, a universal cognitive tool---that makes the pronouncement of ultimate anthropic principles on physical cognition an unwise activity indeed.  

But leaving behind possibilities that can only be guessed at, and concentrating on the immediate here and now, what are the psychological requirements for the successful advancement in the physical sciences of the metageometric endeavour we have here but glimpsed upon or, for that matter, of any other piece of genuinely new physics? Or to put differently, what does it take for human scientists to be capable of highly innovative thought? What kind of mortals are up to such a cyclopean task?

In sketching an answer to this question (and just when we thought we had left evolutionary thoughts behind), a parallel can be drawn with evolutionary theory as suggested by Hardin \citeyear[pp.~258--346]{Hardin:1960}. The key to the successful perpetuation of any species is for its members to differentiate themselves into races, each inhabiting a reproductively \emph{isolated} ecological niche, for then the species as a species enjoys high genetic diversity and is apt to spread into and exploit virgin niches of opportunity (in particular, when confronted with changes of their ecosystems). Species all of whose conservative members are finely adapted only to the peak of ``Mount Tory'' (p.~293) lack the capacity to stray away into virgin land: undiscovered ``Mount Opportunity'' is only available to \emph{some} of the less finely adapted races of ``Mount Risky,'' where each race of the species \emph{evolves in isolation}.  

The same is the case in the sphere of science. The development of truly original ideas requires unorthodox, heretical thought. This kind of highly original thought needed to reach the peak of Mount Opportunity, however, presupposes detachment from the ubiquitous influence of orthodox ideas, from the ebbs and flows of scientific fashions. At some point, the creative spirit must draw his gaze away from his peers and look within himself, making the scientific affair with the natural world a personal matter.\footnote{Charles Darwin, who secluded himself with his family in the English village of Down is, coincidentally, a paradigmatic example of this way of proceeding.} High creativity, and the promise of Mount Opportunity, require thoughts and ideas to \emph{evolve in isolation}. And just like the blind opportunism of evolution can, through the discarding of unfit, wasteful products, create life forms undreamt-of in the drawing board of the most intelligent designer, so can, by the same method, the independent conceptions of myriads of genuinely autonomous men lead to the discovery of ``more things in heaven and earth,'' as Shakespeare's Hamlet would say to today's team-working Horatios, ``than are dreamt of in your philosophy.''

A look at frontier physics reveals that the figure of the intellectually independent man of science is regrettably far from holding court today; creative thought on the basis of collective ideas is not the exception but the rule. Before the approving gaze of the scientific community and of society as a whole, legions of scientists and professional philosophers single-mindedly toil today, among other trends, in the digging of a bottomless hole argument, in the exploration of hidden bends and corners of this universe and the conjuring up of myriads of others, in the knitting of the world out of strings or loops, and---the most irresistible, compelling, and widespread trend of all---in the exploration of the secluded depths of the Planck scale. On the other hand, not to follow such trends and go one's own way is severely looked down upon: not one of the official roads to quantum gravity. But if scientific breakthroughs (as in Kuhnian revolutions) is what we are after---and so it would seem from the grand visions of quantum gravity---these have never been known to arise from the teamwork of hundreds but from the intellect of one creative spirit. Seeing the same old world in a new light is largely a one-man affair. Echoing these thoughts, and even the evolutionary parallelism, we hear the reverberation of Richard Feynman's words: 
\begin{quote}
The only thing that is dangerous is that everybody does the same thing! [\ldots] There are enormous numbers of possibilities\ldots And we have to explore. So we ought to be running around in as many directions as possible.\footnote{In \cite[p.~196]{Davies/Brown:1988}.} 
\end{quote}
Yet, as Hardin \citeyear{Hardin:1960} observed, this independent attitude does not come easy to those already ``breathing an atmosphere of fellowship and togetherness'' (p.~343), i.e.\ to those accustomed to what Smolin \citeyear{Smolin:2001} described as ``the comfort of travelling with a crowd of like-minded seekers'' (p.~10). After all, Man is by nature a gregarious animal.

We recognize in the formulaic, repetitious, and unoriginal character of frontier physics, to borrow Lem's \citeyear[pp.~173--177]{Lem:1984b} words, an ``age of epigonism''\footnote{Epigone, from Greek \emph{epi} ``after'' + \emph{gonos} ``seed'': to be born after; an inferior or undistinguished imitator, follower, or successor.} or ``repetitive fulfillment'' of the geometric physical tradition. Any creative tradition---whether artistic, scientific, or even natural---is based on an underlying space of possibilities determined by its own methods. A painter who only uses a straight-edge will only produce straight-line paintings, just like a physicist who only uses geometric methods will only produce geometric theories of nature. Creation within an budding creative tradition is difficult because, although the field of opportunity is vast, its possibilities have not yet even been mapped out, its scholars resembling more explorers than inventors (cf.\ metageometry). Original creation is easiest when the tradition has been sufficiently explored but its possibilities not even nearly exhausted (cf.\ geometric classical mechanics and the geometric differential calculus). However, when a creative tradition has been nearly thoroughly exploited, the production of new original work within it becomes extremely difficult (cf.\ geometric general relativity), leads sometimes to viable yet malformed new specimens (cf.\ geometric quantum theory), and most of the time reduces to the repetition of older products (cf.\ geometric quantum gravity). 

Returning once more to the evolutionary picture, the best exemplification of repetitive fulfillment within a creative tradition is found in nature itself. When a new taxonomic class of organism (e.g.\ \emph{Insecta}, \emph{Mammalia}) appears, the possible successful species allowed within the creative principle of this class is immense (for \emph{Insecta} of the order of millions). As ecological niches become occupied, however, speciation stops and nature contents itself with an incessant repetition of its earlier successful products. \emph{The creative tradition has become full.} The geometric physical tradition seems now to have reached a similar fulfillment---its fateful age of epigonism.

If forward necessarily means different, how come our present social framework, in which science is cherished so highly, does not encourage the creative development of this discipline? What is the source of this state of affairs? The paradoxical predicament in which science administrators have placed the scientific enterprise is nothing new, as it was noticed by J.~J.~Thomson over 65 years ago (opening quote), but it is correct to say that the situation has only worsened in time. No better analysis of the present predicament of science---and no better thoughts on which to end this dissertation---than in the lucid and uncompromising words of Garrett Hardin:
\begin{quote}
As it became generally realized that an important fraction of the world's research in pure science was done by academic men, administrators defined research as part of the job, and made productivity in research a criterion for advancement. The consequences of this meddling have been about what one would expect. There is now a tendency to choose projects that are pretty sure to give quick results, and to avoid questions on tabooed subjects\ldots As research has become more expensive, the academic man has had to develop a talent for begging. He gets a subsidy from foundations by telling committees what he hopes to accomplish with their money if he gets it. The successful beggar often gives more attention to the committee than he does to the scientific problem. The result: other-directedness is introduced into a realm where it has no business being, the realm of inner-directed science. Orthodoxy is encouraged. This may not be too bad for what we \emph{call} ``science,'' for its fields are almost all freed now of conflict with tradition, and its methods systematized to the point where innumerable and immensely important discoveries can be made by men who are not in the first rank of the heretics. But there is need for the spirit of science to move into fields not now called science, into fields where tradition still holds court. We can hardly expect a committee to acquiesce in the dethronement of tradition. Only an individual can do that, an individual who is not responsible to the mob. Now that the truly independent man of wealth has disappeared, now that the independence of the academic man is fast disappearing, where are we to find the conditions of partial alienation and irresponsibility needed for the highest creativity? \cite[pp.~344--345]{Hardin:1960}
\end{quote}

May the last survivors of these endangered races of men conquer the taboos necessary to unlock the enigmas of the cosmos!

\newpage
\thispagestyle{empty}
\vspace{5cm}
\begin{flushleft}
\begin{tabular}{p{9cm}}
\emph{Geometry! You have accompanied mankind}\\
\emph{For longer than it can remember;}\\
\emph{Well-entrenched within the chambers of its mind,}\\
\emph{You glow as though relentless ember.}

\bigskip

\emph{Anciently, your shapes enticed the Babylonians;}\\
\emph{On clay they would record inscriptions}\\
\emph{Reverenced in all our worldviews: from Newtonian}\\
\emph{To quantum-world ornate descriptions.}

\bigskip

\emph{Theories have been presented in your honour;}\\
\emph{Correctness based upon your clarity;}\\ 
\emph{Ignorance is predicated on the goner}\\
\emph{To do without your familiarity.}

\bigskip

\emph{Presently, despite your triumphs (which are ours),}\\
\emph{Farewell we bid you, \emph{alter ego},}\\ 
\emph{Fearing that you have run out of your powers,}\\
\emph{Disorienting us all, as we go!}

\end{tabular}
\end{flushleft}

\chapter[Appendix A: To picture or to model?]{Appendix A:\\ To picture or to model?}
Throughout this work, the verb ``to picture'' is used instead of the more usual ``to model,'' and likewise ``theoretical picture'' is used instead of ``theoretical model.'' This is because the former usage makes the subjective, theorist-ridden nature of theoretical portrayal in science more explicit by putting it on a par with the equally subjective, artistic renderings of a painter; the meaning of ``to model,'' on the other hand, is ambiguous. 

Moster\'{i}n \citeyear[pp.~131--146]{Mosterin:1984} captured the relationship between the key notions surrounding a representational activity in the canonical expression ``$x$ paints $y$ which represents $z$.'' An application to everyday usage gives ``the painter ($x$) paints the painting ($y$) which represents the model ($z$).'' Here the model---that which is to be subjectively mimicked, emulated, replicated---is the thing (e.g.\ a landscape) or person (e.g.\ a human \emph{model}) which the painter captures---according to his own biases and tools---through and in his painting. Thus, the counterpart of this expression in scientific usage is ``the scientist ($x$) devises the theory ($y$) which represents the section of reality ($z$).'' As a result, Moster\'{i}n noted that the use of the word ``model'' made by natural scientists---to denote a theoretical construct that attempts to explain a section of reality---is diametrically opposed to that of everyday usage, where by ``model'' we rather denote that section of reality which we ``copy,'' i.e.\ that section of reality theorized upon by the scientist or artist.

In everyday usage ``to model'' can also refer to the making of a material object (e.g.\ a maquette of a building or a wooden plane built to scale). Is this meaning opposed to the one just mentioned? Does it make the meaning of ``model'' as ``theory''---both something we make---more plausible? Not at all. When we make a material model, we do not use it as a \emph{theoretical} tool to study something else, but as a scaled-down version of that section of reality we are interested in, and which we thus find more manageable to study. It, the model, is again the object of our theoretical study, not the theoretical tool itself. For example, we now apply the laws about the strength of materials to the maquette of the building, or we place the wooden plane in a wind tunnel to study \emph{its} aerodynamics.  This is what Moster\'{i}n called ``to serve as a model.'' 

Incidentally, the scientific use of ``model'' is also opposed to the meaning given it in the theory of models, which again coincides with that of everyday language. This theory formally studies sections of reality which it calls systems; for every system, it considers a theory that will explain its functioning. If and only if a system works as the theory states, it is called a model of the theory.

Ultimately, one can attribute meaning to words in any way one wishes, and the scientist is free to continue to call his theories models if that is his inclination. But, at least from the perspective explained here, the widely used expression ``theoretical model'' is natural scientists' favourite, unwitting oxymoron.

\chapter[Appendix B: Preparation time]{Appendix B:\\ Preparation time}

\setcounter{chapter}{2}
\renewcommand{\theequation}{\Alph{chapter}.\arabic{equation}}
\setcounter{equation}{0}

\setcounter{chapter}{2}
\renewcommand{\thefigure}{\Alph{chapter}.\arabic{figure}}
\setcounter{figure}{0}

Is there any quantum-mechanical concept of time---or, more precisely, time \emph{instant}---at the geometric level? Could such a concept be connected with the transition function $\langle b|a\rangle$ (geometrically, the inner product)? The idea appears promising because the  transition function represents the inner rearrangement of physical systems at preparation, and change is at the heart of what we understand by time. The concept of time to be derived here, leading to a primitive form of Schr\"{o}dinger's equation, is called \emph{preparation time} after this association.

We start by examining an infinitesimally small change $\delta\langle b|a\rangle$ of $\langle b|a\rangle$ in terms of the interaction between ket and bra vectors $|a\rangle$ and $\langle b|$.\footnote{See \cite[pp.~1550--1551]{Schwinger:1959} for most of the mathematical formalism below but without relation to time.} In order to work with $\delta\langle b|a\rangle$, we need to express it in terms of concepts whose properties and behaviour we know. To this end, we introduce the \emph{first geometric time postulate},
\begin{equation} \label{1GTP}
\delta\langle b|a\rangle=i\langle b|\delta\hat F_{ba}|a\rangle,
\end{equation}
which gives a geometric representation to the infinitesimal change of the transition function as a mediated transition between arrow state vectors. The mediation is performed by operator $\delta\hat F_{ba}$, which works as the intermediary of the interaction between any state vector $|a\rangle$ (in general, $|a_i\rangle$) associated with observable $A$ and any state vector $|b\rangle$ (in general, $|b_i\rangle$) associated with observable $B$. We call $\delta\hat F_{ba}$ the \emph{infinitesimal action operator}. 

Assuming the change of the transition function to be continuous, we find that $\delta$ satisfies the chain rule for a product of transition functions:
\begin{align}\label{chainrule}
\delta (\langle d|c\rangle \langle b|a\rangle)&=
(\langle d|c\rangle + \delta \langle d|c\rangle)(\langle b|a\rangle+\delta\langle b|a\rangle)-\langle d|c\rangle\langle b|a\rangle\nonumber\\
&=\langle d|c\rangle\delta\langle b|a\rangle +\delta\langle d|c\rangle\langle b|a\rangle + \delta\langle d|c\rangle \delta\langle b|a\rangle \nonumber\\
&= \langle d|c\rangle \delta\langle b|a\rangle + \delta\langle d|c\rangle\langle b|a\rangle +O[(\delta \langle b|a\rangle)^2].   
\end{align}
Thus, $\delta$ works mathematically like a differential insofar as the chain rule is concerned, but it must not be confused with a regular differential, since in general $\delta \langle b|a\rangle \neq 0$, where $\langle b|a\rangle$ is a complex number. As it will turn out, and as we could expect, there is no infinitesimal change only for $\langle a_j|a_i\rangle =\delta_{ij} \in \mathbb{R}$, i.e.\ when no transition actually occurs.

Postulate (\ref{1GTP}), decomposition relation (\ref{DR}), and chain rule (\ref{chainrule}) give us enough information about the physical properties of $\delta \hat F_{ba}$. In effect, because
\begin{equation} 
\delta\langle b|a\rangle=\sum_i\Big[ (\delta\langle b|c_i\rangle) \langle c_i|a\rangle+\langle b|c_i\rangle (\delta\langle c_i|a\rangle)\Big],
\end{equation}
we find
\begin{align}
i\langle b|\delta\hat F_{ba}|a\rangle &= \sum_i\left[
i\langle b|\delta\hat F_{bc}|c_i\rangle \langle c_i|a\rangle+
\langle b|c_i\rangle i\langle c_i|\delta\hat F_{ca}|a\rangle\right] 
\nonumber\\
&= i\langle b|\left(\delta\hat F_{bc}\sum_i|c_i\rangle\langle c_i|+\delta\hat F_{ca}\sum_i|c_i\rangle\langle c_i|\right)|a\rangle,
\end{align}
where we have now interpreted the quasigeometric decomposition relation geometrically. This leads to the result
\begin{equation}\label{AOP1}
\delta\hat F_{ba}=\delta\hat F_{bc}+\delta\hat F_{ca} \qquad \mathrm{for\ all}\  a,b,c.	
\end{equation}
This tells us that the interaction between two (geometrically represented) states of a system can be mediated by a third such state in a serial manner. From Eq.~(\ref{AOP1}) we obtain two special results. Setting $c=a$, we find
\begin{equation}
\delta\hat F_{ba}=\delta\hat F_{ba}+\delta\hat F_{aa}, 
\end{equation}
which implies
\begin{equation}\label{AOP2}
\delta\hat F_{aa}=\hat 0 \qquad \mathrm{for\ all}\ a.	
\end{equation}
This means that the infinitesimal action operator is null in the absence of change; i.e.\ in the absence of a transition, as in $\langle a_j|a_i\rangle =\delta_{ij}$, no instant of preparation time is created by its infinitesimal change. On the other hand, the change of a genuine quantum-mechanical transition $\langle b|a\rangle$ does determine an instant of preparation time.

Now, setting $b=a$ in Eq.~(\ref{AOP1}) and using Eq.~(\ref{AOP2}), we find
\begin{equation}
\hat 0 = \delta\hat F_{aa} = \delta\hat F_{ac}+\delta\hat F_{ca}, 
\end{equation}
which implies
\begin{equation}\label{AOP3}
\delta\hat F_{ac}=-\delta\hat F_{ca} \qquad \mathrm{for\ all}\ a,c.	
\end{equation}
This tells us that a change of sign of the action operator indicates a reversal of the direction in which it mediates the transition.

Finally, using Eq.\ (\ref{AOP3}), the transition-function postulate (\ref{TFP}), and an extension to it in the sense that complex conjugation must reverse not only the direction of the transition but also that of a \emph{change} in the transition---$\delta\langle b|a\rangle=(\delta\langle a|b\rangle)^\ast$---we obtain that the action operator is Hermitian. In effect, because
\begin{equation}
i\langle b|\delta\hat F_{ba}|a\rangle=[i\langle a|\delta\hat F_{ab}|b\rangle]^\ast = -i\langle b|\delta\hat F_{ab}^\dag|a\rangle,
\end{equation}
we find
\begin{equation}
\delta\hat F_{ba}=-\delta\hat F_{ab}^\dag =\delta\hat F_{ba}^\dag.
\end{equation}

\begin{figure}
\begin{center}
\includegraphics[width=92mm]{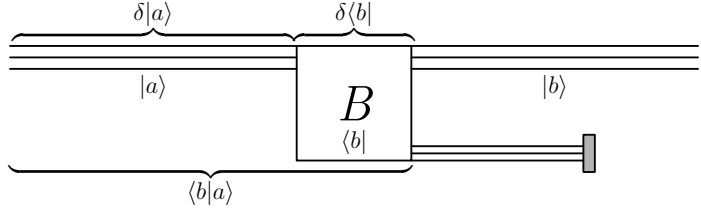}
 \end{center}
\caption[Preparation time]{Schematic representation of preparation time in terms of a transition box and arrow vectors.} 
\label{preptime}
\end{figure}

The introduction of a second geometric time postulate, indicating the simplest possible dependence of $\delta\hat F_{ba}$ on $b$ and $a$, allows us next to recover a primitive form of Schr\"{o}dinger's equation linking infinitesimal changes of $|a\rangle$ and $\langle b|$ with $|a\rangle$ and $\langle b|$ themselves. The \emph{second geometric time postulate} is
\begin{equation} \label{2GTP}
\delta\hat F_{ba}=\hat G(b)-\hat G(a).
\end{equation}
Here the Hermitian operator $\hat G$ works as generator of infinitesimal transformations of $|a\rangle$ and $\langle b|$.
Using this postulate, we find, on the one hand,
\begin{equation}
\delta\langle b|a\rangle =i\langle b|\hat G(b)|a\rangle-i\langle b|\hat G(a)|a\rangle.	
\end{equation}
This is an equation for the infinitesimal change of the transition function in terms of ket and bra vectors. On the other hand, using the geometric decomposition of the transition function\footnote{This is perfectly reasonable because this particular analysis of time is carried out at the geometric level.} and the chain rule, we can find another equation for the infinitesimal change of the transition function:
\begin{equation}
\delta\langle b|a\rangle =(\delta\langle b|)|a\rangle+\langle b|(\delta |a\rangle)
\end{equation}
(Figure \ref{preptime}). The two equations together imply 
\begin{equation} \label{ketSchrodinger}
i\delta |a\rangle=\hat G(a)|a\rangle 
\end{equation}
and
\begin{equation}\label{braSchrodinger}
-i\delta \langle b|=\langle b|\hat G(b), 
\end{equation}
where we dismiss the alternative implication on a physical basis.

Results (\ref{braSchrodinger}) and (\ref{ketSchrodinger}) are \emph{primitive Schr\"{o}dinger equations}, respectively, for the bra vector $\langle b|$ at preparation and for the ket vector $|a\rangle$ just outside preparation. Thus, Hermitian operator $\hat G$, generator of infinitesimal change of $|a\rangle$
and $\langle b|$, can be considered a \emph{primitive ancestor} of the Hamiltonian $\hat H$, itself the generator of translations of a system in time $t$.

\begin{figure}
\begin{center}
\includegraphics[width=90mm]{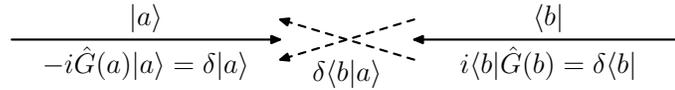}
 \end{center}
\caption[Birth of preparation time]{Birth of preparation time, i.e.\ of a quantum-mechanical geometric time instant, at the borderline of a quantum-mechanical premeasurement process as the combination of the infinitesimal change $\delta\langle b|$ of a bra vector at preparation and the infinitesimal change $\delta|a\rangle$ of a ket vector just outside preparation.} \label{schrpreptime}
\end{figure}

In this geometric analysis, then, the rise of a quantum-mechanical time instant comes about in two parts, namely, the infinitesimal change $\delta\langle b|$ of a bra vector at preparation and the infinitesimal change $\delta|a\rangle$ of a ket vector just outside preparation. Preparation time is by nature inherent in preparation, and it is born at the preparation borderline (Figure \ref{schrpreptime}).

A quantum-mechanical concept of time inspired in a similar idea is that of a \emph{q-tick} proposed by Jaroszkiewicz \citeyear{Jaroszkiewicz:2001}. This time arises qualitatively from the ``irreversible acquisition of information'' that occurrs after a measurement is made on a quantum-mechanical system. In contrast to a q-tick, preparation time arises here from quantum-mechanical transitions (in general, premeasurements) rather than from measurements. Further, the infinitesimal changes of the transition function giving rise to a geometric time instant are governed by the primitive Schr\"{o}dinger equations~(\ref{ketSchrodinger}) and (\ref{braSchrodinger}).

\bibliographystyle{apacite}
\bibliography{fullbibliography}

\begin{thebibliography}{}

\bibitem[\protect\citeauthoryear{%
{Amelino-Camelia}%
}{%
{Amelino-Camelia}%
}{%
{\protect\APACyear{2003}}%
}]{%
Amelino-Camelia:2003}%
\APACinsertmetastar{%
Amelino-Camelia:2003}%
{Amelino-Camelia}, G.%
%
\newblock{}\BBOP{}2003\BBCP{}.
\newblock{}\Bem{\emph{Planck-scale structure of spacetime and some implications
  for astrophysics and cosmology}.}
\newblock{}(arXiv:astro-ph/ 0312014v2)

\bibitem[\protect\citeauthoryear{%
Anandan%
}{%
Anandan%
}{%
{\protect\APACyear{1997}}%
}]{%
Anandan:1997}%
\APACinsertmetastar{%
Anandan:1997}%
Anandan, J.~S.%
%
\newblock{}\BBOP{}1997\BBCP{}.
\newblock{}\BBOQ{}Classical and quantum physical geometry.\BBCQ{}
\newblock{}\BIn{} R.~S. Cohen, M.~Horne\BCBL{}\ \BBA{} J.~Stachel\ (\BEDS),
  \Bem{Potentiality, entanglement and passion-at-a-distance: {Quantum}
  mechanical studies for {Abner} {Shimony}}\ (\BVOL~2, \BPGS\ 31--52).
\newblock{}Dordrecht: Kluwer.
\newblock{}(arXiv:gr-qc/9712015v1)

\bibitem[\protect\citeauthoryear{%
Antonsen%
}{%
Antonsen%
}{%
{\protect\APACyear{1992}}%
}]{%
Antonsen:1992}%
\APACinsertmetastar{%
Antonsen:1992}%
Antonsen, F.%
%
\newblock{}\BBOP{}1992\BBCP{}.
\newblock{}\Bem{Pregeometry}.
\newblock{}Master thesis, University of Copenhagen, Niels Bohr Institute.

\bibitem[\protect\citeauthoryear{%
Ashtekar%
}{%
Ashtekar%
}{%
{\protect\APACyear{2005}}%
}]{%
Ashtekar:2005}%
\APACinsertmetastar{%
Ashtekar:2005}%
Ashtekar, A.%
%
\newblock{}\BBOP{}2005\BBCP{}.
\newblock{}\BBOQ{}Gravity and the quantum.\BBCQ{}
\newblock{}\Bem{New Journal of Physics}, \Bem{7}, 198 (1--33).
\newblock{}(arXiv:gr-qc/0410054v2)

\bibitem[\protect\citeauthoryear{%
Asimov%
}{%
Asimov%
}{%
{\protect\APACyear{1972}}%
}]{%
Asimov:1972}%
\APACinsertmetastar{%
Asimov:1972}%
Asimov, I.%
%
\newblock{}\BBOP{}1972\BBCP{}.
\newblock{}\Bem{Asimov's guide to science}\ (\BVOL~1).
\newblock{}New York: Penguin Books.

\bibitem[\protect\citeauthoryear{%
Auyang%
}{%
Auyang%
}{%
{\protect\APACyear{2001}}%
}]{%
Auyang:2001}%
\APACinsertmetastar{%
Auyang:2001}%
Auyang, S.~Y.%
%
\newblock{}\BBOP{}2001\BBCP{}.
\newblock{}\BBOQ{}Spacetime as a fundamental and inalienable structure of
  fields.\BBCQ{}
\newblock{}\Bem{Studies in History and Philosophy of Modern Physics}, \Bem{32},
  205--215.

\bibitem[\protect\citeauthoryear{%
Baez%
}{%
Baez%
}{%
{\protect\APACyear{2000}}%
}]{%
Baez:2000}%
\APACinsertmetastar{%
Baez:2000}%
Baez, J.~C.%
%
\newblock{}\BBOP{}2000\BBCP{}.
\newblock{}\BBOQ{}Higher-dimensional algebra and {Planck}-scale physics.\BBCQ{}
\newblock{}\BIn{} C.~Callender\ \BBA{} N.~Huggett\ (\BEDS), \Bem{Physics meets
  philosophy at the {Planck} scale}\ (\BPGS\ 177--195).
\newblock{}Cambridge: Cambridge University Press.
\newblock{}(arXiv:gr-qc/9902017v1)

\bibitem[\protect\citeauthoryear{%
Barrow%
}{%
Barrow%
}{%
{\protect\APACyear{1992}}%
}]{%
Barrow:1992}%
\APACinsertmetastar{%
Barrow:1992}%
Barrow, J.~D.%
%
\newblock{}\BBOP{}1992\BBCP{}.
\newblock{}\Bem{Theories of everything: {The} quest for ultimate explanation}.
\newblock{}London: Vintage Books.

\bibitem[\protect\citeauthoryear{%
Bombelli%
, Lee%
, Meyer%
\BCBL{}\ \BBA{} Sorkin%
}{%
Bombelli%
\ \protect\BOthers{.}}{%
{\protect\APACyear{1987}}%
}]{%
Bombelli/Lee/Meyer/Sorkin:1987}%
\APACinsertmetastar{%
Bombelli/Lee/Meyer/Sorkin:1987}%
Bombelli, L.%
, Lee, J.%
, Meyer, D.%
\BCBL{}\ \BBA{} Sorkin, R.~D.%
%
\newblock{}\BBOP{}1987\BBCP{}.
\newblock{}\BBOQ{}Space-time as a causal set.\BBCQ{}
\newblock{}\Bem{Physical Review Letters}, \Bem{59}, 521--524.

\bibitem[\protect\citeauthoryear{%
Borges%
}{%
Borges%
}{%
{\protect\APACyear{1976}}%
}]{%
Borges:1976}%
\APACinsertmetastar{%
Borges:1976}%
Borges, J.~L.%
%
\newblock{}\BBOP{}1976\BBCP{}.
\newblock{}\BBOQ{}Nueva refutaci\'on del tiempo.\BBCQ{}
\newblock{}\BIn{} \Bem{Otras inquisiciones}\ (\BPGS\ 170--187).
\newblock{}Buenos Aires: Alianza.
\newblock{}(http://www.literatura.us/ borges/refutacion.html. Original work
  published 1947.)

\bibitem[\protect\citeauthoryear{%
Boyer%
\ \BBA{} Merzbach%
}{%
Boyer%
\ \BBA{} Merzbach%
}{%
{\protect\APACyear{1991}}%
}]{%
Boyer/Merzbach:1991}%
\APACinsertmetastar{%
Boyer/Merzbach:1991}%
Boyer, C.~B.%
\BCBT{}\ \BBA{} Merzbach, U.~C.%
%
\newblock{}\BBOP{}1991\BBCP{}.
\newblock{}\Bem{A history of mathematics}\ (Second\ \BEd).
\newblock{}New York: Wiley.

\bibitem[\protect\citeauthoryear{%
Brans%
}{%
Brans%
}{%
{\protect\APACyear{1999}}%
}]{%
Brans:1999}%
\APACinsertmetastar{%
Brans:1999}%
Brans, C.~H.%
%
\newblock{}\BBOP{}1999\BBCP{}.
\newblock{}\BBOQ{}Absolute spacetime: {The} twentieth century ether.\BBCQ{}
\newblock{}\Bem{General Relativity and Gravitation}, \Bem{31}, 597--607.
\newblock{}(arXiv:gr-qc/9801029v1)

\bibitem[\protect\citeauthoryear{%
Brassard%
}{%
Brassard%
}{%
{\protect\APACyear{1998}}%
}]{%
Brassard:1998}%
\APACinsertmetastar{%
Brassard:1998}%
Brassard, L.%
%
\newblock{}\BBOP{}1998\BBCP{}.
\newblock{}\Bem{The perception of the image world}.
\newblock{}Doctoral thesis, Simon Fraser University, Vancouver.
\newblock{}(http://www.ensc.sfu.ca/people/ grad/brassard/personal/THESIS)

\bibitem[\protect\citeauthoryear{%
Breuer%
}{%
Breuer%
}{%
{\protect\APACyear{1997}}%
}]{%
Breuer:1997}%
\APACinsertmetastar{%
Breuer:1997}%
Breuer, T.%
%
\newblock{}\BBOP{}1997\BBCP{}.
\newblock{}\BBOQ{}What theories of everything don't tell.\BBCQ{}
\newblock{}\Bem{Studies in History and Philosophy of Modern Physics}, \Bem{28},
  137--143.

\bibitem[\protect\citeauthoryear{%
Bridgman%
}{%
Bridgman%
}{%
{\protect\APACyear{1963}}%
}]{%
Bridgman:1963}%
\APACinsertmetastar{%
Bridgman:1963}%
Bridgman, P.~W.%
%
\newblock{}\BBOP{}1963\BBCP{}.
\newblock{}\Bem{Dimensional analysis}.
\newblock{}New Haven, London: Yale University Press.
\newblock{}(Original revised edition published 1931.)

\bibitem[\protect\citeauthoryear{%
Buccheri%
}{%
Buccheri%
}{%
{\protect\APACyear{2002}}%
}]{%
Buccheri:2002}%
\APACinsertmetastar{%
Buccheri:2002}%
Buccheri, R.%
%
\newblock{}\BBOP{}2002\BBCP{}.
\newblock{}\Bem{\emph{Time and the dichotomy subjective/objective. An
  endo-physical point of view}.}
\newblock{}(http://www.pa.itd.cnr.it/buccheri. Russian Temporology Seminar,
  Moscow, 2002.)

\bibitem[\protect\citeauthoryear{%
Buccheri%
\ \BBA{} Buccheri%
}{%
Buccheri%
\ \BBA{} Buccheri%
}{%
{\protect\APACyear{2005}}%
}]{%
Buccheri/Buccheri:2005}%
\APACinsertmetastar{%
Buccheri/Buccheri:2005}%
Buccheri, R.%
\BCBT{}\ \BBA{} Buccheri, M.%
%
\newblock{}\BBOP{}2005\BBCP{}.
\newblock{}\BBOQ{}Evolution of human knowledge and the endophysical
  perspective.\BBCQ{}
\newblock{}\BIn{} R.~Buccheri, A.~C. Elitzur\BCBL{}\ \BBA{} M.~Saniga\ (\BEDS),
  \Bem{Endophysics, time, quantum and the subjective}\ (\BPGS\ 3--21).
\newblock{}Singapore: World Scientific.

\bibitem[\protect\citeauthoryear{%
Buccheri%
\ \BBA{} Saniga%
}{%
Buccheri%
\ \BBA{} Saniga%
}{%
{\protect\APACyear{2003}}%
}]{%
Buccheri/Saniga:2003}%
\APACinsertmetastar{%
Buccheri/Saniga:2003}%
Buccheri, R.%
\BCBT{}\ \BBA{} Saniga, M.%
%
\newblock{}\BBOP{}2003\BBCP{}.
\newblock{}\BBOQ{}Endo-physical paradigm and mathematics of subjective
  time.\BBCQ{}
\newblock{}\Bem{Frontier Perspectives}, \Bem{12}, 36--40.
\newblock{}(http://www. pa.itd.cnr.it/buccheri)

\bibitem[\protect\citeauthoryear{%
Butterfield%
\ \BBA{} Isham%
}{%
Butterfield%
\ \BBA{} Isham%
}{%
{\protect\APACyear{2000}}%
}]{%
Butterfield/Isham:2000}%
\APACinsertmetastar{%
Butterfield/Isham:2000}%
Butterfield, J.%
\BCBT{}\ \BBA{} Isham, C.%
%
\newblock{}\BBOP{}2000\BBCP{}.
\newblock{}\BBOQ{}Spacetime and the philosophical challenge of quantum
  gravity.\BBCQ{}
\newblock{}\BIn{} C.~Callender\ \BBA{} N.~Huggett\ (\BEDS), \Bem{Physics meets
  philosophy at the {Planck} scale}\ (\BPGS\ 33--89).
\newblock{}Cambridge: Cambridge University Press.
\newblock{}(arXiv:gr-qc/9903072v1)

\bibitem[\protect\citeauthoryear{%
Cahill%
\ \BBA{} Klinger%
}{%
Cahill%
\ \BBA{} Klinger%
}{%
{\protect\APACyear{1996}}%
}]{%
Cahill/Klinger:1996}%
\APACinsertmetastar{%
Cahill/Klinger:1996}%
Cahill, R.~T.%
\BCBT{}\ \BBA{} Klinger, C.~M.%
%
\newblock{}\BBOP{}1996\BBCP{}.
\newblock{}\BBOQ{}Pregeometric modelling of the spacetime phenomenology.\BBCQ{}
\newblock{}\Bem{Physics Letters A}, \Bem{223}, 313--319.
\newblock{}(arXiv:gr-qc/9605018v1)

\bibitem[\protect\citeauthoryear{%
Cahill%
\ \BBA{} Klinger%
}{%
Cahill%
\ \BBA{} Klinger%
}{%
{\protect\APACyear{1997}}%
}]{%
Cahill/Klinger:1997}%
\APACinsertmetastar{%
Cahill/Klinger:1997}%
Cahill, R.~T.%
\BCBT{}\ \BBA{} Klinger, C.~M.%
%
\newblock{}\BBOP{}1997\BBCP{}.
\newblock{}\Bem{\emph{Bootstrap universe from self-referential noise}.}
\newblock{}(arXiv:gr-qc/9708013v1)

\bibitem[\protect\citeauthoryear{%
Chalmers%
}{%
Chalmers%
}{%
{\protect\APACyear{2003}}%
}]{%
Chalmers:2003}%
\APACinsertmetastar{%
Chalmers:2003}%
Chalmers, M.%
%
\newblock{}\BBOP{}2003, November\BBCP{}.
\newblock{}\BBOQ{}Welcome to quantum gravity.\BBCQ{}
\newblock{}\Bem{Physics World}, 27.

\bibitem[\protect\citeauthoryear{%
Clifford%
}{%
Clifford%
}{%
{\protect\APACyear{1886}}%
{\protect\APACexlab{{\protect\BCnt{1}}}}}]{%
Clifford:1886b}%
\APACinsertmetastar{%
Clifford:1886b}%
Clifford, W.~K.%
%
\newblock{}\BBOP{}1886{\protect\BCnt{1}}\BBCP{}.
\newblock{}\BBOQ{}On the nature of things-in-themselves.\BBCQ{}
\newblock{}\BIn{} L.~Stephen\ \BBA{} F.~Pollock\ (\BEDS), \Bem{Lectures and
  essays}\ (\BPGS\ 275--286).
\newblock{}London: Macmillan.
\newblock{}(Original work published 1878.)

\bibitem[\protect\citeauthoryear{%
Clifford%
}{%
Clifford%
}{%
{\protect\APACyear{1886}}%
{\protect\APACexlab{{\protect\BCnt{2}}}}}]{%
Clifford:1886a}%
\APACinsertmetastar{%
Clifford:1886a}%
Clifford, W.~K.%
%
\newblock{}\BBOP{}1886{\protect\BCnt{2}}\BBCP{}.
\newblock{}\BBOQ{}The unseen universe.\BBCQ{}
\newblock{}\BIn{} L.~Stephen\ \BBA{} F.~Pollock\ (\BEDS), \Bem{Lectures and
  essays}\ (\BPGS\ 161--179).
\newblock{}London: Macmillan.
\newblock{}(Original work published 1875.)

\bibitem[\protect\citeauthoryear{%
Cooperstock%
\ \BBA{} Faraoni%
}{%
Cooperstock%
\ \BBA{} Faraoni%
}{%
{\protect\APACyear{2003}}%
}]{%
Cooperstock/Faraoni:2003}%
\APACinsertmetastar{%
Cooperstock/Faraoni:2003}%
Cooperstock, F.%
\BCBT{}\ \BBA{} Faraoni, V.%
%
\newblock{}\BBOP{}2003\BBCP{}.
\newblock{}\BBOQ{}The new {Planck} scale: {Quantized} spin and charge coupled
  to gravity.\BBCQ{}
\newblock{}\Bem{International Journal of Modern Physics D}, \Bem{12},
  1657--1662.

\bibitem[\protect\citeauthoryear{%
{Costa de Beauregard}%
}{%
{Costa de Beauregard}%
}{%
{\protect\APACyear{1977}}%
}]{%
Costa:1977}%
\APACinsertmetastar{%
Costa:1977}%
{Costa de Beauregard}, O.%
%
\newblock{}\BBOP{}1977\BBCP{}.
\newblock{}\BBOQ{}Time symmetry and the {Einstein} paradox.\BBCQ{}
\newblock{}\Bem{Il Nuovo Cimento B}, \Bem{42}, 41--64.

\bibitem[\protect\citeauthoryear{%
Dadi\'{c}%
\ \BBA{} Pisk%
}{%
Dadi\'{c}%
\ \BBA{} Pisk%
}{%
{\protect\APACyear{1979}}%
}]{%
Dadic/Pisk:1979}%
\APACinsertmetastar{%
Dadic/Pisk:1979}%
Dadi\'{c}, I.%
\BCBT{}\ \BBA{} Pisk, K.%
%
\newblock{}\BBOP{}1979\BBCP{}.
\newblock{}\BBOQ{}Dynamics of discrete-space structure.\BBCQ{}
\newblock{}\Bem{International Journal of Theoretical Physics}, \Bem{18},
  345--358.

\bibitem[\protect\citeauthoryear{%
Davidon%
}{%
Davidon%
}{%
{\protect\APACyear{1976}}%
}]{%
Davidon:1976}%
\APACinsertmetastar{%
Davidon:1976}%
Davidon, W.~C.%
%
\newblock{}\BBOP{}1976\BBCP{}.
\newblock{}\BBOQ{}Quantum physics of single systems.\BBCQ{}
\newblock{}\Bem{Il Nuovo Cimento B}, \Bem{36}, 34--40.

\bibitem[\protect\citeauthoryear{%
Davies%
\ \BBA{} Brown%
}{%
Davies%
\ \BBA{} Brown%
}{%
{\protect\APACyear{1988}}%
}]{%
Davies/Brown:1988}%
\APACinsertmetastar{%
Davies/Brown:1988}%
Davies, P. C.~W.%
\BCBT{}\ \BBA{} Brown, J.%
\ (\BEDS).
\newblock{}\BBOP{}1988\BBCP{}.
\newblock{}\Bem{Superstrings: {A} theory of everything?}
\newblock{}Cambridge: Cambridge University Press.
\newblock{}

\bibitem[\protect\citeauthoryear{%
Demaret%
, Heller%
\BCBL{}\ \BBA{} Lambert%
}{%
Demaret%
\ \protect\BOthers{.}}{%
{\protect\APACyear{1997}}%
}]{%
Demaret/Heller/Lambert:1997}%
\APACinsertmetastar{%
Demaret/Heller/Lambert:1997}%
Demaret, J.%
, Heller, M.%
\BCBL{}\ \BBA{} Lambert, D.%
%
\newblock{}\BBOP{}1997\BBCP{}.
\newblock{}\BBOQ{}Local and global properties of the world.\BBCQ{}
\newblock{}\Bem{Foundations of Science}, \Bem{2}, 137--176.
\newblock{}(arXiv:gr-qc/ 9702047v2)

\bibitem[\protect\citeauthoryear{%
Dennett%
}{%
Dennett%
}{%
{\protect\APACyear{1991}}%
}]{%
Dennett:1991}%
\APACinsertmetastar{%
Dennett:1991}%
Dennett, D.~C.%
%
\newblock{}\BBOP{}1991\BBCP{}.
\newblock{}\Bem{Consciousness explained}.
\newblock{}London: Allen Lane.

\bibitem[\protect\citeauthoryear{%
Devitt%
\ \BBA{} Sterelny%
}{%
Devitt%
\ \BBA{} Sterelny%
}{%
{\protect\APACyear{1987}}%
}]{%
Devitt/Sterelny:1987}%
\APACinsertmetastar{%
Devitt/Sterelny:1987}%
Devitt, M.%
\BCBT{}\ \BBA{} Sterelny, K.%
%
\newblock{}\BBOP{}1987\BBCP{}.
\newblock{}\Bem{Language and reality: {An} introduction to the philosophy of
  language}.
\newblock{}Oxford: Basil Blackwell.

\bibitem[\protect\citeauthoryear{%
Dirac%
}{%
Dirac%
}{%
{\protect\APACyear{1958}}%
}]{%
Dirac:1958}%
\APACinsertmetastar{%
Dirac:1958}%
Dirac, P. A.~M.%
%
\newblock{}\BBOP{}1958\BBCP{}.
\newblock{}\Bem{The principles of quantum mechanics}\ (Fourth\ \BEd).
\newblock{}Oxford: Clarendon Press.

\bibitem[\protect\citeauthoryear{%
Drude%
}{%
Drude%
}{%
{\protect\APACyear{1894}}%
}]{%
Drude:1894}%
\APACinsertmetastar{%
Drude:1894}%
Drude, P.%
%
\newblock{}\BBOP{}1894\BBCP{}.
\newblock{}\Bem{Physik des {{\"{A}}thers} auf elektromagnetischer {Grundlage}}.
\newblock{}Stuttgart: Enke.

\bibitem[\protect\citeauthoryear{%
Eakins%
\ \BBA{} Jaroszkiewicz%
}{%
Eakins%
\ \BBA{} Jaroszkiewicz%
}{%
{\protect\APACyear{2002}}%
}]{%
Eakins/Jaroszkiewicz:2002}%
\APACinsertmetastar{%
Eakins/Jaroszkiewicz:2002}%
Eakins, J.%
\BCBT{}\ \BBA{} Jaroszkiewicz, G.%
%
\newblock{}\BBOP{}2002\BBCP{}.
\newblock{}\Bem{\emph{The quantum universe}.}
\newblock{}(arXiv: quant-ph/0203020v1)

\bibitem[\protect\citeauthoryear{%
Eakins%
\ \BBA{} Jaroszkiewicz%
}{%
Eakins%
\ \BBA{} Jaroszkiewicz%
}{%
{\protect\APACyear{2003}}%
}]{%
Eakins/Jaroszkiewicz:2003}%
\APACinsertmetastar{%
Eakins/Jaroszkiewicz:2003}%
Eakins, J.%
\BCBT{}\ \BBA{} Jaroszkiewicz, G.%
%
\newblock{}\BBOP{}2003\BBCP{}.
\newblock{}\Bem{\emph{The origin of causal set structure in the quantum
  universe}.}
\newblock{}(arXiv:gr-qc/0301117v1)

\bibitem[\protect\citeauthoryear{%
Eakins%
\ \BBA{} Jaroszkiewicz%
}{%
Eakins%
\ \BBA{} Jaroszkiewicz%
}{%
{\protect\APACyear{2004}}%
}]{%
Eakins/Jaroszkiewicz:2004}%
\APACinsertmetastar{%
Eakins/Jaroszkiewicz:2004}%
Eakins, J.%
\BCBT{}\ \BBA{} Jaroszkiewicz, G.%
%
\newblock{}\BBOP{}2004\BBCP{}.
\newblock{}\Bem{\emph{Endophysical information transfer in quantum processes}.}
\newblock{}(arXiv:quant-ph/0401006v1)

\bibitem[\protect\citeauthoryear{%
Earman%
}{%
Earman%
}{%
{\protect\APACyear{1989}}%
}]{%
Earman:1989}%
\APACinsertmetastar{%
Earman:1989}%
Earman, J.%
%
\newblock{}\BBOP{}1989\BBCP{}.
\newblock{}\Bem{World enough and space-time: {Absolute} versus relational
  theories of space and time}.
\newblock{}Cambridge, MA: M.I.T. Press.

\bibitem[\protect\citeauthoryear{%
Earman%
, Glymour%
\BCBL{}\ \BBA{} Stachel%
}{%
Earman%
\ \protect\BOthers{.}}{%
{\protect\APACyear{1977}}%
}]{%
Earman/Glymour/Stachel:1977}%
\APACinsertmetastar{%
Earman/Glymour/Stachel:1977}%
Earman, J.%
, Glymour, C.~N.%
\BCBL{}\ \BBA{} Stachel, J.%
\ (\BEDS).
\newblock{}\BBOP{}1977\BBCP{}.
\newblock{}\Bem{Foundations of space-time theories.}
\newblock{}Minneapolis: University of Minnesota Press.
\newblock{}

\bibitem[\protect\citeauthoryear{%
Eddington%
}{%
Eddington%
}{%
{\protect\APACyear{1920}}%
}]{%
Eddington:1920}%
\APACinsertmetastar{%
Eddington:1920}%
Eddington, A.%
%
\newblock{}\BBOP{}1920\BBCP{}.
\newblock{}\Bem{Space, time and gravitation: {An} outline of the general
  relativity theory}.
\newblock{}Cambridge: Cambridge University Press.

\bibitem[\protect\citeauthoryear{%
Einstein%
}{%
Einstein%
}{%
{\protect\APACyear{1935}}%
}]{%
Einstein:1935}%
\APACinsertmetastar{%
Einstein:1935}%
Einstein, A.%
%
\newblock{}\BBOP{}1935\BBCP{}.
\newblock{}\Bem{The world as {I} see it, \emph{A. Harris (Trans.)}}.
\newblock{}London: John Lane.

\bibitem[\protect\citeauthoryear{%
Einstein%
}{%
Einstein%
}{%
{\protect\APACyear{1952}}%
{\protect\APACexlab{{\protect\BCnt{1}}}}}]{%
Einstein:1952b}%
\APACinsertmetastar{%
Einstein:1952b}%
Einstein, A.%
%
\newblock{}\BBOP{}1952{\protect\BCnt{1}}\BBCP{}.
\newblock{}\BBOQ{}The foundation of the general theory of relativity.\BBCQ{}
\newblock{}\BIn{} \Bem{The principle of relativity}\ (\BPGS\ 109--164).
\newblock{}New York: Dover.
\newblock{}(Original work published 1916.)

\bibitem[\protect\citeauthoryear{%
Einstein%
}{%
Einstein%
}{%
{\protect\APACyear{1952}}%
{\protect\APACexlab{{\protect\BCnt{2}}}}}]{%
Einstein:1952a}%
\APACinsertmetastar{%
Einstein:1952a}%
Einstein, A.%
%
\newblock{}\BBOP{}1952{\protect\BCnt{2}}\BBCP{}.
\newblock{}\BBOQ{}On the electrodynamics of moving bodies.\BBCQ{}
\newblock{}\BIn{} \Bem{The principle of relativity}\ (\BPGS\ 35--65).
\newblock{}New York: Dover.
\newblock{}(Original work published 1905.)

\bibitem[\protect\citeauthoryear{%
Einstein%
}{%
Einstein%
}{%
{\protect\APACyear{1961}}%
}]{%
Einstein:1961}%
\APACinsertmetastar{%
Einstein:1961}%
Einstein, A.%
%
\newblock{}\BBOP{}1961\BBCP{}.
\newblock{}\Bem{Relativity: {The} special and the general theory}\ (Fifteenth\
  \BEd).
\newblock{}New York: Three Rivers Press.

\bibitem[\protect\citeauthoryear{%
Einstein%
}{%
Einstein%
}{%
{\protect\APACyear{1982}}%
}]{%
Einstein:1982}%
\APACinsertmetastar{%
Einstein:1982}%
Einstein, A.%
%
\newblock{}\BBOP{}1982\BBCP{}.
\newblock{}\BBOQ{}How {I} created the theory of relativity.\BBCQ{}
\newblock{}\Bem{Physics Today}, \Bem{35}, 45--47.
\newblock{}(Address delivered 14th December, 1922, Kyoto University.)

\bibitem[\protect\citeauthoryear{%
Einstein%
}{%
Einstein%
}{%
{\protect\APACyear{1983}}%
{\protect\APACexlab{{\protect\BCnt{1}}}}}]{%
Einstein:1983a}%
\APACinsertmetastar{%
Einstein:1983a}%
Einstein, A.%
%
\newblock{}\BBOP{}1983{\protect\BCnt{1}}\BBCP{}.
\newblock{}\BBOQ{}Ether and the theory of relativity.\BBCQ{}
\newblock{}\BIn{} \Bem{Sidelights on relativity}\ (\BPGS\ 1--24).
\newblock{}New York: Dover.
\newblock{}(Address delivered 5th May, 1920, University of Leyden.)

\bibitem[\protect\citeauthoryear{%
Einstein%
}{%
Einstein%
}{%
{\protect\APACyear{1983}}%
{\protect\APACexlab{{\protect\BCnt{2}}}}}]{%
Einstein:1983b}%
\APACinsertmetastar{%
Einstein:1983b}%
Einstein, A.%
%
\newblock{}\BBOP{}1983{\protect\BCnt{2}}\BBCP{}.
\newblock{}\BBOQ{}Geometry and experience.\BBCQ{}
\newblock{}\BIn{} \Bem{Sidelights on relativity}\ (\BPGS\ 25--56).
\newblock{}New York: Dover.
\newblock{}(Address delivered 27th January, 1921, Prussian Academy of Sciences,
  Berlin.)

\bibitem[\protect\citeauthoryear{%
Einstein%
}{%
Einstein%
}{%
{\protect\APACyear{1996}}%
{\protect\APACexlab{{\protect\BCnt{1}}}}}]{%
Einstein:1996b}%
\APACinsertmetastar{%
Einstein:1996b}%
Einstein, A.%
%
\newblock{}\BBOP{}1996{\protect\BCnt{1}}\BBCP{}.
\newblock{}\BBOQ{}``{Comments}'' on ``{Outline} of a generalized theory of
  relativity and of a theory of gravitation''.\BBCQ{}
\newblock{}\BIn{} \Bem{The collected papers of {Albert} {Einstein}: {The}
  {Swiss} years}\ (\BVOL~4, \BPGS\ 289--290).
\newblock{}Princeton: Princeton University Press.
\newblock{}(Original work published 1914.)

\bibitem[\protect\citeauthoryear{%
Einstein%
}{%
Einstein%
}{%
{\protect\APACyear{1996}}%
{\protect\APACexlab{{\protect\BCnt{2}}}}}]{%
Einstein:1996a}%
\APACinsertmetastar{%
Einstein:1996a}%
Einstein, A.%
%
\newblock{}\BBOP{}1996{\protect\BCnt{2}}\BBCP{}.
\newblock{}\BBOQ{}On the relativity problem.\BBCQ{}
\newblock{}\BIn{} \Bem{The collected papers of {Albert} {Einstein}: {The}
  {Swiss} years}\ (\BVOL~4, \BPGS\ 306--314).
\newblock{}Princeton: Princeton University Press.
\newblock{}(Original work published 1914.)

\bibitem[\protect\citeauthoryear{%
Ellis%
}{%
Ellis%
}{%
{\protect\APACyear{2005}}%
}]{%
Ellis:2005}%
\APACinsertmetastar{%
Ellis:2005}%
Ellis, J.%
%
\newblock{}\BBOP{}2005, January\BBCP{}.
\newblock{}\BBOQ{}Einstein's quest for unification.\BBCQ{}
\newblock{}\Bem{Physics World}, 56--57.

\bibitem[\protect\citeauthoryear{%
Euclid%
}{%
Euclid%
}{%
{\protect\APACyear{1703}}%
}]{%
Euclid:1703}%
\APACinsertmetastar{%
Euclid:1703}%
Euclid.%
%
\newblock{}\BBOP{}1703\BBCP{}.
\newblock{}\Bem{Opera omnia, \emph{D. Gregory (Ed.)}}.
\newblock{}Oxford: Oxford University Press.

\bibitem[\protect\citeauthoryear{%
Flavin%
}{%
Flavin%
}{%
{\protect\APACyear{2004}}%
}]{%
Flavin:2004}%
\APACinsertmetastar{%
Flavin:2004}%
Flavin, R.%
%
\newblock{}\BBOP{}2004\BBCP{}.
\newblock{}\Bem{\emph{Straight lines: {Selected} reviews}.}
\newblock{}(http://www.flavins corner.com/reviews.htm)

\bibitem[\protect\citeauthoryear{%
Foss%
}{%
Foss%
}{%
{\protect\APACyear{2000}}%
}]{%
Foss:2000}%
\APACinsertmetastar{%
Foss:2000}%
Foss, J.%
%
\newblock{}\BBOP{}2000\BBCP{}.
\newblock{}\Bem{Science and the riddle of consciousness: A solution}.
\newblock{}Boston: Kluwer.

\bibitem[\protect\citeauthoryear{%
Friedman%
}{%
Friedman%
}{%
{\protect\APACyear{1974}}%
}]{%
Friedman:1974}%
\APACinsertmetastar{%
Friedman:1974}%
Friedman, M.%
%
\newblock{}\BBOP{}1974\BBCP{}.
\newblock{}\BBOQ{}Explanation and scientific understanding.\BBCQ{}
\newblock{}\Bem{The Journal of Philosophy}, \Bem{71}, 5--19.

\bibitem[\protect\citeauthoryear{%
Friedman%
}{%
Friedman%
}{%
{\protect\APACyear{1983}}%
}]{%
Friedman:1983}%
\APACinsertmetastar{%
Friedman:1983}%
Friedman, M.%
%
\newblock{}\BBOP{}1983\BBCP{}.
\newblock{}\Bem{Foundations of space-time theories}.
\newblock{}Princeton: Princeton University Press.

\bibitem[\protect\citeauthoryear{%
Galilei%
}{%
Galilei%
}{%
{\protect\APACyear{1957}}%
}]{%
Galilei:1957}%
\APACinsertmetastar{%
Galilei:1957}%
Galilei, G.%
%
\newblock{}\BBOP{}1957\BBCP{}.
\newblock{}\BBOQ{}The assayer.\BBCQ{}
\newblock{}\BIn{} S.~Drake\ (\BED), \Bem{Discoveries and opinions of
  {Galileo}}\ (\BPGS\ 229--280).
\newblock{}New York: Anchor.
\newblock{}(Original work published 1623.)

\bibitem[\protect\citeauthoryear{%
Geroch%
}{%
Geroch%
}{%
{\protect\APACyear{1985}}%
}]{%
Geroch:1985}%
\APACinsertmetastar{%
Geroch:1985}%
Geroch, R.%
%
\newblock{}\BBOP{}1985\BBCP{}.
\newblock{}\Bem{Mathematical physics}.
\newblock{}Chicago: University of Chicago Press.

\bibitem[\protect\citeauthoryear{%
Gibbs%
}{%
Gibbs%
}{%
{\protect\APACyear{1996}}%
}]{%
Gibbs:1996}%
\APACinsertmetastar{%
Gibbs:1996}%
Gibbs, P.%
%
\newblock{}\BBOP{}1996\BBCP{}.
\newblock{}\Bem{\emph{The small scale structure of space-time: {A}
  bibliographical review}.}
\newblock{}(arXiv:hep-th/9506171v2)

\bibitem[\protect\citeauthoryear{%
Goenner%
}{%
Goenner%
}{%
{\protect\APACyear{2004}}%
}]{%
Goenner:2004}%
\APACinsertmetastar{%
Goenner:2004}%
Goenner, H. F.~M.%
%
\newblock{}\BBOP{}2004\BBCP{}.
\newblock{}\BBOQ{}On the history of unified field theories.\BBCQ{}
\newblock{}\Bem{Living Reviews in Relativity}, \Bem{7}, 2.
\newblock{}(http://relativity.livingreviews.org)

\bibitem[\protect\citeauthoryear{%
Gopakumar%
}{%
Gopakumar%
}{%
{\protect\APACyear{2001}}%
}]{%
Gopakumar:2001}%
\APACinsertmetastar{%
Gopakumar:2001}%
Gopakumar, R.%
%
\newblock{}\BBOP{}2001\BBCP{}.
\newblock{}\BBOQ{}Geometry and string theory.\BBCQ{}
\newblock{}\Bem{Current Science}, \Bem{81}, 1568--1575.
\newblock{}(http://www.ias.ac.in/currsci/dec252001/1568.pdf)

\bibitem[\protect\citeauthoryear{%
Gorelik%
}{%
Gorelik%
}{%
{\protect\APACyear{1992}}%
}]{%
Gorelik:1992}%
\APACinsertmetastar{%
Gorelik:1992}%
Gorelik, G.%
%
\newblock{}\BBOP{}1992\BBCP{}.
\newblock{}\BBOQ{}First steps of quantum gravity and the {Planck}
  values.\BBCQ{}
\newblock{}\BIn{} J.~Eisenstaedt\ \BBA{} A.~J. Kox.\ (\BEDS), \Bem{Studies in
  the history of general relativity \emph{({Einstein} {Studies})}}\ (\BVOL~3,
  \BPGS\ 364--379).
\newblock{}Boston: Birkh\"{a}user.

\bibitem[\protect\citeauthoryear{%
Gottstein%
}{%
Gottstein%
}{%
{\protect\APACyear{2003}}%
}]{%
Gottstein:2003}%
\APACinsertmetastar{%
Gottstein:2003}%
Gottstein, K.%
%
\newblock{}\BBOP{}2003\BBCP{}.
\newblock{}\BBOQ{}On the unfathomableness of consciousness by
  consciousness.\BBCQ{}
\newblock{}\BIn{} L.~Castell\ \BBA{} O.~Ischebeck\ (\BEDS), \Bem{Time, quantum
  and information}\ (\BPGS\ 59--72).
\newblock{}Berlin: Springer.
\newblock{}(arXiv:physics/0610011v1)

\bibitem[\protect\citeauthoryear{%
Greene%
}{%
Greene%
}{%
{\protect\APACyear{1999}}%
}]{%
Greene:1999}%
\APACinsertmetastar{%
Greene:1999}%
Greene, B.%
%
\newblock{}\BBOP{}1999\BBCP{}.
\newblock{}\Bem{The elegant universe: {Superstrings}, hidden dimensions, and
  the quest for the ultimate theory}.
\newblock{}London: Jonathan Cape.

\bibitem[\protect\citeauthoryear{%
Hardin%
}{%
Hardin%
}{%
{\protect\APACyear{1960}}%
}]{%
Hardin:1960}%
\APACinsertmetastar{%
Hardin:1960}%
Hardin, G.%
%
\newblock{}\BBOP{}1960\BBCP{}.
\newblock{}\Bem{Nature and {Man's} fate}.
\newblock{}London: Jonathan Cape.

\bibitem[\protect\citeauthoryear{%
Hardin%
}{%
Hardin%
}{%
{\protect\APACyear{1968}}%
}]{%
Hardin:1968}%
\APACinsertmetastar{%
Hardin:1968}%
Hardin, G.%
%
\newblock{}\BBOP{}1968\BBCP{}.
\newblock{}\BBOQ{}The tragedy of the commons.\BBCQ{}
\newblock{}\Bem{Science}, \Bem{162}, 1243--1248.
\newblock{}(http://www.garretthardinsociety.org)

\bibitem[\protect\citeauthoryear{%
Hardin%
}{%
Hardin%
}{%
{\protect\APACyear{1969}}%
}]{%
Hardin:1969}%
\APACinsertmetastar{%
Hardin:1969}%
Hardin, G.%
%
\newblock{}\BBOP{}1969\BBCP{}.
\newblock{}\Bem{Population, evolution, and birth control: {A} collage of
  controversial ideas}.
\newblock{}San Francisco: Freeman.

\bibitem[\protect\citeauthoryear{%
Hardin%
}{%
Hardin%
}{%
{\protect\APACyear{1972}}%
}]{%
Hardin:1972}%
\APACinsertmetastar{%
Hardin:1972}%
Hardin, G.%
%
\newblock{}\BBOP{}1972\BBCP{}.
\newblock{}\Bem{Exploring new ethics for survival: {The} voyage of the
  spaceship {Beagle}}.
\newblock{}New York: Viking.

\bibitem[\protect\citeauthoryear{%
Hardin%
}{%
Hardin%
}{%
{\protect\APACyear{1973}}%
}]{%
Hardin:1973}%
\APACinsertmetastar{%
Hardin:1973}%
Hardin, G.%
%
\newblock{}\BBOP{}1973\BBCP{}.
\newblock{}\Bem{Stalking the wild taboo}.
\newblock{}Los Altos: William Kaufmann.

\bibitem[\protect\citeauthoryear{%
Hardin%
}{%
Hardin%
}{%
{\protect\APACyear{1985}}%
}]{%
Hardin:1985}%
\APACinsertmetastar{%
Hardin:1985}%
Hardin, G.%
%
\newblock{}\BBOP{}1985\BBCP{}.
\newblock{}\Bem{Filters against folly}.
\newblock{}New York: Viking.

\bibitem[\protect\citeauthoryear{%
Hardin%
}{%
Hardin%
}{%
{\protect\APACyear{1993}}%
}]{%
Hardin:1993}%
\APACinsertmetastar{%
Hardin:1993}%
Hardin, G.%
%
\newblock{}\BBOP{}1993\BBCP{}.
\newblock{}\Bem{Living within limits: Ecology, economics, and population
  taboos}.
\newblock{}Oxford: Oxford University Press.

\bibitem[\protect\citeauthoryear{%
Hartle%
}{%
Hartle%
}{%
{\protect\APACyear{2003}}%
}]{%
Hartle:2003}%
\APACinsertmetastar{%
Hartle:2003}%
Hartle, J.~B.%
%
\newblock{}\BBOP{}2003\BBCP{}.
\newblock{}\BBOQ{}Theories of everything and {Hawking}'s wave function of the
  universe.\BBCQ{}
\newblock{}\BIn{} G.~W. Gibons, E.~P.~S. Shellard\BCBL{}\ \BBA{} S.~J. Rankin\
  (\BEDS), \Bem{The future of theoretical physics and cosmology: {Celebrating}
  {Stephen} {Hawking's} 60th birthday}\ (\BPGS\ 38--50).
\newblock{}Cambridge: Cambridge University Press.
\newblock{}(arXiv:gr-qc/0209047v1)

\bibitem[\protect\citeauthoryear{%
Hauser%
}{%
Hauser%
}{%
{\protect\APACyear{2000}}%
}]{%
Hauser:2000}%
\APACinsertmetastar{%
Hauser:2000}%
Hauser, M.~D.%
%
\newblock{}\BBOP{}2000\BBCP{}.
\newblock{}\BBOQ{}What do animals think about numbers?\BBCQ{}
\newblock{}\Bem{American Scientist}, \Bem{88}, 144--151.

\bibitem[\protect\citeauthoryear{%
Heilbron%
}{%
Heilbron%
}{%
{\protect\APACyear{2000}}%
}]{%
Heilbron:2000}%
\APACinsertmetastar{%
Heilbron:2000}%
Heilbron, J.~L.%
%
\newblock{}\BBOP{}2000\BBCP{}.
\newblock{}\Bem{Geometry civilized: {History}, culture, and technique}.
\newblock{}Oxford: Clarendon Press.

\bibitem[\protect\citeauthoryear{%
Hilgevoord%
}{%
Hilgevoord%
}{%
{\protect\APACyear{2002}}%
}]{%
Hilgevoord:2002}%
\APACinsertmetastar{%
Hilgevoord:2002}%
Hilgevoord, J.%
%
\newblock{}\BBOP{}2002\BBCP{}.
\newblock{}\BBOQ{}Time in quantum mechanics.\BBCQ{}
\newblock{}\Bem{American Journal of Physics}, \Bem{70}, 301--306.

\bibitem[\protect\citeauthoryear{%
Hill%
}{%
Hill%
}{%
{\protect\APACyear{1955}}%
}]{%
Hill:1955}%
\APACinsertmetastar{%
Hill:1955}%
Hill, E.~L.%
%
\newblock{}\BBOP{}1955\BBCP{}.
\newblock{}\BBOQ{}Relativistic theory of discrete momentum space and discrete
  space-time.\BBCQ{}
\newblock{}\Bem{Physical Review}, \Bem{100}, 1780--1783.

\bibitem[\protect\citeauthoryear{%
Hilton%
\ \BBA{} Pedersen%
}{%
Hilton%
\ \BBA{} Pedersen%
}{%
{\protect\APACyear{2004}}%
}]{%
Hilton/Pedersen:2004}%
\APACinsertmetastar{%
Hilton/Pedersen:2004}%
Hilton, P.%
\BCBT{}\ \BBA{} Pedersen, J.%
%
\newblock{}\BBOP{}2004\BBCP{}.
\newblock{}\BBOQ{}Proof reading [{Review} of the book \emph{The changing shape
  of geometry: Celebrating a century of geometry and geometry
  teaching}].\BBCQ{}
\newblock{}\Bem{American Scientist}, \Bem{92}, 91--92.

\bibitem[\protect\citeauthoryear{%
Horgan%
}{%
Horgan%
}{%
{\protect\APACyear{1996}}%
}]{%
Horgan:1996}%
\APACinsertmetastar{%
Horgan:1996}%
Horgan, J.%
%
\newblock{}\BBOP{}1996\BBCP{}.
\newblock{}\Bem{The end of science: {Facing} the limits of knowledge in the
  twilight of the scientific age}.
\newblock{}London: Little, Brown and Co.

\bibitem[\protect\citeauthoryear{%
Horwich%
}{%
Horwich%
}{%
{\protect\APACyear{1978}}%
}]{%
Horwich:1978}%
\APACinsertmetastar{%
Horwich:1978}%
Horwich, P.%
%
\newblock{}\BBOP{}1978\BBCP{}.
\newblock{}\BBOQ{}On the existence of time, space and space-time.\BBCQ{}
\newblock{}\Bem{No\^{u}s}, \Bem{12}, 397--419.

\bibitem[\protect\citeauthoryear{%
Horzela%
, Kapu\'{s}cik%
, Kempczy\'{n}ski%
\BCBL{}\ \BBA{} Uzes%
}{%
Horzela%
\ \protect\BOthers{.}}{%
{\protect\APACyear{1992}}%
}]{%
Horzela/Kapuscik/Kempczynski/Uzes:1992}%
\APACinsertmetastar{%
Horzela/Kapuscik/Kempczynski/Uzes:1992}%
Horzela, A.%
, Kapu\'{s}cik, E.%
, Kempczy\'{n}ski, J.%
\BCBL{}\ \BBA{} Uzes, C.%
%
\newblock{}\BBOP{}1992\BBCP{}.
\newblock{}\BBOQ{}On discrete models of space-time.\BBCQ{}
\newblock{}\Bem{Progress in Theoretical Physics}, \Bem{88}, 1065--1071.

\bibitem[\protect\citeauthoryear{%
Isaacson%
\ \BBA{} Isaacson%
}{%
Isaacson%
\ \BBA{} Isaacson%
}{%
{\protect\APACyear{1975}}%
}]{%
Isaacson/Isaacson:1975}%
\APACinsertmetastar{%
Isaacson/Isaacson:1975}%
Isaacson, E.%
\BCBT{}\ \BBA{} Isaacson, M.%
%
\newblock{}\BBOP{}1975\BBCP{}.
\newblock{}\Bem{Dimensional methods in engineering and physics}.
\newblock{}London: Arnold.

\bibitem[\protect\citeauthoryear{%
Isham%
}{%
Isham%
}{%
{\protect\APACyear{1995}}%
}]{%
Isham:1995}%
\APACinsertmetastar{%
Isham:1995}%
Isham, C.%
%
\newblock{}\BBOP{}1995\BBCP{}.
\newblock{}\Bem{Quantum theory: {Mathematical} and structural foundations}.
\newblock{}London: Imperial College Press.

\bibitem[\protect\citeauthoryear{%
Isham%
\ \BBA{} Butterfield%
}{%
Isham%
\ \BBA{} Butterfield%
}{%
{\protect\APACyear{2000}}%
}]{%
Isham/Butterfield:2000}%
\APACinsertmetastar{%
Isham/Butterfield:2000}%
Isham, C.%
\BCBT{}\ \BBA{} Butterfield, J.%
%
\newblock{}\BBOP{}2000\BBCP{}.
\newblock{}\BBOQ{}Some possible roles for topos theory in quantum theory and
  quantum gravity.\BBCQ{}
\newblock{}\Bem{Foundations of Physics}, \Bem{30}, 1707--1735.
\newblock{}(arXiv:gr-qc/9910005v1)

\bibitem[\protect\citeauthoryear{%
Isham%
, Kubyshin%
\BCBL{}\ \BBA{} Renteln%
}{%
Isham%
\ \protect\BOthers{.}}{%
{\protect\APACyear{1991}}%
}]{%
Isham/Kubyshin/Renteln:1991}%
\APACinsertmetastar{%
Isham/Kubyshin/Renteln:1991}%
Isham, C.%
, Kubyshin, Y.~A.%
\BCBL{}\ \BBA{} Renteln, P.%
%
\newblock{}\BBOP{}1991\BBCP{}.
\newblock{}\BBOQ{}Metric space as a model of spacetime: {Classical} theory and
  quantization.\BBCQ{}
\newblock{}\BIn{} J.~D. Barrow, A.~B. Henriques, M.~T. Lago\BCBL{}\ \BBA{}
  M.~S. Longair\ (\BEDS), \Bem{The physical universe: {The} interface between
  cosmology, astrophysics, and particle physics}\ (\BPGS\ 159--173).
\newblock{}Berlin: Springer.

\bibitem[\protect\citeauthoryear{%
Jacobson%
\ \BBA{} Parentani%
}{%
Jacobson%
\ \BBA{} Parentani%
}{%
{\protect\APACyear{2005}}%
}]{%
Jacobson/Parentani:2005}%
\APACinsertmetastar{%
Jacobson/Parentani:2005}%
Jacobson, T.~A.%
\BCBT{}\ \BBA{} Parentani, R.%
%
\newblock{}\BBOP{}2005, December\BBCP{}.
\newblock{}\BBOQ{}An echo of black holes.\BBCQ{}
\newblock{}\Bem{Scientific American}, 69--75.

\bibitem[\protect\citeauthoryear{%
Jammer%
}{%
Jammer%
}{%
{\protect\APACyear{1969}}%
}]{%
Jammer:1969}%
\APACinsertmetastar{%
Jammer:1969}%
Jammer, M.%
%
\newblock{}\BBOP{}1969\BBCP{}.
\newblock{}\Bem{Concepts of space}\ (Second\ \BEd).
\newblock{}Cambridge, MA: Harvard University Press.

\bibitem[\protect\citeauthoryear{%
Jaroszkiewicz%
}{%
Jaroszkiewicz%
}{%
{\protect\APACyear{2001}}%
}]{%
Jaroszkiewicz:2001}%
\APACinsertmetastar{%
Jaroszkiewicz:2001}%
Jaroszkiewicz, G.%
%
\newblock{}\BBOP{}2001\BBCP{}.
\newblock{}\Bem{\emph{The running of the universe and the quantum structure of
  time}.}
\newblock{}(arXiv:quant-ph/0105013v2)

\bibitem[\protect\citeauthoryear{%
Jaroszkiewicz%
}{%
Jaroszkiewicz%
}{%
{\protect\APACyear{2007}}%
}]{%
Jaroszkiewicz:2007}%
\APACinsertmetastar{%
Jaroszkiewicz:2007}%
Jaroszkiewicz, G.%
%
\newblock{}\BBOP{}2007\BBCP{}.
\newblock{}\Bem{\emph{Quantum detector networks: A review of recent
  developments}.}
\newblock{}(arXiv:quant-ph/0709.4198v1)

\bibitem[\protect\citeauthoryear{%
Keller%
}{%
Keller%
}{%
{\protect\APACyear{1933}}%
}]{%
Keller:1933}%
\APACinsertmetastar{%
Keller:1933}%
Keller, H.%
%
\newblock{}\BBOP{}1933\BBCP{}.
\newblock{}\Bem{The world {I} live in}.
\newblock{}London: Methuen.

\bibitem[\protect\citeauthoryear{%
Kolers%
\ \BBA{} {von Gr\"{u}nau}%
}{%
Kolers%
\ \BBA{} {von Gr\"{u}nau}%
}{%
{\protect\APACyear{1976}}%
}]{%
Kolers/vonGrunau:1976}%
\APACinsertmetastar{%
Kolers/vonGrunau:1976}%
Kolers, P.~A.%
\BCBT{}\ \BBA{} {von Gr\"{u}nau}, M.%
%
\newblock{}\BBOP{}1976\BBCP{}.
\newblock{}\BBOQ{}Shape and color in apparent motion.\BBCQ{}
\newblock{}\Bem{Vision Research}, \Bem{16}, 329--335.

\bibitem[\protect\citeauthoryear{%
Kostro%
}{%
Kostro%
}{%
{\protect\APACyear{2000}}%
}]{%
Kostro:2000}%
\APACinsertmetastar{%
Kostro:2000}%
Kostro, L.%
%
\newblock{}\BBOP{}2000\BBCP{}.
\newblock{}\Bem{{Einstein} and the ether}.
\newblock{}Montreal: Apeiron.

\bibitem[\protect\citeauthoryear{%
Kox%
}{%
Kox%
}{%
{\protect\APACyear{1989}}%
}]{%
Kox:1989}%
\APACinsertmetastar{%
Kox:1989}%
Kox, A.~J.%
%
\newblock{}\BBOP{}1989\BBCP{}.
\newblock{}\BBOQ{}{Hendrik} {Antoon} {Lorentz}, the ether, and the general
  theory of relativity.\BBCQ{}
\newblock{}\BIn{} D.~Howard\ \BBA{} J.~Stachel\ (\BEDS), \Bem{{Einstein} and
  the history of general relativity}\ (\BPGS\ 201--212).
\newblock{}Boston: Birkh\"{a}user.
\newblock{}(Original work published 1988.)

\bibitem[\protect\citeauthoryear{%
Lagendijk%
}{%
Lagendijk%
}{%
{\protect\APACyear{2005}}%
}]{%
Lagendijk:2005}%
\APACinsertmetastar{%
Lagendijk:2005}%
Lagendijk, A.%
%
\newblock{}\BBOP{}2005\BBCP{}.
\newblock{}\BBOQ{}Pushing for power.\BBCQ{}
\newblock{}\Bem{Nature}, \Bem{438}, 429.

\bibitem[\protect\citeauthoryear{%
Laughlin%
\ \BBA{} Pines%
}{%
Laughlin%
\ \BBA{} Pines%
}{%
{\protect\APACyear{2000}}%
}]{%
Laughlin/Pines:2000}%
\APACinsertmetastar{%
Laughlin/Pines:2000}%
Laughlin, R.~B.%
\BCBT{}\ \BBA{} Pines, D.%
%
\newblock{}\BBOP{}2000\BBCP{}.
\newblock{}\BBOQ{}The theory of everything.\BBCQ{}
\newblock{}\Bem{Proceedings of the National Academy of Sciences}, \Bem{97},
  28--31.
\newblock{}(http://www.pnas. org/cgi/reprint/97/1/28)

\bibitem[\protect\citeauthoryear{%
Lawrence%
}{%
Lawrence%
}{%
{\protect\APACyear{2003}}%
}]{%
Lawrence:2003}%
\APACinsertmetastar{%
Lawrence:2003}%
Lawrence, P.~A.%
%
\newblock{}\BBOP{}2003\BBCP{}.
\newblock{}\BBOQ{}The politics of publication.\BBCQ{}
\newblock{}\Bem{Nature}, \Bem{422}, 259--261.

\bibitem[\protect\citeauthoryear{%
Lehto%
}{%
Lehto%
}{%
{\protect\APACyear{1988}}%
}]{%
Lehto:1988}%
\APACinsertmetastar{%
Lehto:1988}%
Lehto, M.%
%
\newblock{}\BBOP{}1988\BBCP{}.
\newblock{}\Bem{Simplicial quantum gravity}.
\newblock{}{Ph.D.} thesis, {Research} report 1/1988, Department of Physics,
  University of {Jyv\"{a}skyl\"{a}}.

\bibitem[\protect\citeauthoryear{%
Lehto%
, Nielsen%
\BCBL{}\ \BBA{} Ninomiya%
}{%
Lehto%
\ \protect\BOthers{.}}{%
{\protect\APACyear{1986}}%
{\protect\APACexlab{{\protect\BCnt{1}}}}}]{%
Lehto/Nielsen/Ninomiya:1986b}%
\APACinsertmetastar{%
Lehto/Nielsen/Ninomiya:1986b}%
Lehto, M.%
, Nielsen, H.~B.%
\BCBL{}\ \BBA{} Ninomiya, M.%
%
\newblock{}\BBOP{}1986{\protect\BCnt{1}}\BBCP{}.
\newblock{}\BBOQ{}Diffeomorphism symmetry in simplicial quantum gravity.\BBCQ{}
\newblock{}\Bem{Nuclear Physics B}, \Bem{272}, 228--252.

\bibitem[\protect\citeauthoryear{%
Lehto%
, Nielsen%
\BCBL{}\ \BBA{} Ninomiya%
}{%
Lehto%
\ \protect\BOthers{.}}{%
{\protect\APACyear{1986}}%
{\protect\APACexlab{{\protect\BCnt{2}}}}}]{%
Lehto/Nielsen/Ninomiya:1986a}%
\APACinsertmetastar{%
Lehto/Nielsen/Ninomiya:1986a}%
Lehto, M.%
, Nielsen, H.~B.%
\BCBL{}\ \BBA{} Ninomiya, M.%
%
\newblock{}\BBOP{}1986{\protect\BCnt{2}}\BBCP{}.
\newblock{}\BBOQ{}Pregeometric quantum lattice: {A} general discussion.\BBCQ{}
\newblock{}\Bem{Nuclear Physics B}, \Bem{272}, 213--227.

\bibitem[\protect\citeauthoryear{%
Lem%
}{%
Lem%
}{%
{\protect\APACyear{1984}}%
{\protect\APACexlab{{\protect\BCnt{1}}}}}]{%
Lem:1984}%
\APACinsertmetastar{%
Lem:1984}%
Lem, S.%
%
\newblock{}\BBOP{}1984{\protect\BCnt{1}}\BBCP{}.
\newblock{}\Bem{His master's voice, \emph{M. Kandel (Trans.)}}.
\newblock{}San Diego: Harvest/HBJ.
\newblock{}(Original work published 1968.)

\bibitem[\protect\citeauthoryear{%
Lem%
}{%
Lem%
}{%
{\protect\APACyear{1984}}%
{\protect\APACexlab{{\protect\BCnt{2}}}}}]{%
Lem:1984b}%
\APACinsertmetastar{%
Lem:1984b}%
Lem, S.%
%
\newblock{}\BBOP{}1984{\protect\BCnt{2}}\BBCP{}.
\newblock{}\Bem{Microworlds: {Writings} on science fiction and fantasy}.
\newblock{}San Diego: Harvest/HBJ.
\newblock{}(Original work published 1981.)

\bibitem[\protect\citeauthoryear{%
Lem%
}{%
Lem%
}{%
{\protect\APACyear{1985}}%
}]{%
Lem:1985}%
\APACinsertmetastar{%
Lem:1985}%
Lem, S.%
%
\newblock{}\BBOP{}1985\BBCP{}.
\newblock{}\Bem{The star diaries, \emph{M. Kandel (Trans.)}}.
\newblock{}San Diego: Harvest/HBJ.
\newblock{}(Original work published 1971.)

\bibitem[\protect\citeauthoryear{%
Major%
\ \BBA{} Setter%
}{%
Major%
\ \BBA{} Setter%
}{%
{\protect\APACyear{2001}}%
}]{%
Major/Setter:2001}%
\APACinsertmetastar{%
Major/Setter:2001}%
Major, S.~A.%
\BCBT{}\ \BBA{} Setter, K.~L.%
%
\newblock{}\BBOP{}2001\BBCP{}.
\newblock{}\BBOQ{}On the universality of the entropy-area relation.\BBCQ{}
\newblock{}\Bem{Classical and Quantum Gravity}, \Bem{18}, 5293--5298.
\newblock{}(arXiv:gr-qc/ 0108034v1)

\bibitem[\protect\citeauthoryear{%
Malcolm%
}{%
Malcolm%
}{%
{\protect\APACyear{1972}}%
}]{%
Malcolm:1972}%
\APACinsertmetastar{%
Malcolm:1972}%
Malcolm, N.%
%
\newblock{}\BBOP{}1972\BBCP{}.
\newblock{}\Bem{Ludwig {Wittgenstein}: {A} memoir}.
\newblock{}London: Oxford University Press.

\bibitem[\protect\citeauthoryear{%
Marshack%
}{%
Marshack%
}{%
{\protect\APACyear{1972}}%
}]{%
Marshack:1972}%
\APACinsertmetastar{%
Marshack:1972}%
Marshack, A.%
%
\newblock{}\BBOP{}1972\BBCP{}.
\newblock{}\Bem{The roots of civilisation: {The} cognitive beginnings of
  {Man's} first art, symbol and notation}.
\newblock{}London: Weidenfeld \& Nicolson.

\bibitem[\protect\citeauthoryear{%
Martin%
}{%
Martin%
}{%
{\protect\APACyear{1996}}%
}]{%
Martin:1996}%
\APACinsertmetastar{%
Martin:1996}%
Martin, J.~L.%
%
\newblock{}\BBOP{}1996\BBCP{}.
\newblock{}\Bem{General relativity: A first course for physicists}\ (Second\
  \BEd).
\newblock{}London: Prentice Hall.

\bibitem[\protect\citeauthoryear{%
Misner%
, Thorne%
\BCBL{}\ \BBA{} Wheeler%
}{%
Misner%
\ \protect\BOthers{.}}{%
{\protect\APACyear{1973}}%
}]{%
Misner/Thorne/Wheeler:1973}%
\APACinsertmetastar{%
Misner/Thorne/Wheeler:1973}%
Misner, C.~W.%
, Thorne, K.~S.%
\BCBL{}\ \BBA{} Wheeler, J.~A.%
%
\newblock{}\BBOP{}1973\BBCP{}.
\newblock{}\Bem{Gravitation}.
\newblock{}New York: Freeman.

\bibitem[\protect\citeauthoryear{%
Monk%
}{%
Monk%
}{%
{\protect\APACyear{1997}}%
}]{%
Monk:1997}%
\APACinsertmetastar{%
Monk:1997}%
Monk, N. A.~M.%
%
\newblock{}\BBOP{}1997\BBCP{}.
\newblock{}\BBOQ{}Conceptions of space-time: {Problems} and possible
  solutions.\BBCQ{}
\newblock{}\Bem{Studies in History and Philosophy of Modern Physics}, \Bem{28},
  1--34.

\bibitem[\protect\citeauthoryear{%
Moster\'{i}n%
}{%
Moster\'{i}n%
}{%
{\protect\APACyear{1984}}%
}]{%
Mosterin:1984}%
\APACinsertmetastar{%
Mosterin:1984}%
Moster\'{i}n, J.%
%
\newblock{}\BBOP{}1984\BBCP{}.
\newblock{}\Bem{Conceptos y teor\'{i}as en la ciencia}.
\newblock{}Madrid: Alianza.

\bibitem[\protect\citeauthoryear{%
Nagels%
}{%
Nagels%
}{%
{\protect\APACyear{1985}}%
}]{%
Nagels:1985}%
\APACinsertmetastar{%
Nagels:1985}%
Nagels, G.%
%
\newblock{}\BBOP{}1985\BBCP{}.
\newblock{}\BBOQ{}Space as a ``bucket of dust''.\BBCQ{}
\newblock{}\Bem{General Relativity and Gravitation}, \Bem{17}, 545--557.

\bibitem[\protect\citeauthoryear{%
Newton%
}{%
Newton%
}{%
{\protect\APACyear{1962}}%
}]{%
Newton:1962}%
\APACinsertmetastar{%
Newton:1962}%
Newton, I.%
%
\newblock{}\BBOP{}1962\BBCP{}.
\newblock{}\Bem{Mathematical principles of natural philosophy, \emph{A.~Motte
  \& F.~Cajori (Trans.)}}.
\newblock{}Berkeley: University of {California} Press.
\newblock{}(Original work published 1729.)

\bibitem[\protect\citeauthoryear{%
Ng%
}{%
Ng%
}{%
{\protect\APACyear{2003}}%
}]{%
Ng:2003}%
\APACinsertmetastar{%
Ng:2003}%
Ng, Y.~J.%
%
\newblock{}\BBOP{}2003\BBCP{}.
\newblock{}\BBOQ{}Selected topics in {Planck}-scale physics.\BBCQ{}
\newblock{}\Bem{Modern Physics Letters A}, \Bem{18}, 1073--1097.
\newblock{}(arXiv:gr-qc/0305019v2)

\bibitem[\protect\citeauthoryear{%
North%
}{%
North%
}{%
{\protect\APACyear{1990}}%
}]{%
North:1990}%
\APACinsertmetastar{%
North:1990}%
North, J.~D.%
%
\newblock{}\BBOP{}1990\BBCP{}.
\newblock{}\Bem{The measure of the universe: {A} history of modern cosmology}.
\newblock{}New York: Dover.

\bibitem[\protect\citeauthoryear{%
Ohanian%
}{%
Ohanian%
}{%
{\protect\APACyear{1975}}%
}]{%
Ohanian:1976}%
\APACinsertmetastar{%
Ohanian:1976}%
Ohanian, H.~C.%
%
\newblock{}\BBOP{}1975\BBCP{}.
\newblock{}\Bem{Gravitation and spacetime}.
\newblock{}New York: Norton.

\bibitem[\protect\citeauthoryear{%
Ohanian%
}{%
Ohanian%
}{%
{\protect\APACyear{1989}}%
}]{%
Ohanian:1989}%
\APACinsertmetastar{%
Ohanian:1989}%
Ohanian, H.~C.%
%
\newblock{}\BBOP{}1989\BBCP{}.
\newblock{}\Bem{Physics}\ (Second\ \BEd).
\newblock{}New York: Norton.

\bibitem[\protect\citeauthoryear{%
Oppenheimer%
}{%
Oppenheimer%
}{%
{\protect\APACyear{1973}}%
}]{%
Oppenheimer:1999}%
\APACinsertmetastar{%
Oppenheimer:1999}%
Oppenheimer, J.~R.%
%
\newblock{}\BBOP{}1973\BBCP{}.
\newblock{}\BBOQ{}Prospects in the arts and sciences.\BBCQ{}
\newblock{}\BIn{} L.~Copeland, L.~W. Lamm\BCBL{}\ \BBA{} S.~J. McKenna\
  (\BEDS), \Bem{The world's great speeches}\ (Fourth\ \BEd, \BPGS\ 642--645).
\newblock{}New York: Dover.

\bibitem[\protect\citeauthoryear{%
Pais%
}{%
Pais%
}{%
{\protect\APACyear{1982}}%
}]{%
Pais:1982}%
\APACinsertmetastar{%
Pais:1982}%
Pais, A.%
%
\newblock{}\BBOP{}1982\BBCP{}.
\newblock{}\Bem{Subtle is the lord}.
\newblock{}Oxford: Oxford University Press.

\bibitem[\protect\citeauthoryear{%
Patton%
\ \BBA{} Wheeler%
}{%
Patton%
\ \BBA{} Wheeler%
}{%
{\protect\APACyear{1975}}%
}]{%
Patton/Wheeler:1975}%
\APACinsertmetastar{%
Patton/Wheeler:1975}%
Patton, C.~M.%
\BCBT{}\ \BBA{} Wheeler, J.~A.%
%
\newblock{}\BBOP{}1975\BBCP{}.
\newblock{}\BBOQ{}Is physics legislated by cosmogony?\BBCQ{}
\newblock{}\BIn{} C.~J. Isham, R.~Penrose\BCBL{}\ \BBA{} D.~W. Sciama\ (\BEDS),
  \Bem{Quantum gravity: {An} {Oxford} symposium}\ (\BPGS\ 538--605).
\newblock{}Oxford: Clarendon Press.

\bibitem[\protect\citeauthoryear{%
Peacock%
}{%
Peacock%
}{%
{\protect\APACyear{2006}}%
}]{%
Peacock:2006}%
\APACinsertmetastar{%
Peacock:2006}%
Peacock, J.%
%
\newblock{}\BBOP{}2006\BBCP{}.
\newblock{}\BBOQ{}A universe tuned for life [{Review} of the book \emph{The
  cosmic landscape: String theory and the illusion of intelligent
  design}].\BBCQ{}
\newblock{}\Bem{American Scientist}, \Bem{94}, 168--170.

\bibitem[\protect\citeauthoryear{%
Pedersen%
}{%
Pedersen%
}{%
{\protect\APACyear{1980}}%
}]{%
Pedersen:1980}%
\APACinsertmetastar{%
Pedersen:1980}%
Pedersen, J.%
%
\newblock{}\BBOP{}1980\BBCP{}.
\newblock{}\BBOQ{}Why we still need to teach geometry!\BBCQ{}
\newblock{}\BIn{} M.~Zweng, T.~Green, J.~Kilpatrick, H.~Pollak\BCBL{}\ \BBA{}
  M.~Suydam\ (\BEDS), \Bem{Proceedings of the fourth international congress on
  mathematical education}\ (\BPGS\ 158--160).
\newblock{}Boston: {Birkh\"{a}user}.

\bibitem[\protect\citeauthoryear{%
Penrose%
}{%
Penrose%
}{%
{\protect\APACyear{1998}}%
}]{%
Penrose:1998}%
\APACinsertmetastar{%
Penrose:1998}%
Penrose, R.%
%
\newblock{}\BBOP{}1998\BBCP{}.
\newblock{}\BBOQ{}Mathematical physics in the 20th and 21st centuries?\BBCQ{}
\newblock{}\Bem{DMV-Mitteilungen}, \Bem{2/98}, 56--64.

\bibitem[\protect\citeauthoryear{%
Perez%
}{%
Perez%
}{%
{\protect\APACyear{2006}}%
}]{%
Perez:2006}%
\APACinsertmetastar{%
Perez:2006}%
Perez, A.%
%
\newblock{}\BBOP{}2006\BBCP{}.
\newblock{}\BBOQ{}Loop quantum gravity.\BBCQ{}
\newblock{}\Bem{Europhysics News}, \Bem{37}(3), 17--21.

\bibitem[\protect\citeauthoryear{%
Perez~Bergliaffa%
, Romero%
\BCBL{}\ \BBA{} Vucetich%
}{%
Perez~Bergliaffa%
\ \protect\BOthers{.}}{%
{\protect\APACyear{1998}}%
}]{%
Perez/Romero/Vucetich:1998}%
\APACinsertmetastar{%
Perez/Romero/Vucetich:1998}%
Perez~Bergliaffa, S.~E.%
, Romero, G.~E.%
\BCBL{}\ \BBA{} Vucetich, H.%
%
\newblock{}\BBOP{}1998\BBCP{}.
\newblock{}\BBOQ{}Towards an axiomatic pregeometry of space-time.\BBCQ{}
\newblock{}\Bem{International Journal of Theoretical Physics}, \Bem{37},
  2281--2298.
\newblock{}(arXiv:gr-qc/9710064v1)

\bibitem[\protect\citeauthoryear{%
Planck%
}{%
Planck%
}{%
{\protect\APACyear{1899}}%
}]{%
Planck:1899}%
\APACinsertmetastar{%
Planck:1899}%
Planck, M.%
%
\newblock{}\BBOP{}1899\BBCP{}.
\newblock{}\BBOQ{}\"{U}ber irreversible {Strahlungsvorg\"{a}nge}.\BBCQ{}
\newblock{}\Bem{Sitzungsberichte der Preussischen Akademie der Wissenschaften},
  \Bem{5}, 440--480.

\bibitem[\protect\citeauthoryear{%
Pollard%
}{%
Pollard%
}{%
{\protect\APACyear{1984}}%
}]{%
Pollard:1984}%
\APACinsertmetastar{%
Pollard:1984}%
Pollard, W.~G.%
%
\newblock{}\BBOP{}1984\BBCP{}.
\newblock{}\BBOQ{}Rumors of transcendence in physics.\BBCQ{}
\newblock{}\Bem{American Journal of Physics}, \Bem{52}, 877--881.

\bibitem[\protect\citeauthoryear{%
Pylyshyn%
}{%
Pylyshyn%
}{%
{\protect\APACyear{1979}}%
}]{%
Pylyshyn:1979}%
\APACinsertmetastar{%
Pylyshyn:1979}%
Pylyshyn, Z.%
%
\newblock{}\BBOP{}1979\BBCP{}.
\newblock{}\BBOQ{}Do mental events have durations?\BBCQ{}
\newblock{}\Bem{Behavioral and Brain Sciences}, \Bem{2}, 277--278.

\bibitem[\protect\citeauthoryear{%
Reichenbach%
}{%
Reichenbach%
}{%
{\protect\APACyear{1951}}%
}]{%
Reichenbach:1951}%
\APACinsertmetastar{%
Reichenbach:1951}%
Reichenbach, H.%
%
\newblock{}\BBOP{}1951\BBCP{}.
\newblock{}\Bem{The rise of scientific philosophy}.
\newblock{}Berkeley: University of California Press.

\bibitem[\protect\citeauthoryear{%
Repo%
}{%
Repo%
}{%
{\protect\APACyear{2001}}%
}]{%
Repo:2001}%
\APACinsertmetastar{%
Repo:2001}%
Repo, P.%
%
\newblock{}\BBOP{}2001\BBCP{}.
\newblock{}\Bem{Quantum-mechanical models of black holes}.
\newblock{}{Ph.D.} thesis, {Research} report 12/2001, Department of Physics,
  University of {Jyv\"{a}skyl\"{a}}.

\bibitem[\protect\citeauthoryear{%
Requardt%
}{%
Requardt%
}{%
{\protect\APACyear{1995}}%
}]{%
Requardt:1995}%
\APACinsertmetastar{%
Requardt:1995}%
Requardt, M.%
%
\newblock{}\BBOP{}1995\BBCP{}.
\newblock{}\Bem{\emph{Discrete mathematics and physics on the {Planck} scale}.}
\newblock{}(arXiv:hep-th/9504118v1)

\bibitem[\protect\citeauthoryear{%
Requardt%
}{%
Requardt%
}{%
{\protect\APACyear{1996}}%
}]{%
Requardt:1996}%
\APACinsertmetastar{%
Requardt:1996}%
Requardt, M.%
%
\newblock{}\BBOP{}1996\BBCP{}.
\newblock{}\Bem{\emph{Emergence of space-time on the {Planck} scale described
  as an unfolding phase transition within the scheme of dynamical cellular
  networks and random graphs}.}
\newblock{}(arXiv:hep-th/9610055v1)

\bibitem[\protect\citeauthoryear{%
Requardt%
}{%
Requardt%
}{%
{\protect\APACyear{2000}}%
}]{%
Requardt:2000}%
\APACinsertmetastar{%
Requardt:2000}%
Requardt, M.%
%
\newblock{}\BBOP{}2000\BBCP{}.
\newblock{}\Bem{\emph{Let's call it nonlocal quantum physics}.}
\newblock{}(arXiv:gr-qc/ 0006063v1)

\bibitem[\protect\citeauthoryear{%
Requardt%
\ \BBA{} Roy%
}{%
Requardt%
\ \BBA{} Roy%
}{%
{\protect\APACyear{2001}}%
}]{%
Requardt/Roy:2001}%
\APACinsertmetastar{%
Requardt/Roy:2001}%
Requardt, M.%
\BCBT{}\ \BBA{} Roy, S.%
%
\newblock{}\BBOP{}2001\BBCP{}.
\newblock{}\BBOQ{}({Quantum}) space-time as a statistical geometry of fuzzy
  lumps and the connection with random metric spaces.\BBCQ{}
\newblock{}\Bem{Classical and Quantum Gravity}, \Bem{18}, 3039-3058.

\bibitem[\protect\citeauthoryear{%
Riemann%
}{%
Riemann%
}{%
{\protect\APACyear{1873}}%
}]{%
Riemann:1873}%
\APACinsertmetastar{%
Riemann:1873}%
Riemann, B.%
%
\newblock{}\BBOP{}1873\BBCP{}.
\newblock{}\BBOQ{}On the hypotheses which lie at the bases of geometry.\BBCQ{}
\newblock{}\Bem{Nature, \emph{W. K. Clifford (Trans.)}}, \Bem{VIII (183, 184)},
  14--17, 36, 37.
\newblock{}(http://www.emis.de/classics/Riemann. Address delivered 10th June,
  1854, {G\"{o}ttingen} University.)

\bibitem[\protect\citeauthoryear{%
Rindler%
}{%
Rindler%
}{%
{\protect\APACyear{1977}}%
}]{%
Rindler:1977}%
\APACinsertmetastar{%
Rindler:1977}%
Rindler, W.%
%
\newblock{}\BBOP{}1977\BBCP{}.
\newblock{}\Bem{Essential relativity: {Special}, general, and cosmological}\
  (Second\ \BEd).
\newblock{}New York: Springer.

\bibitem[\protect\citeauthoryear{%
Rovelli%
}{%
Rovelli%
}{%
{\protect\APACyear{1998}}%
}]{%
Rovelli:1998}%
\APACinsertmetastar{%
Rovelli:1998}%
Rovelli, C.%
%
\newblock{}\BBOP{}1998\BBCP{}.
\newblock{}\BBOQ{}Loop quantum gravity.\BBCQ{}
\newblock{}\Bem{Living Reviews in Relativity}, \Bem{1}, 1.
\newblock{}(http://relativity.livingreviews.org)

\bibitem[\protect\citeauthoryear{%
Rovelli%
}{%
Rovelli%
}{%
{\protect\APACyear{2000}}%
}]{%
Rovelli:2000}%
\APACinsertmetastar{%
Rovelli:2000}%
Rovelli, C.%
%
\newblock{}\BBOP{}2000\BBCP{}.
\newblock{}\BBOQ{}Quantum spacetime: {What} do we know?\BBCQ{}
\newblock{}\BIn{} C.~Callender\ \BBA{} N.~Huggett\ (\BEDS), \Bem{Physics meets
  philosophy at the {Planck} scale}\ (\BPGS\ 101--122).
\newblock{}Cambridge: Cambridge University Press.
\newblock{}(arXiv:gr-qc/9903045v1)

\bibitem[\protect\citeauthoryear{%
Rovelli%
}{%
Rovelli%
}{%
{\protect\APACyear{2003}}%
}]{%
Rovelli:2003}%
\APACinsertmetastar{%
Rovelli:2003}%
Rovelli, C.%
%
\newblock{}\BBOP{}2003, November\BBCP{}.
\newblock{}\BBOQ{}Loop quantum gravity.\BBCQ{}
\newblock{}\Bem{Physics World}, 37--41.

\bibitem[\protect\citeauthoryear{%
Rovelli%
}{%
Rovelli%
}{%
{\protect\APACyear{2004}}%
}]{%
Rovelli:2004}%
\APACinsertmetastar{%
Rovelli:2004}%
Rovelli, C.%
%
\newblock{}\BBOP{}2004\BBCP{}.
\newblock{}\Bem{Quantum gravity}.
\newblock{}Cambridge: Cambridge University Press.
\newblock{}(http://www.cpt.univ-mrs.fr/$\sim$rovelli/book.pdf)

\bibitem[\protect\citeauthoryear{%
Rovelli%
}{%
Rovelli%
}{%
{\protect\APACyear{2006}}%
}]{%
Rovelli:2006}%
\APACinsertmetastar{%
Rovelli:2006}%
Rovelli, C.%
%
\newblock{}\BBOP{}2006\BBCP{}.
\newblock{}\Bem{\emph{Unfinished revolution}.}
\newblock{}(arXiv:gr-qc/0604045v2)

\bibitem[\protect\citeauthoryear{%
Saliba%
}{%
Saliba%
}{%
{\protect\APACyear{2002}}%
}]{%
Saliba:2002}%
\APACinsertmetastar{%
Saliba:2002}%
Saliba, G.%
%
\newblock{}\BBOP{}2002\BBCP{}.
\newblock{}\BBOQ{}Greek astronomy and the medieval {Arabic} tradition.\BBCQ{}
\newblock{}\Bem{American Scientist}, \Bem{90}, 360--367.

\bibitem[\protect\citeauthoryear{%
Saniga%
}{%
Saniga%
}{%
{\protect\APACyear{2005}}%
}]{%
Saniga:2005}%
\APACinsertmetastar{%
Saniga:2005}%
Saniga, M.%
%
\newblock{}\BBOP{}2005\BBCP{}.
\newblock{}\BBOQ{}A geometrical chart of altered temporality (and
  spatiality).\BBCQ{}
\newblock{}\BIn{} R.~Buccheri, A.~C. Elitzur\BCBL{}\ \BBA{} M.~Saniga\ (\BEDS),
  \Bem{Endophysics, time, quantum and the subjective}\ (\BPGS\ 245--272).
\newblock{}Singapore: World Scientific.
\newblock{}(http://www.ta3.sk/$\sim$msaniga)

\bibitem[\protect\citeauthoryear{%
Saslow%
}{%
Saslow%
}{%
{\protect\APACyear{1998}}%
}]{%
Saslow:1998}%
\APACinsertmetastar{%
Saslow:1998}%
Saslow, W.~M.%
%
\newblock{}\BBOP{}1998\BBCP{}.
\newblock{}\BBOQ{}A physical interpretation of the {Planck} length.\BBCQ{}
\newblock{}\Bem{European Journal of Physics}, \Bem{19}, 313.

\bibitem[\protect\citeauthoryear{%
Schwinger%
}{%
Schwinger%
}{%
{\protect\APACyear{1959}}%
}]{%
Schwinger:1959}%
\APACinsertmetastar{%
Schwinger:1959}%
Schwinger, J.%
%
\newblock{}\BBOP{}1959\BBCP{}.
\newblock{}\BBOQ{}The algebra of microscopic measurement.\BBCQ{}
\newblock{}\Bem{Proceedings of the National Academy of Sciences}, \Bem{45},
  1542--1553.
\newblock{}(http://www. pnas.org/cgi/reprint/45/10/1542)

\bibitem[\protect\citeauthoryear{%
Schwinger%
}{%
Schwinger%
}{%
{\protect\APACyear{1960}}%
}]{%
Schwinger:1960}%
\APACinsertmetastar{%
Schwinger:1960}%
Schwinger, J.%
%
\newblock{}\BBOP{}1960\BBCP{}.
\newblock{}\BBOQ{}The geometry of quantum states.\BBCQ{}
\newblock{}\Bem{Proceedings of the National Academy of Sciences}, \Bem{46},
  257--265.
\newblock{}(http://www.pnas.org/ cgi/reprint/46/2/257)

\bibitem[\protect\citeauthoryear{%
Shallis%
}{%
Shallis%
}{%
{\protect\APACyear{1982}}%
}]{%
Shallis:1982}%
\APACinsertmetastar{%
Shallis:1982}%
Shallis, M.%
%
\newblock{}\BBOP{}1982\BBCP{}.
\newblock{}\Bem{On time: {An} investigation into scientific knowledge and human
  experience}.
\newblock{}New York: Brunnett Books.

\bibitem[\protect\citeauthoryear{%
Smith%
}{%
Smith%
}{%
{\protect\APACyear{1951}}%
}]{%
Smith:1951}%
\APACinsertmetastar{%
Smith:1951}%
Smith, D.~E.%
%
\newblock{}\BBOP{}1951\BBCP{}.
\newblock{}\Bem{History of mathematics}\ (\BVOL~1).
\newblock{}New York: Dover.

\bibitem[\protect\citeauthoryear{%
Smolin%
}{%
Smolin%
}{%
{\protect\APACyear{1997}}%
}]{%
Smolin:1997}%
\APACinsertmetastar{%
Smolin:1997}%
Smolin, L.%
%
\newblock{}\BBOP{}1997\BBCP{}.
\newblock{}\Bem{The life of the cosmos}.
\newblock{}Oxford: Oxford University Press.

\bibitem[\protect\citeauthoryear{%
Smolin%
}{%
Smolin%
}{%
{\protect\APACyear{2001}}%
}]{%
Smolin:2001}%
\APACinsertmetastar{%
Smolin:2001}%
Smolin, L.%
%
\newblock{}\BBOP{}2001\BBCP{}.
\newblock{}\Bem{Three roads to quantum gravity}.
\newblock{}New York: Basic Books.

\bibitem[\protect\citeauthoryear{%
Smolin%
}{%
Smolin%
}{%
{\protect\APACyear{2003}}%
}]{%
Smolin:2003}%
\APACinsertmetastar{%
Smolin:2003}%
Smolin, L.%
%
\newblock{}\BBOP{}2003\BBCP{}.
\newblock{}\Bem{\emph{How far are we from the quantum theory of gravity?}}
\newblock{}(arXiv:hep-th/0303185v2)

\bibitem[\protect\citeauthoryear{%
Sorkin%
}{%
Sorkin%
}{%
{\protect\APACyear{1991}}%
}]{%
Sorkin:1991}%
\APACinsertmetastar{%
Sorkin:1991}%
Sorkin, R.~D.%
%
\newblock{}\BBOP{}1991\BBCP{}.
\newblock{}\BBOQ{}Spacetime and causal sets.\BBCQ{}
\newblock{}\BIn{} J.~C. D'Olivo, E.~Nahmad-Achar, M.~Rosenbaum, M.~P. Ryan,
  L.~F. Urrutia\BCBL{}\ \BBA{} F.~Zertuche\ (\BEDS), \Bem{Relativity and
  gravitation: {Classical} and quantum}\ (\BPGS\ 150--173).
\newblock{}Singapore: World Scientific.
\newblock{}(http://www.physics.syr. edu/$\sim$sorkin)

\bibitem[\protect\citeauthoryear{%
Sorkin%
}{%
Sorkin%
}{%
{\protect\APACyear{2005}}%
}]{%
Sorkin:2005}%
\APACinsertmetastar{%
Sorkin:2005}%
Sorkin, R.~D.%
%
\newblock{}\BBOP{}2005\BBCP{}.
\newblock{}\BBOQ{}Causal sets: {Discrete} gravity.\BBCQ{}
\newblock{}\BIn{} A.~Gomberoff\ \BBA{} D.~Marolf\ (\BEDS), \Bem{Lectures on
  quantum gravity.}
\newblock{}New York: Springer.
\newblock{}(arXiv:gr-qc/0309009v1)

\bibitem[\protect\citeauthoryear{%
Stachel%
}{%
Stachel%
}{%
{\protect\APACyear{1989}}%
}]{%
Stachel:1989}%
\APACinsertmetastar{%
Stachel:1989}%
Stachel, J.%
%
\newblock{}\BBOP{}1989\BBCP{}.
\newblock{}\BBOQ{}{Einstein's} search for general covariance.\BBCQ{}
\newblock{}\BIn{} D.~Howard\ \BBA{} J.~Stachel\ (\BEDS), \Bem{{Einstein} and
  the history of general relativity}\ (\BPGS\ 63--100).
\newblock{}Boston: Birkh\"{a}user.
\newblock{}(Original work published 1980.)

\bibitem[\protect\citeauthoryear{%
Stapp%
}{%
Stapp%
}{%
{\protect\APACyear{1975}}%
}]{%
Stapp:1975}%
\APACinsertmetastar{%
Stapp:1975}%
Stapp, H.~P.%
%
\newblock{}\BBOP{}1975\BBCP{}.
\newblock{}\BBOQ{}Bell's theorem and world process.\BBCQ{}
\newblock{}\Bem{Il Nuovo Cimento B}, \Bem{29}, 270--276.

\bibitem[\protect\citeauthoryear{%
Stewart%
}{%
Stewart%
}{%
{\protect\APACyear{2002}}%
}]{%
Stewart:2002}%
\APACinsertmetastar{%
Stewart:2002}%
Stewart, I.%
%
\newblock{}\BBOP{}2002\BBCP{}.
\newblock{}\Bem{Does {God} play dice: {The} new mathematics of chaos}.
\newblock{}Malden, MA: Blackwell.

\bibitem[\protect\citeauthoryear{%
Stewart%
}{%
Stewart%
}{%
{\protect\APACyear{2006}}%
}]{%
Stewart:2006}%
\APACinsertmetastar{%
Stewart:2006}%
Stewart, I.%
%
\newblock{}\BBOP{}2006\BBCP{}.
\newblock{}\Bem{Letters to a young mathematician ({Art} of mentoring)}.
\newblock{}New York: Basic Books.

\bibitem[\protect\citeauthoryear{%
Stewart%
\ \BBA{} Golubitsky%
}{%
Stewart%
\ \BBA{} Golubitsky%
}{%
{\protect\APACyear{1993}}%
}]{%
Stewart/Golubitsky:1993}%
\APACinsertmetastar{%
Stewart/Golubitsky:1993}%
Stewart, I.%
\BCBT{}\ \BBA{} Golubitsky, M.%
%
\newblock{}\BBOP{}1993\BBCP{}.
\newblock{}\Bem{Fearful symmetry: {Is} {God} a geometer?}
\newblock{}London: Penguin Books.

\bibitem[\protect\citeauthoryear{%
Stuckey%
}{%
Stuckey%
}{%
{\protect\APACyear{2001}}%
}]{%
Stuckey:2001}%
\APACinsertmetastar{%
Stuckey:2001}%
Stuckey, W.~M.%
%
\newblock{}\BBOP{}2001\BBCP{}.
\newblock{}\Bem{\emph{Metric Structure and Dimensionality over a {Borel} Set
  via Uniform Spaces}.}
\newblock{}(arXiv:gr-qc/0109030v2)

\bibitem[\protect\citeauthoryear{%
Stuckey%
\ \BBA{} Silberstein%
}{%
Stuckey%
\ \BBA{} Silberstein%
}{%
{\protect\APACyear{2000}}%
}]{%
Stuckey/Silberstein:2000}%
\APACinsertmetastar{%
Stuckey/Silberstein:2000}%
Stuckey, W.~M.%
\BCBT{}\ \BBA{} Silberstein, M.%
%
\newblock{}\BBOP{}2000\BBCP{}.
\newblock{}\Bem{\emph{Uniform spaces in the pregeometric modelling of quantum
  non-separability}.}
\newblock{}(arXiv:gr-qc/0003104v2)

\bibitem[\protect\citeauthoryear{%
Synge%
}{%
Synge%
}{%
{\protect\APACyear{1964}}%
}]{%
Synge:1964}%
\APACinsertmetastar{%
Synge:1964}%
Synge, J.~L.%
%
\newblock{}\BBOP{}1964\BBCP{}.
\newblock{}\Bem{Relativity: {The} general theory}.
\newblock{}Amsterdam: North-Holland.

\bibitem[\protect\citeauthoryear{%
Tegmark%
}{%
Tegmark%
}{%
{\protect\APACyear{1998}}%
}]{%
Tegmark:1998}%
\APACinsertmetastar{%
Tegmark:1998}%
Tegmark, M.%
%
\newblock{}\BBOP{}1998\BBCP{}.
\newblock{}\BBOQ{}Is ``the theory of everything'' merely the ultimate ensemble
  theory?\BBCQ{}
\newblock{}\Bem{Annals of Physics}, \Bem{270}, 1--51.
\newblock{}(arXiv:gr-qc/9704009v2)

\bibitem[\protect\citeauthoryear{%
Tegmark%
\ \BBA{} Wheeler%
}{%
Tegmark%
\ \BBA{} Wheeler%
}{%
{\protect\APACyear{2001}}%
}]{%
Tegmark/Wheeler:2001}%
\APACinsertmetastar{%
Tegmark/Wheeler:2001}%
Tegmark, M.%
\BCBT{}\ \BBA{} Wheeler, J.~A.%
%
\newblock{}\BBOP{}2001, February\BBCP{}.
\newblock{}\BBOQ{}100 years of quantum mysteries.\BBCQ{}
\newblock{}\Bem{Scientific American}, 68--75.
\newblock{}(arXiv:quant-ph/0101077v1)

\bibitem[\protect\citeauthoryear{%
Thiemann%
}{%
Thiemann%
}{%
{\protect\APACyear{2003}}%
}]{%
Thiemann:2003}%
\APACinsertmetastar{%
Thiemann:2003}%
Thiemann, T.%
%
\newblock{}\BBOP{}2003\BBCP{}.
\newblock{}\BBOQ{}Lectures on loop quantum gravity.\BBCQ{}
\newblock{}\Bem{Lecture Notes in Physics}, \Bem{631}, 41--135.
\newblock{}(arXiv:gr-qc/0210094v1)

\bibitem[\protect\citeauthoryear{%
Tulving%
}{%
Tulving%
}{%
{\protect\APACyear{1989}}%
}]{%
Tulving:1989}%
\APACinsertmetastar{%
Tulving:1989}%
Tulving, E.%
%
\newblock{}\BBOP{}1989\BBCP{}.
\newblock{}\BBOQ{}Remembering and knowing the past.\BBCQ{}
\newblock{}\Bem{American Scientist}, \Bem{77}, 361--367.

\bibitem[\protect\citeauthoryear{%
Turing%
}{%
Turing%
}{%
{\protect\APACyear{1950}}%
}]{%
Turing:1950}%
\APACinsertmetastar{%
Turing:1950}%
Turing, A.%
%
\newblock{}\BBOP{}1950\BBCP{}.
\newblock{}\BBOQ{}Computing machinery and intelligence.\BBCQ{}
\newblock{}\Bem{Mind}, \Bem{59}, 433--460.

\bibitem[\protect\citeauthoryear{%
{van der Waerden}%
}{%
{van der Waerden}%
}{%
{\protect\APACyear{1983}}%
}]{%
vanderWaerden:1983}%
\APACinsertmetastar{%
vanderWaerden:1983}%
{van der Waerden}, B.~L.%
%
\newblock{}\BBOP{}1983\BBCP{}.
\newblock{}\Bem{Geometry and algebra in ancient civilizations}.
\newblock{}Berlin: Springer.

\bibitem[\protect\citeauthoryear{%
Volovich%
}{%
Volovich%
}{%
{\protect\APACyear{1987}}%
}]{%
Volovich:1987}%
\APACinsertmetastar{%
Volovich:1987}%
Volovich, I.~V.%
%
\newblock{}\BBOP{}1987\BBCP{}.
\newblock{}\Bem{\emph{Number theory as the ultimate physical theory}.}
\newblock{}(Preprint CERN-TH., 4781--4787.)

\bibitem[\protect\citeauthoryear{%
{von Borzeszkowski}%
\ \BBA{} Treder%
}{%
{von Borzeszkowski}%
\ \BBA{} Treder%
}{%
{\protect\APACyear{1988}}%
}]{%
vonBorzeszkowski/Treder:1988}%
\APACinsertmetastar{%
vonBorzeszkowski/Treder:1988}%
{von Borzeszkowski}, H.%
\BCBT{}\ \BBA{} Treder, H.%
%
\newblock{}\BBOP{}1988\BBCP{}.
\newblock{}\Bem{The meaning of quantum gravity}.
\newblock{}Dordrecht: Reidel.

\bibitem[\protect\citeauthoryear{%
Weinberg%
}{%
Weinberg%
}{%
{\protect\APACyear{1993}}%
}]{%
Weinberg:1993}%
\APACinsertmetastar{%
Weinberg:1993}%
Weinberg, S.%
%
\newblock{}\BBOP{}1993\BBCP{}.
\newblock{}\Bem{Dreams of a final theory}.
\newblock{}London: Vintage Books.

\bibitem[\protect\citeauthoryear{%
Weinberg%
}{%
Weinberg%
}{%
{\protect\APACyear{1999}}%
}]{%
Weinberg:1999}%
\APACinsertmetastar{%
Weinberg:1999}%
Weinberg, S.%
%
\newblock{}\BBOP{}1999, December\BBCP{}.
\newblock{}\BBOQ{}A unified physics by 2050?\BBCQ{}
\newblock{}\Bem{Scientific American}, 36--43.

\bibitem[\protect\citeauthoryear{%
Wertheimer%
}{%
Wertheimer%
}{%
{\protect\APACyear{1912}}%
}]{%
Wertheimer:1912}%
\APACinsertmetastar{%
Wertheimer:1912}%
Wertheimer, M.%
%
\newblock{}\BBOP{}1912\BBCP{}.
\newblock{}\BBOQ{}Experimentelle {Studien} \"uber das {Sehen} von
  {Bewegung}.\BBCQ{}
\newblock{}\Bem{Zeitschrift f\"ur {Psychologie}}, \Bem{61}, 161--265.

\bibitem[\protect\citeauthoryear{%
Weyl%
}{%
Weyl%
}{%
{\protect\APACyear{1949}}%
}]{%
Weyl:1949}%
\APACinsertmetastar{%
Weyl:1949}%
Weyl, H.%
%
\newblock{}\BBOP{}1949\BBCP{}.
\newblock{}\Bem{Philosophy of mathematics and natural science}.
\newblock{}Princeton: Princeton University Press.

\bibitem[\protect\citeauthoryear{%
Wheeler%
}{%
Wheeler%
}{%
{\protect\APACyear{1964}}%
}]{%
Wheeler:1964}%
\APACinsertmetastar{%
Wheeler:1964}%
Wheeler, J.~A.%
%
\newblock{}\BBOP{}1964\BBCP{}.
\newblock{}\BBOQ{}Geometrodynamics and the issue of the final state.\BBCQ{}
\newblock{}\BIn{} C.~De~Witt\ \BBA{} B.~S. De~Witt\ (\BEDS), \Bem{Relativity,
  groups and topology}\ (\BPGS\ 317--520).
\newblock{}New York: Gordon and Breach.

\bibitem[\protect\citeauthoryear{%
Wheeler%
}{%
Wheeler%
}{%
{\protect\APACyear{1980}}%
}]{%
Wheeler:1980}%
\APACinsertmetastar{%
Wheeler:1980}%
Wheeler, J.~A.%
%
\newblock{}\BBOP{}1980\BBCP{}.
\newblock{}\BBOQ{}Pregeometry: {Motivations} and prospects.\BBCQ{}
\newblock{}\BIn{} A.~R. Marlov\ (\BED), \Bem{Quantum theory and gravitation.}
\newblock{}New York: Academic Press.

\bibitem[\protect\citeauthoryear{%
Whitehead%
}{%
Whitehead%
}{%
{\protect\APACyear{1948}}%
}]{%
Whitehead:1948}%
\APACinsertmetastar{%
Whitehead:1948}%
Whitehead, A.~N.%
%
\newblock{}\BBOP{}1948\BBCP{}.
\newblock{}\Bem{Science and the modern world}.
\newblock{}New York: Mentor.

\bibitem[\protect\citeauthoryear{%
Whittaker%
}{%
Whittaker%
}{%
{\protect\APACyear{1951}}%
}]{%
Whittaker:1951}%
\APACinsertmetastar{%
Whittaker:1951}%
Whittaker, E.~T.%
%
\newblock{}\BBOP{}1951\BBCP{}.
\newblock{}\Bem{A history of the theories of aether and electricity}\ (\BVOLS\
  1--2).
\newblock{}London, New York: Tomash Publishers \& {American} Institute of
  Physics.

\bibitem[\protect\citeauthoryear{%
Whorf%
}{%
Whorf%
}{%
{\protect\APACyear{1956}}%
}]{%
Whorf:1956}%
\APACinsertmetastar{%
Whorf:1956}%
Whorf, B.~L.%
%
\newblock{}\BBOP{}1956\BBCP{}.
\newblock{}\BBOQ{}Science and linguistics.\BBCQ{}
\newblock{}\BIn{} J.~B. Carroll\ (\BED), \Bem{Language, thought and reality}\
  (\BPGS\ 207--219).
\newblock{}Cambridge, MA: M.I.T Press.
\newblock{}(Original work published 1940.)

\bibitem[\protect\citeauthoryear{%
Wilson%
}{%
Wilson%
}{%
{\protect\APACyear{1985}}%
}]{%
Wilson:1985}%
\APACinsertmetastar{%
Wilson:1985}%
Wilson, R.~J.%
%
\newblock{}\BBOP{}1985\BBCP{}.
\newblock{}\Bem{Introduction to graph theory}\ (Third\ \BEd).
\newblock{}Essex: Longman Group.

\bibitem[\protect\citeauthoryear{%
Wittgenstein%
}{%
Wittgenstein%
}{%
{\protect\APACyear{1922}}%
}]{%
Wittgenstein:1922}%
\APACinsertmetastar{%
Wittgenstein:1922}%
Wittgenstein, L.%
%
\newblock{}\BBOP{}1922\BBCP{}.
\newblock{}\Bem{Tractatus logico-philosophicus, \emph{C.~K.~Ogden (Trans.)}}.
\newblock{}London: Routledge \& Kegan.

\bibitem[\protect\citeauthoryear{%
Woit%
}{%
Woit%
}{%
{\protect\APACyear{2002}}%
}]{%
Woit:2002}%
\APACinsertmetastar{%
Woit:2002}%
Woit, P.%
%
\newblock{}\BBOP{}2002\BBCP{}.
\newblock{}\BBOQ{}Is string theory even wrong?\BBCQ{}
\newblock{}\Bem{American Scientist}, \Bem{90}, 110--112.

\end{thebibliography}

\end{document}